\documentclass{styles/IDR}
\usepackage{url}
\usepackage{bbold}
\usepackage[english]{babel}
\usepackage{subfigure}
\usepackage{graphicx}
\usepackage{multirow}
\usepackage{amsmath}
\usepackage{amssymb}
\usepackage{amsfonts}
\usepackage{slashed}
\usepackage{bbm}
\usepackage{ctable,longtable}
\usepackage{cancel}
\newcommand\Dmq{\Delta m^2}
\newcommand\eps{\varepsilon}
\newcommand\eVq{\text{eV}^2}
\def\gev{\ensuremath{\,\text{Ge}\text{V\/}}}
\newcommand\ldm{\Delta m_{31}^2}
\newcommand\sdm{\Delta m_{21}^2}
\newcommand\stheta{\sin^2 2\theta_{13}}
\def\nubarmu{\ensuremath{\overline{\nu}_{\mu}}}

\begin{document}
%
%
\selectlanguage{english}
\makeatletter

\pagestyle{empty}
\thispagestyle{empty}
%
\begin{figure}
  \vspace{-1cm}
  \begin{center}
    \includegraphics[width=\textwidth]{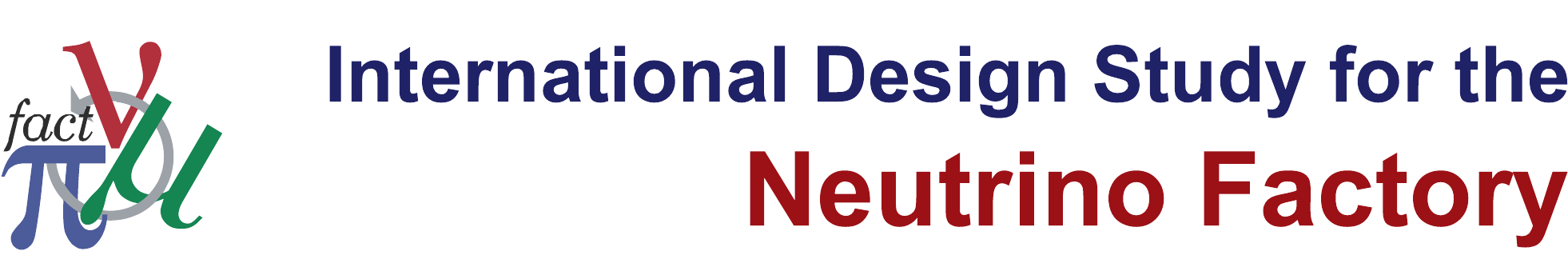}
  \end{center}
  \rightline{IDS-NF-020}
  \rightline{BNL-96453-2011}
  \rightline{CERN-ATS-2011-216}
  \rightline{EUROnu-WP1-05}
  \rightline{FERMILAB-PUB-11-581-APC}
  \rightline{RAL-TR-2011-018}
  \vspace{2cm}
\end{figure}
\begin{center}
  {\bf\LARGE 
    Interim Design Report \\
  }
  \vspace{0.75cm}
  {\bf\it\Large The IDS-NF collaboration} \\
  \vspace{2cm}
\end{center}

\vspace{3cm}

\begin{center}
  \vspace{3cm}
  19 October 2011
\end{center}
\vfill

\cleardoublepage
%
\cleardoublepage
{\setlength\parindent{0em}
  S.~Choubey, R.~Gandhi, S.~Goswami
  \\{\it
    Harish-Chandra Research Institute, Chhatnag Road, Jhunsi,
    Allahabad, 211019, India
  }
  \par \filbreak
  J.S.~Berg, R.~Fernow, J.C.~Gallardo, R.~Gupta, H.~Kirk, N.~Simos, N.~Souchlas
  \\{\it
    Brookhaven National Lab, P.O. Box 5000, Upton, NY 11973-5000, 
    USA
  } 
  \par \filbreak
  M.~Ellis\footnote{Now at Westpac Institutional Bank, Sydney, NSW, 
  Australia},
  P.~Kyberd
  \\{\it
    Brunel University West London, Uxbridge, Middlesex UB8 3PH,
    UK
  } 
  \par \filbreak
  E.~Benedetto, E.~Fernandez-Martinez, I.~Efthymiopoulos, R.~Garoby, 
  S.~Gilardoni, M.~Martini, G.~Prior
  \\{\it
    European Organization for Nuclear Research, CERN CH-1211, 
    Geneva 23, Switzerland 
  } 
  \par \filbreak
  D.~Indumathi, N.~Sinha
  \\{\it
   The Institute of Mathematical Sciences, IV Cross Road, 
   CIT Campus, Taramani, Chennai 600 113, Tamil Nadu, India
  } 
  \par \filbreak
  P.~Ballett, S.~Pascoli
  \\{\it
    Institute for Particle Physics Phenomenology, Department of
    Physics, University of Durham, Science Laboratories, South Rd,
    Durham, DH1 3LE, UK
  } 
  \par \filbreak
  A.~Bross, S.~Geer, C.~Johnstone, J.~Kopp, N.~Mokhov,
  J.~Morfin, D.~Neuffer, S.~Parke, 
  M.~Popovic, J.~Strait, S.~Striganov
  \\{\it
    Fermilab, P.O. Box 500, Batavia, IL 60510-5011, USA
  } 
  \par \filbreak
  A.~Blondel, F.~Dufour
  \\{\it
    University de Geneve, 24, Quai Ernest-Ansermet, 1211 Geneva 4,
    Suisse
  } 
  \par \filbreak
  A.~Laing, F.J.P.~Soler
  \\{\it
    School of Physics and Astronomy, Kelvin Building, 
    University of Glasgow, Glasgow G12 8QQ, Scotland, UK
  } 
  \par \filbreak
  M.~Lindner, T.~Schwetz
  \\{\it
    Max Planck Institut f\"ur Kernphysik, PO Box 103980, 
    69029 Heidelberg, Germany
  } 
  \par \filbreak
  A.~Alekou, M.~Apollonio, M.~Aslaninejad, C.~Bontoiu, P.~Dornan, 
  R.~Eccleston, A.~Kurup, K.~Long, J.~Pasternak, J.~Pozimski
  \\{\it
    Physics Department, Blackett Laboratory, Imperial College London,
    Exhibition Road, London, SW7 2AZ, UK
  } 
  \par \filbreak
  A.~Bogacz, V.~Morozov, Y.~Roblin
  \\{\it
    Jefferson Laboratory, 12000 Jefferson Avenue, Newport News, 
    VA 23606, USA
  } 
  \par \filbreak
  S.~Bhattacharya, D.~Majumdar
  \\{\it
    Saha Institute of Nuclear Physics, Sector-I, Block-AF, Bidhannagar, 
    Kolkata 700064, India
  } 
  \par \filbreak
  Y.~Mori, T.~Planche
  \\{\it
    Kyoto University, Research Reactor Institute, 2,Asashiro-Nishi, 
    Kumatori-cho, Sennan-gun, Osaka 590-0494 Japan
  } 
  \par \filbreak
  M.~Zisman
  \\{\it
    Lawrence Berkeley National Laboratory, 1 Cyclotron Road,
    Berkeley, CA 94720, USA
  } 
  \par \filbreak
  D.~Cline, D.~Stratakis, X.~Ding
  \\{\it
    Department of Physics and Astronomy, University of California, 
    Los Angeles, CA 90095, USA
  } 
  \par \filbreak
  P.~Coloma,
  A.~Donini\footnote{Also at Universidad de Valencia and Instituto de 
  Fisica Corpuscular, UV/CSIC, Apart. 22085, E-46071, Valencia, Spain}, 
  B.~Gavela, J.~Lopez Pavon, M.~Maltoni
  \\{\it
    Universidad Autonoma de Madrid and Instituto de Fisica Teorica,  
    UAM/CSIC, Cantoblanco, E-28049, Madrid, Spain;
} 
  \par \filbreak
  C.~Bromberg
  \\{\it
    Michigan State University, 150 Administration Building,
    East Lansing, Michigan 48824, USA
  } 
  \par \filbreak
  M.~Bonesini
  \\{\it
    Sezione INFN Milano Bicocca, Dipartimento di Fisica G. Occhialini,
    Piazza Scienza 3, 20126 Milano, Italy
  }
  \par \filbreak
  T.~Hart
  \\{\it
    The University of Mississippi, Department of Physics and Astronomy,
    108 Lewis Hall, PO Box 1848, Oxford, Mississippi 38677-1848, USA
  }
  \par \filbreak
  Y.~Kudenko
  \\{\it
    Institute for Nuclear Research of Russian, Academy of Sciences,
    7a, 60th October Anniversary prospect, Moscow 117312, Russia
  } 
  \par \filbreak
  N.~Mondal
  \\{\it
    Tata Institute of Fundamental Research, School of Natural Sciences,
    Homi Bhabha Rd., Mumbai 400005, India
  }
  \par \filbreak
  S.~Antusch, M.~Blennow, T.~Ota
  \\{\it
    Max Planck Institut f\"ur Physik, Werner Heisenberg Institut 
    f\"ur Physik, Fohringer Ring 6, D-80805 Munich, Germany
  } 
  \par \filbreak
  R.J.~Abrams, C.M.~Ankenbrandt, K.B.~Beard, M.A.C.~Cummings, 
  G.~Flanagan, R.P.~Johnson, T.J.~Roberts, C.Y.~Yoshikawa
  \\{\it
    Muons Inc., 552 N. Batavia Avenue, Batavia, 
    IL 60510, USA
  } 
  \par \filbreak
  P.~Migliozzi, V.~Palladino
  \\{\it
    Universita di Napoli Federico II, Dipartimento di Scienze Fisiche,
    Complesso Universitario di Monte S. Angelo, via Cintia, 
    I-80126 Napoli, Italy
  }
  \par \filbreak
  A.~de~Gouvea
  \\{\it
    Northwestern University, Dept. of Physics and Astronomy, 
    2145 Sheridan Road, Evanston, Illinois 60208-3112 USA
  } 
  \par \filbreak
  V.B.~Graves
  \\{\it
    Oak Ridge National Laboratory, P.O. Box 2008, Oak Ridge, 
    TN 37831, USA
  } 
  \par \filbreak
  Y.~Kuno
  \\{\it
    Osaka University, Graduate School, School of Science, 
    1-1 Machikaneyama-cho, Toyonaka, Osaka 560-0043, Japan
  } 
  \par \filbreak
  J.~Peltoniemi
  \\{\it
    Puulinnankatu 12 F 50, FI-90570 Oulu, Finland
  } 
  \par \filbreak
  V.~Blackmore, J.~Cobb, H.~Witte
  \\{\it
    Particle Physics Department, The Denys Wilkinson Building,
    Keble Road, Oxford, OX1 3RH, UK
  } 
  \par \filbreak
  M.~Mezzetto, S.~Rigolin
  \\{\it
    Dipartimento di Fisica, Universit´a di Padova and INFN Padova, 
    Via Marzolo 8, I-35131, Padova, Italy
  } 
  \par \filbreak
  K.T.~McDonald
  \\{\it
    Princeton University, Princeton, NJ, 08544, USA
  } 
  \par \filbreak
  L.~Coney, G.~Hanson, P.~Snopok
  \\{\it
    Department of Physics and Astronomy, University of California, 
    Riverside, CA 92521, USA
  } 
  \par \filbreak
  L. Tortora
  \\{\it
    Sezione INFN Roma Tre, Dipartimento di Fisica E. Amaldi,
    Via della Vasca Navale 84, 00146 Roma, Italy
  }
  \par \filbreak
  C.~Andreopoulos, J.R.J.~Bennett, S.~Brooks, O.~Caretta, T.~Davenne, 
  C.~Densham, R.~Edgecock\footnote{also at University of Huddersfield, 
  Queensgate, Huddersfield, HD1 3DH, UK}, D.~Kelliher, P.~Loveridge, 
  A.~McFarland, S.~Machida, C.~Prior, G.~Rees, C.~Rogers, J.W.G.~Thomason
  \\{\it
    STFC Rutherford Appleton Laboratory, Chilton, Didcot, 
    Oxfordshire, OX11 0QX, UK
  } 
  \par \filbreak
  C.~Booth, G.~Skoro
  \\{\it
    University of Sheffield, Dept. of Physics and Astronomy, 
    Hicks Bldg., Sheffield S3 7RH, UK
  } 
  \par \filbreak
  Y.~Karadzhov, R.~Matev, R.~Tsenov
  \\{\it
    Department of Atomic Physics, St. Kliment Ohridski 
    University of Sofia, 5 James Bourchier Boulevard, 
    BG-1164 Sofia, Bulgaria
  } 
  \par \filbreak
  R.~Samulyak
  \\{\it
    Physics and Astronomy Department,
    Stony Brook University,
    Stony Brook, NY  11794-3800, USA
  }
  \par \filbreak
  S.R.~Mishra, R.~Petti
  \\{\it
    Department of Physics and Astronomy, 
    University of South Carolina, 
    Columbia SC 29208, USA
  }
  \par \filbreak
  M.~Dracos
  \\{\it
    IPHC, Universit\'e de Strasbourg, CNRS/IN2P3, F-67037 Strasbourg, 
    France
  } 
  \par \filbreak
  O. Yasuda
  \\{\it
    Department of Physics, Toyko Metropolitan University,
    1-1 Minami-Osawa, Hachioji-shi, Toyko, Japan 192-0397
  }
  \par \filbreak
  S.K.~Agarwalla, A. Cervera-Villanueva, J.J.~Gomez-Cadenas, 
  P.~Hernandez, T.~Li, J.~Martin-Albo 
  \\{\it
    Instituto de Fisica Corpuscular (IFIC), Centro Mixto CSIC-UVEG,
    Edificio Investigacion Paterna, Apartado 22085, 46071 Valencia, 
    Spain
  } 
  \par \filbreak
  P.~Huber
  \\{\it
    Virginia Polytechnic Inst. and State Univ., Physics Dept.,
    Blacksburg, VA 24061-0435
  }
  \par \filbreak
  J.~Back, G.~Barker, P.~Harrison
  \\{\it
    Department of Physics, University of Warwick, Coventry,
    CV4 7AL, UK
  } 
  \par \filbreak
  D.~Meloni, J.~Tang, W.~Winter
  \\{\it
    Fakult\"at f\"ur Physik und Astronomie, Am Hubland, 
    97074 W\"urzburg, Germany
  } 
  \par \filbreak
}

\cleardoublepage
%
%
\parindent 10pt
\pagenumbering{roman}                   
\setcounter{page}{1}
\thispagestyle{plain}
\pagestyle{plain}
%
\section*{Foreword}

The International Design Study for the Neutrino Factory (the IDS-NF)
\cite{IDSNFWWWSite} was established by the community at the ninth
``International Workshop on Neutrino Factories, super-beams, and
beta-beams'' which was held in Okayama in August 2007 \cite{NuFact07}.
The IDS-NF mandate is to deliver the Reference Design Report (RDR) for
the facility on the timescale of 2012/13 \cite{Ref:IDS-NF-Spec}.
The RDR is conceived as the document upon which requests for the
resources to carry out the first phase of the Neutrino Factory project
can be made.
The RDR will include: the physics performance of the Neutrino Factory
and the specification of each of the accelerator, diagnostic, and
detector systems; an estimate of the cost of the facility; and an
estimate of the schedule for its implementation. 
The RDR will also include a discussion of the remaining technical
risks and an appropriate risk-mitigation plan.

The mandate for the study \cite{Ref:IDS-NF-Spec} requires an Interim
Design Report to be delivered midway through the project as a step on
the way to the RDR.
This document, the IDR, has two functions: it marks the point in the
IDS-NF at which the emphasis turns to the engineering studies required to
deliver the RDR and it documents baseline concepts for the
accelerator complex, the neutrino detectors, and the instrumentation
systems. 
The IDS-NF is, in essence, a site-independent study.
Example sites, CERN, FNAL, and RAL, have been identified to allow
site-specific issues to be addressed in the cost analysis that will be
presented in the RDR.
The choice of example sites should not be interpreted as implying a
preferred choice of site for the facility.

This document is organised as follows. 
The physics case for the Neutrino Factory is reviewed and the
performance of the IDS-NF baseline for the facility are presented in
section \ref{Sect:PPEG}.
The baseline concepts for the accelerator facility and the neutrino
detectors are reviewed in sections \ref{Sect:AccWG} and
\ref{Sect:DetWG} respectively. 
The steps that the IDS-NF collaboration plans to take to prepare the
RDR are presented in section \ref{Sect:FuturePlans} together with a
summary of the accelerator and detector R\&D programmes that are
required to complete the detailed specification of the facility and to
mitigate technical risks that have been identified.

\cleardoublepage
%
\tableofcontents
\cleardoublepage
\pagenumbering{arabic}                   
\setcounter{page}{1}
%
%
\section*{Executive summary}

\subsection*{Introduction}

The starting point for the International Design Study for the Neutrino
Factory (the IDS-NF \cite{IDSNFWWWSite}) was the output of the earlier
International Scoping Study for a future Neutrino Factory and
super-beam facility (the ISS)
\cite{Bandyopadhyay:2007kx,Apollonio:2009,Abe:2007bi}.
The accelerator facility described in section \ref{Sect:AccWG}
incorporates the improvements that have been derived from the
substantial amount of work carried out within the Accelerator Working
Group. 
Highlights of these improvements include:
\begin{itemize}
  \item Initial concepts for the implementation of the proton driver
    at each of the three example sites, CERN, FNAL, and RAL;
  \item Detailed studies of the energy deposition in the target area;
  \item A reduction in the length of the muon beam phase-rotation and
    bunching systems;
  \item Detailed analyses of the impact of the risk that stray
    magnetic field in the accelerating cavities in the ionisation
    cooling channel will reduce the maximum operating gradient.
    Several alternative ionisation-cooling lattices have been
    developed as fallback options to mitigate this technical risk;
  \item Studies of particle loss in the muon front-end and the
    development of strategies to mitigate the deleterious effects of
    such losses;
  \item The development of more complete designs for the muon linac and
    re-circulating linacs;
  \item The development of a design for the muon FFAG that
    incorporates insertions for injection and extraction; and 
  \item Detailed studies of diagnostics in the decay ring.
\end{itemize}
Other sub-systems have undergone a more ``incremental'' evolution; an
indication that the design of the Neutrino Factory has achieved a
degree of maturity.
The design of the neutrino detectors described in section
\ref{Sect:DetWG} has been optimised and the Detector Working Group has
made substantial improvements to the simulation and analysis of the
Magnetised Iron Neutrino Detector (MIND) resulting in an improvement
in the overall neutrino-detection efficiency and a reduction in the
neutrino-energy threshold. 
In addition, initial consideration of the engineering of the MIND
has generated a design that is feasible and a finite element analysis
of the toroidal magnetic field to produce a realistic field map has
been carried out.
Section \ref{Sect:DetWG} also contains, for the first time, a
specification for the near-detector systems and a demonstration that
the neutrino flux can be determined with a precision of 1\% through
measurements of inverse muon decay at the near detector.

The performance of the facility, the work of the Physics and
Performance Evaluation Group, is described in section
\ref{Sect:PPEG}.
The effect of the improved MIND performance is to deliver a discovery
reach for CP-invariance violation in the lepton sector, the
determination of the mass hierarchy, and of $\theta_{13}$ that extends
down to values of $\sin^2 2 \theta_{13} \sim 5 \times 10^{-5}$ and is
robust against systematic uncertainties. 
In addition, the improved neutrino-energy threshold has allowed an
indicative analysis of the kind of re-optimisation of the facility
that could be carried out should $\theta_{13}$ be found close to the
current upper bound.
The results presented in section \ref{Sect:PPEG} demonstrate
that the discovery reach as well as the precision with which the
oscillation parameters can be measured at the baseline Neutrino
Factory is superior to that of other proposed facilities for all
possible values of $\sin^2 2 \theta_{13}$.

\subsection*{Motivation}

The phenomenon of neutrino oscillations, arguably the most significant
advance in particle physics over the past decade, has been established
through measurements on neutrinos and anti-neutrinos produced in the
sun, by cosmic-ray interactions in the atmosphere, nuclear reactors,
and beams produced by high-energy particle accelerators
\cite{Amsler:2008zzb}.
In consequence, we know that the Standard Model is incomplete and must
be extended to include neutrino mass, mixing among the three neutrino
flavours, and therefore lepton-flavour non conservation.
These observations have profound implications for the ultimate theory
of particle interactions and for the description of the structure and
evolution of the Universe.  
In particular:
\begin{itemize}
  \item Mixing among the three massive neutrinos admits the
    possibility that the matter-antimatter (CP) symmetry is violated
    via the neutrino-mixing matrix.  
    Such an additional source of CP-invariance violation may hold the
    key to explaining how the antimatter created in the Big Bang
    was removed from the early Universe; 
  \item If a neutrino is to be distinguished from its antineutrino
    counterpart it is necessary to assign a conserved lepton number
    to the neutrino.
    At present there is no theoretical justification for such a
    conserved quantum number.
    If lepton number is not conserved, then a neutrino is
    indistinguishable from an antineutrino, i.e. the neutrino is a
    Majorana particle; a completely new state of matter.
    If this is so, then it is conceivable that the ``see-saw''
    mechanism may give an explanation of why the neutrino mass
    is tiny compared to the other matter fermions and may help to
    explain why the neutrino mixing angles are so large compared to
    those of the quarks; and
  \item The neutrino abundance in the Universe is second only to that
    of the photon and so, even with a tiny mass, the neutrino may make
    a significant contribution to the dark matter which is known to
    exist.
    Therefore, the neutrino may play an important role in determining
    the structure of the Universe.
\end{itemize}
These exciting possibilities justify an energetic and far reaching
programme.

An essential part of such a programme is to make precision
measurements of the oscillation parameters. 
Assuming the three flavours and the unitary neutrino-mixing matrix
that is presently favoured, oscillation measurements can be used to
determine the three mixing angles and the phase parameter that can
provide a new source of CP-invariance violation.  
Neutrino-oscillation measurements can also be used to determine the
two (signed) mass differences. 
This programme is similar to the long-standing investigations of quark
mixing via the CKM matrix and it would now seem to be clear that an
understanding of the flavour problem will definitely necessitate
precision measurements in both quark and lepton sectors.  

Not all the properties of the neutrino can be determined by
oscillation experiments. 
Equally important is the determination of the Majorana or Dirac nature
of the neutrino which requires a totally different experiment. 
Currently, the search for neutrinoless double beta decay is the most
promising. 
In addition, although oscillation measurements determine the mass
differences, they are insensitive to the absolute mass, $m_1$, of the 
lightest mass state. 
The determination of $m_1$ requires a very precise measurement of the
end-point of the electron spectrum in beta decay.
The mass would also follow from the observation of neutrinoless double
beta decay. 

The Neutrino Factory, described in this report, is capable of
generating the intense, high-energy neutrino and anti-neutrino beams
which are required to make the exquisitely sensitive oscillation
measurements:
\begin{itemize}
  \item The deviation of $\theta_{23}$ from $\pi/4$;
  \item The deviation of $\theta_{13}$ from $0$;
\end{itemize}
and, for $\sin^2 2 \theta_{13} \gtrsim 5 \times 10^{-5}$:
\begin{itemize}
  \item The degree to which CP-invariance is violated in the (Dirac)
    lepton sector; and
  \item The neutrino mass hierarchy.
\end{itemize}
The intense beams will be equally valuable should there be either an
extended neutrino sector that includes, for example, a fourth
generation or sterile-neutrino state or new non-standard
interactions. 
The large data sets that will be collected will have a unique
potential to throw light on the physics of flavour and hence the
ultimate theory of particle physics.  

\subsection*{Neutrino oscillations and the Neutrino Factory}

Neutrino oscillations are described by a unitary ``mixing matrix''
that rotates the mass basis into the flavour basis. 
The mixing matrix is parametrised by three mixing angles
($\theta_{12}$, $\theta_{23}$, and $\theta_{13}$) and one phase
parameter ($\delta$)
\cite{Pontecorvo:1957cp,Pontecorvo:1957qd,Maki:1962mu,Bilenky:2001rz}. 
If $\delta$ is non-zero (and not equal to $\pi$), then CP-invariance
violation in the neutrino sector will occur so long as 
$\theta_{13} > 0$.
Measurements of neutrino oscillations in vacuum can be used to
determine the moduli of the mass-squared differences 
$\Delta m^2_{31} = m^2_3 - m^2_1$ and 
$\Delta m^2_{21} = m^2_2 - m^2_1$ (where the $m_i$ are the masses of 
the neutrino mass eigenstates) and, with the aid of interactions with
matter, also the sign. 
The bulk of the measurements of neutrino oscillations to date have
been collected using the dominant, ``disappearance'', channels 
$\nu_e \rightarrow \nu_e$ and $\nu_\mu \rightarrow \nu_\mu$.  
These data have yielded values for two of the three mixing
angles ($\theta_{12}$ and $\theta_{23}$), the magnitude of the
mass-squared differences $\Delta m^2_{31}$ and $\Delta m^2_{21}$, and
that $m_2 > m_1$ (i.e. that $\Delta m^2_{21} > 0$).
The challenge to the neutrino community, therefore, is to determine
the sign of $\Delta m^2_{31}$ (the ``mass hierarchy''), to measure
$\theta_{13}$ and $\delta$, and to improve the accuracy with which
$\theta_{23}$ is known.

Over the next few years, the T2K, NO$\nu$A, Double Chooz,
Daya Bay, and RENO experiments will exploit the sub-leading, 
$\nu_\mu \rightarrow \nu_e$ and  $\bar{\nu}_e \rightarrow \bar{\nu}_x$
channels to improve significantly the precision with which
$\theta_{13}$ is known. 
The NO$\nu$A long-baseline experiment may also be able to determine
the mass hierarchy if $\theta_{13}$ is close to the present upper
limit.
However, it is very unlikely that either T2K or NO$\nu$A will be able
to discover CP-invariance violation, i.e. that $\delta \ne 0,\pi$.

The most effective channel for a precision measurement of
$\theta_{13}$ and the search for CP-invariance violation is the
sub-leading $\nu_e \rightarrow \nu_\mu$ oscillation.
The determination of the mass hierarchy relies on the measurement of
the oscillation frequency of neutrinos passing through the earth,
hence the sensitivity increases as the distance the neutrino beam
propagates through the earth increases.
The best sensitivity to CP-invariance violation is also found at large
source-detector distances as long as the neutrinos have a sufficiently
large energy, $E_\nu$.  
Therefore, optimum sensitivity to the parameters of the Standard
Neutrino Model (S$\nu$M) can be achieved at a facility that provides
intense, high-energy $\nu_e$ and $\bar{\nu}_e$ beams.

In the Neutrino Factory, beams of $\nu_e$ and $\bar{\nu}_\mu$ 
($\bar{\nu}_e$, $\nu_\mu$) are produced from the decays of $\mu^+$
($\mu^-$) circulating in a storage ring.
As the muon charge-to-mass ratio is large the neutrinos carry away
a substantial fraction of the energy of the parent muon, hence, high
neutrino energies can readily be achieved.  
Time-dilation is also beneficial, allowing sufficient time to produce
a pure, collimated beam.
Charged-current interactions induced by ``golden channel'', 
$\nu_e \rightarrow \nu_\mu$, oscillations produce muons of charge
opposite to those produced by the $\bar{\nu}_\mu$ in the beam and thus
a magnetised detector is required.   
The additional capability to investigate the ``silver'' 
($\nu_e \rightarrow \nu_\tau$) and ``platinum'' 
($\nubarmu \rightarrow \bar \nu_e$) channels makes the Neutrino
Factory the ideal place to look for oscillation phenomena that are
outside the standard, three-neutrino-mixing paradigm.
It is thus the ideal facility to serve the precision era of neutrino
oscillation measurements. 
The performance of the Neutrino Factory is described in detail in
section \ref{Sect:PPEG}.

\subsection*{The IDS-NF baseline}

The optimisation of the Neutrino Factory baseline to maximise the
performance of the facility for the discovery of CP-invariance
violation, the mass hierarchy, and the determination of
$\theta_{13}$ is described in section \ref{Sect:PPEG}.
The optimum requires two distant detectors. 
One at the ``magic baseline'', $7\,000$--$8\,000$\,km from the
source, will have excellent sensitivity to the mass hierarchy and 
$\sin^2 2 \theta_{13}$.
The second source-detector distance in the range
$2\,500$--$5\,000$\,km will give the best sensitivity to CP-invariance
violation but requires a stored-muon energy in excess of 20~GeV. 
The sensitivity to non-standard interactions at the Neutrino Factory
improves as the stored-muon energy is increased, reaching a plateau at
around $\sim25$~GeV \cite{Kopp:2007mi}.   
A baseline stored muon energy of $25$~GeV has therefore been adopted.

The baseline accelerator facility described in section
\ref{Sect:AccWG} provides a total of $10^{21}$ muon decays per year
split between the two distant neutrino detectors.  
The process of creating the muon beam begins with the bombardment of a
pion-production target with a 4\,MW, pulsed proton beam.
The target must be sufficiently heavy to produce pions copiously, yet
not so large as to cause a significant rate of interaction of the
secondary pions within the target material. 
In addition, the target must withstand the substantial beam-induced
shock.
The IDS-NF baseline calls for a free-flowing, liquid-mercury-jet
target operating in a solenoid-focusing, pion-capture channel.  
This is followed by a solenoidal transport channel in which the pions
decay to muons.
The emerging muon beam is then bunched and rotated in phase space to
produce a beam with small energy spread.
At this point, the muon beam occupies a large volume of
phase space which must be reduced, ``cooled'', before it can be injected
into the acceleration sections.  
As a consequence of the short muon lifetime, traditional cooling
techniques are inappropriate, so the required phase-space reduction is
achieved by means of ionisation cooling.
This involves passing the muon beam through a material in which
it loses energy through ionisation and then re-accelerating the beam
in the longitudinal direction to replace the lost energy.   
The IDS-NF baseline calls for lithium-hydride absorbers embedded in a
solenoidal transport channel with re-acceleration achieved using
201\,MHz cavities at a gradient of 16\,MV/m.  
Subsequent muon acceleration must be rapid, especially at low muon
energy. 
For the baseline, muons are accelerated to 0.9\,GeV in a
superconducting linac and then to 12.6\,GeV in a sequence of two
re-circulating linear accelerators (RLAs). 
The final stage of acceleration, from 12.6\,GeV to the stored-muon
energy of 25\,GeV, is provided by a fixed-field alternating-gradient
(FFAG) accelerator. 
 
\begin{figure}
  \begin{center}
    \includegraphics[width=0.31\textwidth]{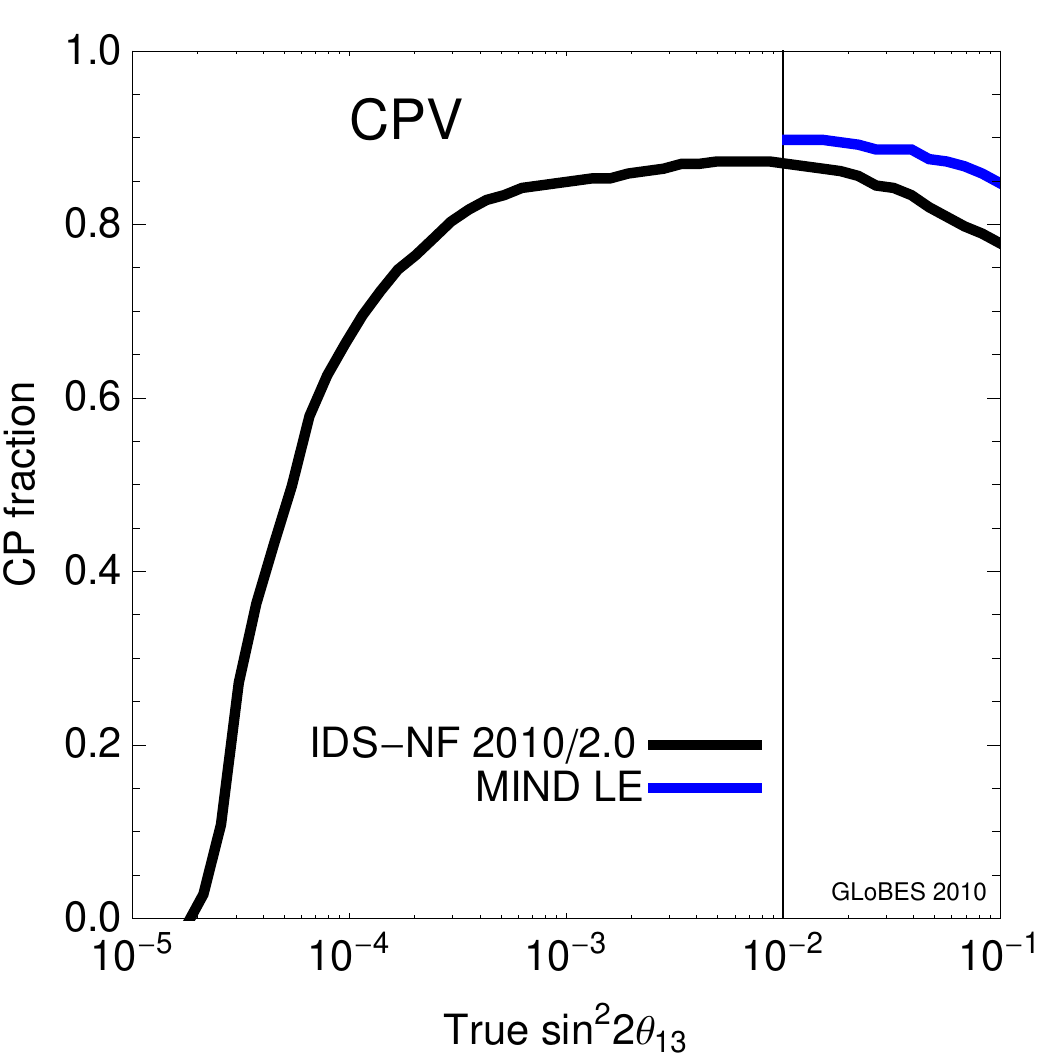}%
    \includegraphics[width=0.31\textwidth]{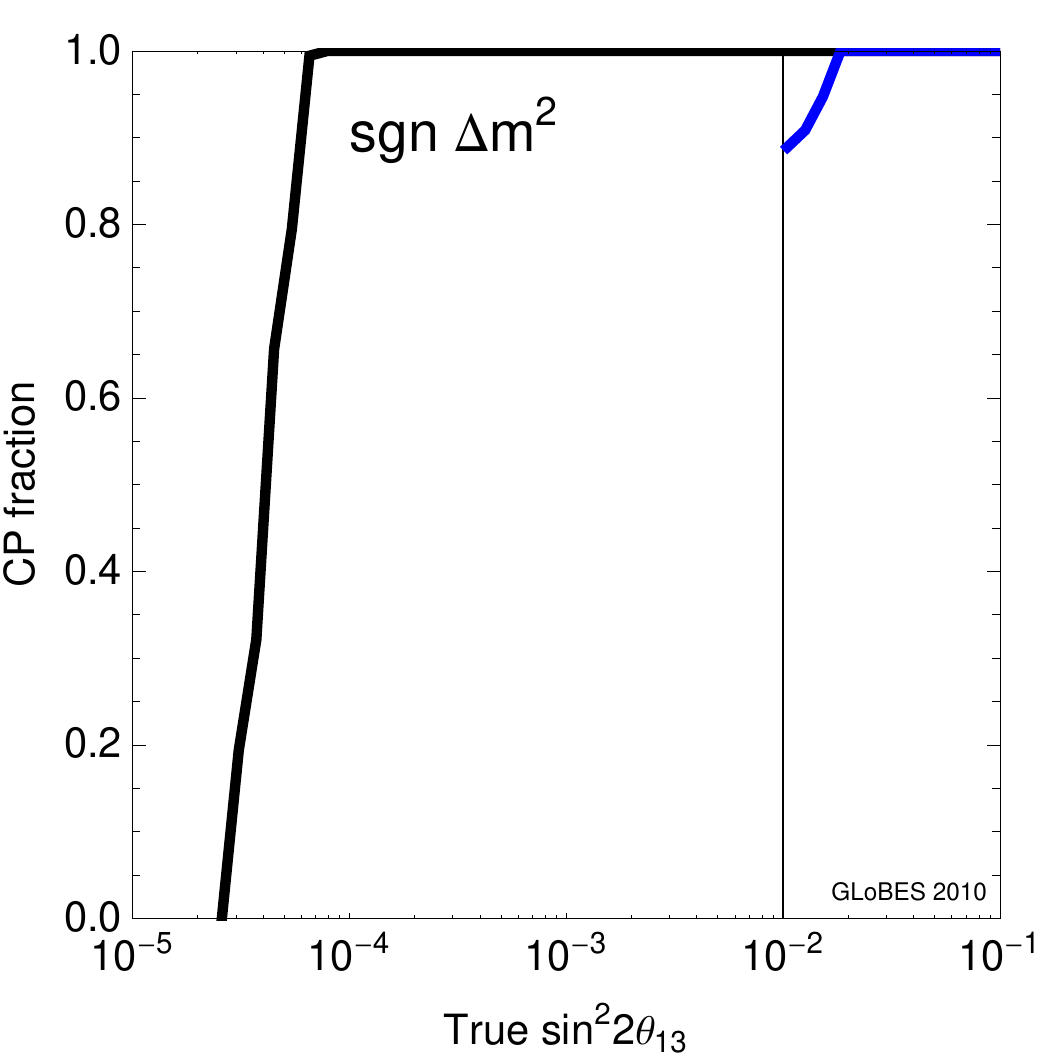}%
    \includegraphics[width=0.31\textwidth]{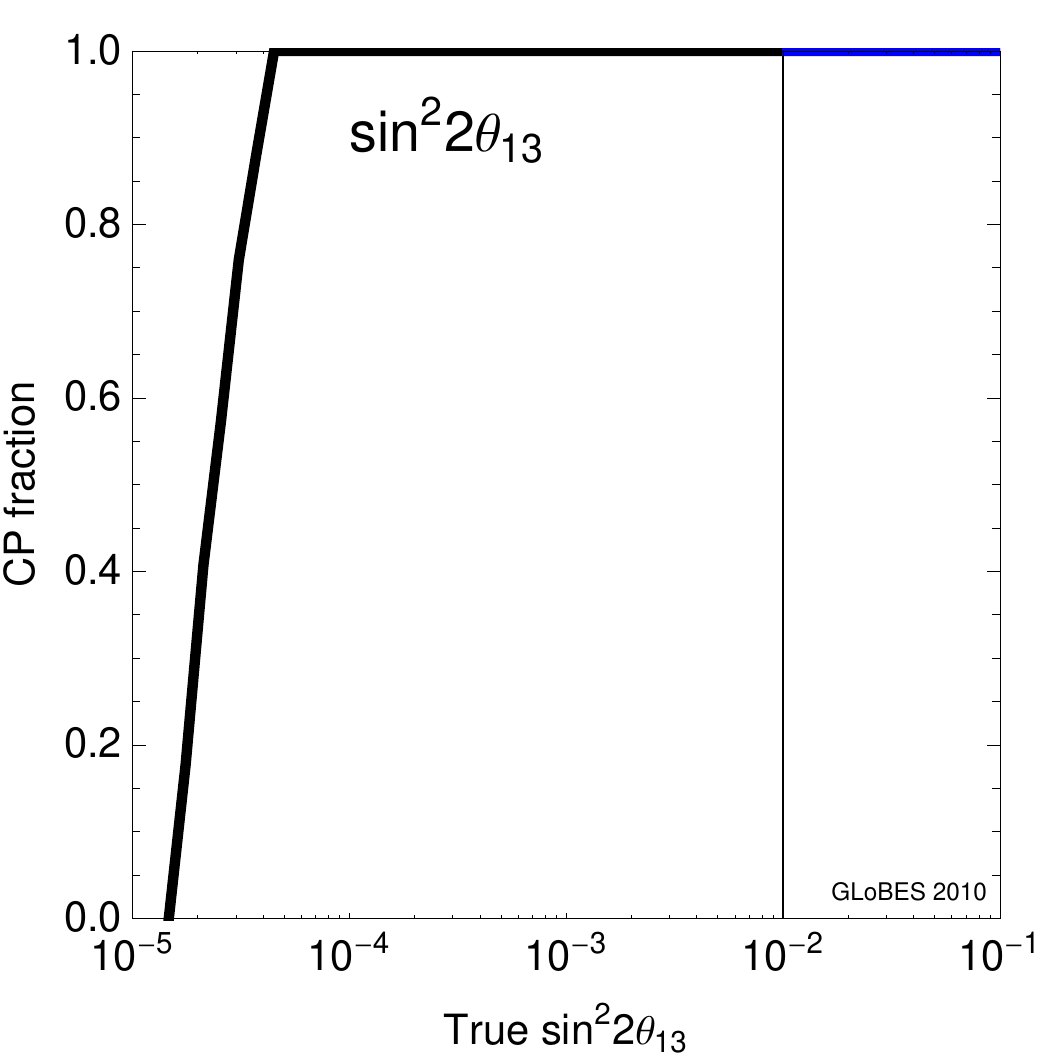}
  \end{center}
  \caption{
    The discovery potential at $3 \sigma$ for CP violation (left
    panel), the mass hierarchy (central panel), and 
    $\sin^2 2 \theta_{13}$ (right panel).
    The discovery reach is plotted in terms of the CP fraction as a
    function of $\sin^2 2 \theta_{13}$.  
    The performance of the IDS-NF baseline Neutrino Factory is shown
    as the black solid line. 
    The Neutrino Factory re-optimised for large $\sin^2 2 \theta_{13}$ 
    is shown as the blue solid line. 
  }
  \label{Fig:Performance} 
\end{figure}

The baseline neutrino detector described in \ref{Sect:DetWG} is a
revision of the Magnetised Iron Neutrino Detector (MIND) developed
within the ISS \cite{Abe:2007bi}.
MIND is a MINOS-like iron-scintillator sandwich calorimeter with a
sampling fraction optimised for the Neutrino Factory beam
\cite{Cervera:2010rz}.
The baseline calls for a fiducial mass of 100~kTon to be placed at
the intermediate baseline and a 50\,kTon detector at the magic
baseline. 
The performance of the IDS-NF baseline Neutrino Factory in terms of
the $3 \sigma$ discovery reach for CP violation, the mass hierarchy,
and $\theta_{13}$ is shown in figure \ref{Fig:Performance}.  
The discovery reach is presented in terms of the fraction of all
possible values of $\delta$ (the ``CP fraction'') and plotted as a
function of $\sin^2 2 \theta_{13}$.
As described in detail in section \ref{Sect:PPEG}, the discovery reach
of the Neutrino Factory is significantly better than the realistic
alternative facilities, particularly for small values of
$\theta_{13}$. 
Should $\theta_{13}$ be discovered to be large, i.e. close to the
present upper bound, a re-optimisation of the baseline Neutrino
Factory will still yield superior performance.

\subsection*{The Neutrino Factory as part of the muon-physics
             programme}

The properties of the muon make it a unique tool for particle
physics. 
In addition to the decays that provide intense, high-energy $\nu_e$
and $\bar{\nu}_e$ beams, the great precision with which properties
such as $g-2$ can be calculated using the Standard Model makes it an
ideal tool in the search for new phenomena. 
Furthermore, the observation of charged lepton flavour violation (cLFV)
in muon decay, predicted in many models, would revolutionise current
theories, whilst the muon's comparatively large mass and point-like
nature make it an appealing candidate to provide multi-TeV
lepton-antilepton collisions at a Muon Collider. 

Neutrino oscillations involve processes in which lepton flavour is not
conserved.
Therefore, processes such as $\mu \rightarrow e \gamma$, 
$\mu \rightarrow eee$, and muon to electron conversion in the field of
the nucleus ($\mu N \rightarrow e N$) will occur.
Rates for such cLFV processes can be calculated in the Standard Model
extended to take into account neutrino oscillations but they are
minuscule (of the order of $10^{-54}$) and so the observation of such
processes would be a clear signal of new physics. 
To achieve the requisite sensitivities, intense muon beams are
required and the techniques proposed for the Neutrino Factory, such as
high power, pulsed proton beams with short ($\lesssim 10\,{\rm ns}$)
bunches, pion capture in solenoidal fields, the manipulation of muon
beams of large emittance, and FFAG acceleration are of great
relevance.

A Muon Collider \cite{Alsharoa:2002wu} offers crucial advantages over
an $e^+e^-$ collider of the same centre-of-mass energy and luminosity
because the muon mass is roughly 200 times that of the electron.
The large muon mass leads to a relatively low rate of synchrotron
radiation, making it possible to design a circular machine in which
100\% of multi-TeV $\mu^+ \mu^-$ collisions occur within $\sim 0.1$\%
of the nominal centre-of-mass energy of the collider.
In addition, should the Higgs boson be discovered and have the
expected coupling proportional to mass, the Muon Collider could be
used as a ``Higgs Factory'' at which detailed studies of its
properties may be carried out.

A conceptual design for the Muon Collider has been proposed
\cite{Palmer:2009zzb} in which the systems that make up the Neutrino
Factory form the ``front-end'' of the Muon Collider.
Indeed, should the techniques required to realise the Neutrino
Factory be demonstrated, the principal accelerator-system challenge
that would remain for the Muon Collider would be the development of an 
ionisation-cooling system that could reduce all six phase-space
dimensions of the muon beams such that a luminosity in excess of
$10^{34}$\,cm$^{-2}$ could be achieved.
Therefore, the implementation of the Neutrino Factory is desirable to
mitigate the technical risks presented by the Muon Collider. 

The Neutrino Factory is the facility of choice for the study of
neutrino oscillations: it has excellent discovery reach and offers the
best precision on the mixing parameters, particularly in the difficult
region of small $\theta_{13}$.
The ability to vary the stored-muon energy and perhaps the detector
technology can provide the flexibility to respond to developments in
our understanding and the discovery of new phenomena.
The R\&D programme required to make the Neutrino Factory a reality
will directly benefit the development of the Muon Collider and
experiments that seek to discover cLFV.
A comprehensive muon-physics programme is compelling indeed. 

\cleardoublepage
%
\section{Physics at the Neutrino Factory}
\label{Sect:PPEG}

\subsection{The physics case for advanced neutrino experiments}
\label{subsec:physicscase}

The unambiguous observation of neutrino oscillations is arguably the
most significant development in particle physics of the past two
decades \cite{GonzalezGarcia:2007ib}.  
It revealed, beyond reasonable doubt, that neutrinos have mass and
that leptons mix.  
Neutrino masses are not accounted for in the Standard Model of
particle physics.  
It is not hard to modify the Standard Model Lagrangian to accommodate
neutrino masses \cite{Mohapatra:2005wg}.
Indeed, one can write down several different ``Standard Neutrino
Model'' (S$\nu$M) Lagrangians that render neutrinos massive that are
consistent with the available data.
The candidates for the S$\nu$M introduce various new
degrees of freedom at a variety of energy scales (from below the
electron-Volt scale all the way to the Grand Unification scale).  
Part of the mission of particle-physics experiments in the next few
decades will be to elucidate the mechanism by which neutrino mass is
generated. 
The next generation of neutrino experiments will represent a moderate
extension of existing techniques.
This makes them technologically attractive, since the technical risks
are relatively low, but also limits their physics sensitivity. 
To exploit to the full the discoveries made to date in neutrino
physics, advanced neutrino experiments, based on novel technologies,
are required. 
The Neutrino Factory is the ultimate advanced neutrino oscillation
facility.
In this section, the scientific case for advanced neutrino oscillation
experiments in general and of the Neutrino Factory in particular is
reviewed.

While the mechanism behind neutrino-mass generation remains unknown, a 
very successful phenomenological description exists.  
It considers the existence of three massive neutrinos 
$\nu_1$, $\nu_2$, and $\nu_3$ with masses $m_1$, $m_2$, and $m_3$,
respectively.
These mass eigenstates are linear combinations of the ``active''
neutrino flavours, $\nu_e$, $\nu_{\mu}$, and $\nu_{\tau}$, labelled
according to the way they interact with the $W$-boson and the
charged-leptons, $e$, $\mu$, and $\tau$.
One can pick a weak-basis in which neutrinos with a well-defined
flavour are related to neutrinos with a well-defined mass via the
unitary lepton mixing matrix $U$: $\nu_{\alpha}=U_{\alpha i}\nu_i$
($\alpha=e,\mu,\tau$, $i=1,2,3$).  
It is customary to define the
neutrino masses as follows: $m_1<m_2$ while $m_3$ is either the
heaviest or the lightest neutrino.  
To identify $m_3$, $|m_3^2-m^2_1|$ and $|m_3^2-m^2_2|$ are evaluated
and the smaller combination is chosen such that it is larger than
$m_2^2-m_1^2$. 
With the masses labelled in this way, $m_3>m_2$ if $m_3^2-m_1^2$ is
positive and $m_3<m_1$ if $m_3^2-m_1^2$ is negative.  
Note that
$\Delta m^2_{21}=m_2^2-m_1^2$ is a positive-definite quantity, while
$\Delta m^2_{31}=m_3^2-m_1^2$ is allowed to have either sign.  
It is
also customary to refer to the spectrum $m_3>m_2>m_1$ as ``normal'',
in which case $\Delta m^2_{31}>0$, while the $m_2>m_1>m_3$ spectrum is
referred to as ``inverted,'' in which case $\Delta m^2_{31}<0$.  
Once the neutrino mass eigenstates are properly defined, it is also
customary to parametrise $U$ as prescribed in the Particle Data Book
\cite{Nakamura:2010zzi}.
The neutrino mixing angles ($\theta_{ij}$) are related to the
mixing-matrix elements by:
\begin{equation}
  \frac{|U_{e2}|^2}{|U_{e1}|^2}\equiv \tan^2\theta_{12};
  ~~~~\frac{|U_{\mu3}|^2}{|U_{\tau3}|^2}\equiv \tan^2\theta_{23}; {\rm~and}
  ~~~~U_{e3}\equiv\sin\theta_{13}e^{-i\delta};
\end{equation}
where $\delta$ is the CP-odd, ``Dirac'', phase.
Two other CP-odd phases might be required to complete the
parametrisation of $U$ if the neutrinos are Majorana fermions. 
However, these ``Majorana'' phases can not be observed in
neutrino oscillations and will henceforth be ignored.  
Neutrino-oscillation data are sensitive to the neutrino mass-squared
differences, the values of the mixing angles $\theta_{12}$,
$\theta_{13}$, $\theta_{23}$, and $\delta$.  
Two of the three mixing angles are large, in stark contrast to the
quark-mixing matrix, in which all mixing parameters are either small 
($\sin\theta_c\sim 0.23$, where $\theta_c$ is the Cabibbo angle) or
very small ($|V_{ub}|\sim 0.004$).  
In the neutrino sector, three parameters remain unknown or only poorly
constrained: the sign of $\Delta m^2_{31}$; $\delta$; and the mixing
angle $\theta_{13}$.  
The remaining parameters are known at the $\sim 5$--$10$\% level
\cite{GonzalezGarcia:2010er,Mezzetto:2010zi}.  

If $\theta_{13}$ is ``large'', the current generation of experiments
(which includes Double Chooz \cite{Ardellier:2004ui}, RENO
\cite{Ahn:2010vy}, Daya Bay \cite{Guo:2007ug}, T2K \cite{Itow:2001ee}
and No$\nu$a \cite{Ayres:2004js}) may determine that $\theta_{13}$ is
not zero and No$\nu$a may provide a non-trivial hint regarding the
sign of $\Delta m^2_{31}$. 
It is also expected that the precision with which $|\Delta m^2_{31}|$
and $\sin^22\theta_{23}$ are known will improve significantly.  
Even under these conditions, it is widely anticipated that by the end
of the current decade we shall not know the sign of 
$\Delta m^2_{31}$ and will have no information regarding whether
CP-invariance is violated in neutrino interactions 
($\delta\neq 0,\pi$).   
On the other hand, if $\theta_{13}$ is ``small'', not withstanding a
measurement of $|\Delta m^2_{31}|$ with significantly improved
precision, a more precise measurement of $\sin^22\theta_{23}$, and a
more stringent bound on $\theta_{13}$, our understanding of the
neutrino-oscillation parameters might not be too different from what
it is today.

To complete our knowledge of the parameters that describe neutrino
oscillations, the goals of advanced experiments must be:
\begin{itemize}
  \item 
    To search for CP-invariance violation in neutrino oscillations by
    seeking to measure the CP-odd phase $\delta$;
  \item
    To establish whether the neutrino-mass spectrum is normal or
    inverted by determining the sign of $\Delta m^2_{31}$;
  \item
    To determine $\theta_{13}$ with good precision (or constrain it to
    be very small indeed); and
  \item
    To measure as precisely as possible all the oscillation
    parameters.
\end{itemize}
The search for CP-invariance violation in neutrino oscillations
provides a unique opportunity to further our understanding of
CP-invariance.  
Experiments have demonstrated that the CP-odd phase of the quark
mixing matrix, $\delta_{\rm CKM}$, controls all CP-invariance
violation in the quark sector.
The past generation of neutrino-oscillation experiments has revealed
that there is at least one more CP-violating parameter in the lepton
mixing matrix, $\delta$. 
In practise, $\delta$ can only be observed in neutrino oscillation
experiments and an advanced neutrino experiment is required to explore
this new window into CP-invariance violation.

To determine how precisely the oscillation parameters must be
measured, it is important to assess the precision at which our
understanding would change qualitatively.
Attempts to understand the underlying physics behind the pattern of
lepton mixing provide some guidance 
\cite{Altarelli:2004za,Mohapatra:2005wg,Albright:2006cw}).  
Such attempts often make predictions regarding the specific values for
the different parameters or, perhaps more generically, make predictions
regarding relations among different oscillation parameters. 
One guiding principle for precision measurements of neutrino
oscillations should be the ability to provide unambiguous tests of
these relations. 
An example is given by the tri-bimaximal mixing pattern
($\sin^2\theta_{23}=1/2$, $\sin^2\theta_{12}=1/3$, and
$\sin^2\theta_{13}=0$) \cite{Harrison:2002er}; a popular, zeroth-order
ansatz that guides theoretical research in neutrino flavour. 
Different models make different predictions regarding deviations from
this ansatz, which are usually small and proportional to some other
small parameters in flavour physics. 
Neutrino physics offers one known small (but non-zero) parameter: the
ratio of the two mass-squared differences, 
$\Delta m^2_{12}/|\Delta m^2_{31}|\equiv\epsilon\sim 0.03$
\cite{deGouvea:2004gr}. 
Current data constrain
$\sin^2\theta_{13}\lesssim\epsilon$, $\sin^2\theta_{12}-1/3\lesssim
\epsilon$, while $\sin^2\theta_{23}-1/2\lesssim \sqrt{\epsilon}$.  
An important next-to-next-generation goal would be to test whether
$\sin^2\theta_{13}\lesssim\epsilon^2$, $\sin^2\theta_{12}-1/3\lesssim
\epsilon^2$, and $\sin^2\theta_{23}-1/2\lesssim\epsilon$ (or better).

High-precision measurements of neutrino oscillations are required to
test whether the origin of neutrino mass is also the origin of
the baryon asymmetry of the Universe, through leptogenesis
\cite{Fukugita:1986hr,Davidson:2008bu}.  
Several indirect generic predictions of leptogenesis can be verified
with neutrino experiments: neutrinos are expected to be Majorana
fermions, and it is expected, quite generically, that neutrino
oscillations violate CP-invariance, even if no model independent
relation between ``high-energy'' and ``low-energy'' CP-invariance
violation exists.   
Quantitative tests are all model dependent and will rely on very
precise measurements in neutrino oscillations and elsewhere.  
For example, precise measurements of oscillation parameters could
validate a specific flavour model, which allows one to relate
``high-energy'' and ``low-energy'' parameters.  
Furthermore, discoveries at collider experiments and searches for
charged-lepton flavour violation may provide other hints that render
leptogenesis either ``very likely'' or ``most improbable''
\cite{Buckley:2006nv}.
We are very far from testing leptogenesis conclusively, but precise
measurements of all neutrino oscillation parameters -- far beyond
where we are now -- are a {\it conditio sine qua non}.

Precision measurements of the oscillation parameters are also required
to confirm or refute with confidence the current three-active-neutrino
formalism.
Ultimately, one aims not only at constraining the mixing-parameter
space, but also at {\it over-constraining} it.  
Several important questions need to be addressed: 
\begin{itemize}
  \item
    Are there really only three light neutrinos and is $U$ a unitary
    matrix? 
  \item
    Are there other neutrino interactions?  And, 
  \item
    Is there only one source of CP-invariance violation in the
    neutrino sector?  
\end{itemize}
While our understanding of the neutrino has increased tremendously
over the past decade, we are far from providing a satisfactory answer
to any of these questions.   
In this respect, our understanding of the lepton-flavour sector is far
behind our understanding of the quark-flavour sector.

Some manifestations of new physics are best investigated with neutrino
oscillation experiments.  
For example, the search for light sterile neutrinos which may be
related to the origin of neutrino masses.
Sterile neutrinos can be detected via the observation of new
oscillation frequencies and mixing angles. 
Tests of the unitarity of the lepton-mixing matrix may also point to
new ``neutrino'' degrees of freedom that are too heavy to be seen in
oscillation experiments. 
Neither of these phenomena can be studied outside of neutrino
oscillation experiments.
Other new-physics ideas to which measurements of neutrino oscillation
are uniquely sensitive are related to physics at the electroweak
scale, including new four-fermion neutrino interactions (of the
current--current type, 
$\propto(\bar{\nu}_{\alpha}\Gamma\nu_{\beta})(\bar{f}\Gamma' f)$,
where $f$ is a charged fermion).  
Some of these possibilities will be discussed in more detail later in
this section. 
Indeed, non-standard neutrino interactions are often used as proxies
for the discussion of the sensitivity of neutrino-oscillation
experiments to new physics. 

In many candidate new-physics scenarios, a combination of different
experimental probes will be required in order to piece together a more
fundamental description of how nature works at the smallest distance
scales.  
In addition to studies of neutrino flavour-change, these include
collider experiments (for example, the LHC and a next-generation
lepton collider), searches for charged-lepton flavour violation in the
muon and tau sectors, searches for the permanent electric dipole
moments of fundamental particles, including the electron and the muon,
searches for lepton-number violation, especially searches for
neutrinoless double-beta decay, and direct and indirect searches for
dark matter.  
In many new-physics scenarios, in particular when it
comes to identifying the physics responsible for neutrino masses,
advanced neutrino oscillation experiments are guaranteed to play a
leading role.

Neutrino experiments have proved, over the past few
decades, that our ability to predict what will be detected, and to
identify what are the important questions, is limited at best.  
It is
safe to state, however, that a Neutrino Factory, combined with a
multi-kTon detector at an underground facility, offers a unique and 
powerful tool for the study of fundamental physics.  
In order to prepare for the unexpected, it is vital that advanced
set-ups be versatile and multifaceted.  
The Neutrino Factory fits the bill.
In addition to providing the neutrino beams required for the
definitive, precision neutrino-oscillation programme, the Neutrino
Factory also provides an ideal environment in which to study a variety
of other phenomena.   
The well-characterised neutrino beam from the muon storage ring allows
a programme of extremely precise neutrino scattering measurements to
be carried out at a near detector, including studies of neutrino
flavour-change at very short baselines and precision measurements of
neutrino scattering on nucleons \cite{Mangano:2001mj} and electrons
\cite{deGouvea:2006cb}.
The latter allow for uniquely sensitive tests of the electroweak
theory.

Activities not directly related to neutrino physics can also be
addressed at a Neutrino Factory complex.  
The availability
of a large number of muons allows one to consider new set-ups for
searching for rare muon processes \cite{Aysto:2001zs}, especially
$\mu\to e$ conversion in nuclei, and for measurements of the
electroweak properties of the muon, including the muon electric and
magnetic dipole moments.  
The availability of a large number of protons---used to make the muons
for the storage ring---allows one to consider a suite of hadronic
experiments including, for example, those required to study very rare
kaon phenomena
($K\to\pi\bar{\nu}\nu$, $K\to\pi \mu^{\pm} e^{\mp}$, $K^+\to\pi^- \ell^+\ell'^+$, 
etc).   
At the opposite end of the neutrino beam, the very large detector
complex also serves many purposes.  
Depending on the location and composition, very large detectors can be
used to study naturally occurring neutrinos---especially the
atmospheric neutrinos and, perhaps, neutrinos from Supernova
explosions. 
Finally, the large
instrumented volumes can be used for searching for proton decay.
Indeed, the Kamiokande and IMB experiments were originally constructed
to look for proton decay, stumbled upon atmospheric neutrino
oscillations, and observed neutrinos produced in super-nov\ae along
the way.

\subsection{Neutrino oscillation update} 

A reasonable understanding of the parameters describing three--flavour
neutrino oscillations has been developed from the results of a variety
of experiments involving solar and atmospheric neutrinos, as well as
man-made neutrinos from nuclear power plants and accelerators.
Table~\ref{tab:summary} summarises the results of two recent global
fits to world neutrino data from~\cite{GonzalezGarcia:2010er}
and~\cite{Schwetz:2008er}, both updated as of November 2010.
Details of another recent analysis can be found 
in~\cite{Fogli:2008ig,Fogli:2009ce}.
\begin{table}
  \caption{
    Determination of three--neutrino
    oscillation parameters from 2010 global data~\cite{Schwetz:2008er,
      GonzalezGarcia:2010er}. For $\Delta m^2_{31}$ and
    $\sin^2\theta_{13}$ the upper (lower) row corresponds to inverted
    (normal) neutrino mass hierarchy.
  }
  \label{tab:summary} 
  \centering
  \begin{tabular}
    {|@{\quad}l@{\quad}|@{\quad}c@{\quad}@{\quad}c@{\quad}
     |@{\quad}c@{\quad}@{\quad}c@{\quad}|}
    \hline
    & \multicolumn{2}{c|@{\quad}}{Ref.~\cite{GonzalezGarcia:2010er} (updated)}
    & \multicolumn{2}{c|}{Ref.~\cite{Schwetz:2008er} (updated)}
    \\
    parameter
    & best fit$\pm 1\sigma$ & 3$\sigma$ interval
    & best fit$\pm 1\sigma$ & 3$\sigma$ interval
    \\
    \hline
    $\Dmq_{21} ~[10^{-5}~\eVq]$
    & $7.60 \pm 0.20$ & 6.98--8.18
    & $7.59_{-0.17}^{+0.20}$  & 7.09--8.19
    \\[1mm]
    $\Dmq_{31} ~[10^{-3}~\eVq]$ &
    \begin{tabular}{c}
      $-2.35 \pm 0.09$ \\[-1mm]
      $+2.44 \pm 0.09$
    \end{tabular}
    &
    \begin{tabular}{c}
      $-$(2.07--2.65) \\[-1mm]
      $+$(2.17--2.74)
    \end{tabular}
    &
    \begin{tabular}{c}
      $-(2.34_{-0.09}^{+0.10})$ \\[-1mm]
      $+2.45\pm0.09$
    \end{tabular}
    &
    \begin{tabular}{c}
      $-(2.08-2.64)$\\[-1mm]
      $+(2.18-2.73)$
    \end{tabular}
    \\
    $\sin^2\theta_{12}$
    & $0.317_{-0.016}^{+0.017}$ & 0.27--0.37
    & $0.318_{-0.016}^{+0.019}$ & 0.27--0.38
    \\[1mm]
    $\sin^2\theta_{23}$
    & $0.45_{-0.05}^{+0.09}$ & 0.34--0.65
    & $0.51\pm{0.06}$ & 0.39--0.64
    \\[1mm]
    $\sin^2\theta_{13}$
    & $0.009_{-0.008}^{+0.015}$ & $\leq$ 0.048
    &
    \begin{tabular}{c}
      $0.017\pm0.010$ \\[-1mm]
      $0.012^{+0.010}_{-0.007}$
    \end{tabular}
    &
    \begin{tabular}{c}
      $\leq$ 0.048 \\[-1mm]
      $\leq$ 0.042
    \end{tabular}
    \\
    \hline
  \end{tabular}
\end{table}

Spectral information from the KamLAND reactor
experiment~\cite{KamLAND:2008ee, Gando:2010aa} leads to an accurate
determination of $\Dmq_{21}$ with the remarkable precision of 7\% at
$3\sigma$.
The sign of $\Dmq_{21}$ being determined using solar neutrino data
from SNO (see for example \cite{Aharmim:2011vm}).
KamLAND data also start to
contribute to the lower bound on $\sin^2\theta_{12}$, whereas the upper
bound is dominated by solar data.
From the Sudbury Neutrino Observatory (SNO) we use the data of its
final phase, where the neutrons produced in the neutrino neutral
current (NC) interaction with deuterium are detected mainly by an
array of $^3$He NC detectors (NCD)~\cite{Aharmim:2008kc}, as well as
the recent joint re-analysis of data from Phase I and Phase II
(the pure D$_2$O and salt phases)~\cite{Aharmim:2009gd}. 
In this
analysis, an effective electron kinetic energy threshold of 3.5~MeV
has been used (Low Energy Threshold Analysis, LETA), and the
determination of the total neutrino flux has been improved by about a
factor two.  
These improvements have been possible thanks mainly to
the increased statistics, in particular the NC event sample in the
LETA is increased by about 70\% as a result of the decrease of the
thresholds of 5~MeV and 5.5~MeV used in the Phase~I in Phase~II
analyses respectively. 
Furthermore, energy resolution, background suppression, and
systematic uncertainties have been improved.
Data from SNO are combined with the global data on solar neutrinos
\cite{Cleveland:1998nv,Altmann:2005ix,Hosaka:2005um} including updates
on Gallium experiments \cite{Abdurashitov:2009tn, Kaether:2010ag} as
well as results from Borexino \cite{Arpesella:2008mt} on $^7$Be
neutrinos yielding an improved determination of the atmospheric
parameters.

The MINOS experiment~\cite{Adamson:2008zt} is searching for $\nu_\mu$
disappearance with a baseline of 735~km. 
The latest data were presented at the Neutrino 2010
Conference~\cite{vahle} for the neutrino (7.2$\times10^{20}$ p.o.t.)
and the anti-neutrino (1.71$\times10^{20}$ p.o.t.) running modes. 
The data confirm the energy-dependent disappearance of $\nu_\mu$,
showing significantly fewer events than expected in the case of no
oscillations in the energy range $\lesssim 6$~GeV, whereas above $\sim
6$\,GeV the spectrum is consistent with the no-oscillation
expectation. 
\begin{figure}
  \centering
  \includegraphics[width=0.8\textwidth]{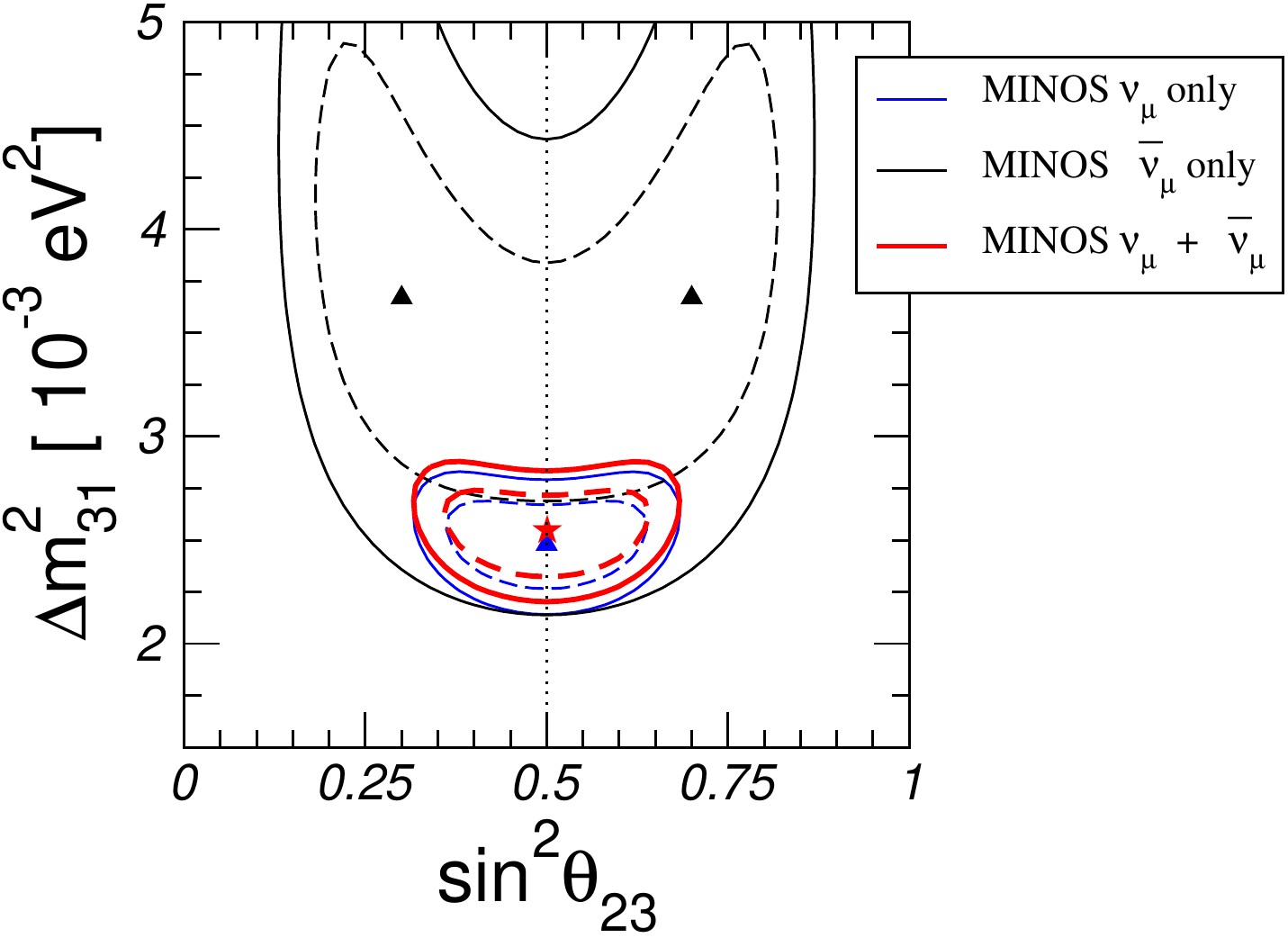}
  \caption{
    Allowed regions from recent MINOS data, using neutrinos-only,
    anti-neutrinos-only, and in combination.
  }
  \label{fig:minos} 
\end{figure}

Figure~\ref{fig:minos} shows a re-analysis of MINOS neutrino and
anti-neutrino data. One can see that there is a slight tension between the
neutrino and anti-neutrino results: there is no overlap of the allowed
regions at less than 90\%~CL. However, at $3\sigma$ the results are fully
consistent. We find the following $\chi^2$ minima and goodness-of-fit (GOF)
values:
\begin{equation}
  \begin{array}{rcc}
    \nu:     & \chi^2_{\text{min},\nu} = 24.4/(27-2)     & \text{GOF} = 49.6\%; \\
    \bar\nu: & \chi^2_{\text{min},\bar\nu} = 15.0/(13-2) & \text{GOF} = 18.4\%; \\
    \nu+\bar\nu: & \chi^2_\text{min,tot} = 46.1/(40-2)   & \text{GOF} = 17.3\%.
  \end{array}
\end{equation}
Hence the combined neutrino and anti-neutrino fit still provides a
very good GOF. 
Using the consistency test from reference~\cite{Maltoni:2003cu}
yields $\chi^2_\text{PG} = \chi^2_\text{min,tot} -
\chi^2_{\text{min},\nu} - \chi^2_{\text{min},\bar\nu} = 6.6$. 
The
value of $\chi^2_\text{PG}$ has to be evaluated for 2 degrees of
freedom, which implies that neutrino and anti-neutrino data are
consistent with a probability of 3.7\%. 
This number indicates a slight
tension between the sets, at the level of about $2.1\sigma$. 
In the
global analysis~\cite{Schwetz:2008er} only neutrino data are used from
MINOS, whereas in the analysis in \cite{GonzalezGarcia:2010er} both
neutrino and anti-neutrino data are included. 
It is clear from figure \ref{fig:minos} that the improvement in the
quality of the fit when the anti-neutrino data are added is
negligible.

We combine the long-baseline accelerator data from MINOS as well as
from K2K~\cite{Aliu:2004sq} with atmospheric neutrino measurements
from Super-Kamiokande I+II+III~\cite{Wendell:2010md}.
The determination of $|\Dmq_{31}|$ is dominated by spectral data from
the MINOS experiment, where the sign of $\Dmq_{31}$ (\textit{i.e.},
the neutrino mass hierarchy) is undetermined by present data. 
The
measurement of the mixing angle $\theta_{23}$ is still largely
dominated by atmospheric neutrino data from Super-Kamiokande with a
best-fit point close to maximal mixing. 
Small deviations from maximal
mixing due to sub-leading, three-flavour effects (not included in the
analysis of~\cite{Schwetz:2008er}) have been found in
references~\cite{Fogli:2005cq,GonzalezGarcia:2007ib,GonzalezGarcia:2010er}, 
see table~\ref{tab:summary}. 
A comparison
of these subtle effects can be found in reference~\cite{snow}. 
While an excess of sub-GeV $e$-like data provides an explanation for
the deviations from maximality obtained
in~\cite{Fogli:2005cq,GonzalezGarcia:2010er}, these results are not
statistically significant, and are not confirmed by a recent analysis
including $\Delta m^2_{21}$ effects from the
Super-Kamiokande collaboration \cite{Wendell:2010md}. 
\begin{figure}
  \centering
  \includegraphics[width=0.8\textwidth]{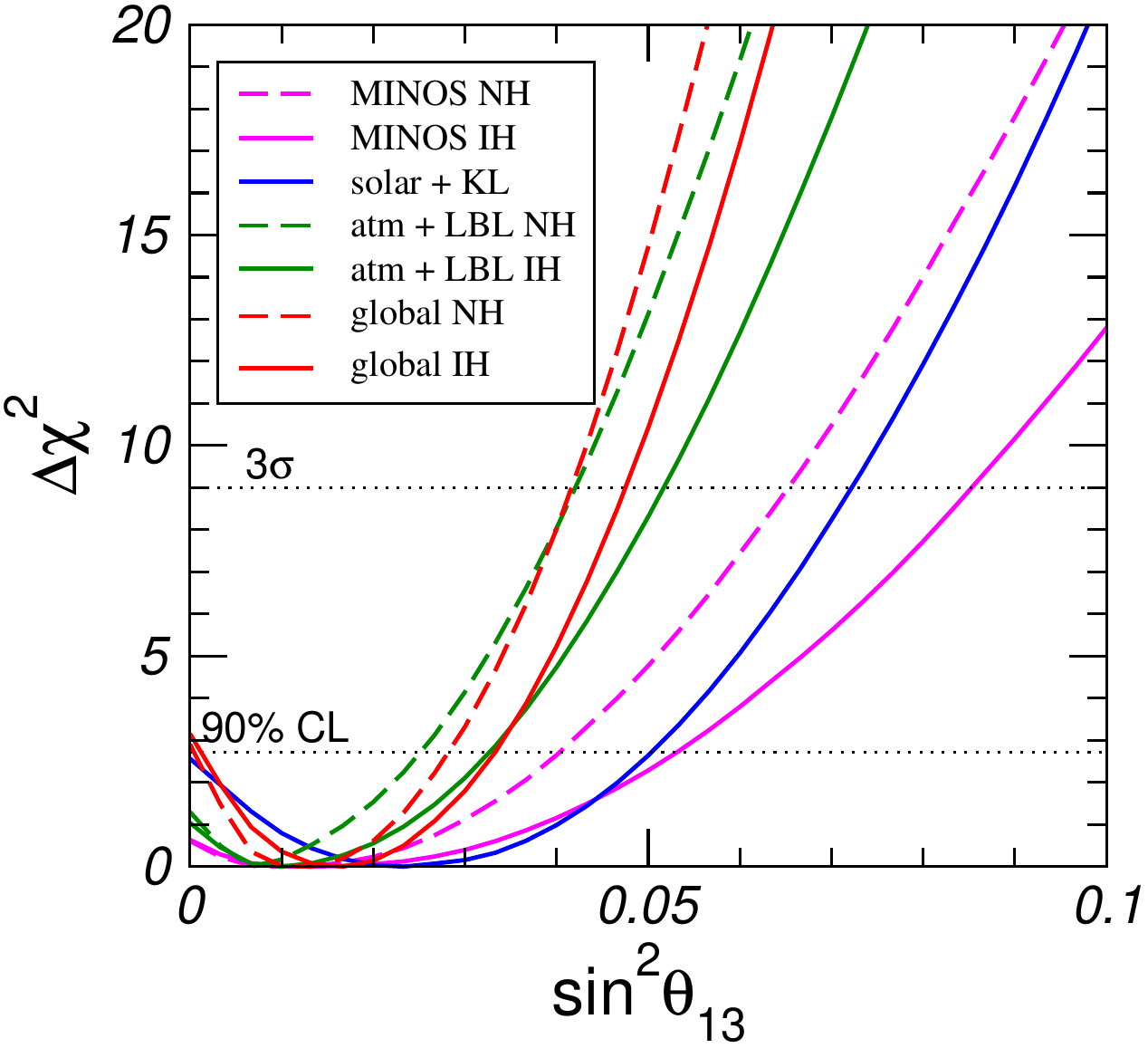}
  \caption{
    Dependence of $\Delta\chi^2$ on $\sin^2\theta_{13}$ from various
    data sets, shown for NH (solid) and IH (dashed). 
    The curves labelled ``MINOS'' include disappearance and appearance
    data, and the curves labelled ``atm + LBL'' include Super-K
    atmospheric data + MINOS (disappearance and appearance) + K2K +
    CHOOZ.
  } 
  \label{fig:the13} 
\end{figure}

The third mixing angle $\theta_{13}$ is of crucial importance for
future oscillation experiments.
Figure~\ref{fig:the13} summarises the information on $\theta_{13}$ from
present data, which emerges from an interplay of different data
sets. 
An important contribution to the bound comes from
the CHOOZ reactor experiment~\cite{Apollonio:2002gd} combined with the
determination of $|\Dmq_{31}|$ from atmospheric and long-baseline
experiments.
Using this set of data, a possible hint for a non-zero $\theta_{13}$
from atmospheric data has been found in
references~\cite{Fogli:2005cq,Fogli:2008jx}. 
The origin of such a hint has been investigated in
some detail in reference~\cite{Maltoni:2008ka}, and more recently in
\cite{Fogli:2009ce, GonzalezGarcia:2010er}.
From these results one may conclude that the statistical relevance of
the hint for non-zero $\theta_{13}$ from atmospheric data depends
strongly on the details of the rate calculations and of the $\chi^2$
analysis. 
Furthermore, the origin of this effect might be traced back
to a small excess (at the $1\sigma$ level) in the multi-GeV
electron-like ($e$-like) data sample in SK-I, which however is no
longer present in the combined $\text{SK-I} + \text{SK-II}$ data and
is extremely weak in the combined 
$\text{SK-I} + \text{SK-II} + \text{SK-III}$ data set.

A recent analysis (neglecting sub-leading $\Dmq_{21}$ effects) from the
Super-Kamiokande collaboration finds no evidence of such a
hint~\cite{Wendell:2010md}. 
The results of this analysis have also been
used in the global fit of~\cite{Schwetz:2008er}. 
However, in the
combination of these data with MINOS disappearance and appearance data
a slight preference for $\theta_{13} > 0$, with $\Delta\chi^2 = 1.6
(1.9)$ at $\theta_{13}=0$ for normal hierarchy (inverted hierarchy). 
This happens due to a small
mismatch of the best fit values for $|\Delta m^3_{31}|$ at
$\theta_{13}=0$, which can be resolved by allowing for non-zero values
of $\theta_{13}$.

The MINOS Collaboration has recently reported new data from the search
of $\nu_\mu \to \nu_e$ transitions in the Fermilab NuMI
beam~\cite{Adamson:2010uj}. 
The new data are based on a total exposure
of 7$\times$10$^{20}$ protons-on-target, more than twice the size of
the previous data release~\cite{Adamson:2009yc}. 
The new MINOS far
detector data consists of 54 electron-neutrino events, while,
according to the measurements in the MINOS Near Detector, $49.1 \pm
7.0 (stat) \pm 2.7 (syst)$ background events were expected. 
Hence the
observed number of events is in agreement with background expectations
within 0.7$\sigma$ and the hint for a non-zero value of $\theta_{13}$
present in previous data has largely disappeared.

An important piece of information on $\theta_{13}$ comes from solar
and KamLAND data. 
The relevant survival probabilities are given by:
\begin{equation}\label{eq:Pee}
  P_{ee} \approx
  \begin{cases}
    \cos^4\theta_{13} \left(1- \sin^22\theta_{12}\right \langle \sin^2\phi\rangle)
    & \text{solar, low energies / KamLAND}
    \\
    \cos^4\theta_{13} \, \sin^2\theta_{12}
    & \text{solar, high energies}
  \end{cases}
  \, ;
\end{equation}
where $\phi = \Dmq_{21} L / 4E$ and $\langle \sin^2\phi\rangle \approx
1/2$ for solar neutrinos.  
Equation~(\ref{eq:Pee}) implies an
anti-correlation of $\sin^2\theta_{13}$ and $\sin^2\theta_{12}$ for
KamLAND and low-energy solar neutrinos. 
In contrast, for the high-energy part of the spectrum, which undergoes
the adiabatic MSW conversion inside the sun and which is constrained
by the SNO CC/NC measurement, a positive correlation of
$\sin^2\theta_{13}$ and $\sin^2\theta_{12}$ emerges. 
As discussed
in~\cite{Maltoni:2004ei, Goswami:2004cn}, this complementarity leads
to a non-trivial constraint on $\theta_{13}$ and allows the hint for a
non-zero value of $\theta_{13}$ to be understood, which helps 
to reconcile the slightly different best fit points for $\theta_{12}$
as well as for $\Dmq_{21}$ for solar data and data from KamLAND
separately~\cite{Balantekin:2008zm,Goswami:2004cn,Fogli:2008jx,Maltoni:2003da,GonzalezGarcia:2007ib}.  
We found that the inclusion
of the new solar data, and in particular of the SNO-LETA results, tends
to lower the statistical significance of $\theta_{13}\neq 0$ from this
data set. 
There is a minor dependence of the hint
for $\theta_{13}$ on the solar model, as discussed for example in
\cite{GonzalezGarcia:2010er}. 

From the global data a significance for $\theta_{13} > 0$ of
$1.8\sigma$ for the inverted hierarchy and $1.7\sigma$ for the normal
hierarchy (NH) is obtained in~\cite{Schwetz:2008er} and $1.3\sigma$
in~\cite{GonzalezGarcia:2010er}. 
We find that the inverted hierarchy (IH)
gives a slightly better fit, however, with only $\Delta\chi^2 = 0.7$
\cite{Schwetz:2008er} with respect to the best fit in normal hierarchy
and even less ($\Delta\chi^2 = 0.1$) in \cite{GonzalezGarcia:2010er}.

\begin{figure}
  \includegraphics[width=0.8\textwidth]{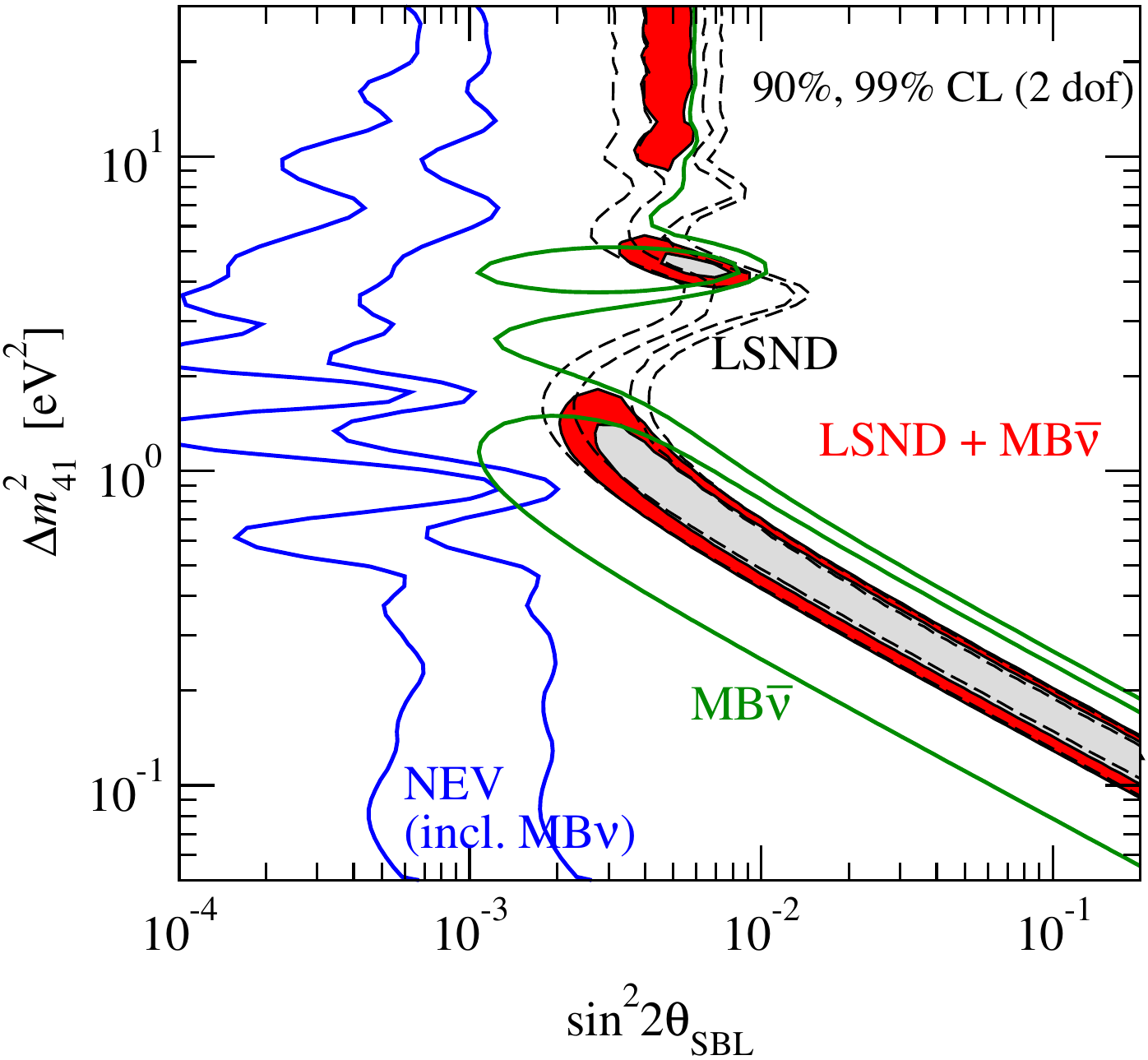}
  \caption{\label{fig:mboone} Constraint from no-evidence data (NEV)
  compared to the combined allowed regions from LSND and MiniBooNE $\bar\nu$
  data (shaded) at 90\% and 99\% CL for (3+1) oscillations. We also show the
  individual regions from LSND and MiniBooNE $\bar\nu$ data.}
\end{figure}

Recently the MiniBooNE collaboration announced updated results of their
search for $\bar\nu_\mu\to\bar\nu_e$
transitions~\cite{AguilarArevalo:2010wv}. 
In the full energy range from 200~MeV to 3~GeV they find an excess of
$43.2\pm22.5$ events over the expected background (the error includes
statistical and systematic uncertainties). 
In the oscillation-sensitive region of 475~MeV to 1250~MeV the
background-only hypothesis has a probability of only
0.5\%~\cite{AguilarArevalo:2010wv}. 
This result is consistent with the
evidence for $\bar\nu_\mu \to \bar\nu_e$ transitions reported by
LSND~\cite{Aguilar:2001ty} if interpreted in terms of neutrino oscillations,
see figure~\ref{fig:mboone}. 
Any explanation of these hints for
$\bar\nu_\mu \to \bar\nu_e$ transitions at the scale of $E/L \sim 1$~eV$^2$
has to satisfy strong constraints from various experiments. 
First, no
evidence for transitions has been found in MiniBooNE neutrino data above
475~MeV~\cite{AguilarArevalo:2007it}. 
This suggests that CP (or even CPT)
violation has to be invoked to reconcile neutrino and anti-neutrino data.
Second, severe constraints exist for $\bar\nu_e$ \cite{Declais:1994su,
Apollonio:2002gd} and $\nu_\mu, \bar\nu_\mu$ \cite{Dydak:1983zq,
AguilarArevalo:2009yj, Adamson:2010wi, Bilenky:1999ny} disappearance at this
scale, which have to be respected by any explanation of the
$\bar\nu_\mu\to\bar\nu_e$ excesses.

The standard approach to the LSND problem is to introduce one or more
sterile neutrinos at the eV scale. 
Adding one sterile neutrino one obtains
the so-called (3+1) mass scheme. 
In this framework there is no CP violation
at short baselines, and disappearance experiments strongly disfavour an
explanation of the appearance signals, see for
example~\cite{Maltoni:2002xd}. 
The latest situation is summarised in
figure~\ref{fig:mboone} (taken from \cite{Akhmedov:2010vy}), which shows
the well known tension in the (3+1) oscillation fit: the region indicated
from LSND combined with MiniBooNE anti-neutrino data (shaded region) and the
constraint from all other experiments seeing no positive signal (blue
curves) touch each other at $\Delta\chi^2 = 12.7$, which corresponds to
99.8\%~CL for 2 degrees of freedom.
If two neutrino mass states at the eV scale are present \cite{Peres:2000ic,
Sorel:2003hf} ((3+2) scheme), the possibility of CP violation opens
up~\cite{Karagiorgi:2006jf}, which allows the LSND and MiniBooNE
neutrino data to be reconciled \cite{Maltoni:2007zf}. 
However, constraints from disappearance
data still impose a challenge to the fit, and the overall improvement with
respect to the (3+1) case is not significant~\cite{Maltoni:2007zf,
Karagiorgi:2009nb}, see \cite{Akhmedov:2010vy} for updated (3+2) results.

Apart from sterile-neutrino oscillations, various more exotic explanations
of the LSND signal have been proposed, among them, sterile-neutrino
decay~\cite{Ma:1999im, PalomaresRuiz:2005vf, Gninenko:2010pr}, CPT
violation~\cite{Murayama:2000hm, Barenboim:2002ah, GonzalezGarcia:2003jq,
Barger:2003xm}, violation of Lorentz symmetry~\cite{Kostelecky:2004hg,
deGouvea:2006qd, Katori:2006mz}, quantum decoherence~\cite{Barenboim:2004wu,
Farzan:2008zv}, mass-varying neutrinos~\cite{Kaplan:2004dq, Barger:2005mh},
shortcuts of sterile neutrinos in extra dimensions~\cite{Pas:2005rb},
sterile neutrino oscillations with a non-standard energy
dependence~\cite{Schwetz:2007cd}, and sterile neutrinos plus non-standard
interactions~\cite{Akhmedov:2010vy,Nelson:2010hz}. 
Some of these proposals
involve very speculative physics, and many of them fail to describe all data
consistently.

Unfortunately the LSND puzzle is far from solved. 
The statistical
significance of MiniBooNE data is not high enough to settle this issue
unambiguously, and recent anti-neutrino data added more confusion to
the topic. 
Moreover, the event excess in the low energy region observed at about
$3\sigma$ in MiniBooNE neutrino data \cite{AguilarArevalo:2007it} provides
another puzzle, which might or might not be related to the LSND one.
Currently only global analyses relying on various different data sets can
address the LSND issue, with disappearance experiments playing a crucial
role. 
In view of future high precision oscillation experiments it would be
highly desirable to understand better whether there is interesting neutrino
physics happening at the scale $E_\nu/L \sim 1$~eV$^2$.

\subsection{Expectations ahead of the Neutrino Factory}

If a decision were to be taken after the completion of the Reference
Design Report (RDR), say, in 2013, it would be necessary to evaluate
the knowledge of the parameters that govern neutrino oscillations at
this time and to assess the scientific risks for the Neutrino
Factory.
In anticipation, therefore, we describe the expectations for the
status of measurements of $\theta_{13}$, leptonic CP violation, and
the mass hierarchy that will pertain after the Interim Design
Report (IDR), but ahead of any decision to initiate the Neutrino
Factory project.

\subsubsection{Between IDR and RDR}

\begin{figure}
\begin{center}
\includegraphics[width=0.85\textwidth]{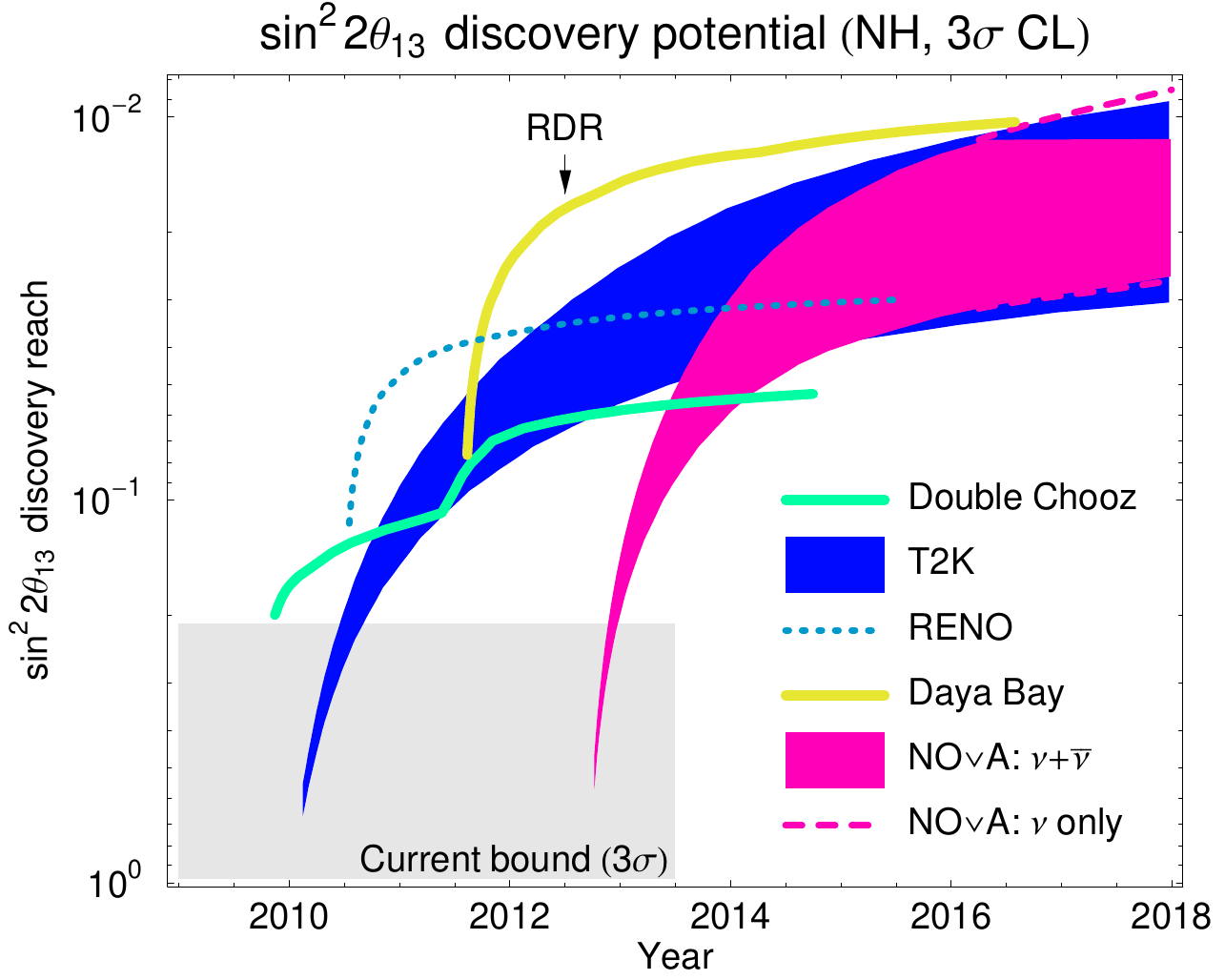}
\end{center}
\caption{\label{fig:evoldisc} Evolution of the $\theta_{13}$ discovery
  potential as a function of time ($3\sigma$~CL), i.e., the smallest
  value of $\theta_{13}$ that can be distinguished from zero at
  $3\sigma$.  The bands reflect the (unknown) true value of
  $\delta$. Normal hierarchy assumed. Figure taken from
  reference \cite{Huber:2009cw}.}
\end{figure}

Here we discuss the likely evolution of measurements of neutrino
oscillations in the period between the completion of the IDR and the
RDR, i.e., the time from now until the end of 2012.  
The most important
question is whether $\theta_{13}>0$ will have been discovered over
this period.  
To address this question, we show in figure
\ref{fig:evoldisc} the discovery potential as a function of
$\theta_{13}>0$ for those experiments in operation or under
construction during 2009 which may provide useful information before
the time that the RDR will be prepared (marked by the arrow).  
It may be
seen from the figure that the key player is expected to be the Daya Bay
experiment, which has a $\theta_{13}$ discovery reach of $\stheta
\gtrsim 0.02$ at the $3 \sigma$ CL ($\stheta \gtrsim 0.01$ at the 90\%
CL).  
Even if the anticipated timescale is not met, it is clear from the
figure that T2K and NO$\nu$A will discover $\theta_{13}$ for the
median $\delta$ within a few years of the publication of the RDR.
Therefore, we assume that around the time of the RDR, we will know
whether $\stheta \gtrsim 0.01$ (``$\theta_{13}$ discovered'') or
$\stheta \lesssim 0.01$ (``$\theta_{13}$ not discovered'').
\begin{figure}
\begin{center}
\includegraphics[width=0.45\textwidth]{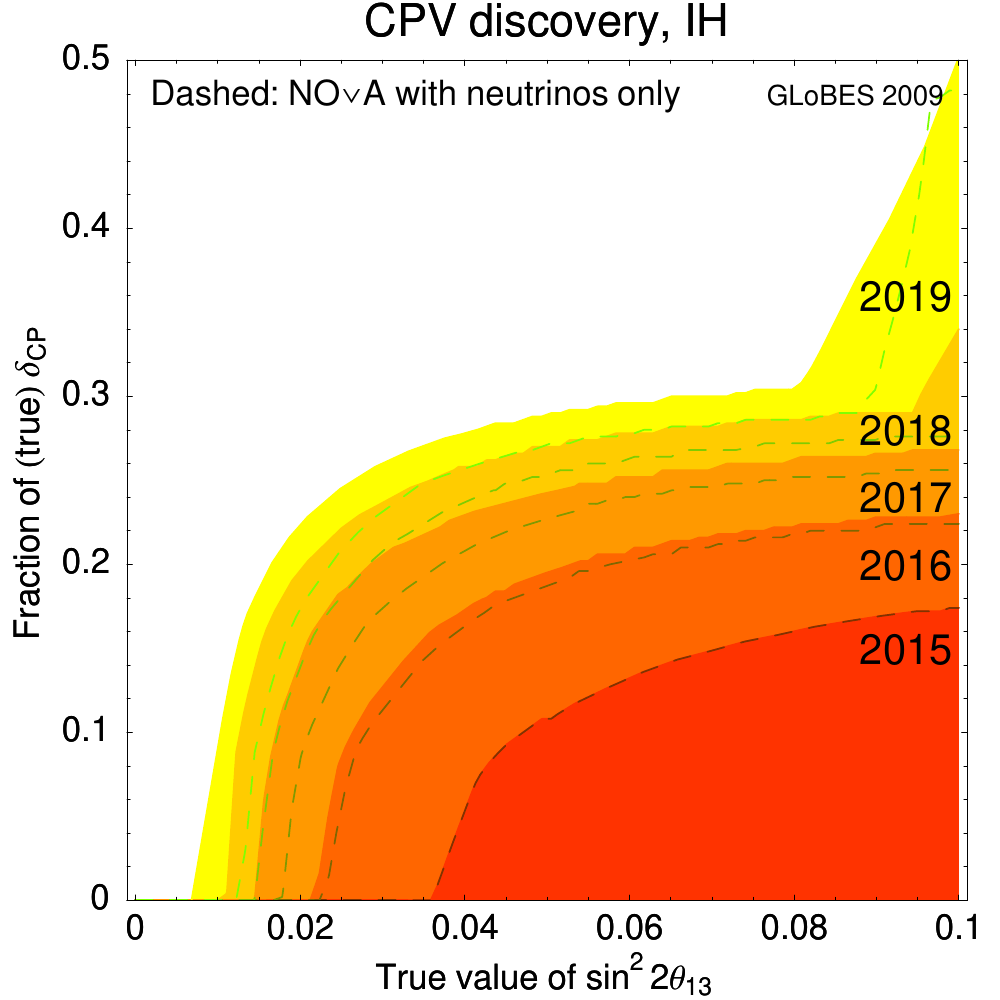}
\includegraphics[width=0.45\textwidth]{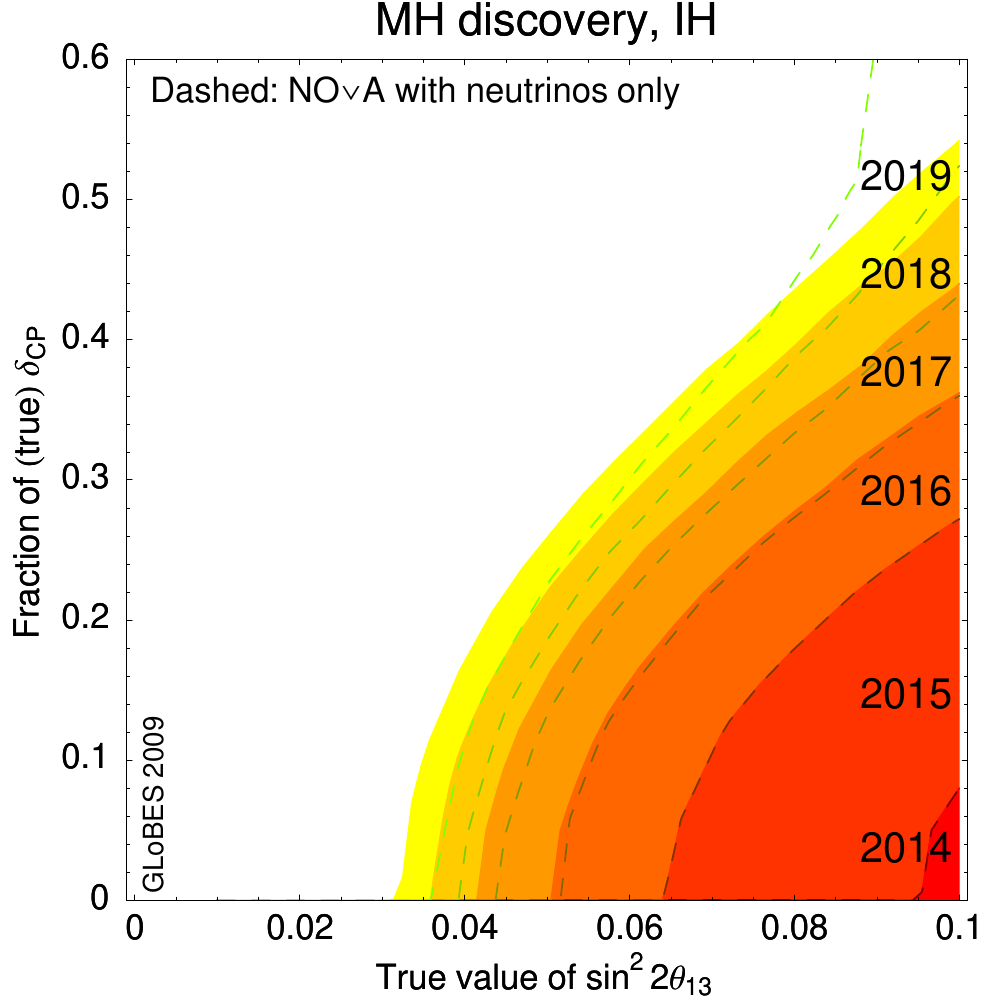}
\end{center}
\caption{\label{fig:tsl} CP violation (left panel) and mass hierarchy
  (right panel) discovery potentials as a function of true $\stheta$
  and fraction of true $\delta$ at the 90\%~CL from T2K, NO$\nu$A
  (reactor experiments, Double Chooz and Daya Bay, are included in the
  evaluation to constrain $\theta_{13}$).  The dashed curves
  refer to NO$\nu$A with neutrino running only, whereas the shaded
  contours refer to the nominal NO$\nu$A neutrino-anti-neutrino plan.
  If no contour is shown for a specific year, there is no sensitivity.
  Note the different scales on the vertical axes. Inverted hierarchy
  assumed. Figure taken from reference \cite{Huber:2009cw}.}
\end{figure}

The most interesting question is whether CP violation and the mass
hierarchy can be measured by the time the RDR is finished.  We show in
figure \ref{fig:tsl} the potential for the discovery of CP violation
(left panel) and the mass hierarchy (right panel) as a function of
$\stheta$ and the fraction of true $\delta$ at the 90\%~CL from T2K,
NO$\nu$A and including data from reactor experiments (Double
Chooz and Daya Bay) to improve the constraint on $\theta_{13}$. 
The figure 
shows that even under optimistic conditions, such as at the 90\% CL,
for very large $\stheta$ close to the current bound, and for the
inverted mass ordering (for which the combined performance is better),
there cannot be any information on these performance indicators before
the RDR even from a combined fit to all data.  Since no next
generation experiment beyond the ones discussed here are under
construction at the moment, we do not expect any information on CP
violation and the mass hierarchy from measurements of neutrino
oscillations before the RDR is prepared.

\subsubsection{Between RDR and data taking at the Neutrino Factory}

After the RDR has been delivered, the precision with which $\stheta$
is known will improve rather slowly.
Indeed, the ultimate precision will only be a factor of two or so
better than the precision achieved at the time of the RDR since
there is no experiment under construction which could improve this
bound significantly (see figure \ref{fig:evoldisc}). 
However, at the time the RDR is prepared, there may be a decision in
favour of an alternative next-generation experiment, such as a
super-beam upgrade, which may start data taking earlier or on a
timescale similar to that of the Neutrino Factory.  
Therefore, we discuss the results to be expected from the experiments
that will be in operation at the time the RDR is prepared and the
results from potential upgrades separately. 
We assume that data taking at the Neutrino  Factory may start as early
as 2020

If $\stheta \gtrsim 0.01$ is discovered before the RDR is delivered,
there is a chance that CP violation may be observed and the mass
hierarchy determined.
This chance is quantified in figure \ref{fig:tsl}. 
If one believes the hint for $\stheta \simeq 0.06$ in reference
\cite{Fogli:2008jx}, CP violation and the mass hierarchy might be
discovered for about 30\% of all values of $\delta$ at the 
90\% CL for the inverted hierarchy. 
Therefore, it is clear that a high-confidence-level discovery in
advance of the start of data taking at the Neutrino Factory 
is extremely unlikely and could only come from a combination of
different data \cite{Huber:2009cw}. 
The next generation of experiments will be needed to confirm
this result and to determine the oscillation parameters with greater
precision.

In summary, if $\theta_{13} > 0$ has not been discovered before
the RDR is prepared, it is quite unlikely to be discovered until data
from the experiments that are now under construction becomes
available. 
Further, the experiments that are either operating or being
constructed now will not be able to access CP violation or the mass
hierarchy for $\stheta \lesssim 0.01$ (see figure \ref{fig:tsl}). 
Therefore, an advanced, high-sensitivity experiment will be needed.

\subsection{The baseline configuration}
\label{sec:henf}

Here we describe the optimisation and performance of the Neutrino
Factory baseline setup.  Two different optimisation strategies are
possible depending on the available information on $\stheta$: if
$\stheta$ is not discovered prior to the start of the Neutrino Factory
project, then the prime goal of a Neutrino Factory would be to
determine $\stheta$ \emph{and} to be able to address the mass
hierarchy and CP violation for the largest range of $\stheta$
feasible.  We will refer to this as the ``small $\stheta$'' case.  On
the other hand, if $\stheta$ has been measured prior to the start of
the Neutrino Factory project, then the goal is to obtain the most
definitive measurements of the CP phase and the atmospheric
parameters.  We will refer to this as the ``large $\stheta$'' case.
Note that, for large $\stheta$, the mass hierarchy would be measured
by NO$\nu$A at less than $3\,\sigma$ for the vast majority of CP
phases and that the precision on $\stheta$ obtained by Daya Bay would
not be likely to be improved by any accelerator-based experiment.  For
practical purposes, the limiting value of $\stheta$ that separates the
large and small $\stheta$ regimes is given by the sensitivity of Daya
Bay which will eventually reach $\stheta\simeq0.01$ at the $3\,\sigma$
level~\cite{Huber:2009cw}.  The Long Baseline Neutrino Experiment
(LBNE)~\cite{LBNE-physics} at FNAL has the potential to improve upon
the sensitivity of Daya Bay by a factor of between 1 and 5.  However,
this limit would be approached only towards the end of the 10 years of
LBNE running and thus would be too late to inform the Neutrino Factory
optimisation process.  For a more detailed discussion of these
possible optimisation strategies see reference \cite{Tang:2009wp}.
The optimisation of the Neutrino Factory has been studied in great
detail, for instance, in references
\cite{Barger:1999fs,Cervera:2000kp,Burguet-Castell:2001ez,Freund:2001ui,Huber:2003ak,Donini:2005db,Huber:2006wb,Gandhi:2006gu,Kopp:2008ds,Agarwalla:2010xx}.
Here we will give a brief summary of the considerations which led to
the IDS-NF baseline setup. All results in this section have been
computed using the General Long Baseline Experiment Simulator (GLoBES)
software package~\cite{Huber:2004ka,Huber:2007ji}.

\subsubsection{Optimisation of a single baseline Neutrino Factory}
\label{sec:opti1}

\begin{figure}[t!]
\begin{center}
\includegraphics[width=0.48\textwidth]{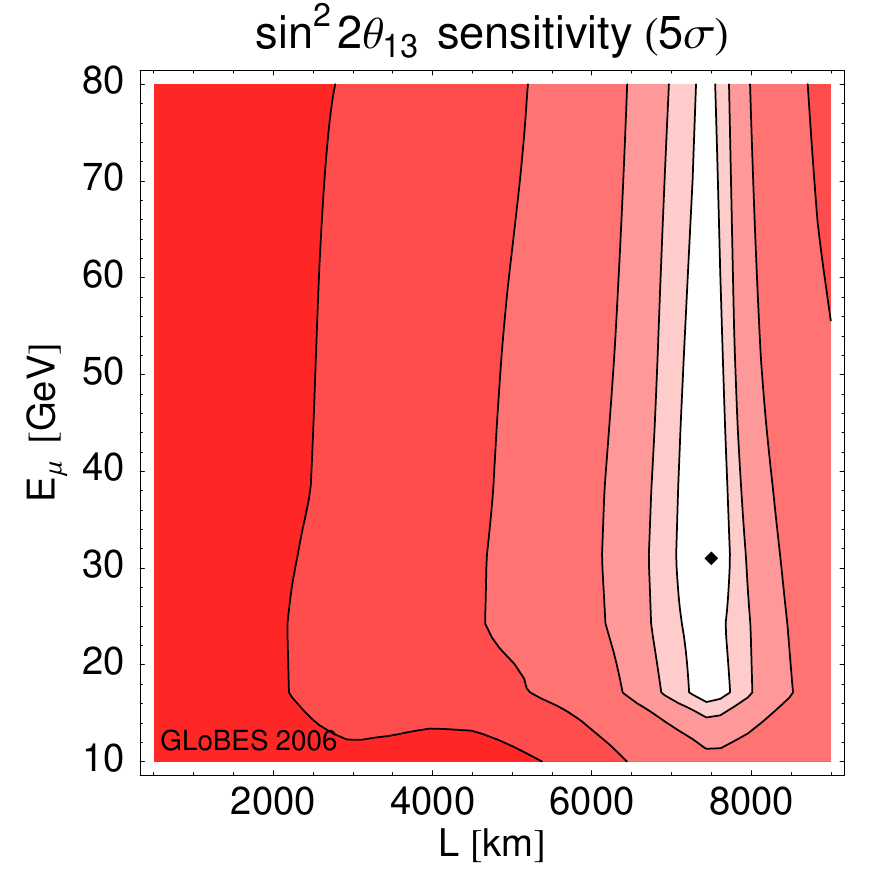} %
\includegraphics[width=0.48\textwidth]{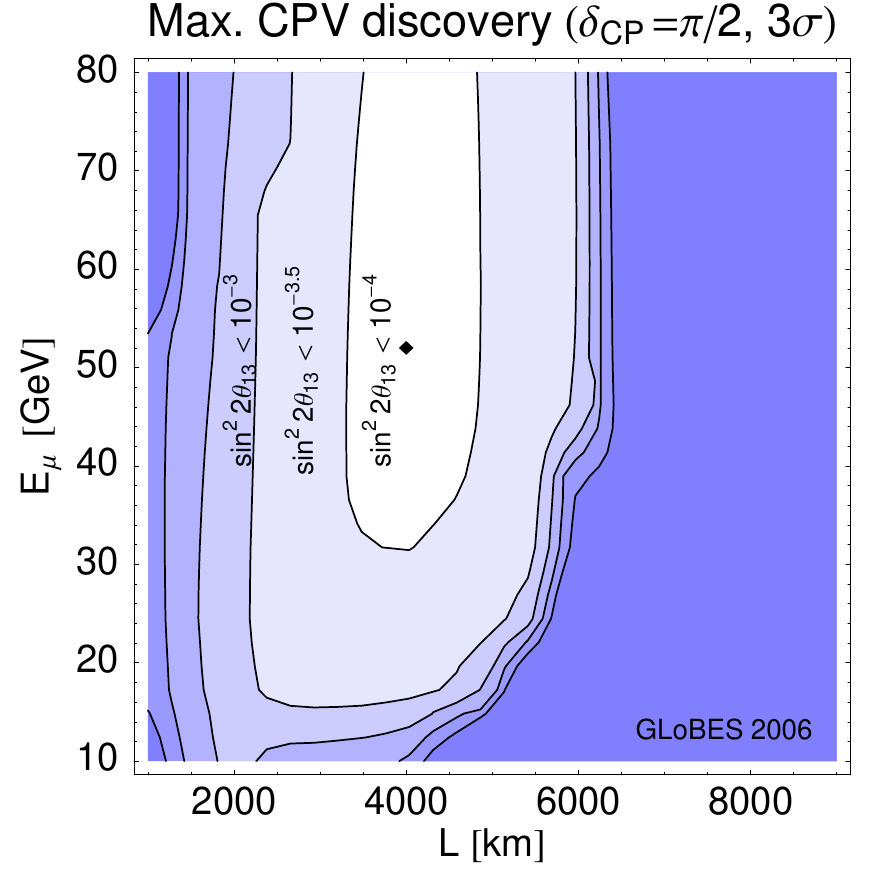}
\end{center}
\caption{\label{fig:elopt} Optimisation of a single baseline Neutrino
  Factory as a function of baseline $L$ and muon energy $E_\mu$. The
  left panel shows the $\stheta$ sensitivity (exclusion limit) with
  the optimum sensitivity $\stheta \simeq 5.0 \times 10^{-4}$ (diamond).
  The contours show the regions within a factor of 0.5, 1, 2, 5, and
  10 of the optimal sensitivity. The right panels shows the
  discovery reach for maximal CP violation, $\delta=\pi/2$, in terms
  of $\stheta$. Figure taken from reference \cite{Huber:2006wb}.  }
\end{figure}

In figure~\ref{fig:elopt} we show the sensitivity reach in $\stheta$
and $\delta$ as a function of the baseline, $L$, and stored-muon
energy, $E_\mu$.  The left panel shows the $\stheta$ sensitivity
(exclusion limit) including correlations and degeneracies while the
right panel shows the CP violation discovery reach for one particular
value of maximal CP violation.  
The results shown in figure \ref{fig:elopt} have been used to optimise
the facility for the largest possible reach in $\stheta$, in other
words, for the small $\stheta$ case.  
The optimum discovery reach in $\stheta$ is for the $L$ and $E_\mu$
combinations in the white regions of the plots; for details, see
reference \cite{Huber:2006wb}.  The excellent sensitivity to $\stheta$
at about 7\,500\,km can be easily understood in terms of the magic
baseline \cite{Huber:2003ak,Smirnov:2006sm}.  At this baseline, which
only depends on the matter-density profile, the effect of the CP phase
vanishes, which results in a clean signal for $\stheta$ and the mass
hierarchy.  The optimum baseline for the measurement of CP-violation
is found for $L$, $E_\mu$ combinations in the white region of the
right panel.  It turns out that, in terms of the $\stheta$ reach, a
baseline between 2\,500\,km and 5\,000\,km is close to optimal.  Therefore,
the IDS-NF baseline setup specifies two long-baseline detectors, one
between 2\,500~km to 5\,000~km and the second at between 7\,000~km
and 8\,000~km.

As far as the muon energy dependence is concerned, there are two
competing factors.  The oscillation maximum is at relatively low
energies in the Neutrino Factory spectrum, therefore a low energy
threshold, has been identified as the main priority for the
optimisation of the Magnetised Iron Emulsion Detector (MIND)
\cite{Huber:2006wb}.
The low energy threshold provided by the MIND (see section
\ref{Sect:DetWG}) allows for a reduction of $E_\mu$ to values as low
as about $\sim 20$~GeV without significant loss of sensitivity
\cite{Huber:2006wb}. 
Therefore, the IDS-NF baseline setup with a MIND detector uses
$E_\mu=25$~GeV.  On the other hand, matter effects play an important
role, such as in the determination of the mass hierarchy.  Assuming an
averaged matter density $\rho=3.4 \, \mathrm{g/cm^3}$, the
matter-resonance energy for a baseline of 4\,000~km is at 9.3~GeV.
One can already see from the left panel in figure \ref{fig:elopt} that
this leads to the requirement that the peak of the neutrino flux
exceeds this energy, which is equivalent to $E_\mu \gtrsim 15$~GeV for
optimal performance at the very long baseline.  Therefore, for the
baseline Neutrino Factory, $E_\mu$ between about 15~GeV and 25~GeV can
be considered optimal, where the lower end of this range is determined
by Earth matter effects, and the upper end by detector performance and
cost.
These results, which are valid for the small $\stheta$ case, are fully
confirmed by a recent analysis \cite{Agarwalla:2010xx} based on the
updated MIND performance as described in section~\ref{sec:MIND}.

On the one hand, the results on the optimisation of a single baseline
serve as a starting point for the optimisation of the two baseline
setup.  As we will discuss later, two baselines are essential to
ensure uniformly robust performance throughout the parameter space.
On the other hand, a single baseline Neutrino Factory may be feasible,
or even appropriate, under certain conditions, for example large
$\theta_{13}$.

A Neutrino Factory with a single baseline of $\sim 1\,000\,\mathrm{km}$
was first proposed in references \cite{lenf1,lenf2} 
and was termed the ``low-energy Neutrino Factory'' (LENF), since the
associated muon beam energies were much lower than the usual $25$~GeV.
At the LENF baseline, the oscillation spectrum at energies below
$5$~GeV is very rich, potentially allowing for very precise
measurements of the unknown oscillation parameters $\theta_{13}$,
$\delta$ and the mass hierarchy.  Since the initial studies
\cite{lenf1,lenf2}, developments of the accelerator and detector
designs have enabled the experimental simulations to be refined and
more detailed optimisation studies to be performed
\cite{LENF,mythesis}.  The key finding of these studies is that,
given sufficiently high statistics and a detector with excellent
detection efficiency at low neutrino energy ($\lesssim 1$\,GeV), an
optimised LENF can have excellent sensitivity to the standard
oscillation parameters, competitive with the usual two baseline, high
energy setup for $\sin^{2}2\theta_{13}\gtrsim 10^{-2}$.

In order to exploit the rich signatures at low energies, alternative
detector types with a lower neutrino energy threshold and a better
energy resolution compared to the MIND are necessary.  Two possible
alternative detector technologies have been considered: either a
20\,kTon totally active scintillating detector (TASD) \cite{ISS} or a
100\,kTon liquid argon (LAr) detector \cite{LAr}, both of which would
need to be magnetised; a challenging requirement.  
These detectors would be capable of detecting and
identifying the charges of both electrons and muons, providing access
to multiple oscillation channels: the $\nu_{\mu}$ ($\bar{\nu}_{\mu}$)
disappearance channels, as well as the golden
($\nu_{e}\rightarrow\nu_{\mu}$ and
$\bar{\nu}_{e}\rightarrow\bar{\nu}_{\mu}$)~\cite{Cervera:2000kp} and platinum
channels ($\nu_{\mu}\rightarrow\nu_{e}$ and
$\bar{\nu}_{\mu}\rightarrow\bar{\nu}_{e}$).  A detailed description of
the assumptions used to simulate the detectors is given in reference
\cite{LENF}.  The main features of the TASD that were assumed were an
energy threshold of 0.5 GeV with a detection efficiency of $94\%$
above 1~GeV and $74\%$ below 1~GeV for $\nu_{\mu}$ and
$\bar{\nu}_{\mu}$ with a background of $10^{-3}$.  For $\nu_{e}$ and
$\bar{\nu}_{e}$ an efficiency of $47\%$ above 1~GeV and $37\%$ below
1~GeV with a background of $10^{-2}$ were assumed.  The main sources of
background were assumed to arise from charge mis-identification and
neutral current events.  The energy resolution was assumed to be
$10\%$ for all channels.  The possibility of using non-magnetised TASD
or LAr detectors was explored in reference \cite{nonmagnetic}; however
the reference LENF setup assumes that magnetisation will be possible.
Recent interest in the LENF has also prompted the authors of reference
\cite{LENF_LENA} to study a setup with a 50~kTon liquid-scintillator
detector.  Each of the alternative detector technologies considered in
the LENF studies is essentially fully active throughout its volume
and therefore offers, for the first time, the possibility of observing
the \emph{platinum} channels in addition to the golden channels.
Indeed, there were early indications that such detector technologies
might be capable of detecting and distinguishing between $e^{-}$ and
$e^{+}$ as well as $\mu^{-}$ and $\mu^{+}$.  It is found that the
complementarity between the platinum and golden channels can be of
great benefit if statistics are limited, improving the sensitivity to
the standard oscillation parameters.  The inclusion of the platinum
channels also turns out to be crucial for resolving degeneracies
between the oscillation parameters and non-standard
effects~\cite{mynsi,mythesis} in the absence of a second baseline.
\begin{figure}[tp!]
  \begin{center}
    \setlength\subfigcapskip{4mm}
    \subfigure[$\sin^22\theta_{13}=10^{-1}$]{\includegraphics[angle=270,width=0.47\linewidth]{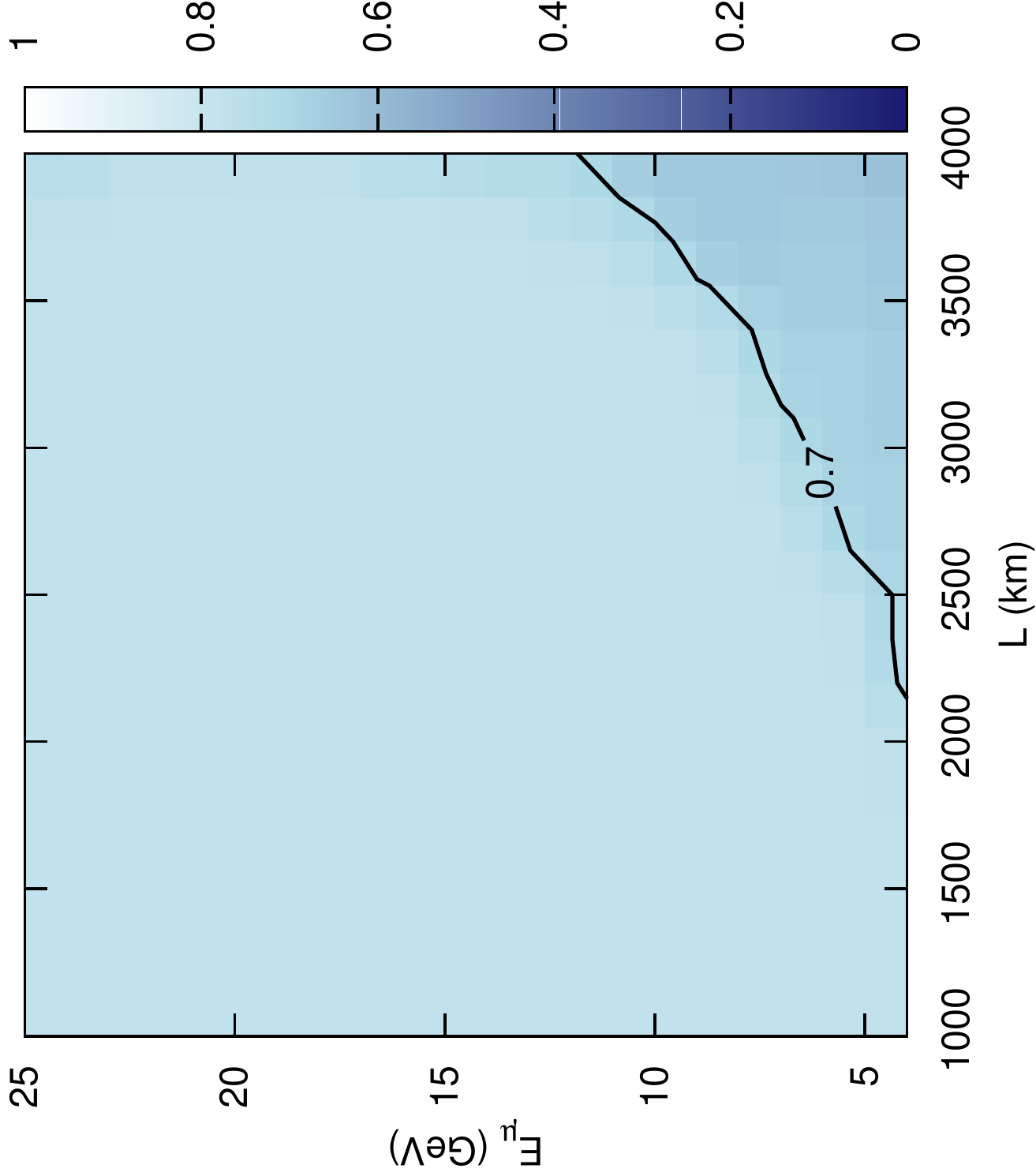}}%
    \hspace{0.04\linewidth}%
    \subfigure[$\sin^22\theta_{13}=10^{-2}$]{\includegraphics[angle=270,width=.47\linewidth]{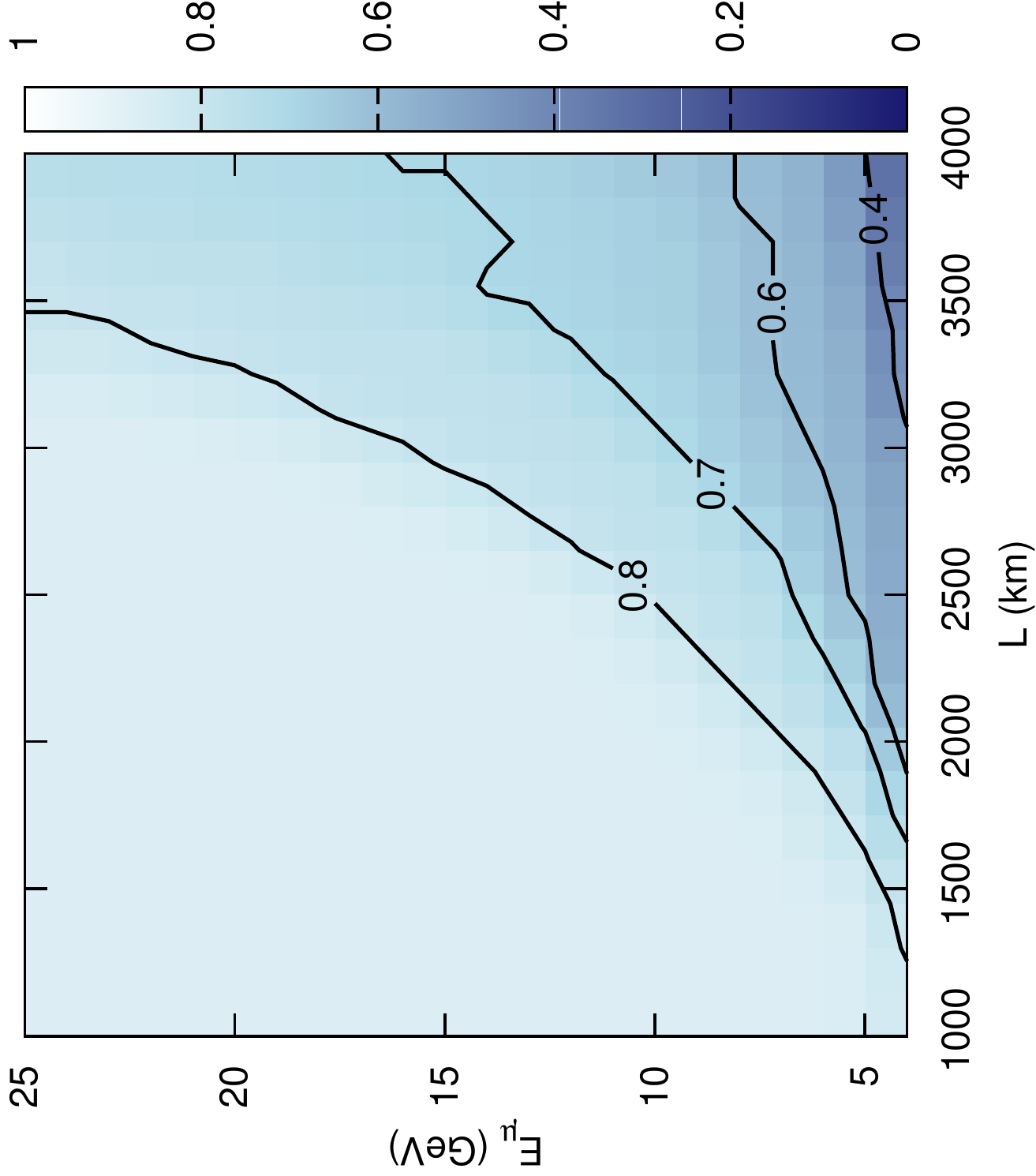}}\\
    \subfigure[$\sin^22\theta_{13}=10^{-3}$]{\includegraphics[angle=270,width=.47\linewidth]{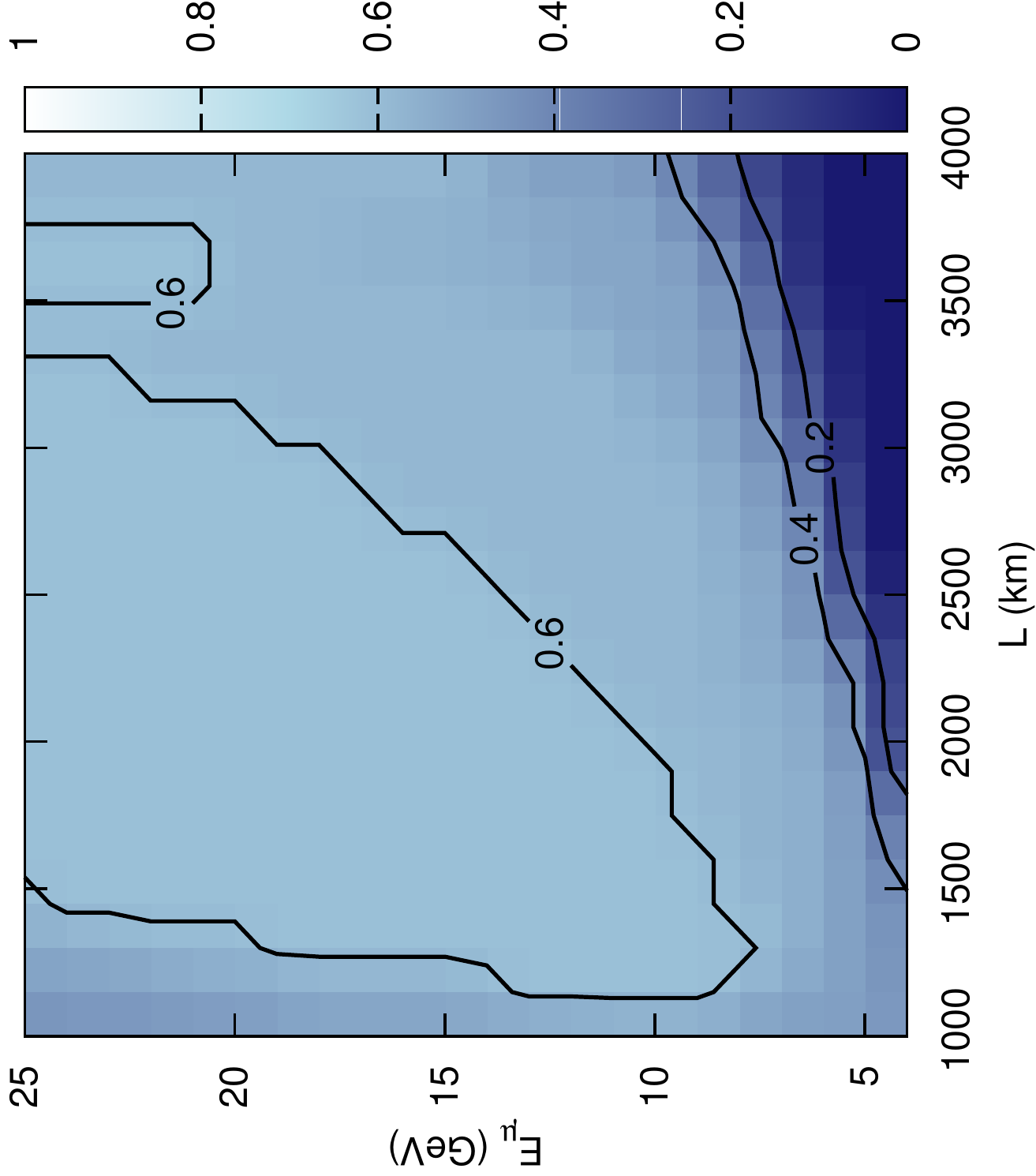}}%
    \hspace{0.04\linewidth}%
    \subfigure[$\sin^22\theta_{13}=10^{-4}$]{\includegraphics[angle=270,width=.47\linewidth]{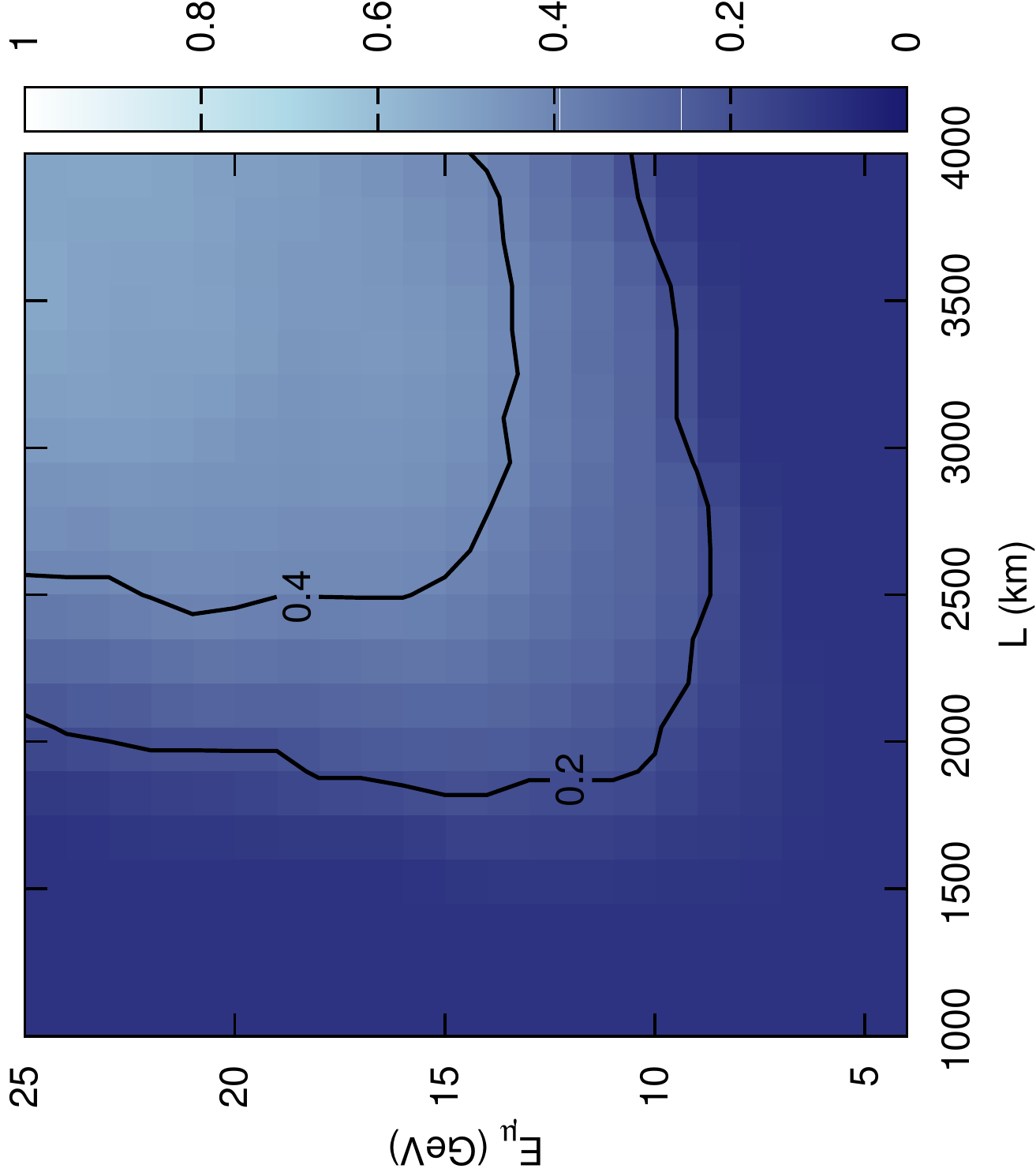}}
  \end{center}
  \caption{CP fraction as a function of baseline, $L$, and stored muon
    energy, $E_\mu$.  Results are shown for a $20\,\mathrm{kTon}$ TASD
    using the default muon intensity of $5\times10^{21}$ $\mu^+$ and
    $5\times10^{21}$ $\mu^-$ for four different values of
    $\theta_{13}$: 
    (a) $\sin^2 2 \theta_{13} = 10^{-1}$;
    (b) $\sin^2 2 \theta_{13} = 10^{-2}$;
    (c) $\sin^2 2 \theta_{13} = 10^{-3}$; and
    (d) $\sin^2 2 \theta_{13} = 10^{-4}$.
  }
\label{fig:plots1}
\end{figure}

In previous work \cite{Pascoli-Li2010, Pascoli2008}, the LENF was
configured with a stored muon energy of between $4$~GeV and $5$~GeV
and a baseline of around $1\,300$ to $1\,500\,$km, with $1\,480\,$km
corresponding to the configuration Fermilab to Henderson mine and
$1300\,$km to the combination Fermilab to Homestake
mine\footnote{Homestake is the host site the proposed Deep Underground
  Science and Engineering Laboratory (DUSEL).}.  
Here we 
focus on optimising a LENF for the purpose of measuring CP-violation.
To this end, we compute the ability, under a range of experimental
scenarios, of ruling out all CP-conserving values of $\delta$. For
each detector, a CP-violation discovery-potential-plot was first
produced for each baseline and muon energy combination.  
Using a Poissonian $\chi^2$ function and marginalising over all
parameters except $\delta$, plots showing the range of values of
``true'' $\theta_{13}$ and $\delta$ for which CP-violation could be
detected at the $3\sigma$ significance were produced.
The simulations assumed a normal mass hierarchy and the hierarchy
degeneracy was taken into account during the marginalisation.  
The simulated oscillation parameters were chosen 
to be $\sin^22\theta_{12}=0.3$, $\theta_{23}=\pi/4$, $\Delta
m^2_{12}=8.0\times10^{-5}\,\mathrm{eV}^2$ and 
$|\Delta m^2_{13}|=2.5\times10^{-3}\,\mathrm{eV}^2$ with an
uncertainty of $4\,$\% and $10\,$\% on the solar and atmospheric
parameters respectively. 
For a few illustrative values of true $\theta_{13}$, the CP fraction
was then calculated for each choice of energy and baseline.  
The CP fraction is defined for each true $\theta_{13}$ as the fraction
of true $\delta$ values which fall inside the CP discovery potential
curve. Using these values, a contour plot was then produced of how the
CP fraction varies as a function of baseline and muon energy for a
given true $\theta_{13}$, as shown in figure \ref{fig:plots1}.

From figure~\ref{fig:plots1}, we see that for large values of
$\theta_{13}$, around $\sin^22\theta_{13}\gtrsim10^{-2}$, CP fractions
of $70\,$\% to $90\,$\% for most energies and baselines can be
obtained. 
The only deviation from this is in the region of the lowest energies
but longest baselines where we see a worse performance at around
$40\,$\% to $80\,$\%.  
This poor performance may partially be explained by
considering the uncertainty in $\theta_{13}$: as this parameter is
being marginalised over, the ability to measure the CP phase is
diminished if $\theta_{13}$ cannot be constrained.  The best
constraints on $\theta_{13}$ are found near the first
oscillation maximum; in the poor-performance region, the majority of
neutrinos have baseline-energy ratios not matching this condition.
This region is also affected by the decrease of the neutrino flux at
large baselines which compounds the poor performance.

\begin{figure}[tp]
 \centering
 \includegraphics[width=0.8\textwidth]{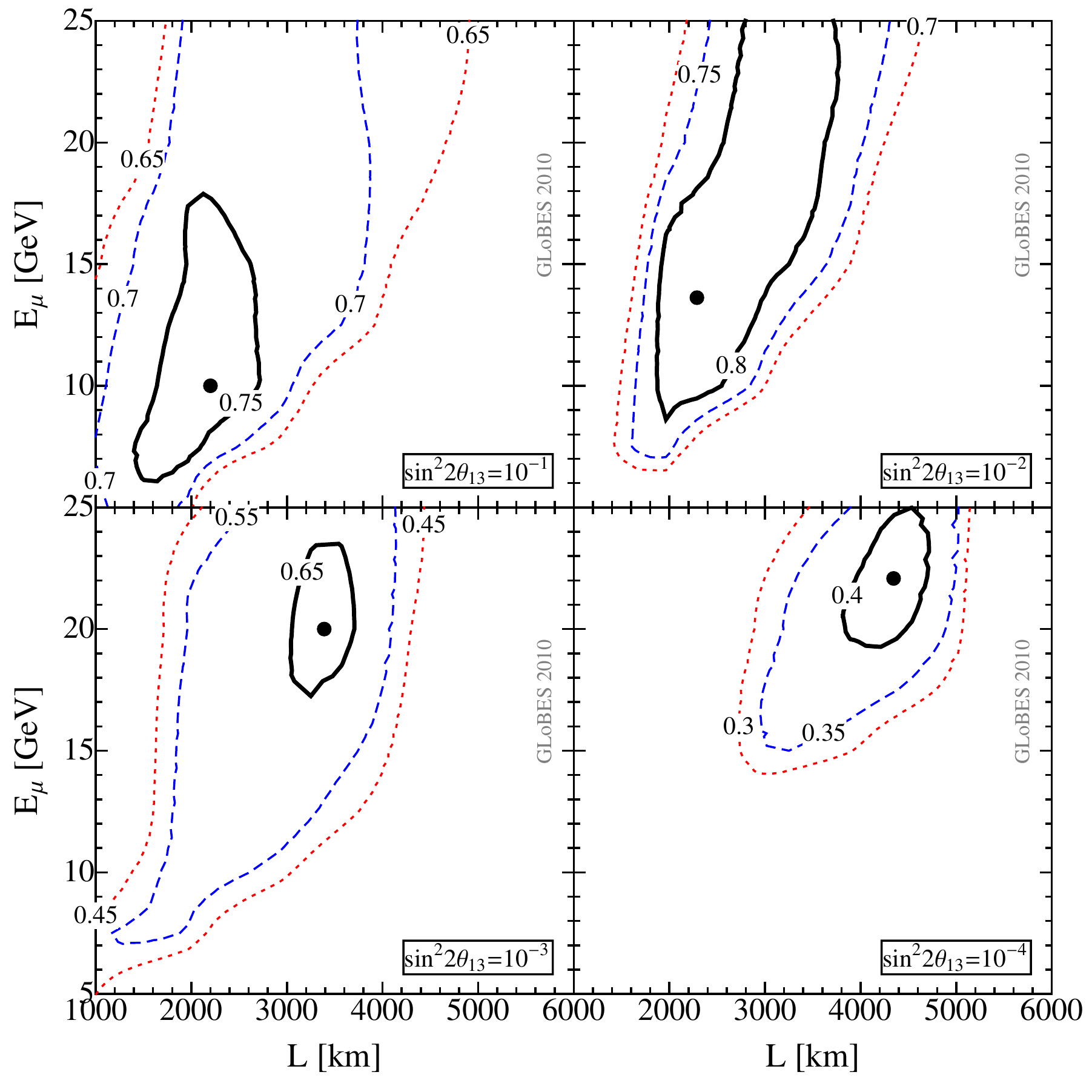}
 \caption{ \label{fig:LvsE} Fraction of $\delta$ for which CPV will
   be discovered ($3 \sigma$ CL) as a function of $L$ and $E_\mu$ for
   the single baseline Neutrino Factory.  The different panels
   correspond to different true values of $\stheta$, as given there.
   Here SF=1 ($2.5\times10^{20}$ muons per year and polarity) is used
   with a 50 kt detector. The optimal performance is marked by a dot:
   (2200,10.00), (2288,13.62), (3390,20.00) and (4345,22.08). Figure
   and caption taken from reference~\cite{Agarwalla:2010xx}.}
\end{figure}

Looking at smaller values of $\theta_{13}$, around
$\sin^22\theta_{13}\approx10^{-3}$ we see a general decrease in CP
fraction to a range of $50\,$\% to $60\,$\% and the region of poor
performance which was mentioned earlier worsens to $0\,$\% to
$40\,$\%. 
We also see another region in which the performance is significantly
reduced; the region of highest-energies but shortest-baselines.  This
drop in performance is attributed to a decrease in statistics: this
region has the smallest baseline-energy ratio which means it is the
furthest away from the oscillation maxima.  When these effects are
combined with the drop in statistics that arises from the lowering of
$\theta_{13}$, experiments in this region start to be severely
compromised by the number of events.  The same behaviour is seen for
the updated MIND detector \cite{Agarwalla:2010xx} and in figure
\ref{fig:LvsE}.

As the true value of $\theta_{13}$ drops to
$\sin^22\theta_{13}\lesssim10^{-4}$, the performance decreases rapidly
and we generally see CP fractions in the range of $0\,$\% to
$50\,$\%.
The two poorly performing regions continue to worsen and are mostly in
the range $0\,$\% to $30\,$\%.  
Eventually, only the regions of $E\gtrsim10$\,GeV and
$L\gtrsim2\,000$\,km have any appreciable CP fractions. 

The high-energy regions of our CP fraction plots overlap with part of
a previous optimisation study that considered the baseline Neutrino
Factory \cite{Huber:2006wb,Agarwalla:2010xx}.  In
figure~\ref{fig:LvsE} we show the same type of result as in
figure~\ref{fig:plots1}, but now for a $50\,\mathrm{kTon}$ MIND detector
corresponding to the one described in~\ref{sec:MIND}.  Qualitatively
the results in figures~\ref{fig:plots1} and~\ref{fig:LvsE} are very
similar and demonstrate that the decision between shorter baseline and
lower energy versus longer baseline and higher energy is more driven
by the magnitude of $\stheta$ than by the detector technology alone.
Although the two experimental simulations have many differences, we
see similar generic features: higher energies require longer baselines
and as $\theta_{13}$ decreases the low energy, short baseline setups
become increasingly disfavoured. 
Also, not surprisingly, when using a MIND
somewhat larger energies at large $\stheta$ are favoured in order to
avoid the effects from the energy threshold of the detector. For large
$\theta_{13}$, the baseline Neutrino Factory has a CP fraction around
$80\,$\%~(see figure~\ref{fig:heperf}) and the LENF has CP fractions
of $70\,$\% to $80\,$\% over a comparable region.  But also, a MIND
detector with 10~GeV muon energy has a similar performance of 80\%,
see figure~\ref{fig:LvsE}.  Thus, low energy setups are preferable at
large $\stheta>0.01$ for \emph{both} MIND and TASD detectors. Any
perceived performance advantage of one detector type with respect to
the other strongly depends on the relative detector sizes and assumed
beam luminosities.

\paragraph*{The platinum channel}

\begin{figure}[tp]
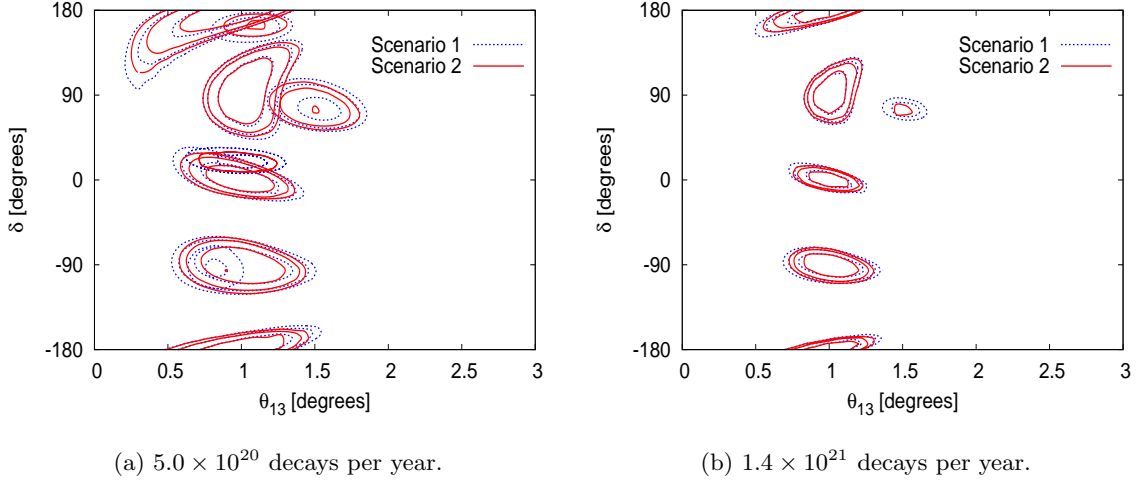

   \centering
   \subfigure[~$5.0\times10^{20}$ decays per year.]{
         \includegraphics[width=7.5cm,height=6cm]{01-PPEG/Figures/./plat_lo_2}}
   \subfigure[~$1.4\times10^{21}$ decays per year.]{
         \includegraphics[width=7.5cm,height=6cm]{01-PPEG/Figures/./plat_hi_2}}
       \caption{~Sensitivity to $\theta_{13}$ and $\delta$ without the
         platinum channel (Scenario 1 - dashed blue lines) and with
         the platinum channel with a $10^{-2}$ background (Scenario 2
         - solid red lines), in the case of a) low statistics
         ($5.0\times10^{20}$ decays per year and b) high statistics
         ($1.4\times10^{21}$ decays per year, significantly larger
         than the present IDS-NF baseline), for
         $\theta_{13}=1^{\circ}$. From reference \cite{LENF}.}
\label{fig:plat}
\end{figure}

The most significant addition relative to the initial LENF studies
would be the inclusion of the platinum channels,
$\nu_{\mu}\rightarrow\nu_{e}$ and
$\bar{\nu}_{\mu}\rightarrow\bar{\nu}_{e}$. 
We find that, if backgrounds
are at a negligibly low level, the platinum channels always give a
significant improvement to the setup. 
However, once realistic
background levels of at least $\sim10^{-2}$ on the platinum channels
are included, much of this benefit is lost. 
There is still some improvement if statistics are limited to below the
$5.0\times10^{20}$ decays per year per polarity assumed in references
\cite{lenf1,lenf2}, but virtually nothing is gained in the case of
higher statistics.
This is illustrated in figure \ref{fig:plat} where the $68\%$, $90\%$
and $95\%$ contours in the $\theta_{13}-\delta$ plane are shown for a
true simulated value of $\theta_{13}=1^{\circ}$ and $\delta=0$,
$\pm90^{\circ}$ and $\pm180^{\circ}$, comparing the results from a
setup without the platinum channel (`Scenario 1' - blue dashed lines)
and a setup which does include the platinum channel with a background
of $10^{-2}$ (`Scenario 2' - solid red lines). 
Figure \ref{fig:plat}a shows the
results in the low statistics case and figure \ref{fig:plat}b the high
statistics case.

\subsubsection{Optimisation of a two baseline Neutrino Factory}
\label{sec:opti2}

\begin{figure}[tp]
  \begin{center}
    \includegraphics[width=\textwidth]{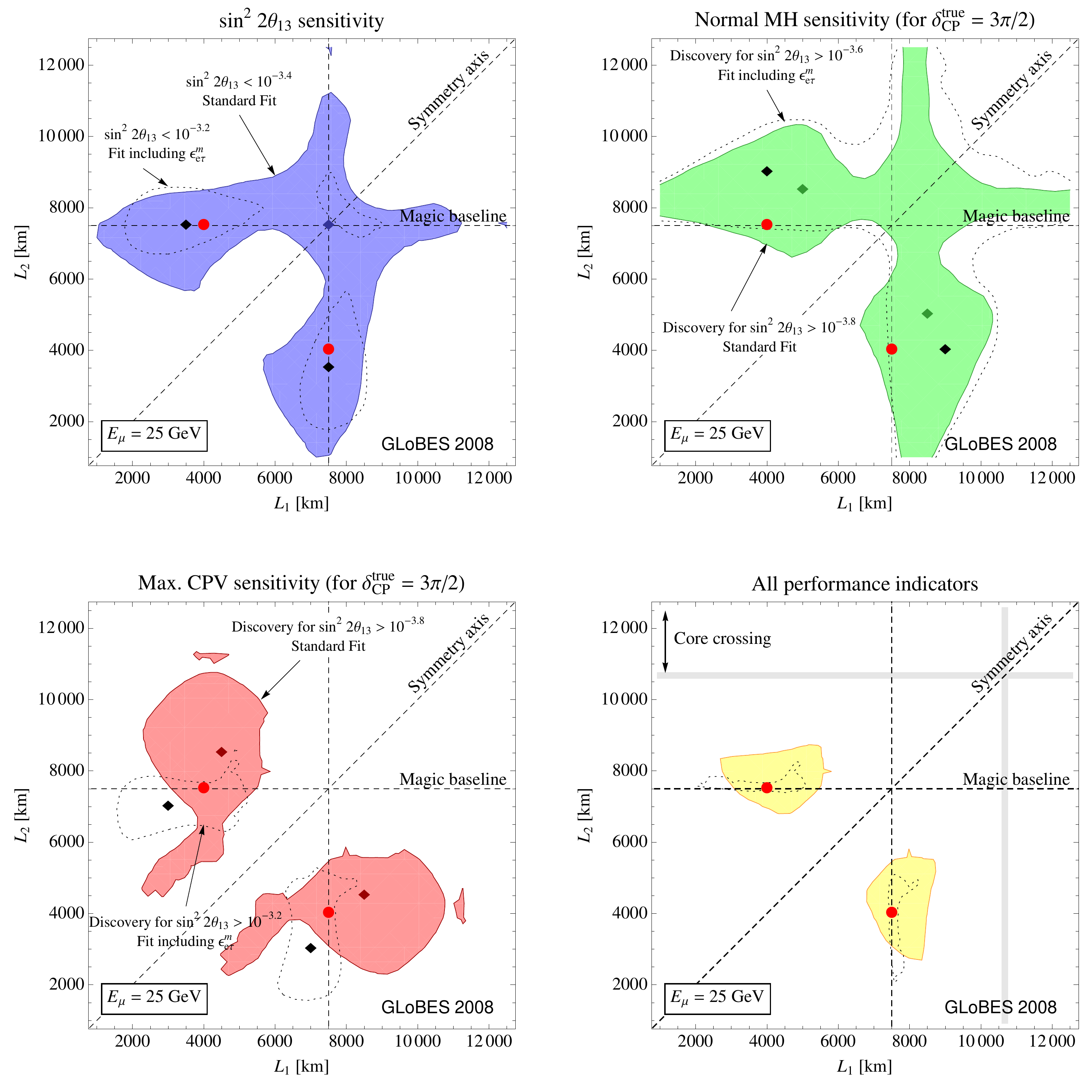}
  \end{center}
  \caption{ \label{fig:allopt} Two-baseline optimisation of the
    Neutrino Factory for the standard oscillation performance
    indicators. The upper left panel shows the (shaded) region where
    the sensitivity to $\stheta$ better than $10^{-3.4}$ ($5 \sigma$),
    the upper right panel shows the (shaded) region where the
    sensitivity to the normal mass hierarchy (MH) is given for all
    $\stheta \ge 10^{-3.8}$ ($5 \sigma$, $\delta_{\rm CP}^{\rm true} =
    3\pi/2$), the lower left panel shows the (shaded) region where the
    sensitivity to maximal CP violation (CPV) is given for all
    $\stheta \ge 10^{-3.8}$ ($5 \sigma$, $\delta_{\rm CP}^{\rm true} =
    3\pi/2$), and the lower right panel shows the intersection of the
    three regions as the shaded region.  The dotted curves have been
    obtained from a fit including a non-standard interaction parameter
    $\epsilon_{e \tau}$ marginalised (for the $\stheta$ ranges given
    in the plots).  The diamonds show the setups with optimal
    sensitivities (coloured/grey for the shaded contours, black for the
    dotted contours), whereas the circles correspond to the {\sf
      IDS-NF} standard choices $L_1 = 4 \, 000$~km and $L_2 = 7 \,
    500$~km. Figure taken from reference \cite{Kopp:2008ds}.}
\end{figure}

The simultaneous optimisation of two Neutrino Factory baselines has
been studied in reference \cite{Kopp:2008ds}, see figure \ref{fig:allopt}
(shaded regions). 
One easily recovers the behaviour discussed above: $\stheta$
(upper left panel) and mass hierarchy (upper right panel) sensitivity
prefer a long baseline, but the combination with a shorter baseline
is not significantly worse.
For CP violation (lower left panel), however, one
baseline has to be considerably shorter. 
If all performance indicators
are combined (lower right panel), the optimum at 4\,000~km plus
7\,500~km is obtained. 
A very interesting study is performed in the dashed
curves: these curves illustrate the effect of a potential non-standard
interaction $\epsilon_{e \tau}$, which is known to harm especially the
appearance channels. 
While the absolute performance for all
performance indicators decreases, the position of the optimum in two
baseline space remains qualitatively unchanged. 
Therefore, the two
baseline setup is not only optimal for standard physics, but also
robust with respect to unknown effects (see also
reference \cite{Ribeiro:2007ud}).

All this optimisation is based on the assumption that the experiment
is optimised in the $\stheta$ direction. On the other hand, it has
been demonstrated in the literature that intermediate values of
$\stheta$ require the magic baseline as a degeneracy resolver (see
reference \cite{Barger:2001yr} for degeneracy issues). Therefore, the
optimisation of the Neutrino Factory does change qualitatively.
However, note that for larger values of $\stheta$, which
may be found in a first low energy phase of the Neutrino Factory,
somewhat shorter baselines than 3000~km can be used; cf., figure 8 in
reference \cite{Tang:2009wp}.

\paragraph*{Robustness of performance}

\begin{figure}
  \begin{center}
    \includegraphics[width=1\textwidth]{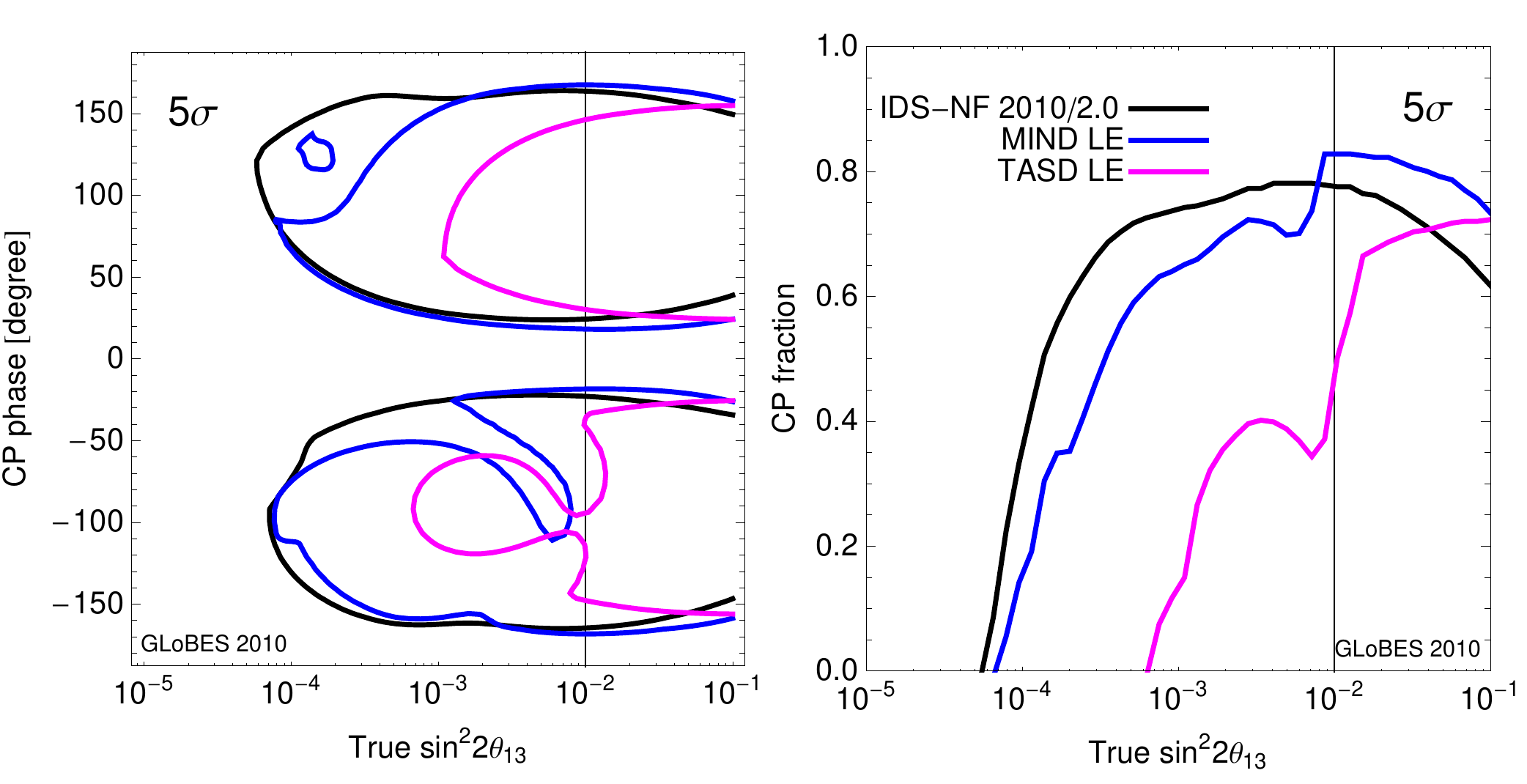}
  \end{center}
  \caption{ 
    CP violation discovery reach (left hand panel) and CP fraction
    (right hand panel) comparing possible single-baseline setups with
    the two-baseline default configuration (IDS-NF 2010/2.0)
    TASD LE stands for a $20\,\mathrm{kTon}$ TASD detector at
    $1\,300$~km baseline with $E_\mu=5$~GeV and MIND LE stands for a
    $100$~kTon MIND at $2\,000$~km with $E_\mu=10$~GeV.  
    Note, that the results are shown at the $5\,\sigma$ confidence
    level.
  }
  \label{fig:robustness}
\end{figure}

Given the excellent performance of single baseline configurations over
a wide range of $\stheta$ values, the question arises whether a
second baseline is strictly required.  The answer is a resounding yes,
as the results shown in figure~\ref{fig:robustness} decisively
demonstrate.  In this figure, we show the sensitivity reach for CP
violation as a function of $\stheta$ in the left hand panel for three
different configurations.  ``IDS-NF 2010/2.0'' refers to a the
reference setup defined in section~\ref{sec:performance}, it employs
two baselines of $4\,000$~km with a $100\,\mathrm{kTon}$ MIND and
$7\,500$\,km with a $50\,\mathrm{kTon}$ MIND.  The muon energy is
$25$~GeV.  ``MIND LE'' uses a single baseline of $2\,000$~km with a
$100$~kTon MIND and a muon energy of $10$~GeV.  ``TASD LE'' uses a
single baseline of $1\,300$~km with a $20$~kTon TASD and a muon energy 
of $5$~GeV.  The MIND performance is described in
section~\ref{sec:MIND} and the TASD performance is taken from
reference \cite{LENF}.  All setups are assumed to run for 10 years
with $10^{21}$ useful muon decays per year summed over both
polarities.  For a two baseline setup, this means that each detector
receives only one half of this flux.  The systematics for all setups
is the same: 1\% on signal and 20\% on background.  The results are
shown at the $5\,\sigma$ level, which helps to emphasise the crucial
difference.  For values of $\stheta>0.01$ (thin vertical line) all
three setups show a qualitatively very similar behaviour, but
nonetheless there is a distinct differences in performance; the 
single-baseline setups perform better at very large values of
$\stheta$.   
For values of $\stheta<0.01$ the picture changes
dramatically and the single-baseline setups suffer from a phenomenon
called $\pi$-transit resulting in large regions of parameter space
where sensitivity is lost.  The reason for the occurrence of the
$\pi$-transit was described in reference \cite{Huber:2002mx}: for
certain combinations of the true values of the CP phase and $\stheta$,
a CP violating true solution can be fitted with the wrong mass
hierarchy and $\delta=\pi$.  The location of the $\pi$-transit depends
sensitively on the baseline and energy, and therefore MIND~LE and
TASD~LE show this effect for slightly different values of $\stheta$,
whereas IDS-NF~2010/2.0, the two baseline setup, is immune against
this effect due the ability of the magic baseline to identify
unambiguously the mass hierarchy.  This is a rather general
conclusion: all single-baseline setups we have studied suffer from the
$\pi$-transit problem due to their relatively weak ability to
determine the mass hierarchy.

We conclude that single-baseline setups, whether they use a MIND or
TASD, all suffer from problems with degeneracies, also, in the context
of new physics as will be discussed later in this section, for values
of $\stheta\lesssim 0.01$. The actual extent of the problem and the
precise boundary in $\stheta$ depend crucially on the detector
performance and available statistics. However, once these factors are
known it is straightforward to derive the limiting values of $\stheta$
above which a given single baseline setup is safe from degeneracies.
In those cases, single baseline setups typically achieve a higher
precision for the CP phase than our standard two baseline
configuration. Hence, for sufficiently large values of $\stheta$, a
single baseline configuration offers a cost-effective way to improve
the ability to measure the leptonic CP phase. Fortunately, the typical
magnitude of the delimiting value of $\stheta$ is within the
sensitivity reach of Daya Bay and therefore, sufficient data for a
informed decision will be available.

\subsubsection{Performance of the Neutrino Factory}
\label{sec:performance}

\begin{figure}
  \begin{center}
    \includegraphics[width=0.3\textwidth]{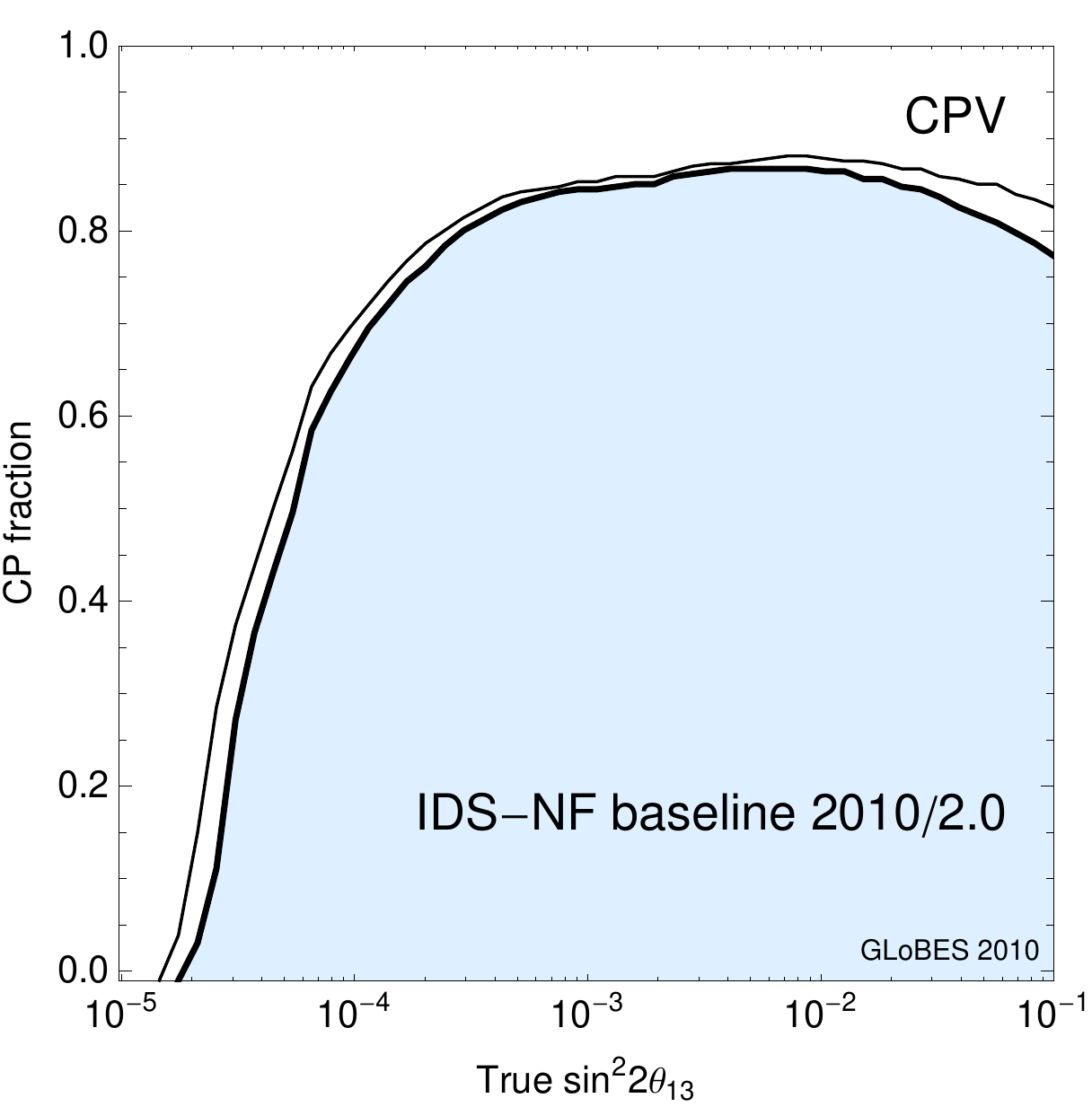}%
    \includegraphics[width=0.3\textwidth]{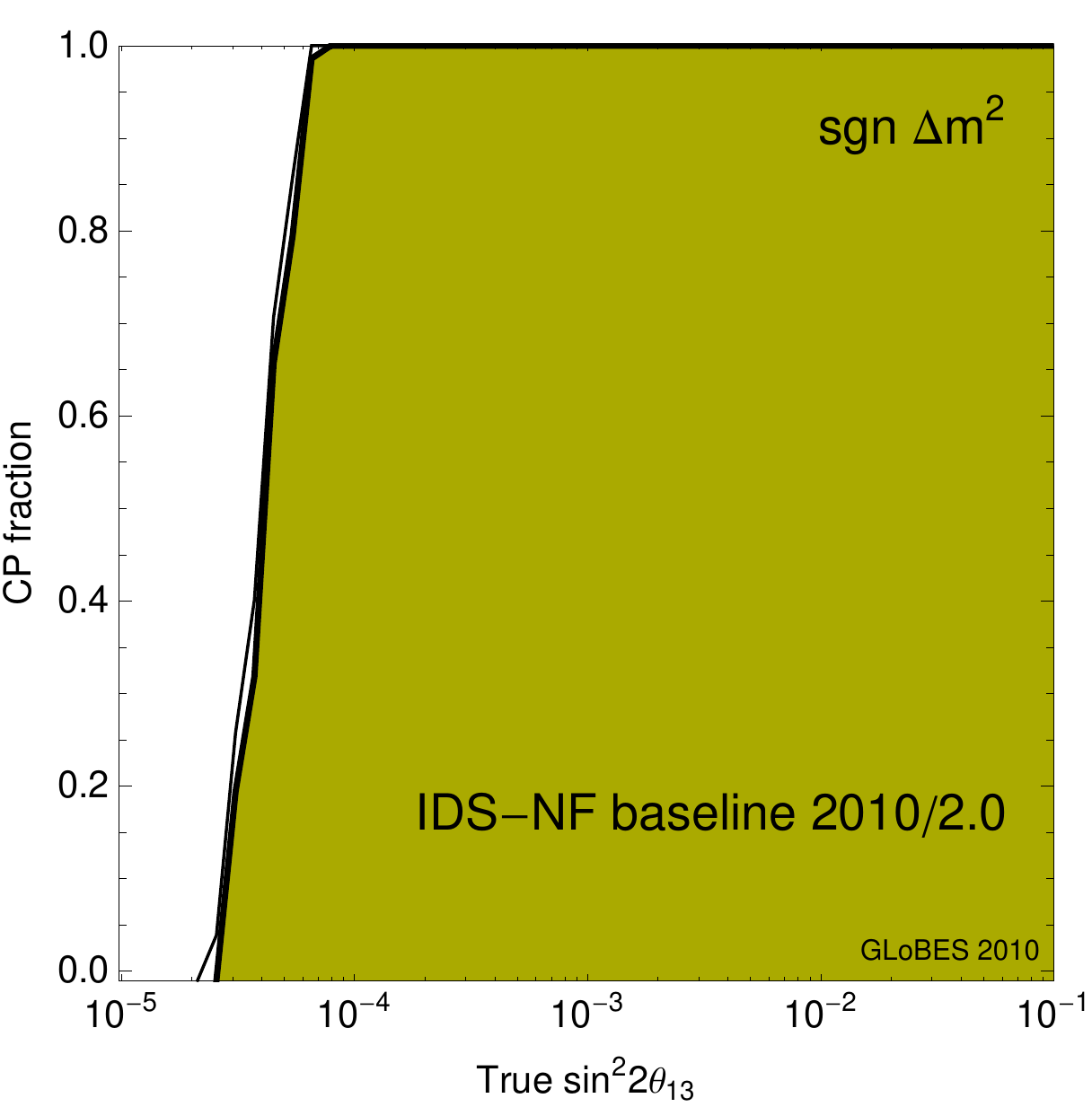}%
    \includegraphics[width=0.3\textwidth]{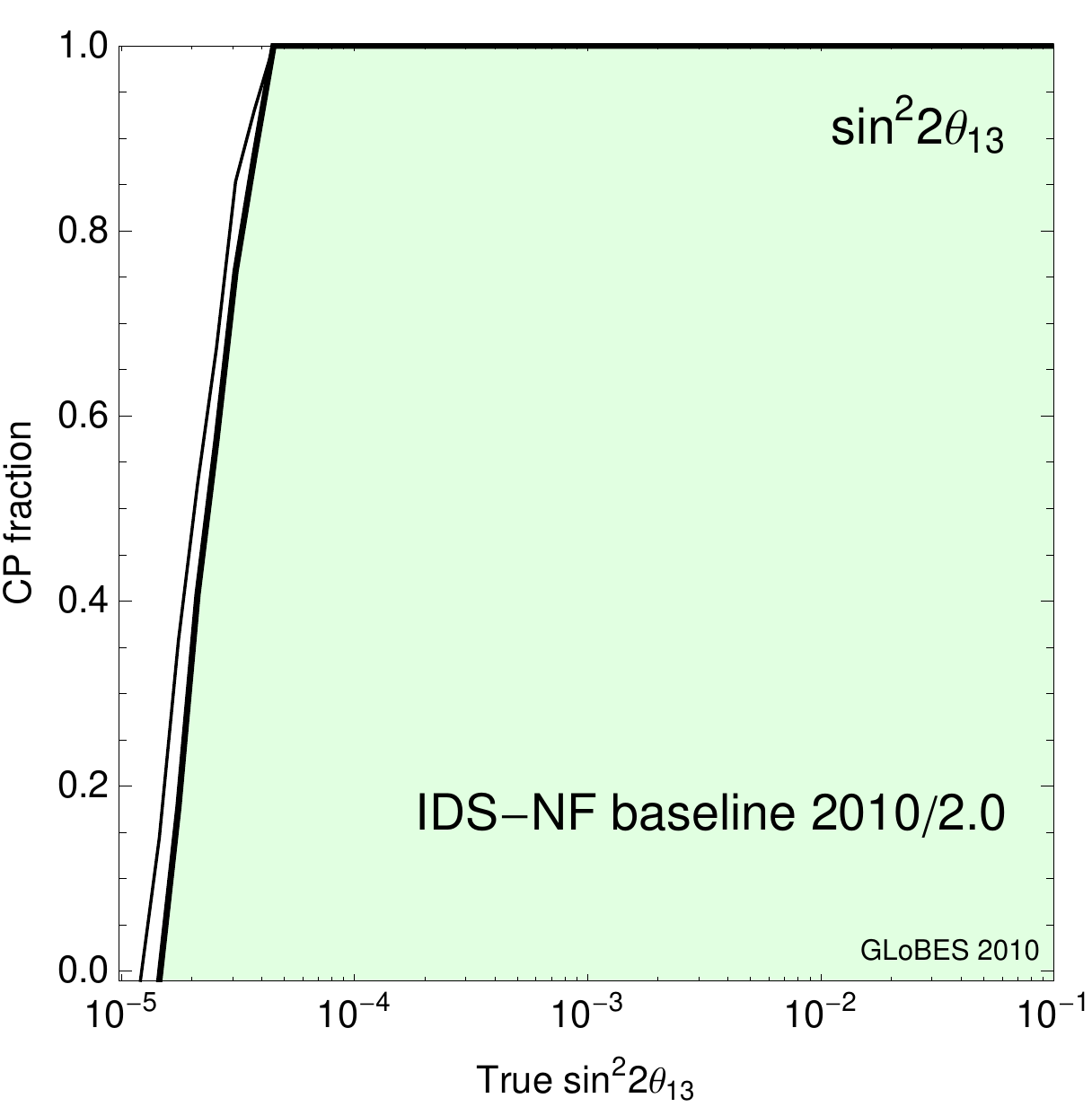}
  \end{center}
  \caption{
    Performance of the baseline Neutrino Factory (100~kt at 4\,000~km
    plus 50~kt at 7\,500~km, updated MIND) in terms of the
    $\theta_{13}$, CP violation, and mass hierarchy discovery reach at
    $3\sigma$ CL. 
    The thick black line assumes default systematics, whereas the thin
    black line is for the case with no beam or detector systematics. 
  }
  \label{fig:heperf} 
\end{figure}

Here we discuss the performance of the high energy Neutrino Factory
for standard oscillation physics.  We show in figure \ref{fig:heperf}
the performance of the high energy Neutrino Factory (100~kt at 4000~km
plus 50~kt at 7500~km) in terms of the $\theta_{13}$, CP violation,
and mass hierarchy discovery reach at $3\sigma$ CL using the updated
MIND performance.  The figure shows that for all performance
indicators discovery reaches for $\stheta$ between $10^{-5}$ and
$10^{-4}$ can be achieved, which corresponds to oscillation amplitudes
about four orders of magnitude smaller than the current bounds.
Non-zero $\theta_{13}$ and the mass hierarchy are discovered almost
independently of the true value of $\delta$.  The CP-violation
discovery reach is very stable over about two orders of magnitude
$10^{-3.5} \lesssim \stheta \lesssim 10^{-1.5}$.  Only for large
values of $\stheta$, the fraction of $\delta$ (leftmost panel)
decreases.  This effect comes from correlations among matter density
uncertainty (see references~\cite{Ohlsson:2003ip}), oscillation
parameters (see, e.g., reference \cite{Huber:2002mx}), and other
systematics.  In addition, the explicit treatment of near detectors
affects this range~\cite{Tang:2009na}.  To illustrate the effect of
beam and detector systematics we show the thin black line, which is
computed for zero beam and detectors systematics.  Even with the
updated MIND the baseline setup is still found to be optimal for
$\stheta<0.01$ and the impact of tau contamination on the standard
performance indicators is found to be
negligible~\cite{Agarwalla:2010xx}.  This issue is discussed in detail
in section~\ref{sec:tau}.

\begin{figure}
  \begin{center}
  \begin{tabular}{cc}
\raisebox{6.3cm}{\includegraphics[angle=-90,width=0.55\textwidth]{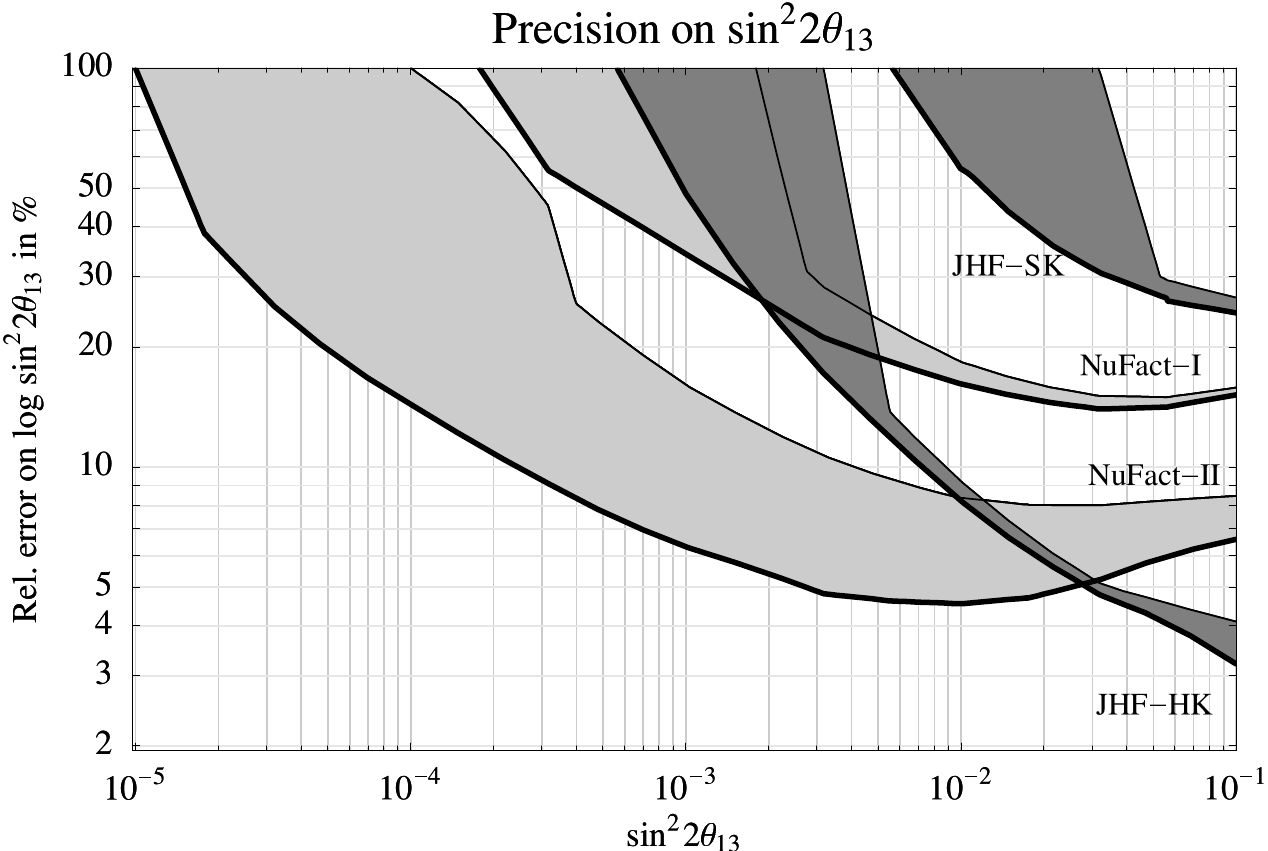}} & 
\includegraphics[width=0.40\textwidth]{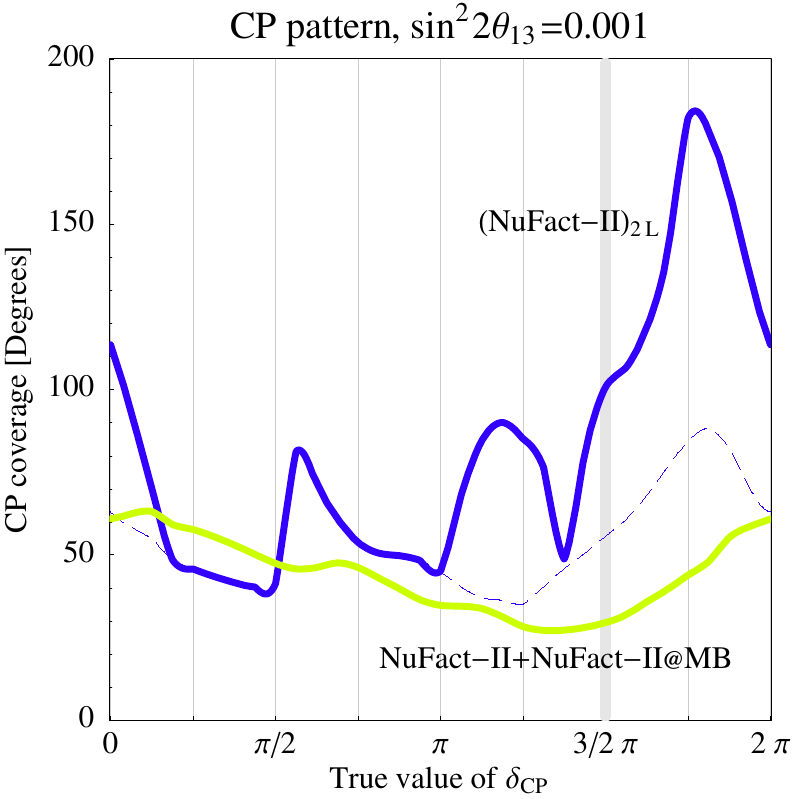} 
\end{tabular}
\end{center}
\caption{\label{fig:cppattern} Performance indicators for the
  precision of $\theta_{13}$ (left panel) and $\delta$ (right panel)
  measurement at the Neutrino Factory. The precision of $\theta_{13}$
  is shown as a function of the true $\stheta$ for a single baseline
  Neutrino Factory ($90\%$\,CL); taken from reference \cite{Huber:2002mx}.
  The bands reflect the unknown true value of $\delta$. The
  precision of $\delta$ (``CP coverage'') is shown as a function of
  the true $\delta$ for a true $\stheta=0.001$ ($3 \sigma$ CL);
  taken from reference \cite{Huber:2004gg}. The different curves correspond
  to a single baseline version with all muons in one storage ring
  (dark/blue solid curve) and a two baseline version (light/yellow
  solid curve). The dashed curve corresponds to not taking 
  degeneracies into account.}
\end{figure}

Apart from the discovery reaches, for which canonical performance
indicators representing the whole parameter space exist, the precision
measurements of $\theta_{13}$ or $\delta$ strongly depend on the true
values of $\theta_{13}$ and $\delta$ themselves.  This is illustrated
in figure \ref{fig:cppattern} for two performance indicators.  In the
left panel, the precision of $\theta_{13}$ is shown as a function of
the true $\stheta$ for a single baseline Neutrino Factory
``NuFact-II''.  The bands reflect the unknown true value of $\delta$.
One can easily see from figure \ref{fig:cppattern} that different
experiments can be compared in this representation.  In the right
panel, the precision of $\delta$ (``CP coverage'') is shown as a
function of the true $\delta$.  Note that a CP coverage of 360$^\circ$
means that all values of $\delta$ fit the chosen true value, i.e., no
information on $\delta$ can be obtained (for details, see reference
\cite{Huber:2004gg}).  The maximum in the figure of 180$^\circ$ means
that about 50\% of all values of $\delta$ can be excluded.  Comparing
the two solid curves, one can easily see that the magic baseline
combination acts as a risk minimiser: the curve becomes relatively
flat with a $3\sigma$ (full width) error of 30$^\circ$ to 60$^\circ$,
corresponding to a $1\sigma$ (half width) error of 5$^\circ$ to
10$^\circ$.  This corresponds to a precision similar to the quark
sector.  Note that the right panel is shown for a particular true
value of $\stheta$.  
Another performance indicator that is often used is obtained by
fitting for $\theta_{13}$ and $\delta$ simultaneously at several
selected sets of true $\{ \theta_{13}, \delta \}$.  
While this result corresponds most closely with what would be expected
from the Neutrino Factory, the analysis has yet to be repeated for the
revised Neutrino Factory baseline presented here.
Figure \ref{fig:cppattern} is therefore included to provide examples
of the precision with which $\theta_{13}$ and $\delta$ would be
measured at representative facilities. 

\begin{figure}
\begin{center}
\includegraphics[width=\textwidth]{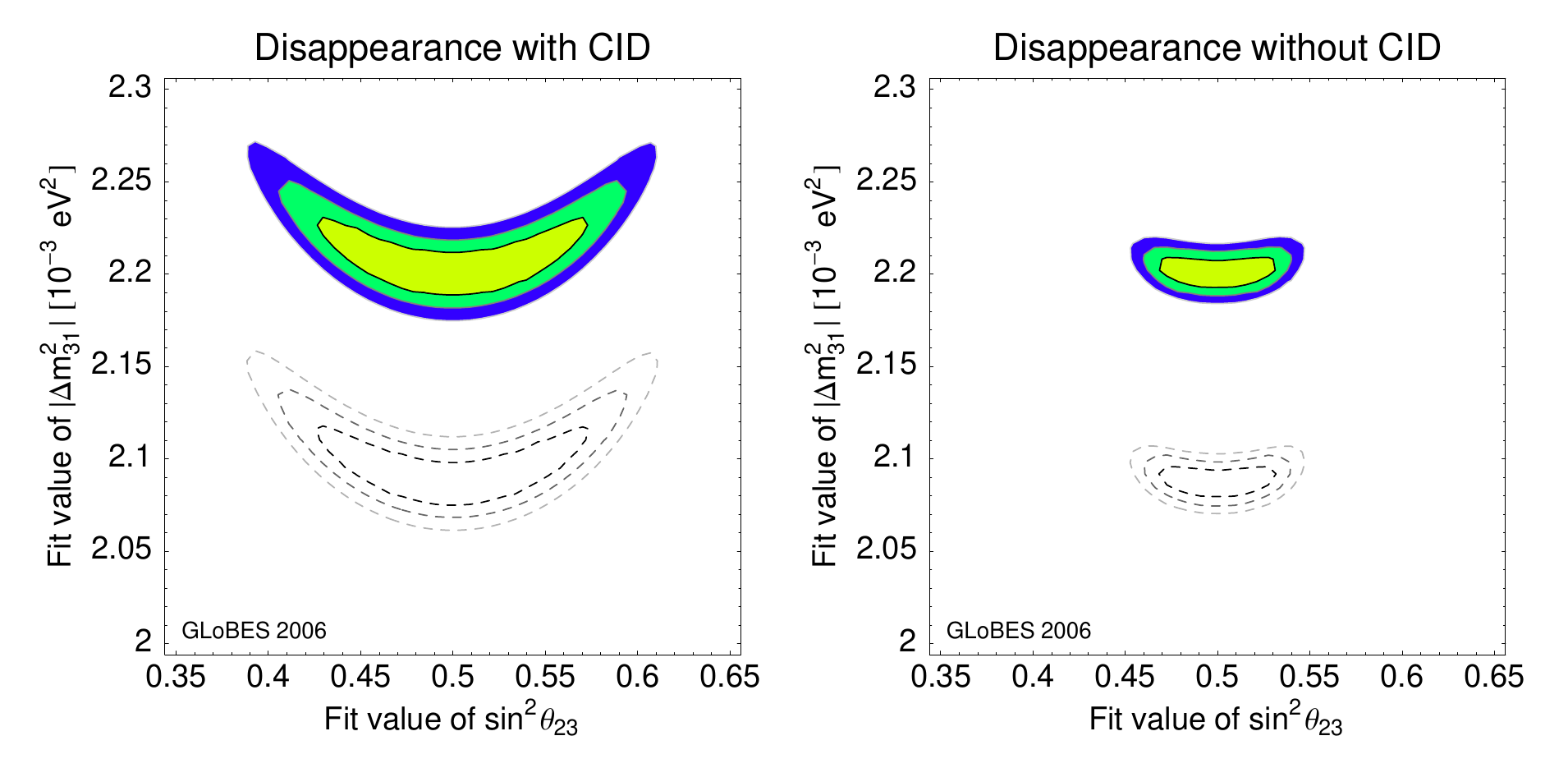}
\end{center}
\caption{\label{fig:cid} 
  Comparison of $\ldm$-$\theta_{23}$-precision in an analysis with
  charge identification (CID, left panel) and without CID (right
  panel) in the disappearance channel including all correlations
  ($1\sigma$, $2 \sigma$, $3\sigma$, 2 d.o.f., $\stheta=0$). 
  The appearance
  information is added as usual with CID. Dashed curves correspond to
  the inverted hierarchy solution. Figure taken from reference
  \cite{Huber:2006wb}.} 
\end{figure}

There are other measurements in standard oscillation physics which can
be performed at the Neutrino Factory and which are not related to the
performance indicators discussed above. 
For example, we show in figure \ref{fig:cid} the
combined precision in $\ldm$ and $\theta_{23}$ for maximal atmospheric
mixing. The different shaded and dashed regions correspond to the
normal and inverted hierarchy fit (for the true normal hierarchy),
which are separated by $2 \cdot \Delta m_{21}^2 \cdot \cos^2
\theta_{12}$ (see, e.g., reference \cite{deGouvea:2005mi}). Obviously, one
can measure $\ldm$ with a precision better than $\sdm$.
The splitting of the measured region is a relic of choosing 
$\Delta m_{31}^2$ instead of an appropriate intermediate value between
$\Delta m_{31}^2$ and $\Delta m_{32}^2$, which results in a different
mass-squared splitting for normal and inverted hierarchy for fixed
$|\Delta m_{31}^2|$. 
It does not mean that one can measure the mass hierarchy using this
effect \cite{deGouvea:2005mi}.
By comparing the left and right panels, it can be seen that it is
important to use a data sample without charge identification (CID) for
the analysis of the disappearance channels, because the overall
efficiency and threshold are much better and the CID background is
comparatively small \cite{Huber:2006wb}. 
In reference \cite{Huber:2006wb} it also has been demonstrated that
the very long baseline clearly helps to constrain $\theta_{23}$ even
further.

Other potential measurements within the S$\nu$M framework
at the Neutrino Factory include deviations from maximal
mixings~\cite{Antusch:2004yx}, the high-confidence check of the MSW
effect in Earth matter, even for $\stheta=0$, by the solar term of the
appearance channel~\cite{Winter:2004mt}, the percent-level
determination of the average matter density along the
baseline~\cite{Winter:2005we,Minakata:2006am,Gandhi:2006gu}, and the
mass hierarchy determination if
$\stheta=0$~\cite{deGouvea:2005hk,deGouvea:2005mi}. For these
measurements, a very long baseline is typically the key ingredient.

\subsubsection{The problem of $\tau$-contamination}
\label{sec:tau}

The problem of $\tau$-contamination was first studied in reference
\cite{Indumathi:2009hg} in the context of precision measurements of
the atmospheric parameters $(\Delta m_{32}^2,\theta_{23})$ using the
$\nu_\mu \to \nu_\mu$ channel at a Neutrino Factory.  It was shown in
section \ref{sec:performance}, figure \ref{fig:cid}, that it is useful
to add all muons in the final state without CID.  The improved
efficiency in the low-energy part of the neutrino spectrum, however,
has the drawback that a previously irrelevant background now becomes
potentially harmful.  Oscillations into $\nu_\tau$, unless suppressed
by low efficiency at low energy, enhance both the right- and
wrong-sign muon samples.  In previous analyses, the poor efficiency of
the detector below 10~GeV allowed $\tau$-contamination to be neglected
\cite{Cervera:2000vy,Cervera:2000kp}.  In the case of the detection of
the platinum channel, $\tau$-contamination also affects the electron
sample.  However, detectors capable of measuring the platinum channel
are usually assumed to be capable of distinguishing the electron from
tau decay from the true platinum-electron sample.  Oscillations of
$\nu_e, \nu_\mu \to \nu_\tau$ will produce $\tau$s through $\nu_\tau
N$ CC interactions within the detector that will, eventually, decay
into muons (approximately 17\% of them). These muons from taus will,
therefore, ``contaminate'' the ``direct'' muon samples (coming from
$\nu_e,\nu_\mu \to \nu_\mu$ oscillations). Notice that muons from taus
are not background but as good a signal as the direct muons.  Muons
from $\tau$-decay accumulate in low-energy muon bins, since the
associated neutrinos produced in $\tau$-decay result in a
``secondary'' muon which has, on average, 1/3 of the $\tau$
energy \cite{Indumathi:2009hg}.

It is very hard to remove the muons from tau decay using kinematic
cuts in a MIND-like detector.  Any cuts that attempt to do so
drastically reduce the direct muon events as well and hence worsen the
sensitivity to the oscillation parameters.  They escape essentially
all filters designed to kill the dominant backgrounds and directly add
to the direct muon sample, see reference \cite{Indumathi:2009hg}.  On
the other hand, neglect of the tau contribution will lead to an
incorrect conclusion about the precision achievable at the Neutrino
Factory on a given observable.  The ``$\tau$-contamination'' must be
added to the signal and it must be studied together with it.

\paragraph{Impact on the atmospheric parameters measurement}
\label{sec:rightsigntaus}

\begin{figure}
  \begin{tabular}{cc}
    \includegraphics[width=0.50\textwidth]%
      {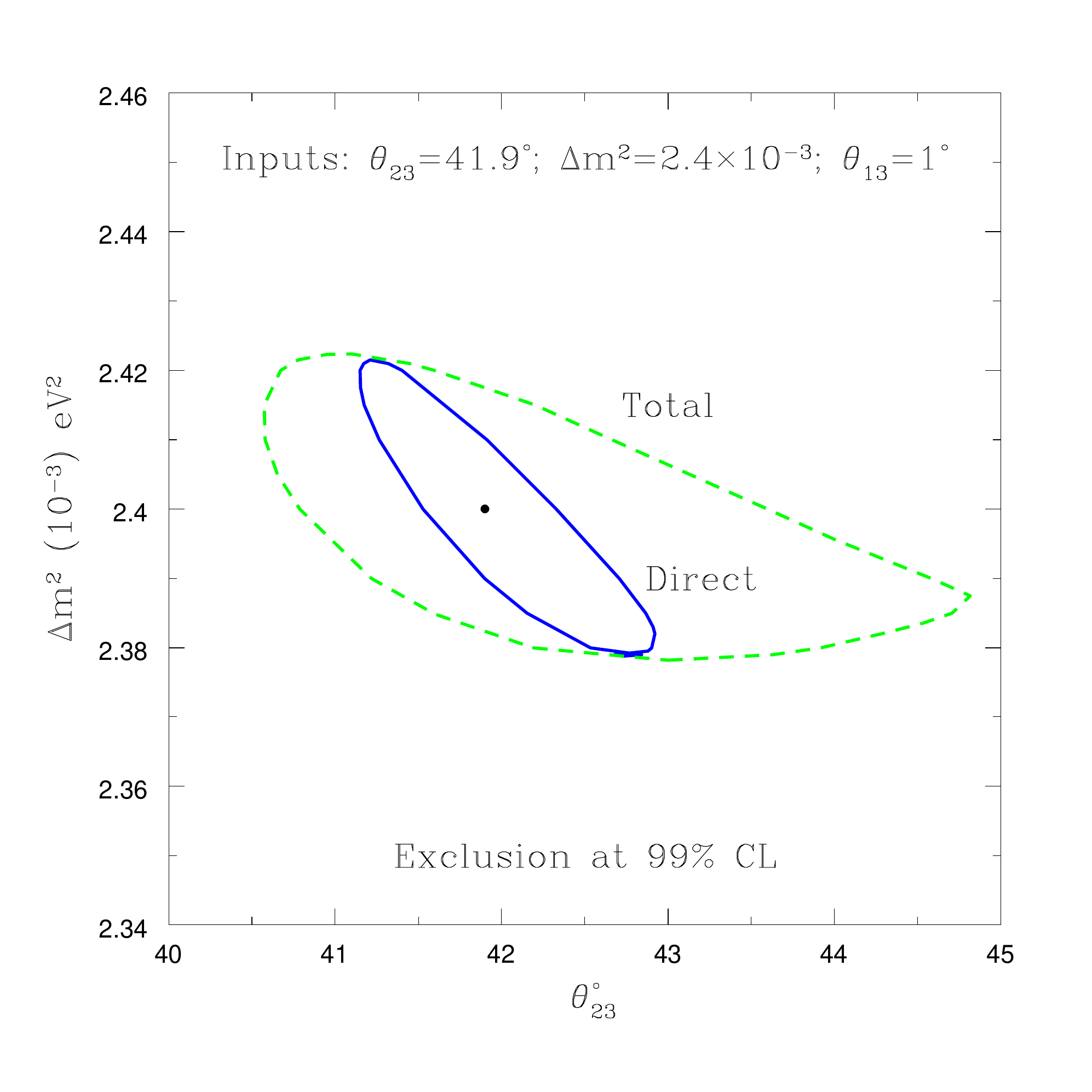} & 
    \includegraphics[width=0.50\textwidth]%
      {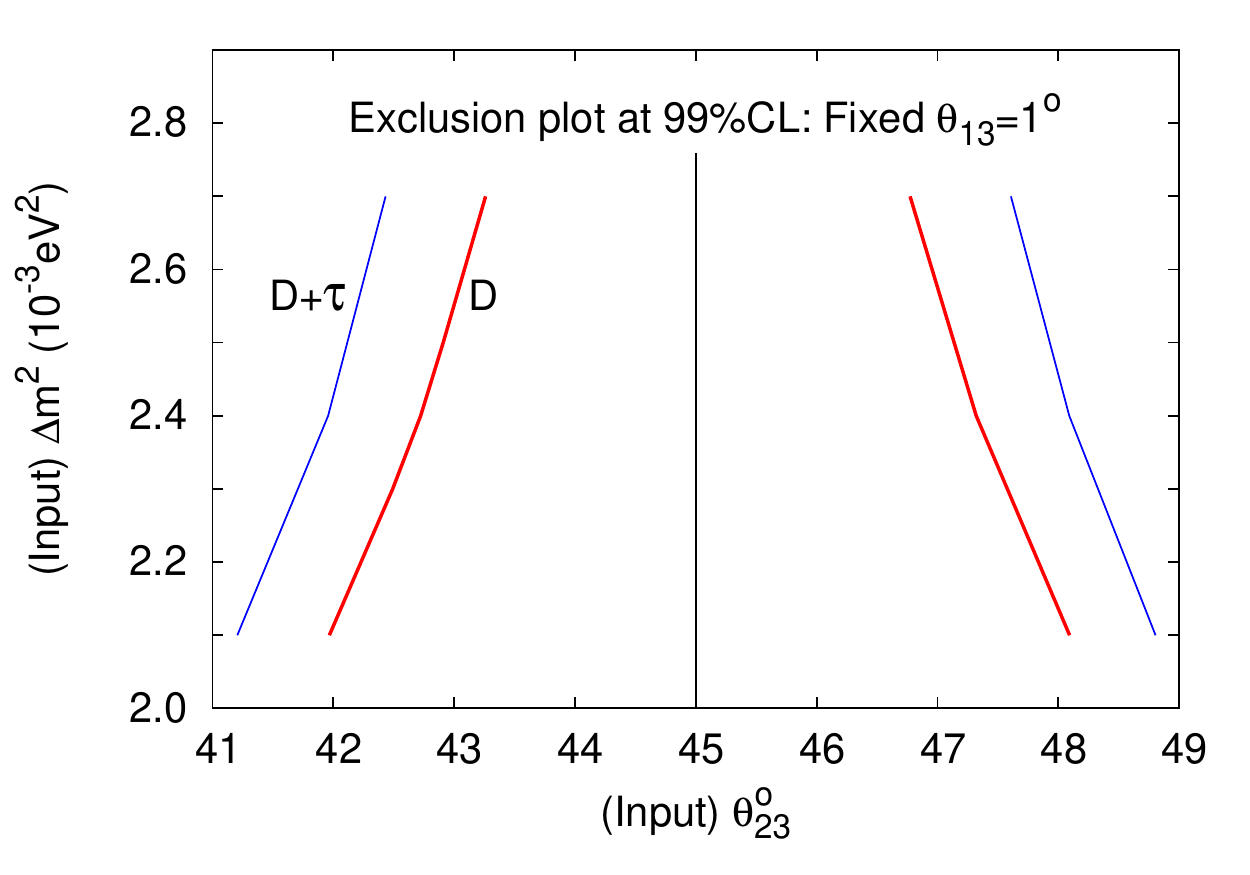} 
  \end{tabular}
  \caption{ Left: 99\% CL contours in $\Delta m^2$--$\theta_{23}$ from
    CC events directly producing muons (solid line) and with the
    inclusion of muons from tau decay as well (dashed line). The input
    parameters are: $\Delta m^2 = \Delta m^2_{31} = 2.4\times 10^{-3}$ eV$^2$,
    $\theta_{23} = 41.9^\circ$.  From reference
    \cite{Indumathi:2009hg}.  Right: The largest (smallest) true value
    of $\theta_{23}$ in the first (second) octant that can be
    discriminated at 99\% CL from $\theta_{23} = \pi/4$ as a function
    of $\Delta m^2$ when only Direct ($D$) and total ($D + \tau$)
    events are considered.  In both panels, $\theta_{13}$ has been
    fixed to $1^\circ$ and $\delta = 0$.  From
    reference~\cite{Indumathi:2009hg}.  }
  \label{fig:contt13fix}
\end{figure}

The fit of the atmospheric parameters for a typical sample set of
inputs 
($\Delta m^2 = \Delta m^2_{31} = 2.4\times 10^{-3}$ eV$^2$, 
$\theta_{23} = 41.9^\circ; \theta_{13} = 1^\circ; \delta = 0$) is
shown in figure \ref{fig:contt13fix}\,(left), from reference
\cite{Indumathi:2009hg}. 
The contours represent the allowed region in $\Delta
m^2$--$\theta_{23}$ parameter space at 99\% CL ($\Delta \chi^2 =
9.21$).  The solid line corresponds to direct muon events alone and
the dashed line to the total right-sign muon sample, including those
from tau decay.  It can be seen that the allowed region is much more
constrained with direct muons only than when including muons from
taus.  In particular, the inclusion of the tau contribution worsens
the precision with which $\theta_{23}$ and its deviation from
maximality can be measured: a spread of $\sim 2^\circ$ if tau events
are removed turns into $\sim 4.5^\circ$ when they are included.  The
reason is the following: since the $\theta_{23}$-dependent terms come
with opposite sign in $P_{\mu\mu}$ and $P_{\mu\tau}$, and the
statistics of $\tau$-events is significant, the combination of muons
from direct production and from tau decays marginally decreases the
sensitivity of the event rates to this angle.  On the other hand, the
$\tau$-contamination does not affect the determination of (the modulus
of) $\Delta m^2$.  The largest true value of $\theta_{23}$ that can be
discriminated at 99\% CL from maximal $\theta_{23}$ is shown in figure
\ref{fig:contt13fix}\,(right) as a function of $\Delta m^2$.  While
there is a distinct but mild dependence on $\Delta m^2$, it is seen
that $\tau$-contamination worsens the ability to discriminate
$\theta_{23}$ from maximality, thus making this measurement harder
than originally expected.

\paragraph{Impact on the $\theta_{13}$ and $\delta$ measurement}
\label{sec:wrongsigntaus}

The problem of $\tau$-contamination of the $\nu_e \to \nu_\mu$ channel
was studied in detail in reference \cite{Donini:2010xk}.  
A good signal-to-background ratio is crucial to determine
simultaneously, and with good precision,
$\theta_{13}$ and $\delta$, since in this channel we have a sample
of tens of events at most.  
In order to separate high-energy charged
currents from the low-energy dominant neutral current background good
energy reconstruction is crucial. 
For this reason, in the standard
MIND analysis at the Neutrino Factory, the neutrino energy is
reconstructed by adding the energy of the muon and that of the
hadronic jet.  
This operation, however, yields a biased result when the
muon comes from a tau decay and it is detected in an iron calorimeter
such as MIND, since no additional information regarding the missing
energy arising from the neutrinos in the $\tau \to \nu_\tau \bar
\nu_\mu \mu^-$ decay can be provided.
This would not be the case in an emulsion cloud chamber or
liquid-argon detector capable of separating the $\nu_e \to \nu_\tau$
signal from $\nu_e \to \nu_\mu$ and of measuring precisely the
kinematics of the process under study.
In this case the neutrino energy could also be reconstructed 
for $\nu_e \to \nu_\tau \to \tau \to \mu$ transitions. 
The sample of wrong-sign muons from the decay of wrong-sign taus will
be distributed erroneously in neutrino energy bins, thus contaminating
the wrong-sign muon sample by events in which the parent neutrino
energy reconstruction is biased.

Consider a $\nu_\tau$ of energy $E_{\nu_\tau}$, interacting in MIND
and producing a wrong-sign $\tau$ of energy $E_\tau$ together with a
hadronic jet of energy $E_h$. After $\tau$-decay, $E_{\nu_\tau} =
E_\tau + E_h = (E_\mu + E_{miss}) + E_h$, where $E_{miss}$~is the
missing energy carried away by the two neutrinos in the $\tau$-decay.
Experimentally, we observe the secondary muon and a hadronic jet, a
signal essentially indistinguishable from that of a wrong-sign muon
from CC $\nu_\mu$ interactions. However, in the latter case, the
addition of the (primary) muon energy $E_\mu$ and of the hadronic jet
energy $E_h$ results in the correct parent $\nu_\mu$ energy,
$E_{\nu_\mu} = E_\mu + E_h$. On the other hand, in the former case the
addition of the (secondary) muon energy $E_\mu$ and of the hadronic
jet energy $E_h$ results in the wrongly reconstructed fake neutrino
energy $E_{fake} = E_\mu + E_h = E_{\nu_\tau} - E_{miss}$. If we
divide the $\tau$ three-body decay energy distribution in discrete
fake neutrino energy bins, we find that for a monochromatic $\nu_\tau$
beam of energy $E_{\nu_\tau}$, the final muon will be assigned to a
given fake neutrino energy bin of energy $E_{\mu,j}$ with probability
$V_j(E_{\nu_\tau})$, where $j = 1, \dots, N^\mu_{bin}$.  We can
compute the distribution of $\nu_\tau$ of a given energy
$E_{\nu_\tau}$ and divide them into $\nu_\tau$ energy bins of energy
$E_{\tau,i}$, where $i = 1, \dots, N^\tau_{bin}$.  The ensemble of the
probability vectors $V_j (E_{\tau.i})$, for $i$ and $j$ running over all
the $\nu_\mu$ and $\nu_\tau$ energy bins, is represented by the
migration matrix $M_{ij}$. After having computed $M_{ij}$, the number
of total wrong-sign muons in a given neutrino energy bin is given by:
\begin{equation}
\label{eq:ntot}
N_i (\theta_{13},\delta) = \sum_{i = 1, N_{bin}} \left [ N^\mu_i (\theta_{13},\delta) + \sum_{j = 1, N_{bin}} M_{ij} N^\tau_j (\theta_{13},\delta) \right ] \, .
\end{equation}
The resulting fraction of wrong sign muons which stem from
$\tau$-decay is found to be as large as 60\% in the energy range from
$5-10\,\mathrm{GeV}$ but only a few percent for higher energies.
\begin{figure}
  \begin{tabular}{cc}
    \includegraphics[width=0.45\textwidth]%
      {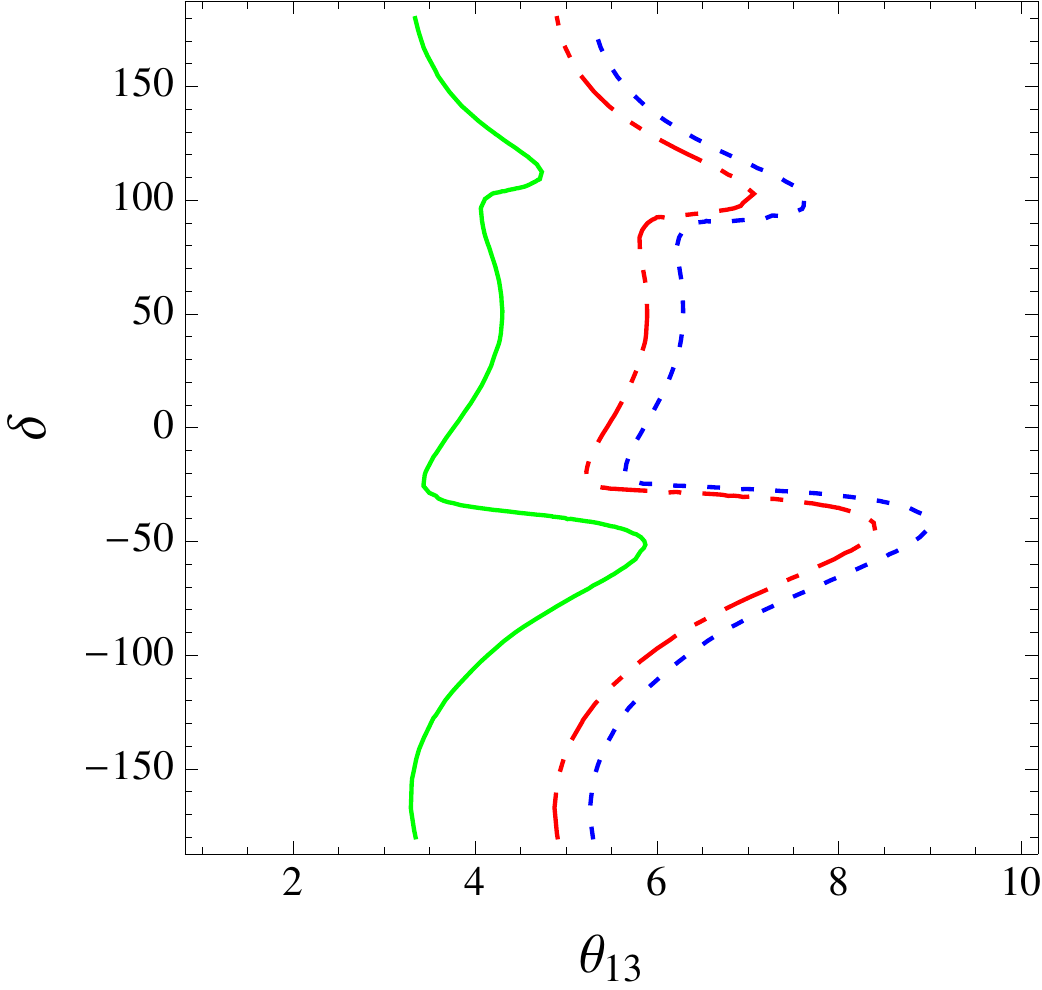} & 
    \includegraphics[width=0.45\textwidth]%
      {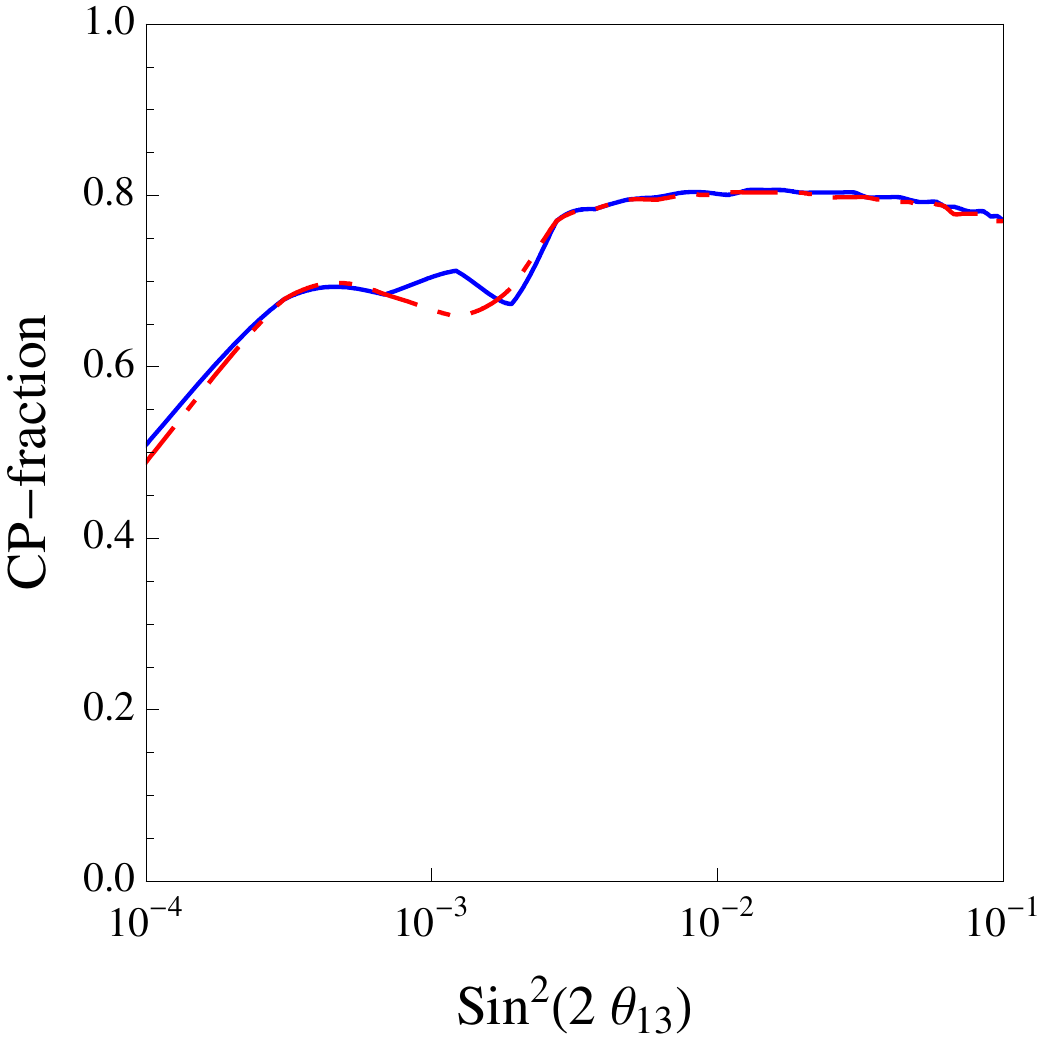} 
  \end{tabular}
  \caption{ 
    Left: Test of the hypothesis that a
    simulation of the data that includes the effect of the
    $\tau$-contamination at $L = 4\,000$~km can be fitted with the
    golden muon theoretical distribution.  
    In the regions to the right of the
    contour lines, the hypothesis can be rejected at 1, 2 or 3$\sigma$
    (from left to right), assuming the goodness-of-fit statistics
    follows the $\chi^2$ distribution with $n =8$ dof. From
    reference \cite{Donini:2010xk}.  
    Right: Comparison of the CP-fraction
    achievable at the IDS-NF baseline Neutrino Factory when
    the $\tau$-contamination is correctly taken into account (dashed
    red line) with respect to the ideal one in which no muons from
    taus are considered (solid blue line).  
  }
  \label{fig:taucontapp}
\end{figure}

If not properly treated, the $\tau$-contamination introduces an
intolerable systematic error, in particular for large $\theta_{13}$.
Figure \ref{fig:taucontapp}~(left), from reference \cite{Donini:2010xk}, shows
that the test of the hypothesis that simulated data including the
$\tau$-contamination, $N_i (\theta_{13},\delta)$, can be fitted using
the direct wrong-sign muon distribution, $N^\mu_i
(\theta_{13},\delta)$, fails at more than 3$\sigma$ for $\theta_{13}
\geq 5^\circ$.  For $\theta_{13} \in [1^\circ,5^\circ]$, even if
$N^\mu_i (\theta_{13},\delta)$ can fit the $\tau$-contaminated data
(albeit with a relatively poor $\chi^2$), the error in the joint
measurement of $\theta_{13}$ and $\delta$ can be so large that it
could actually obviate the use of the Neutrino Factory as a precision
facility (see, again, reference \cite{Donini:2010xk} for a detailed
analysis of the errors introduced by a wrong treatment of the
$\tau$-contamination). On the other hand, once $M_{ij}$ has been
statistically computed, experimental data distributed in reconstructed
neutrino energy bins, can be fitted using the complete wrong-sign
muons distribution $N_i (\theta_{13},\delta)$, properly taking into
account the $\tau$-contamination of the golden-muon sample. Using this
procedure, the systematic error introduced by the muons from taus is
completely removed. The remaining error is the statistical error of
the migration matrix elements, which is under control.

It is worth noting that the $\tau$-contamination of the wrong-sign
muon sample, once properly treated, does not worsen the measurement of
$\theta_{13}$ and $\delta$, as was the case for the measurement of the
atmospheric parameters.  This is illustrated in figure
\ref{fig:taucontapp}~(right) where the CP fraction achieved using the
golden-muon sample only and the total wrong-sign muon sample are
compared.  It can be seen that the only difference between the two
lines is a slight displacement of the wiggles at $\sin^2 2 \theta_{13}
\sim 10^{-3}$ ($\theta_{13} \sim 1^\circ$).  The wiggles are a
consequence of the loss of sensitivity to CP violation introduced by
the so-called ``sign clones'' for negative $\delta$ (a phenomenon
known as $\pi$-transit, \cite{Huber:2002mx}).  Since the location of
the clones in the two samples differs, a small difference in the
location of the wiggles is found when the two lines are compared.  We
can see, however, that once the $\tau$-contamination is properly
treated, no (significant) loss in the CP-fraction is found anywhere
else.  This result differs from that shown in section
\ref{sec:rightsigntaus}.  The main difference between the two channels
is the statistical weight of the two signals.  In the case of the
$\theta_{23}$-measurement through $\nu_\mu \to \nu_\mu$ disappearance,
the right-sign muon direct sample is represented by tens of thousands
of events, and the corresponding $\tau$-contamination by $\sim$ 10\%
of the signal.  This is still a huge number of events, and it affects
the precision with which we are able to measure $\theta_{23}$ and its
deviation from maximality.  On the other hand, when dealing with the
golden channel measurement of $\theta_{13}$ and $\delta$, the signal
is represented by tens of events.  Once the problem of the wrong
assignment of muons from taus into reconstructed energy bin is solved
by means of the migration matrix approach, the residual statistical
impact of the $\tau$-contamination for $\theta_{13} \leq 10^\circ$ is
small.

\subsection{Comparison with the physics performance of alternative experiments}
\label{sec:competition}

A Neutrino Factory is not the only facility that has been proposed for
the study of neutrino oscillations with great accuracy.  Several other
approaches, in particular beta-beams and super-beams, are currently
being examined.  Given the existence of several options, each of which
comes with its own advantages and disadvantages, a critical comparison
of the various facilities is called for. However, this report is
about the feasibility and physics reach of the Neutrino Factory and,
therefore, we feel that we have neither the space, the expertise, nor
a mandate to provide an in-depth critical comparison.  At the same
time, it would be a disservice to the reader to shun the comparison
altogether and thus, we will present a comparison of the physics
sensitivities only, which in turn are based on the outcome of studies
of the facilities that are presently under way.  All results in this
section have been computed using the GLoBES
software~\cite{Huber:2004ka,Huber:2007ji}.

\subsubsection{Super-Beams}

There are two large super-beam studies under way, one is for the SPL
at CERN as part of EUROnu and the other is in the context of LBNE in the
US.  Both experiments will employ a horn-focused pion beam as the
neutrino source and use baselines of $130$~km (SPL) and $1\,300$\,km
(LBNE).  In both experiments the primary proton-beam power will be at
the MW level, however the proton energies will be very different,
4~GeV for the SPL and 120~GeV for LBNE.  The SPL will use a water
Cherenkov detector, with $440$~kTon fiducial volume \cite{wp2}; the
detector performance is taken from
references~\cite{Campagne:2004wt,Campagne:2006yx}.  Note that the SPL
beam has been studied in quite some detail as part of EUROnu and there
it was found that the neutrino fluxes can be optimised by changing the
target and horn configuration~\cite{wp2}, which improves the
anticipated performance compared with that obtained in earlier
studies.

LBNE will use either a water Cherenkov detector or a liquid-Argon
detector with a 6-times smaller mass.  
It has been shown \cite{LBNE-physics,Huber:2010dx} that
physics-wise the two detector configurations are equivalent and
therefore we will consider only the water-Cherenkov option.  
The default fiducial volume for the
water Cherenkov detector in LBNE is $200$~kTon and the baseline is from
Fermilab to DUSEL, which is $1\,300$\,km.  The implementation of both
the beam spectrum and detector performance follows
reference~\cite{LBNE-physics}. We also consider a possible luminosity
upgrade as provided by Project X.

\subsubsection{Beta-beams}

Beta-beams have been proposed in reference~\cite{Zucchelli:2002sa} and
the idea is to exploit the $\beta$-decay of relatively short-lived
radio-isotopes with half-lives of around $1$~s which are
ionised, accelerated to relativistic speeds and put into a storage
ring, where they eventually decay. 
As a result of the high Lorenz
$\gamma\simeq 50-500$, the (anti-)neutrino emission will be directed
in the forward direction with an opening angle of $1/\gamma$.
Neutrino beams can be produced using $\beta^+$ emitters.  
The maximum neutrino energy that is available, $E_\text{max}$, is
given by the end-point of the $\beta$-spectrum, $Q_\beta$, and the
Lorenz boost, $E_\text{max}=\gamma Q_\beta$.  
Given the fact that $2 Q_\beta$ lies in
the range of $1-10\,\mathrm{MeV}$, it follows that neutrino beam
energies in the range from $0.1-5\,\mathrm{GeV}$ are possible.
Currently, four isotopes are being studied: $^{18}$Ne and $^8$B for
neutrino production and $^6$He and $^8$Li for anti-neutrino
production; their properties are listed in table~\ref{tab:ions}.
Clearly, the ions with a large $Q_\beta$ will yield higher neutrino
energies for a given Lorenz $\gamma$, but for a given energy they will
yield less flux since the flux is proportional to $\gamma^{-2}$. 
All
of these isotopes are too short-lived to have significant natural
abundance and thus they need to be produced artificially. 
It turns out that the
achievable production rates ultimately limit the luminosity of these
kinds of neutrino beams. 
The intensities considered to be the goals are
$1.1 \times 10^{18}$ ions per year for $^{18}$Ne and $2.9 \times 10^{18}$
ions per year for $^{6}$He~\cite{EURISOL}. 
For some time it was not quite clear whether these ion intensities
are feasible, but recent developments within EUROnu indicate that, for
$^{18}$Ne, the required intensity can be achieved
\cite{Wildner:2010zz}. 
\begin{table}
\caption{\label{tab:ions} $A/Z$, half-life and end-point energies for
  two $\beta^+$-emitters ($^{18}$Ne and $^8$B) and two
  $\beta^-$-emitters ($^6$He and $^8$Li).Table adapted
  from~\cite{Bernabeu:2010rz}.}
\begin{center}
\begin{tabular}{|c|c|c|c|c|} \hline \hline
   Element  & $A/Z$ & $T_{1/2}$ (s) & $Q_\beta$  [MeV] & Decay Fraction \\ 
\hline
  $^{18}$Ne &   1.8 &     1.67      &        3.41         &      92.1\%    \\
           
  $^{8}$B   &   1.6 &     0.77      &       13.92         &       100\%    \\
\hline
 $^{6}$He   &   3.0 &     0.81      &        3.51         &       100\%    \\ 
 $^{8}$Li   &   2.7 &     0.83      &       12.96         &       100\%    \\ 
\hline
\hline
\end{tabular}
\end{center}
\end{table}

In the phenomenological literature it has been pointed out that large
Lorenz boosts, around 350, possibly combining data from all four ion
species, can lead to physics sensitivities which are on par with that
of the Neutrino Factory
\cite{Burguet-Castell:2003vv,Burguet-Castell:2005pa,Agarwalla:2005we,%
  Huber:2005jk,Donini:2006tt,Agarwalla:2006vf,Agarwalla:2008ti,%
  Agarwalla:2008gf,Coloma:2010wa}.  However, there are currently no
attempts to perform a feasibility study of any of these so called
``high-$\gamma$'' beta-beams.  As a consequence, it it is not possible
to assess the cost and technological risks associated with the
high-$\gamma$ options.  Therefore, we will not consider them here and
instead will focus on those options which are currently studied, at
a machine level, within the EUROnu Beta-Beam work package; i.e., only
the $\gamma=100$ option will be considered.  
For this reason, we also do
not consider $^8$Li or $^8$B because they require higher values of
$\gamma$ to be useful, see, for example, \cite{Coloma:2010wa}.

\subsubsection{Comparison}

\begin{figure}[tp]
  \begin{center}
    \includegraphics[width=0.49\textwidth]%
      {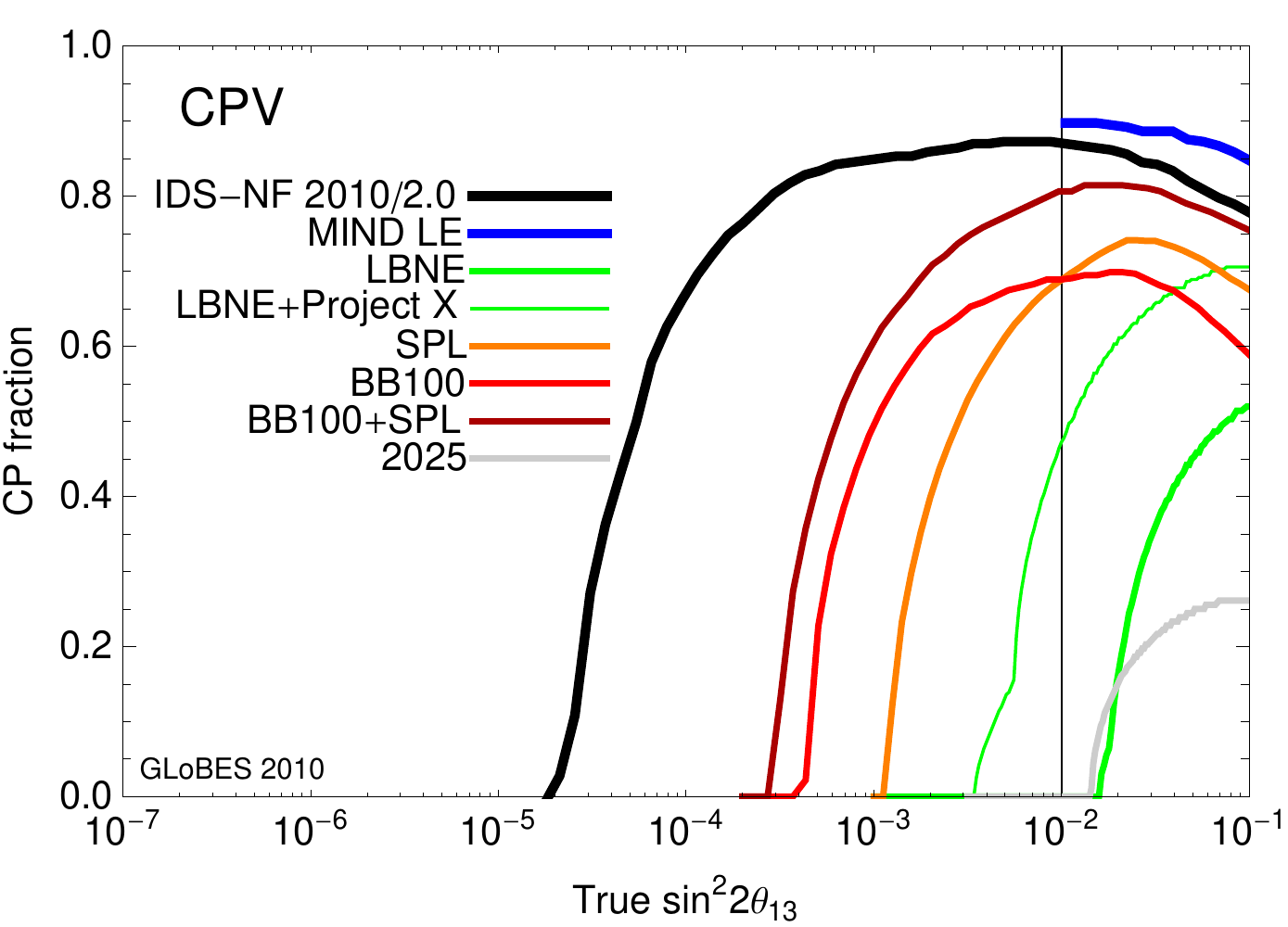} %
    \includegraphics[width=0.49\textwidth]%
      {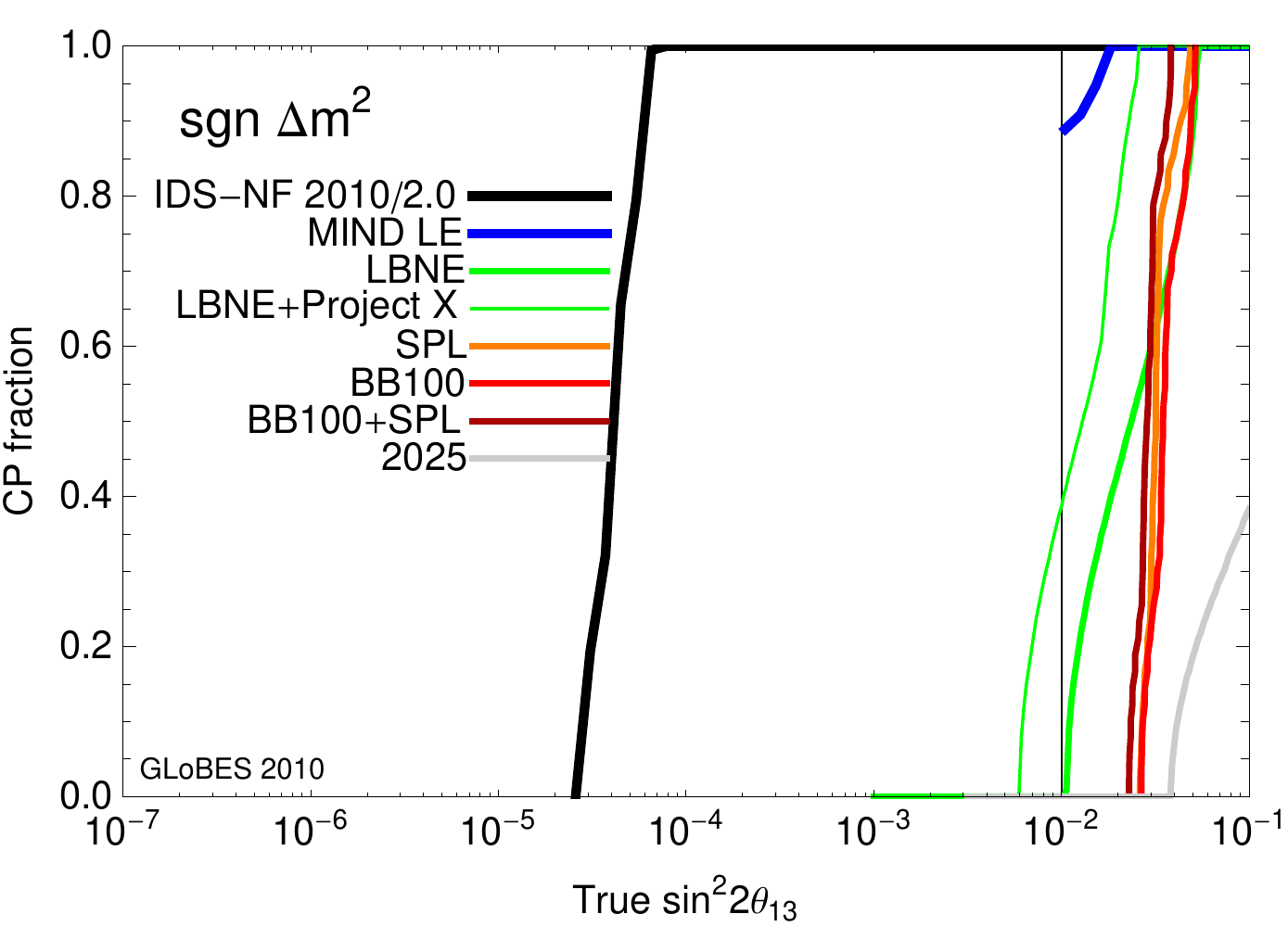}
    \includegraphics[width=0.49\textwidth]%
      {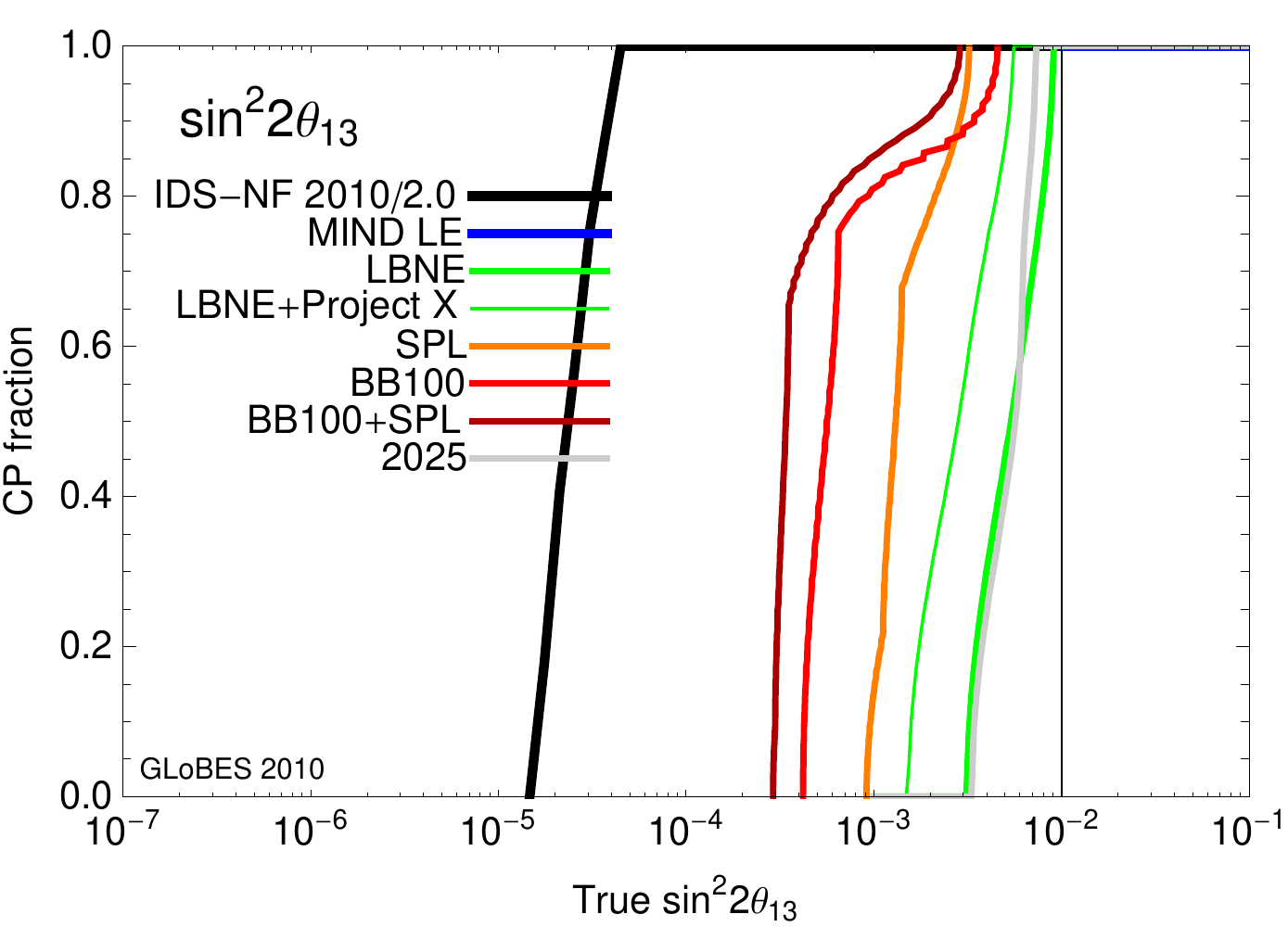} %
  \end{center}
  \caption{ Comparison of the physics reach of different future
    facilities for the discovery of CP violation (top left panel), the
    mass hierarchy (top right panel), and $\sin^2 2 \theta_{13}$
    bottom panel.  The sensitivities of the SPL super-beam are taken
    from \cite{Campagne:2006yx}.  The beta-beam curves are also taken
    from \cite{Campagne:2006yx}, however with the ion intensities
    reduced to the EURISOL values \cite{EURISOL}.  Curves for LBNE are
    taken from \cite{Huber:2010dx} and correspond to the results in
    \cite{LBNE-physics}.  The $\theta_{13}$ sensitivities expected
    from current experiments are shown as vertical lines
    \cite{Huber:2009cw}.  MIND LE is a single-baseline Neutrino
    Factory optimised for large $\stheta>0.01$, see also
    section~\ref{sec:opti1}.  }
  \label{fig:euronu2009}
\end{figure}

It is possible that a next-generation super-beam (or combined
super-beam/beta-beam) experiment could be constructed on the same
timescale as the Neutrino Factory, or might even start earlier. 
Possible
candidates are the LBNE and SPL experiments
shown in figure \ref{fig:euronu2009} with their $\theta_{13}$, CP
violation, and mass hierarchy discovery potentials.

If $\stheta \gtrsim 0.01$ is discovered before the RDR is prepared, CP
violation can be measured at the $3 \sigma$ CL, depending on the
option and exact value of $\stheta$, for between 60\% and 75\% of all
values of $\delta$.  For the current hint $\stheta=0.06$, a discovery
can be made for up to 70\% of all possible values of $\delta$.  The
mass hierarchy discovery is, in this case, guaranteed by the LBNE
option; for the alternatives the discovery reach strongly depends on
$\stheta$ in this range.  For comparison, in figure
\ref{fig:euronu2009} the $3 \sigma$ curves from the existing
experiments in 2025 are also shown (curves ``2025''). As we have
demonstrated, the Neutrino Factory can do significantly better for the
CP violation discovery, and, of course, also measure the mass
hierarchy in that range. Very importantly, the CP fraction in figure
\ref{fig:euronu2009} can be related to the precision measurement of
$\delta$ for the CP conserving values. Therefore, the higher the CP
fraction, the higher the precision. In either case, a Neutrino Factory
will provide a more precise measurement than any other experiment,
especially if a single baseline configuration, optimised for the, by
then known, value of $\stheta$ were employed.

If the data indicate that $\stheta \lesssim 0.01$ at the time the RDR
is in preparation, alternatives to the Neutrino Factory have a
discovery potential for $\stheta$ significantly exceeding the existing
experiments, in some cases by up to an order of magnitude.  
Depending on $\stheta$, CP violation may be discovered for up to 65\%
of possible values of $\delta$ (SPL \& LBNE).  
The mass hierarchy can only be accessed by LBNE in a small fraction
of the parameter space.  The CP violation plot demonstrates that these
experiments have limited potential for $\stheta \lesssim 0.01$, since
the small data samples that can be expected will cut off the
sensitivity at some value of $\sin^2 2 \theta_{13}$.  Figure
\ref{fig:euronu2009} shows that the Neutrino Factory can do
significantly better.  It is also interesting from figure
\ref{fig:euronu2009} that the alternatives are either optimised for
the CP violation (SPL/BB100) or the mass hierarchy discovery (LBNE).
No option other than the Neutrino Factory can do all these
measurements equally well.

In summary, even if $\theta_{13} > 0$ is discovered by the generation
of experiments currently under construction, it is likely that the
discovery of CP violation and precision measurement of the CP phase
require data from advanced experiments, like the Neutrino Factory.
If $\theta_{13}$ is not discovered by the generation of experiments
presently under construction, there will be no further information on
CP violation and mass hierarchy from these experiments, and an advanced
experiment is mandatory to make further progress.  In this case, the
super-beam and beta-beam based alternatives will be strongly limited
by the size of the data sample that can be collected and only the
Neutrino Factory provides a robust sensitivity to all observables
throughout the accessible parameter space.

%
\subsection{Non-standard neutrino interactions}
\label{sec:nsi}

Since the Neutrino Factory is the most ambitious concept available for
advancing our knowledge of lepton flavour physics, the search for new and
exotic phenomena in the lepton sector will be an important part of its physics
program.  In this section, we will outline the results that can be obtained on
non-standard neutrino interactions, non-unitarity in the lepton mixing matrix,
and sterile-neutrino scenarios.


\subsubsection{Effective theory of new physics in the neutrino sector}

Without any knowledge of possible high-energy extensions of the Standard
Model, an effective theory approach is well-suited to parameterising the
effects of such extensions on neutrino-oscillation experiments.  Non-standard
interactions (NSI) mediated by effective higher-dimensional operators can lead
to additional, possibly flavour-violating, contributions to the MSW potential
that neutrinos experience when travelling through matter, and they can affect
the neutrino production and detection processes~\cite{Wolfenstein:1977ue,
Grossman:1995wx}.  The neutrino oscillation probability in the presence of NSI
can be written:
\begin{align}
  P(\nu^s_\alpha \rightarrow \nu^d_\beta)
    &= \Big| \big[ \big( 1 + \eps^d \big)^T \,\, e^{-i H_{\rm NSI} L} \,\,
                   \big( 1 + \eps^s \big)^T \big]_{\beta\alpha} \Big|^2 \,,
  \label{eq:P-NSI}
\end{align}
with the Hamiltonian:
\begin{align}
  H_{\rm NSI} &= U \begin{pmatrix}
    0 &                      & \\
    & \Delta m_{21}^2 / 2E & \\
    & & \Delta m_{31}^2 / 2E
                   \end{pmatrix} U^\dag
  + V_{\rm MSW} (\mathbb{1} + \eps^m) \,.
\end{align}
Here, $U$ is the lepton-mixing matrix, $V_{\rm MSW}$ is the
Mikheyev-Smirnov-Wolfenstein (MSW) potential corresponding to the matter
traversed by the neutrinos, and $\eps^d$, $\eps^s$, and $\eps^m$ are complex $3
\times 3$ matrices parameterising the relative strength of the non-standard
effects compared to Standard Model weak interactions. If $\eps^s_{\alpha\beta}$
is non-zero, neutrinos produced in interactions with charged leptons of flavour
$\alpha$ will have a non-standard component of flavour $\beta$. If
$\eps^d_{\beta\alpha} \neq 0$, neutrinos of flavour $\beta$ can produce a
charged lepton of flavour $\alpha$ in the detector. The Hermitian matrix
$\eps^m$ is responsible for non-standard matter effects.  Note that in most
concrete models, the elements of $\eps^d$, $\eps^s$, and $\eps^m$ are not all
independent.

Model-independent experimental constraints on the $\eps^{s,d,m}_{\alpha\beta}$
range from $10^{-4}$ to $\mathcal{O}(1)$~\cite{Fornengo:2001pm,
Davidson:2003ha, GonzalezGarcia:2007ib, Biggio:2009kv, Biggio:2009nt}, while
model-dependent limits are often much tighter~\cite{Antusch:2008tz,
Gavela:2008ra} since, in concrete models, NSI in the neutrino sector are
usually accompanied by non-standard effects in the charged-lepton sector that
are constrained, for example, by rare decay searches or measurements of the
weak mixing angle.

If NSI are induced by dimension-6 operators that arise when heavy mediator fields
are integrated out, we expect $\eps^{s,d,m}_{\alpha\beta} \sim g^2 M_W^2 /
g_{\rm NSI}^2 M_{\rm NSI}^2$, where $g$ is the weak gauge coupling, $g_{\rm
NSI}$ is the coupling constant of the heavy mediators, and $M_{\rm NSI}$ is
their mass.  If $g_{\rm NSI} \sim g$ and $M_{\rm NSI} \sim \text{few} \times
100$~GeV, we naively expect $\eps^{s,d,m}_{\alpha\beta} \sim 0.01$. Most
dimension-6 operators of this magnitude are, however, already ruled out by
several orders of magnitude by constraints from the charged-lepton sector. 

Such constraints are avoided if NSI are generated by dimension-8
operators such as:
\begin{equation}
  (\bar{L}_\alpha H) \gamma^\rho (\bar{H}^\dag L_\beta) 
  (\bar{E}_\gamma \gamma_\rho E_\delta)\,,
\end{equation}
where $L$, $H$, and $E$ are a left-handed lepton field, the Standard
Model Higgs field, and a right-handed charged-lepton field,
respectively, and $\alpha$, $\beta$, $\gamma$, $\delta$ are flavour
indices.  
When $H$ acquires a
non-zero vacuum expectation value $v$, this operator leads to
$SU(2)$-violating four-fermion contact interactions, so that NSI in
the neutrino sector can be generated without inducing
phenomenologically-problematic four-charged-lepton couplings or other
harmful effects. On the other hand, this procedure leads to an
additional suppression of order $v2/M2$ and requires some cancellation
conditions at tree level~\cite{Gavela:2008ra}. Larger non-standard
effects could be generated in theories involving \emph{light}
mediators ($\ll M_W$) with very small couplings. Models containing
light new particles have recently received a lot of interest in the
context of Dark Matter searches (see for example references
\cite{Fitzpatrick:2010em, Hooper:2010uy, Arina:2010rb, Lavalle:2010yw,
  Cumberbatch:2010hh, Kuflik:2010ah}), but possible effects in the
neutrino sector are less well explored in the literature.  For very
light mediators, processes with small or no momentum transfer, such as
coherent forward scattering leading to non-standard neutrino matter
effects, will be enhanced compared to processes with large momentum
transfer from which most NSI constraints are derived.

Finally, there are more exotic scenarios of new interactions in the neutrino
sector that could be tested at the Neutrino Factory. One example for such a
scenario is long-range leptonic forces~\cite{Joshipura:2003jh, Grifols:2003gy,
GonzalezGarcia:2006vp, Bandyopadhyay:2006uh, Samanta:2010zh, Heeck:2010pg},
which are phenomenologically equivalent to NSI even though the theoretical
motivation is very different.

\subsubsection{Neutrino Factory sensitivities to non-standard interactions}

\begin{figure}
  \begin{center}
    \includegraphics[width=13cm]{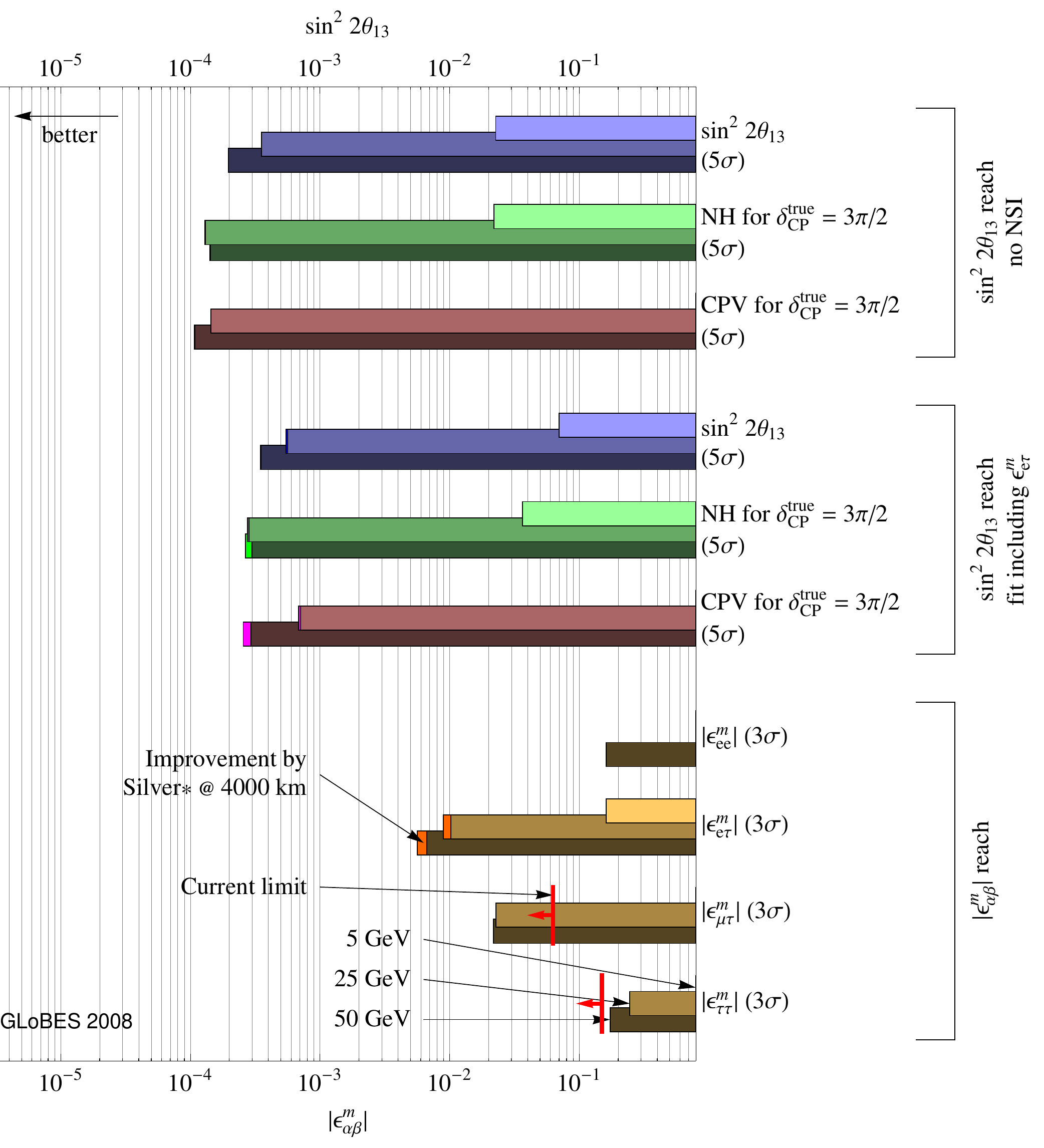}
  \end{center}
  \vspace{-0.5cm}
  \caption{Summary of the Neutrino Factory performance with and
    without the presence of non-standard interactions. The IDS-NF
    setup with two 50~kTon magnetised iron detectors (MIND) at
    baselines of 4\,000~km and 7\,500~km was used, and the ``true''
    parameter values $\sin^2 2\theta_{13} = 0.001$ and 
    $\delta_{\rm CP} = 3 \pi/2$ were assumed (the detector mass used
    differs from the present IDS-NF baseline).
    The plot shows that sensitivities are poor at
    $E_\mu = 5$~GeV (light bars) due to the high energy threshold of
    the MIND detector, but increase dramatically at $E_\mu = 25$~GeV
    (medium light bars). The benefit from increasing $E_\mu$ further
    to 50~GeV (dark bars) is only marginal, as is the benefit from
    including a silver channel $\nu_\tau$ detector at 4\,000~km
    (``Silver*''). Current model-independent limits on NSI are taken
    from~\cite{Davidson:2003ha, Biggio:2009nt, GonzalezGarcia:2007ib}.
    No strong limits exist for $\epsilon^m_{ee}$ and
    $\epsilon^m_{e\tau}$. Figure taken from reference~\cite{Kopp:2008ds}.}
  \label{fig:NSI-discreach}
\end{figure}

The far detectors of the Neutrino Factory will offer a unique opportunity to
constrain or measure the elements of $\eps^m$, corresponding to non-standard
matter effects. This is shown in figure \ref{fig:NSI-discreach} under the
assumption that only one of the $\eps^m_{\alpha\beta}$ is non-zero at a time.
We see that existing model-independent constraints can be significantly
improved, and that the Neutrino Factory would begin to probe the parameter
region $\eps^m_{\alpha\beta} \lesssim 0.01$ that is most interesting from the
model-building point of view. We also compare the sensitivity
to standard oscillation observables ($\theta_{13}$, CP violation, MH)
in a conventional fit to the sensitivities that may be achieved if one
NSI parameter ($\epsilon^m_{e\tau}$) is allowed to be non-zero. 
It turns out that allowing for the possibility of non-standard effects
reduces the sensitivity to standard oscillation observables.

For NSI affecting the neutrino production and detection mechanisms,
even tighter constraints can often be obtained from the Neutrino
Factory \emph{near} detector.  To demonstrate this, we consider
non-standard contributions to the charged-current neutrino production
and detection processes~\cite{Grossman:1995wx}. New contributions to
$\nu_e$ and $\nu_\mu$ production or detection are difficult to observe
because they are degenerate with systematic biases in the neutrino
cross section calculations or in the normalisation of the beam flux. A
near detector with charge identification capabilities, however, would
be sensitive to flavour-violating processes such as $\mu \to e \nu_e
\nu_e$, $\mu \to e \nu_\mu \nu_\mu$, $\nu_e + N \to \mu + X$, and
$\nu_\mu + N \to e + X$~\cite{Tang:2009na}.  
Another interesting possibility is
the construction of a detector sensitive to $\nu_\tau$ to look for
processes such as $\mu \to e \nu_{\alpha} \nu_\tau$ or $\nu_\alpha + N
\to \tau + X$ for $\alpha = e, \mu$. This possibility has been
considered in references
\cite{Tang:2009na,Antusch:2009pm,Meloni:2009cg,MINSIS2009Madrid}.  In
table~\ref{tab:cc-nsi-bounds}, we summarise some expected bounds on CC
NSI operators for the Neutrino Factory with and without a near
detector capable of $\nu_\tau$ detection.  We see that some bounds can
be improved by almost two orders of magnitude compared to their
current values, and that the Neutrino Factory would probe the
parameter region expected if NSI are induced by
$\mathcal{O}(\text{TeV})$ mediators.
\begin{table}
  \caption{Expected sensitivities and current bounds for the coupling constants
    (relative to Standard Model weak interactions) of some NSI operators
    relevant to the Neutrino Factory experiment.  Numbers in
    parenthesis are for a model with charged-singlet
    scalar mediation, while numbers without parenthesis are model-independent.}
  \begin{center}
    \begin{tabular}{cccc}
    \hline \hline
      & Sensitivity w/o $\nu_\tau$ ND & Sensitivity w/ $\nu_\tau$ ND
      & Current bound~\cite{Tang:2009na,Antusch:2008tz} \\
      \hline
      $\epsilon^{s}_{e\tau}$ ($\mu \to e \nu_\mu \nu_\tau$)
          & $ 4 \cdot 10^{-3}$
          & $ 7 \cdot 10^{-4}$
          & $ 3.2 \cdot 10^{-2}$ ($ 1.8 \cdot 10^{-3}$) \\
      $\epsilon^{s}_{\mu\tau}$ ($\mu \to e \nu_e \nu_\tau$)
          & $ 4 \cdot 10^{-1}$ ($3\cdot 10^{-3}$)
          & $ 6 \cdot 10^{-4}$ ($6 \cdot 10^{-4}$)
          & $ 3.2 \cdot 10^{-2}$ ($ 1.9 \cdot 10^{-3}$) \\
      \hline \hline
    \end{tabular}
  \end{center}
  \label{tab:cc-nsi-bounds}
\end{table}

\subsubsection{Non-standard interactions in a low-energy Neutrino Factory}

Let us now comment on the NSI sensitivity of a low-energy Neutrino Factory
(LENF).  We have analysed the LENF setup defined in
\cite{FernandezMartinez:2010zza}, using the MonteCUBES (Monte Carlo
Utility Based Experiment Simulator) software package
\cite{Blennow:2009pk,MCBhome}. 
The LENF has leading order sensitivity to 
$\varepsilon^m_{e\mu}$ and $\varepsilon^m_{e\tau}$ from the golden and platinum
channels. We have found that the sensitivity is limited by the correlations
between the standard oscillation and non-standard parameters, and thus
that increasing statistics does \emph{not} increase the sensitivity of the
setup, in direct contrast to standard oscillation measurements. Instead, it is
necessary to include information from a second baseline, or from a
complementary channel such as the platinum channel. With the performance of the
platinum channel as assumed in \cite{FernandezMartinez:2010zza} ($47\%$
efficiency, $10^{-2}$ background), the platinum channel has the greatest impact
for \emph{large} values of $\theta_{13}$. However, we have found that a
platinum channel with hypothetically perfect performance ($\sim100\%$
efficiency, negligible background) has an effect even for small values of
$\theta_{13}$, indicating that there is a critical signal-to-background
threshold that must be overcome in order for it to become effective.

\begin{figure}
\includegraphics[width=7.5cm,height=6cm]{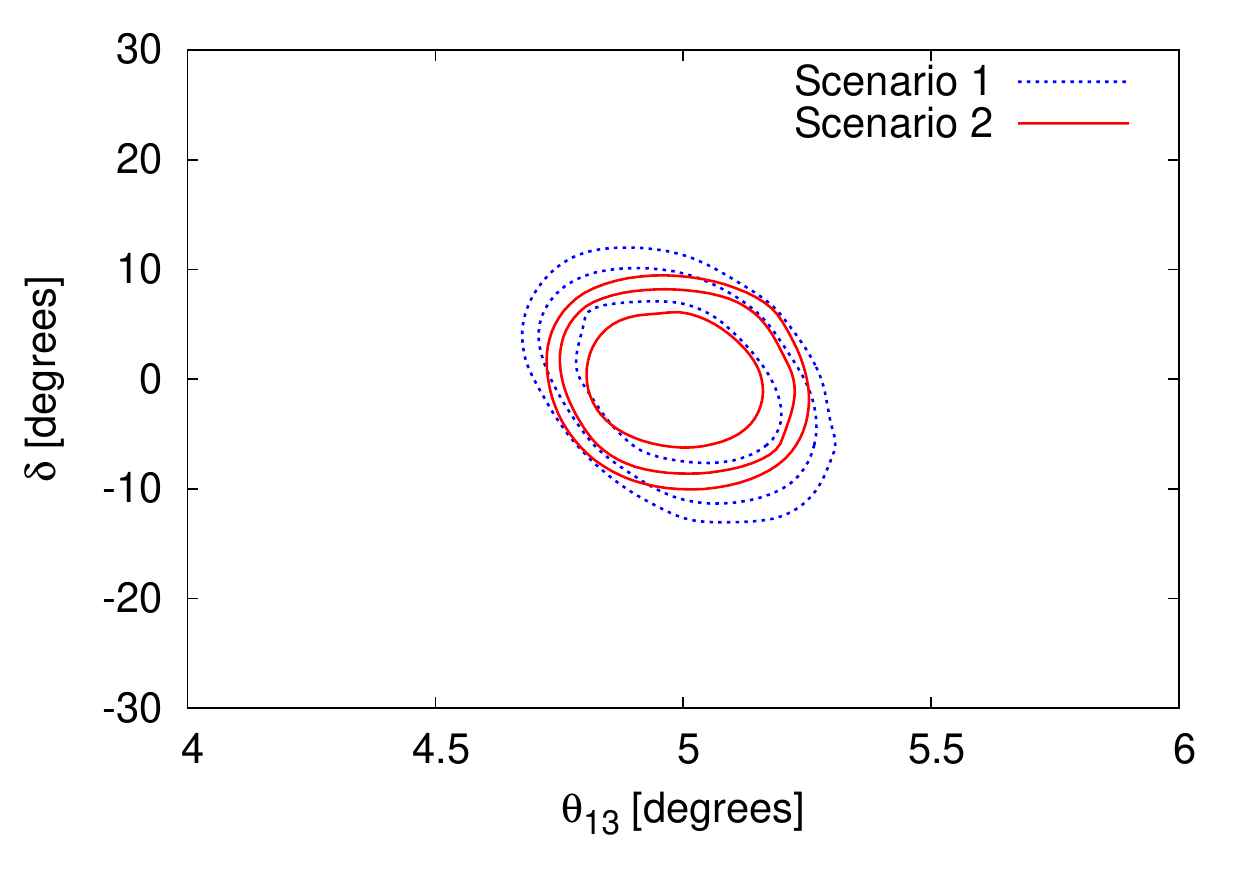}
\includegraphics[width=7.5cm,height=6cm]{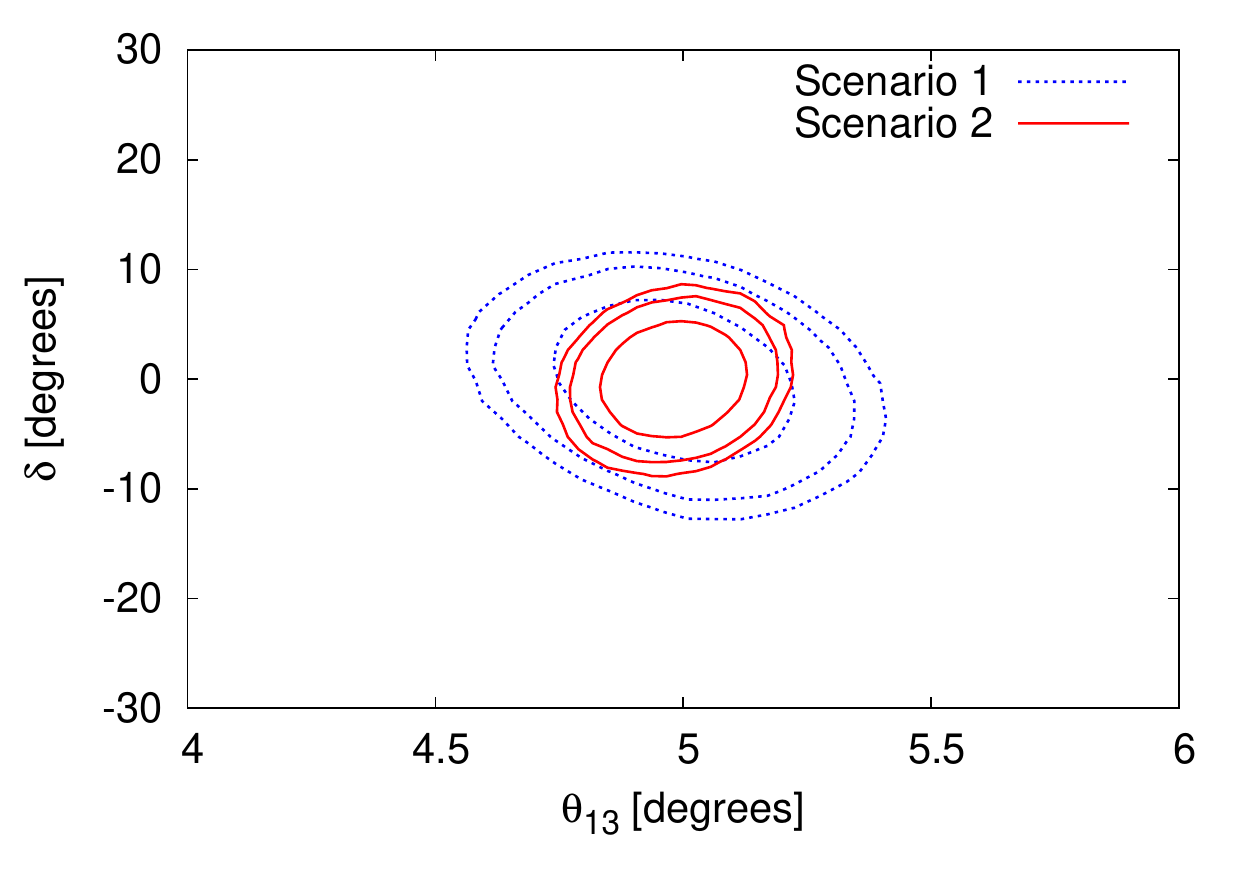}
\caption{$68\%$, $90\%$ and $95\%$ regions in the $\theta_{13}-\delta$ plane
  when marginalising over standard oscillation parameters only (left) and when
  marginalising over all standard oscillation and NSI parameters (right), for
  true values of $\theta_{13}=5^{\circ}$, $\delta=0$ and
  $\varepsilon^m_{e\mu}=\varepsilon^m_{e\tau}=0$. Scenario 1 is a LENF setup without
  platinum channel; Scenario 2 is the LENF with platinum channel. Figure taken from reference~\cite{FernandezMartinez:2010zza}.}
  \label{fig:th13d}
\end{figure}

In figure \ref{fig:th13d} we show how the platinum channel helps to maintain
sensitivity to the standard oscillation parameters, when marginalisation over
all NSI parameters, as well as the oscillation parameters is performed: we show
the $68\%$, $90\%$ and $95\%$ regions obtained in the $\theta_{13}-\delta$
plane when only the golden channel and $\nu_{\mu} / \bar{\nu}_{\mu}$
disappearance channels are used (`Scenario 1' - dotted blue lines), and when
combined with the platinum channel (`Scenario 2' - solid red lines). The
left-hand figure shows the results when \emph{only} the standard oscillation
parameters are marginalised over, and the right-hand figure the results when
all oscillation \emph{and} NSI parameters are taken into account. This shows
that the sensitivity of Scenario 1 to $\theta_{13}$ and $\delta$ is diminished
by taking into account the possibility of non-zero NSIs, but that Scenario 2
is almost unaffected---the addition of the platinum channel
makes the setup more robust.

\begin{figure}
  \includegraphics[width=7.5cm,height=6cm]{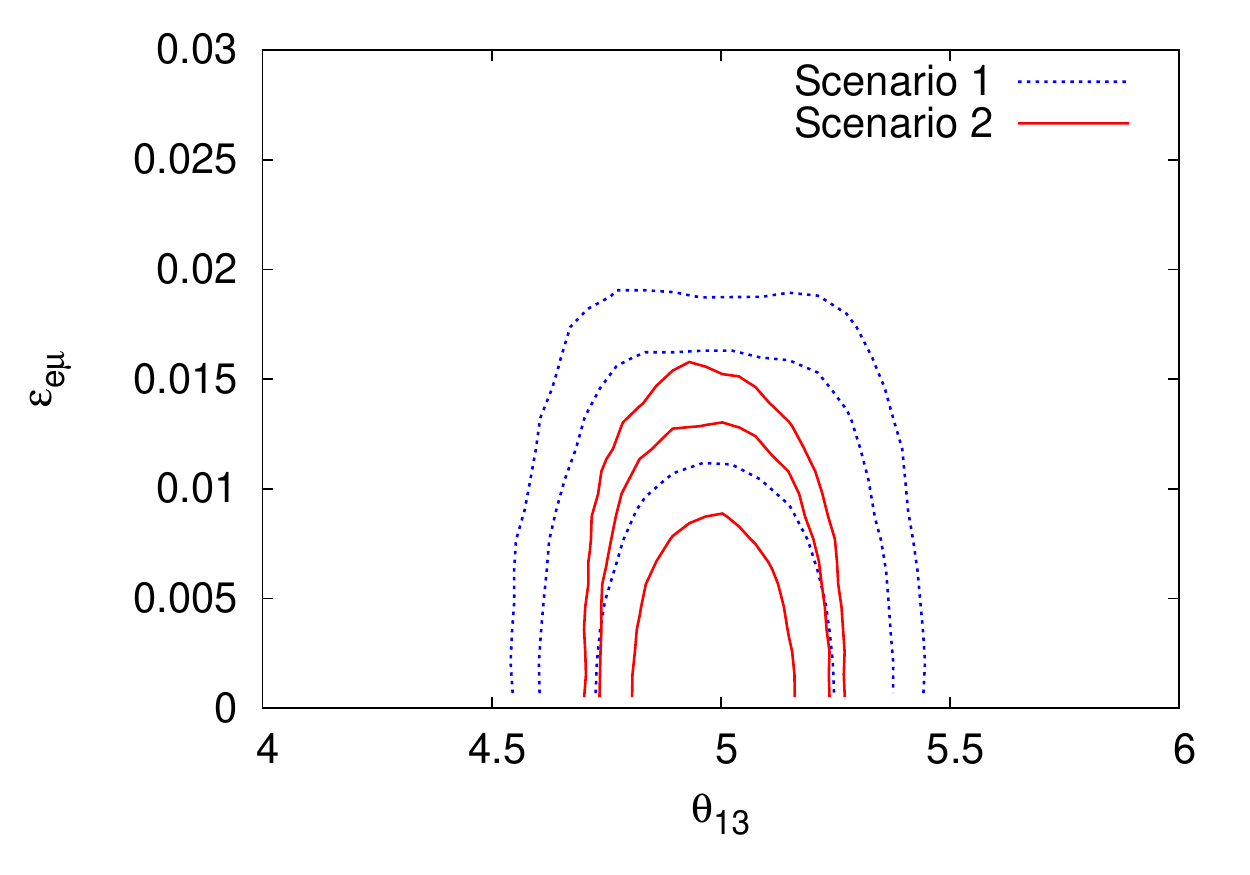}
  \includegraphics[width=7.5cm,height=6cm]{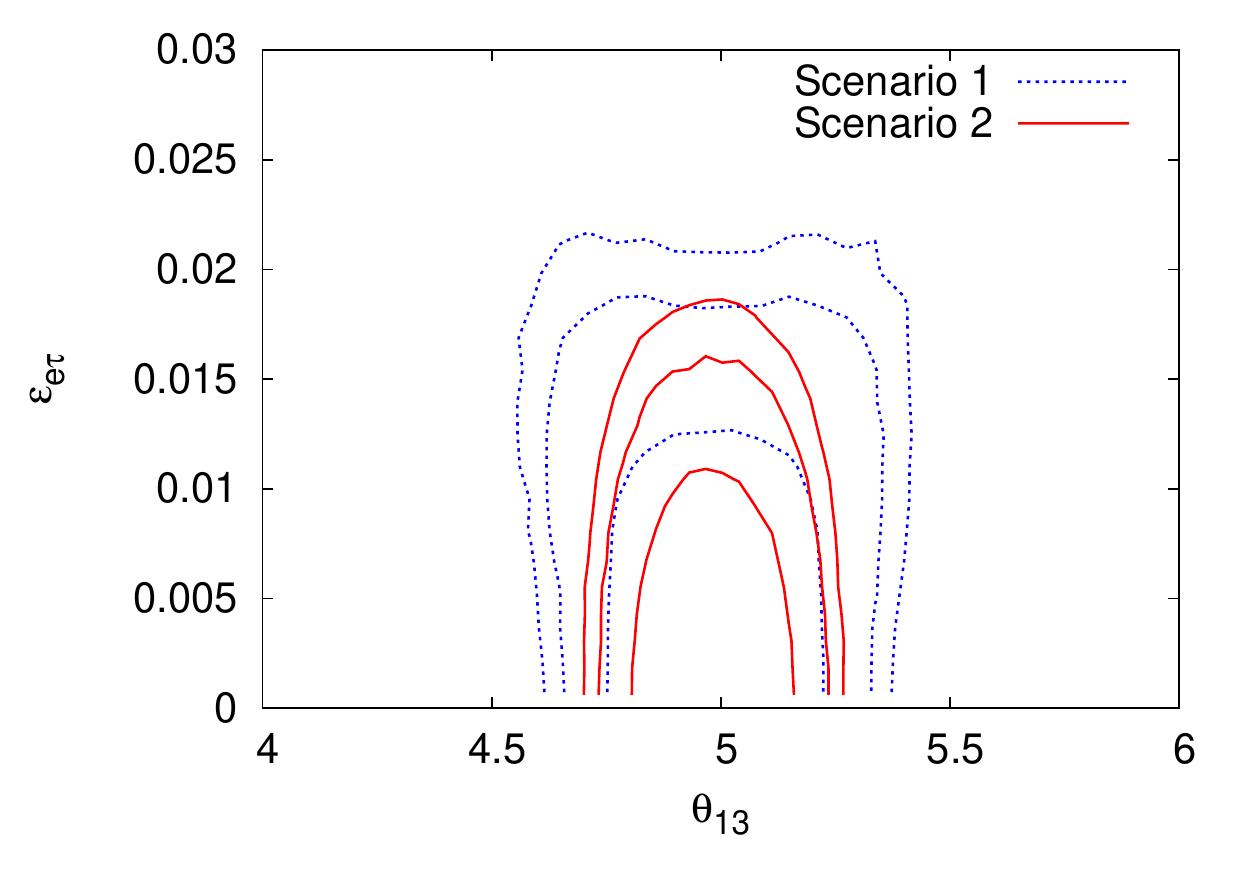}
  \caption{$68\%$, $90\%$ and $95\%$ regions in the
    $\theta_{13}-\varepsilon^m_{e\mu}$ (left) and
    $\theta_{13}-\varepsilon^m_{e\tau}$ (right) plane for
    $\theta_{13}=5^{\circ}$, $\delta=0$, and
    $\varepsilon^m_{e\mu}=\varepsilon^m_{e\tau}=0$.  Figure taken from
    reference~\cite{FernandezMartinez:2010zza}.}
  \label{fig:eps}
\end{figure}

For measuring the NSI parameters themselves, the LENF running for 5 years per
polarity (i.e., 10 years running are assumed, the same assumption that
has been made in evaluating the performance of the other long-baseline
experiments) has sensitivity to $\varepsilon^m_{e\mu}$ and
$\varepsilon^m_{e\tau}$ down to $\sim10^{-2}$ (figure \ref{fig:eps}). 
If the running time is doubled, no significant improvement
is gained, indicating that the limiting factor is not statistics. 
These bounds are a significant improvement upon the current bounds 
\cite{Biggio:2009kv,Biggio:2009nt}, but not as strong as those which
the Neutrino Factory can obtain.

\subsubsection{Non-unitarity of the leptonic mixing matrix}

A non-unitary mixing matrix in the charged-current coupling of
neutrinos and charged leptons is a general prediction of extensions 
of the Standard Model (SM) that include new fermion degrees of freedom that can
mix with the SM leptons \cite{Langacker:1988ur}. Indeed, the full mixing matrix
including the extra degrees of freedom would thus be larger than the standard
three-by-three matrix. 
While the full mixing matrix should be unitary, such a
constraint does not apply to the $3 \times 3$ sub-matrix accessible at low
energies. In particular, many extensions of the SM that try to accommodate the
evidence for neutrino masses and mixings from neutrino-oscillation experiments
introduce such extra leptonic degrees of freedom. This is the case for example
of the type-I and type-III seesaw models. The study of deviations from
unitarity of the leptonic mixing matrix can therefore provide a tool to explore
the origin of neutrino masses beyond the SM~\cite{Broncano:2002rw,Broncano:2003fq,
Antusch:2009gn,Gavela:2009cd,Antusch:2009gn,Ohlsson:2010ca}.

In a completely general way, the non-unitary leptonic mixing matrix
$N$ can be parametrised as the product of a Hermitian matrix,
$1 + \eta$, times a unitary matrix $U$
\cite{FernandezMartinez:2007ms}: 
\begin{equation}
N=(1+\eta)U\;.
\label{eq:param}
\end{equation}
Such a modification of the SM charged-current interactions would affect lepton
universality tests, the measurement of $G_F$ from muon decay compared to other
measurements, rare lepton decays and the invisible width of the $Z$,
and strong constraints can be derived on the allowed size of
$\eta$~\cite{Langacker:1988ur,Nardi:1994iv,Tommasini:1995ii,Antusch:2006vwa,Antusch:2008tz}:
$\eta_{ee}<2.0 \times 10^{-3}$, $\eta_{e \mu}<5.9 \times 10^{-5}$, $\eta_{e
\tau}<1.6 \times 10^{-3}$, $\eta_{\mu \mu}<8.2 \times 10^{-4}$, $\eta_{\mu
\tau}<1.0 \times 10^{-3}$ and $\eta_{\tau \tau}<2.6 \times 10^{-3}$, at the
$90~\%$\,CL.
Up to order $\eta$ effects, the matrix $U$ can be identified with
the PMNS matrix as extracted from experimental data without taking
non-unitarity into account. Since $\eta$ is much smaller than the experimental
uncertainties on the PMNS matrix elements, the corrections are negligible.

The effects of a non-unitary mixing at the Neutrino Factory have been
studied in references \cite{FernandezMartinez:2007ms, Holeczek:2007kk,
Goswami:2008mi, Antusch:2009pm, Meloni:2009cg}.  In particular,
in~\cite{Antusch:2009pm} this study was performed in detail for a setup very
similar to the IDS-NF baseline Neutrino Factory (muon energy 25~GeV,
two 50~kt magnetised iron detectors at baselines of 4\,000~km and
7\,500~km) and considering simultaneously the effect of all extra
non-unitarity parameters in the fit.  
To 
perform a fit in this high-dimensional parameter space, the Markov Chain Monte
Carlo tool MonteCUBES, a plug-in for the widely used GLoBES
software~\cite{Huber:2004ka,Huber:2007ji}, has been used.  It was found that,
given the very stringent present limits on the allowed deviations from
unitarity, the Neutrino Factory could only improve constraints on the 
$\eta_{e \tau}$ and $\eta_{\mu \tau}$ elements of the matrix $\eta$.
Indeed, it is
easier to search for off diagonal elements since they imply lepton flavour
violation and the present bounds on $\eta_{e \mu}$ from $\mu \to e \gamma$ are
too strong to improve through measurements of neutrino oscillations. 
The sensitivity of the Neutrino Factory to these parameters could be
significantly improved if a near detector capable of $\tau$ detection
is included, since the zero-distance effects caused by the unitarity
deviations could be tested.  
\begin{figure}
 \begin{center}
  \includegraphics[width=0.49\textwidth]{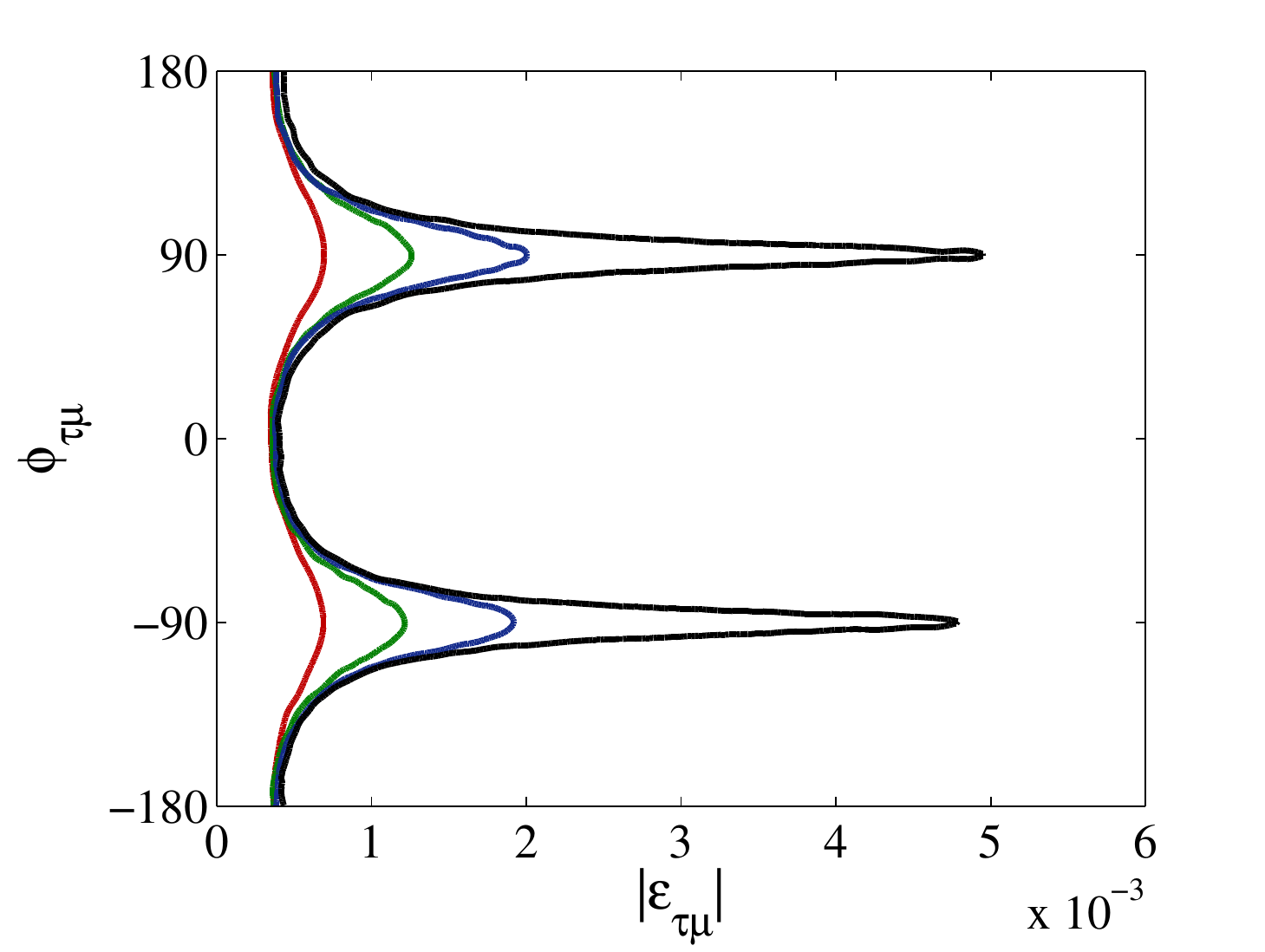}
  \includegraphics[width=0.49\textwidth]{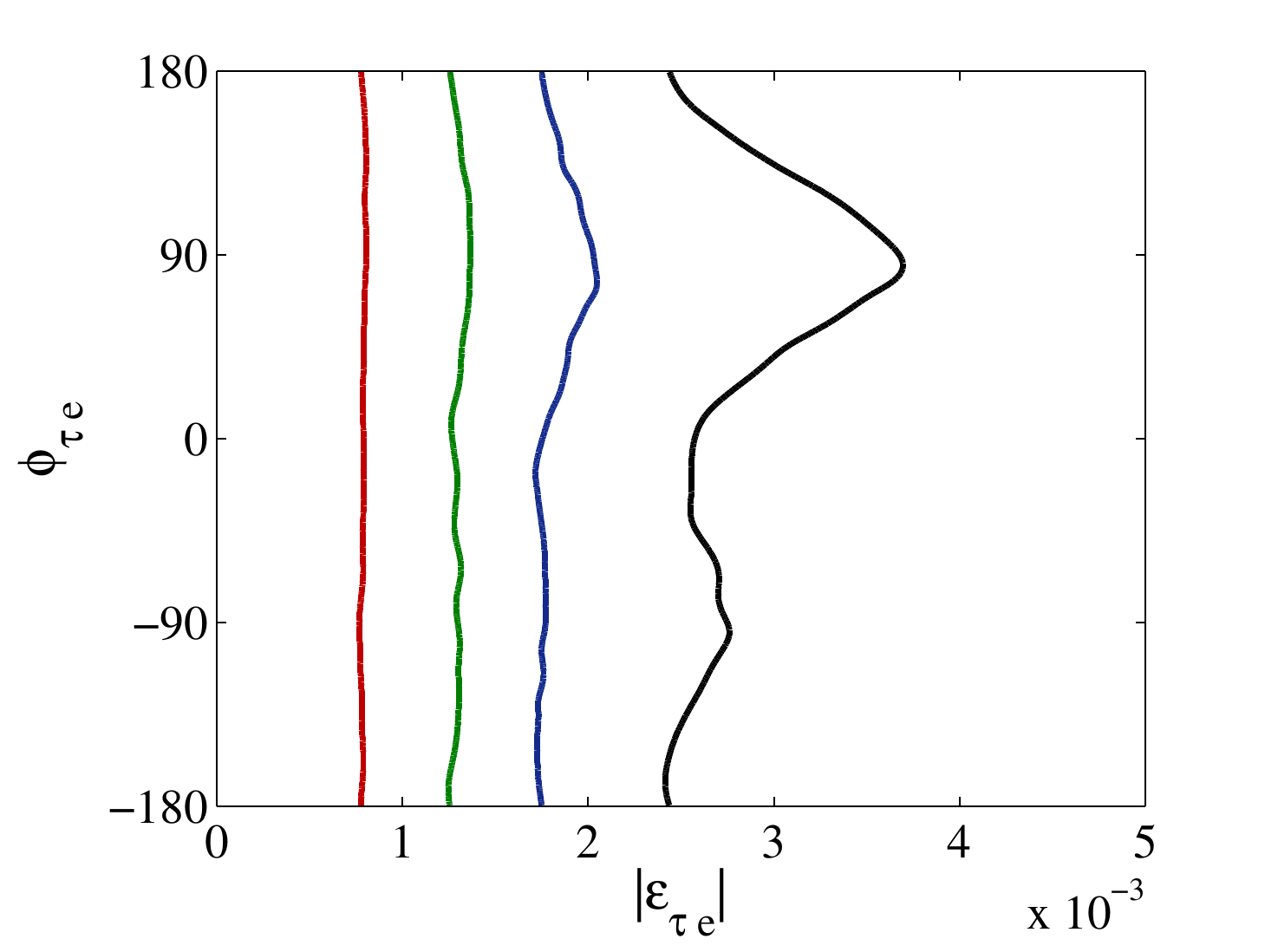}
 \end{center}
 \caption{ The 90~\% confidence level sensitivity of the IDS Neutrino
   Factory to the unitarity violating parameters $\eta_{\tau\mu} =
   |\varepsilon_{\tau\mu}|e^{i\phi_{\tau\mu}}$ (left) and $\eta_{\tau
     e} = |\varepsilon_{\tau e}|e^{i\phi_{\tau e}}$ (right). The
   different curves correspond to different sizes of the near $\tau$
   detector, from left to right, 10~kTon, 1~kTon, 100~ton, no near
   detector.  Figure taken from reference~\cite{Antusch:2009pm}.}
 \label{fig:sens}
\end{figure}

In figure \ref{fig:sens} the sensitivity limits on the parameters 
$\eta_{\tau\mu} = |\varepsilon_{\tau\mu}|e^{i\phi_{\tau\mu}}$ and 
$\eta_{\tau e} = |\varepsilon_{\tau e}|e^{i\phi_{\tau e}}$ are
depicted. 
The different curves correspond to different sizes of the $\tau$
detector close to the muon storage ring; from left to right, 
10~kTon, 1~kTon, 100~ton, no near detector. 
As can be seen, the sensitivity to
the real part of $\eta_{\mu \tau}$ is remarkable, down to $\sim 5 \times
10^{-4}$ without the inclusion of any near detector. This stems from the fact
that matter effects in the $\nu_\mu$ disappearance channel depend linearly on
this parameter. The inclusion of near detectors able to detect $\nu_\tau$
appearance with increasing size improves also the sensitivity to the imaginary
part of $\eta_{\mu \tau}$, since the zero-distance effect only depends on the
modulus, but the sensitivity is not as good, reaching $\sim 10^{-3}$ for a
1~kTon detector, since the effect is second order. 
A larger detector would thus
be required to improve over present limits. 
Regarding $\eta_{e \tau}$, the
sensitivity of the Neutrino Factory in the absence of a near
detector sensitive to $\tau$ production is much poorer, down to 
$\sim 2 \times 10^{-3}$ for some values of the phase. 
This mainly stems
from second-order terms in the $\nu_\mu$ appearance channel. The inclusion of
near detectors to search for $\tau$ appearance again improves the sensitivity.
However, as for the $\eta_{\mu \tau}$ element, detectors larger than 1~kTon
would be needed to improve over present bounds.  

\subsubsection{Discovery reach for sterile neutrinos}

Sterile neutrinos (Standard Model singlet fermions) are a very common
prediction in many models of new physics and, if they are light and mix with
the three standard neutrinos, it is possible to search for them at neutrino
oscillation experiments. Sterile neutrinos have received a lot of attention in
the context of seemingly anomalous results from the LSND and MiniBooNE
experiments~\cite{Aguilar:2001ty, AguilarArevalo:2010wv}.
Here, we will discuss
the discovery reach for sterile neutrinos at the Neutrino Factory.  We consider
the simplest scenario in which only one sterile neutrino is added to the three
active neutrinos of the Standard Model in the so-called 3+1 scheme, which
recovers the standard picture in the case of small active-sterile
mixings.  
For the sake of simplicity, we focus only on the scheme where the
fourth state is the heaviest and the normal hierarchy is assumed in
the standard sector. 

\paragraph{Theoretical formalism/notation}
\label{sec:ana}
The numerical results on the discovery reach can be understood from the
analytical expressions of the transition probabilities.  Instead of using a
parametrisation-independent approach, where all the probabilities are directly
expressed in terms of matrix elements of the $4 \times 4$ unitary mixing matrix
$U$, we rely on a particular  parametrisation which allows the
standard PMNS matrix to be recovered for small ``new'' mixing angles: 
\begin{equation}
    \label{equ:3+1param1}
    U =
    R_{34}(\theta_{34} ,\, 0) \; R_{24}(\theta_{24} ,\, 0) \;
    R_{14}(\theta_{14} ,\, 0) \;
    R_{23}(\theta_{23} ,\, \delta_3) \;
    R_{13}(\theta_{13} ,\, \delta_2) \; 
    R_{12}(\theta_{12} ,\, \delta_1) \,,
\end{equation}
where $R_{ij}(\theta_{ij},\ \delta_l)$ are the complex rotation matrices in the
$ij$-plane.  In the short-baseline limit $|\Delta_{41}|= \Delta m^2_{41}L/4 E
\sim \mathcal{O}(1) \gg |\Delta_{31}|$,  the matter effects can be safely
ignored, and the relevant probabilities read:
\begin{align}
&\mathcal{P}_{e\mu}= \mathcal{P}_{\mu e} = 4 c_{14}^2 s_{14}^2 s_{24}^2 \sin ^2 \Delta_{41} \label{equ:pem2}\,;  \\
&\mathcal{P}_{ee}=1-\sin ^2\left(2 \theta _{14}\right)\sin ^2 \Delta_{41}  \label{equ:pee2}\,;\\
&\mathcal{P}_{\mu \tau}= 4 c_{14}^4 c_{24}^2 s_{24}^2 s_{34}^2 \sin ^2
  \Delta_{41}\,; {\rm and} \\
&\mathcal{P}_{\mu\mu}=1-c_{14}^2 s_{24}^2  \left[3+2 c_{14}^2 \cos \left(2 \theta _{24}\right)-\cos \left(2 \theta _{14}\right)\right]\sin ^2 \Delta_{41}  \label{equ:pmm2}\,;
\end{align}
where we used the short-hand notation $\Delta_{ij}= \Delta m^2_{ij} L/4 E$.
From these probabilities we can see that $\theta_{24}$ can be measured by
$\mathcal{P}_{\mu \mu}$  and  $\theta_{14}$  by   $\mathcal{P}_{e e}$. On the
other hand, $\theta_{34}$ only shows up in combination with the other small
mixing angles.  For long baselines,  some of the relevant features of the
probability transitions can be well understood using simple perturbative
expansions: for $\Delta_{31} = \mathcal{O}(1) \ll \Delta_{41}$ and up to the
second order in
$
s_{13},  s_{14},  s_{24},  s_{34}, \hat s_{23}=\sin\theta_{23}-\frac{1}{\sqrt{2}}
$,
and considering
$
\Delta_{21}
$
as small as $s_{ij}^2$,
we obtain:
\begin{eqnarray}
\mathcal P_{\mu\mu} &=& \cos^2 (\Delta_{31}) (1 - 2 s_{24}^2) + 8 \hat s_{23}^2\sin^2(\Delta_{31}) + c_{12}^2 \Delta_{12}   \sin(2\Delta_{31}) +  \nonumber \\
&&   2 s_{24} s_{34} \cos \delta_3 \Delta_n  \sin(2\Delta_{31}) - \label{pmumulo}\\ 
&& 2 s_{13}^2 \Delta_{31}\cos( \Delta_{31})\,\frac{(\Delta_{31}-\Delta_e) \Delta_e \sin (\Delta_{31})-\Delta_{31}\sin\left(\Delta_{31}-\Delta_e \right)\sin (\Delta_e )}{(\Delta_{31}-\Delta_e)^2} \, ;
 and \nonumber \\
\mathcal P_{\mu\tau} &=&\sin^2 (\Delta_{31})(1-8 \hat s_{23}^2-s_{24}^2-s_{34}^2)-c_{12}^2 \Delta_{12} \sin (2\Delta_{31})-\nonumber \\
&& s_{24}s_{34}\sin (2\Delta_{31})\left[2 \Delta_n \cos \delta_3-\sin\delta_3\right] - \label{pmutaulo}  \\ 
&& s_{13}^2 \Delta_{31}\sin \Delta_{31} \,\frac{\Delta_{31}
\left\{\sin (\Delta_{31}-\Delta_e) +\sin (\Delta_e )\right\}-2 (\Delta_{31}-\Delta_e) 
\Delta_e \cos (\Delta_{31})}{(\Delta_{31}-\Delta_e)^2} \, ;\nonumber 
\end{eqnarray}
from which we learn that with long baselines, $\theta_{24}$ is best
accessed via $\mathcal{P}_{\mu \mu}$ with the first term proportional to
$\cos^2 (\Delta_{13})$.  The leading sensitivity to $\theta_{34}$ can be
expected from $\mathcal{P}_{\mu \tau}$  (the {\it discovery channel} as introduced
in \cite{Donini:2008wz}).  Notice also the dependence on the phase $\delta_3$
in both probabilities, which makes them useful to check whether other sources of
CP violation beside the standard one ($\delta_2$ in our parametrisation) can
be tested at the Neutrino Factory.

\paragraph{Results for the IDS-NF baseline}

\begin{figure}
\includegraphics[width=0.32\textwidth]{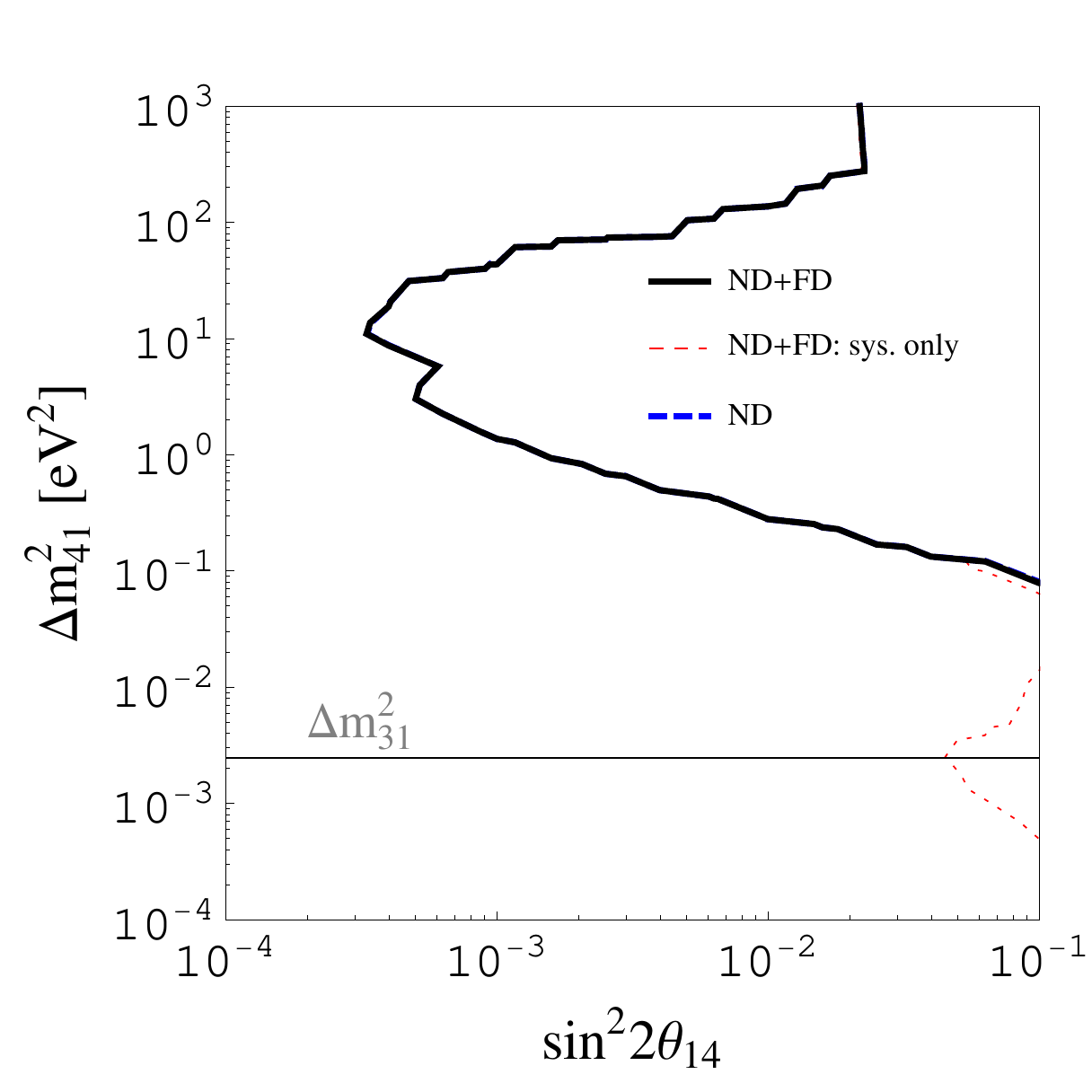}
\includegraphics[width=0.32\textwidth]{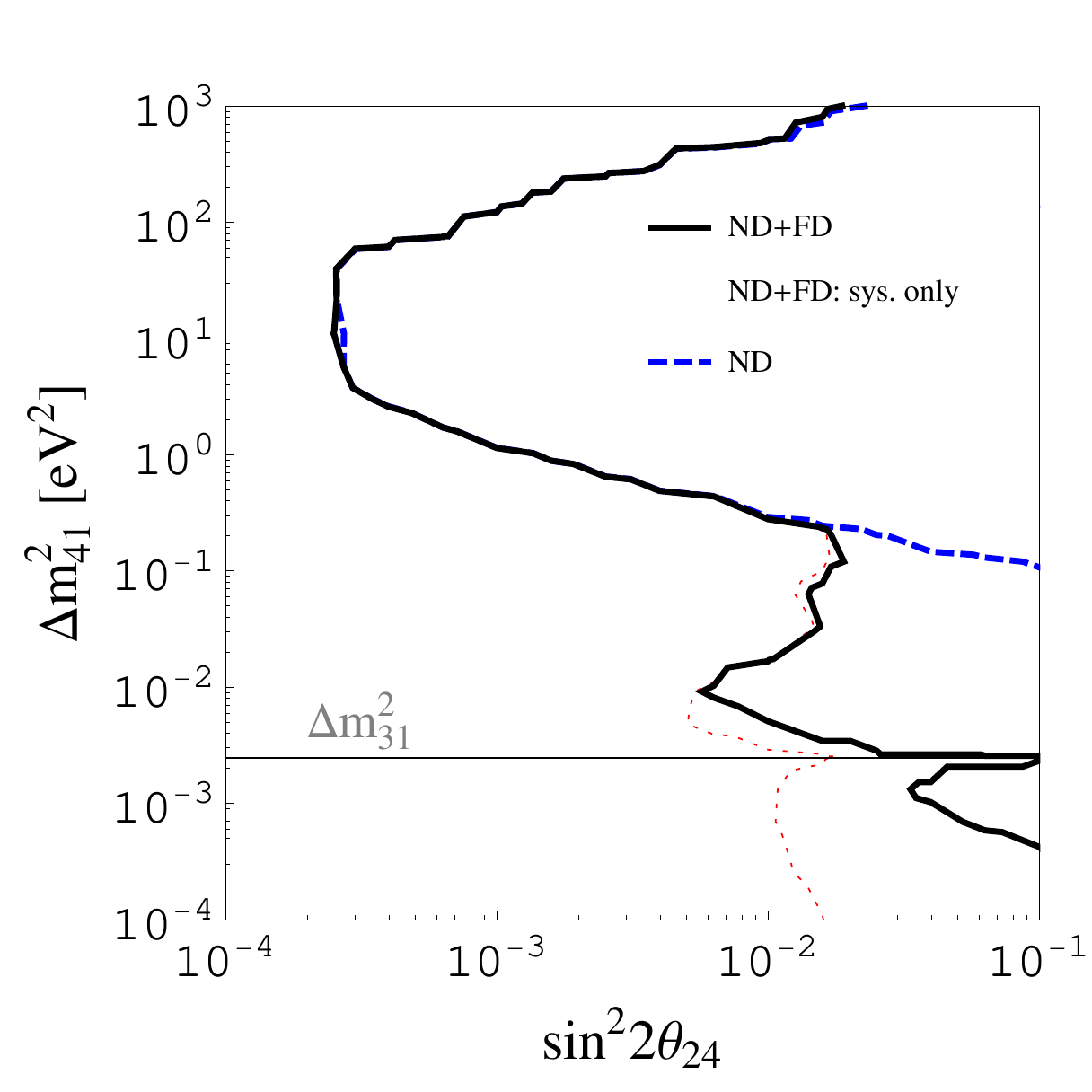}
\includegraphics[width=0.32\textwidth]{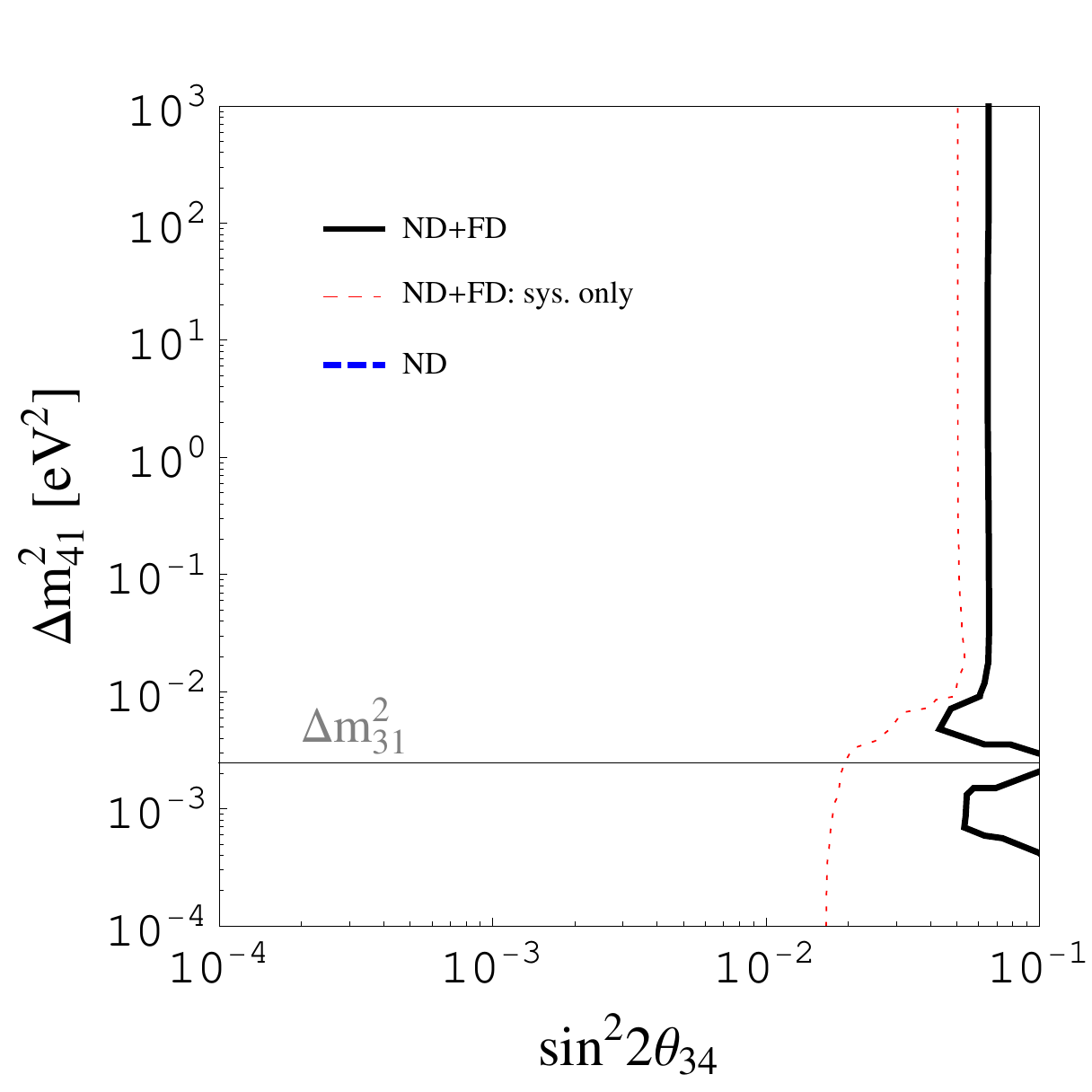}
\caption{
  The exclusion limit at 90\% CL (2 d.o.f) for
  $\sin^22\theta_{i4}$--$\Delta m_{41}^2 (i=1,2,3)$ (region on r.h.s.\ of
  curves excluded). We have assumed a Neutrino Factory with two 50\,kTon
  magnetised iron far detectors (FD) at  $4\,000$~km and $7\,500$~km, and with
  32\,Ton near detectors (ND) at $\sim 2$~km. The thick solid black curve is for
  the full ND + FD setup, while the dashed blue curve shows the impact of the
  near detectors separately. The thin red dotted curve illustrates how the
  sensitivity would improve if parameter correlations and degeneracies were
  absent, so that only statistical and systematical uncertainties would
  contribute. Figure taken from reference~\cite{Meloni:2010zr}.}
\label{fig:theta-mass}
\end{figure}

We first discuss general constraints on the new mixing angles $\theta_{14}$,
$\theta_{24}$, and $\theta_{34}$,  and on the additional mass squared
difference $\Delta m_{41}^2$ without any additional assumptions
\cite{Meloni:2010zr}.  
The oscillation channels considered, simulated with a
modified version of the GLoBES software, are electron-to-muon neutrino
oscillations (appearance channels) and muon-to-muon neutrino
oscillations (disappearance channels) (see \cite{Meloni:2010zr} for
all details of the simulations).  
In addition to the IDS-NF baseline setup, we have also considered
near detectors with a fiducial mass of $32$\,Ton and a distance of
$d=2$~km from the end of the decay straight, which corresponds to an
effective baseline of 2.28~km. 
We show the 
resulting exclusion limits in the $\theta_{i4}$---$\Delta m_{41}^2$ planes in
figure \ref{fig:theta-mass}.

The main sensitivity is obtained at about $\Delta m_{41}^2 \simeq 10 \,
\mathrm{eV}^2$, which comes from the distance chosen for the near detectors.
Since the efficiencies for muon-neutrino detection are typically better than
those for $\nu_e$, the sensitivity to $\theta_{24}$ is slightly better than
that to $\theta_{14}$ for our assumptions. As expected, there is no sensitivity
to $\theta_{34}$ coming from the near detectors, because the $\nu_\tau$
disappearance channel does not exist.  For the effect at long baselines, it is
useful to consider first the thin dashed curves in figure \ref{fig:theta-mass},
which correspond to a simulation including only statistical and systematic
errors, but no parameter correlations. In all three panels, the sensitivity
changes as a function of $\Delta m_{41}^2$ in the region where $\Delta m_{41}^2
\sim \Delta m_{31}^2$. This is due to the fact that the Neutrino Factory is
sensitive to the atmospheric oscillation frequency, whereas for $\Delta
m_{41}^2 \sim \Delta m_{21}^2$, no particular additional effects from the solar
frequency can be found. As expected (see equation (\ref{pmumulo})), the main
sensitivity is found for $\theta_{24}$.  However, there is also some
sensitivity to $\theta_{14}$, which vanishes after the marginalisation, and
some sensitivity to $\theta_{34}$, which is even present for $\Delta
m_{41}^2=0$ for systematics only. After we include parameter correlations by
marginalising over all free oscillation parameters (thick solid curves), only
the sensitivities to $\theta_{24}$ and $\theta_{34}$ remain in the $\Delta
m_{41}^2$ regions close to the atmospheric $\Delta m_{31}^2$ and above, where
the effects of $\Delta m_{41}^2$ average out. Very interestingly, note that
mixing-angle correlations destroy the sensitivities for $\Delta m_{41}^2=\Delta
m_{31}^2$, where $m_4=m_3$ and no additional $\Delta m_{41}^2$ is observable,
leading to small gaps (see horizontal lines).  We have tested that the
sensitivity to $\theta_{34}$ is a matter-potential-driven, statistics-limited,
higher-order effect present in the muon-neutrino disappearance channels.  We
see from the three panels of figure \ref{fig:theta-mass} that it is not easy to
disentangle the parameters for arbitrarily-massive sterile neutrinos.  Parameter
correlations lead to a pollution of the exclusion limit of a particular mixing
angle with $\Delta m_{41}^2$. In addition, there is a competition between
$\Delta m_{41}^2$ and $\Delta m_{31}^2$ at the long baseline. Near detectors,
on the other hand, have very good sensitivities to $\theta_{14}$ and
$\theta_{24}$ but cannot measure $\theta_{34}$. Nevertheless, the absolute
values of the sensitivities are quite impressive. Especially, $\theta_{24}$ can
be very well constrained close to the atmospheric mass squared difference
range. This indicates that sterile neutrino bounds in that range should be also
obtainable from current atmospheric neutrino oscillation experiments.  Note
that while one may expect some effect from $\mathcal P_{\mu \tau}$, we have
tested the impact of $\nu_\tau$ detectors at all baselines, and we have not
found any improvement of the sensitivities.  The reason is that the
$\theta_{34}$ effect at the long baseline comes with the same energy dependence
as the $\theta_{24}$ effect, which means that one cannot disentangle these, as
discussed in section \ref{sec:ana}.

\paragraph{Optimisation issues and sensitivity of alternative setups}

Although a similar analysis for CP-violating signals has not been carried out,
the authors of reference \cite{Donini:2008wz} have presented a first
analysis of the precision achievable in the simultaneous measurement
of mixing angles and CP-violating phases (but see
\cite{Donini:1999jc,Kalliomaki:1999ii,Donini:2001xp,Donini:2001xy,Dighe:2007uf,Donini:2008wz,Goswami:2008mi,Giunti:2009en,Yasuda:2010rj}
for preliminary analyses). The experimental setup 
adopted there is a Neutrino Factory with 50 GeV stored muons, with two
detectors of the hybrid-MIND type (a magnetised emulsion cloud chamber
followed by a magnetised iron tracking calorimeter), located at 
$L = 3\,000$~km and $7\,500$~km. 
The $\nu_\tau$ 
identification turns out to be of primary importance for the study of CP
violation. As an example, for the simultaneous measurement of $\theta_{34}$ and
$\delta_3$ it has been shown that,  using the $\nu_\mu$ disappearance channel only,
we are able to measure a non-vanishing $\delta_3$ for values of $\theta_{34}$
above $\sin^2 2 \theta_{34} \geq 0.4\ (\theta_{34} \geq 18^\circ)$,  and that
the detector at $L = 3\,000$~km has no $\delta_3$-sensitivity
whatsoever.  
Adding 
the $\nu_\mu \to \nu_\tau$ discovery channel data, the $L=3000$ km detector is
no longer useless for the measurement of $\delta_3$: spikes of
$\delta_3$-sensitivity for particular values of $\delta_3$ can be
observed, in some cases outperforming the far-detector results. 
However, it is in the combination of the two
baselines that a dramatic improvement in the $\delta_3$-discovery potential
is achieved.  When the $\nu_\mu \to \nu_\tau$ data are included, a
non-vanishing $\delta_3$ can be measured for values of $\theta_{34}$ as small
as $\sin^2 2 \theta_{34} = 0.06\ (\theta_{34} = 7^\circ)$ for $\theta_{24} =
5^\circ$ and $\sin^2 2 \theta_{34} = 0.10\ (\theta_{34} = 9^\circ)$ for
$\theta_{24} = 3^\circ$.

\subsubsection{New physics summary}

We have shown that, in addition to its unique capability of clarifying the
flavour structure of neutrinos in the standard framework, the Neutrino Factory
is also able to probe possible modifications of the oscillation pattern due to
new physics. For example, it is very sensitive to non-standard matter effects,
non-standard charged current interactions, non-unitarity in the leptonic mixing
matrix, and sterile neutrinos. Especially for non-standard effects modifying
the neutrino-production and detection processes and, for short-baseline
oscillations into sterile neutrinos, a crucial role is played by the near
detector.  A very important aspect of new physics searches at a Neutrino
Factory is the possibility to identify new sources of CP violation by studying
their effect on the oscillation pattern. Many other electroweak precision
experiments do not rely on interference phenomena and are therefore insensitive
to CP-violating phase factors.

Quantitatively, we have shown that the constraints the Neutrino Factory can set
on non-unitarity in the mixing matrix and on active-sterile neutrino mixing are
comparable to or slightly better than the existing ones, while current
model-independent bounds on non-standard neutrino interactions can be
significantly improved. The Neutrino Factory could probe interactions
100 times weaker than Standard Model weak interactions, and the low energy
option, though less sensitive, would still allow for improvements compared to
existing bounds.

The NSI sensitivity of the Neutrino Factory naively translates into
sensitivity to new physics scales of several 100~GeV, implying that
the Neutrino Factory could provide measurements complementary to those
expected from the LHC.  On the other hand, in many concrete models,
non-standard neutrino interactions large enough to be seen at the
Neutrino Factory are already ruled out by other experiments probing
correlated effects in the charged-lepton sector.  Therefore, from a
model-builder's point of view the new physics discovery reach of the
Neutrino Factory is limited.  However, as neutrinos have proved
theoretical prejudices to be wrong in the past, the versatility and
precision of a Neutrino Factory will still make new physics searches a
worthwhile and integral part of its physics program.

\subsection{Physics Summary}
\label{sec:physics-summary}

Neutrino oscillations and the associated large leptonic family mixing
are arguably the first clear-cut discovery of physics beyond the
Standard Model that is not obtained from cosmology. Large mixing
angles came as a surprise and, while we have been able to formulate a
plethora of theories which can accommodate these results, we are far
from an actual understanding of mixing in particular and flavour in
general. This lack of understanding is also exemplified by the
non-observation of new physics in B-factories.  In some models
of neutrino mass generation, there is a connection between neutrinos
and Grand Unification. In this class, called see-saw models, neutrinos
can be ultimately responsible for the observed matter-antimatter
asymmetry of the Universe. In the first place, this requires neutrinos
themselves to violate matter-antimatter symmetry. The origin of flavour
is one of the profound open questions in particle physics and studies
of neutrino oscillations will play a crucial part in providing the
data, which is necessary to make progress towards an answer.

The Neutrino Factory will be the ultimate neutrino physics facility
and provides a superior performance for CP violation, the mass
hierarchy and $\theta_{13}$, see figure~\ref{fig:euronu2009}. At the
same time, it provides the necessary precision to test rigorously the
three-flavour oscillation framework and thus, also has the potential
to discover new physics.  
The IDS-NF baseline configuration for the Neutrino Factory consists of
one 100\,kTon MIND at a baseline of between 2\,500\,km and 5\,000\,km,
to  determine the CP phase, and one 50\,kTon MIND at a baseline of
between 7\,000\,km and 8\,000\,km to determine $\theta_{13}$ and the
mass hierarchy. 
The stored muon energy is 25\,GeV and
$10^{21}$ useful muon decays per year are available. This baseline
configuration has been derived under the assumption that
$\stheta<0.01$, i.e., $\theta_{13}\neq0$ has not been discovered
prior to the start of the Neutrino Factory project. In this case, the
second, longer baseline is necessary to ensure a robust physics
performance throughout the parameter space, see
figure~\ref{fig:robustness}, and to provide the best possible
sensitivity to new physics.  On the other hand, if $\stheta>0.01$,
then it is very likely that $\stheta$ will be measured before we
embark onto the Neutrino Factory project and hence we would use this
information to optimise the design of the facility. 
Thus, in this case, we would use a
single baseline between 1\,500---2\,500\,km and a stored muon energy
around 10\,GeV, see top left panel of figure~\ref{fig:elopt}.  
The detector
can still be a 100\,kTon MIND, since recent updates of the MIND
analysis, in particular the inclusion of quasi-elastic events, have
improved the low-energy performance significantly as explained in
detail in section~\ref{sec:MIND}.

In terms of new physics sensitivities, we have studied sterile
neutrinos and non-standard interactions as generic examples and find
the Neutrino Factory, in its baseline configuration, provides good
sensitivity to physics that cannot be studied at the LHC or
elsewhere. In addition, a Neutrino Factory, in most cases, would be
able to disentangle a new physics contribution from the usual
oscillations. 

We, as a community, have been searching for new physics for many
decades in vain. The discovery of neutrino mass is the first tangible
sign for physics beyond the Standard Model. It seems therefore, natural
and compelling to follow up on this initial discovery by a dedicated,
long term science program with the goal to establish the flavour
physics of neutrinos at a level of precision comparable to that at
which we have tested the CKM description of quark flavour. The results
presented in this section are supported by a large body of literature
and demonstrate clearly that a Neutrino Factory is the ultimate
tool to study neutrinos and their flavour transitions with the highest
precision.

\clearpage
%
\section{The Neutrino Factory accelerator complex}
\label{Sect:AccWG}
\subsection{Overview}
\label{Sect:AccOview}

\begin{figure}
  \begin{center}
    \includegraphics[width=0.9\linewidth]{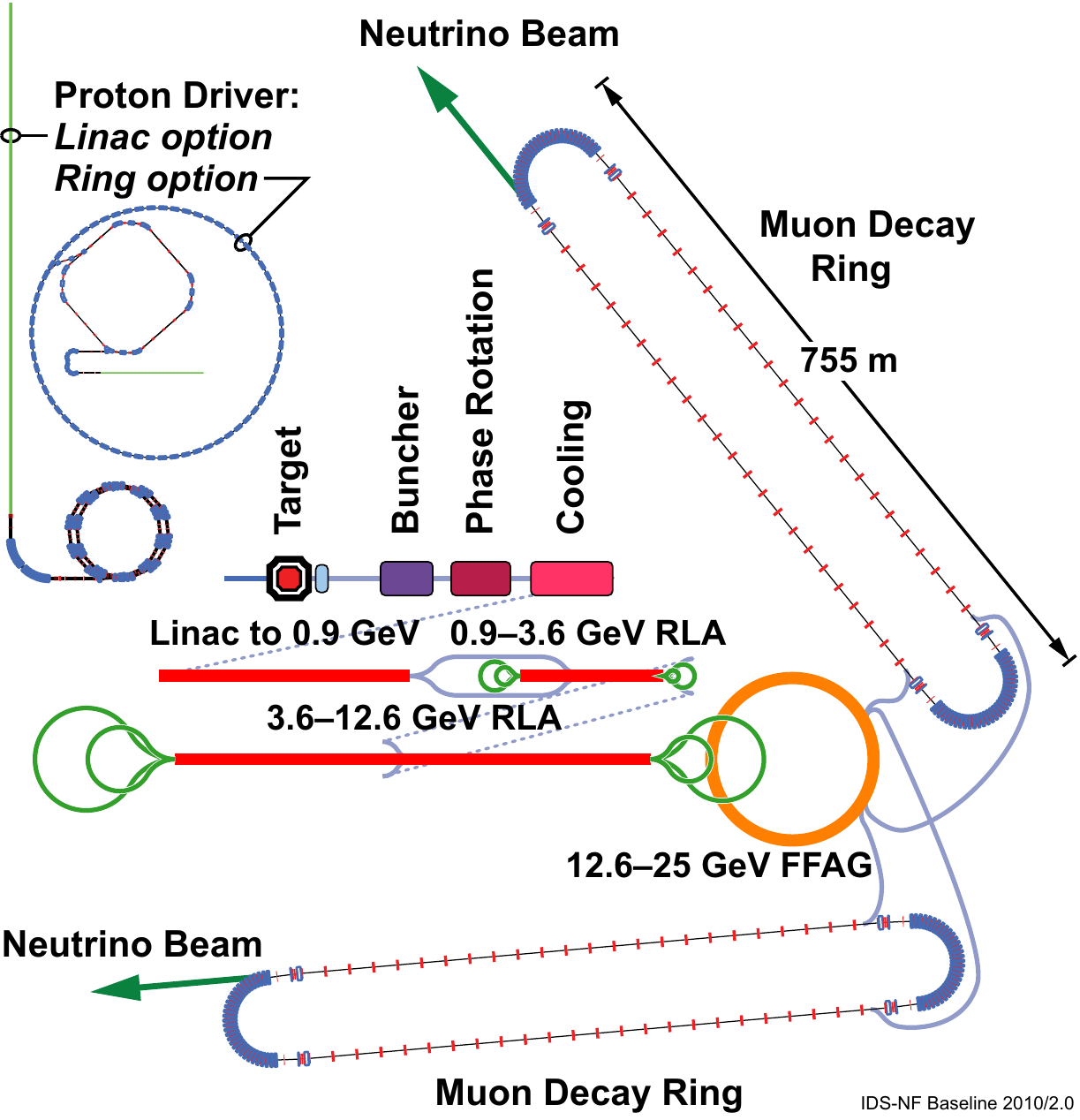}
  \end{center}
  \caption{Schematic drawing of the IDS-NF accelerator complex.}
  \label{Fig:AccNFSchematic}
\end{figure}

The Neutrino Factory accelerator systems are required to produce a
high-intensity, high-energy neutrino beam with a very well-known
energy spectrum and without the contamination of unwanted neutrino flavours.  
The energy spectrum is well-defined since it arises
from the decay of muons confined within a storage ring with a
relatively small energy spread. 
The muon beam is accelerated to high energy thereby creating a
high-energy neutrino spectrum.
The accelerator facility (shown schematically in figure
\ref{Fig:AccNFSchematic} produces a high-intensity neutrino beam by 
creating large numbers of muons, capturing as many of them as
possible, and minimising the loss of these muons before they can be
directed toward the detector.   
Table \ref{tab:acc:muons} quantifies these characteristics: it gives
the muon beam energy and the number of muon decays in the direction of
the neutrino detector.
The uncertainty in the neutrino flux is limited by the angular
divergence of the stored muon beam \cite{Crisan:2000gc}.
Table \ref{tab:acc:muons}, therefore, also specifies the maximum
angular divergence of the muon beam.
Two storage rings are required to serve simultaneously both the
detector at the intermediate baseline and that at the long baseline.
The decay rings have straight sections inclined so that the straight
sections point at the distant detectors.  
The lengths of the straight sections form a significant fraction of
the decay ring circumference so that the fraction of the stored muons
decaying in the direction of the detectors is maximised.  
\begin{table}[hb]
  \caption{
    Parameters characterising the muon beam produced by the
    accelerator facility.  Muon decays are a total for all
    signs and detector baselines.
  }
  \label{tab:acc:muons}
  \begin{tabular}{|l|r|}
    \hline
    {\bf Parameter}                                                & {\bf Value}       \\
    \hline
    Muon total energy                                              & 25 GeV            \\
    Production straight muon decays in $10^7$ s                    & $10^{21}$          \\
    Maximum RMS angular divergence of muons in production straight & $0.1/\gamma$      \\
    Distance to intermediate baseline detector                     & 2\,500--5\,000 km \\
    Distance to long baseline detector                             & 7\,000--8\,000 km \\
    \hline
  \end{tabular}
\end{table}

\subsubsection{Principal subsystems}

Muon beams at the Neutrino Factory are produced from the decays in
flight of pions produced by the bombardment of a target by a
high-power, pulsed proton beam.
The muons produced in this way must be captured and manipulated in
6-D phase space such that the maximum number are transported into the
phase-space region that is required for acceleration and
storage in the decay ring.  
After the phase-space manipulation, the muon beam energy is increased
to 25\,GeV in a number of acceleration stages.
Finally, the beam is injected into the storage ring, where it
circulates as the muons decay into neutrinos. 

The number of muons created is approximately proportional to the power
in the proton beam, defined as the product of the number of protons
per bunch, the energy of each proton, and the rate at which bunches
hit the target.  
The power of the proton beam is chosen to be sufficient to produce the
desired number of muons (see table~\ref{tab:acc:muons}).
Other parameters are chosen such that the number of muons produced per
unit power is nearly optimal (see table~\ref{tab:acc:pd:parm}) given
the method we are using to capture the muons.  
This report will not define a specific proton driver for the Neutrino
Factory.
Rather, we describe three designs that could be implemented by
appropriately developing infrastructure at three of the world's
proton-accelerator laboratories (CERN, FNAL, and RAL).
These laboratories have been chosen as example sites to allow
practical considerations to be evaluated.
The range of proton-driver technologies being studied is
matched to the constraints of the example sites and covers the
technologies most likely to be employed.
We believe that this strategy will allow us to demonstrate that it is
feasible to deliver the required beam parameters from a number of
possible sites and using a variety of technologies.

The target that the proton beam hits is a liquid mercury jet.  
The liquid
jet avoids issues of structural damage from the beam that a solid
target faces.  
The proton beam hits the target inside a 20~T solenoid,
which tapers down to 1.5\,T in a distance of 15\,m.  
The high field and the taper give a large angular acceptance for the
pions at the target. 
The solenoid channel allows both muon signs to be captured.

The pion beam from the target decays into muons in a long decay channel,
and the resulting muon beam has a large energy spread as well as an
energy-time correlation, the latter arising because the pions are not
highly relativistic.
A sequence of RF cavities, the frequency of which decreases with
distance, then turns this distribution into a train of bunches. 
A subsequent sequence of cavities, with frequency decreasing down
the channel, then changes the energies of the individual bunches so that
all the bunches have the same energy.
These ``bunching'' and ``phase-rotation'' sections are followed by an
ionisation-cooling channel, which reduces the transverse emittance of
the bunches, thereby increasing the number of muons that can be
transmitted into the acceptance of the acceleration systems that
follow.  

Acceleration of the muon beam occurs in a number of stages.  
The first stage is a linac, which accelerates the beam to 0.9~GeV total
energy. 
This is followed by two recirculating linear accelerators (RLAs),
taking the beam to 3.6~GeV and 12.6~GeV, respectively.  
RLAs are preferable to a series of single-pass linacs since the beam
makes multiple passes through the RF cavities (4.5 passes in our
design), consequently reducing the cost of the acceleration system.
The RLA technique cannot be applied at lower energies, however,
since the beam has a low velocity which, combined with its large
relative-energy spread and large geometric emittance, would make a
linac phased for a higher energy inefficient at the lower, initial
energy.
The final stage of acceleration is performed in a fixed-field
alternating-gradient (FFAG) accelerator, which permits even more
passes (11) through the RF cavities.
The switchyard in an RLA will not permit such a large number of passes,
while the FFAG cannot achieve this large number number of turns at
lower energies.  
Each acceleration stage is thus chosen to use the most efficient
acceleration technique for its energy range.

The full-energy beam is next injected into two racetrack-shaped decay
rings.  
Each decay ring points toward one of the far detectors (the
``intermediate'' baseline or the ``long'' baseline).  
Each decay ring is capable of storing both muon signs simultaneously.

\subsubsection{Progress within the IDS-NF accelerator study to date}

The IDS-NF adopted, with slight modifications, the specification for
the Neutrino Factory accelerator complex developed by the Accelerator
Working Group of the International Scoping Study of a future Neutrino
Factory and super-beam facility (the ISS) \cite{Apollonio:2009}.
The initial baseline (referred to as 2007/1.0) is defined in
\cite{IDS-NF-002}.
The specification of the accelerator facility presented below
(referred to as 2010/1.0) is the
result of a substantial amount of work and constitutes a revision of
the initial baseline \cite{IDS-NF-020}.
The principal developments that have resulted from the work of the
Accelerator Working Group of the IDS-NF or that have informed the
choice of baseline are summarised below.

The MERIT experiment, a test of a liquid-mercury-jet target,
has been completed, and the
data have been analysed~\cite{McDonald:2009zz}.
The mercury jet was in a high-field solenoid and was hit by a proton
beam from the CERN PS.
The experiment demonstrated that such a target is able to withstand
the high power-density that will be present at the Neutrino Factory.
MERIT also showed that if two bunches hit the target in rapid succession,
the disruption of the mercury jet caused by the first bunch does not
reduce the number of pions produced by the second bunch unless the
bunches are separated by more than 350~$\mu$s (because the disruption
increases with time).  
This result is important for the specification of the time structure
of the bunches coming from the proton driver when beam loading in the
RF cavities in the muon acceleration system is considered. 
In addition, the observed disruption length combined with the speed of
the jet allow operation at repetition rates of up to 70\,Hz.

The geometry of the proton beam and mercury jet has been optimised as
a function of energy \cite{Ding:2009zz,Ding:2010zzd}.  
The performance of the target was improved as a result, and the energy
dependence of the pion-production rate was computed more precisely.
The effect of a non-zero beam divergence was computed.
As a result of this calculation, a requirement for the proton-beam
emittance was determined (see table \ref{tab:acc:pd:parm}).
Comparisons were made between
MARS~\cite{Mokhov:1995wa,Mokhov:2000ih,Mokhov:2003jq,Mokhov:2004aa,mars}
(version 1507 is used throughout this document)
and FLUKA~\cite{Fluka1,Fluka2}
simulations for this
target system~\cite{Back1:2010}.

Very detailed studies were made of the energy deposition in the target
region~\cite{Ding:2010zzd,Back2:2010}.  
The studies indicate that the energy deposition in the superconducting
solenoids in the target region is too large,
so the solenoids in the target region
will be redesigned to allow more room for shielding.

In the muon front-end, the bunching and phase-rotation sections have
been shortened, creating a system that produces the same number of
muons in a shorter bunch train~\cite{Ankenbrandt:2009zza,Neuffer:2003zz}.  
Studies of the performance of RF cavities in magnetic fields have
continued~\cite{Huang:2009zzn} and the MICE experiment continues to
make progress towards a demonstration of the ionisation cooling
technique~\cite{Alekou:2010zz}.  
Indications that the maximum cavity-gradient may be reduced in
magnetic fields have led to the consideration of front-end lattices
that give adequate performance should the RF cavities not
reach the gradients specified for the magnetic field configuration
of the baseline lattice for the IDS-NF muon front end.
The performance of the baseline lattices, along with that
of the modified lattices, has been
computed~\cite{Gallardo:2010td,Stratakis:2010zz,Rogers:2009zzc,Alekou:2010zza,Neuffer:2009zz,Neuffer:2003zz}.
Particle losses and energy deposition in the front end have been computed,
and methods to reduce their impact are beginning to be
studied.

Tracking studies with more realistic fields
have been done on the first acceleration linac,
resulting in slight modifications to the lattice design.
The RLA designs have been improved by modifying the quadrupole
gradient profile in the linacs to increase
performance \cite{Bogacz:2009zzc}.  
A chicane has been designed for the locations where the arcs cross
each other \cite{Ankenbrandt:2009zza}.
In the arc-crossing and injection chicanes, chromatic
corrections have been incorporated and preliminary tracking results have
indicated that chromatic correction in these regions alone is
sufficient to achieve acceptable performance.

A new FFAG design has been developed that has a high average RF
gradient to reduce the effect of non-linear coupling of transverse
into longitudinal motion \cite{Berg:2009zz}.  
Injection and extraction schemes have been developed for the
FFAG ring \cite{Pasternak:2010zza} and we have begun to study the
designs of the injection and extraction kickers and septa.  
These preliminary studies indicated that longer drifts were necessary,
and as a result the FFAG design was modified.
Chromaticity correction in the FFAG has been studied more extensively
\cite{Berg:2009zz}.  

We have begun studies of diagnostics in the decay ring.  
We have a preliminary design for a polarimeter to be used for
beam-energy measurement \cite{Apollonio:2010zz}. 
In addition, a helium gas Cherenkov detector has been
considered for angular divergence measurement \cite{Piteira:2001}.
Other methods of beam-divergence measurement are now being studied as
the Cherenkov solution proved difficult.

\subsection{Proton driver}
\label{Sect:ProtonDriver}

The proton driver at the Neutrino Factory is required to deliver a
proton-beam of 4~MW at a repetition rate of 50 Hz to the
pion-production target.
The proton-beam energy must be in the multi-GeV range in order to
maximise the pion yield.  
In addition, the Neutrino Factory specifies a particular time structure
consisting of three very short bunches separated by about 120\,$\mu$s. 
To allow the muon beam to be captured efficiently, short, 1--3\,ns rms,
bunches are required. 
Each bunch from the proton driver will become a separate muon bunch train.
The bunch separation is constrained by beam loading in the downstream
muon accelerator systems and by the time scale for
disruption of the mercury-jet target.  
The proton beam parameters necessary to produce the desired number of
muons in the storage rings of the Neutrino Factory are listed in table
\ref{tab:acc:pd:parm}.
In order to achieve such short bunches, a dedicated bunch compression
system must be designed to deal with the very strong space-charge
forces.
Several proton-driver schemes fulfilling these requirements have been
proposed (see below and appendices).  
Typically they consist of an H$^-$-ion source followed by a
radio-frequency quadrupole (RFQ),
a chopper, and a linear accelerator.  
In some cases
the final energy of the proton driver is delivered by the linac. 
In these linac-based scenarios, the beam time structure must be obtained
with the help of charge-exchange injection into an accumulator ring
followed by fast phase rotation in a dedicated compressor ring. 
\begin{table}
  \caption{
    Proton driver requirements.
    A proton kinetic energy in the range 5~GeV to 15~GeV has been shown
    to provide adequate performance.
    The number of protons, beam radius, $\beta^*$, and geometric emittance
    (see section~\ref{Sect:Trgt})
    correspond to the values
    for an 8\,GeV proton beam.
  }
  \label{tab:acc:pd:parm}
  \centering
  \begin{tabular}{|l|r|}
    \hline
    {\bf Parameter}         & {\bf Value}                      \\
    \hline
    Kinetic energy          & 5--15 GeV                        \\
    Average beam power      & 4 MW                             \\
                            & ($3.125\times 10^{15}$ protons/s) \\
    Repetition rate         & 50 Hz                            \\
    Bunches per train       & 3                                \\
    Total time for bunches  & 240 $\mu$s                       \\
    Bunch length (rms)      & 1--3 ns                          \\
    Beam radius             & 1.2 mm (rms)                     \\
    Rms geometric emittance & $<5~\mu$m                        \\
    $\beta^*$ at target     & $\geq 30$~cm                     \\
    \hline
  \end{tabular}
\end{table}

\begin{figure}
  \begin{center}
    \includegraphics[angle=270,width=0.5\linewidth]{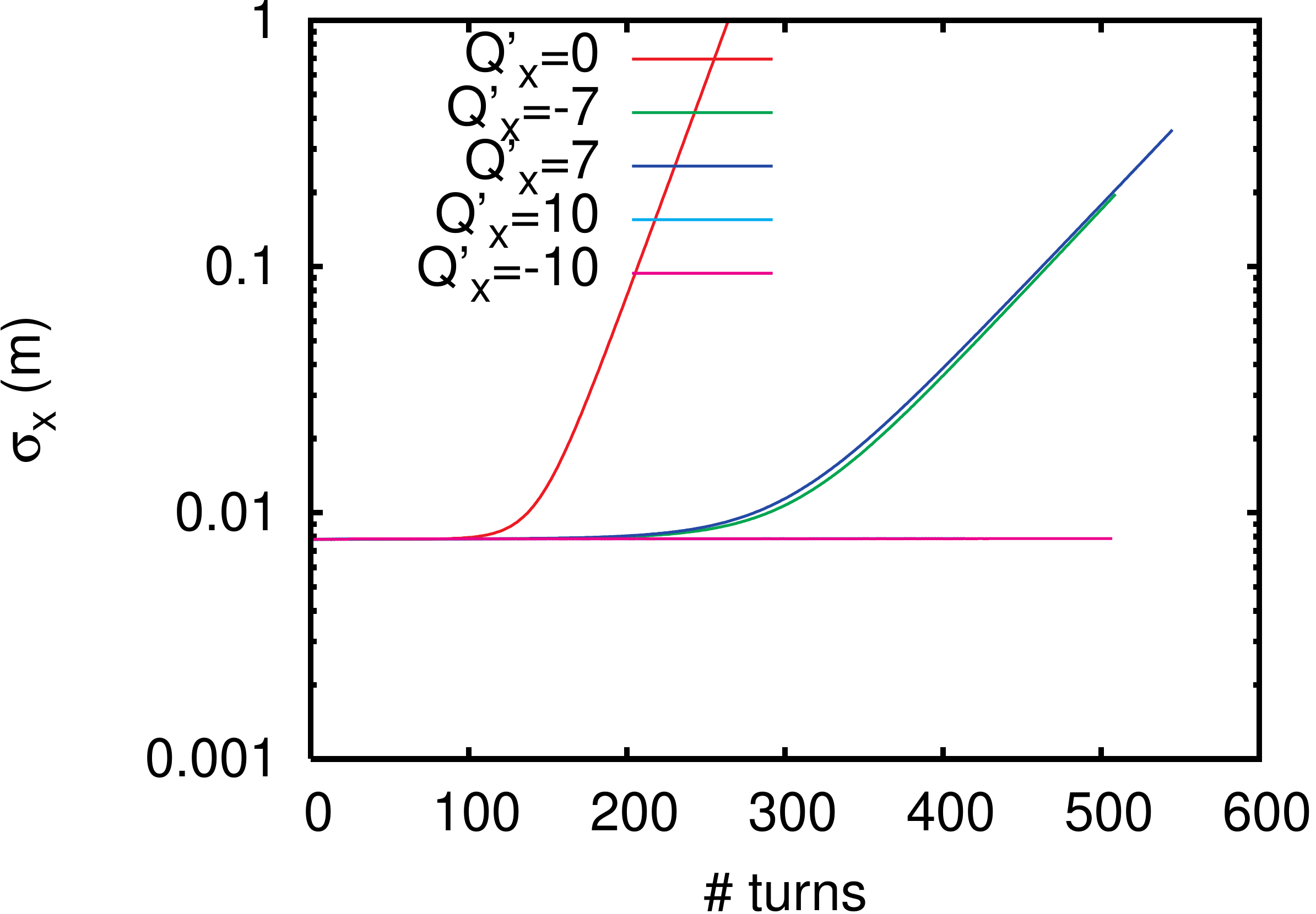}%
    \includegraphics[angle=270,width=0.5\linewidth]{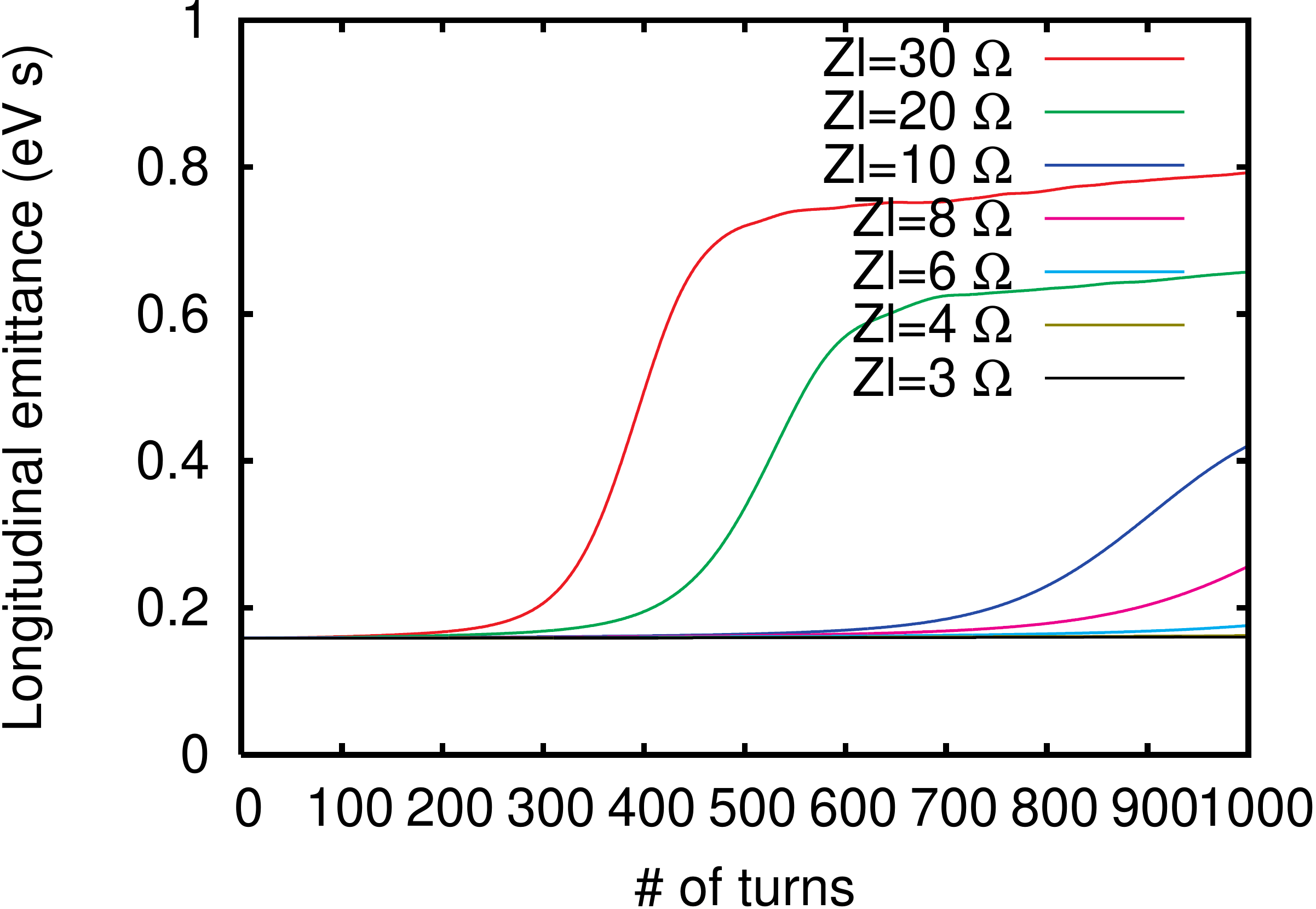}
  \end{center}
  \caption{Horizontal beam size evolution, for various values of
    chromaticity, assuming a transverse impedance of 1~M$\Omega$/m, $Q_R$=1,
    $f_R$=1~GHz (left).  Longitudinal emittance evolution for different values
    of $Z_{\|}/n$ (right). The lines for 3 and 4~$\Omega$ lie on top of each
    other in the figure.}
  \label{fig:acc:pd:cern1}
\end{figure}

\begin{figure}
  \begin{center}
    \includegraphics[width=0.5\linewidth]{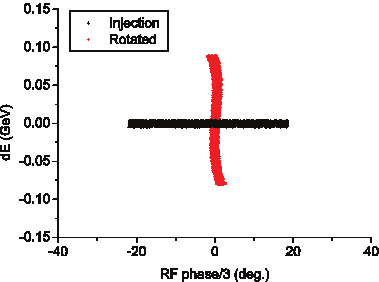}%
    \includegraphics[width=0.5\linewidth]{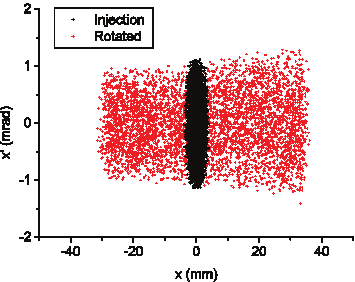}
  \end{center}
  \caption{Phase space plots before and after bunch rotation.}
  \label{fig:acc:pd:cern2}
\end{figure}

\begin{figure}
  \begin{center}
    \includegraphics[width=0.6\linewidth]{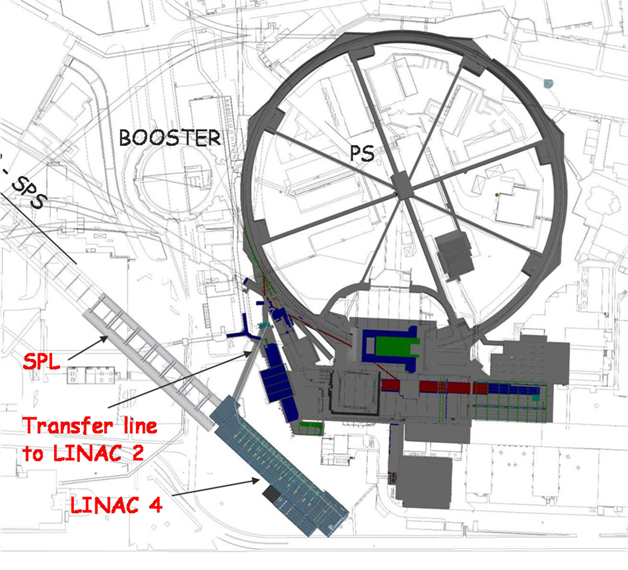}
  \end{center}
  \caption{The CERN Linac4 with respect to the rest of the site.}
  \label{fig:acc:pd:cernsite}
\end{figure}

Such a linac-based solution was adopted for the CERN Neutrino Factory
scenario, which would be based on the proposed 5\,GeV, high-power version of
the Superconducting Proton Linac (SPL)~\cite{Gerigk:2006qi},
which can deliver $10^{14}$
protons at the repetition rate of 50~Hz \cite{Garoby:2009b}. 
In the recent past, the Superconducting Proton Linac study
evolved into an international collaboration whose aim is the
optimisation of the architecture of a pulsed superconducting
high-power proton linac.  The most recent design of the SPL and the
description of the goals of the collaboration, can be found
in~\cite{Gerigk:2010}.
In the CERN scenario,
the chopped beam from the SPL would be injected into an isochronous
accumulator ring in which 120~ns long bunches are formed without the
need for an RF system. 
The absence of synchrotron motion in the accumulator ring makes it
important to study the stability of the beam in the presence of
space-charge.
As presented in~\cite{Benedetto:2009EB},
transverse stability can be obtained with a suitable
choice of chromaticity as shown in figure~\ref{fig:acc:pd:cern1} (left
panel) and longitudinal stability can be achieved by limiting the
longitudinal broad-band impedance to a few ohms as shown in
figure~\ref{fig:acc:pd:cern1} (right panel)~\cite{Benedetto:2009EB}.  
Two-dimensional phase-space painting is used in the stripping injection
into the accumulator ring, allowing the temperature of the stripping
foil to be kept below 2000~K. 
The beam parameters after accumulation are obtained as a compromise
between the competing requirements of minimising the heating of the
injection foil, maximising the aperture, and adequate compensation of
the space-charge forces and are set to allow for RF phase-rotation
in the downstream compressor ring.  
The size of the two rings is determined by the requirement that
successive bunches must arrive at the correct location in the
compressor ring.
The compressor ring has a large phase slip factor, which is needed for
the fast phase rotation.
Tracking simulations in the compressor ring have been performed using
the ORBIT code~\cite{orbit}.
The good performance of the compressor ring is demonstrated in
figure~\ref{fig:acc:pd:cern2} (left panel).
The simulations have also been used to investigate the transverse
phase space. Figure~\ref{fig:acc:pd:cern2} (right panel)
shows that the transverse space charge can be tolerated due
to the limited number of turns of the beam in the compressor ring and
the relatively large dispersion, which effectively lowers the tune
shift by enlarging the beam size.
The parameters of the accumulator and compressor rings are listed in
table \ref{tab:acc:pd:cern1}.
More details of the CERN proton driver scenario can be found in
\cite{Garoby:2009a}.
The low energy normal-conducting part of the SPL is currently under
construction and should become operational in the following few years
as part of the LHC injector chain. The existing proton linac, the
Linac2, will be replaced soon by the more modern Linac4~\cite{Arnaudon:2006jt}
that
will accelerate H- up to 160 MeV, before injecting them in the PSB.
The linac performances will match the requirements of the program of
increasing the LHC luminosity.  The status of the construction of the
Linac4 in 2010 can be found in~\cite{Vretenar:2010}; in particular,
the construction of the new
Linac building was recently concluded.
Figure~\ref{fig:acc:pd:cernsite}
shows an overview of the Linac integration with respect to the
existing accelerator complex.
\begin{table}
  \caption{
    Parameters of the accumulator and compressor rings for the CERN
    proton driver scenario.
  }
  \label{tab:acc:pd:cern1}
  \centering
  \begin{tabular}{|l|r|}
    \hline
    {\bf Parameter}               & {\bf Value}               \\
    \hline
    \multicolumn{2}{|c|}{Accumulator ring}                    \\
    \hline
    Circumference                 &185 m                      \\
    No. of turns for accumulation & 640                       \\
    Working point (H/V)           & 7.37/5.77                 \\
    Total bunch length            & 120 ns                    \\
    RMS momentum spread           & 0.863 $\times 10^{-3}$\\
    \hline
    \multicolumn{2}{|c|}{Compressor ring}                     \\
    \hline
    Circumference                 & 200 m                     \\
    No. of turns for compression  & 86                        \\
    RF voltage                    & 1.7 MV                    \\
    Gamma transition              & 2.83                      \\
    Working point                 & 4.21/2.74                 \\
    \hline
  \end{tabular}
\end{table}

A proton driver for a Neutrino Factory situated at Fermilab would be
based upon an upgrade of the proposed Project~X linac, as described in
Appendix \ref{sec:acc:fnal}. 
Fermilab is currently designing a high-intensity proton source that
will deliver beam at 3~GeV and 8~GeV. 
As presently conceived, the Project~X linac will deliver only $\approx
10\%$ of the proton-beam power needed for the Neutrino Factory at
8~GeV. 
However, Project~X is designed such that it can be upgraded to deliver
the full beam power (4~MW at 8~GeV) required for the Neutrino Factory. 
Just as in the CERN scheme, additional accumulator and compressor
rings will be needed to provide the correct time structure.

Project~X will accelerate H$^{-}$-ions in two superconducting linacs.
A CW linac will accelerate beam to 3~GeV, where the majority of the
beam will be used for experiments. 
A small portion of the beam will be directed into a pulsed linac to be
further accelerated to 8~GeV. 
At 8~GeV, the beam will be accumulated in the Recycler ring before
being transferred to the Main Injector.   
The proton beam will then be accelerated to higher energy for the
long-baseline neutrino program.  

The CW linac will be operated with an average current of 1~mA.
Less than 5\% of the CW-linac beam will be directed to the pulsed
linac. 
Initially, the pulsed-linac beam will provide $\approx 350$~kW of beam
power at 8~GeV.
If the pulsed linac is capable of operating 50\% of the time,
then the required beam power of 4~MW could be reached.  
The option of converting the pulsed linac into a CW linac is not being
considered.
Instead, the average current of the CW linac can be raised to 4-5~mA.
While designing Project~X, provision is being made to allow the CW
linac to be upgraded to accelerate the increased beam current that
will be required to serve the Neutrino Factory.

For Project~X to serve the Neutrino Factory, an accumulation ring will
be put at the end of the second linac.
At injection into the accumulation ring there will be a stripping
system, foil or laser based, to convert the H$^{-}$-ions to protons.
The front-end of Project~X will have a programmable chopper so that
beam will be injected into three RF buckets in the accumulation ring.
After injection is complete, the RF bucket voltage will be increased
to shorten the bunches. 
The accumulated protons will then be transferred to a separate
bunch-shortening (compressor) ring.
Bunch rotation will be used to achieve the final bunch length
(2~ns). 
The three bunches will be extracted with the proper spacing
(120~$\mu$s) to the target station.
The accumulation, bunch shortening, and targeting will be done at 50~Hz,
as required for the Neutrino Factory. 

Project~X will be located within the Tevatron ring at Fermilab.
The accumulation ring will be situated near the end of the pulsed
linac.
Once the beam has been converted from H$^{-}$-ions to protons, the
beam can be transported some distance to the bunch-shortening ring.
Extraction from the bunch shortening ring to the target sets the
orientation of the ensuing muon collection and acceleration.
All components of the Neutrino Factory, including a decay ring, fit
within the Tevatron ring footprint.  

\begin{figure}
  \begin{center}
    \parbox{0.48\linewidth}{\includegraphics[width=\linewidth]{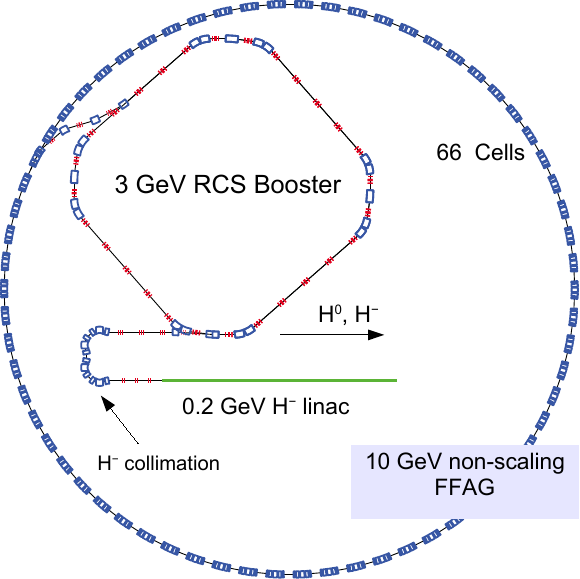}}\hspace{0.04\linewidth}%
    \pdfpxdimen=1in%
    \divide\pdfpxdimen by96%
    \parbox{0.48\linewidth}{\includegraphics*[viewport=144px 192px 912px 608px,width=\linewidth]{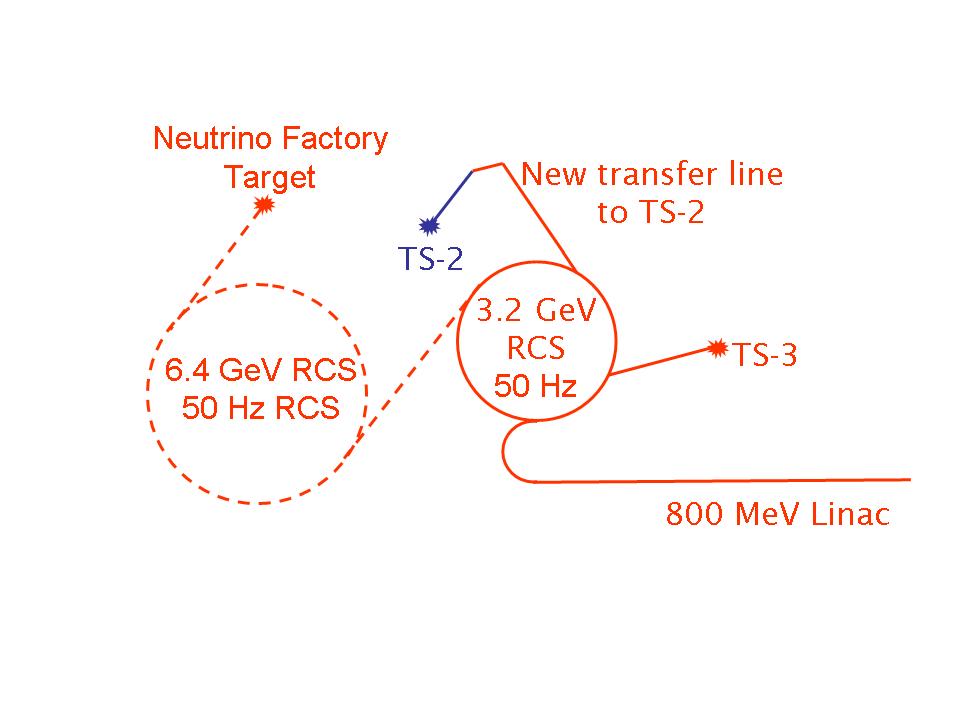}}
  \end{center}
  \caption{Left panel: Layout of the ISS proton driver based on an RCS and
    an FFAG. Right panel: Layout of the common proton driver for ISIS and
    the Neutrino Factory at RAL.}
  \label{fig:acc:pd:ral}
\end{figure}

A Neutrino Factory sited at the Rutherford Appleton Laboratory (RAL)
would be served by a proton driver based on an upgrade to the ISIS
pulsed-proton source.
In this scenario, a chain of circular accelerators, typically
Rapid Cycling Synchrotrons (RCSs), provides an alternative to the
linac-based options outlined above.
Here, bunch compression is accomplished adiabatically in
the RCS or, alternatively, in an FFAG ring as proposed in the ISS
study \cite{Apollonio:2009}.   
Recently the attractive idea of a common proton driver for the
spallation neutron source and the Neutrino Factory was proposed in the
framework of the ongoing ISIS megawatt-upgrade programme.
In such a scenario, the proton drivers for both facilities would share the
same source, chopper, linac, accumulator, and acceleration up to
3.2~GeV.  
After extraction, a number of bunches would be sent directly to the
neutron-spallation target while three others would be injected into a
second RCS or FFAG where, after acceleration to somewhere between
6.4 and 10.3~GeV followed by bunch compression, the beam
would be extracted towards the Neutrino Factory pion-production
target. 
The layout of the proton drivers using the RCS machines can be seen
in figure \ref{fig:acc:pd:ral}.
The left panel shows the solution based on an RCS booster and FFAG
ring developed during the ISS \cite{Apollonio:2009}.
The right panel shows the layout of the common proton driver for the
spallation-neutron source and the Neutrino Factory that is presently
under development.
For the ISIS megawatt upgrade to be compatible with the Neutrino
Factory, an 800~MeV H$^-$ linac has been designed, candidate
lattices for the 3.2~GeV booster RCS have been identified, and
preliminary parameters for the final RCS ring have been proposed.
More details of the Neutrino Factory proton driver option based on an
upgrade of facilities at RAL are presented in Appendix~\ref{sec:acc:ral}.

\subsection{Target}
\label{Sect:Trgt}

\subsubsection{Introduction to the target system}

\begin{figure}
  \includegraphics[width=0.8\textwidth]{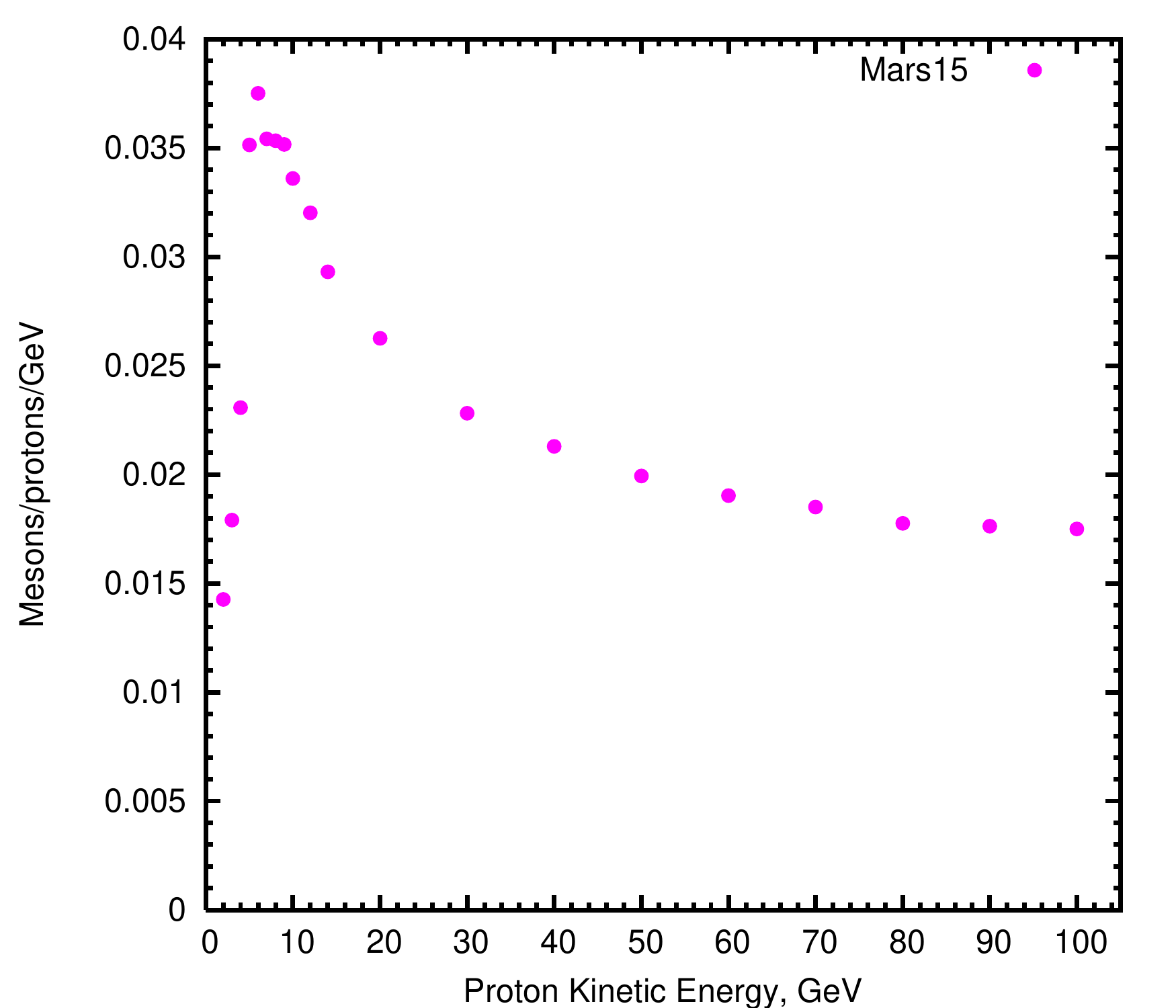}
  \caption{The meson production efficiency, normalised to the beam power,
    as a function of incoming
    proton beam \cite{Ding:2009zz}. Production
    efficiency is the number of mesons 50~m downstream from the target
    with kinetic energy in the range of 40--180~MeV. This criterion
    has been found to have a good correlation with the muons that are
    transmitted through the front end.}
  \label{fig:acc:tgt:1a}
\end{figure}

\begin{figure}
  \includegraphics[width=0.8\textwidth]{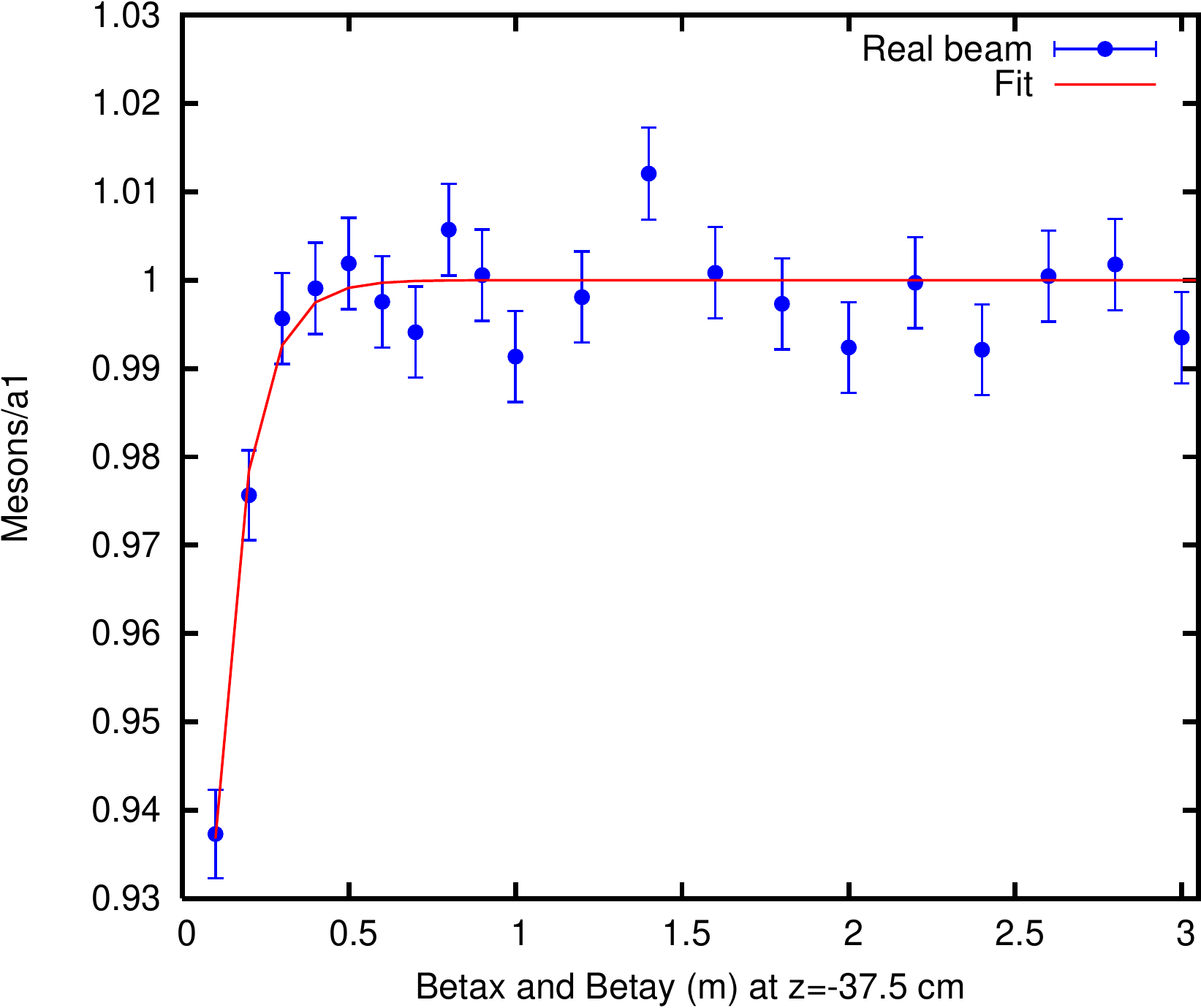}
  \caption{The meson production efficiency as a function of the
    $\beta^*$ of the incoming proton beam~\cite{Ding:2010zzd}. The
    data are fit to the function $y=a_1-a_2e^{-kx}$, with $x$ being
    the beta function of the beam distribution at $z=-37.5\text{ cm}$
    (the centre of the beam-target crossing)
    and $y$ being the production efficiency. The results are scaled to
    $a_1$, the limit of the fit as the beta function goes to
    infinity.}
  \label{fig:acc:tgt:1b}
\end{figure}

The requirements for the proton driver at the Neutrino Factory
\cite{Alsharoa:2002wu} (summarised in table \ref{tab:acc:pd:parm})
call for a target capable of intercepting and surviving a 4~MW pulsed
proton beam with a repetition rate of 50~Hz.
The assumption of 8~GeV for the proton-beam energy arises from an
optimisation study (performed with
MARS~\cite{Mokhov:1995wa,Mokhov:2000ih,Mokhov:2003jq,Mokhov:2004aa,mars})
in which the target parameters were evaluated as a
function of the kinetic energy of the incoming proton beam.
The resulting meson (``meson'' in this document denotes muons, charged pions,
and charged kaons)
production-efficiency is shown in figure
\ref{fig:acc:tgt:1a}.
The peak production-efficiency occurs for proton kinetic energies of
6--8~GeV.
The $\beta^*$ requirement (at the centre of the beam-target crossing)
was determined by studying the production efficiency for the target
geometry defined in table \ref{tab:acc:tgt:parm} for a proton beam of
8\,GeV and 1.2\,mm rms radius as a function of $\beta^*$.
The result, shown in figure \ref{fig:acc:tgt:1b}, lead to the
requirement that $\beta^*$ should be at least 30\,cm.

\begin{figure}
  \includegraphics[width=0.8\textwidth]{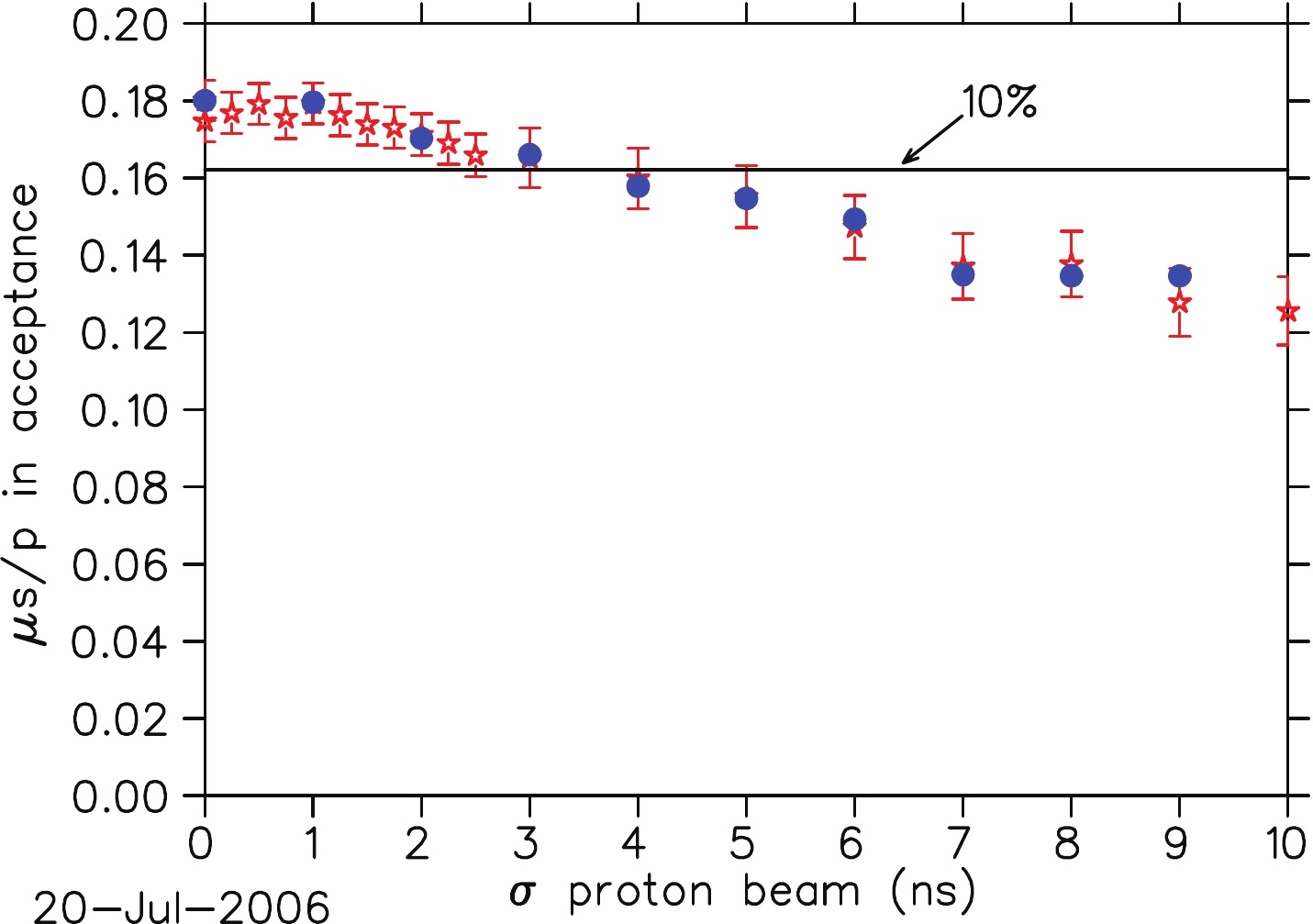}
  \caption{The final muon throughput of the front end as a function of
    the incoming proton beam bunch length~\cite{Gallardo:2006zz}.
    Solid blue dots and red
    stars are data for two equivalent methods of creating the beam
    distribution in time.}
  \label{fig:acc:tgt:1c}
\end{figure}

The proton-beam bunch-length requirement is derived from a calculation
in which the front end of a Neutrino Factory complex is simulated with
the code ICOOL~\cite{Fernow:2005cq}.
The muon throughput is obtained as a function of the mesons produced
at the target for proton beams with various bunch
lengths~\cite{Gallardo:2006zz}.  
The results are shown in figure \ref{fig:acc:tgt:1c}.

The proposed target system simultaneously captures charged pions of
both signs, hence the use of solenoid magnets in the target system
(rather than toroidal magnets that primarily capture particles of only
one sign, as is typical in target systems for ``conventional''
neutrino beams).   

The target, the proton beam dump, and a shield/heat exchanger are to
be located inside a channel of superconducting solenoid magnets that
capture, confine and transport secondary pions and their decay muons
to the muon front-end (see section
\ref{Sect:FrontEnd}). 
In the present baseline configuration, most of the 4~MW beam power is
dissipated within a few meters of the target, inside the
solenoid channel, which presents severe engineering challenges.

Maximal production of low-energy pions is obtained with a proton beam
of 1--1.5~mm (RMS) radius and a target of radius three times this.
The target parameters have been determined by extensive simulations
using the MARS modelling code \cite{Ding:2010zzd}.

The effect of beam-induced shock in solid targets has been studied
using a high-current pulse with a rapid rise time to deliver a high
power-density in tungsten and tantalum wires \cite{Simos:2008zz}.
The results indicate that a system in which tungsten bars are
exchanged between beam pulses can withstand the beam-induced shock.
However, schemes for a set of moving solid targets have yet to be
shown to be compatible with the confined environment provided by the
solenoid magnets that form the pion-capture system.
Hence, the baseline target concept is for a free liquid-mercury jet.
A free-flowing jet is chosen because the intense beam-induced pressure
waves in the liquid target would damage or lead to the failure of
any pipe containing the liquid in the interaction region.  
The parameters of the present target-system baseline are summarised in
table \ref{tab:acc:tgt:parm}.
\begin{table}
  \caption{Baseline target system parameters.}
  \label{tab:acc:tgt:parm}
  \centering
  \begin{tabular}{|l|r|}
    \hline
    {\bf Parameter}                                      & {\bf Value}      \\
    \hline
    Target type                                          & Free mercury jet \\
    Jet diameter                                         & 8 mm             \\
    Jet velocity                                         & 20 m/s           \\
    Jet/solenoid axis angle                              & 96 mrad          \\
    Proton beam/solenoid axis angle                      & 96 mrad          \\
    Proton beam/jet angle                                & 27 mrad          \\
    Capture solenoid (SC-1) field strength               & 20~T             \\
    Front-end $\pi/\mu$ transport channel field strength & 1.5 T            \\
    Length of transition between 20 T and 1.5 T          & 15 m             \\
    \hline
  \end{tabular}
\end{table}

The concept of a mercury jet target within a high-field solenoid
has been validated by R\&D over the
past decade, culminating in the MERIT
experiment~\cite{Bennett:2004} that ran in the fall of 2007 at the
CERN PS.
The experiment benefited from the intensity of the beam pulses (up to
$30\times10^{12}$ protons per pulse) and the flexible beam structure
available for the extracted PS proton beam.  
Key experimental results include the demonstration that
\cite{McDonald:2009zz}:
\begin{itemize}
\item 
  The magnetic field of the solenoid greatly mitigates both the
  extent of the disruption of the mercury and the velocity of the
  mercury ejected after interception of the proton beam.  
  The disruption of a 20~m/s mercury jet in a 20~T field is
  sufficiently limited that a repetition rate as high as 70~Hz
  is feasible without
  loss of secondary particle production;
\item 
  Individual beam pulses with energies up to 115~kJ can safely be 
  accommodated;
\item 
  Subsequent proton-beam pulses separated by up to 350~$\mu$s have
  the same efficiency for secondary particle production as does the
  initial pulse; and
\item 
  The disruption of the mercury jet caused by the second of two beam
  pulses separated by more than 6~$\mu$s is unaffected by the presence
  of the first beam pulse.
\end{itemize}

\begin{figure}
  \includegraphics[width=0.8\textwidth]{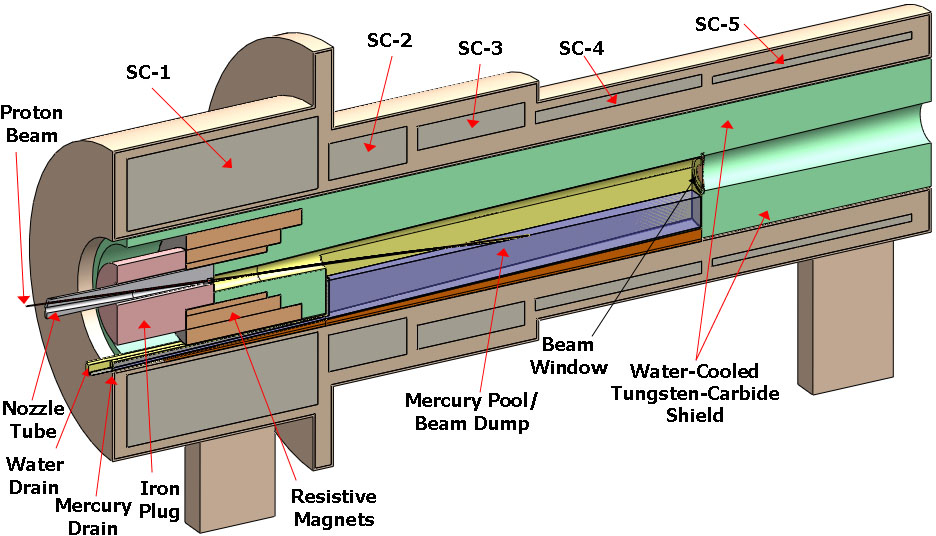}
  \caption{Baseline target system concept, with small changes from
    Neutrino Factory Study~II~\cite{Ozaki:2001}.  SC-$n$ are the
    superconducting magnets.}
  \label{fig:acc:tgt:all}
\end{figure}

In the Neutrino Factory target system,
the mercury jet is collected in a pool, inside the solenoid magnet
channel, that also serves as the proton beam-dump, as sketched in
figure \ref{fig:acc:tgt:all}.  
Disruption of this pool by the mercury
jet (equivalent to a mechanical power of 3\,kW) and by the non-interacting
part of the proton beam is nontrivial, and needs further study.

The superconducting magnets of the target system must be shielded
against the heat and radiation damage caused by secondary
particles issuing from the target (and beam dump).  A high-density shield is
favoured to minimise the inner radii of the magnets.

The baseline shield concept is a stainless-steel vessel of complex
shape (see figure \ref{fig:acc:tgt:all}) containing water-cooled
tungsten-carbide beads.  
Calculations using MARS in an MCNP mode (MCNP mode gives more accurate
treatment of low-energy neutrons)
show that for an 8~GeV, 4~MW incoming proton beam, the radiation
penetrating the shielding (as configured in Study~II~\cite{Ozaki:2001})
results in 50~kW
of power being deposited in the superconducting coils that surround
the target~\cite{Souchlas:2010}.
This would present a severe operational burden
on the cryo-system and hence is not considered
practical. 
Effort to produce a new
baseline for the shielding is under way.

The solenoid magnets of the target system vary in strength from 20~T
(SC-1 in figure \ref{fig:acc:tgt:all}) down to 1.5~T in the subsequent
constant-field transport channel.
Table \ref{tab:acc:tgt:capturecoils} gives the solenoid parameters as
established in Study~II \cite{Ozaki:2001}.
The interface between the target station and the muon front-end is
taken to be the point at which the constant-field capture channel
begins, 15~m downstream of magnet SC-1 in the present baseline.
The initial superconducting coil (SC-1), which is responsible for
generating a 14~T field at the site of the target, has the most
challenging mechanical requirements.  The stored energy of this magnet
is 600~MJ, and the radial and hoop stresses are 110~MPa and 180~MPa,
respectively~\cite{Loveridge:2010}.
\begin{table}[htb]
\begin{center}
\caption[The Study 2 Solenoid coil parameters ]
{Solenoid coil and iron plug geometric parameters.
  The centre of the beam-target
  crossing is at $-37.5$~cm.}
\label{tab:acc:tgt:capturecoils}
\vspace{2.5mm}
\begin{tabular}{|c|c|c|c|c|c|c|c|}
\hline
  & $\bf{z}$ & $\bf{\Delta z}$ & $\bf{R_i}$ & $\bf{\Delta R}$ & $\bf{I/A}$ & $\bf{n I}$ & $\bf{n I l}$\\ 
& (m) & (m) & (m) & (m) &  (A/mm$^2$) & (A) & (A-m) \\ 
\hline
Fe
& 0.980 &  0.108 &  0.000 &  0.313 &   0.00 &  0.00 &  0.00  \\
& 1.088 &  0.312 &  0.000 &  0.168 &   0.00 &  0.00 &  0.00  \\
\hline
Cu coils 
& 1.288 &  0.749 &  0.178 &  0.054 & 24.37 &  0.98 &  1.26  \\
& 1.288 &  0.877 &  0.231 &  0.122 & 19.07 &  2.04 &  3.74  \\
& 1.288 &  1.073 &  0.353 &  0.137 & 14.87 &  2.18 &  5.78  \\
\hline
SC coils
& 0.747 &  1.781 &  0.636 &  0.642 & 23.39 & 26.77 & 160.95  \\
& 2.628 &  0.729 &  0.686 &  0.325 & 25.48 &  6.04 & 32.23  \\
& 3.457 &  0.999 &  0.776 &  0.212 & 29.73 &  6.29 & 34.86  \\
& 4.556 &  1.550 &  0.776 &  0.107 & 38.26 &  6.36 & 33.15  \\
& 6.206 &  1.859 &  0.776 &  0.066 & 49.39 &  6.02 & 30.59  \\
& 8.000 &  0.103 &  0.416 &  0.051 & 68.32 &  0.36 &  1.00  \\
& 8.275 &  2.728 &  0.422 &  0.029 & 69.27 &  5.42 & 14.88  \\
&11.053 &  1.749 &  0.422 &  0.023 & 75.62 &  3.00 &  8.18  \\
&12.852 &  1.750 &  0.422 &  0.019 & 77.37 &  2.61 &  7.09  \\
&14.652 &  1.749 &  0.422 &  0.017 & 78.78 &  2.30 &  6.22  \\
&16.451 &  1.750 &  0.422 &  0.015 & 79.90 &  2.07 &  5.59  \\
&18.251 &  2.366 &  0.422 &  0.013 & -0.85 &  2.53 &  6.80  \\
\hline
\end{tabular}
\end{center}
\end{table}

A 20~T field is beyond the capability of Nb$_3$Sn, and magnet SC-1 is
proposed as a hybrid of a 14~T superconducting coil with a 6~T
hollow-core copper solenoid insert.  A 45~T solenoid with this type of
construction (but a significantly smaller bore of 32~mm diameter)
has been operational since 2000 at the National High
Magnetic Field Laboratory (Florida, USA)~\cite{NHMFL:45T}.  A 19~T resistive magnet
with a 16~cm bore at the Grenoble High Magnetic Field
Laboratory~\cite{GHMFL:M8} was used in an earlier
phase~\cite{Fabich:2002} 
of the Neutrino Factory R\&D program.  
A topic for further study is possible
fabrication of SC-1 as a high-T$_{\text{C}}$ magnet with no
copper-solenoid insert, which could provide more space for internal
shielding of SC-1 and/or permit operation at a higher field for a
potential increase in the meson-capture efficiency.

The target system (and also the subsequent $\pi$/$\mu$ solenoid transport
channel) will be subjected to considerable activation, such that once
beam has arrived on target all subsequent maintenance must be
performed by remote-handling equipment.  
The infrastructure associated
with the target hall, with its remote-handling equipment, and
hot-cells for eventual processing of activated materials, may be the
dominant cost of the target system.

\subsubsection{Sub-systems}

\paragraph{Target}
\subparagraph{Baseline Concept}

The target itself is a free liquid mercury jet ($Z = 80$, $A = 200.6$,
density $\rho = 13.5\text{ g/cm}^3$, nuclear interaction
length $\lambda_I \approx 15\text{ cm}$) of
diameter $d = 8\text{ mm}$, flowing at $v = 20 \text{ m}/\text{s}$.
The volume-flow rate is 1.0~L/s and the mechanical power in the
flowing jet is 2.7~kW.  
The flow speed of 20~m/s ensures that the gravitational curvature of
the jet over the two nuclear interaction lengths (30~cm) through which the
proton beam passes is negligible compared to its diameter and that
more than two nuclear interaction lengths of new target material are presented
to the beam every 20~ms, the time between beam pulses at a repetition
rate of 50~Hz. 

According to a MARS15 simulation \cite{Ding:2010zzd,Ding:2010}, 
$\sim 11\%$ of the beam energy is deposited in the target, corresponding to
9~kJ/pulse at 50~Hz.  
This energy is deposited over two nuclear interaction lengths along the
jet (30~cm, 15~cm$^3$), so, noting that the specific heat of mercury
is $\sim 4.7$~J/cm$^3$/K, the temperature rise of the mercury during a
beam pulse is about 130~K.  
The boiling point of mercury is 357~$^\circ$C, so the mercury jet, which
enters the target volume at room temperature, is not vaporised at
50-Hz operation.

Although the mercury jet is not vaporised, it will be disrupted and
dispersed by the pressure waves induced by the pulsed energy
deposition.  
The MERIT experiment \cite{McDonald:2010zz} showed that for pulses
equivalent to 50~Hz operation at 4~MW beam power, this disruption
results in droplets with peak velocity of $\sim 50$~m/s in a 
15~T field.
Extrapolating these results to 20~T field at the Neutrino Factory
target yields a maximum droplet-velocity of $\sim 30$~m/s.

The optimal production of pions in the acceptance of the muon
front-end is achieved by appropriate tilts of the mercury jet and
proton beam with respect to the magnetic axis.  
These tilts depend slowly on the proton-beam energy (as does the
optimum radius of the jet), and the current best values are based on
MARS15 simulations and are given in table \ref{tab:acc:tgt:parm}.

The Reynolds number of the mercury flow in the jet is
$R=\rho vd/\eta\approx 1400$. 
Noting that the viscosity of mercury is
$\eta=1.5$~mPa$\cdot$s,  the flow can be turbulent.  
Hence, the
quality of the jet is an issue, although operation in a high 
magnetic field damps surface perturbations~\cite{Fabich:2002}. 
The nozzle will need to be
as close as feasible to the interaction region.  
High velocity mercury may generate erosion of the nozzle, e.g., by
erosion-corrosion or by cavitation; this part of the system would need to be
carefully designed but may be studied off-line using a suitable mercury flow
loop.
Detailed design work of the nozzle is under way.

\subparagraph{Possible Target Alternatives}

Alternatives to the liquid-mercury jet target that are under 
consideration are: a liquid-metal jet using a metal that is solid at room
temperature; a helium-cooled metallic low-\textit{Z} static packed bed;
a metal-powder-jet; and a system of solid tungsten bars
that are exchanged between beam pulses.

A eutectic alloy of
lead and bismuth (melting point 124~$^\circ$C)
would have similar performance
to a mercury target, with the challenge of operating
the target flow-loop at temperatures above the boiling point of water,
with that flow-loop in thermal contact with the water-cooled shield of
the superconducting magnets.  
The activation products from a lead-bismuth target are more
troublesome than those of a mercury target. 
A helium cooled low-\textit{Z} static packed bed of beryllium (or possibly
titanium) would give a low-\textit{Z} (the advantages of which are described
below) alternative to graphite
that is considerably more tolerant to radiation damage and therefore
would have a greater lifetime than the graphite,
provided that sufficient cooling can be
achieved.
The powder-jet and solid alternatives are discussion in Appendices
\ref{App:PowderJet} and \ref{App:SolidTrgt}.

\paragraph{Proton Beam Dump}

The target system requires the proton beam dump to be inside the
superconducting magnet channel, only $\approx 1\text{ m}$ from the
target.  
The baseline design is to use the pool that collects the
mercury from the target jet as the beam dump.  
The mercury is to be
drained from the upstream end of the pool, in a passage that is a
sector of the annular space between the resistive magnet and
superconducting magnet SC-1.

The dump must dissipate the $\sim 3$~kW of mechanical power in the
mercury jet, as well as $\sim 20$\,kW of power in the attenuated
proton beam. The beam dump must also be able to withstand one or two
full intensity beam pulses before the beam can be tripped in the event
that the mercury jet fails, or the beam misses the jet.
The vessel that contains the mercury pool will be subject to
substantial radiation damage and heating by the secondary particles
from the target, and must be replaced periodically.

\paragraph{Beam Windows}
\label{sec:acc:tgt:bw}

The volume that contains the target and mercury pool beam dump is the
primary containment vessel for the mercury.  
This containment vessel
includes a small window upstream through which the proton beam enters
and a larger window on the downstream face of the vessel, currently
specified to be at the boundary between superconducting magnets SC-4
and SC-5, through which the desired secondary pions and muons (as well
as other particles) pass.  
The containment vessel is to be operated with helium gas (plus mercury
vapour) at atmospheric pressure.

\begin{figure}
  \centering
  \includegraphics[width=0.29\linewidth]{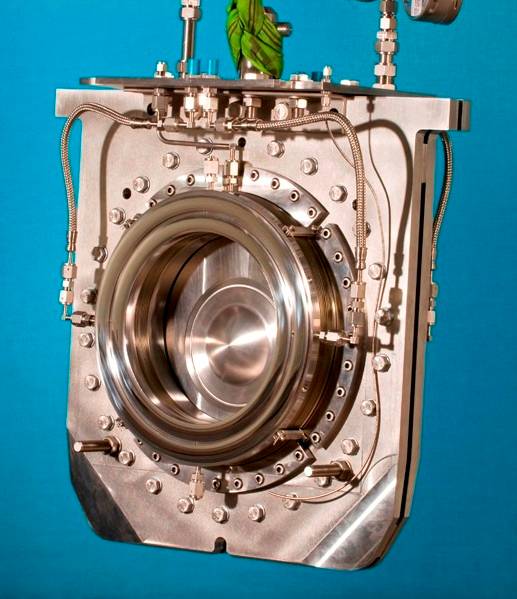}%
  \hspace{0.04\linewidth}%
  \includegraphics[width=0.67\linewidth]{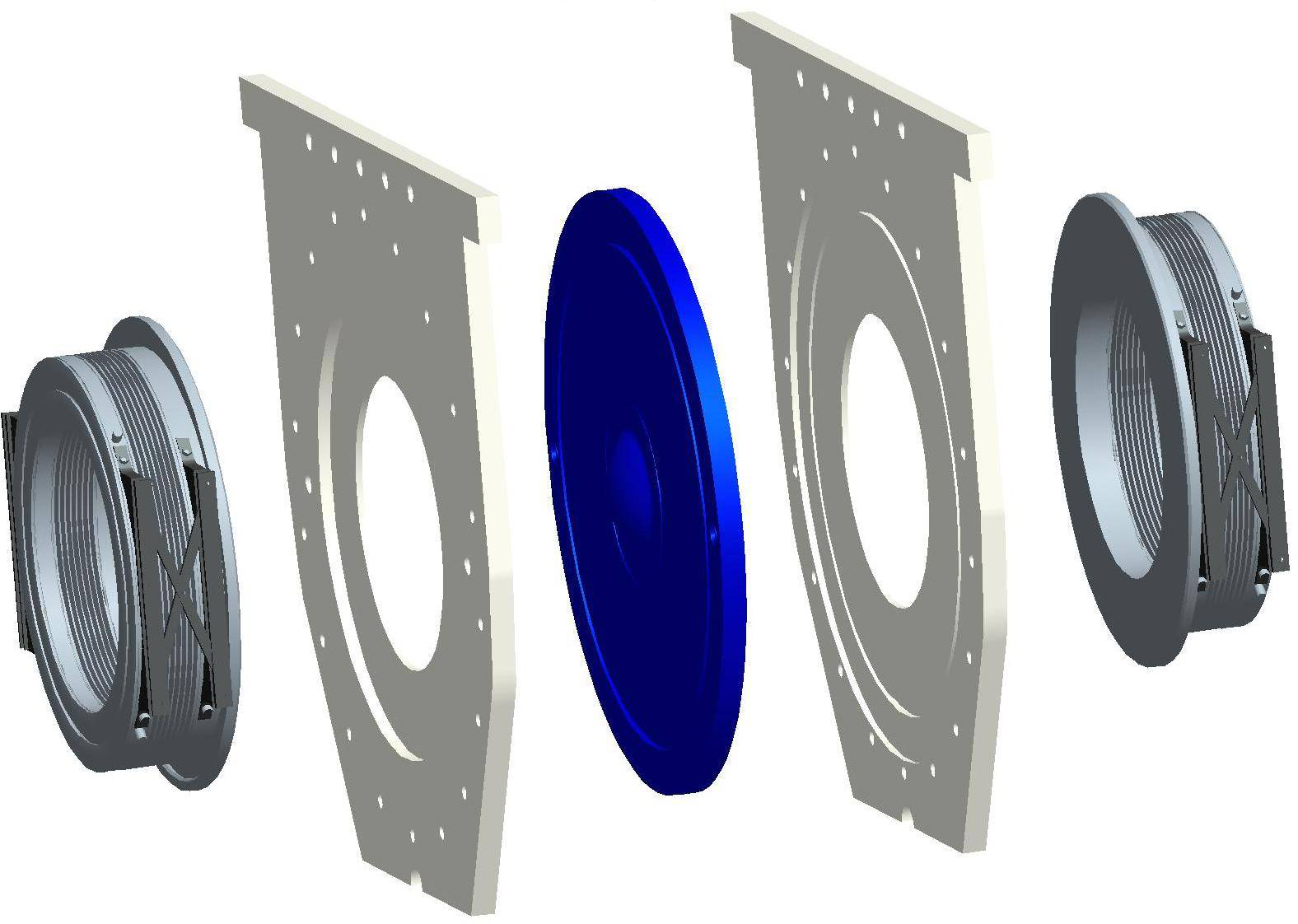}
  \caption{Photograph and schematic of the T2K beam window~\cite{Rooney:2008},
    which was designed for a 0.75~MW beam.}
  \label{fig:acc:tgt:t2k}
\end{figure}

The upstream proton-beam window has yet to be specified, either as to
design or location. 
The larger exit window is specified to be made of beryllium.  
It will be a double window, such that the volume between
the two windows can support a flow of helium gas to cool the
window, and which can be monitored for indications of window
failure.
The complexity of such a window system is indicated in figure
\ref{fig:acc:tgt:t2k} which shows the beam window in use on the
T2K target at J-PARC~\cite{Rooney:2008}.
The window system is a replaceable item, and will be sealed to the
downstream face of the primary containment vessel and to the upstream
face of the pion-decay-channel vessel via inflatable ``pillow''
seals. 

\paragraph{The Internal Shield}
\label{sec:acc:tgt:shield}

A major challenge of the target system is the dissipation of the 4~MW
of beam power inside the superconducting magnet string without
quenching the magnets or extreme shortening of their operational
lifetime due to radiation damage.  
Most of the beam power will be dissipated in an internal shield
composed of a high-\textit{Z} material, which will have to extend well beyond
the downstream end of the target system into the muon front-end.

The baseline scenario is for a shield of tungsten-carbide beads cooled
by water.  Pure tungsten beads would provide better shielding, but
tungsten corrodes in water in a high radiation environment~(see, for
example, \cite{Li:2007}).  Random packing of spherical beads of a
single radius will result in a configuration with about 63\% by volume
of tungsten-carbide, and 37\% water~\cite{Song:2008}.  The flow path
of the coolant is not presently specified; multiple inlets and outlets
will be appropriate for a shield of total length 30\,m---50\,m.
One inlet is required at the very upstream end of the shield where the
heat load is the largest.

All high-\textit{Z} target systems proposed so far rely on recirculating target
material from the target interaction region to be externally cooled.
This requires either a radial or longitudinal gap to be engineered between
the solenoid coils and shielding.

The outer radius of the shield was specified to be 63~cm in
Feasibility Study~II \cite{Ozaki:2001}, but subsequent MARS15
simulations indicate that this would imply a load of 
$\sim 25$~kW in magnet SC-1~\cite{Ding:2010zzd}.  
A low-\textit{Z} target material would considerably reduce
neutron production compared with a high-\textit{Z} material,
which could potentially reduce this energy deposition.
Interim values of up to 75~cm for the outer
radius of the shield are being considered.  Studies are under way to determine
acceptable criteria for the total power deposition and the peak local
power deposition for the various magnets to be stable against
quenching and radiation damage.  
The details of the present baseline concept should therefore be regarded as
preliminary.

\paragraph{The solenoid magnets}
\label{sec:acc:tgt:sols}

An early concept~\cite{Palmer:1994km} for a muon collider
assumed separate targets for production/collec\-tion of positive and
negative particles.  It was soon realised that the use of solenoid magnets
would permit a single channel to operate with both
signs~\cite{Palmer:1995jy}, and that the initial capture in a
high-field solenoid followed by a series of solenoids each of slightly
lower field than the last exchanges transverse for longitudinal
momentum~\cite{Green:1995vg}.  
Another advantage is that the solenoid-magnet coils would be farther
from the high radiation due to secondary particles from the
target than would toroidal coils \cite{Mokhov:1996tp}.

The design of the first magnet, with baseline field of 20~T is
challenging.  
The use of a 6~T water-cooled, hollow-core-copper
solenoid insert is required if the superconducting outsert is
made from Nb$_3$Sn.  This copper magnet receives a very high radiation
dose (while acting as a partial shield for the superconducting outsert)
and is anticipated to be a replaceable component with a lifetime of 4
years or less.  If the presence of this copper magnet leads to a
requirement for thicker shielding and consequently larger inner
diameter for the superconducting outsert, such that the latter is
untenable, we must consider the option of only a 14~T Nb$_3$Sn magnet,
or development of a large-bore high-T$_{\text{C}}$ magnet (or more simply, a
high-T$_{\text{C}}$-Nb$_3$Sn hybrid~\cite{Gupta:2010a}; tests of YBCO
indicate that it has good resistance to radiation
damage~\cite{Gupta:2010b}).
The impact on muon yield of any such revision will also need to be
addressed. 

Another issue is the very large axial forces between the various
magnets of the target system.  A further complication is the
requirement that the axial field profile in the beam-jet interaction
region be smooth, such that the mercury jet is minimally perturbed as
it enters this field.  The baseline scenario calls for an iron plug at
the upstream end of the first magnet, through which the proton beam
and mercury jet enter.  The presence of this plug adds considerable
complexity to the mechanical design of the system and is an important
technical issue.

\subsubsection{Particle Production Simulations}

Many of the studies above were performed using the
MARS
simulation code.  It is essential to understand
the dependence of particle production
on target parameters (geometry, proton energy, material, etc.), and to have
the proper values for the production rates.  This is necessary both
for particles giving the desired muons and for particles that lead
to undesired energy deposition in the surrounding system.

\begin{figure}
  \includegraphics[width=\linewidth]{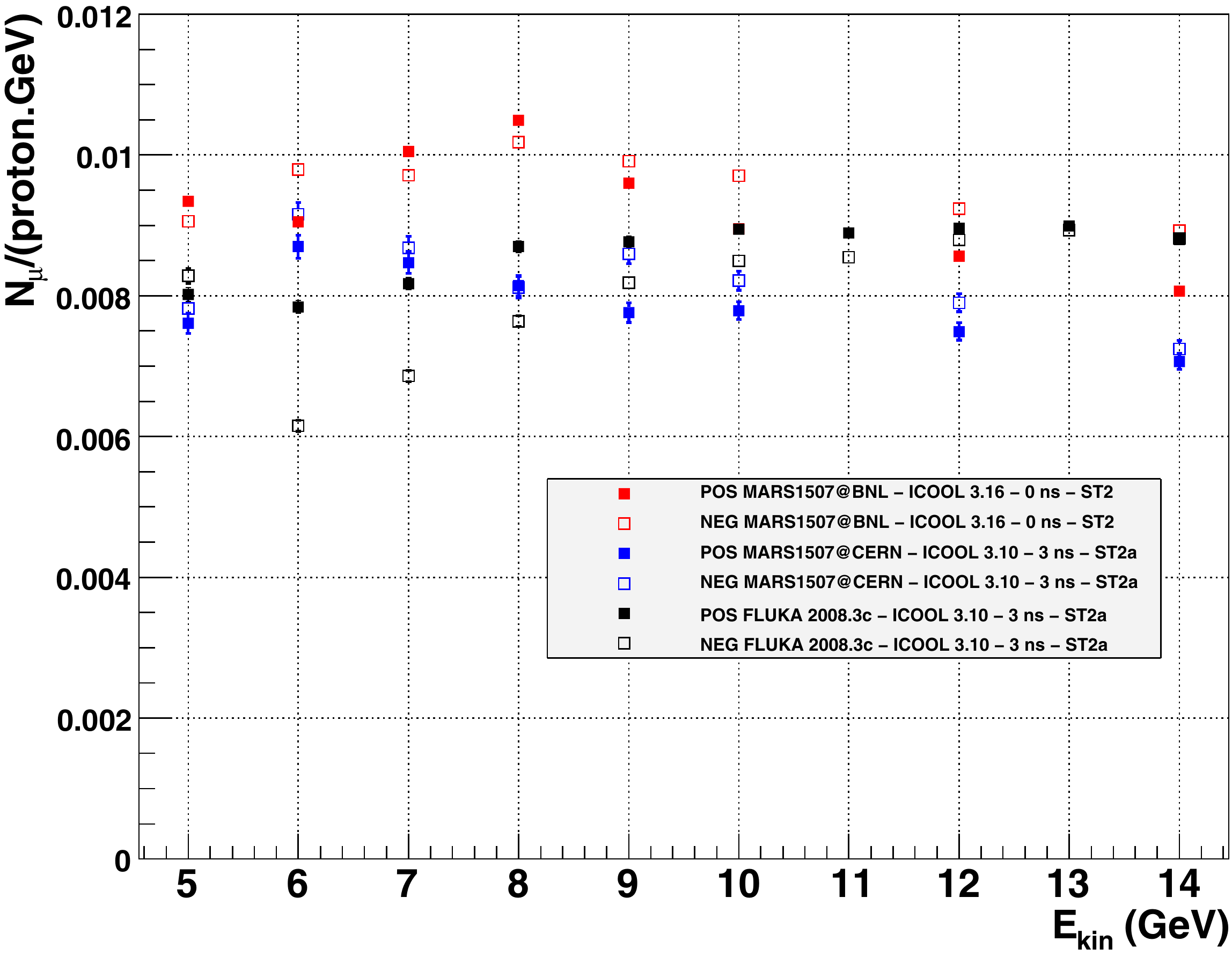}
  \caption{The muon yield, normalised to the beam power.  An initial
    beam is generated by MARS as installed at BNL (red) and CERN
    (blue), and FLUKA 2008.3c (black), with an effective bunch length
    of 0~ns (red), 3~ns (blue and black).  Mesons are then tracked
    through the front end using ICOOL 3.10 (blue) and ICOOL 3.16
    (red).  For the black points, an
    acceptance map approximating the effect of passing the particle
    through ICOOL~3.10 has been applied.  After this, an acceptance
    cut is performed for a longitudinal momentum range of
    100--400~MeV/$c$, normalised longitudinal acceptance of 150~mm,
    and normalised transverse acceptance of 30~mm using
    ECALC9F~\cite{Fernow:2003zz}.  ST2
    and ST2a denote different field tapering parameters, with the
    field going from 20~T to 1.25~T and 1.75~T respectively.}
  \label{fig:acc:tgt:gersende}
\end{figure}

Computations of particle production by both MARS and
FLUKA~\cite{Fluka1,Fluka2,Back2:2010}
have been performed.  Summaries of some simulations
comparing results from different codes and lattices
are shown in figure \ref{fig:acc:tgt:gersende}.  
A study~\cite{Soler:2010} of the
dependence of muon yield on target material has been performed
using G4beamline~\cite{Roberts:2008zzc} for different hadronic models.
The results from this study give a flat distribution for high-\textit{Z}
materials for a beam energy between 5 and 8~GeV, whereas low-\textit{Z}
materials give a strong dependence of the muon yield on beam energy,
favouring low energy.

A recent study~\cite{Strait:2010zzb} has been made, using HARP
data~\cite{Catanesi:2008uv,Bolshakova:2009if}
convolved with the acceptance of the front-end channel of the Neutrino
Factory and correcting for the phase space not covered as well as
for thick-target effects not accounted for in the cross-section data.  The
simulations were performed with a more recent version (1509) of the MARS code.
Results from this study show that the dependence
of the muon yield on proton beam energy is relatively flat, with the
yield being within 10\% of the maximum value (at 7\,GeV) in an energy
range of 4\,GeV---11\,GeV.  These results are in basic agreement with those
from figure \ref{fig:acc:tgt:gersende}, but show a somewhat
flatter distribution than those of the pure simulation studies.
It will be important to cross-validate these simulations against newer
versions of the simulation codes as they become available.

\subsection{Muon front-end}
\label{Sect:FrontEnd}

The Neutrino Factory muon front-end consists of a pion decay channel
and longitudinal drift, followed by an adiabatic buncher,
phase-rotation system, and ionisation-cooling channel.

The present design is based on the lattice presented in the Neutrino
Factory Study 2A report~\cite{Albright:2004iw} and subsequently
developed in the ISS \cite{Apollonio:2009} with several modifications:
the taper from the target solenoid has been adjusted; the
solenoid-field strength in the drift, buncher, and phase rotation
sections has been reduced from 1.75 T to 1.5 T; the whole system has
been shortened; and the thickness of the lithium hydride absorbers in the
cooling section has been increased. 
These changes result in the same muon-capture performance
in a shorter bunch train, reducing requirements on some systems
downstream of the muon front-end.

\subsubsection{Decay and longitudinal drift}

Downstream of the target solenoid, the magnetic field is adiabatically
reduced from 20 T to 1.5 T over a distance of 15~m.
Over the same distance, the beam pipe radius increases from 0.075~m to
0.3~m.  
This arrangement captures within the 1.5~T decay channel a
secondary-pion beam with a large energy spread. 

The initial proton bunch is relatively short (between 1~ns and 3~ns rms,
see section \ref{Sect:ProtonDriver}) resulting in a short pion
bunch. 
As the secondary pions travel from the target they drift
longitudinally, following $ct = s/\beta_z + ct_0$, where $s$ is
distance along the transport line and $\beta_z = v_z/c$ is the
relativistic longitudinal velocity. Hence, downstream of the target,
the pions and their daughter muons develop a position-energy
correlation in this RF-free decay channel.  In the present baseline,
the longitudinal drift length $L_D$ = 57.7 m, and at the end of the
decay channel there are about 0.4 muons of each sign per incident
8~GeV proton.

\subsubsection{Buncher}

The drift channel is followed by a buncher section that uses RF
cavities to form the muon beam into a train of bunches and a
phase-energy rotating section that decelerates the leading high-energy
bunches and accelerates the late low energy bunches, so that each
bunch has the same mean energy.  
The baseline design delivers a bunch train
that is less than 80~m long. This is an improvement over the version
of the design developed for the ISS~\cite{Apollonio:2009} which
delivered a 120 m long bunch train containing the same number of
muons. 

A shorter bunch train makes some downstream systems easier
to design. 
For example, one of the constraints on the minimum length of the decay
rings is the total length of the bunch train. 
By making the bunch train shorter, it may be possible to make the
decay rings shorter. 
Also, the FFAG ring has a rather demanding kicker system, mostly
driven by the total circumference of the ring but also influenced by
the length of the bunch train.
A shorter bunch train makes these kickers slightly easier to
construct.

To determine the required buncher parameters, we consider reference
particles (0, $N_B$) at $p_0 = 233$ MeV/c and $p_{N_B} = 154$ MeV/c,
with the intent of capturing muons from an initial kinetic energy
range of 50 to 400 MeV.  The RF cavity frequency, $f_{RF}$, and phase
are set to place these particles at the centre of bunches while the RF
voltage increases along the channel.
These conditions can be
maintained if the RF wavelength, $\lambda_{RF}$, increases along the
buncher, following:
\begin{equation}
  N_B \lambda_{RF}(s) = N_B \frac{c}{f_{RF}(s)} =
  s\left(\frac{1}{\beta_{N_B}}-\frac{1}{\beta_0}\right) \, ;
\end{equation}
where $s$ is the total distance from the target, $\beta_0$ and
$\beta_{N_B}$ are the velocities of the reference particles, and $N_B$
is an integer. For the present design, $N_B$ is chosen to be 10, and
the buncher length is 31.5~m. 
With these parameters, the RF cavities
decrease in frequency from 320~MHz ($\lambda_{RF} = 0.94$ m) to
230~MHz ($\lambda_{RF} = 1.3$~m) over the length of the buncher. 

The initial geometry for the placement of the RF cavities uses
$0.4-0.5$~m long cavities placed within 0.75~m long cells.  
The 1.5~T solenoid focusing of the decay region is continued through
the buncher and the rotator section which follows.  
The RF gradient is increased from cell to
cell along the buncher, and the beam is captured into a string of
bunches, each of which is centred about a test particle position,
with energies determined by the spacing from the initial test particle
such that the $i^{th}$ reference particle has velocity:
\begin{equation}
  1/\beta_i = 1/\beta_0 + \frac{i}{N_B}
  \left(\frac{1}{\beta_{N_B}} - \frac{1}{\beta_0}\right).
\end{equation}
In the initial design, the cavity gradients, $V_{RF}$, follow a linear
increase along the buncher:
\begin{equation}
\label{eq:linear_ramp}
  V_{RF}(z) \approx 9 \frac{z}{L_B} MV/m \, ;
\end{equation}
where $z$ is distance along the buncher and $L_B$ is the length of
the buncher.  The gradient at the end of
the buncher is 9 MV/m.  This gradual increase of the bunching voltage
enables a somewhat adiabatic capture of the muons into separated
bunches, which minimises phase-space dilution.
\begin{table*}
  \caption{Summary of front-end RF requirements. The total installed RF voltage
  is 1184~MV.}
\begin{center}
\begin{tabular}{|l|c|c|c|c|c|c|}
\hline
        & {\bf Length} & {\bf Number}      & {\bf Frequencies}     & {\bf Number}         & {\bf Peak gradient} & {\bf Peak power}    \\
        & {\bf [m]}    & {\bf of cavities} & {\bf [MHz]}           & {\bf of frequencies} & {\bf [MV/m]}        & {\bf requirements}  \\
\hline
Buncher & 33.0         & 37                & 319.6 to 233.6        & 13                   & 3 to 9.71           & 1--3.5\,MW/freq.    \\ 
Rotator & 42.0         & 56                & 230.2 to 202.3        & 15                   & 13                  & 2.5\,MW/cavity      \\ 
Cooler  & 97.5         & 130               & 201.25                & 1                    & 15                  & 4\,MW/cavity        \\ 
\hline
Total   & 172.5        & 219               & 319.6 to 201.25       & 29                   &                     & 562\,MW             \\ 
\hline
\end{tabular}
\label{tab:acc:fe:rf_settings}
\end{center}
\end{table*}

In the practical implementation of the buncher concept, this linear
ramp of cavity frequency is approximated by a sequence of
RF cavities that decrease in frequency along the 33 m beam transport
allotted to the buncher. The number of different RF frequencies is limited to a
more manageable 13 (1--4 RF cavities per frequency).  The linear ramp in
gradient described by equation \ref{eq:linear_ramp} is approximated
by the placement and gradient of the cavities in the buncher.  
Table \ref{tab:acc:fe:rf_settings} shows a summary of the RF cavities
that are needed in the buncher, rotator, and cooling sections. 

\subsubsection{Rotator}

\begin{figure}
  \begin{center}
    \includegraphics[width=0.9\textwidth]%
      {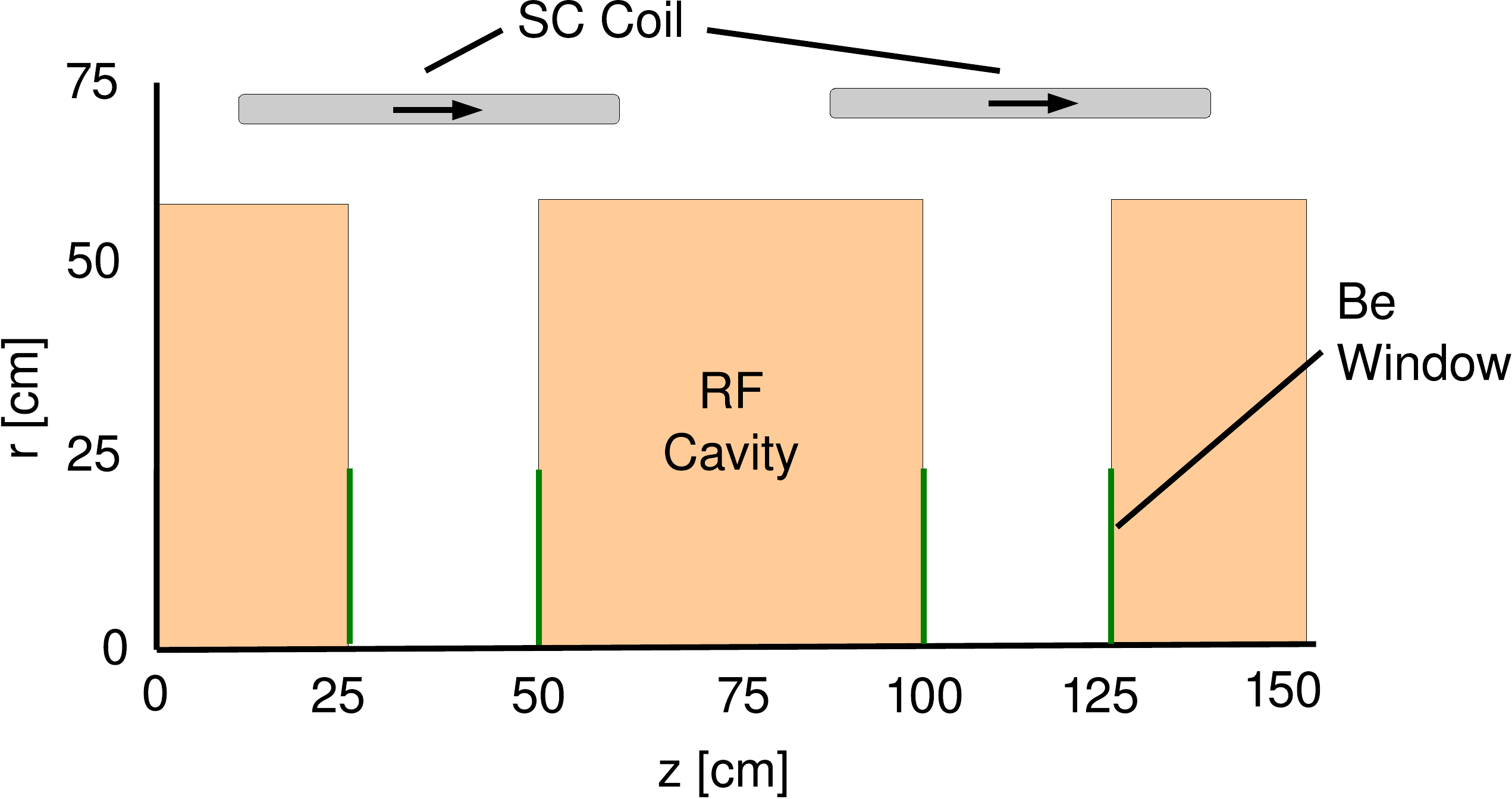}
  \end{center}
  \caption{Schematic radial cross section of a rotator cell. }
  \label{fig:acc:fe:rotator_schematic}
\end{figure}

In the rotator section, the RF bunch-spacing between the reference
particles is shifted away from the integer, $N_B$, by an increment,
$\delta N_B$, and phased so that the high-energy reference particle is
stationary and the low-energy one is uniformly accelerated to arrive
at the same energy as the first reference particle at the end of the
rotator.  For the baseline, $\delta N_B =$ 0.05 and the bunch spacing
between the reference particles is $N_B + \delta N_B = 10.05$.  This
is accomplished using an RF gradient of 12 MV/m in 0.5 m long RF
cavities within 0.75 m long cells.  The RF frequency decreases from
230.2 MHz to 202.3 MHz along the length of the 42 m long rotator
region. 
A schematic of a rotator cell is shown in figure
\ref{fig:acc:fe:rotator_schematic}. 

The RF frequency is set by requiring that the trajectories of the
reference particles be spaced in $ct$ by $(N_B + \delta N_B)$
wavelengths. 
In a practical implementation, a continuous change in frequency from
cavity to cavity is replaced by grouping adjacent sets of cavities
into the same RF frequency. 
The 42\,m long RF rotator, then contains 56
RF cavities grouped into 15 frequencies.

\begin{figure}
  \begin{center}
    \includegraphics[width=0.63\textwidth]%
      {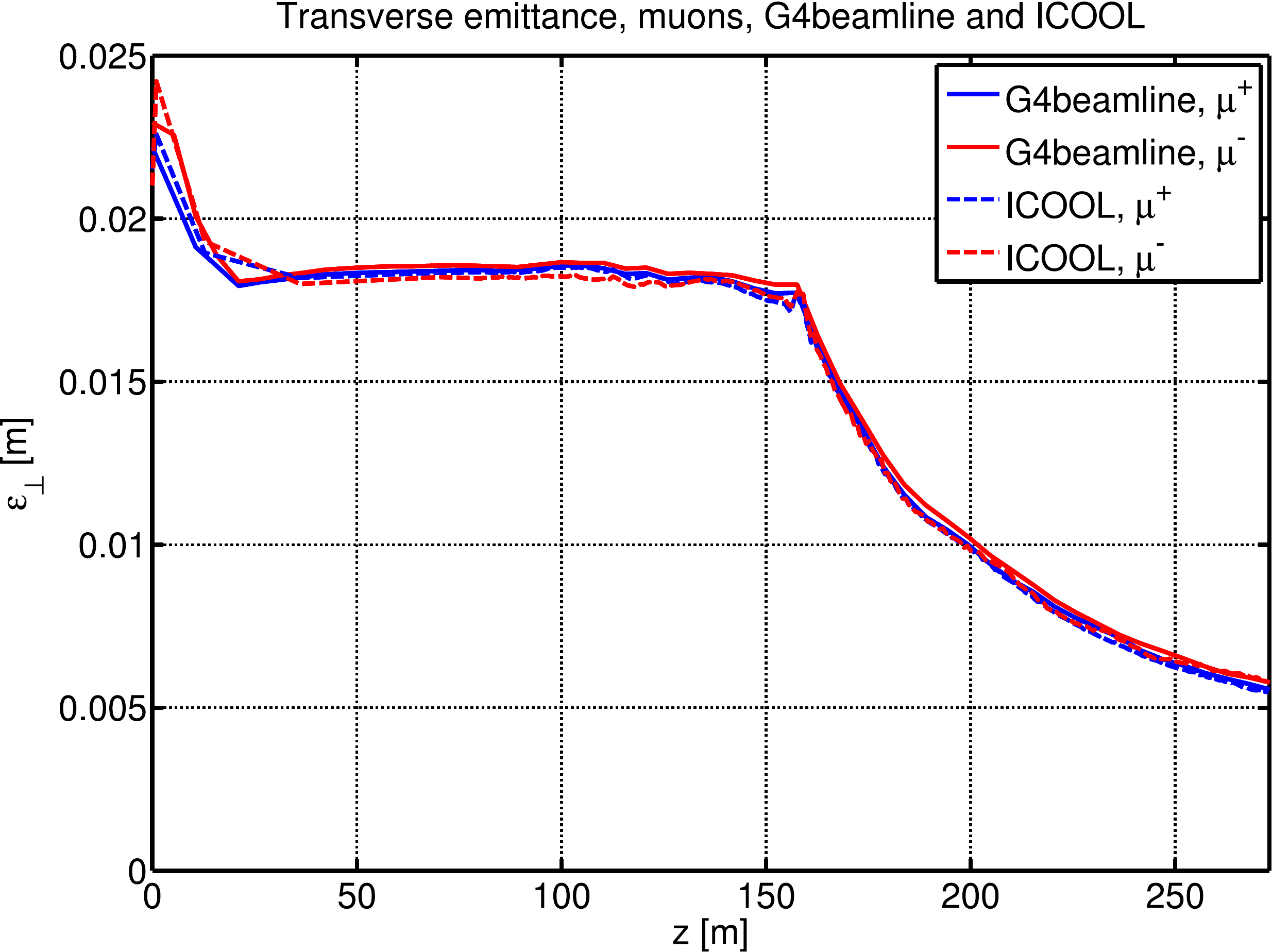}
    \includegraphics[width=0.63\textwidth]%
      {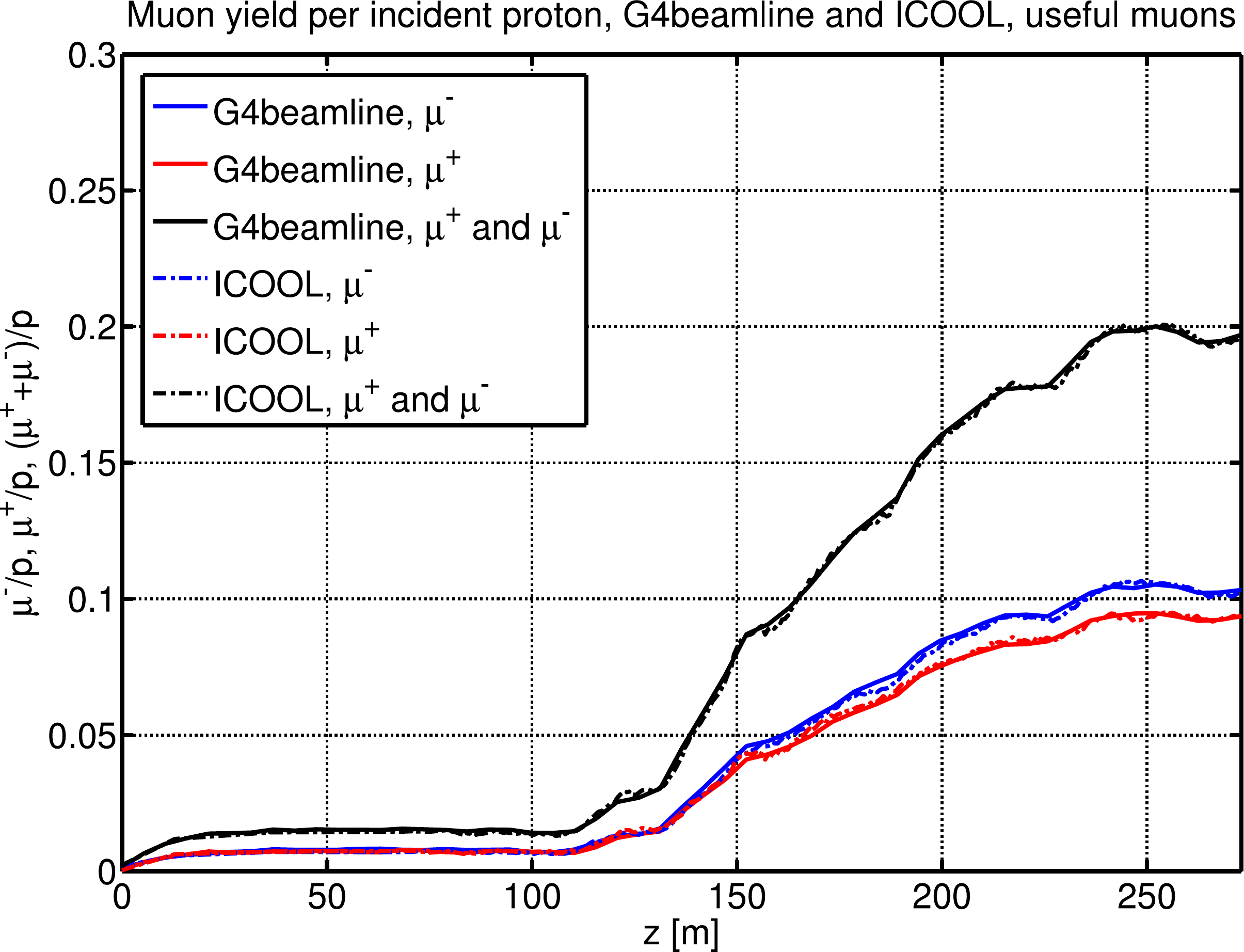}
  \end{center}
\caption{
  Performance of the bunching and cooling channel as a function
  of distance along the channel, as simulated using the ICOOL
  code~\cite{Fernow:2005cq} and the G4beamline
  code~\cite{Roberts:2008zzc}. (top) The evolution of the rms
  transverse emittance (computed over all bunches).
  (bottom) The evolution of the number of
  muons within a reference acceptance (muons within 201.25 MHz RF
  bunches with momentum in the range 100--300 MeV/c, transverse
  amplitude squared less than 0.03\,m and longitudinal amplitude
  squared less than 0.15\,m). The cooling section starts at $s = 155$
  m, where the rms transverse emittance is 0.018 m and 0.08 $\mu$ per
  proton are in the reference acceptance. The capture performance is
  shown for a cooling channel extending to $s = 270$~m although in
  this design the cooling channel extends only to 230~m.  
  Acceptance is maximal at 0.20~$\mu$ per initial 8~GeV proton at 
  $s = 240$~m (85~m of cooling) and the RMS transverse emittance is
  7\,mm.
  At $s = 230$~m (75~m of cooling) the number of $\mu$ per proton
  is 0.19 and the transverse emittance is 7.5\,mm.
  }
  \label{fig:acc:fe:front_end_performance}
\end{figure}

Within the rotator, as the reference particles are accelerated to the
central energy (at $p =$ 233 MeV/c) at the end of the channel, the
beam bunches formed before and after the central bunch are decelerated
and accelerated respectively, obtaining at the end of the rotator a
string of bunches of equal energy for both muon species.
At the end of the rotator the RF frequency matches into the RF
frequency of the ionisation cooling channel (201.25 MHz).  The average
momentum at the rotator is 230 MeV/c. The performance of the bunching
and phase rotation channel, along with the subsequent cooling channel,
is displayed in figure \ref{fig:acc:fe:front_end_performance}, which
shows, as a function of the distance down the channel, the number of
muons within a reference acceptance. 
The phase rotation increases the ``accepted'' muons by a factor of
four.

\begin{figure}
  \begin{center}
    \includegraphics[width=0.9\textwidth]%
      {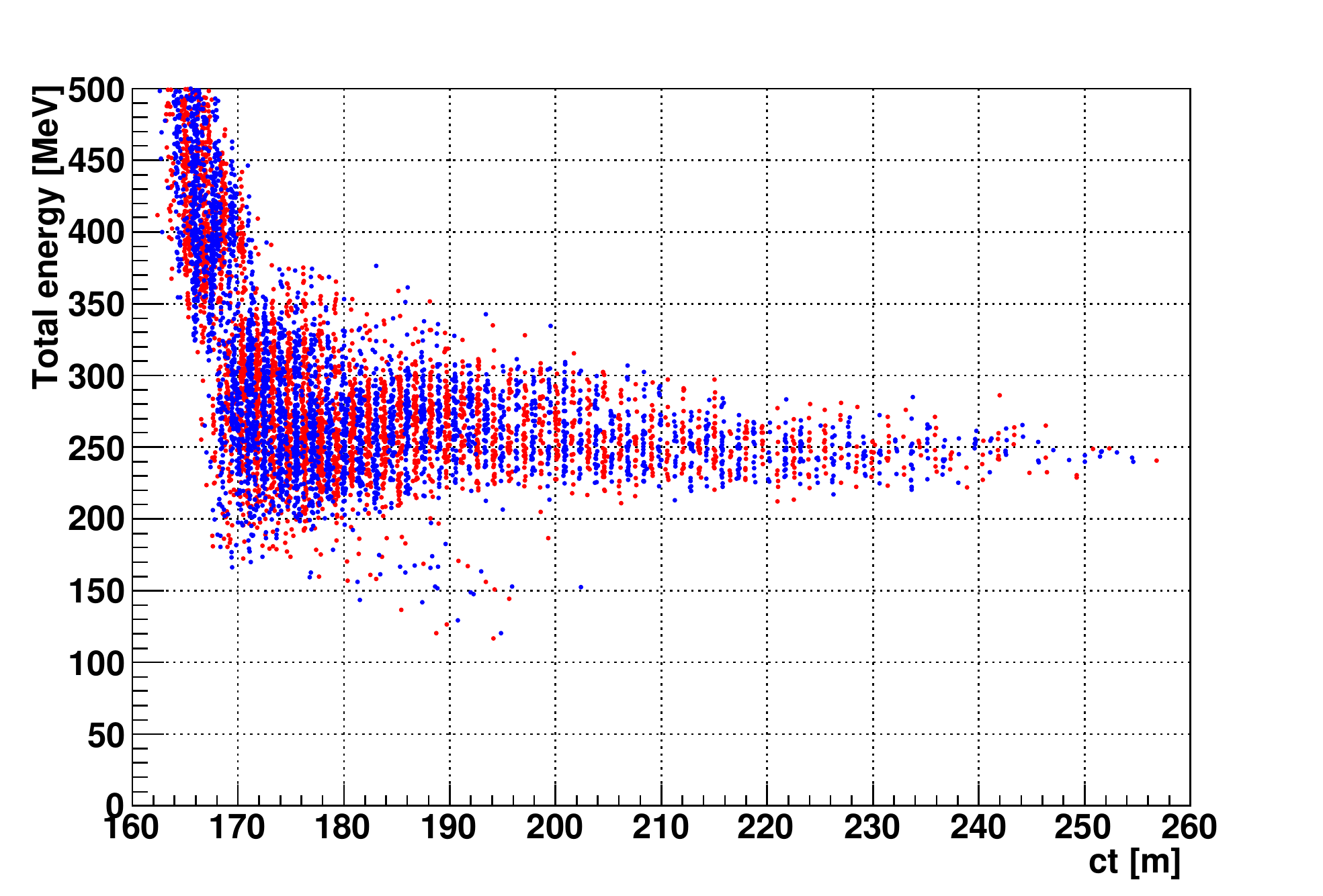}
  \end{center}
  \caption{
    Distribution of particles in longitudinal phase-space at the
    phase rotation end. $\mu+$ are shown in red and $\mu-$ are shown in
    blue.
  }
  \label{fig:acc:fe:neg_and_pos_figure}
\end{figure}

A critical feature of the muon production, collection, bunching, and
phase rotation system is that it produces bunches of both signs
($\mu^+$ and $\mu^-$) at roughly equal intensities. 
This occurs because the focusing systems are solenoids which focus
both signs, and the RF systems have stable acceleration for both
signs, separated by a phase difference of $\pi$. 
The distribution of muons in longitudinal phase
space for particles of both signs at the end of the rotator is shown
in figure \ref{fig:acc:fe:neg_and_pos_figure}.

\subsubsection{Cooling channel}

\begin{figure}
  \begin{center}
    \includegraphics[width=0.9\textwidth]%
      {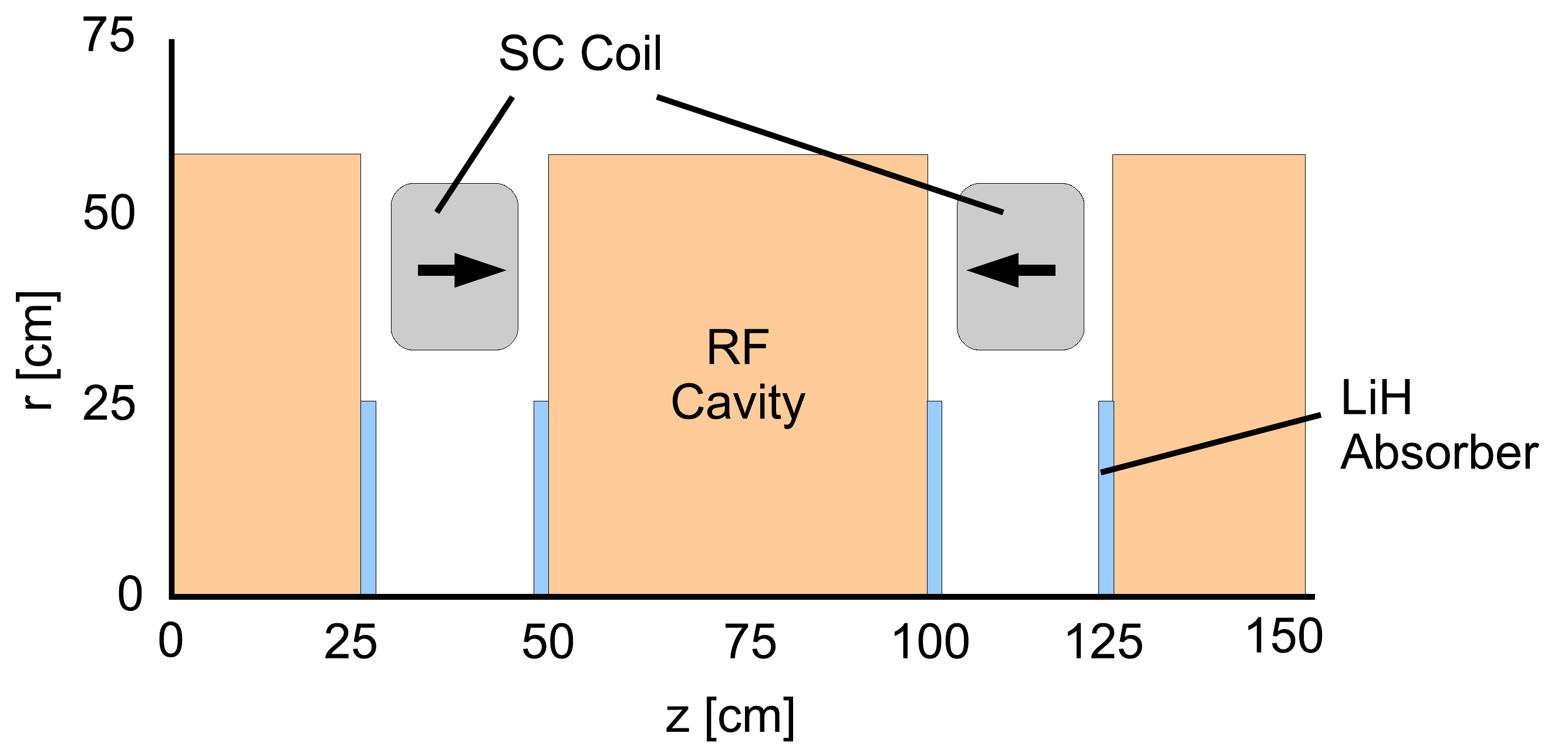}
  \end{center}
  \caption{Schematic radial cross section of a cooling cell.}
  \label{fig:acc:fe:cooling_schematic}
\end{figure}

The baseline cooling-channel design consists of a sequence of
identical 1.5 m long cells (figure
\ref{fig:acc:fe:cooling_schematic}).  
Each cell contains two 0.5 m-long RF cavities, with 1.1 cm thick LiH
discs at the ends of each cavity (4 per cell) and a 0.25 m spacing
between cavities. 
The LiH discs provide the energy-loss material for
ionisation cooling.  
The cells contain two solenoid coils with opposite polarity.
The coils produce an approximately sinusoidal variation of the
magnetic field in the channel with a peak value on-axis of 2.8 T,
providing transverse focusing with $\beta_\perp = 0.8$~m.  
The currents in the first two
cells are perturbed from the reference values to provide matching from
the constant-field solenoid in the buncher and rotator sections.  The
total length of the cooling section is 75~m (50 cells).  Based on the
simulation results shown in figure \ref{fig:acc:fe:front_end_performance}, the
cooling channel is expected to reduce the rms transverse normalised
emittance from $\epsilon_{N} = 0.018$~m to 
$\epsilon_{N} = 0.0075$~m.  
The rms longitudinal emittance is $\epsilon_{L}= 0.07$~m/bunch.

The effect of the cooling can be measured by counting the number of
simulated particles that fall within a reference acceptance that
approximates the expected acceptance of the downstream accelerator.

The squared amplitude $A_\perp^2$ is given by:
\begin{equation}
  A_\perp^2 = p_z/m \big[ \beta_\perp(x^{\prime2}+y^{\prime2})+
  \gamma_\perp(x^2+y^2)+2\alpha_\perp(xx^\prime+yy^\prime)+
  2(\beta_\perp\kappa - \mathcal{L})(xy^\prime-yx^\prime) \big] \, ;
\end{equation}
where $\beta_\perp$, $\alpha_\perp$, $\gamma_\perp$ are solenoidal
equivalents of the Twiss parameters, $\kappa$ is the solenoidal
focusing strength, and $\mathcal{L}$ is the dimensionless kinetic
angular momentum~\cite{Fernow:2003zz,Penn:2000mt}.

For longitudinal motion, the variables $t_c = ct$ (phase lag in periods
within a bunch multiplied by RF wavelength) and $\Delta E$ (energy
difference from centroid) are used rather than $(z, z^\prime)$. The
longitudinal squared amplitude is given by:
\begin{equation}
  A_L^2=\frac{c}{m_\mu} \left[ \frac{t_c^2}{\delta} +
    \delta \left( \Delta E - \frac{\alpha_L t_c}{\delta}\right)^2
    \right] \, ;
\end{equation}	
where $\delta$ is defined by:
\begin{equation}
  \delta= \frac{c \left<t_c^2\right>}{m_\mu\epsilon_L} \, ;
\end{equation}
$\epsilon_L$ is a normalised longitudinal emittance:
\begin{equation}
  \epsilon_L= \frac{c}{m_\mu}\sqrt{\left<t_c^2\right>\left<\Delta
    E^2\right>-\left<t_c\Delta E\right>^2} \, ;
\end{equation}
and $\alpha_L$ is a correlation factor:
\begin{equation}
  \alpha_L= \frac{c}{m_\mu \epsilon_L}\left<t_c\Delta E\right> \, .
\end{equation}
Following criteria developed using the ECALC9F program (distributed with
ICOOL), a particle is
considered to be within the acceptance of the machine if the
transverse amplitude squared $A_\perp^2$ is less than 0.03~m and the
longitudinal amplitude squared is less than 0.15\,m. Note that the
transverse and longitudinal notations are not the same and
transverse-longitudinal amplitude correlations are not included.
This is a crude first approximation to the muon accelerator
acceptance, but is used in the present tables for consistent
comparison of simulations.

Using the output from our re-optimised buncher and rotator, we have
tracked particles through the cooling channel, and obtain, within
the reference acceptances, 0.19~$\mu$ per 8~GeV incident proton. 
The acceptance criteria remove larger amplitude particles from the
distribution and the rms emittance of the accepted beam is therefore
much less than that of the entire beam.  The rms transverse emittance, 
$\epsilon_\perp$, of the accepted beam is 0.004 m and the rms
longitudinal emittance is 0.036 m.

At the end of the cooling channel, there are interlaced trains of
positive and negative muon bunches.  The trains of usable muon
bunches are 80~m long (50 bunches), with 70$\%$ of the muons in the
leading 20 bunches (30 m).  The bunch length is 0.16~m in $ct$ for
each bunch, with a mean momentum of 230\,MeV/c and an rms width 
$\delta p$ of $28$~MeV/c.  For the accepted beam, the rms bunch width
is 3.8~cm and the rms transverse momentum is 10~MeV/c.  

\subsubsection{Simulation codes}

Two independently-developed codes have been used for tracking
simulations of the muon front-end by the Monte Carlo method: ICOOL
version 3.20 \cite{Fernow:2005cq}; and G4beamline version 2.06
\cite{Roberts:2008zzc}.

ICOOL is under active development at the Brookhaven National
Laboratory. It is a 3-dimensional tracking program that was
originally written to study ionisation cooling of muon beams. The
program simulates particle-by-particle propagation through materials
and electromagnetic fields. The physics model is most accurate for
muons in the kinetic energy range of 50\,MeV to 1\,GeV, but tracking of
electrons, pions, kaons, and protons is also possible. ICOOL includes
a number of custom models for particle decay, delta-ray production,
multiple Coulomb scattering, ionisation energy loss, and energy
straggling.

G4beamline is a particle tracking and simulation program
under active development by Muons, Inc. Physics processes are modelled
using the Geant4 toolkit with the QGSP\_BERT physics package~\cite{Agostinelli:2002hh}
and it is specifically designed to simulate beam lines, and other
systems using single-particle tracking.

Both codes use semi-analytic procedures to compute electromagnetic
fields. 
Solenoid fields are generated as a sum of elliptic integrals
calculated using the solenoid-coil geometry~\cite{Fernow:2003}. 
RF cavities are modelled using a Bessel function radially and a
sinusoid in time for the ideal field produced by a cylindrical pillbox
cavity. 

Good agreement is shown in the muon yield and the yield of other
particle species from the two codes
(see figures \ref{fig:acc:fe:front_end_performance} and
\ref{fig:acc:fe:power_deposition}). Note that different versions of
ICOOL have been shown to disagree at the level of a few percent
\cite{Rogers:2010}.

Analysis of results is performed using ECALC9F version 2.07 in
addition to custom scripts written in Python and MATLAB. The beam has
been generated using MARS 15.07. We expect a significant systematic
error on the overall rate owing to uncertainty in the models used to
generate the input beam.

\subsubsection{RF Requirements and Design}

The RF cavities in this design are all normal conducting cavities having
29 frequencies in the range 201.25 MHz to 320 MHz. The cavities are
50\,cm long with peak field gradients in the range 4\,MV/m to
15\,MV/m, with the highest voltage required for the 201 MHz cavities.
The power consumption of these cavities has been estimated
semi-analytically using standard formul\ae, and the results are listed
in tables 
\ref{tab:acc:fe:buncher_rf_settings} and
\ref{tab:acc:fe:rot_cool_rf_settings},  together with the RF cavities
required. The position and phase of every cavity is listed in Table
\ref{tab:acc:fe:rot_rf_detailed_list}. 
\begin{table*}
\caption{Front-end RF requirements for the buncher system.}
\begin{center}
\begin{tabular}{|c|c|c|c|c|c|}
\hline
{\bf Frequency}   &{\bf ${\bf V_{\text{Tot}}}$ (per}      &{\bf Number of}  &{\bf Length} &{\bf Gradient} & {\bf Peak RF Power} \\
{\bf\,[MHz]}\,         &{\bf frequency) [MV]}   &{\bf cavities}
&{\bf [m]}          &{\bf [MV/m]}         & {\bf (per frequency) [MW]}
\\ 
\hline
319.63   &1.37    & 1   &0.4    &3.42  &0.2  \\
305.56   &3.92    & 2   &0.4    &4.894 &0.6  \\
293.93   &3.34    & 2   &0.45   &4.17  &0.5  \\
285.46   &4.8     & 2   &0.45   &5.34  &1    \\
278.59   &5.72    & 2   &0.45   &6.36  &1.25 \\
272.05   &6.66    & 3   &0.45   &4.94  &1.5  \\
265.8    &7.57    & 3   &0.45   &5.61  &1.5  \\
259.83   &8.48    & 3   &0.45   &6.3   &2    \\
254.13   &9.41    & 3   &0.45   &6.97  &2.3  \\
248.67   &10.33   & 4   &0.45   &7.65  &2.3  \\
243.44   &11.23   & 4   &0.45   &8.31  &2.5  \\
238.42   &12.16   & 4   &0.45   &9.01  &3    \\
233.61   &13.11   & 4   &0.45   &9.71  &3.5  \\
Total    &98.1    &37   &       &      &22   \\
\hline
\end{tabular}
\label{tab:acc:fe:buncher_rf_settings}
\end{center}
\end{table*}
\begin{table*}
  \caption{Front-end RF requirements for the rotator and cooler systems.
    $f_{\text{RF}}$ is the RF frequency, 
    $V_{\text{tot}}$ is the total voltage required at that frequency, 
    $n_{\text{cav}}$ is the number of cavities at that frequency, 
    $E_{\text{peak}}$ is the peak gradient in each cavity,
    and $P_{\text{peak}}$ is the peak RF power required per cavity.}
\begin{center}
\begin{tabular}{|c|c|c|c|c|}
\hline
{\bf $f_{\text{RF}}$ [MHz]}&{\bf $V_{\text{tot}}$ [MV]}&{\bf $n_{\text{cav}}$}&{\bf $E_{\text{peak}}$ [MV/m]}&{\bf $P_{\text{peak}}$ [MW]}\\
\hline
Rotator&       &   &      &       \\
230.19 & 19.5  & 3 & 13.0 &   2.2 \\
226.13 & 19.5  & 3 & 13.0 &   2.2 \\
222.59 & 19.5  & 3 & 13.0 &   2.3 \\
219.48 & 19.5  & 3 & 13.0 &   2.4 \\
216.76 & 19.5  & 3 & 13.0 &   2.4 \\
214.37 & 19.5  & 3 & 13.0 &   2.5 \\
212.48 & 19.5  & 3 & 13.0 &   2.5 \\
210.46 & 19.5  & 3 & 13.0 &   2.6 \\
208.64 & 26.0  & 4 & 13.0 &   2.6 \\
206.9 & 26.0   & 4 & 13.0 &   2.7 \\
205.49 & 26.0  & 4 & 13.0 &   2.7 \\
204.25 & 32.5  & 5 & 13.0 &   2.7 \\
203.26 & 32.5  & 5 & 13.0 &   2.8 \\
202.63 & 32.5  & 5 & 13.0 &   2.8 \\
202.33 & 32.5  & 5 & 13.0 &   2.8 \\
Total  & 364.0 & 56 &     & 144.8 \\
\hline
Cooler &       &    &     &       \\
201.25 & 880   & 130& 16  &   4.3 \\
\hline
\end{tabular}
\label{tab:acc:fe:rot_cool_rf_settings}
\end{center}
\end{table*}
\begin{table*}
\caption{Full list of RF cavities. Cavities are grouped by
  frequency. Position is the position of the upstream edge of the
  first cavity in the group. $z$-Separation is the distance between the
  centres of each cavity in the group. 0$^\circ$ is bunching mode
  while 5$^\circ$ and 35$^\circ$ are partially accelerating modes.} 
\begin{center}
\begin{tabular}{|c|c|c|c|c|c|c|}
\hline
{\bf $z$ Position}  &{\bf Phase}      &{\bf Peak gradient}  & {\bf Frequency} & {\bf Length} & {\bf Number} & {\bf $z$-Separation} \\
{\bf \,[m]\,}       &{\bf [deg.]}     &{\bf [MV/m]}         & {\bf [MHz]}     & {\bf [m]}    &              & {\bf [m]}            \\ 
\hline
81.27 & 0.0 & 3.42 & 319.63 & 0.40 & 1 &  \, \\    
85.02 & 0.0 & 4.894 & 305.56 & 0.40 & 2 & 0.40 \\
88.77 & 0.0 & 4.17 & 293.93 & 0.40 & 2 & 0.40 \\
91.77 & 0.0 & 5.34 & 285.46 & 0.45 & 2 & 0.45 \\
94.02 & 0.0 & 6.36 & 278.59 & 0.45 & 2 & 0.45 \\
96.27 & 0.0 & 4.94 & 272.05 & 0.45 & 3 & 0.45 \\
98.52 & 0.0 & 5.61 & 265.8 & 0.45 & 3 & 0.45 \\ 
100.77 & 0.0 & 6.3 & 259.83 & 0.45 & 3 & 0.45 \\
103.02 & 0.0 & 6.97 & 254.13 & 0.45 & 3 & 0.45 \\
105.27 & 0.0 & 7.65 & 248.67 & 0.45 & 3 & 0.45 \\
107.52 & 0.0 & 8.31 & 243.44 & 0.45 & 3 & 0.45 \\   
109.77 & 0.0 & 9.01 & 238.42 & 0.45 & 3 & 0.45 \\   
112.02 & 0.0 & 9.71 & 233.61 & 0.45 & 3 & 0.45 \\   
114.27 & 5.0 & 13.0 & 230.19 & 0.5 & 3 & 0.75 \\   
116.52 & 5.0 & 13.0 & 226.13 & 0.5 & 3 & 0.75 \\   
118.77 & 5.0 & 13.0 & 222.59 & 0.5 & 3 & 0.75 \\   
121.02 & 5.0 & 13.0 & 219.48 & 0.5 & 3 & 0.75 \\   
123.27 & 5.0 & 13.0 & 216.76 & 0.5 & 3 & 0.75 \\   
125.52 & 5.0 & 13.0 & 214.37 & 0.5 & 3 & 0.75 \\   
127.77 & 5.0 & 13.0 & 212.48 & 0.5 & 3 & 0.75 \\   
130.02 & 5.0 & 13.0 & 210.46 & 0.5 & 3 & 0.75 \\   
132.27 & 5.0 & 13.0 & 208.64 & 0.5 & 4 & 0.75 \\   
135.27 & 5.0 & 13.0 & 206.90 & 0.5 & 4 & 0.75 \\    
138.27 & 5.0 & 13.0 & 205.49 & 0.5 & 4 & 0.75 \\   
141.27 & 5.0 & 13.0 & 204.25 & 0.5 & 5 & 0.75 \\   
145.02 & 5.0 & 13.0 & 203.26 & 0.5 & 5 & 0.75 \\   
148.77 & 5.0 & 13.0 & 202.63 & 0.5 & 5 & 0.75 \\   
152.52 & 5.0 & 13.0 & 202.33 & 0.5 & 5 & 0.75 \\   
154.6 & 35.0 & 16.0 & 201.25 & 0.5 & 130 & 0.75 \\    
\hline
\end{tabular}
\label{tab:acc:fe:rot_rf_detailed_list}
\end{center}
\end{table*}

Several RF cavities have been constructed to support the muon
accelerator design effort~\cite{Huang:2009zzn}. A 43~cm long, 201~MHz
RF cavity has been constructed and operated at peak gradients up
to 21~MV/m and 10 more RF cavities are under construction as part of
the international Muon Ionisation Cooling Experiment
(MICE)~\cite{Alekou:2010zz}. 
Additionally, several 805~MHz cavities have been constructed and
operated at gradients up to 40~MV/m. 
Design and construction of cavities with intermediate
frequencies is not expected to present any additional difficulties.

\subsubsection{Effect of magnetic field on RF gradient}

\begin{figure}
  \begin{center}
    \includegraphics[width=0.9\textwidth]%
      {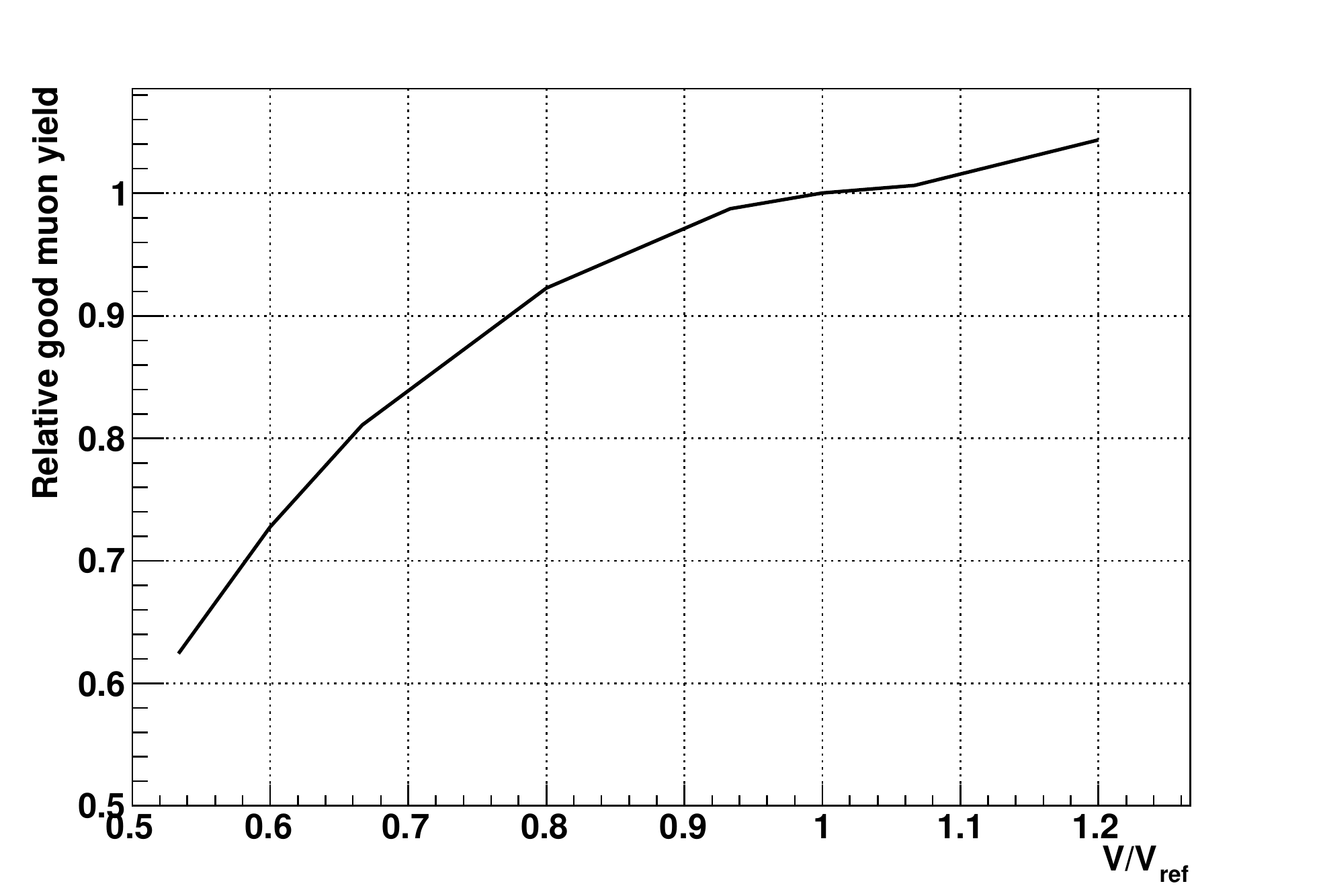}
  \end{center}
  \caption{
    Relative (good) muon yield within the nominal acceptance
    as a function of the peak field gradient, simulated in
    ICOOL. $V_{\text{ref}}$ is the baseline gradient; all cavities were scaled
    by the constant $V/V_{\text{ref}}$. 
  }
  \label{fig:acc:fe:reduced_rf}
\end{figure}

There is empirical evidence that magnetic fields
overlapping RF cavities, as present in the muon front-end,
may induce breakdown in the cavities~\cite{Moretti:2005zz,Palmer:2009zza}.
The performance of the muon front-end using a
reduced field has been explored using ICOOL. In
figure \ref{fig:acc:fe:reduced_rf}, the muon transmission is shown as
a function of fractional change in RF gradient in the buncher,
rotator, and cooling channel.
The simulations indicate that around the nominal gradient, muon
transmission is rather insensitive to the peak gradients. 
If the achievable RF gradient falls dramatically below the nominal
value, there is a significant effect on muon transmission. 
In order to mitigate this technical risk, several alternative
lattices have been developed and are described briefly in Appendix
\ref{sec:acc:fe:FrontEnd_Alternatives}.  

\subsubsection{Magnet requirements and design}

Pion capture in the target system is achieved using a series of
solenoids in which the field tapers from 20~T to 1.5~T over 15~m (see
section \ref{Sect:Trgt}).
Here we describe solenoids in the drift, buncher, and rotator that
produce a constant magnetic field of 1.5~T and solenoids in the cooler
that produce the alternating solenoid field configuration. 
These coils are summarised in table \ref{tab:acc:fe:magnet_settings}.
\begin{table*}
\caption{Summary of front-end magnet requirements.}
\begin{center}
\begin{tabular}{|l|c|c|c|c|c|}
\hline
                  & {\bf Length} & {\bf Inner radius} & {\bf Radial thickness} & {\bf Current density} & {\bf Number} \\
                  & {\bf [m]}    & {\bf [m]}          & {\bf [m]}              & {\bf [A/mm$^2$]}      &  \\
\hline
Initial transport & 0.5    & 0.68         &  0.04     & 47.5      & 180 \\
Cooling channel   & 0.15   & 0.35         &  0.15     & $\pm$107  & 130 \\
\hline
\end{tabular}
\label{tab:acc:fe:magnet_settings}
\end{center}
\end{table*}

The 1.5~T solenoids must accommodate the beam pipe, with a 30 cm
radius. Within the buncher and rotator, they must also accommodate RF
cavities with radii of 60 cm.  This can be achieved using coils with
an inner radius of 68 cm and a conductor radial thickness of 4\,cm, so
that the cavities fit entirely within the coils. A coil length of 50 cm
spaced at 75 cm intervals leaves a gap of 25 cm between coils,
matching the periodicity of the cooling channel and enabling access
for room temperature services such as vacuum and RF power feeds.  The
required current for these coils is 47.5 A/mm$^2$ to give a total
current of 0.95 MA-turns. The coils are therefore large enough to
accommodate the beam pipe, RF and diagnostics, and added shielding. A
smaller radius could be used in the first 60 m, which has no RF. The
135 m transport requires 180 such magnets.

The cooling system requires strong alternating-sign coils that are
placed between RF cavities, fitting within the 25 cm inter-cavity
spaces (figure \ref{fig:acc:fe:cooling_schematic}).  The coils are 15 cm long
with inner radius 35 cm, radial thickness 15 cm and current density of
$\pm107\text{ A/mm}^2$ to give a total current of 2.4~MA-turns.  The coil
currents alternate in direction from coil to coil.  These coils
produce an on-axis solenoid field that varies from $+$2.8~T to
$-$2.8~T over a 1.5~m period, following an approximately sinusoidal 
dependence.  Maximum fields in the cooling cell volume are 5~T near
the coil surfaces. 100 such coils are needed in a 75~m cooling system.

\subsubsection{Beam Losses}

There are significant particle losses along the beam line and these
will result in a large energy deposition in superconducting magnets and
other equipment. 
Two main risks have been identified: energy deposition by all
particles may cause superconducting equipment to quench; and energy
deposition by hadrons and other particles may activate equipment
preventing hands-on maintenance. 

\begin{figure}
  \begin{center}
    \includegraphics[width=0.82\textwidth]%
      {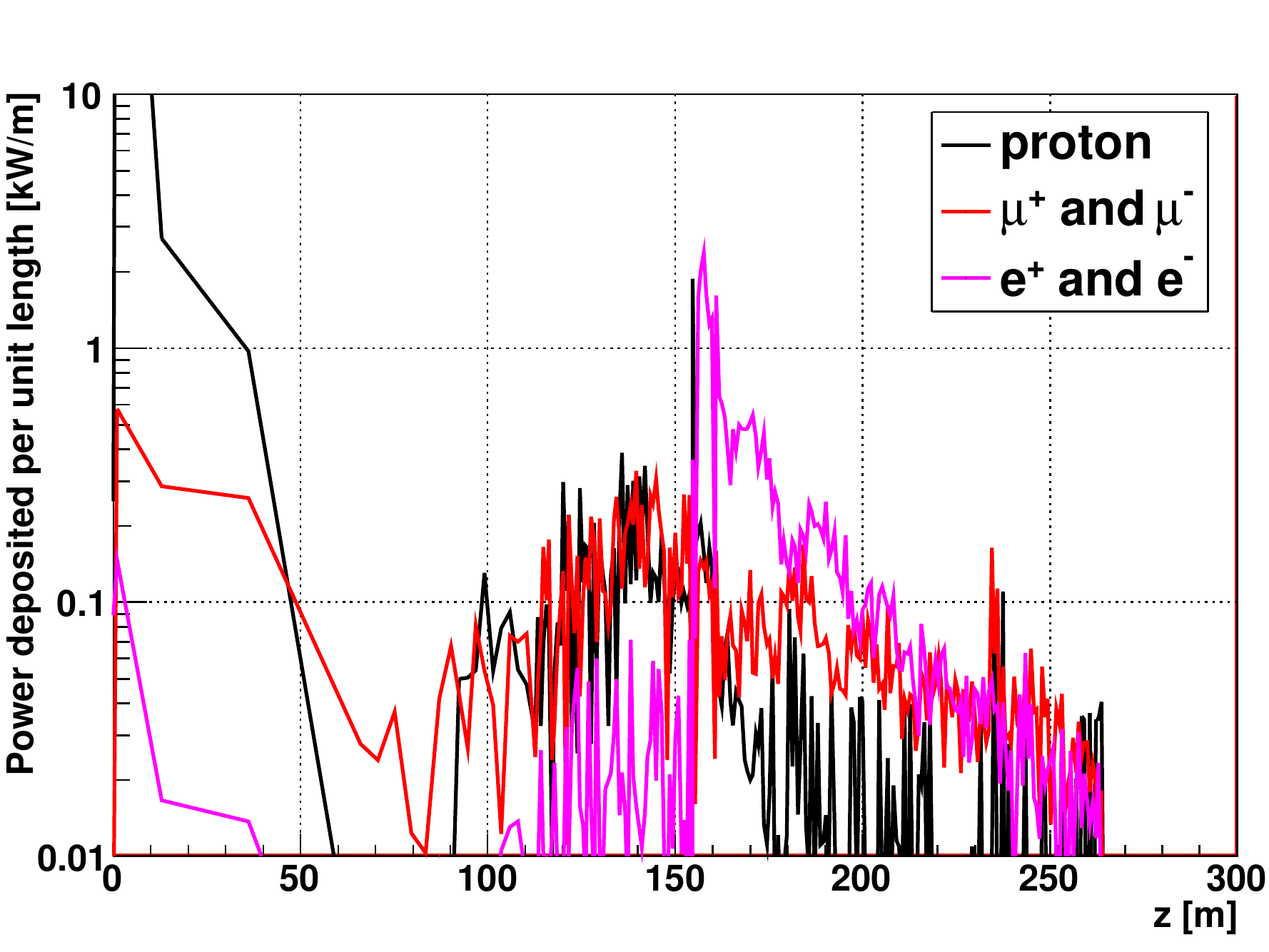}
    \includegraphics[width=0.78\textwidth]%
      {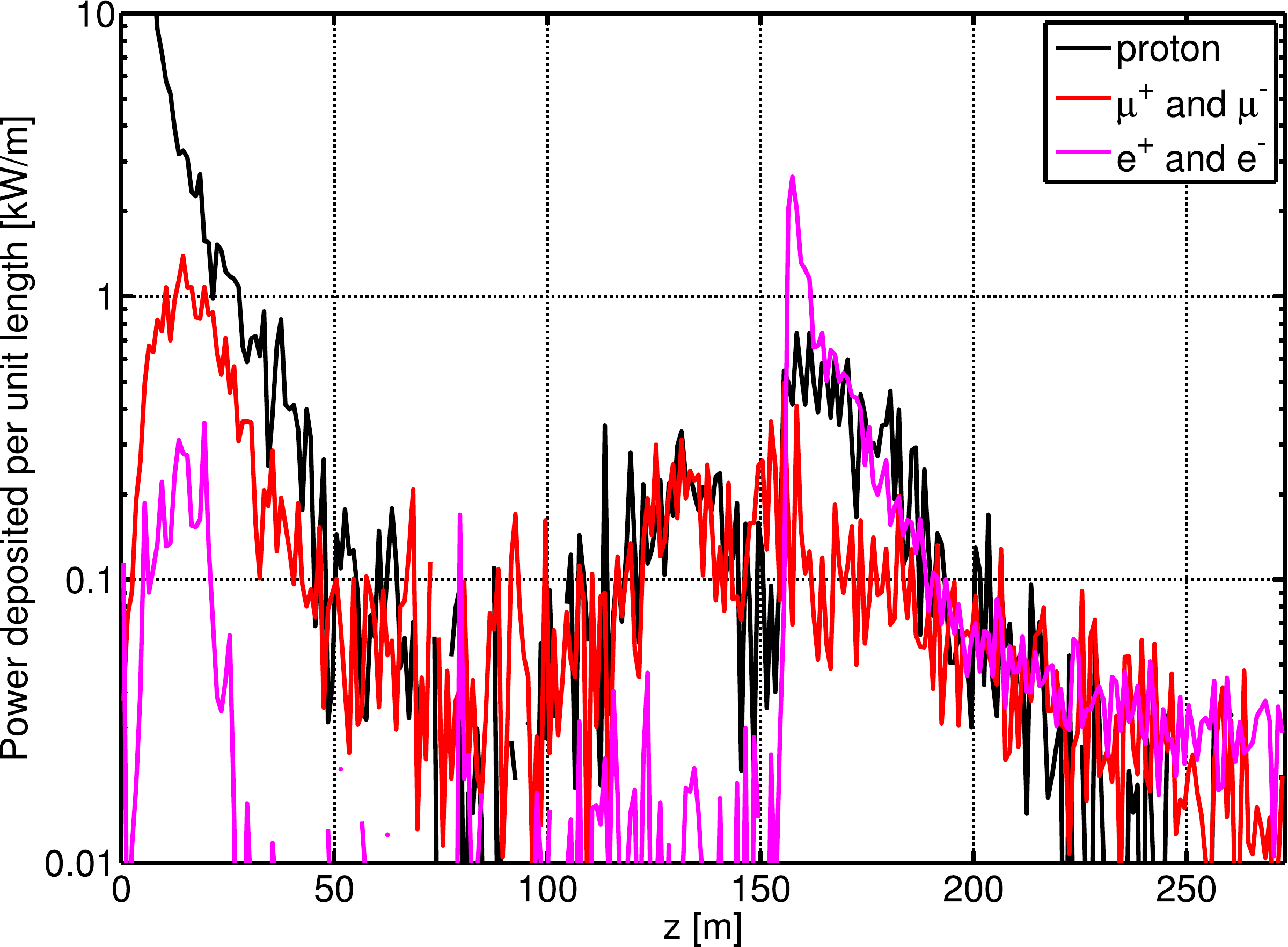}
  \end{center}
  \caption{
    Power deposited by transmission losses of various particle
    species in the surrounding equipment simulated in (top) ICOOL and
    (bottom) G4Beamline.
  }
  \label{fig:acc:fe:power_deposition}
\end{figure}

In figure \ref{fig:acc:fe:power_deposition}, the power deposited by
transmission losses per unit length from various particle species is
shown as a function of distance along the channel. 
Note that energy deposition in RF
windows and absorbers is not included in this calculation. 
It is expected that this equipment will absorb several kilowatts of
beam power from each particle species.

In currently operating accelerators, uncontrolled hadronic losses must
be less than $\approx$1~W/m to enable ``hands-on'' maintenance without
additional time, distance, or shielding constraints.
Magnets are expected to
quench with beam losses above a few tens of W/cm$^3$. 
Several schemes are envisaged to control the beam losses and reduce
them below these values. 

Three devices are under study for reducing the transmission losses in
the front-end:
\begin{itemize}
\item 
  Low momentum protons may be removed by passing the bean through a
  low-$Z$ ``proton absorber'' . 
  This device takes advantage of the different stopping
  distance of protons compared with other particles in material;
\item 
  Particles with a high momentum, outside of the acceptance of the
  front-end, may be removed using a pair of chicanes. 
  Dispersion is induced in the beam by means of bending
  magnets or tilted solenoids in a chicane arrangement and
  high-momentum particles are passed onto a beam dump. 
  A design for the chicane is being developed that can accommodate the
  large momentum spread.
  In order to retain both muon species, one chicane is required for
  each sign; and
\item 
  Particles with transverse amplitude outside of the acceptance of
  the front-end may be removed using transverse collimators.
\end{itemize}

\subsubsection{Summary}
The Neutrino Factory muon front-end captures a substantial proportion
of the muons produced by the Neutrino Factory target. Longitudinal
capture is achieved using a buncher and energy-time phase-rotation
system while transverse capture is achieved using a high field
solenoid adiabatically tapered to 1.5 T and enhanced by ionisation
cooling. 
Technical risks to the muon front-end are presented by the requirement
for high peak RF fields in the presence of intense magnetic fields and
irradiation of the accelerator hardware due to uncontrolled particle
losses.
Strategies have been outlined by which these risks can be mitigated.
Overall, the muon front-end increases the capture rate of muons in the
nominal accelerator acceptance by a factor 10.

\subsection{Linac and RLA}

\begin{figure}
  \begin{center}
    \includegraphics[width=0.95\textwidth]%
      {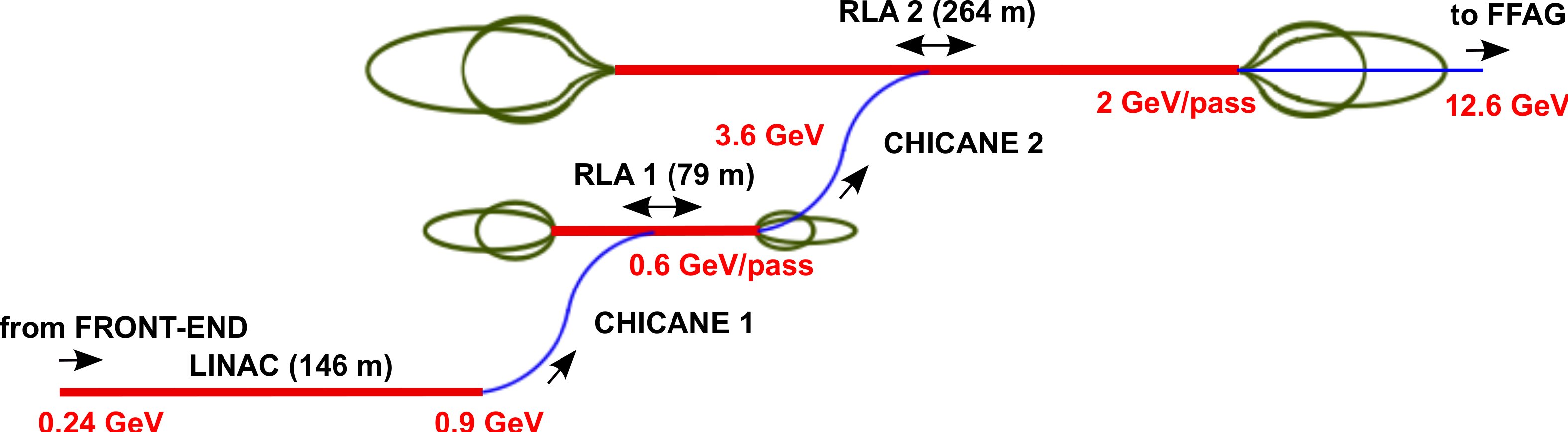}
  \end{center}
  \caption{
    Layout of the linac and recirculating linacs connected by
    chicanes. 
  }
  \label{fig:acc:rla:sketch}
\end{figure}

The muon acceleration process involves a complex chain of accelerators
including a (single-pass) linac, two recirculating linacs (RLAs) and an
FFAG ring~\cite{Bogacz:2008zzb}.  
This section will discuss the linac, the two RLAs and the chicanes,
shown schematically in figure \ref{fig:acc:rla:sketch}.

Acceleration starts after ionisation cooling at 230\,MeV/c and proceeds
to 12.6 GeV. 
The pre-accelerator captures a large muon phase space and accelerates
muons to relativistic energies, while adiabatically decreasing the
phase-space volume so that effective acceleration in the RLA is
possible.
The RLA further compresses and shapes the longitudinal and
transverse phase spaces, while increasing the energy. 
An appropriate choice of multi-pass linac optics based on FODO
focusing assures a large number of passes in the RLA. 
The proposed ``dog-bone'' configuration facilitates simultaneous 
acceleration of both $\mu^+$ and $\mu^-$ species through the
requirement of mirror-symmetric optics in the return ``droplet''
arcs. 

The linac consists of superconducting RF cavities and iron-shielded
solenoids grouped in cryo-modules \cite{Bogacz:2003uk},
while the recirculating linacs RLA~I and RLA~II consist of
superconducting RF cavities and quadrupoles. 
The linac is required to accelerate 0.22~GeV/c muons coming from the
muon front-end to 0.9~GeV/c and, given these relatively low energies,
solenoidal transverse focusing has been chosen so that the beam
preserves its initial horizontal-vertical phase-space coupling.
The transfer to RLA~I is performed through the double chicane,
``chicane~1'', which consists of a vertical dipole spreader (at the
beginning), horizontal bending magnets, a vertical dipole combiner (at
the end), and quadrupoles for transverse focusing
\cite{Bogacz:2009zzc}.
In this manner both positive and negative muons can be transferred,
while keeping RLA~I at the same height as the linac, a decision taken
to simplify the civil engineering.
With the beam now being relativistic, quadrupole focusing in a
FODO-lattice is preferred for chicane~1.
In addition, a number of sextupoles must be inserted at positions
where the dispersion caused by bending has a maximum. 
Entering RLA~I, the beam performs 4.5 passes at an average gain of
0.6~GeV/pass with an excursion along the return arcs after each pass.
Single-cell, superconducting cavities provide acceleration along RLA~I
while transverse focusing is achieved by quadrupoles in a FODO
arrangement along RLA~I and its arcs. 
Similarly, through another double chicane, ``chicane~2'', the 3.6\,GeV
beam is transferred to RLA~II where it again performs 4.5 passes but
now at an average gain of 2~GeV/pass. 
Eventually, the beam is extracted and injected into the FFAG at
12.6~GeV.
Section \ref{sec:acc:rla:track} reports the latest results of
muon-beam tracking and the work which has been performed to allow
tracking using realistic field maps for both RF cavities and
solenoids.

\subsubsection{Overall lattice description }
\paragraph{Linear Pre-Accelerator}

\begin{figure}
  \begin{center}
    \includegraphics[width=0.98\textwidth]%
      {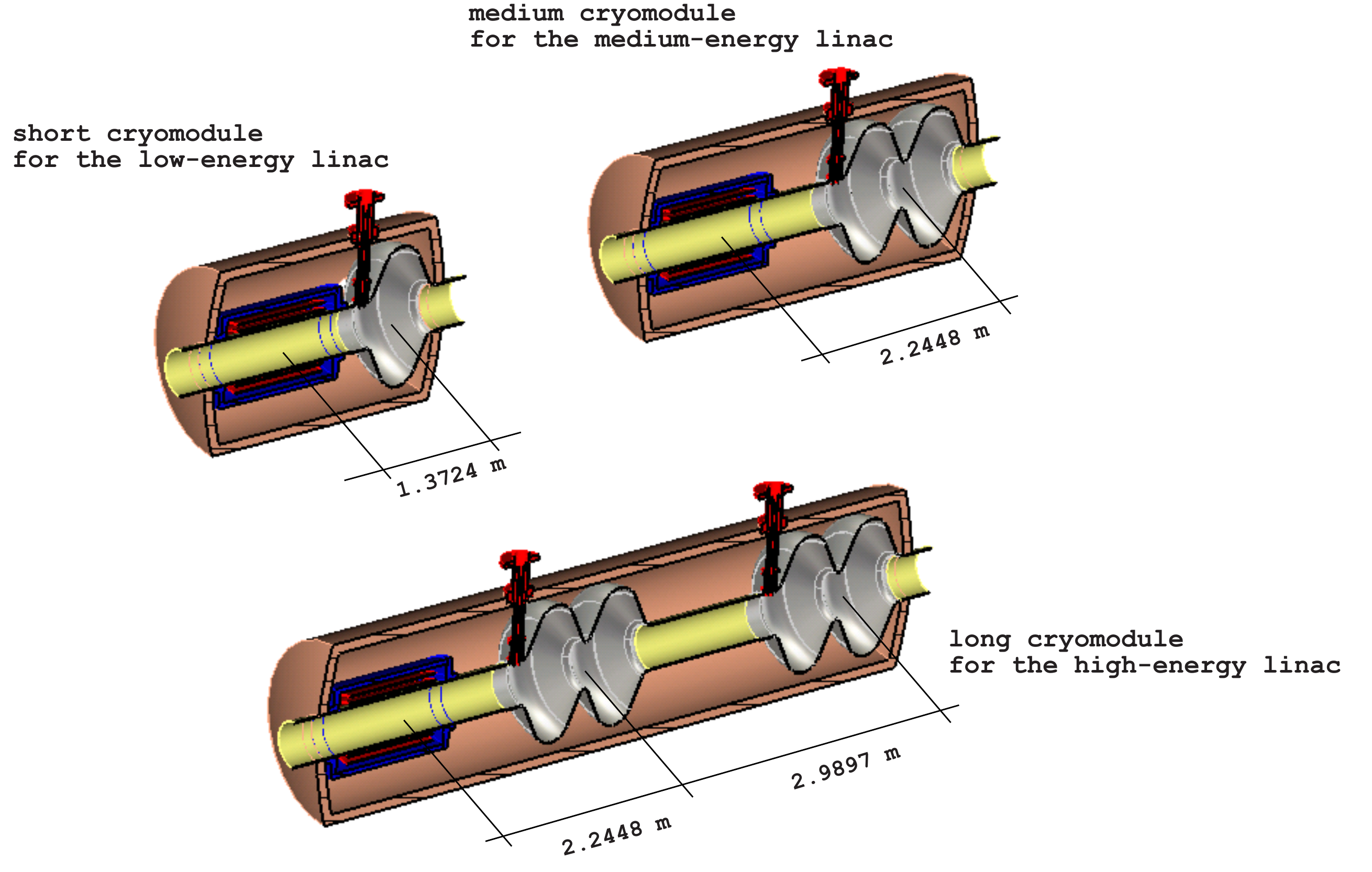}
  \end{center}
  \caption{
    Layout of the three cryo-module types. 
    The modules share the same solenoid and cavity type throughout in this
    study.
  } 
  \label{fig:acc:rla:modules}
\end{figure}

\begin{figure}
  \begin{center}
    \includegraphics[width=0.95\textwidth]%
      {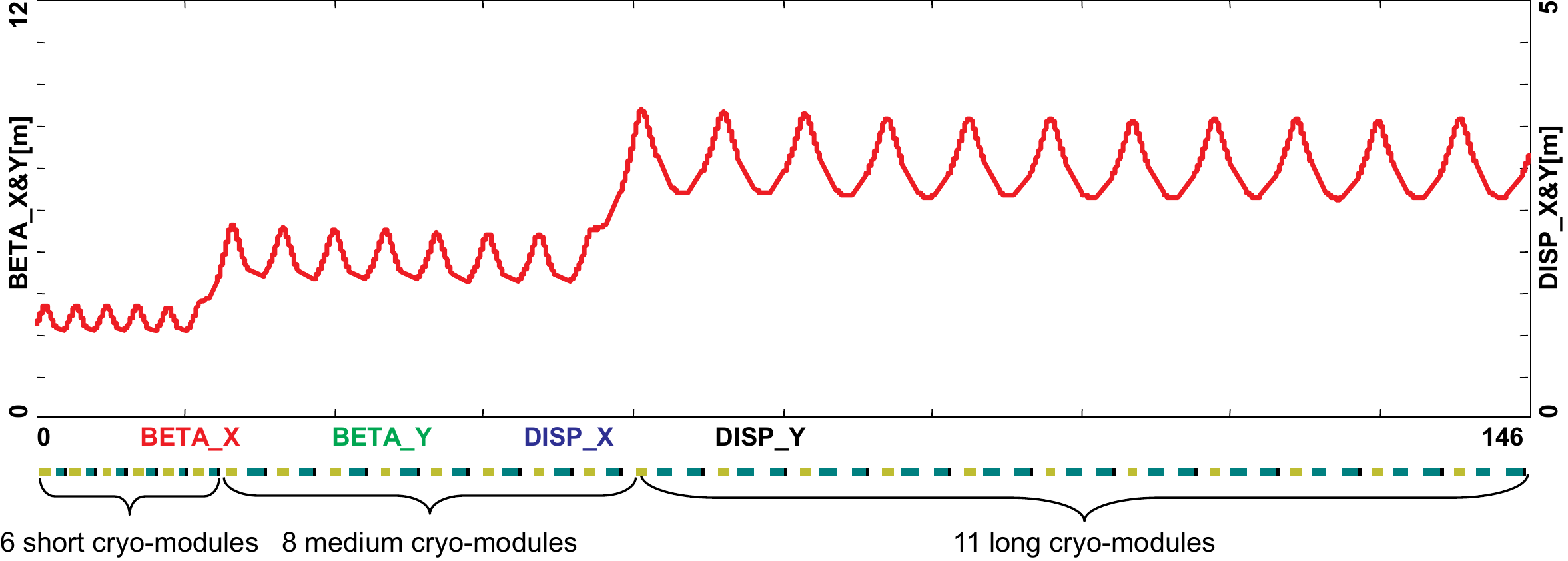}
  \end{center}
  \caption{Transverse FOFO optics of the entire linac; the upper,
    middle and lower periodic sections uniformly matched at the
    junctions.}
  \label{fig:acc:rla:transv}
\end{figure}

A single-pass linac raises the total energy from 0.244 GeV to 0.9 GeV.
This makes the muons sufficiently relativistic to facilitate further
acceleration in the RLA. 
The initial phase-space of the beam, as delivered by the muon
front-end, is characterised by significant energy spread; the linac
has been designed so that it first confines the muon bunches in
longitudinal phase-space, then adiabatically super-imposes
acceleration over the confinement motion, and finally boosts the
confined bunches to 0.9~GeV. 
To achieve a manageable beam-size in the front-end of the linac, short
focusing cells are used for the first six cryo-modules
\cite{Bogacz:2005zr}.
The beam size is adiabatically damped with acceleration, allowing the
short cryo-modules to be replaced with intermediate length
cryo-modules, and then with 11 long cryo-modules. 
Consequently, the linac was split into three consecutive sections
(referred as: the ``upper'', ``middle'' and ``lower'' linac sections)
each section being built of a particular type of cryo-module as shown
in figure \ref{fig:acc:rla:modules}. 
The important parameters for each section are summarised in
table \ref{tab:acc:rla:linacsumm}.
Each linac section is configured with periodic FOFO 
cells, matched at the section junctions, as illustrated in 
figure \ref{fig:acc:rla:transv} \cite{Berg:2005ut}.
Periodicity within each section is maintained by scaling the solenoid
fields in consecutive cryo-modules linearly with increasing momentum as
summarised in table \ref{tab:acc:rla:table1}. 
\begin{table}
  \caption{Summary of important linac parameters. Energy gain per cavity
    indicates which (half-)cavity in table~\ref{tab:acc:rla:cav} is being used.}
  \label{tab:acc:rla:linacsumm}
  \begin{tabular}{|l|r|r|r|}
    \hline
    {\bf Linac section} & {\bf upper} & {\bf middle} & {\bf lower}  \\
    \hline
    Cell length (m)     & 3           & 5            & 8            \\
    Cryo-modules        & 6           & 8            & 11           \\
    Maximum $\beta$ function (m) & 2.90 & 4.93       & 8.25         \\
    RF cavities/lattice cell & 1           & 1            & 2            \\
    RF cells/cavity     & 1           & 2            & 2            \\
    Maximum energy gain/cavity (MeV)&11.25&25.5&25.5\\
    \hline
  \end{tabular}
\end{table}
\begin{table}
  \caption{For each cryo module, the kinetic energy at the exit of the
    module, the solenoid field (for a 1~m solenoid), and the RF phase
    (zero is on-crest). Taken from the OptiM model.}
  \label{tab:acc:rla:table1}
  {\bf Short cryo-modules} \\[0.2ex]
  \begin{tabular}{|l|r|r|r|r|r|r|}
    \hline
    Kin. energy (MeV)&141.6&145.1&149.0&153.1&157.6&162.4\\
    Field (T)&$-1.04$&1.06&$-1.07$&1.09&$-1.12$&0.89\\
    RF phase (deg.)&$-73.3$&$-71.6$&$-69.8$&$-68.1$&$-66.4$&$-64.6$\\
    \hline
  \end{tabular}\\[0.2ex]
  {\bf Medium cryo-modules} \\[0.2ex]
  \begin{tabular}{|l|r|r|r|r|r|r|r|r|}
    \hline
    Kin. energy (MeV)&174.3&187.3&201.3&216.5&232.6&249.7&267.7&286.6\\
    Field (T)&$-0.99$&0.98&$-1.03$&1.08&$-1.14$&1.21&$-1.28$&1.03\\
    RF phase (deg.)&$-62.11$&$-59.13$&$-56.22$&$-53.31$&$-50.40$&$-47.52$&$-44.67$&$-41.85$\\
    \hline
  \end{tabular}\\[0.2ex]
  {\bf Long cryo-modules} \\[0.2ex]
  \begin{tabular}{|l|r|r|r|r|r|r|r|r|r|r|r|}
    \hline
    Kin. energy (MeV)&326.4&368.6&412.6&458.3&505.5&553.7&602.8&652.6&702.8&753.3&803.9\\
    Field (T)&$-1.13$&1.17&$-1.29$&1.41&$-1.54$&1.68&$-1.81$&1.95&$-2.09$&2.23&$-2.37$\\
    RF phase (deg.)&$-38.1$&$-33.8$&$-29.5$&$-25.5$&$-21.5$&$-17.8$&$-14.2$&$-10.8$&$-7.6$&$-4.5$&$-1.7$\\
    \hline
  \end{tabular}
\end{table}

\begin{figure}
  \begin{center}
    \includegraphics[width=0.95\textwidth]%
      {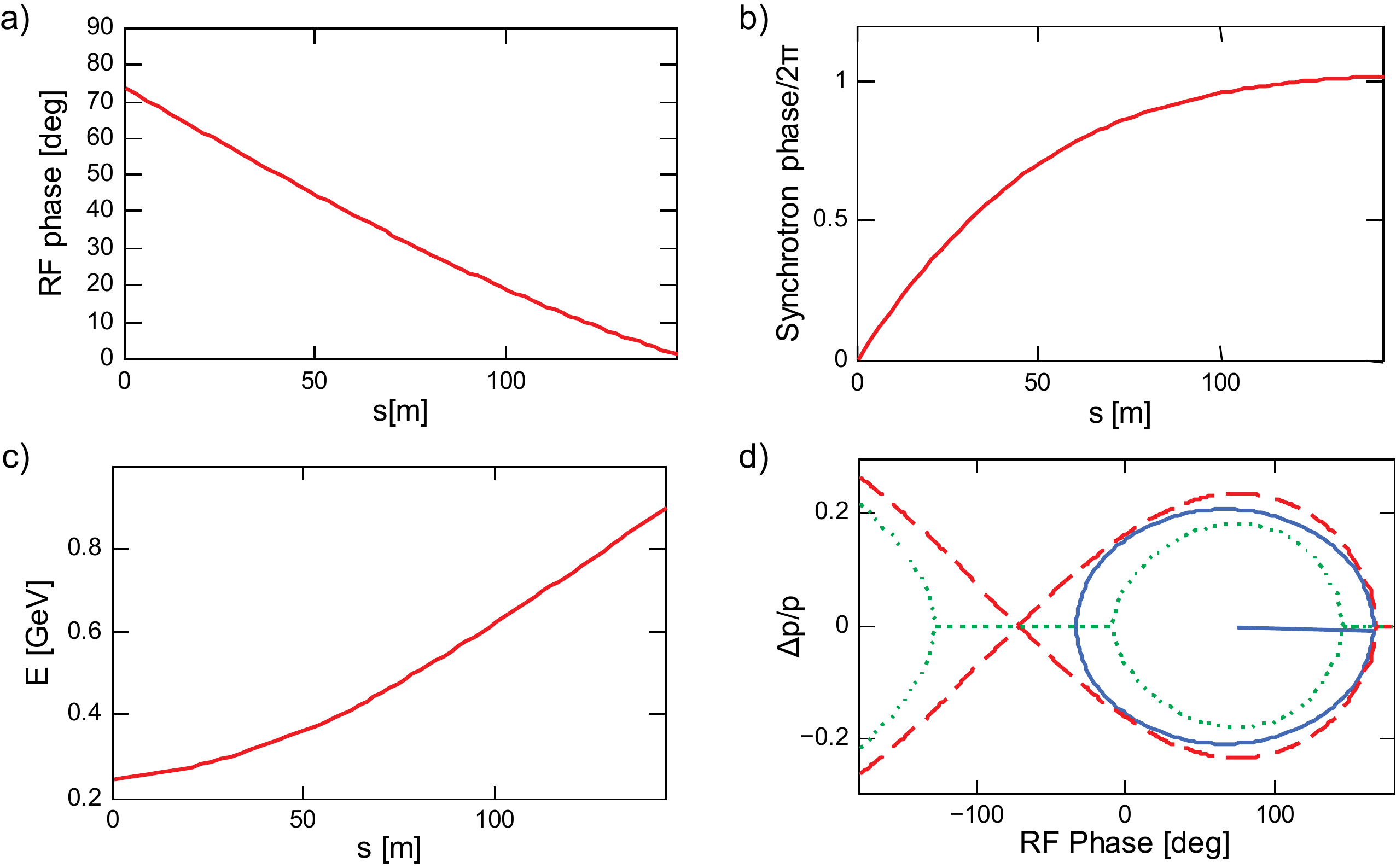}
  \end{center}
  \caption{Longitudinal matching of a single-pass linac: a) the bunch
    starts far off-crest to capture the beam at low energy and then
    moves to the crest as the energy increases and the longitudinal
    acceptance
    improves; b) the synchrotron phase advances by a full period
    from the beginning to the end of the linac; c) the rate of
    energy gain increases as the bunch moves closer to crest.
    d) Longitudinal acceptance matched
    inside the separatrix and optimised for `full bucket'
    acceleration.}
  \label{fig:acc:rla:long}
\end{figure}

One of the main requirements of the single-pass pre-accelerator linac
is to compress adiabatically the longitudinal phase-space volume in the 
course of acceleration. 
The initial longitudinal acceptance of the linac (chosen to be
2.5~$\sigma$) calls for ``full bucket acceleration''; with an initial
momentum acceptance of $\Delta p/p = \pm$17\% and a bunch length of  
$\Delta\phi$= $\pm$102$^\circ$ (in RF phase). 
To perform adiabatic bunching one needs to drive rather strong
synchrotron motion along the linac. 
The profile of the RF-cavity phases is organised so that the phase of
the first cavity is shifted by 73$^\circ$ (off crest) and then the
cavity phase is gradually changed to zero by the end of the linac, see
table \ref{tab:acc:rla:table1} and figure \ref{fig:acc:rla:long}a.  
In the initial part of the linac, when the beam is still not relativistic, 
the far-off-crest acceleration induces rapid synchrotron motion (one full 
period, see figure \ref{fig:acc:rla:long}b), which allows bunch
compression in both bunch-length and momentum spread as illustrated in
figure~\ref{fig:acc:rla:long}.

\begin{figure}
  \begin{center}
    \mbox{\vrule height1mm width0mm}\\
    \raisebox{-\height}[0mm][\totalheight]{\large\textsf{a})}%
    \raisebox{-\height}[0mm][\totalheight]{\includegraphics[height=62mm]{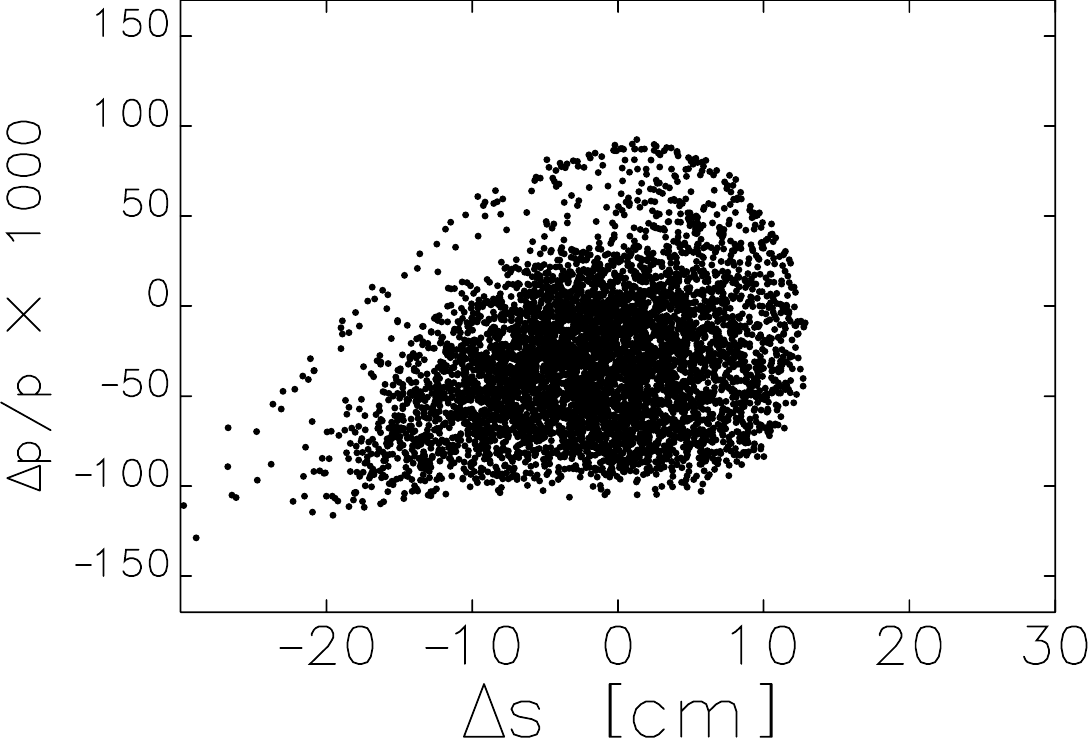}}%
    \hspace{3mm}%
    \raisebox{-\height}[0mm][\totalheight]{\large\textsf{b})}%
    \raisebox{-\height}[0mm][\totalheight]{\includegraphics[height=61mm]{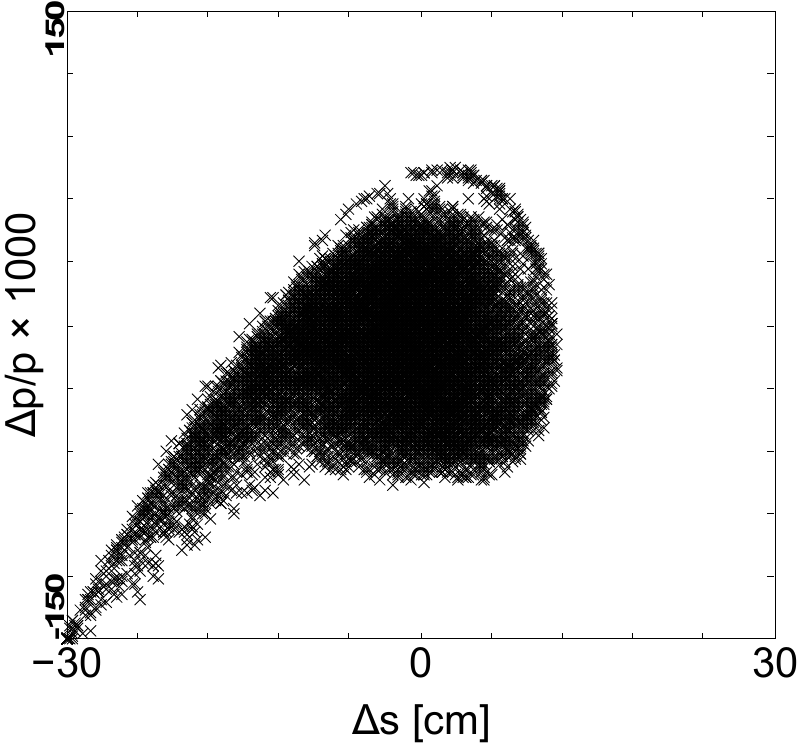}}
  \end{center}
  \caption{Longitudinal phase-space compression at the linac end as
    simulated by ELEGANT~\cite{Borland:2000,elegant} (left) and
    OptiM~\cite{optim} (right).}
  \label{fig:acc:rla:dpp}
\end{figure}

To maximise the longitudinal acceptance, the initial position of the bunch
is shifted relative to the centre of the bucket, to keep the
beam boundary inside the separatrix~\cite{Bogacz:2003uk}, as
illustrated in figure~\ref{fig:acc:rla:long}d. 
The synchrotron motion also suppresses the sag in acceleration for the
bunch head and tail. 
In our tracking simulation we have assumed a particle distribution
that is Gaussian in 6D phase space with the tails of the distribution
truncated at 2.5 $\sigma$, which corresponds to the beam acceptance. 
Despite the large initial energy 
spread, the particle tracking simulation through the linac does not predict 
any significant emittance 
growth~\cite{Bogacz:2009zzc}.
Results of the simulation are illustrated in figure
\ref{fig:acc:rla:dpp} which shows the longitudinal phase-space at the
end of the linac as simulated by ELEGANT~\cite{Borland:2000,elegant};
tracking through the individual field maps of the linac's RF cavities
and solenoids (figure \ref{fig:acc:rla:dpp}a) and by a simple matrix
based code OptiM \cite{optim} (figure \ref{fig:acc:rla:dpp}b).

\paragraph{RLA I}

\begin{figure}
  \begin{center}
    \mbox{\vrule height12mm width0mm}\\
    \includegraphics[width=\linewidth]%
    {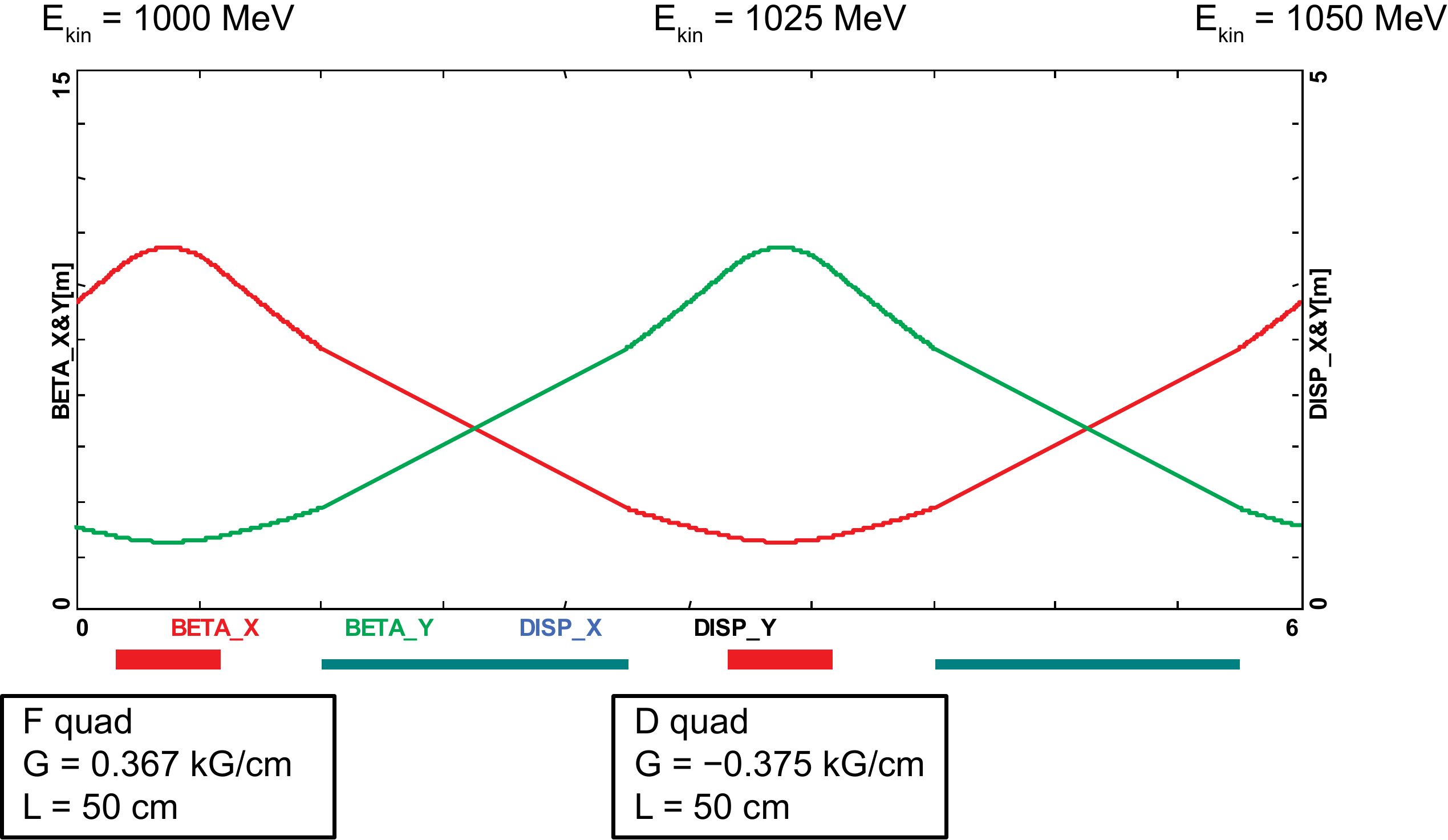}
  \end{center}
  \caption{Periodic 90$^\circ$ FODO 6~m linac cell. A pair
    of 2-cell cavities is interleaved between quadrupoles.  The cavity
    length is 1 RF wavelength and the elements are equally spaced.}
  \label{fig:acc:rla:fodo}
\end{figure}

The dog-bone RLA I is designed to accelerate simultaneously the 
$\mu^+$ and $\mu^-$ beams from 0.9 GeV to 3.6\,GeV. 
The injection energy into the 
RLA and the energy gain per RLA linac (0.6 GeV) were chosen so that 
a tolerable level of RF phase slippage along the linac could be 
maintained ($\sim20^\circ$ in RF phase). To suppress chromatic effects,
90$^\circ$ FODO optics was used as a building block for both the
linac and the return arcs. 
The layout and optics of the linac's periodic cell is shown in figure
\ref{fig:acc:rla:fodo}. The cavity from table~\ref{tab:acc:rla:cav}
with the higher energy gain is used for the linac.

\begin{figure}
  \begin{center}
    \includegraphics[width=0.94\textwidth]%
    {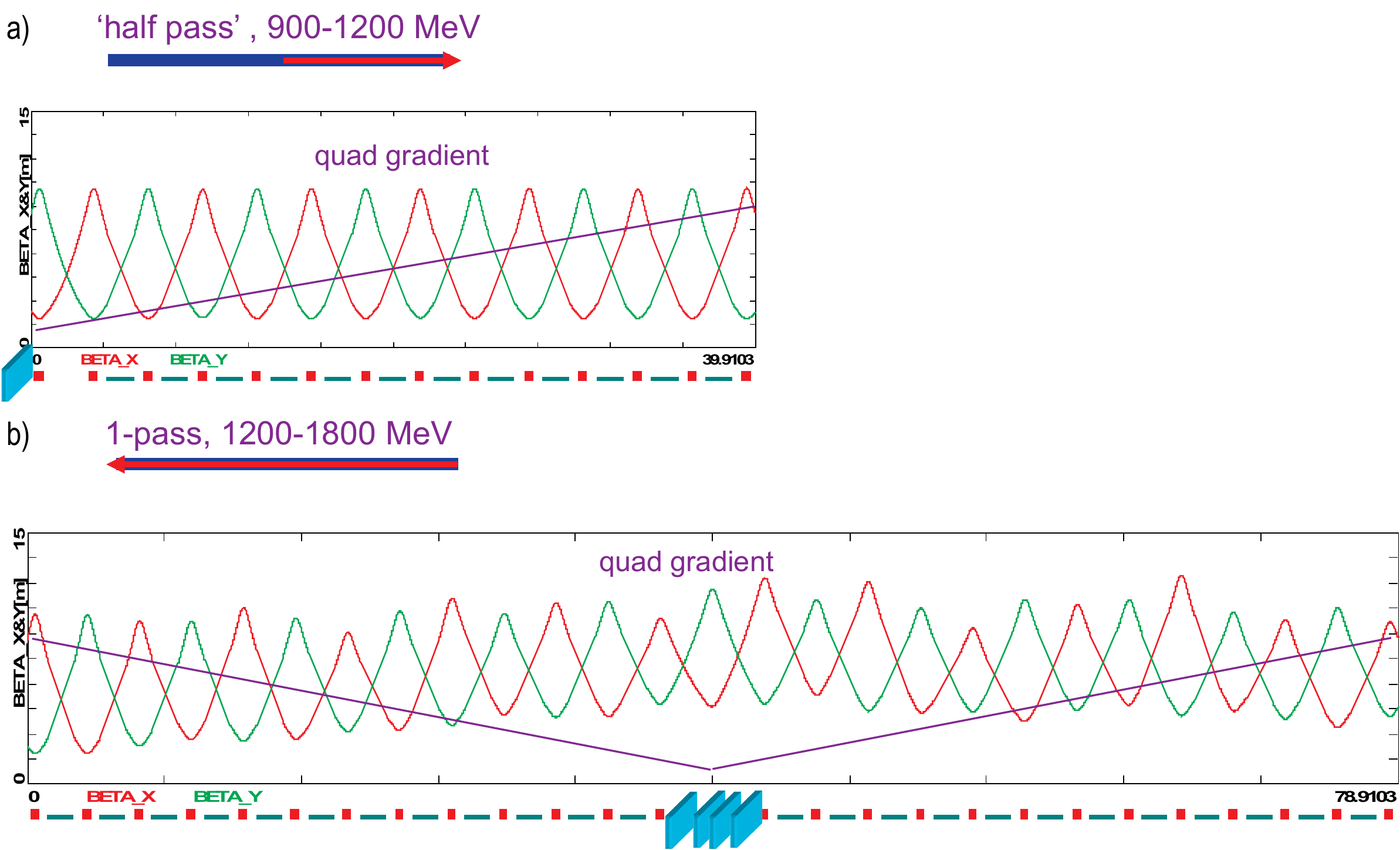}
  \end{center}
  \caption{FODO-based multi-pass linac optics. The quadrupole
    gradients are scaled up with momentum for the first half of the
    linac, then they are mirror reflected in the second half. The
    injection chicane magnets are located at the middle of the linac
    (marked in blue).}
  \label{fig:acc:rla:multipass}
\end{figure}

The focusing profile along the linac was chosen so that beams with a
large energy spread could be transported within the given
aperture. 
Since the beam is traversing the linac in both directions, a
`bisected' focusing profile \cite{Ankenbrandt:2009zza} was chosen for
the multi-pass linac.
Here, the quadrupole gradients scale up with momentum to maintain
90$^\circ$ phase advance per cell for the first half of the linac
(figure \ref{fig:acc:rla:multipass}a), and then are mirror reflected
in the second half, as illustrated in figure
\ref{fig:acc:rla:multipass}b.  
The complete focusing profile along the entire linac is summarised in
table \ref{tab:acc:rla:table2}.
\begin{table}
  \caption{Gradient, cryo-module by cryo-module along the
    0.6 GeV linac.  The quadrupole centres start at longitudinal position
    46~cm with a spacing between quadrupole centres of 3~m.  The FODO lattice
    starts at one end of the linac with the first F quad in the table,
    and is reflection symmetric about the
    centre of the last D quad shown in the table.
    There are RF cavities
    between each of the quadrupoles, except there are no cavities between
    the central D magnet and its adjacent F quadrupoles. In the first full
    pass, each cavity causes
    the beam to gain 25~MeV, starting with a kinetic energy of
    1103.9~MeV at the beginning of that pass.}
  \label{tab:acc:rla:table2}
  \begin{tabular}{|l|r|r|r|r|r|r|r|}
    \hline
    F Gradient (kG/cm)&0.402&0.385&0.368&0.352&0.335&0.318&0.301\\
    D Gradient
    (kG/cm)&$-0.394$&$-0.377$&$-0.360$&$-0.343$&$-0.325$&$-0.308$&$-0.301$ \\
    \hline
  \end{tabular}
\end{table}%

\begin{figure}
  \begin{center}
    \includegraphics[width=\textwidth]%
      {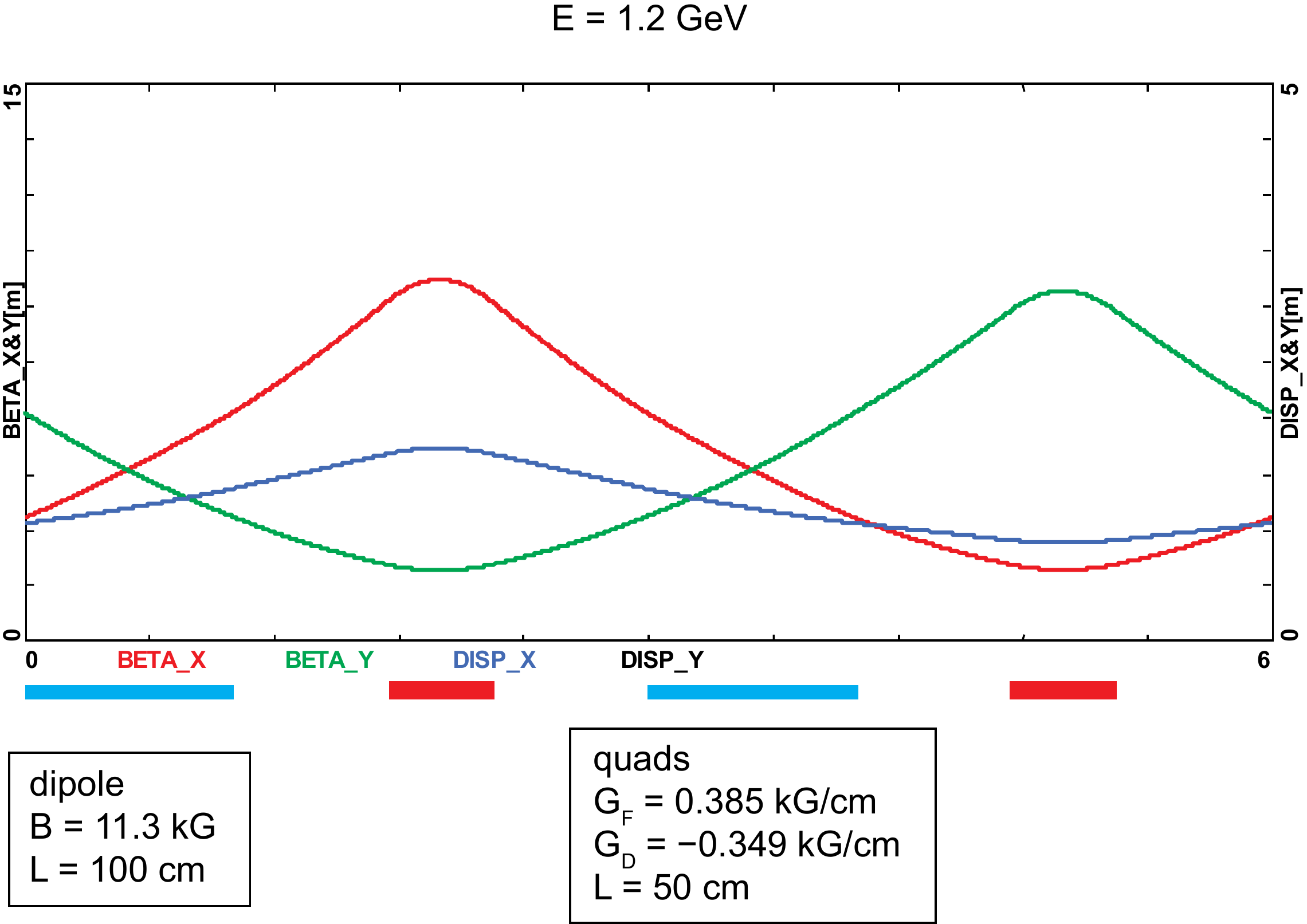}
  \end{center}
  \caption{Periodic 90$^\circ$ FODO 6~m inward-bending cell, including
    a pair of rectangular bends interleaved between quads.  All
    elements are equally spaced.}
  \label{fig:acc:rla:inward}
\end{figure}

At the ends of the RLA linacs, the beams need to be directed into the
appropriate energy-dependent (pass-dependent) droplet arc for
recirculation \cite{Bogacz:2005zr}. 
The entire droplet-arc architecture is based on 90$^\circ$
phase-advance cells with periodic beta functions, as shown in
figure \ref{fig:acc:rla:inward}.  

\begin{figure}
  \begin{center}
    \includegraphics[width=\textwidth]%
      {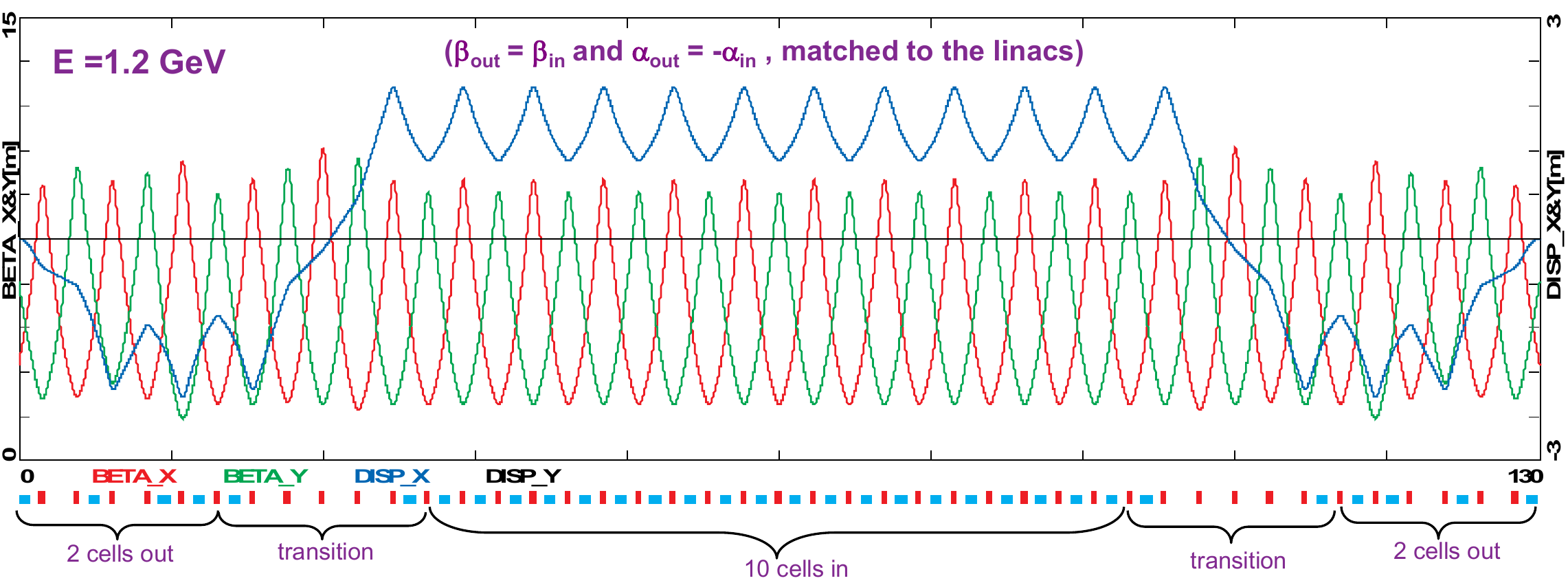}
  \end{center}
  \caption{Compact droplet arc optics, featuring uniform periodicity
    of beta functions and dispersion, which is suppressed geometrically
    via `missing dipoles' at the arc ends.}
  \label{fig:acc:rla:compact}
\end{figure}

For practical reasons, horizontal rather than vertical beam separation
has been chosen.
Rather than suppressing the horizontal dispersion created by the
spreader, it has been matched to that of the outward
60$^\circ$ arc.  This is partially
accomplished by removing one dipole (the one furthest
from the spreader) from each of
the two cells following the spreader.
To switch from outward to inward bending, three transition cells are used,
wherein the four central dipoles are removed.  The two remaining dipoles
at the ends bend the same direction as the dipoles to which they are closest.
To facilitate simultaneous acceleration of both $\mu^+$
and $\mu^-$ bunches, a mirror symmetry is imposed on the
droplet-arc optics (oppositely charged bunches move in opposite
directions through the arcs)~\cite{Bogacz:2008zzb}.
This puts a constraint on the exit/entrance
Twiss functions for two consecutive linac passes, 
namely 
$\beta_{out}^n = \beta_{in}^{n+1}$ and 
$\alpha_{out}^n = -\alpha_{in}^{n+1}$, 
where $n = 0, 1, 2 ...$ is the pass index. 
Complete droplet arc optics for the lowest energy arc (1.2 GeV) are
shown in figure \ref{fig:acc:rla:compact}.

The momentum compaction is relatively large (6.5 m), which guarantees
significant rotation in longitudinal phase-space as the beam
passes through the arc.
This effect, combined with off-crest acceleration in the subsequent
linac, yields further compression of the longitudinal phase-space as
the beam is accelerated \cite{Bogacz:2009zzc}.

All higher arcs are based on the same bends (1~m, 1.1~T dipoles) and
they are configured via extending Arc 1 by additional inward and
outward cells as summarised in table \ref{tab:acc:rla:table3}.  The transition
regions always to have three cells.
The quadrupole strengths in the arcs are scaled up linearly with
momentum to preserve the 90$^\circ$ FODO lattice, as summarised in
table \ref{tab:acc:rla:table3}.  Since the linac does not have a
90$^\circ$ phase advance per cell for later passes, the quadrupoles
near the spreader must be modified to achieve matching, as described
in Table~\ref{tab:acc:rla:arcmatch}.  The matching designs will
need to be slightly modified in future work to ensure that common
magnets (namely the F magnet at the end of the linac and the ends
of all the arcs) all have the same parameters.
\begin{table}
  \caption{Higher arcs optics architecture; arc circumference and 
    quadrupole strength scaled up with momentum.  $E$ is the arc energy,
    $p/p_1$ is the ratio of the momentum to the first arc momentum,
    $n_{\text{out}}$ is the number of outward bending cells in each outward
    bend (the total is twice this), $n_{\text{in}}$ is the number of inward
    bending cells, $L$ is the arc length, and $G_F$ and $G_D$ are the
    focusing and defocusing quadrupole gradients, respectively.}
  \label{tab:acc:rla:table3}
  \begin{tabular}{|r|r|r|r|r|r|r|}
    \hline
    {\bf $E$} (GeV) & {\bf $p/p_1$} & {\bf $n_{\text{out}}$} & {\bf $n_{\text{in}}$} & {\bf $L$} (m)
    & {\bf $G_F$} (kG/cm)& {\bf $G_D$} (kG/cm)\\
    \hline
    1.2&1&2&10&130&0.385&$-0.349$\\
    1.8&3/2&3&15&172&0.578&$-0.524$\\
    2.4&2&4&20&214&0.770&$-0.698$\\
    3.0&5/2&5&25&256&0.962&$-0.873$\\
    \hline
  \end{tabular}
\end{table}
\begin{table}
  \caption{Strengths of matching quadrupoles in arc~2 of RLA~I.  Values
    for other arcs have not yet been computed.}
  \label{tab:acc:rla:arcmatch}
  \begin{tabular}{|l|rrrrrr|}
    \hline
    Gradient (kG/cm)&0.5248&$-0.5260$&0.6321&$-0.6878$&0.6151&$-0.5463$\\
    \hline
  \end{tabular}
\end{table}
\paragraph{RLA II}

\begin{figure}
  \begin{center}
    \includegraphics[width=0.77\textwidth]%
    {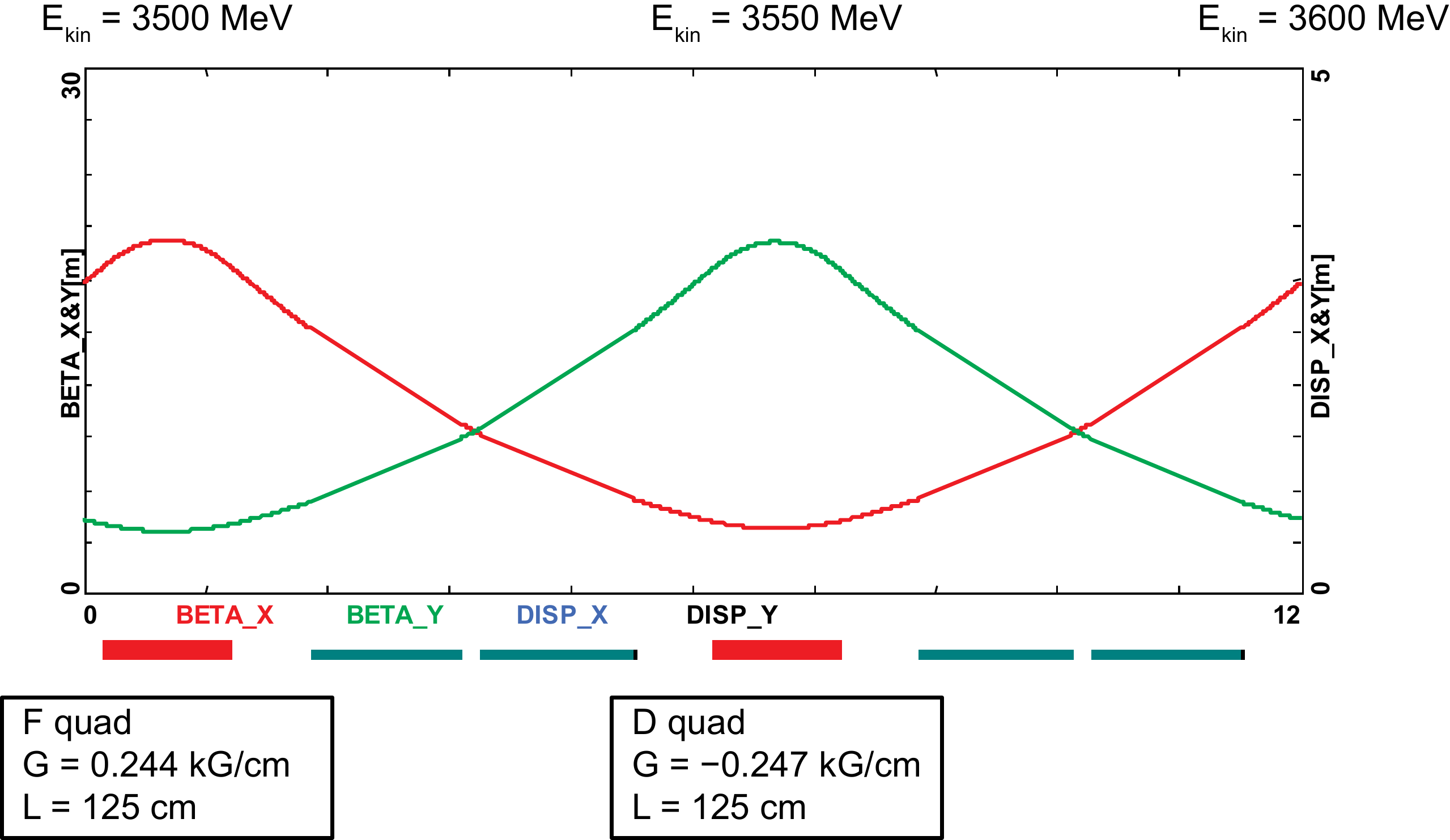}
  \end{center}
  \caption{Periodic 90$^\circ$ FODO linac cell (12~m), including
    two pairs of 2-cell cavities interleaved between quadrupoles.}
  \label{fig:acc:rla:fodo12}
\end{figure}

RLA~II is designed to accelerate the $\mu^+$ and $\mu^-$ beams from
3.6\,GeV to 12.6\,GeV. 
The injection energy and the energy gain per linac (2~GeV) were chosen
so that a tolerable level of RF-phase slippage along the linac could  
be maintained. 
The optics configuration is almost identical to that of RLA~I. It is  
again based on a 90$^\circ$ FODO lattice and a bisected multi-pass
linac. 
The main difference is that the fundamental cell-length was extended to
12~m (cf.\ 6~m for RLA~I).  
The layout and optics of the linac's periodic cell is shown in figure
\ref{fig:acc:rla:fodo12}.
The cavity from table~\ref{tab:acc:rla:cav}
with the higher energy gain is used for the linac.

\begin{figure}
  \begin{center}
    \includegraphics[width=0.95\textwidth]%
      {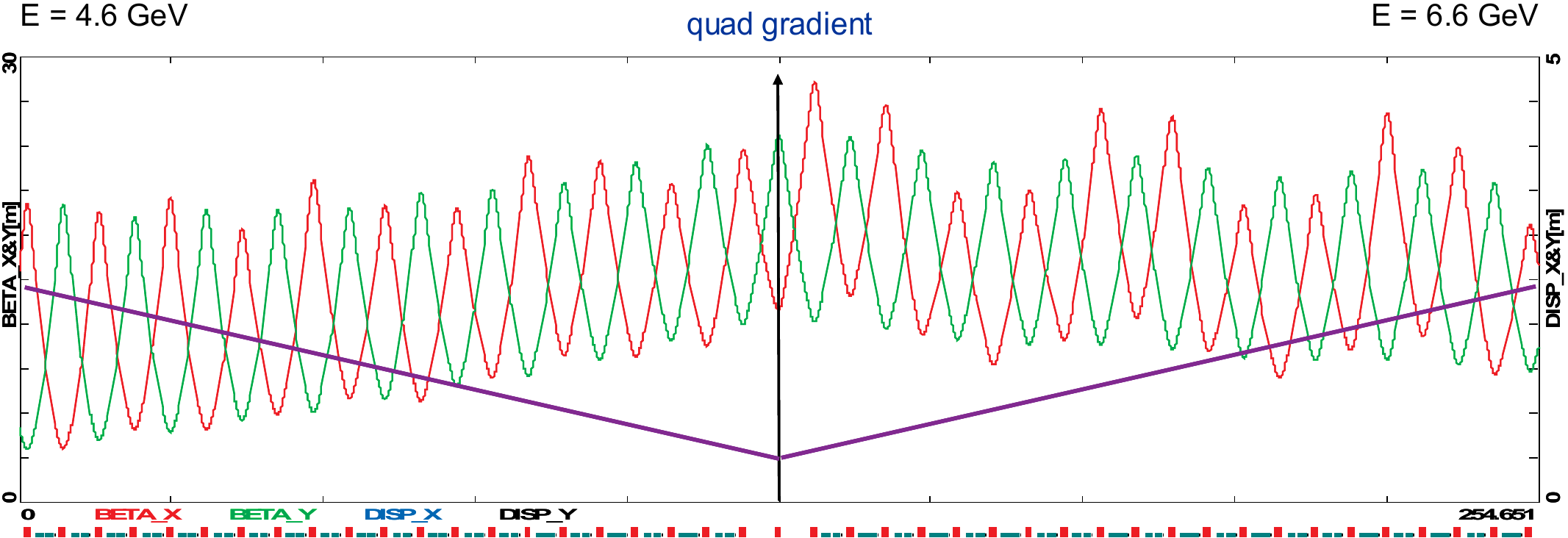}
  \end{center}
  \caption{Bisected multi-pass linac optics. The quadrupole gradients
    are scaled up with momentum for the first half of the linac and
    mirror reflected in the second half.}
  \label{fig:acc:rla:bisect}
\end{figure}

As for RLA I, a bisected focusing profile was chosen for the
multi-pass linac in RLA~II, with the quadrupole gradients
scaled up with momentum in the first half of the linac and mirror
reflected in the second half, as illustrated in
figure \ref{fig:acc:rla:bisect}. 
The complete focusing profile along the entire linac is summarised in
table \ref{tab:acc:rla:table4}.
\begin{table}
  \caption{Gradient and energy profile, cryo-module by cryo-module along
    the 2 GeV linac.  As in table~\ref{tab:acc:rla:table2}, except that
    the spacing between quadrupole centres is 6~m, the first quad centre is
    at 133~cm, the energy gain between quadrupoles is 50~MeV, and
    the kinetic energy is 2894~MeV at the beginning of the first full pass.}
  \label{tab:acc:rla:table4}
  \begin{tabular}{|l|r|r|r|r|r|r|}
    \hline
    Sequence&1&2&3&4&5&6\\
    F grad. (kG/cm)&0.308   &0.305   &0.298   &0.291   &0.285   &0.278   \\
    D grad. (kG/cm)&$-0.313$&$-0.306$&$-0.299$&$-0.292$&$-0.285$&$-0.278$\\\hline
    Sequence       &7       &8       &9       &10      &11      & \\
    F grad. (kG/cm)&0.271   &0.264   &0.257   &0.251   &0.244   & \\
    D grad. (kG/cm)&$-0.271$&$-0.264$&$-0.257$&$-0.250$&$-0.244$& \\
    \hline
  \end{tabular}
\end{table}

All higher arcs are based on the same bends (2~m, 2.1~T dipoles) and
they are configured via extending Arc 1 by additional inward and
outward cells as summarised in table \ref{tab:acc:rla:table5}. 
Magnets to achieve these dipole fields will likely need to be superconducting.
The quadrupole strengths in the arcs are scaled up linearly with 
momentum to preserve the 90$^\circ$ FODO lattice, as summarised in
table \ref{tab:acc:rla:table5}. 
\begin{table}
  \caption{As in table~\ref{tab:acc:rla:table3}, but for arcs in RLA~II.
    $B$ is field for the dipoles.}
  \label{tab:acc:rla:table5}
  \begin{tabular}{|r|r|r|r|r|r|r|r|}
    \hline
    {\bf $E$} (GeV) & {\bf $p/p_1$} & {\bf $n_{\text{out}}$} & {\bf $n_{\text{in}}$} & {\bf $L$} (m)
    & {\bf $G_F$} (kG/cm)& {\bf $G_D$} (kG/cm)&$B$ (T)\\
    \hline
    4.6 &1  &2&10&260&1.45&$-1.42$&2.10\\
    6.6 &1.43&3&15&344&2.08&$-2.04$&2.01\\
    8.6 &1.87&4&20&428&2.71&$-2.66$&1.96\\
    10.6&2.30&5&25&512&3.34&$-3.27$&1.94\\
    \hline
  \end{tabular}
\end{table}%

\paragraph{Injection double chicane}

\begin{figure}
  \includegraphics[width=0.95\linewidth]{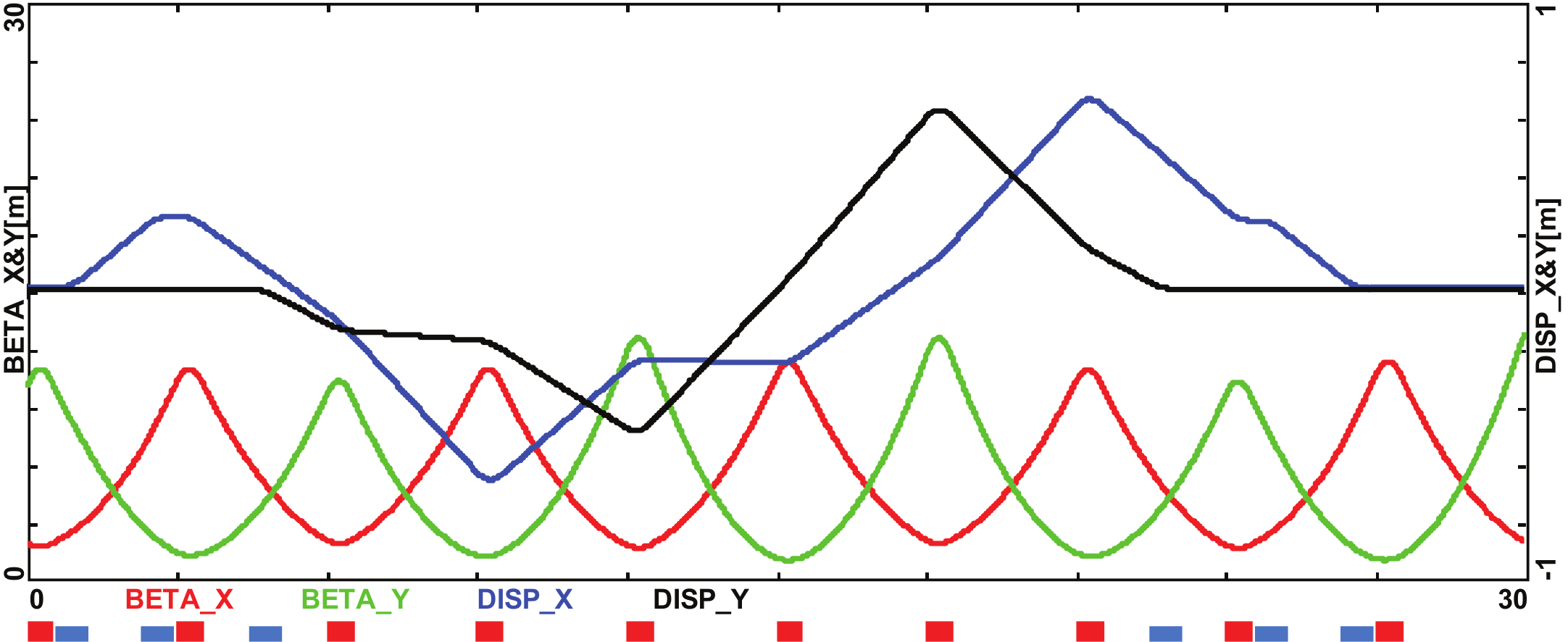}
  \caption{Diagram of the injection chicane lattice.}
  \label{fig:acc:rla:chicane}
\end{figure}

To transfer both $\mu^+$ and $\mu^-$ bunches from one accelerator to
the next, which is located at a different vertical elevation, we
use a compact double chicane \cite{Bogacz:2008zzb}.
A FODO lattice is used with a 90$^\circ$ phase advance horizontally
and a 120$^\circ$ phase advance vertically.
Each ``leg'' of the chicane involves four horizontal and two vertical 
bending magnets, forming a double achromat in the horizontal and
vertical planes, while preserving periodicity of the beta functions
\cite{Bogacz:2009zzc}.  The lattice for RLA~I is described in
table~\ref{tab:acc:rla:chicane} and figure~\ref{fig:acc:rla:chicane}.
The injection chicane
for RLA~II will be similar, but has yet to be completed.
The chicanes will have sextupoles to correct vertical emittance
dilution, located at the peaks of the vertical dispersion
(see figure~\ref{fig:acc:rla:chicane}).
Their integrated strengths (in the first injection
chicane) will be approximately 0.01\,kG/cm.
\begin{table}
  \caption{Lattice of injection chicane into RLA~I.  The left hand table
    gives the dipole parameters, the right hand table gives the quadrupole
    parameters.  The dipoles are 60~cm long, the quadrupoles are 50~cm
    long.  The chicane is 30~m long and designed for a kinetic energy
    of 803.877\,MeV.  $s$ is the longitudinal position of the end of
    the magnet, $B$ is the
    magnetic field of the dipole (H for horizontal, V for vertical),
    $\theta$ is the bend angle of the
    dipole, $G$ is the quadrupole gradient, and $\varepsilon_{\text{in}}$
    and $\varepsilon_{\text{out}}$ are the entrance and exit angles,
    respectively.  Zero entrance and exit angles would correspond to a sector
    bend; an edge angle with the same sign as the bend angle rotates the
    edge such that a vector perpendicular to the edge and pointing out of the
    magnet rotates away from the centre of curvature (MAD convention).}
  \label{tab:acc:rla:chicane}
  \begin{tabular}[c]{|crrrrr|}
    \hline
    Type&$s$ (cm)&$B$ (kG)&$\theta$ (deg.)&$\varepsilon_{\text{in}}$ (deg.)&$\varepsilon_{\text{out}}$ (deg.)\\\hline
    H& 120&7.024468&8.01381&0&9\\
    H& 290&$-7.024468$&$-8.01381$&$-9$&0\\
    V&504&$-4.682978$&$-5.34254$&0&$-6$\\
    V&2306&4.682978&5.34254&6&0\\
    H&2520&$-7.024468$&$-8.01381$&$-9$&0\\
    H&2690&7.024468&8.01381&0&9\\
    \hline
  \end{tabular}
  \begin{tabular}[c]{|rr|}
    \hline
    $s$ (cm)&$G$ (kG/cm)\\\hline
    50&$-0.3682620$\\
    350&0.3090135\\
    650&$-0.3623229$\\
    950&0.3121510\\
    1250&$-0.3641247$\\
    1550&0.3121510\\
    1850&$-0.3641247$\\
    2150&0.3090135\\
    2450&$-0.3623229$\\
    2750&0.3172734\\
    \hline
  \end{tabular}
\end{table}

\begin{figure}
  \begin{center}
    \includegraphics[width=0.95\textwidth]%
    {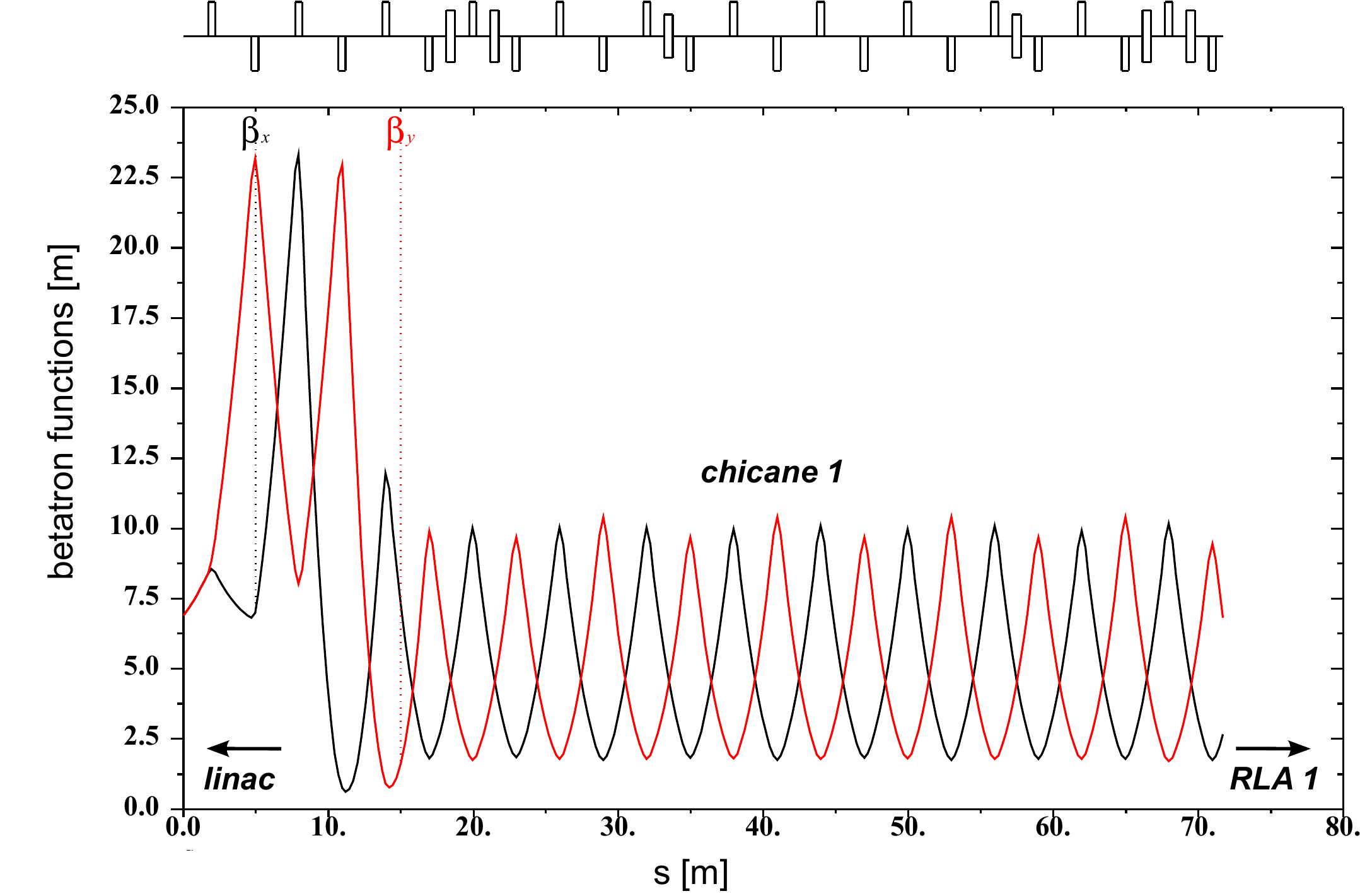}
  \end{center}
  \caption{Horizontal and vertical betatron functions as the beam
    passes through chicane 1, calculated with MADX.}
  \label{fig:acc:rla:beta_chicane1}
\end{figure}

The chicane and RLA lattices have not yet been simulated with 
realistic field maps but only with usual beam optics codes such as
MADX \cite{mad-x}, and OptiM \cite{optim}. 
As an example, the horizontal and vertical betatron functions of
chicane~1 are shown in figure \ref{fig:acc:rla:beta_chicane1}. 

\paragraph{Arc crossings}

\begin{figure}
  \begin{center}
    \includegraphics[width=\linewidth]{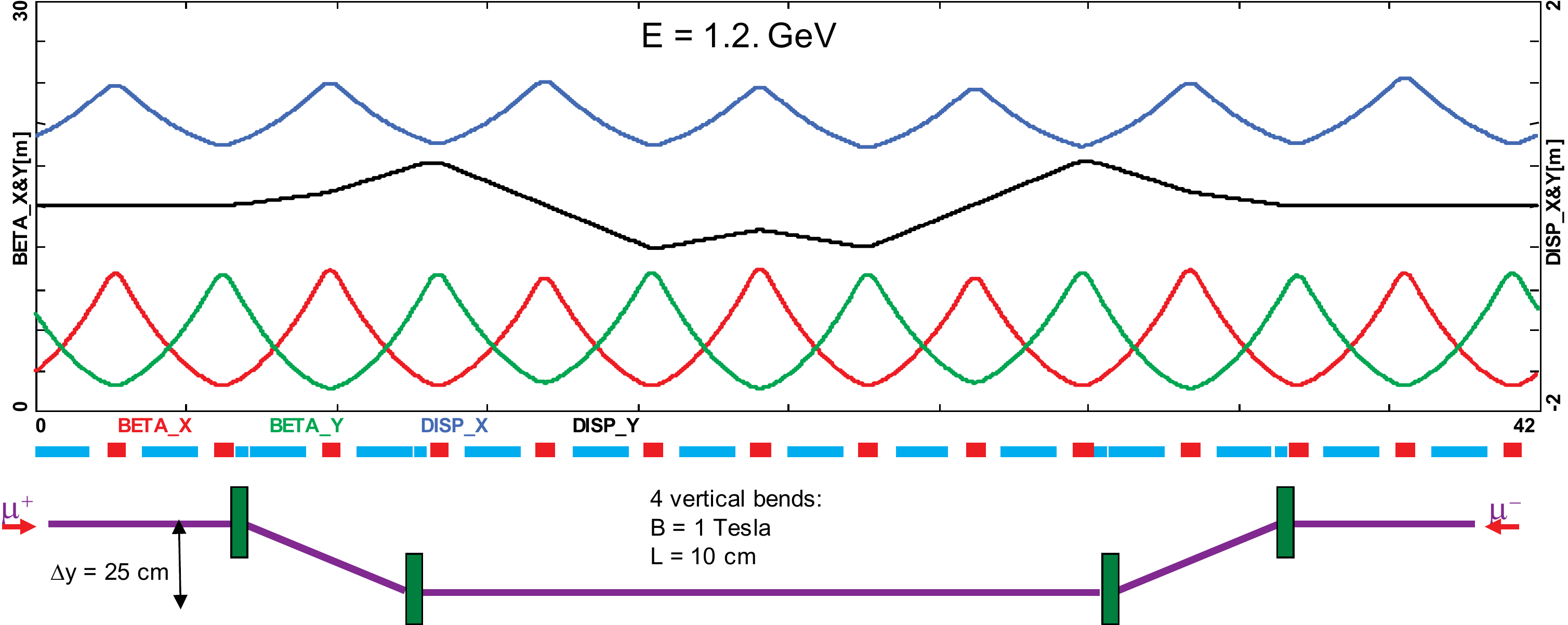}\\[-1mm]
  \end{center}
  \caption{Diagram of the arc bypass lattice.  The vertical dipoles are
    10 cm long, centred in the gap between the adjacent dipole and quadrupole.}
  \label{fig:acc:rla:bypass}
\end{figure}

One unfortunate feature of the dogbone lattice is that the arcs will cross
each other.  To address this, the arcs will have a vertical bypass in the
region of the crossing, one arc going up and the other down, as shown
in figure~\ref{fig:acc:rla:bypass}.

\subsubsection{Magnets}

\begin{figure}
  \begin{center}
    \includegraphics[width=0.35\textwidth]%
      {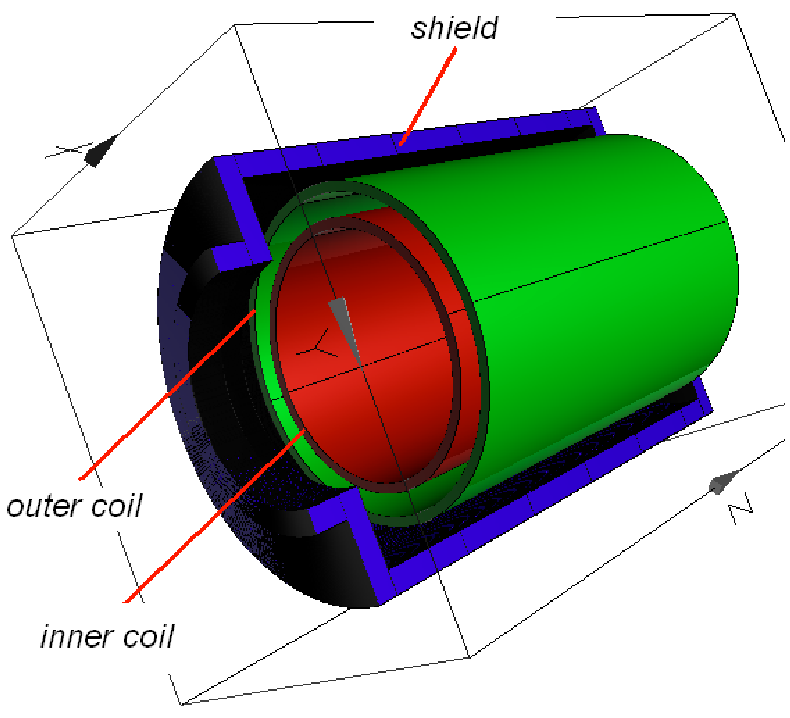}\hspace{0.4cm}
    \includegraphics[width=0.6\textwidth]%
      {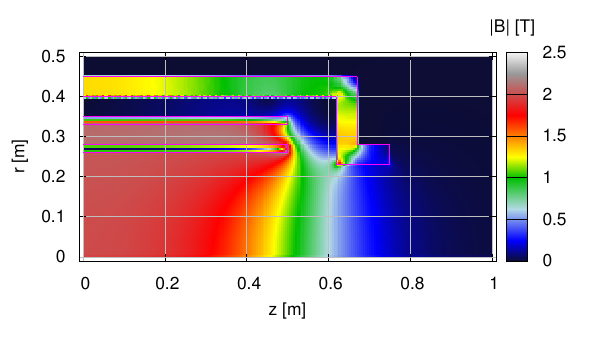}
  \end{center}
  \caption{Layout of the shielded superconducting solenoid (left) and its 2D
    magnetic field map across the axial plane (right) for a peak
    magnetic field of 2~T.}
  \label{fig:acc:rla:solenoid}
\end{figure}

The most important magnet types of the linac and RLA sections are:
superconducting solenoids (linac) and normal conducting dipoles,
combined dipole/sextupoles and quadrupoles (RLAs and chicanes).
A detailed numerical simulation has been performed for the
superconducting solenoids, while the design of normal conducting
magnets is in progress.  
As shown in figure \ref{fig:acc:rla:solenoid}, a configuration 
of two superconducting coils with opposite currents shielded by a 5~cm
thick iron wall is good enough to prevent longitudinal field leakage
which may cause quenching of the neighbouring superconducting
cavities. 
Also, it has been shown that the coils are far from the quench limit
and may be made thinner.  

\subsubsection{RF system}

The lattice designs for the acceleration systems are based on the Study~II
RF-cavity designs \cite{Ozaki:2001}, the parameters of which are summarised
in table \ref{tab:acc:rla:cav}.
\begin{table}
  \caption{Important parameters for the Study~II superconducting RF cavity
    designs~\cite{Ozaki:2001}.  When single-cell cavities are used,
    appropriate parameters are cut in half.}
  \label{tab:acc:rla:cav}
  \begin{tabular}{|l|r|r|}
    \hline
    RF frequency (MHz)&201.25&201.25\\
    Cells per cavity&2&2\\
    Aperture diameter (mm)&300&460\\
    Energy gain/cavity (MeV)&25.5&22.5\\
    Stored energy/cavity (J)&2008&1932\\
    Input power/cavity (kW)&1016&980\\
    RF on time (ms)&3&3\\
    Loaded $Q$&$10^6$&$10^6$ \\
    \hline
  \end{tabular}
\end{table}

\begin{figure}
  \begin{center}
    \begin{tabular}{l}
      length = 0.7448 m\\
      radius = 0.6854 m\\
      resonance freq. = 201.247 MHz\\
      quality fact. = 24.67$\times 10^9$\\
      transit time factor = 0.650 - 0.690\\
      peak $\hat{E}$ =  26.17 MV/m\\
      $|E|^{max}_{surf}$ = 21.70 MV/m\\
      $|H|^{max}_{surf}$ = 48.06 kA/m\\
      stored energy = 712 J
    \end{tabular}\hspace{4em}
    \raisebox{-0.5\height}{\includegraphics[width=0.3\textwidth]{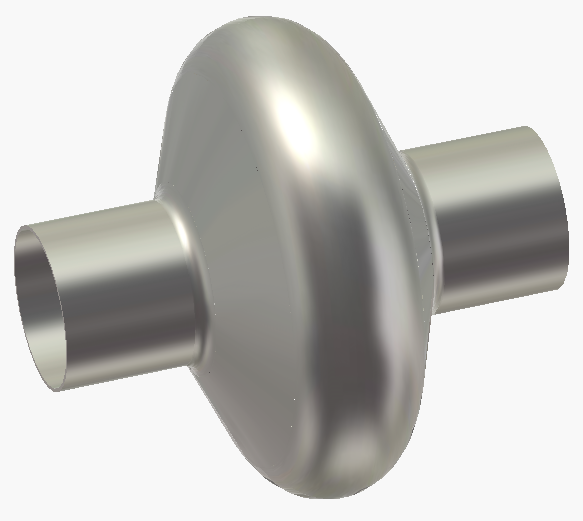}}\\
    \includegraphics[width=0.55\textwidth]{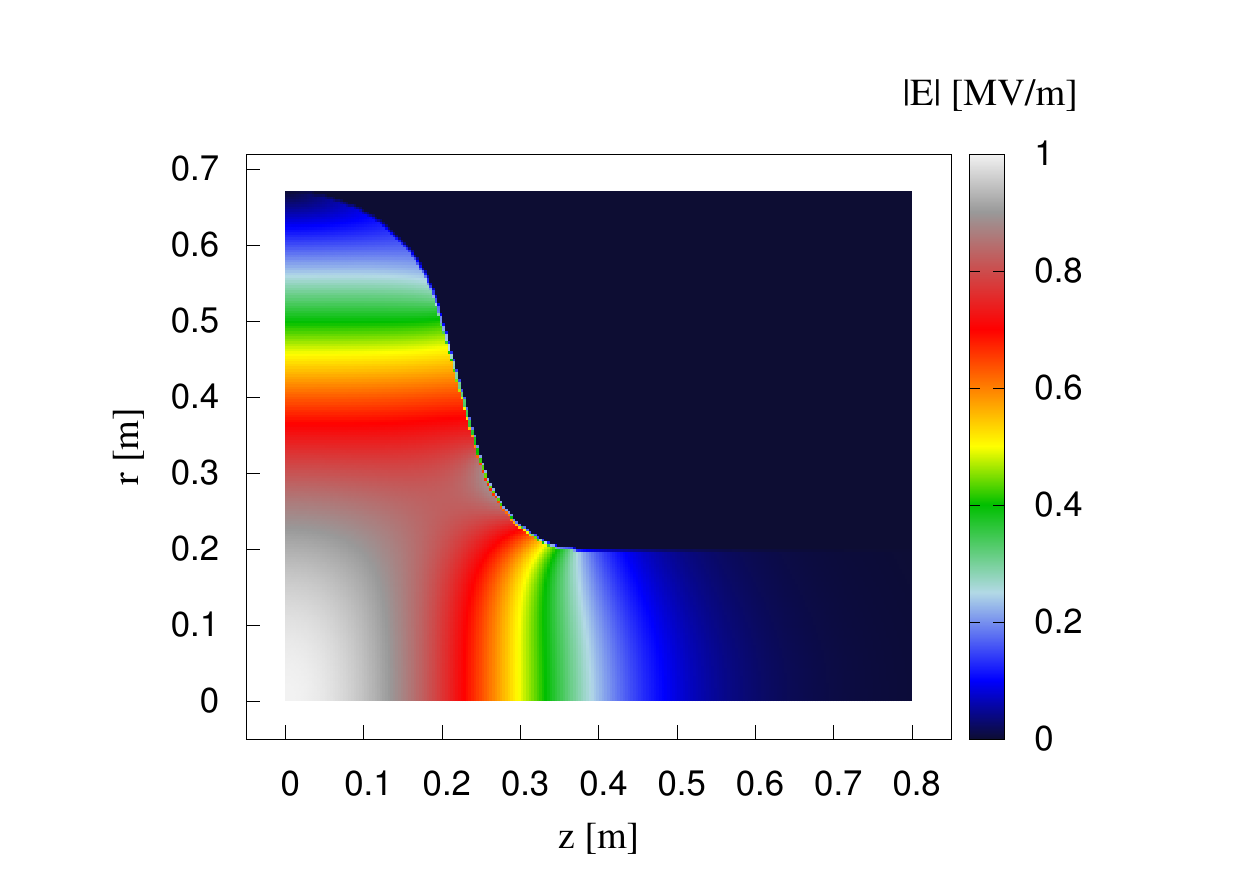}%
    \includegraphics[width=0.45\textwidth]{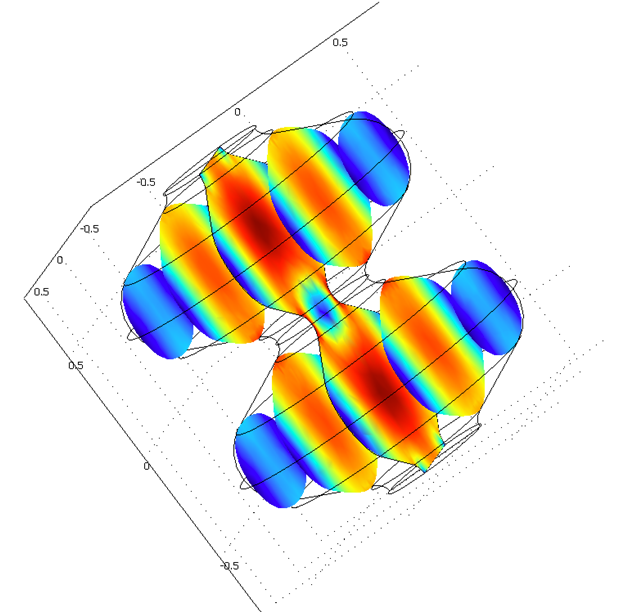}
  \end{center}
\caption{Main parameters of the superconducting RF cell and its electric field maps.}
\label{fig:acc:rla:cavity}
\end{figure}

We have studied the optimisation of the cavity design.
The layout of the standing wave superconducting RF cell has been
optimised for a number of parameters.
The target energy gain over the whole 66 cells of the linac is 
$\sim 665$~MeV, or a little more than 10 MeV/cell as shown in figure
\ref{fig:acc:rla:cavity}.  
Both single-cell and double-cell ($\pi$-mode) cavities have been
simulated with specialised codes including Superfish
\cite{superfish,Halbach:1976cp}, CST-MWS \cite{cst-mws} and Comsol
\cite{comsol} in order to obtain the electric- and magnetic-field
maps necessary for particle tracking.  

The peak electric fields are 23.09~MV/m in the upper linac and
26.17~MV/m the middle and lower linacs.  
In previous work~\cite{Ozaki:2001}, the references to 15~MV/m,
17~MV/m, and 17~MV/m cavities included the transit time factor for a
$\beta=1$ particle ($26.17\text{ MV/m}\times 0.65 = 17$~MV/m, 0.65 is the
transit time factor).   
Here $0.90\leq\beta \leq 0.99$ and the cavity phases were offset to
introduce $\sim 1$ synchrotron period over the whole linac.  

\subsubsection{Beam tracking}
\label{sec:acc:rla:track}

\begin{figure}
  \begin{center}
    \includegraphics[width=0.85\textwidth]{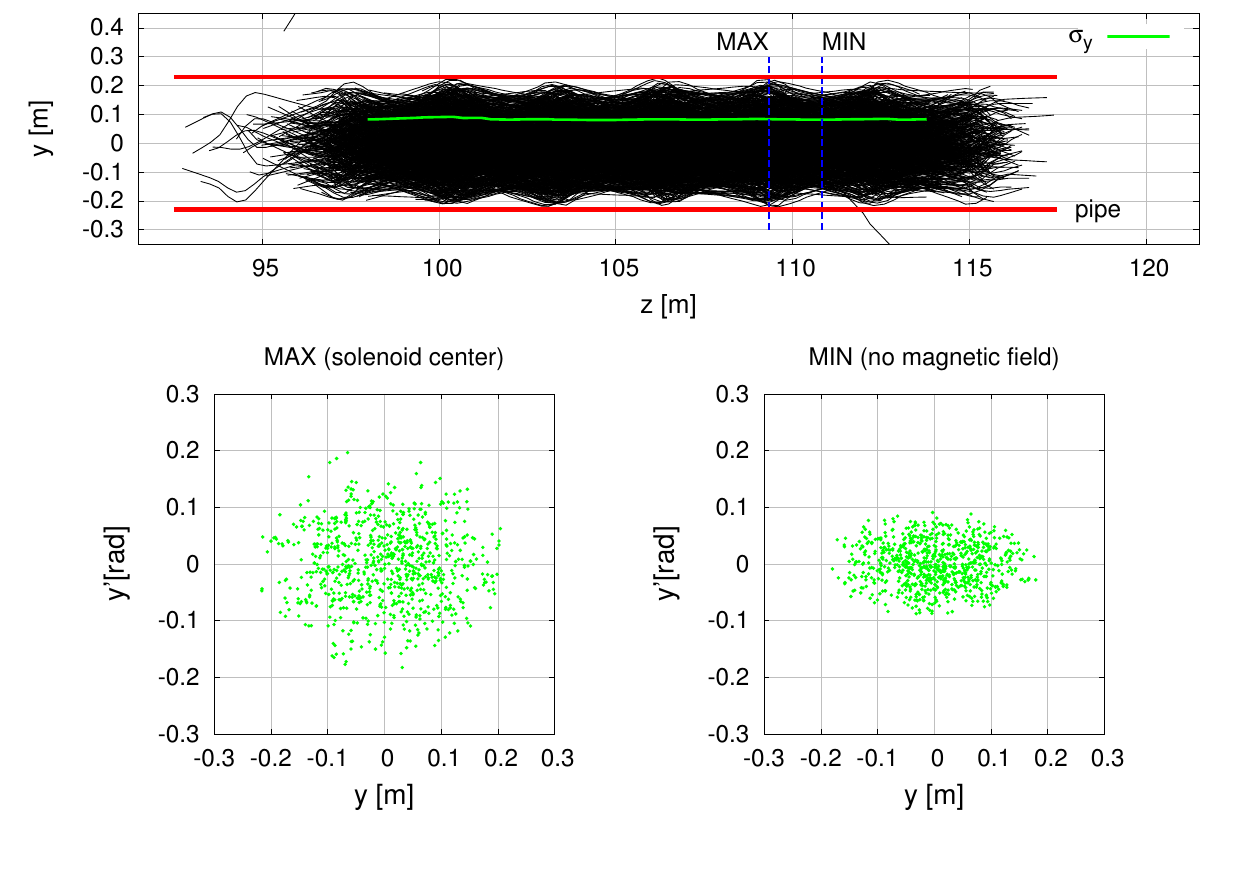}\\    
    \begin{tabular}{|c| c | l| l| l| l| l| l| l|}
      \hline
      \multicolumn{2}{|c|}{linac}            & $\bar{\varepsilon}_{x}$ & $\sigma_{x}$ & $\sigma_{x'}$ & $\bar{\varepsilon}_{y}$ & $\sigma_{y}$ & $\sigma_{y'}$  \\
      \multicolumn{2}{|c|}{section}          &      [mm rad]           & [mm]         & [mrad]        & [mm rad]                & [mm]         & [mrad]         \\ \hline
      \textit{\textbf{\textit{upper}}} & MIN &  5.7                    & 78           & 43            & 5.7                     &  82          & 40             \\
      \textit{\textbf{\textit{middle}}}& MIN &  3.4                    & 79           & 26            &  3.3                    &  79          & 25             \\
      \textit{\textbf{\textit{lower}}} & MIN &  2.2                    & 87           & 15            &  2.1                    &  88          & 16 \\
    \hline
    \end{tabular}
  \end{center}
  \caption{Stable trajectories in the upper linac 36 cells downstream
    from the injection point (top), the \textit{y}-phase space at the two
    positions indicated (middle) and values of the normalised
    transverse emittances and standard deviations (bottom).}
  \label{fig:acc:rla:au}
\end{figure}

First, using only solenoid field maps (no-acceleration), Gaussian muon
bunches with very large emittances have been tracked through many
cells of the three linac cell types in order to obtain
the transverse acceptances.  
A significant number of muons are lost in the early part of the
linac.
Later in the linac, the beam is transported without additional loss
making it possible to determine the rms transverse emittances shown in
figure \ref{fig:acc:rla:au}.

\begin{figure}
  \begin{center}
    \includegraphics[width=0.84\textwidth]%
      {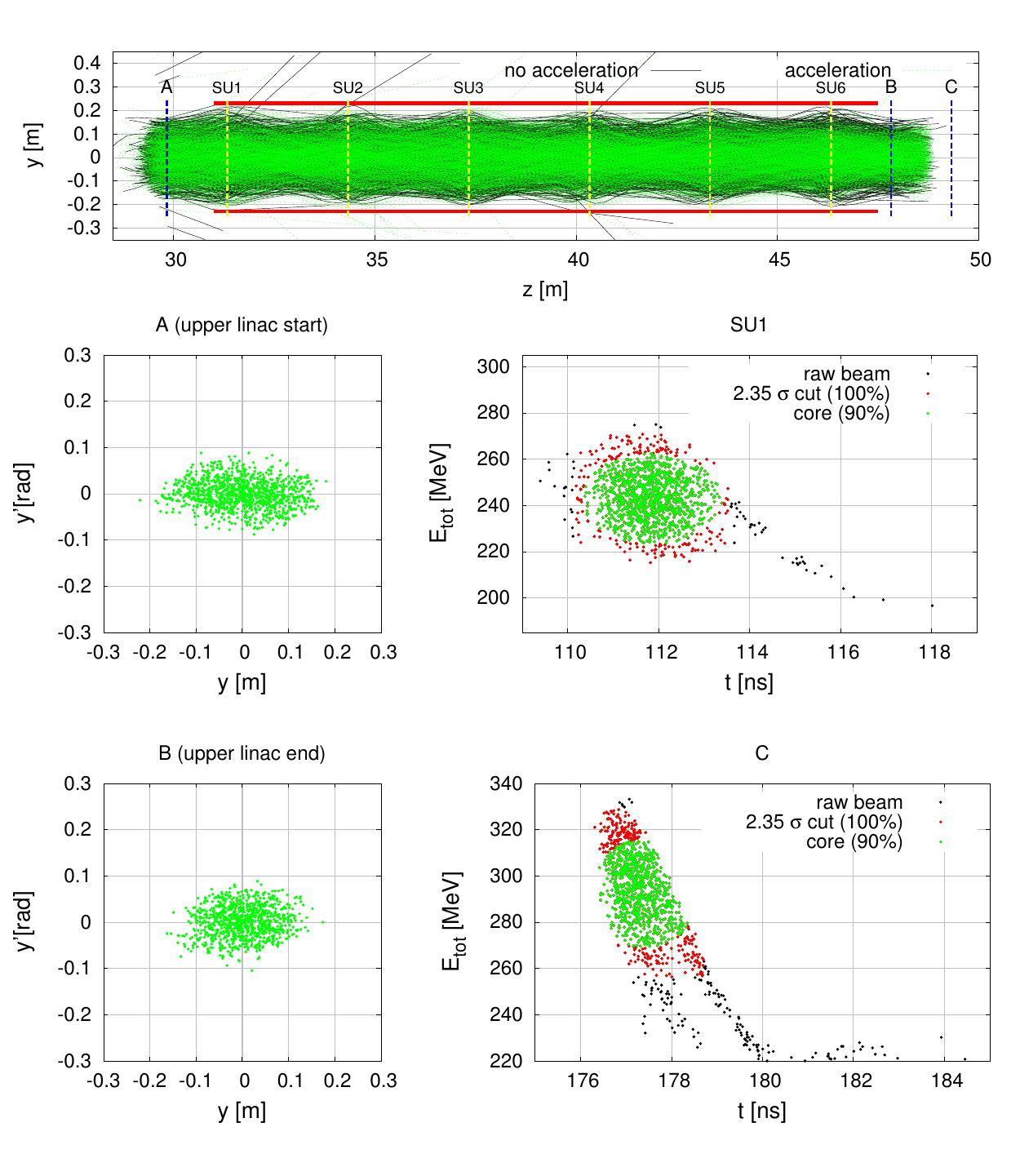}\\
    \begin{tabular}{|c |c | l |l |l |l |l |l |l |}
      \hline
      \multicolumn{2}{|c|}{\textit{phase}}         &$\bar{\varepsilon}_{x}$ & $\sigma_{x}$ & $\sigma_{x'}$ & $\bar{\varepsilon}_{y}$ & $\sigma_{y}$ & $\sigma_{y'}$  \\
      \multicolumn{2}{|c|}{\textit{space}}         & [mm rad]               & [mm]         & [mrad]        & [mm rad]                & [mm]         & [mrad]     \\ \hline
      \multirow{2}*{\textit{\textbf{transverse}}}  & A & 4.9                & 81           & 34            &  5.0                    & 87           & 31         \\
                                                   & B & 4.5                & 61           & 33            &  4.6                    & 62           & 36 \\ 
      \hline
      \multicolumn{2}{|c}{\textit{}}                & \multicolumn{2}{|l|}{(1/$\beta\gamma$)$\times$ $\bar{\varepsilon}_{||}$}& \multicolumn{2}{|l|}{$\sigma_{t}$} &\multicolumn{2}{|l|}{$\sigma_{E_{tot}}$}              \\
      \multicolumn{2}{|c}{\textit{}}                  & \multicolumn{2}{|l|}{[ms eV]}             & \multicolumn{2}{|l|}{[ns]}           &  \multicolumn{2}{|l|}{[MeV]}         \\ \hline
      \multirow{2}*{\textit{\textbf{longitudinal}}}& SU1& \multicolumn{2}{|l|}{2.7}                & \multicolumn{2}{|l|}{0.72}        &   \multicolumn{2}{|l|}{13}     \\ 
      & C  & \multicolumn{2}{|l|}{1.7}                & \multicolumn{2}{|l|}{0.55}        &   \multicolumn{2}{|l|}{15}  \\ \hline
    \end{tabular}
  \end{center}
  \caption{GPT particle trajectories through the upper linac (top),
    the \textit{y}-phase space at its beginning/end (middle) and figures
    of the normalised emittances and standard deviations (bottom).}
  \label{fig:acc:rla:trup}
\end{figure}

\begin{figure}
  \begin{center}
    \includegraphics[width=0.87\textwidth]%
      {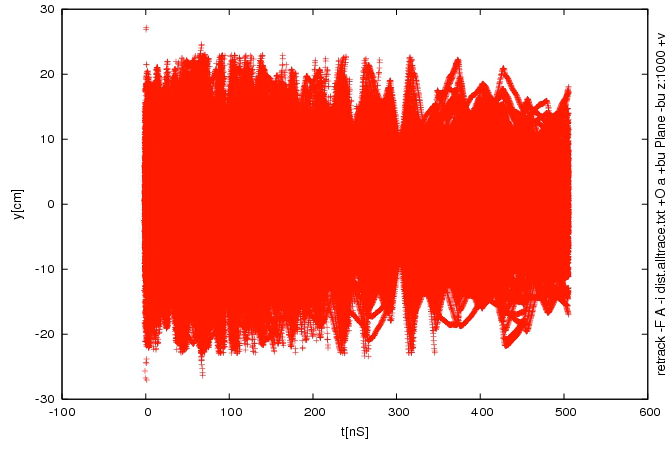}\\
    \includegraphics[width=0.87\textwidth]%
      {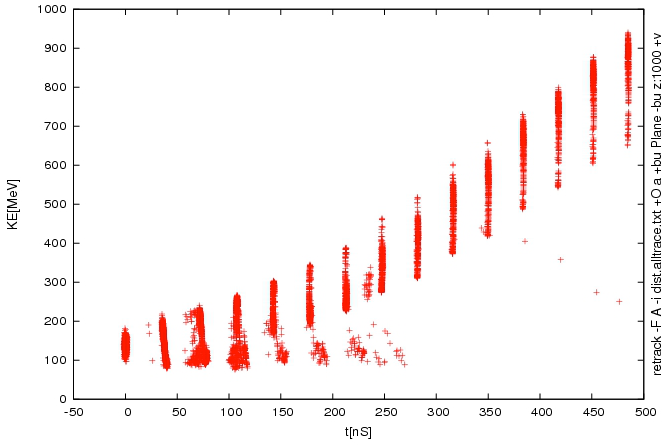}
  \end{center}
  \caption{
    G4beamline particle tracking showing vertical projections of the
    trajectories (top) and kinetic energy (bottom) along the entire
    linac.
  }
  \label{fig:acc:rla:g4bl}
\end{figure}

End-to-end, particle-tracking studies are in progress for the 
linac using the codes GPT~\cite{gpt}, G4beamline~\cite{Roberts:2008zzc},
OptiM and ELEGANT.  
As an example, figure \ref{fig:acc:rla:trup} shows the longitudinal and
transverse phase-space evolution for the upper linac while figure
\ref{fig:acc:rla:g4bl} shows the particle trajectories and
kinetic-energy increase along the entire linac. 

As figure \ref{fig:acc:rla:trup} shows, there is noticeable (adiabatic)
reduction of the transverse beam-size as the beam undergoes
acceleration,
although the phasing of the RF cavities is not yet fully optimised.
Future studies will focus on the longitudinal phase space and on
controlling the longitudinal-transverse phase-space coupling with the
aim of transporting the beam along the whole linac with minimum
losses.

\subsection{Linear non-scaling FFAG}

In the linac and RLAs described above, the cost of the RF systems 
dominates the cost of the beam lines that accelerate the beam.
A better balance between magnet and RF costs
would result by making more passes through the RF cavities.
Unfortunately, the switchyard precludes one from making a large
number
of passes in the RLA linac.
It would thus be advantageous to eliminate the switchyard and use a single
arc for all energies.  
This could be accomplished by using
a machine that allows a large energy range
within a single fixed-field arc, i.e.,   
a fixed field alternating gradient (FFAG) accelerator.

To keep the machine cost low and to allow the use of 201.25~MHz RF
cavities, we use a design that differs from the original type of FFAG
(referred to as a ``scaling'' FFAG).  
This new type of machine is called a
linear ``non-scaling'' FFAG.  
Its important features are that it is
isochronous near the middle of the energy range, limiting the
phase slippage and thus permitting the use of high RF frequencies, and
that its magnets are more compact than those of a comparable scaling
FFAG. 

\begin{figure}
  \centering
  \includegraphics[width=0.75\linewidth]{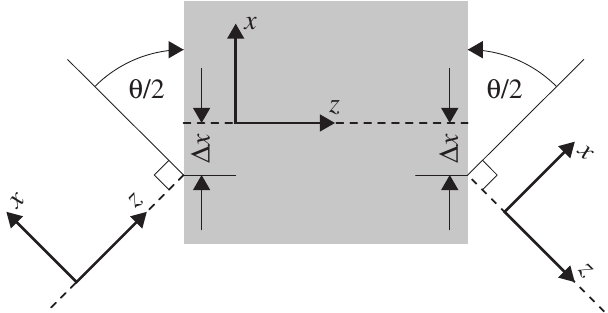}
  \caption{Geometry at a magnet.  $\theta$ is the angle given in
    table~\ref{tab:acc:ffag:ring}, and $\Delta x$ is the
    shift given in that same table.}
  \label{fig:acc:ffag:geom}
\end{figure}

We have optimised the design of a linear non-scaling FFAG
\cite{Berg:2004wc}, and the resulting parameters are given in tables
\ref{tab:acc:ffag:goals}--\ref{tab:acc:ffag:rf}.
The important criteria for this optimisation were that the long drift
length would be 5.0~m (to accommodate the septum, see below) and that
the time-of-flight be a half integer number of RF periods per cell.  
With the exception of the lattice cells containing the injection and
extraction insertions, the ring consists of identical cells, each
containing an FDF triplet. 
Most cells contain a superconducting cavity, the parameters of
which are given in table \ref{tab:acc:ffag:rf}.
The magnet aperture is increased by 30\% from what is required
for the beam to ensure that the beam is in a region of the magnet with
good field quality.  The magnet angle and shift are defined as follows
(see figure \ref{fig:acc:ffag:geom}):
upon entering the magnet, the coordinate axis bends toward the centre
of the ring by half the given angle.  The centre of the magnet is
then shifted away from the centre of the ring by the shift value.  Upon
exiting the magnet, the coordinate axis bends toward the centre of the ring
by the second half of the bend angle.
\begin{table}
  \caption{Design requirements for the FFAG accelerator.
    Energies are total energy. Transverse acceptance is
    defined as $a^2p/(\beta_\bot mc)$,
    where $a$ is the maximum beam radial displacement,
    $\beta_\bot$ is the Courant-Snyder beta
    function, $p$ is the total momentum, $m$ is the muon mass, and $c$ is the
    speed of light.  Longitudinal acceptance is $\Delta E\Delta t/mc$,
    where $\Delta E$ is the maximum energy half-width and $\Delta t$
    is the maximum half-width in time (assuming an upright longitudinal
    ellipse).}
  \label{tab:acc:ffag:goals}
  \begin{tabular}{|l|r|l|r|}
    \hline
    Injection energy &12.6 GeV&Transverse acceptance&30 mm\\
    Extraction energy&25.0 GeV&Longitudinal acceptance&150 mm\\
    Muons per train&$2.4\times10^{12}$&Muon trains&3\\
    Time between trains&120 $\mu$s&Repetition rate&150 Hz\\
    \hline
  \end{tabular}
\end{table}
\begin{table}
  \caption{Magnetic lattice parameters of the linear non-scaling FFAG.
    The short drift is between each D and F in the triplet, the
    long drift is between the F magnets.  The angle and shift are described
    in the text.  The maximum magnet field is at the magnet aperture.}
  \label{tab:acc:ffag:ring}
  \centering
  \begin{tabular}[c]{|l|r|}
    \hline
    Cells            &64        \\
    Long drift       &5.0 m     \\
    Short drift      &0.5 m     \\
    Circumference    &667 m     \\
    \hline
  \end{tabular}
  \hfil
  \begin{tabular}[c]{|l|r|r|}
    \hline
                        &D        &F       \\
    \hline
    Length (m)          &2.251117 &1.086572\\
    Angle (mrad)        &156.837  &-29.331 \\
    Shift (mm)          &41.003   &13.907  \\
    Field (T)           &4.20784  &-1.39381\\
    Gradient (T/m)      &-13.55592&18.04570\\
    Aperture radius (mm)&137      &163     \\
    Maximum field (T)   &6.1      &4.3     \\
    \hline
  \end{tabular}
\end{table}
\begin{table}
  \caption{RF parameters and derived performance of the linear non-scaling
    FFAG.  Cavity voltage is the maximum energy gain of a speed of light
    muon passing through the cavity~\cite{Ozaki:2001}.}
  \label{tab:acc:ffag:rf}
  \centering
  \begin{tabular}[c]{|l|r@{\qquad}|l|r|}
    \hline
    RF frequency     &201.25 MHz&Cavity voltage  &25.5 MV\\
    Aperture diameter&30 mm     &Cells per cavity&2      \\
    Input power      &1 MW      &RF pulse length &3 ms   \\
    Stored energy    &2008 J    &                &       \\
    Cavities         &48        &Total voltage   &1214 MV\\
    Turns            &11.6      &Decay loss      &6.7\%  \\
    \hline
  \end{tabular}
\end{table}

\subsubsection{Injection and Extraction}

Injection and extraction occur at opposite sides of the ring. Each
system will have reflection symmetry to accommodate both positively
and negatively charged muons. There will be two septa in each system
with kicker magnets between them.  All the kickers in each system have
the same strength. The septum and kicker magnets have length 4.4~m,
i.e., there is 30~cm separation between each of these elements and the
adjacent main
magnets. Parameters for the injection/extraction magnets are given in
table~\ref{tab:acc:ffag:inj}.
\begin{table}
  \caption{Injection/Extraction system parameters.  Pattern contains a
    \texttt{+} for an outward kick, \texttt{-} for an inward kick,
    and a \texttt{O} for an empty drift. A
    septum is located at both ends of the kicker system.}
  \label{tab:acc:ffag:inj}
  \centering
  \begin{tabular}{|l|r|r|}
    \hline
    &Injection&Extraction\\
    \hline
    Kickers&2&4\\
    Pattern&\texttt{-O-}&\texttt{++OO++}\\
    Kicker field (T)&0.089&0.067\\
    Septum field (T)&0.92&1.76 \\
    Kicker/septum length (m) & 4.4 & 4.4 \\
    \hline
  \end{tabular}
\end{table}

A septum field below 2~T is chosen in order to minimise the
effect of the septum stray-field on the circulating beam.  
In order to
estimate the clearance needed so that the extracted or injected beam
avoids the magnet immediately following the septum, the dimensions of
the superconducting combined-function magnet (SCFM) in the J-PARC
neutrino beam line is used. The width of the J-PARC SCFM from the
inner radius of the coils to the outer radius of the cold mass is
19.8~cm~\cite{Ogitsu:2009}. This distance is used, together with the size
of the muon beam, in calculating the required septum bending
angle. Taken together, these constraints require the long drift
in the triplet to be at least 5~m in length.

\begin{figure}
  \begin{center}
    \includegraphics[width=0.5\linewidth]{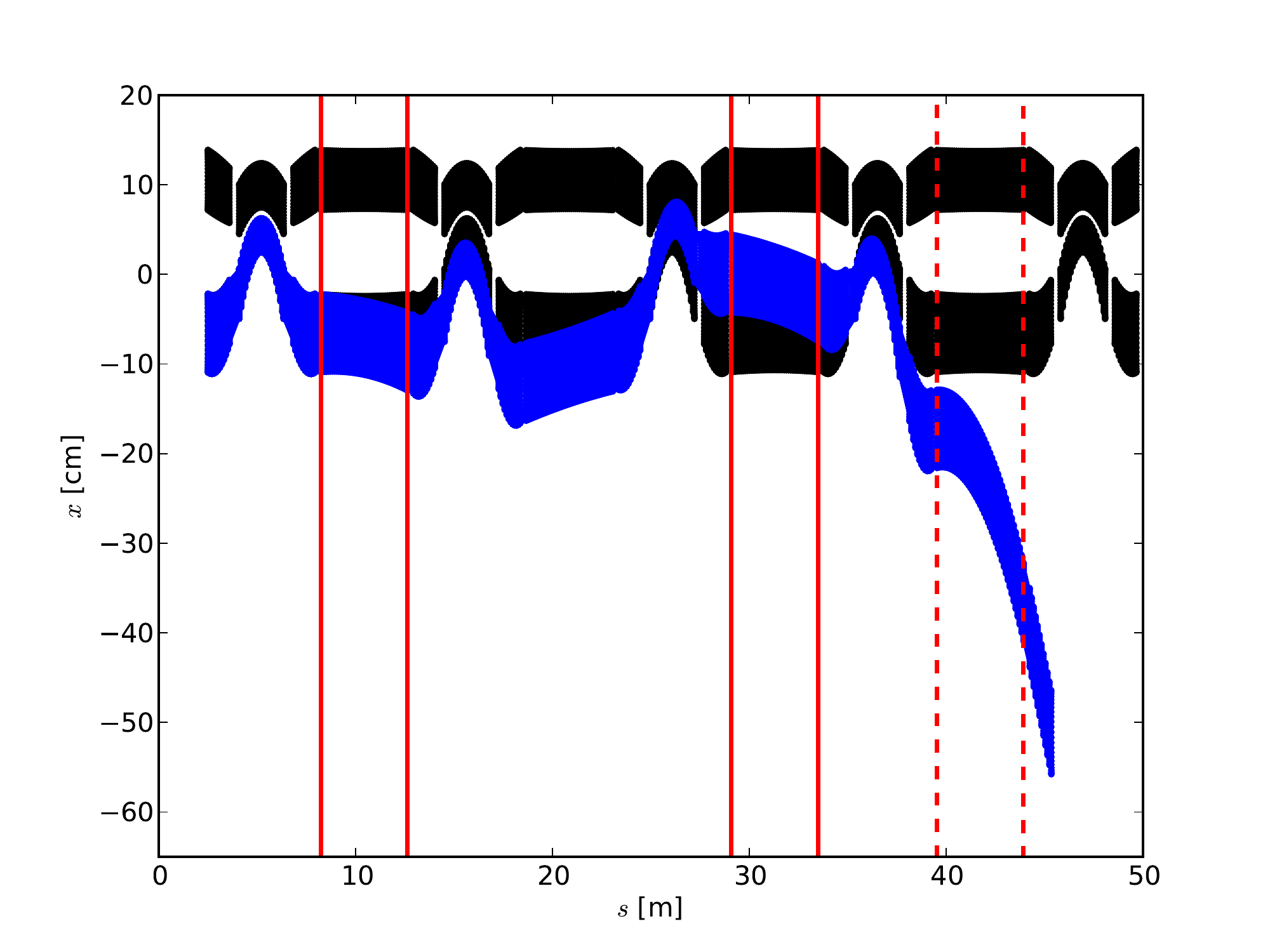}%
    \includegraphics[width=0.5\linewidth]{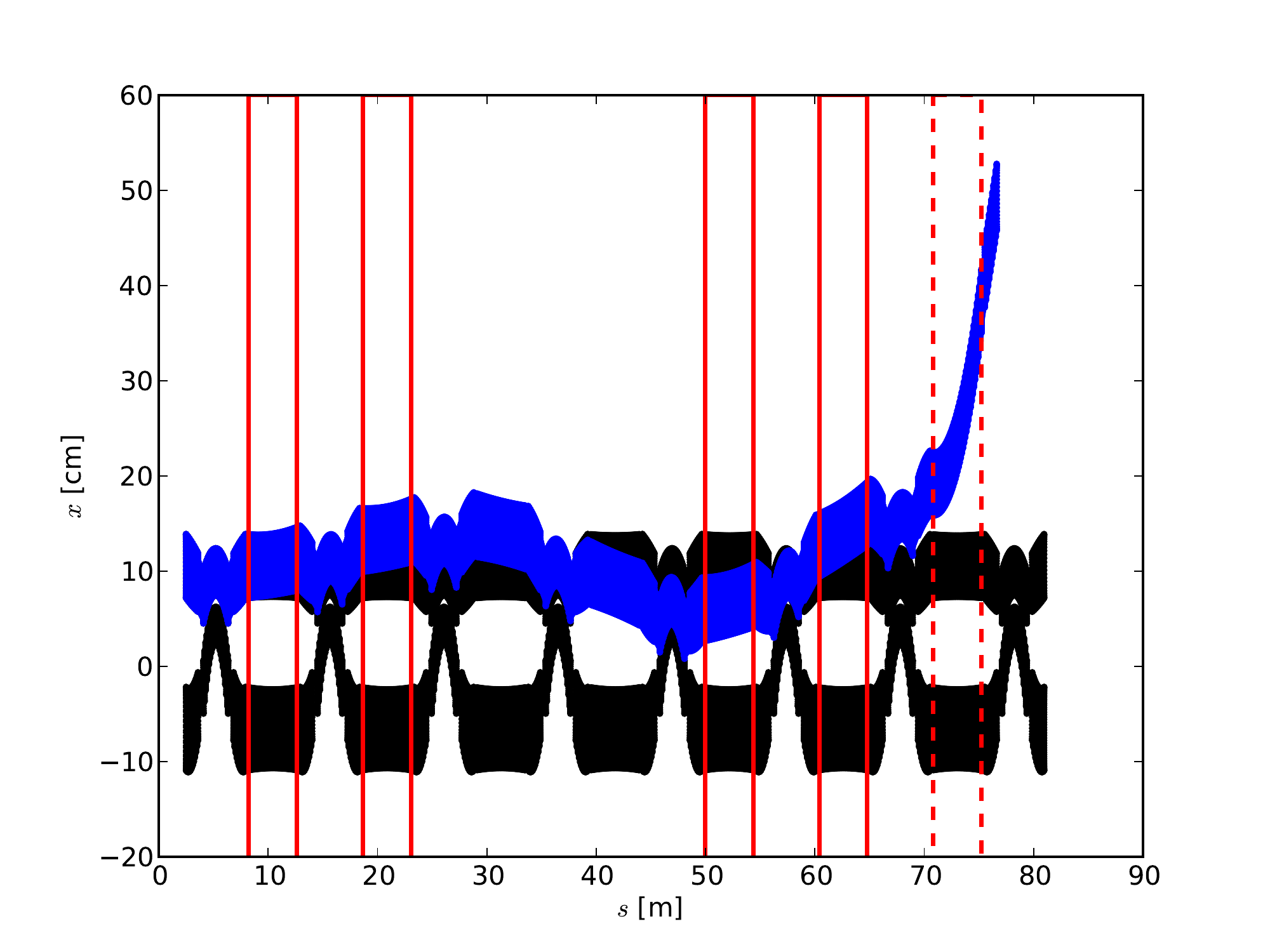}
  \end{center}
  \caption{Injection scheme (left) and extraction scheme (right) for
    the FFAG ring. The full-acceptance circulating beam at injection and
    extraction
    momenta are shown in black and the injected and extracted beams in
    blue. The red vertical lines represent the location of the kicker
    magnets (solid) and septum (dashed).}
  \label{fig:acc:ffag:injext}
\end{figure}

In order to be feasible from an engineering perspective, the
kicker-magnet peak fields are constrained to be less than 0.1~T. The
separation between the kicked and circulating beams at the
entrance of the septum is 2~cm. 
Two kicker magnets are required for injection. Due
to the relatively high horizontal phase advance at this energy and the
requirement of mirror symmetry, the kickers need to be separated by a
single cell, resulting in an empty long drift at the centre of the
injection system. At extraction four kickers are required, i.e., two
pairs separated by two empty long drifts. Tracking results of the muon
beam through the injection and extraction systems are shown in
figure \ref{fig:acc:ffag:injext}. As can be seen from the figure,
injection is into the inside of the ring whereas the beam is extracted
to the outside.
\begin{table}
  \caption{Magnet apertures (i.e., inner radius of magnet) in the
    Injection/Extraction system. In the rest of the ring, the magnet apertures
    are 16.3 cm and 13.7 cm for the F and D magnets, respectively.}
  \label{tab:acc:ffag:aper}
  \centering
  \begin{tabular}{|l|r|r|r|}
    \hline
    &Magnet&Number&Aperture (cm)\\
    \hline
    Injection&F&4&20.8\\
    &D&4&16.1\\
    Extraction&F&8&19.8\\
    &D&2&15.5 \\
    \hline
  \end{tabular}
\end{table}

\begin{figure}
  \begin{center}
    \includegraphics[width=0.75\linewidth]{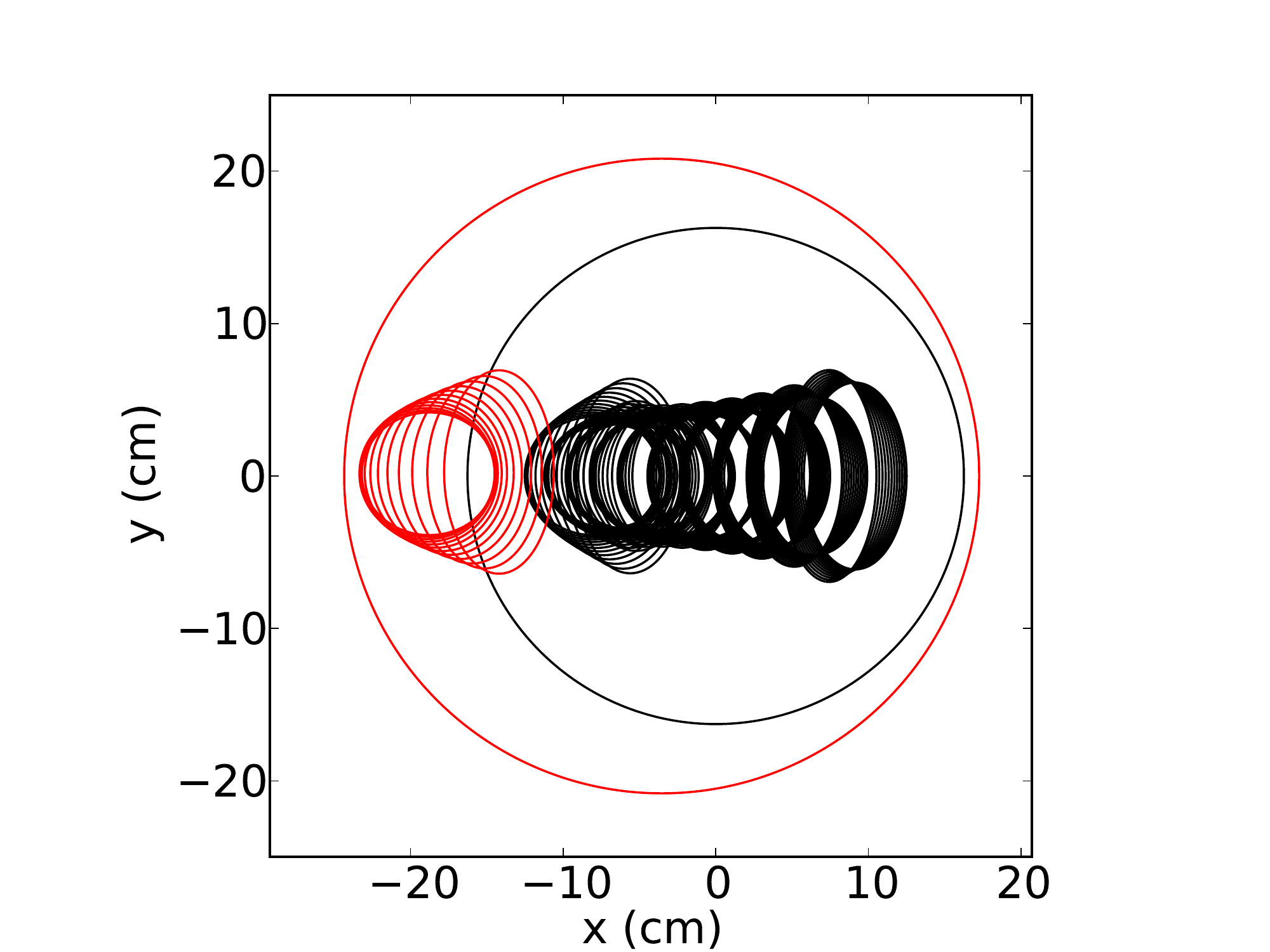}
  \end{center}
  \caption{The circulating beam (black ellipses)
    enclosed by the normal F magnet aperture (black circle) and the
    additional aperture (red circle) required to accommodate the kicked
    beam (red ellipses, shown
    just before the septum at injection.}
  \label{fig:acc:ffag:aperture}
\end{figure}

The distribution of kickers over several cells means that a number of
large-aperture main magnets are required. The apertures are calculated
by finding the smallest circle that encloses the kicked and
circulating beam and adding an additional 1~cm clearance (for example
see figure \ref{fig:acc:ffag:aperture}). The number of these magnets and
their apertures are listed in table~\ref{tab:acc:ffag:aper}.

\subsubsection{Kickers and Septa for the FFAG}

\begin{figure}
  \includegraphics[width=0.8\textwidth]%
    {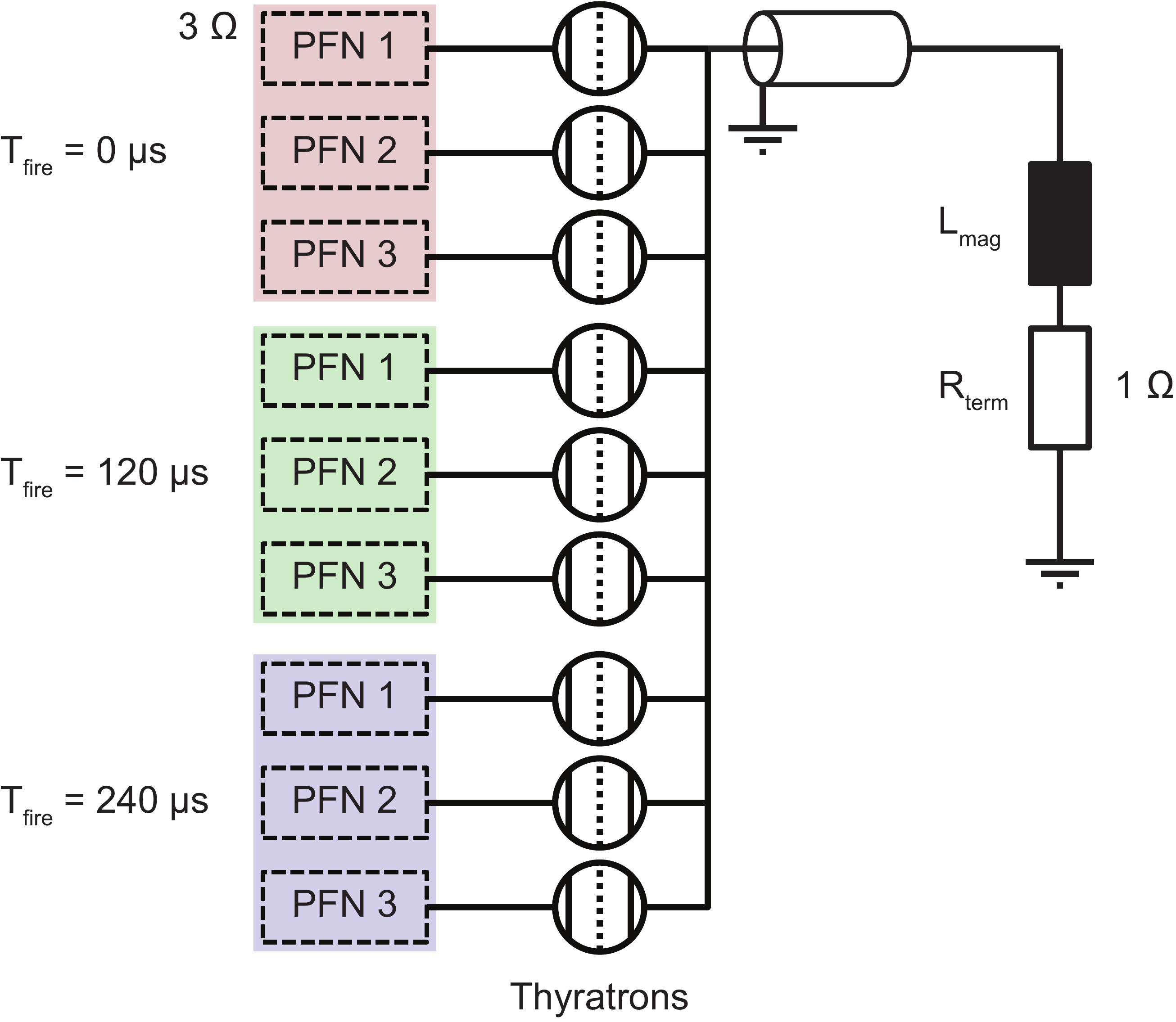}
    \caption{Schematic of the FFAG injection/extraction kicker system
      operating with 3 muon bunch trains. The
    above system is needed for every
    sub-kicker unit.}
  \label{fig:acc:ffag:kicksys}
\end{figure}

Travelling-wave kicker magnets are proposed both for
injection and extraction.  The kicker system consists of a power
supply charging pulse forming networks (PFNs), which are connected via
fast thyratron switches and coaxial cables to the kicker
magnet and terminated via a matching resistor.  A CX1925X~\cite{cx1925}
thyratron switch
and RG192 coaxial cables connected in
parallel could be used.  A current of about 30 kA is necessary to
produce the required 0.1~T magnetic field in an aperture of about
$0.3\text{ m}\times0.3\text{ m}$.
The same design is proposed for injection and
extraction, as the requirements with respect to rise/fall time are
very similar.  This dictates the need for careful impedance matching
to avoid any reflections, which is essential, especially for
the injection case.  In order to keep the rise- and fall-times low, the
kicker magnet must be sub-divided into four or five separate units.
The kicker uses a window-frame geometry, with the yoke
made of ferrite to enhance the field level, limit the
field leakage and improve the field quality.  The ferrite Ferroxcube
8C11~\cite{8C11} could be used for the kicker core.  The detailed
design of the kicker, including the matching capacitance scheme to
minimise reflections, remains
to be addressed in future studies.  A voltage of 60 kV is assumed,
which is compatible with existing fast thyratron switches and sets
the impedance for the magnet and termination resistor at 1~$\Omega$.
The thyratron CX 1925X is limited to 10\,kA peak current, which
dictates the use of three 3~$\Omega$ PFNs connected in parallel for each
sub-kicker.  As the baseline muon pulse consists of three muon 
bunch-trains separated by 120~$\mu$s, a total of 9 PFNs for each
sub-kicker is needed to avoid requiring an extremely demanding power
supply to recharge the PFNs between the individual bunch trains.  The
total number of PFNs and switches per kicker unit installed in a
single drift is thus 36 (assuming that the kicker is subdivided into
four sub-kickers).  Power supplies of $\approx 2.5$~MW peak power
are needed for every kicker and although it is expected that the
majority of the power will be dumped into the termination resistor,
thermal issues for the kicker magnet remain to be studied.  
A schematic of the kicker system is shown in figure
\ref{fig:acc:ffag:kicksys} and some of its parameters are collected in
table \ref{tab:acc:ffag:kick}.
\begin{table}
  \caption{Parameters of the kicker system}
  \label{tab:acc:ffag:kick}
  \centering
  \begin{tabular}{|l|r|}
    \hline
    Kicker total aperture (h$\times$v)&0.3$\times$0.3 m\\
    Kicker length &4.4 m\\
    Rise/fall time (5-95$\%$)&1.9 $\mu$s\\
    Kicker max field& $\approx$0.1 T\\
    Kicker pulse duration at the top & 0.3 $\mu$s\\
    Charging voltage & 60 kV\\
    Peak current in the magnet& 30 kA\\
    Kicker inductance & 5.1 $\mu$H\\ 
    Kicker impedance &  1 $\Omega$\\
    Peak current at switch & 10 kA\\
    Repetition rate &     50 Hz\\
    Number of sub-kickers&  4-5\\
    Number of PFNs per micro-pulse per sub-kicker& 3\\
    Total number of PFNs &   36 (for 4 sub-kickers)\\
    Total averaged power per kicker &  $\approx$1.25 MW\\
    Total peak power per kicker &  $\approx$2.5 MW\\
    \hline
  \end{tabular}
\end{table}

\begin{figure}
  \includegraphics[width=0.8\textwidth]{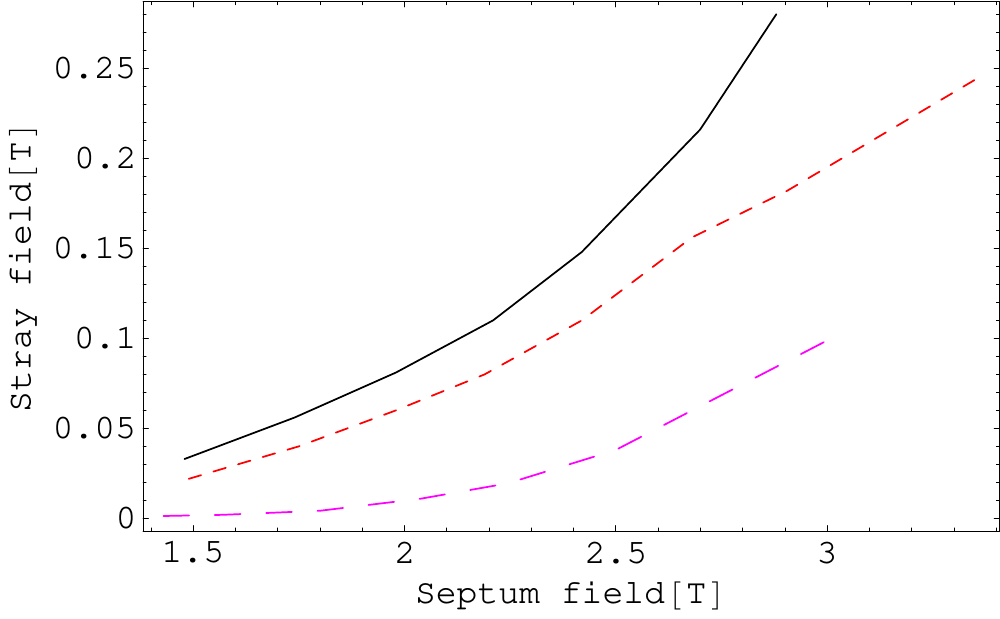}
  \vspace*{8pt}
  \caption{Stray fields in the superconducting
    septum using standard soft iron (black, solid line), using
    advanced material-VACOFLUX 50 (red line, short dash) and
    introducing a chamfer (magenta, long dash). The simulations
    were performed using the COMSOL 4.0
    package.}
  \label{fig:acc:ffag:septum1}
\end{figure}

\begin{figure}
  \includegraphics[width=0.8\textwidth]{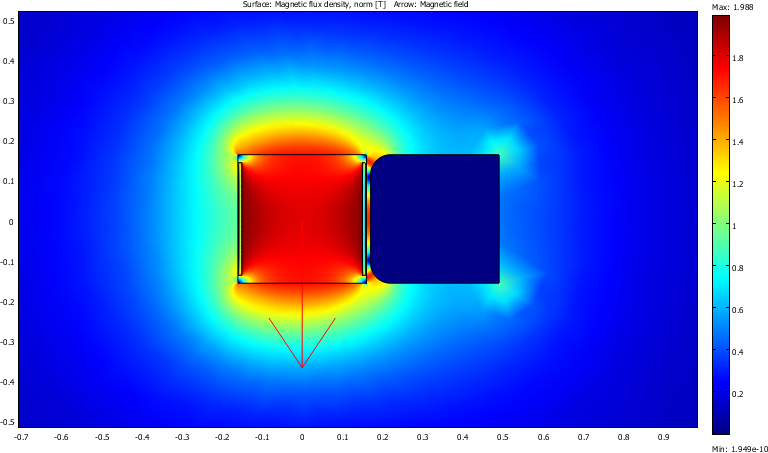}
  \vspace*{8pt}
  \caption{2D section of the superconducting
    septum for a muon FFAG.} 
  \label{fig:acc:ffag:septum2}
\end{figure}

Kicker magnets are used to suppress (create) the separation between
the injected (extracted) and the circulating beam, but in order
for the beam to clear
the external aperture of the main magnets,
septa are required.
The magnetic fields of septa for both injection and
extraction are shown in table~\ref{tab:acc:ffag:inj}. As their aperture
is large ($\approx$0.3$\times$0.3 m total), a rather high current-turn
product of
450 kA-turns is required.  Although the septum design could be based
on normal conducting technology, a DC solution would be rather
challenging due to problems with cooling the coils.  This cooling
problem could be reduced by pulsing the septum, but that would introduce
eddy currents and variable stray fields in close proximity to the
main superconducting magnets.  Moreover, a rather substantial power supply
would be
needed.  To avoid these challenges we propose to use superconducting
technology for the septa, both for injection and extraction.  As the
extraction septum is the more challenging, design work has focused on
this case. An identical device could also be used for injection.  By
using superconducting technology, high magnetic fields can easily be
obtained using very thin conductor (8 mm) with a modest current density of
200~A/mm$^2$.  The most challenging issue is the shielding of the
circulating beam region from stray fields.  In order to reduce the
field leakage, COMSOL 4.0 \cite{comsol} simulations have been
performed using various yoke geometries.  The current design is based
on a window frame magnet with the yoke extending all around both
extracted and circulating beam areas.  The septum itself, which is 2
cm thick, is subdivided into the conductor part and the yoke part. Two
yoke magnetic materials have been studied: a ``standard'' soft iron
and the soft magnetic cobalt-iron alloy VACOFLUX
50~\cite{vacoflux}.  In addition, a chamfer was introduced on
the side facing the circulating beam in order to further limit the field
leakage.  Figure \ref{fig:acc:ffag:septum1} shows the simulations of the
stray fields using different yoke materials and with and without the
chamfer.  The leakage field at the boundary of the septum on the side
of the circulating beam is shown and was used to evaluate various
designs.  The 2D design of the extraction septum is shown in
figure \ref{fig:acc:ffag:septum2}.  Based on the electromagnetic
simulations, the septum field will be held to less than 2~T, which
dictates the length of the septum necessary to permit the beam to
clear the adjacent main magnets.  
This turns out to be an important constraint in the FFAG
lattice definition.  The detailed cryogenic design and quench limit of
the superconducting septum remain to be addressed. 

\subsubsection{Main Ring Magnets}

\begin{figure}
  \includegraphics[width=0.95\textwidth]{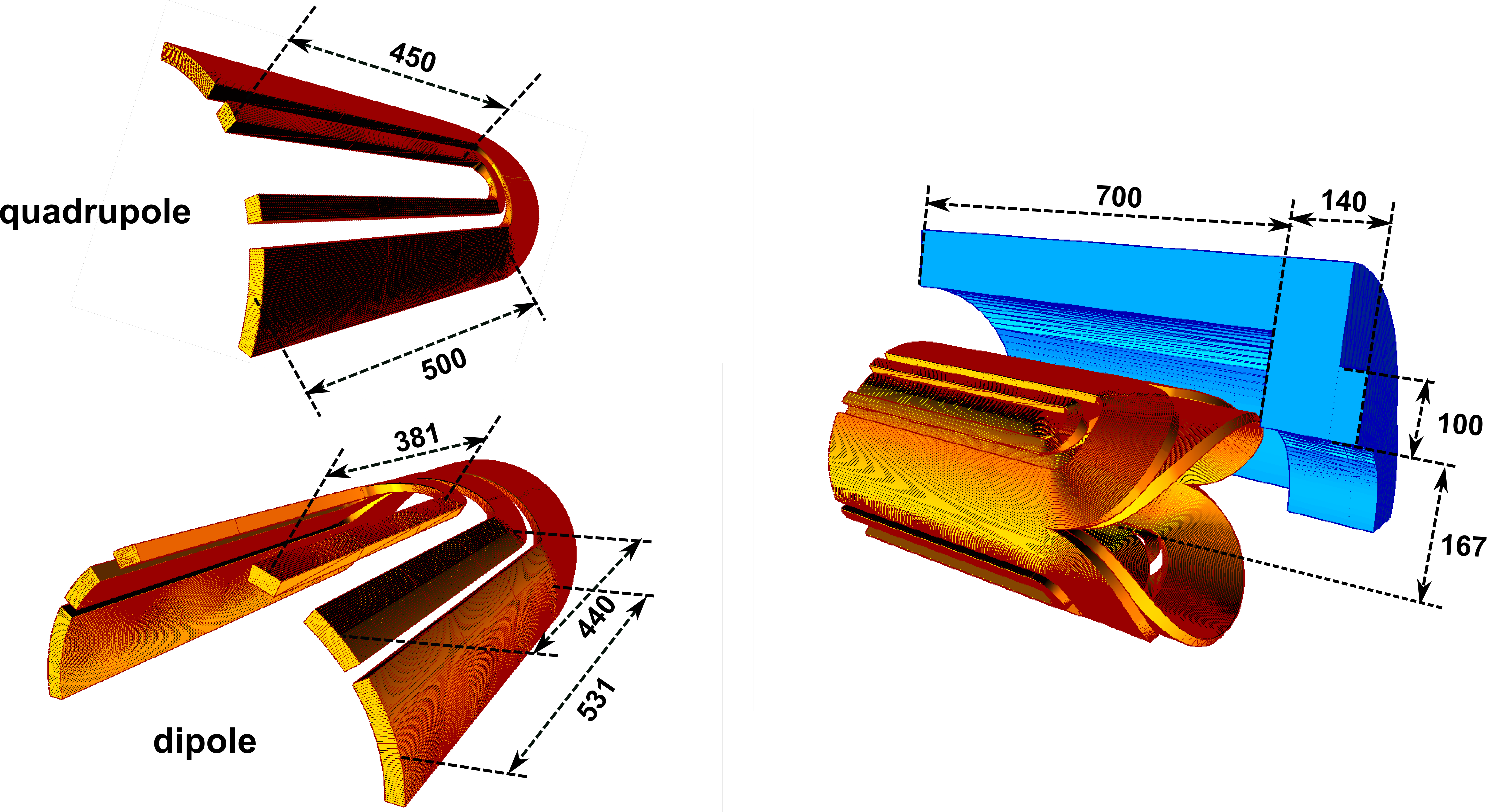}
  \caption{Geometry of the design of half of the
    F magnet for the FFAG ring.}
  \label{fig:ffag:magnet}
\end{figure}

\begin{figure}
  \includegraphics[width=0.8\textwidth]{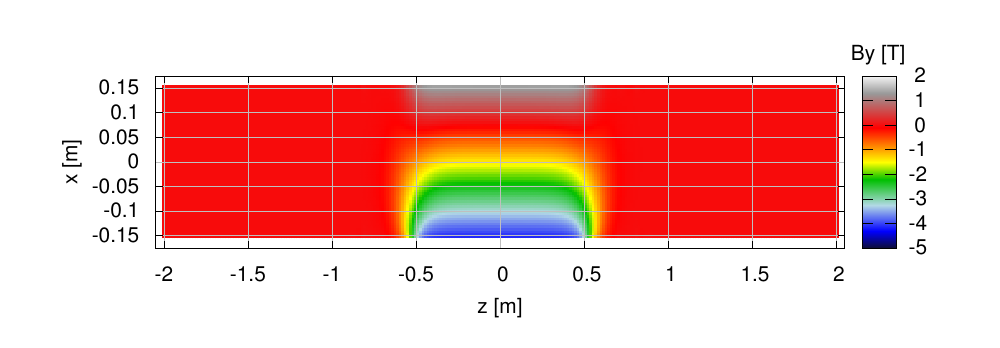}
  \caption{Vertical component of the magnetic
    field on the median plane of the F magnet obtained with
    the ROXIE code.}
  \label{fig:ffag:magnetfield}
\end{figure}

The preliminary design of the main superconducting combined-function
magnets for the linear non-scaling FFAG has been performed using the
CERN ROXIE code~\cite{roxie}. 
As the design of the focusing (F) magnet is more advanced,  
we present only this element. 
The design uses the well-established technology of Nb-Ti 
superconducting magnets fabricated using Rutherford cable. 
In order to simplify the geometry and allow for flexible 
beam optics tuning, the magnet design is based on separate coils for
dipole and quadrupole field components, which are 
assembled according to the conventional ``$\cos\theta$'' geometry. 
The inner layer creates a dipole field using 3 conductor blocks,
and the outer layer creates a quadrupole field with 2 conductor
blocks. 
For the needs of the current design the dipole and quadrupole cables
have been generated in ROXIE based on the standard Large Hadron
Collider (LHC) main dipole inner cable~\cite{Bruning:2004ej} at CERN. 
The cables consist of 28 strands and have a trapezoidal geometry. 
They would be constructed using the same filaments as the LHC magnet.
An iron yoke made of soft magnetic steel is placed beyond the
quadrupole layer. 
The magnet is closed with a clamp in order to 
limit field leakage in the long straight section, where it could
affect hardware components like superconducting RF cavities or
kickers. 
The geometry of coils and yoke for half of the F magnet is shown in
figure~\ref{fig:ffag:magnet}. 
The dipole field and gradient on axis have been reproduced according
to the lattice design specifications with an accuracy of
$\sim10^{-4}$. 
The vertical field component on the median plane of the F magnet is
shown in figure~\ref{fig:ffag:magnetfield}. 
The field quality off-axis still needs to be improved, which will
require an update of the magnet geometry. 
A quench analysis has been performed using ROXIE, and calculations of
the temperature margins suggest stable magnet operation.
The principal parameters of the focusing (F) FFAG magnet are collected
in table~\ref{tab:ffag:magnet}.
Future studies will address the field quality, the possible addition
of higher multipoles (mainly sextupole) for chromaticity correction,
and the cryogenic design. 
\begin{table}
  \caption{Parameters of the main focusing FFAG magnet.}
  \label{tab:ffag:magnet}
  \centering
  \begin{tabular}{|l|r|}
  \hline
    Strand diameter & 1.065 mm\\
    Number of strands per cable (dipole and quadrupole)& 28\\
    Cable height (dipole and quadrupole)&15~mm\\
    Dipole cable inner/outer width & 1.58/1.75~mm \\
    Quadrupole cable inner/outer width & 1.83/1.98~mm \\
    Number of conductors in dipole blocks& 52, 25, 13\\
    Number of conductors in quadrupole blocks& 41, 10\\
    Inner radius of dipole blocks& 163~mm\\
    Inner radius of quadrupole blocks& 179~mm\\
    Inner radius of the yoke& 300~mm\\
    Yoke thickness & 100~mm\\
    Half of the yoke length (with clamp) & 840 mm\\
    Dipole current& 3190~A\\
    Quadrupole current& 8490~A\\
    Dipole current density& 332~A/mm$^2$\\
    Quadrupole current density& 885~A/mm$^2$\\
    Peak field in the conductors & 5.75 T\\
    Minimal temperature margin to quench at 2.2 K& 3.84 K\\
    \hline
  \end{tabular}
\end{table}

\subsubsection{Beam loading}

Beam loading will cause the different bunches to gain different
amounts of energy. 
There are two time scales to consider: different energy gains within a
train, and different energy gains for different trains.

On each pass, a bunch train extracts 9.7~J from a cavity.  The last
bunch will therefore gain 62~keV less energy per cavity pass than the
first bunch.  Through the entire acceleration cycle, this amounts to
an energy difference from head to tail of 30~MV.  This is small compared
to the energy spread expected in the bunches.

Each proton driver cycle accelerates three bunches, which will
hit the target in relatively rapid succession.  Each proton driver bunch
creates a separate bunch train in the muon accelerator.
After a bunch train has completed its acceleration cycle, it has extracted
116~J from the cavities.  If no power is added to the cavities, then the
next bunch train will gain 360~MV less energy, which is too big a difference.
To replace the extracted
energy with an input power limitation of 1~MW per cavity requires 120~$\mu$s
between bunch trains and therefore between proton driver bunches.

\subsubsection{Risk Mitigation}

The use of a linear non-scaling FFAG is expected to be a more
cost-effective alternative to acceleration with an RLA.  
There are potential issues with this design, however.
It is known that the large transverse emittance
can lead to an effective longitudinal emittance growth in such a
design~\cite{Machida:2006sj}.
If necessary, this effect could be reduced by increasing the average RF gradient
and/or correcting the chromaticity, even partially~\cite{Berg:2007}.
The former can be accomplished by
increasing the drift length to 5.5~m and putting a second RF cavity in
each drift,
but this would increase the machine cost by around 25\%.  Chromaticity
correction is briefly discussed in Appendix~\ref{App:FFAG}.

One could eliminate the FFAG, and extend the energy range of the linac
and the RLAs so as to accelerate to 25~GeV.  While early estimates
led us to believe this is significantly more costly than accelerating
with an FFAG, it remains to be demonstrated in detail.  Furthermore,
the injection and extraction hardware is technologically challenging,
and likely very costly.
Should the FFAG not be found to give a
significant cost benefit, this alternative acceleration scenario without
an FFAG could be used.

\subsection{Decay ring}

In the Neutrino Factory \cite{Apollonio:2009}  neutrinos are generated
from muon decays according to:
\begin{equation}
  \mu^- \rightarrow e^- \bar{\nu}_e \nu_{\mu} \, ; \text{ and~}
  \mu^+ \rightarrow e^+ \nu_e \bar{\nu}_{\mu} \, .
\end{equation}
Neutrinos are aimed at far detectors by collecting muons in storage rings
with long straight sections pointing at the distant experimental
facilities.
The storage rings dip into the ground with angles of 18$^{\circ}$ and
36$^{\circ}$, for baselines of 4000 and 7500~km, respectively.
 
\begin{figure}
  \begin{center}
    \includegraphics*[width=0.95\textwidth]%
      {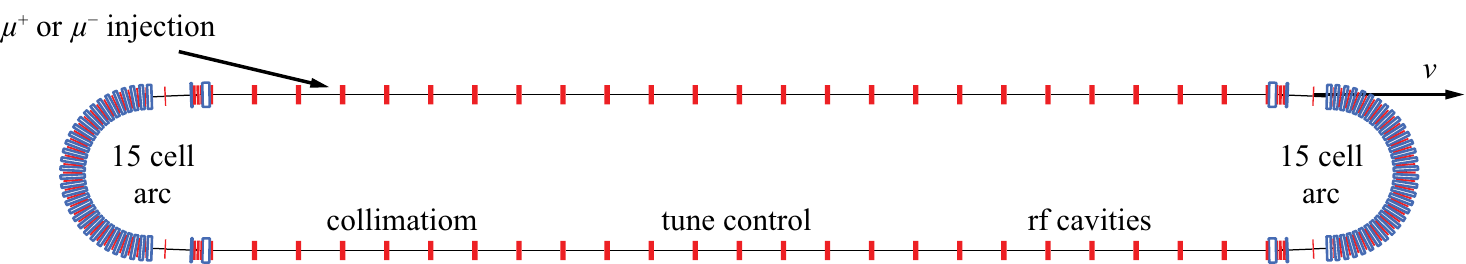}
  \end{center}
  \caption{Racetrack design for the Neutrino Factory storage rings.}
  \label{fig:acc:ring:RaceTrack}
\end{figure}

\begin{figure}
  \centering
    \includegraphics*[width=0.5\textwidth]%
      {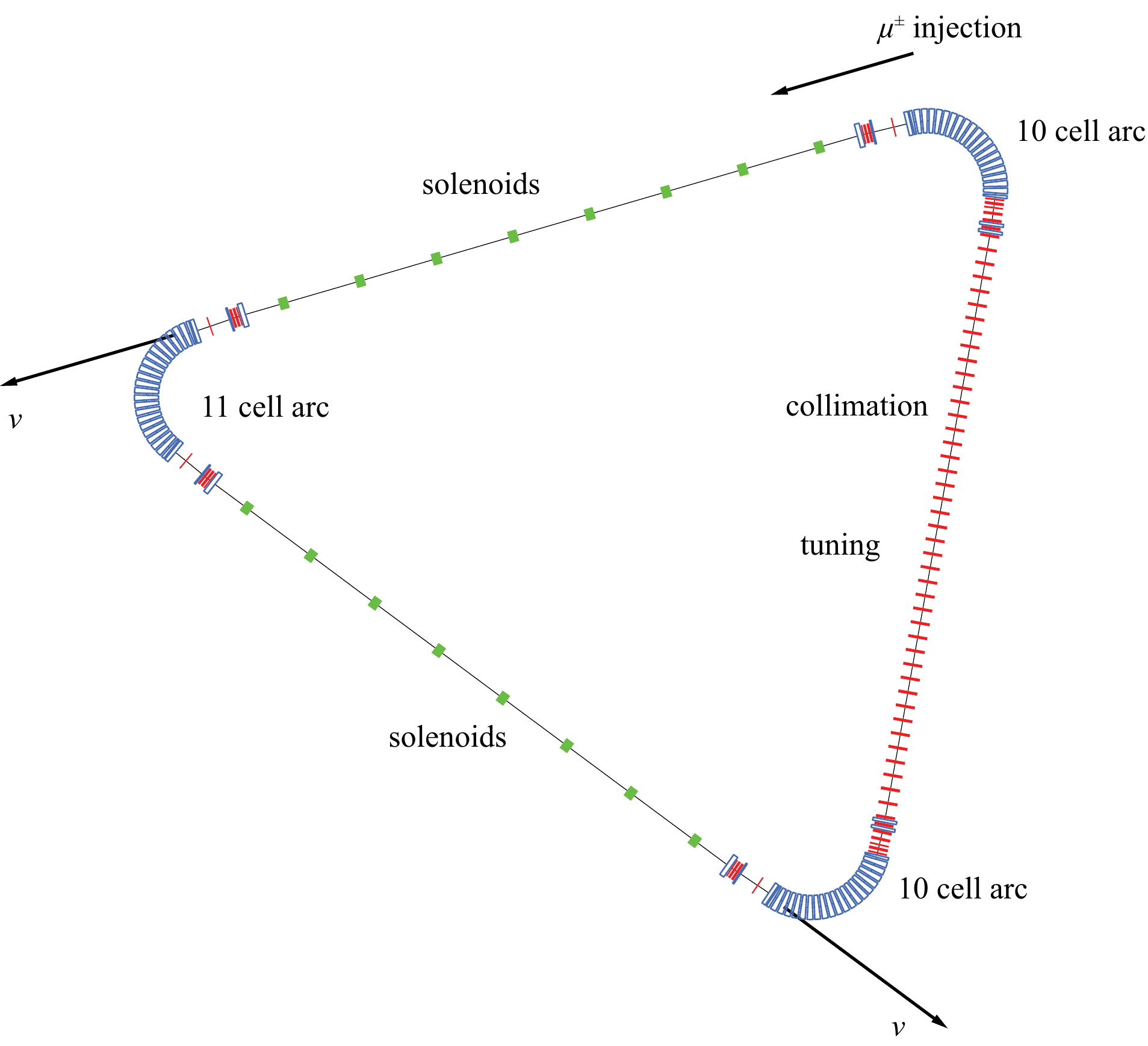}
  \caption{
     A triangular muon storage ring able to send neutrinos to
     detectors in two different directions.
  }
  \label{fig:acc:ring:Triangle}
\end{figure}

Two geometries of ring have been considered.
The geometry shown in figure \ref{fig:acc:ring:RaceTrack} is
designed to store either $\mu^+$ or $\mu^-$ with a single straight
pointing into the ground. 
The return straight is used for collimation, RF and tune control. 
A development of this concept can accommodate counter-rotating muons
of both signs.  
An alternative is a triangular lattice (figure
\ref{fig:acc:ring:Triangle}) with two production straights pointing in
different directions sending neutrinos to combination of detectors
dictated by the apex angle \cite{Prior:2009zz}.
Since the beam must travel in a unique direction,
two triangular rings would be built side by side in the same tunnel,
one containing $\mu^+$ and the other $\mu^-$.

Some parameters driving the choice of the muon storage-ring geometry
are the efficiency, defined as the ratio of the total length of
neutrino production straights to the total circumference, and the
depth of the tunnel, which implies geological and cost considerations.
There will be at most three $\mu^+$ and three $\mu^-$ bunch trains
in the ring.
Based on the proton driver described in section
\ref{Sect:ProtonDriver}, a decay ring of 1.6~km in circumference can
accommodate the equally-spaced, 250~ns long bunch trains and can allow
time intervals of at least 100~ns between the neutrino bursts arising
from the $\mu^+$ and the $\mu^-$ bunches.
The maximal tunnel depths for rings of this size are 444~m for the
racetrack and 493~m for the triangle. 

The optical properties of these rings are challenging. 
In order to keep the divergence of the neutrino beam sufficiently
small, the rms divergence of the muon beam
in the production straight section
must not exceed 10\% of the
natural angular divergence of $1/\gamma$ that arises from the
kinematics of muon decay.
This requirement translates into high values for the $\beta$-functions
in the production straights ($\sim 150$~m) which need to be matched into the low
values of $\sim 14$~m in the arcs. 
At the ends of each straight, small bending magnets are included to
ensure that neutrinos created in the matching section
(where the muon beam still has a large divergence angle)
cannot be seen by the detector.
The production straights for the racetrack design are 599.4~m long,
providing an efficiency of 37.2\% for single-sign muon beams.
The corresponding figures for the triangle lattice is
398.5~m (2 straights, 24.8\% efficiency each).
 
The racetrack design has been chosen as the baseline.
It is the most flexible possibility for directing neutrinos to any
combination of far detector locations, and can direct neutrinos
resulting from the decays of both muon charges to the same far
detector if desired.
In the following paragraphs we focus on the performances of the
race-track rings and on the issue of muon-beam monitoring. 
 
\subsubsection{Performance}

\begin{figure}
  \centering
    \includegraphics*[width=0.95\textwidth]%
      {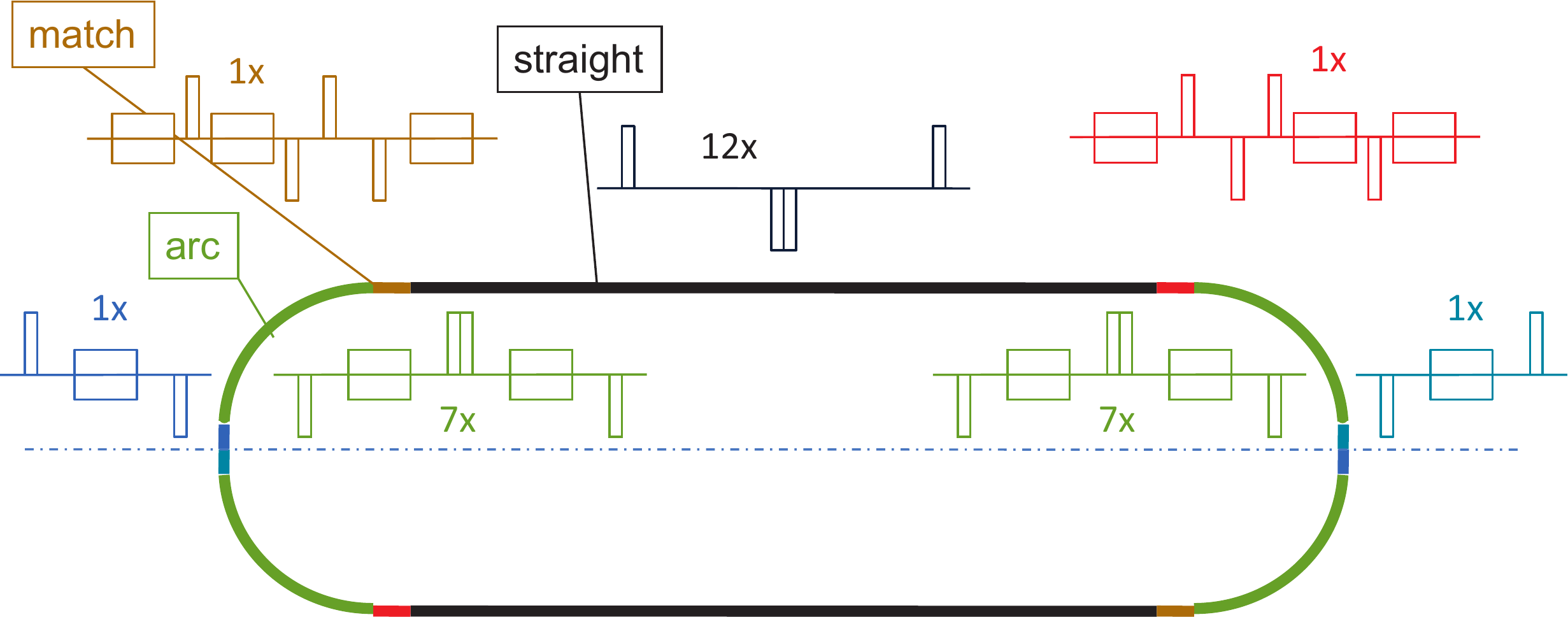}
  \caption{
    Schematic diagram of the racetrack decay ring illustrating its
    linear optics elements: (black) decay straights, (green/blue/cyan)
    arc sections, (red/brown) matching sections.
  }
  \label{fig:acc:ring:ring}
\end{figure}

\begin{figure}
  \begin{center}
    \includegraphics[angle=-90,width=0.48\linewidth]{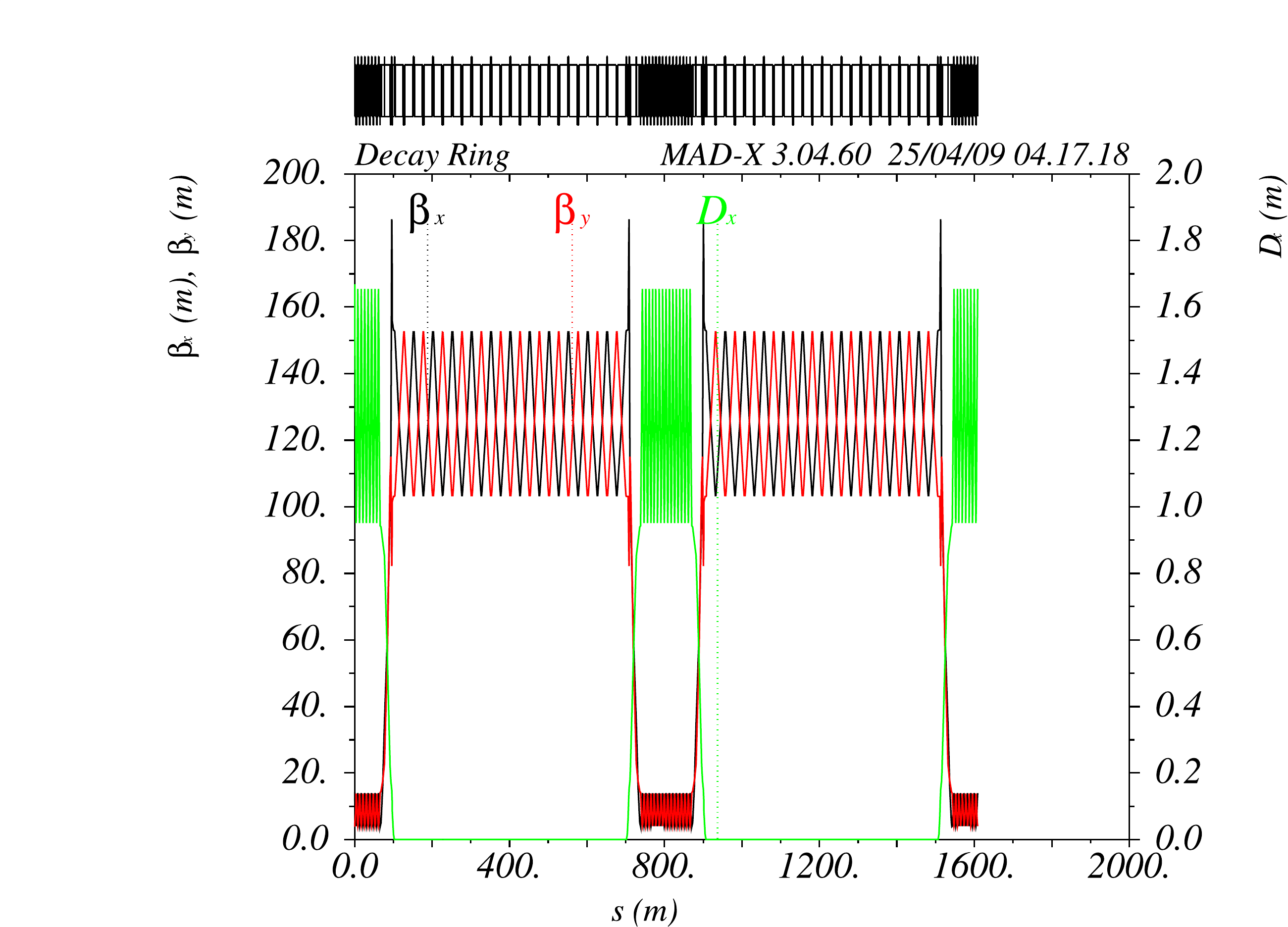}%
    \hspace{0.04\linewidth}%
    \includegraphics[angle=-90,width=0.48\linewidth]{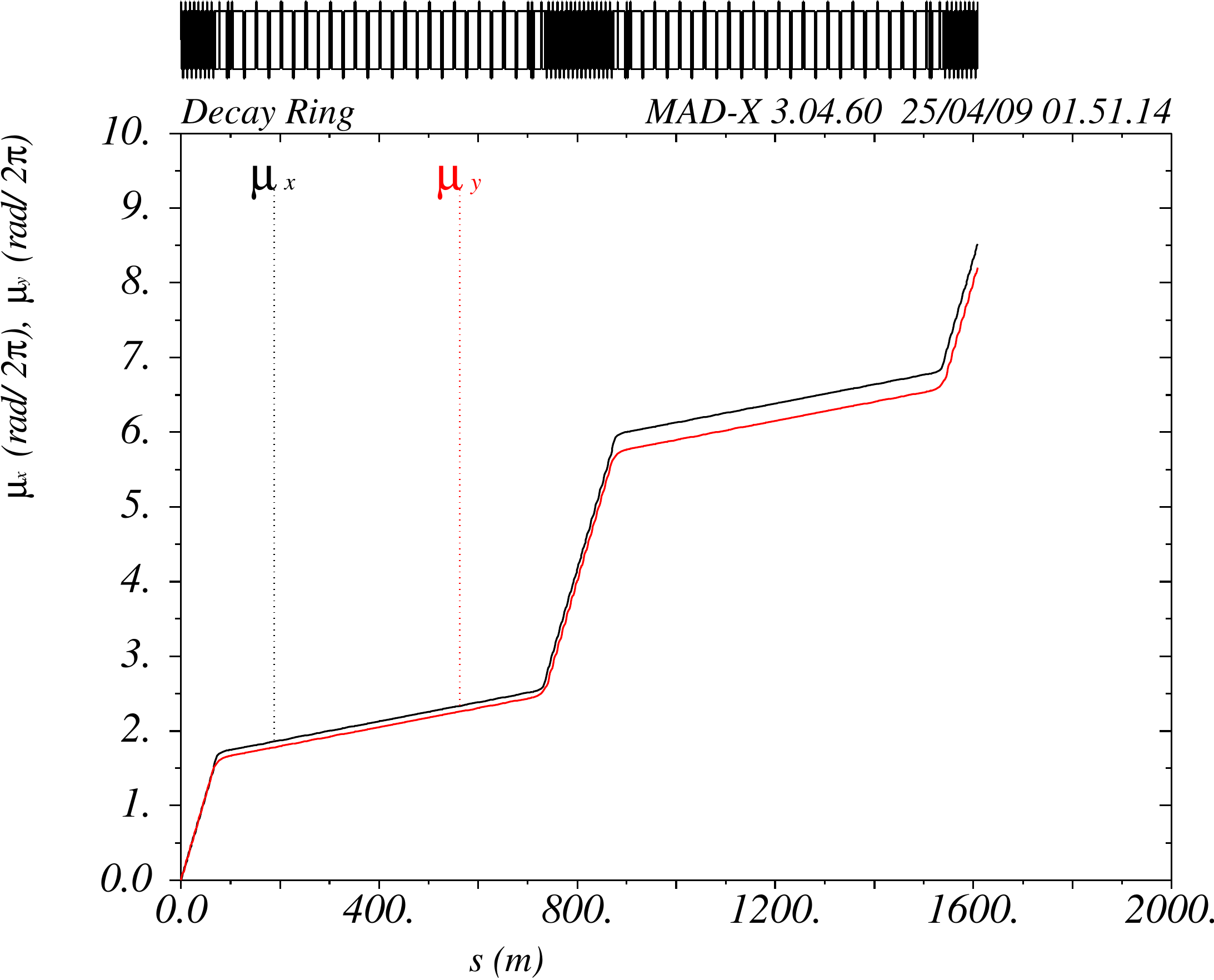}
  \end{center}
\caption{(Left): $\beta_x$ (black) and $\beta_y$ (red) functions for the decay rings. Their values 
  are deliberately chosen high in the straight sections to reduce the divergence of the beam. Dispersion in the arcs is also shown (green).
  (Right): phase advance over one period in the decay ring with a fast increase in the arcs due the small beta functions and a slow increase in the straights where $\beta_{x,y}$ are large.}
\label{fig:acc:ring:twiss}
\end{figure}

Starting from the initial design described in \cite{Prior:2009zz},
MAD-X~\cite{mad-x} has been used to determine the basic machine
properties: Twiss parameters, working point, and dynamic aperture
(DA).
The structure  of the ring is illustrated in figure~\ref{fig:acc:ring:ring}, 
which shows the three main elements of the
optics: straight sections; arcs; and matching sections. 
The central momentum is 25~GeV/c and the total length of this machine
is 1608.8~m. 
Figure \ref{fig:acc:ring:twiss} (left) shows the optics of the
decay-ring lattice, while the lattice, subdivided into its
three main parts (straights, arcs, and matching sections), is described
in tables \ref{tab:acc:ring:seq}--\ref{tab:acc:ring:arc}.
\begin{table*}
  \caption{
    Sequence of regions and their lengths for the decay ring.
  }
  \label{tab:acc:ring:seq}
  \centering
  \begin{tabular}{|l|r|}
    \hline
    {\bf Section} & {\bf Length} (m)  \\
    \hline
    1/2 straight  & 299.7             \\
    Matching cell &  36.5             \\
    Arc           & 132.0             \\
    Matching cell &  36.5             \\
    1/2 straight  & 299.7             \\
    1/2 straight  & 299.7             \\
    Matching cell &  36.5             \\
    Arc           & 132.0             \\
    Matching cell &  36.5             \\
    1/2 straight  & 299.7             \\
    \hline
    Total length  & 1608.8            \\
    \hline
  \end{tabular}
\end{table*}
\begin{table*}
  \caption{
    Lattice for straight cell.  
    There are 12 such cells in each of the two straights.
    The magnet type (superconducting, SC, or normal, NC) is given in
    the column headed ``Type''.
  }
  \label{tab:acc:ring:str}
  \centering
  \begin{tabular}{|l|r|l|l|}
    \hline
    {\bf Element}   & {\bf Length} (m)  & {\bf Field/Gradient} (T/Tm$^{-1}$) & {\bf Type} \\
    \hline
    QF              &    1.500          & 0.462                             & NC         \\
    Drift           &   21.975          &                                   &            \\
    QD              &    1.500          & $-0.462$                            & NC         \\
    QD              &    1.500          & $-0.462$                            & NC         \\
    Drift           &   21.975          &                                   &            \\
    QF              &    1.500          & 0.462                             & NC         \\
    \hline
    Cell length     &   49.950          &                                   &            \\
    Straight length &  599.400          &                                   &            \\
    \hline
    Total length    & 1198.800          &                                   &            \\
    \hline
  \end{tabular}
\end{table*}
\begin{table*}
  \caption{Lattice for matching cell.  There is one such cell in each of the
    four matching sections.
    The magnet type (superconducting, SC, or normal, NC) is given in
    the column headed ``Type''.
  }
  \label{tab:acc:ring:match}
  \centering
  \begin{tabular}{|l|r|l|l|}
    \hline
    {\bf Element} & {\bf Length} (m) & {\bf Field/Gradient} (T/Tm$^{-1}$) & {\bf Type} \\
    \hline
    Drift         &   0.41           &                                   &            \\
    Dipole        &   4.00           & $-0.64$                           & NC         \\
    Drift         &   0.41           &                                   &            \\
    QD            &   0.80           & $-9.17$                           & SC         \\
    Drift         &   0.50           &                                   &            \\
    QF            &   1.60           & 11.5                              & SC         \\
    Drift         &   0.45           &                                   &            \\
    QD            &   1.60           & $-7.62$                           & SC        \\
    Drift         &   0.50           &                                   &            \\
    Dipole        &   0.60           & $-1.9$                            & SC         \\
    Drift         &  14.25           &                                   &            \\
    QF            &   0.80           & 4.00                              & SC         \\
    Drift         &   7.28           &                                   &            \\
    Dipole        &   2.40           & 0.35                              & NC         \\
    Drift         &   0.90           &                                   &            \\
    \hline 
    Cell length   &  36.50           &                                   &            \\
    \hline
    Total length  & 146.00           &                                   &            \\
    \hline
  \end{tabular}
\end{table*}
\begin{table*}
  \caption{
    Lattice for arc cell.  
    There are 15 such cells in each of two arcs.
    The magnet type (superconducting, SC, or normal, NC) is given in
    the column headed ``Type''.
  }
  \label{tab:acc:ring:arc}
  \centering
  \begin{tabular}{|l|r|l|l|}
    \hline
    {\bf Element} & {\bf Length} (mm) & {\bf Field/Gradient} (T/Tm$^{-1}$) & {\bf Type} \\
    \hline
    QD            &   0.5 & $-23.64$ & SC\\
    Drift         &   0.7 & &\\
    Dipole        &   2.0 & $-4.27$ & SC\\
    Drift         &   0.7 & &\\
    QF            &   0.5 & 24.05 & SC\\
    QF            &   0.5 & 24.05 & SC\\
    Drift         &   0.7 & &\\
    Dipole        &   2.0 & $-4.27$ & SC\\
    Drift         &   0.7 & &\\
    QD            &   0.5 & $-23.64$ & SC\\
    \hline 
    Cell length   &   8.8 & &\\
    Arc length    & 132.0 & &\\
    \hline
    Total length &  264.0 & & \\
    \hline
  \end{tabular}
\end{table*}

Beta functions are kept low in the arcs to reduce the size of the beam and
maximise transmission. Conversely the beam envelope is increased in
the straight sections, where keeping a small divergence
is the primary requirement.
The merging of these two opposite criteria is accomplished by the
matching sections, which also eliminate the high dispersion in
the arcs. 

\begin{figure}
  \centering
    \includegraphics[width=0.95\textwidth]%
      {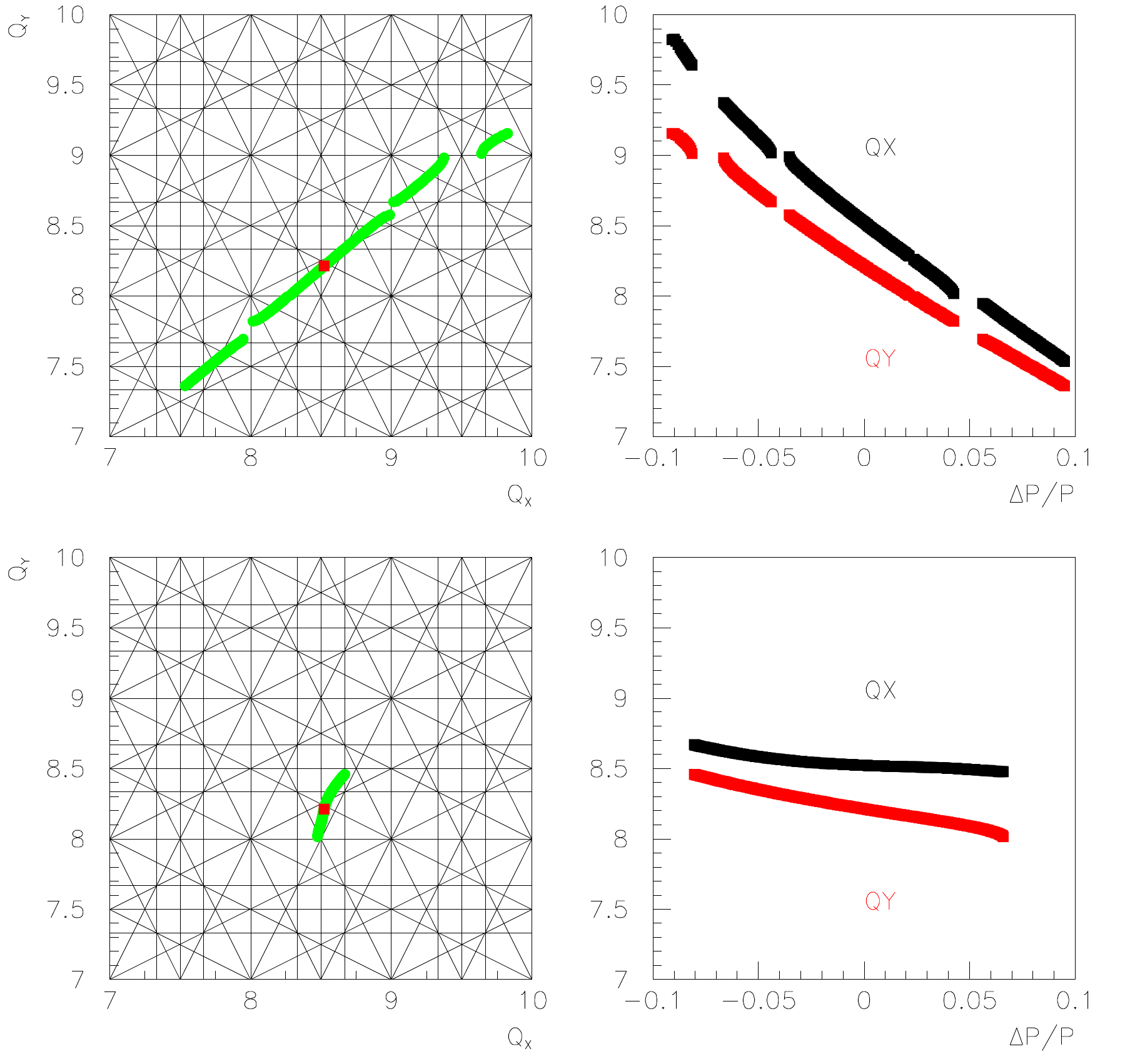}
  \caption{
    Upper section: linear optics configuration, showing the crossing
    of resonances (left) as a function of momentum spread (green
    line): the red dot shows the working point. 
    The natural chromaticity of the beam is shown on the right for
    both the transverse planes. 
    Lower section: the effect of sextupoles is shown. 
    Chromaticity is strongly reduced (right) and so is the crossing of
    potentially dangerous resonances.
  }
  \label{fig:acc:ring:reso}
\end{figure}

Figure \ref{fig:acc:ring:twiss} (right) illustrates the phase advance along
the ring: the working point is ($Q_x = 8.5229$, $Q_y = 8.2127$). 
Without sextupoles, a $\pm 10$\% change in momentum causes the working point to 
move substantially in the ($Q_x$, $Q_y$) plane, crossing resonances that could
be detrimental for the stability of the beam. To mitigate this effect,
sextupoles can be introduced in the dispersive arcs. 
Chromaticity correction could be motivated by the desire for large
momentum acceptance, the need for which is not yet fully known due to
concerns about 
longitudinal emittance growth in the FFAG, which is presently under
study~\cite{Berg:2009zz}. 
A visual summary of these results is shown in figure
\ref{fig:acc:ring:reso} where resonance diagrams and chromaticity
plots are displayed.  

\paragraph{Particle tracking and dynamic aperture}

\begin{figure}
  \centering
    \includegraphics[width=0.94\textwidth]%
    {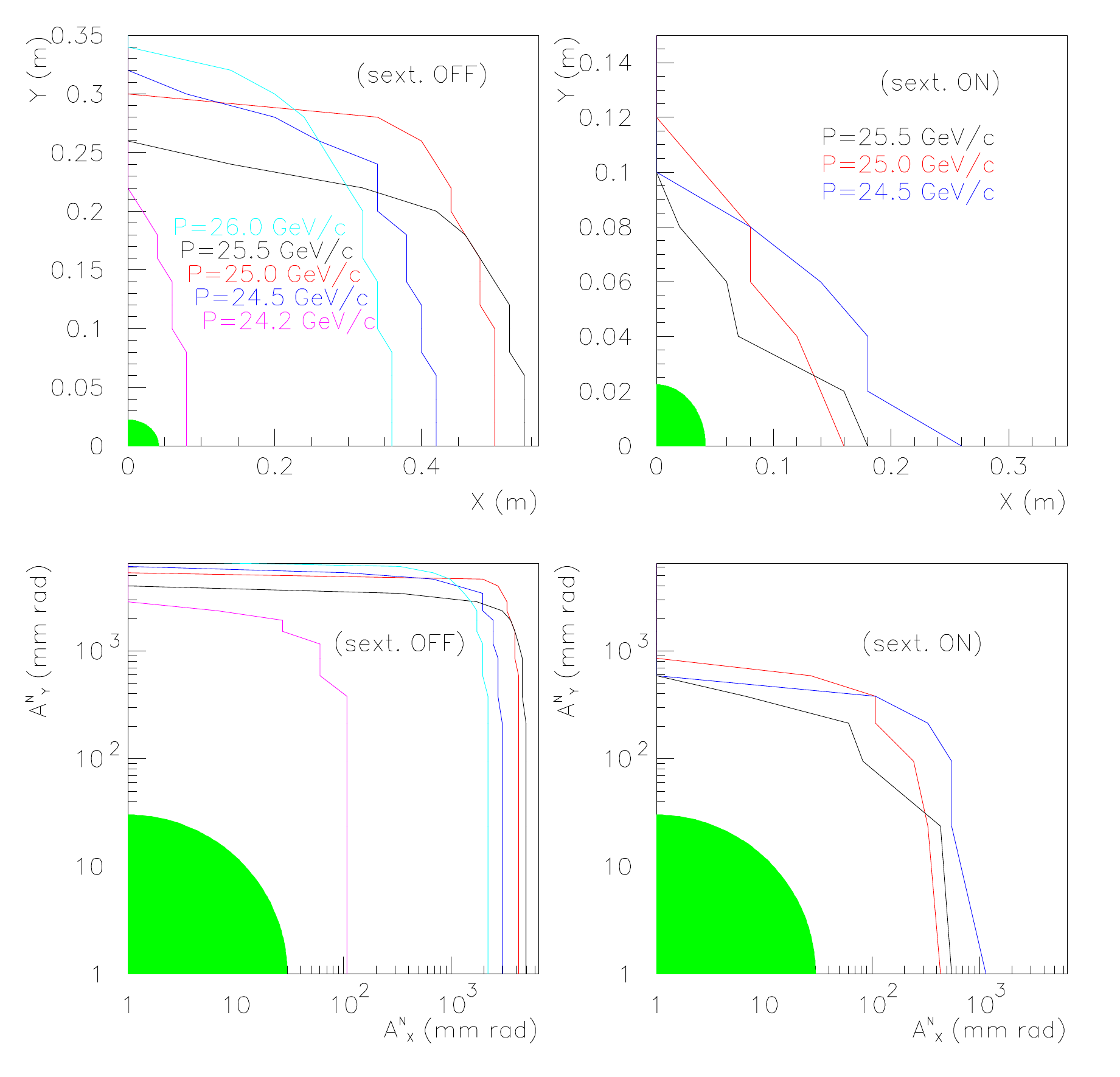}
  \caption{
    Dynamic aperture for the decay ring: (left column) linear optics,
    (right column) sextupoles introduced in the lattice. 
    (Top) contour plots defining the maximal radius for a particle
    injected at the midpoint of an arc before being lost in the
    lattice. 
    In off-momentum cases this radius is generally reduced. 
    The green area represents the beam size corresponding to the
    nominal acceptance (30 mm rad). 
    Points on the contour plots $(R_x,R_y)$ are used to define the
    amplitudes in the two transverse planes ($A^N_x$,$A^N_y$), where
    $(A^N_i=\frac{p}{m_{\mu}}R^2_i/\beta_{i})_{i=x,y}$ (bottom). 
    In this case the green area marks the 30 mm rad fiducial
    acceptance.
  }
  \label{fig:acc:ring:da}
\end{figure}

The introduction of non-linear elements along the ring is advocated to
mitigate resonance-crossing effects, potentially catastrophic for a
storage ring. 
However, two things should be stressed in the case of a
muon ring: (a) at 25~GeV/c the average muon-lifetime corresponds to
fewer than 100 turns in the 1608.8~m long ring, 
(b) the sextupoles introduced in the optics create coupling between
the transverse planes and reduce the dynamic aperture of the
beam. 
Concerning point (a), the question naturally arises whether it is
important to avoid resonance crossing, given the relatively short 
beam-lifetime required.
The second aspect has been investigated to produce a more quantitative
answer and a summary of the dynamic aperture  calculations is shown in
figure \ref{fig:acc:ring:da}.
In the absence of errors, the dynamic aperture comfortably
accommodates the nominal 30~mm~rad acceptance.
Studies of the dynamic aperture in the presence of errors will be
carried out.

\subsubsection{Decay ring instrumentation}

In order to ensure the precise determination of the neutrino flux and
momentum spectrum, continuous measurements of a number of the muon
beam parameters are required.
The critical parameters are: the muon energy (and polarisation), the
beam divergence, and the beam current. 
Efforts are presently focused on the measurement of the first two
parameters which are considered the more challenging.
The beam current can be measured with sufficient precision using
standard beam-current monitors.
The following sections outline the principles of operation of the
devices being considered to measure the beam energy (and polarisation)
and the beam divergence.

\paragraph{Muon beam energy determination using spin polarisation
           measurements} 
\label{Sect:Polar}

\begin{figure}
  \centering
  \includegraphics*[width=0.9\textwidth]%
    {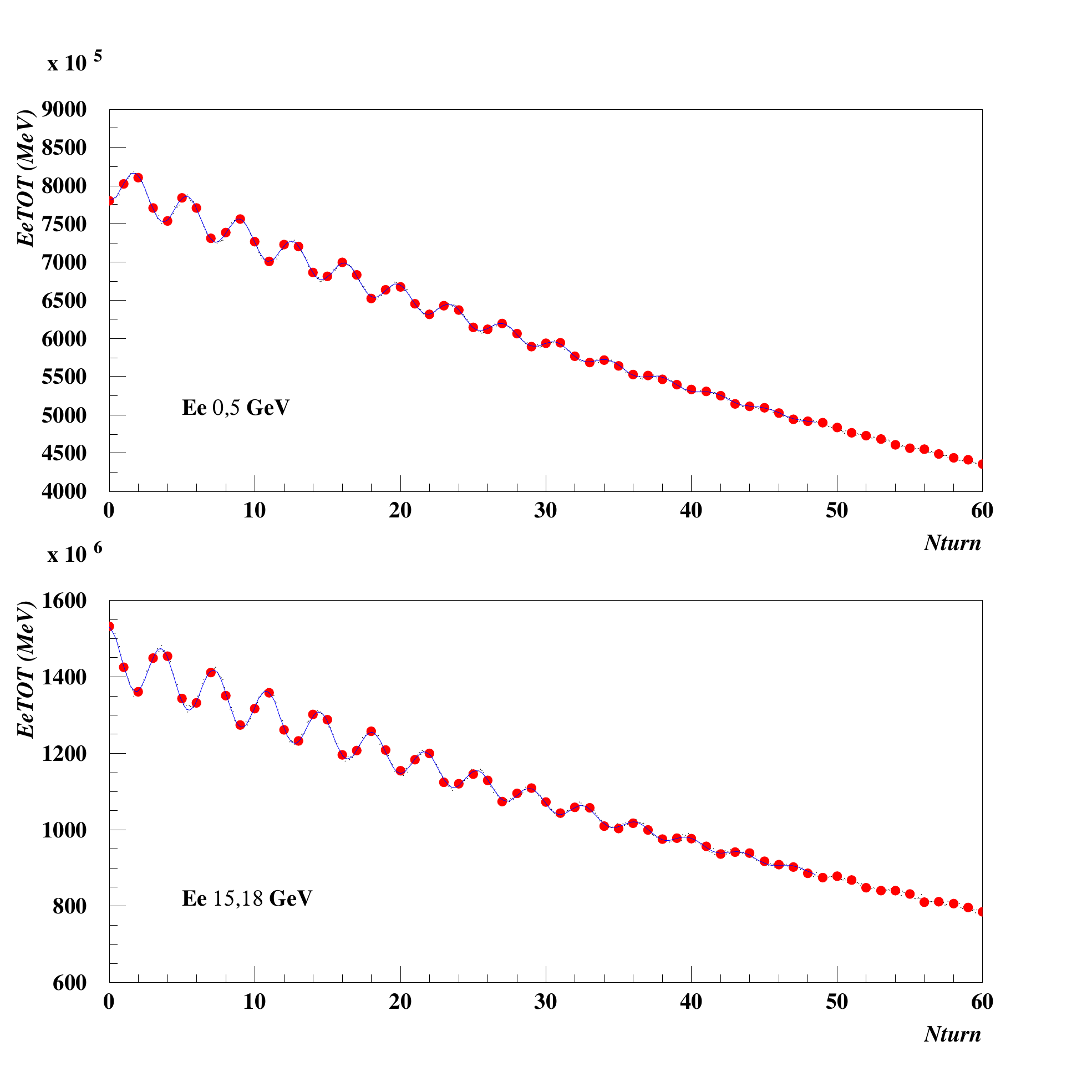}
  \caption{
     Total energy from electrons belonging to the [0,5]~GeV (top) and
     [15,18]~GeV (bottom) GeV intervals of the spectrum as a function
     of the number of turns. 
     The red dots are samples taken at every turn, the superimposed
     curve is a fit to the function defined in equation
     \ref{eq:acc:ring:fit}.
  }
\label{fig:acc:ring:Ee_vs_turn}
\end{figure}

Muons from pion decay are produced with a longitudinal polarisation
of $-100$\% in the decay rest frame. 
In the laboratory frame this number is expected to become $-18$\% for
pions between 200~MeV/c and 300~MeV/c \cite{Blondel:2000vz}.
This residual beam-polarisation can be used to determine the central
energy of the beam \cite{Raja:1997ep}. 
Electrons from muon decays can be collected at the end of a straight
section, exploiting the bending power of the dipoles (e.g., in the
matching sections or in the arcs) and channelling the spectral
components to counting devices \cite{Autin:1999ci}. 
Only part of the electron spectrum can be sampled (since electrons at high
energy travel nearly parallel to the muon beam), however 
a general expression for the expected signal can be
written as \cite{Autin:1999ci}:
\begin{equation}
 \mathcal{E}(t) = \mathcal{E}_0 e^{-\alpha t} \left[ 1 + \frac{\beta}{7} e^{-\frac{1}{2}
     \left( \omega \frac{\Delta E}{E} t \right)^2} \cdot \mathcal{P}
   cos(\phi+\omega t)  \right] \, ;
\label{eq:acc:ring:fit}
\end{equation}
where $\mathcal{E}_0$ is the total energy of the electrons at turn 0 and
the Gaussian term parametrises the beam energy spread. 
If it is assumed that all the decay electrons can be collected, then
$\mathcal{P}$ is the polarisation of the beam.
If this is not the case, the $\mathcal{P}$ is a free parameter that
can be related to the beam polarisation with further study. 
A fit to $\mathcal{E}(t)$ allows both the beam energy and the energy
spread to be determined. 
An example of such a fit is shown in figure
\ref{fig:acc:ring:Ee_vs_turn} for two energy bins, [0,5]~GeV and
[15,18]~GeV.
The precision with which $E_{\mu}$ and $\Delta E_{\mu}/E_\mu$ can be
measured is a function of the number of captured electrons and of
the number of sampled turns. 
With $\sim 3\times 10^5$ electrons per turn reaching the device and
50 turns sampled, we expect a statistical precision 0.2\% on the central
muon energy and 4\% on the energy spread for each machine fill. Since 
the physics requires only that the average energy distribution be known over
significant periods of running, with 50 fills of the ring per second,
the statistical uncertainties on the flux will be negligible.
The systematic uncertainties of this method are also expected to be small.
The central energy measurement affects the flux as $E^3$, but the
precision achievable will be sufficient to make this source of flux uncertainty
negligible. The measurement of the polarisation itself is more difficult,
because it relies on knowing the collection efficiency of the decay
electrons, which in turn requires detailed knowledge of the geometry
of both the ring and the polarimeter. In a ring geometry, however,
the polarisation precesses and thus its effect on the flux averages
to zero with sufficient precision~\cite{Blondel:2000vz}.
These preliminary numbers need a more realistic study of the
systematics of the measurement, but they give an idea of the
limits of the method.
The relation between the oscillation excursions in the total recorded
energy and the polarisation provides a measure of the latter.  

\begin{figure}
  \begin{center}
    \includegraphics[width=0.5\textwidth]{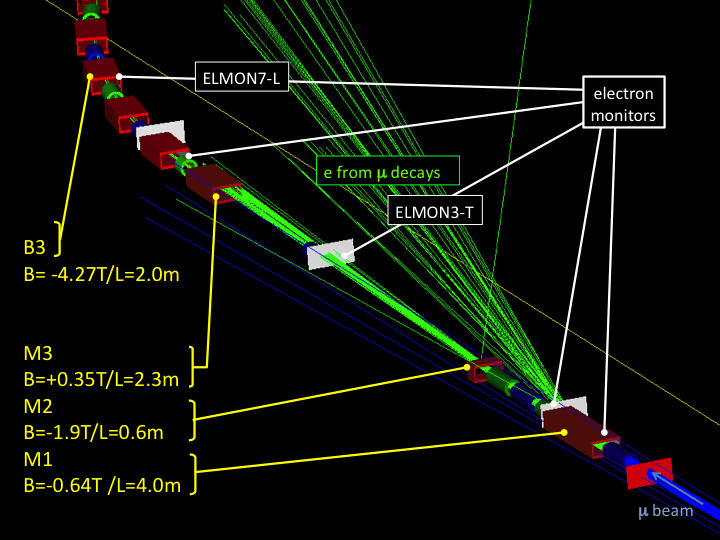}%
    \includegraphics[width=0.5\textwidth]{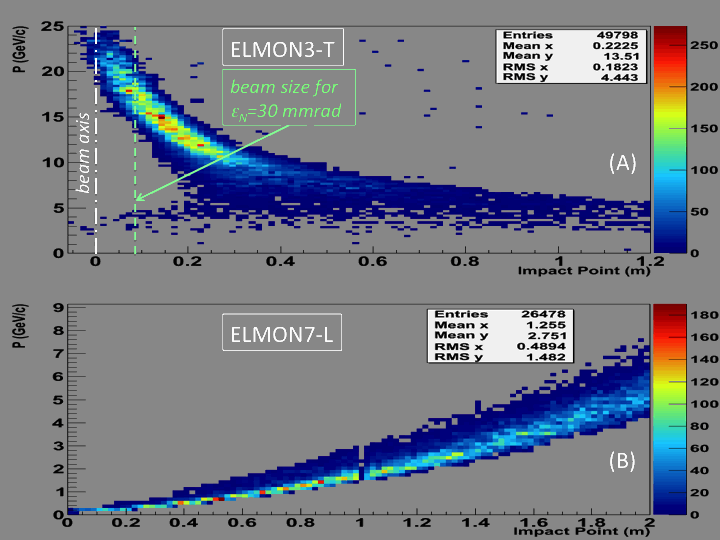}
  \end{center}
\caption{(Left): G4Beamline rendering of the region at the end of a straight section 
and beginning of the arc-section. In the middle the magnets of a
  matching section are visible. Locations for electron monitors are
  suggested. Blue tracks represent muons, green tracks the electrons from decays.
(Right) Momentum as a function of the impact point position for: (A)
  a monitor placed transversely to the beam 
  after a drift of 13 m past a dipole in the matching section. (B) a
  monitor parallel to the beam trajectory. In case (A) the
  position of the beam and its maximum width (for a normalised
  emittance of 30 mm rad) are shown.}
\label{fig:acc:ring:DR-monitors}
\end{figure}

Once the storage-ring lattice has been designed, a suitable location
for the electron calorimeter can be determined.
In the studies presented here, the electron calorimeter is taken to
refer to a ``generic'' device that will eventually be designed to
measure the electron energy.
Figure \ref{fig:acc:ring:DR-monitors} shows some of the possible
locations for electron monitors, together with the main features of the
dipole magnets used as spectrometers to separate electrons of
different energies. 
When choosing a particular location, one must consider both the spectral
power and the purity of the signal. Ideally, the device should collect
electrons produced in the proximity of the spectrometer in order to
avoid spurious effects from other magnets. 
With this in mind, we focused on two main cases:
\begin{itemize}
\item (A): a monitor placed downstream of the second bending magnet in
  the matching section, just before a quadrupole and orthogonal to the
  direction of the beam; and
\item (B): a monitor placed in the third bending element of the arc
  section, parallel to the beam direction.
\end{itemize}
In case (A) the device sits transversely to the nominal
beam orbit at a distance of about 13\,m from a dipole
($B=1.9\text{ T}$, $L_{\text{eff}}=0.6$ m) with the long drift
allowing a good spectral power 
despite the relatively low magnetic field. The advantage of this
location is that no special magnet is needed, which simplifies
the engineering. Figure~\ref{fig:acc:ring:DR-monitors} (Case A)
shows that at this
location a device placed at $>$10 cm from the beam axis can
intercept a fraction of the spectrum between 0~GeV and 18~GeV
without disrupting the muon beam. In this case we simulated muon
decays along the whole 13 m of drift to check the uniformity of the signal.
Case (B) is similar to the idea for the polarimeter proposed in
\cite{Autin:1999ci}. 
The advantage here is the good spectral power of the
bending magnet due to its effective length and high field 
($B=4.3$~T, $L_{\text{eff}}=2$~m). 
However, the requirement to modify the shape of the magnet,
introducing apertures through the superconducting element, would pose 
engineering challenges. 
The graph shown in figure~\ref{fig:acc:ring:DR-monitors} is a superposition of
decays that occur in the 2.4~m path between dipole B2 and dipole B3
(where the electron monitor is located). As in case (A), a good
signal uniformity is found.

\paragraph{Muon beam divergence}

The decay rings have been designed in order to keep a low rms
divergence in the straight sections. This should be of the order of
10\% of the natural divergence originating from muon decays into
neutrinos ($1/\gamma=4.0$ mrad). 
In order to measure this quantity, a device has
to be inserted into the beam.  The material budget of the device needs to be
small so as to reduce the effects of multiple scattering and energy
straggling. 
The original idea of a Cherenkov detector has been re-analysed in a
critical way, while a new concept based on optical transition
radiation is under study. 

\subparagraph*{Cherenkov Radiator}

This concept dates back to the first proposals for a Neutrino Factory
and consists of a tank with thin walls containing He at high pressure
($>10$~atm).  
Photons from Cherenkov radiation reach a mirror telescope located
at about 20 m from the source and bounce back to a collector with
sensors. The device is sensitive to the beam divergence, such that an
inclination of the beam translates into a change in the image
point. 
According to \cite{Piteira:2001} this device should be very precise
(0.04~mrad resolution), however a review of this proposal makes it
impractical for a number of reasons.
The geometry and size of the device is related to the Cherenkov
angle, which is a function of the helium-gas pressure. A pressure
of the order of 15~atm is needed in order to fulfil the geometrical constraints
of the telescope. 
This implies the use of a thick transparent window to contain the 
gas and to let the light travel towards the optical system. 
A thickness of 200~$\mu$m, which may be optimistically thin, causes
the initial divergence of the beam of $0.4$~mrad to grow to
0.46~mrad after 200 turns.  
The Cherenkov telescope is not particularly compact, due to the
requirement to exploit a long lever arm.
It should be noted that any layer of material placed across the beam
will heat up due to energy loss, reaching more than 100~$^\circ$C in 
$\sim 10$ seconds.
A heat-dissipation system is therefore needed to maintain an
appropriate working temperature. 

\subparagraph*{Optical transition radiation device}

In order to overcome some of the aforementioned issues, alternative
solutions have been examined that rely on the principle of optical transition
radiation (OTR)~\cite{Qiu:1994}.
When a charged particle crosses the boundary between vacuum and
conductor (dielectric), electromagnetic radiation is produced both in
the forward and backward directions (generated by the image charge
within the conductor). 
A particular configuration is obtained when the 
conductor forms an angle of 45$^{\circ}$ with respect to the incident beam. In this
case the ``backward'' radiation is orthogonal to the incident beam and
can be collected out of the beam pipe. The shape of this radiation
shows a typical two-lobed distribution with an opening angle of
$1/\gamma$. 
Depending on the optics chosen, one can infer the beam
divergence in two ways:
a) by collecting the OTR pattern in the focal plane with one foil; or
b) by reconstructing the beam in the image plane with three stations
and then infer the divergence of the beam by means of the usual
transport-matrix techniques.
In case (a) the OTR pattern definition is a function of the beam
divergence, so the divergence can be found by studying the ratio
between minimum and maximum intensities of the OTR figure. 
From the literature~\cite{Qiu:1995xz,Fiorito:1993zu},
we learn that in our case a resolution of 0.5~mrad
should be achievable. This is not ideal, though further studies
should be done in order to understand the real potential of this
technique with a realistic muon beam, an optical system and the best
sensors available today.
Case (b) relies on the reconstruction of beam size at three different
points. The spatial resolution is dominated by the optical system and
the sensors used. At CEBAF a beam spot of 100 $\mu$m can be resolved
\cite{Denard:1997zz}, which makes this method interesting for our
case, even if the uncertainty of the divergence measurement still has
to be evaluated.
The proposed OTR devices all rely on very thin (order of 50~$\mu$m)
metal foils, which makes the technique interesting as it has the
potential to offer a low material budget and has the capability of
dissipating heat.  

In conclusion, we identified three possible ways of measuring
the divergence, one based on Cher\-en\-kov radiation and the other two
relying on OTR techniques.  
The first one seems impractical, while the OTR techniques necessitate
further studies to understand the level of precision reachable in our case.

\clearpage
%
\section{Neutrino Detectors for the Neutrino Factory}
\label{Sect:DetWG}

\subsection{Introduction}

\subsubsection{Baseline description for the far detectors}

The IDS-NF baseline for the Neutrino Factory has been optimised as 
described in section \ref{Sect:PPEG}.
The optimum strategy to measure $\delta$, $\theta_{31}$, and the mass
hierarchy (the sign of $\Delta m^2_{13}$) includes having two
Magnetised Iron Neutrino Detectors (MIND), one with a fiducial mass of
100~kTon at $\sim 4\,000$~km and another with a fiducial mass 50\,kTon
at $\sim 7\,500$\,km.   
The latter has been termed the ``magic baseline'', since matter
effects cancel the effect of CP violation at this distance.
The detector is optimised to carry out the detection of the ``golden
channel'' ($\nu_e\rightarrow \nu_\mu$) through the wrong-sign muon
signature. 
This strategy is more efficient for resolving degeneracies in the
neutrino-oscillation formul\ae and provides better sensitivity than,
for example, measuring the golden and the ``silver'' channel
($\nu_e\rightarrow \nu_\tau$) simultaneously.
   
The original golden channel at a Neutrino Factory analysis
\cite{Cervera:2000kp} assumed a cylindrical geometry with a
cross-sectional radius of 10~m, with iron plates 6~cm thick,
scintillator planes 2~cm thick and a 1~T solenoidal field operating at
a 50~GeV Neutrino Factory. 
The International Scoping Study (ISS)
\cite{CerveraVillanueva:2008zz,Abe:2007bi} assumed a cuboidal geometry
of $14\times 14$~m$^2$ with 4~cm thick iron and 1~cm thick
scintillator and a 1~T dipole field, while operating at a 25 GeV
Neutrino Factory. 

For the most recent studies we have adopted a baseline cuboidal geometry with a cross-sectional area of $15\times15$~m$^2$ and length of either 63~m or 125~m, depending on the mass of the detector. The thickness of each plane of iron is 3~cm, followed by two planes of scintillator, each with 1~cm thickness. The three planes form a module of thickness 5~cm. The lateral resolution requirement is 1~cm, which is provided by having co-extruded scintillator bars 15~m long and 3.5~cm wide, read out using optical fibres and silicon photo-multipliers (SiPMT). The magnetic field assumed for this baseline is a 1~T dipole field. Table \ref{tb:MIND_par} lists the key parameters of the two far detectors. The purpose of studying this geometry was to  perform a direct comparison with previous results and to determine the influence of a full reconstruction in conjunction with a modern likelihood analysis to eliminate background.

\begin{figure}
  \begin{center}
    \includegraphics[width=0.9\textwidth]%
      {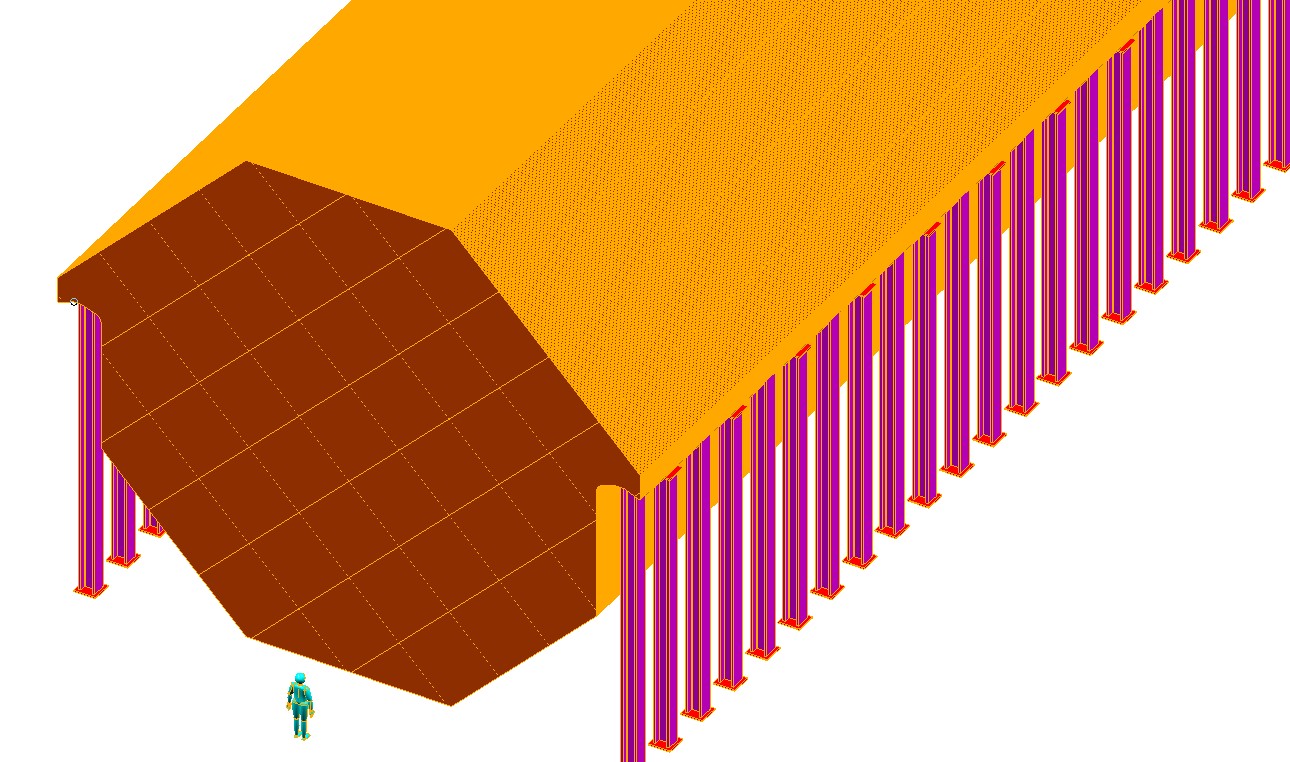}
  \end{center}
  \caption{
    Schematic drawing of the Magnetised Iron Neutrino Detector (MIND). 
  }
  \label{fig:MIND}
\end{figure}

While we adopted this simplifying geometry and magnetic field to
compare with previous simulations, we are aware that a 1\,T dipole
field is not practical from an engineering point of view. 
However, as described in section~\ref{sec:MIND_conceptual_design} we
have now studied a more realistic octagonal geometry (14~m octagonal
iron plates as shown in figure \ref{fig:MIND}), with a toroidal field
between 1\,T and 2.2\,T over the whole fiducial area.
These parameters can be achieved with a 100~kA/turn current traversing
the centre of the MIND plates and are shown to be feasible to
manufacture. 
This more realistic geometry has not been simulated yet but we do not
expect that the performance will differ greatly from what has been
achieved, and will be described in section~\ref{sec:MIND}.  
\begin{table*}
  \caption{
    Baseline parameters for the Magnetised Iron Neutrino Detectors
    at a Neutrino Factory. MIND~1 refers to the detector at $\sim
    4\,000$~km and MIND~2 refers to the detector at $\sim 7\,500$~km. 
  }
  \begin{center}
    \begin{tabular}{|l|r|r|}
      \hline
      {\bf Parameter}                         & {\bf MIND 1} & {\bf MIND 2}\\
      \hline
      Distance (km)                           & 3000-5000 & 7000-8000     \\
      Fiducial mass (kT)                  & 100  & 50   \\
      Size iron plates (cm$^3$)               & 1500$\times$1500$\times$3 & 1500$\times$1500$\times$3 \\ 
      Length detector (m)                     & 125  & 62.5 \\
      Number iron plates               & 2500  & 1250  \\
      Dimensions scintillator bars (cm$^3$)        & 1500$\times$3.5$\times$1  & 1500$\times$3.5$\times$1   \\
      Number scintillator bars per plane      & 429 &  429 \\
      Total number of scintillator bars      & $2.14\times 10^6$ &  $1.07\times 10^6$ \\
      Total number of readout channels  & $4.28\times 10^6$ &  $2.14\times 10^6$   \\
      Photon detector & SiPMT & SiPMT \\
      Magnetic field (T)   & $>$~1 & $>$~1  \\
      \hline
    \end{tabular}
  \end{center}
  \label{tb:MIND_par}
\end{table*}

\subsubsection{Baseline description for the near detectors}
\label{sec:Baseline_near}

The baseline for the Neutrino Factory includes one or more near
detectors.
It is necessary to have one near detector for each of the straight
sections of the storage ring at each of the two polarities, so four
near detectors designed to carry out measurements essential to the
oscillation-physics programme are required.
The near-detector measurements that are essential for the neutrino
oscillation analysis are:
\begin{itemize}
  \item Determination of the neutrino flux through the measurement
        of neutrino-electron scattering;
  \item Measurement of the neutrino-beam properties that are
        required for the flux to be extrapolated with accuracy to the
        far detectors;
  \item Measurement of the charm production cross sections (charm
        production in far detectors is one of the principal
        backgrounds to the oscillation signal); and
  \item Measurement of the neutrino-nucleon deep inelastic,
        quasi-elastic, and resonant-scattering cross sections.
\end{itemize}

The intense neutrino beam delivered by the Neutrino Factory makes it
possible to carry out a unique neutrino-physics programme at the near
detectors. 
This programme includes fundamental electroweak and QCD physics, such
as measurements of parton distribution functions as a function of
$Q^2$ and Bjorken $x$, QCD sum rules, nuclear re-interaction effects,
strange particle production, a precise measurement of $\sin^2\theta_W$.
The near detector must also be capable of searching for new physics,
for example by detecting tau-leptons which are particularly sensitive
probes of non-standard interactions at source and at detection.
Tau neutrino detection is also important in the search for sterile
neutrinos.

Here we itemise the general design features of the near detector:
\begin{itemize} 
\item A detector with micron-scale resolution for charm and tau
  identification (either a silicon vertex or an emulsion-based
  detector); 
\item A low-$Z$, high-resolution target for flux and
  $\nu_\mu$- and $\nu_e$-leptonic cross-section measurements (i.e., a
  scintillating-fibre tracker or a straw-tube tracker);  
\item A magnetic field for charged particle momentum measurement (with
  $\delta p/p \sim 1\%$ (for $p \sim 2-3$\,GeV);
\item A muon catcher for muon identification;
\item Electron identification capabilities;
\item Excellent energy resolution for flux extrapolation: this needs
  to be better than for the far detector, so the goal is to achieve
  $\delta E/E \sim 1\%$; and
\item A variety of nuclear targets to measure cross-sections in iron and
  as a function of the nuclear target mass number $A$. 
\end{itemize}
There are two options currently being considered: one which includes a
high resolution scintillating fibre tracker and the other includes a
transition-radiation straw-tube tracker. 
Both of these options will be studied to determine their capabilities.

\subsection{Far Detectors}

\subsubsection{Magnetised Iron Neutrino Detector Performance}
\label{sec:MIND}

\paragraph{Introduction}
\label{par:MindIntro}

Early papers on the physics outcomes of a Neutrino Factory
concentrated on the sub-dominant $\nu_e \rightarrow \nu_\mu$
oscillation \cite{DeRujula:1998hd} in which a muon of opposite charge
to that stored in the storage ring (wrong-sign muon) would be produced
in a far detector by the charge current (CC) interactions of the
oscillated $\nu_\mu$.
The first analysis of the capabilities of a large Magnetised Iron
Neutrino Detector to detect the wrong-sign muon signature (the golden
channel) was discussed in~\cite{Cervera:2000kp}, where it was
demonstrated that this combination was capable of extracting the
remaining unknown parameters in the three-by-three neutrino mixing
matrix.
This analysis was carried out assuming a Neutrino Factory with 50~GeV
muons, and was optimised for high energy using a detector with 4~cm
thick iron plates and 1~cm scintillator planes. 
Hence, the ability to reconstruct low energy muons was not part of the
optimisation.

The International Scoping Study for a future Neutrino Factory and super-beam facility (the ISS) \cite{Bandyopadhyay:2007kx,Apollonio:2009,CerveraVillanueva:2008zz,Abe:2007bi} adopted a muon energy of 25~GeV, so the analysis was re-optimised for low-energy neutrino interactions. This study focused on the topology and kinematics of neutrino events in the detector, assuming perfect pattern recognition and smearing of the kinematic variables of the scattered muon and hadronic shower. Using a combination of cuts on the relative length of the two longest particles in the event and the momentum and isolation of this candidate, it was demonstrated that high signal-identification efficiency and background suppression could be achieved. 

The work that has been carried out since the ISS, and is included in this Interim Design Report, has focused on testing these assumptions, by performing a full simulation and reconstruction of the candidate neutrino events, allowing a full characterisation of the detector response. Incorrect charge assignment (charge mis-identification) of non-oscillated $\overline{\nu}_\mu$ CC interactions, background from meson decays in the hadronic shower, and misidentification of neutral current (NC) and $\nu_e$ CC events present the most significant background for the Neutrino Factory beam.

The present study first focused on demonstrating the performance of reconstruction algorithms using deep inelastic scattering (DIS) neutrino events generated using LEPTO61~\cite{Ingelman:1997Cp} and simulated using GEANT3~\cite{geant3wu}, as in the study performed for the ISS. Pattern recognition and reconstruction of candidate muons were carried out using a Kalman filter (RecPack \cite{CerveraVillanueva:2004kt}) and a Cellular Automaton algorithm, particularly useful in extracting low energy neutrino events \cite{Emeliyanov_otr/itr-cats:tracking}. Preliminary results were shown in~\cite{Cervera:2008nf} and were further developed in \cite{Cervera:2010rz}. 

In conjunction with the development of reconstruction algorithms for MIND, an entirely new simulation was developed based on the NUANCE event generator \cite{Casper:2002sd} and GEANT4~\cite{Apostolakis:2007zz}. This simulation, named G4MIND, will ultimately allow for a full optimisation of MIND in terms of segmentation, technology options and analysis algorithms. The work undertaken to develop the simulation and digitisation as well as the application and re-optimisation of the reconstruction algorithm introduced in \cite{Cervera:2010rz} is described in this report. These results were used to determine the response matrices for this detector system (presented in Appendix~\ref{app:response}), which has been used to deduce the expected sensitivity of the combination of the MIND and a 25~GeV Neutrino Factory to key oscillation parameters.

\begin{figure}
  \begin{center}
    \includegraphics[width=0.56\textwidth]{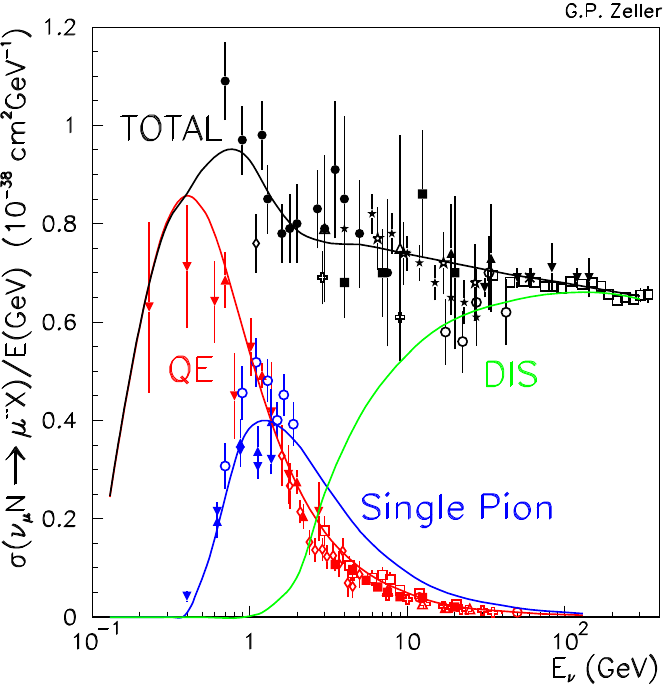}
  \end{center}
  \caption{Experimentally measured cross-sections for quasi-elastic, single pion and deep inelastic $\nu_\mu$ CC interactions with the predictions made by NUANCE~\cite{Zeller:2008zz}.}
  \label{fig:intmuCC}
\end{figure}

\paragraph{Simulation using NUANCE and GEANT4}
\label{sec:evGen}
In previous studies, only deep inelastic scattering (DIS) events
generated with LEPTO61 were considered. However, at energies below
5~GeV there are large contributions from quasi-elastic (QE), single
pion production (1$\pi$) and other resonant production (DIS) events
(see figure \ref{fig:intmuCC}). 
A QE event may be accurately described by considering the interaction
of the neutrino to take place with a nucleon as a whole
($\nu_\mu+n\rightarrow\mu^-+p$). 
The energy of such an event may readily be reconstructed from the
momentum of the muon and the angle it makes with the beam direction. 
In addition, the low multiplicity of the event makes muon
reconstruction simpler. 
1$\pi$ events should also improve the purity of the low-energy event
sample due to their low multiplicity, although there may be increased
background from pion decay. 
Other nuclear resonant events, producing 2 or 3 pions, as well as
diffractive and coherent production, have much smaller contributions.

\begin{figure}
  \begin{center}$
    \begin{array}{cc}
      \includegraphics[width=0.45\textwidth]{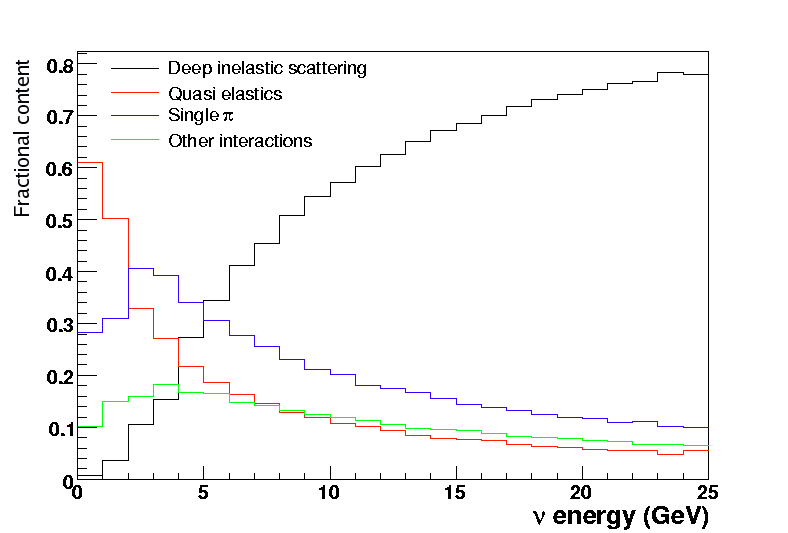} &
      \includegraphics[width=0.45\textwidth]{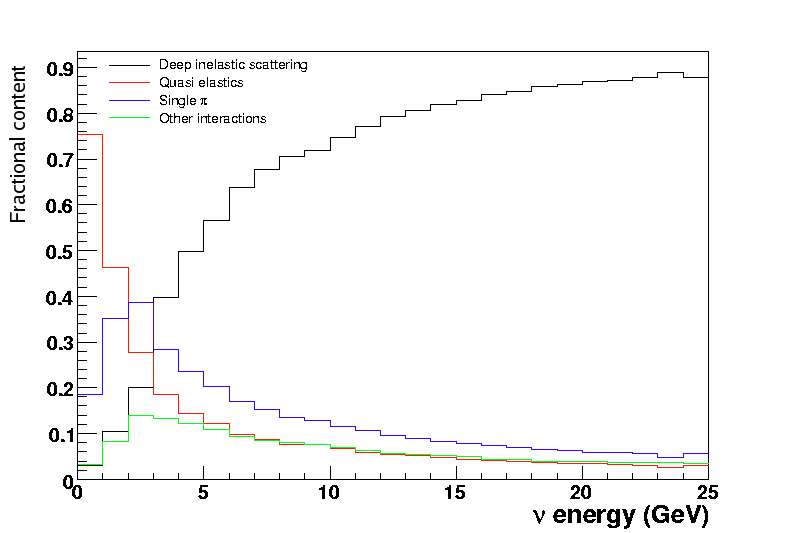}\\
      \includegraphics[width=0.45\textwidth]{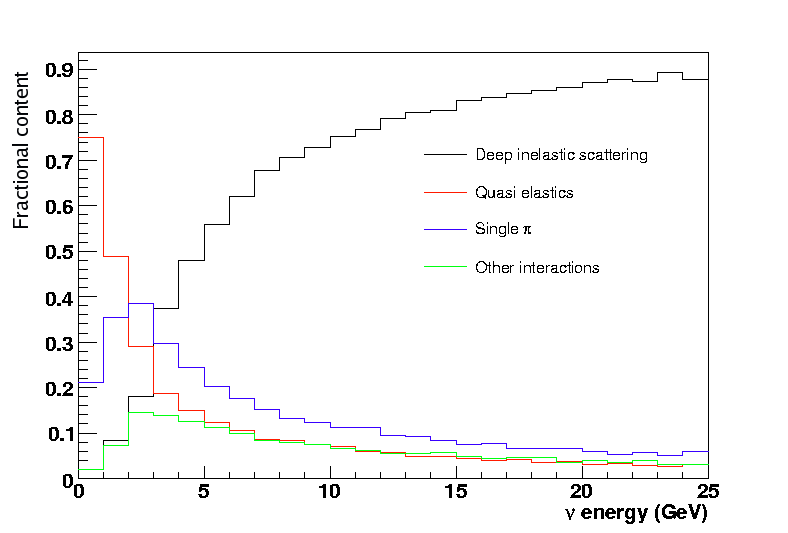} &
      \includegraphics[width=0.45\textwidth]{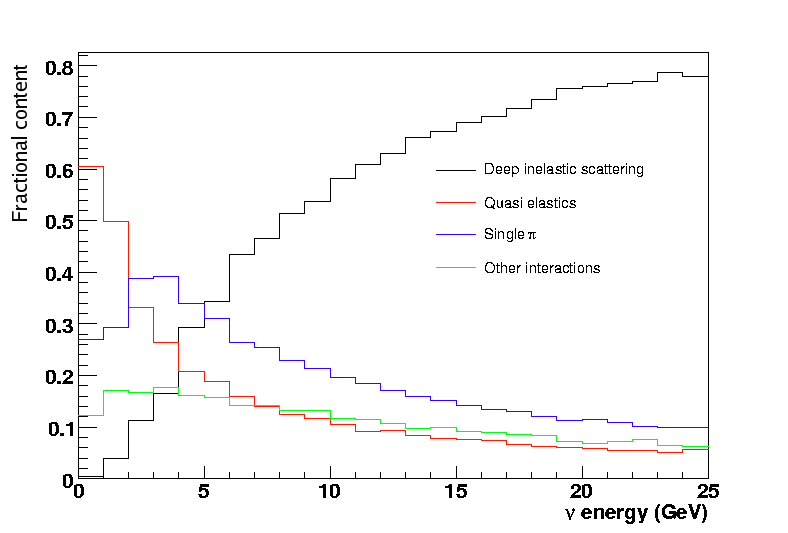}\\
      \includegraphics[width=0.45\textwidth]{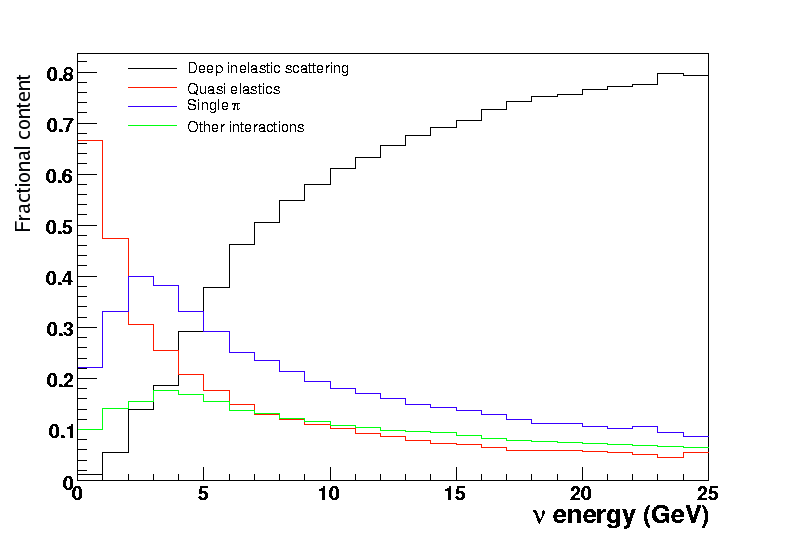} &
      \includegraphics[width=0.45\textwidth]{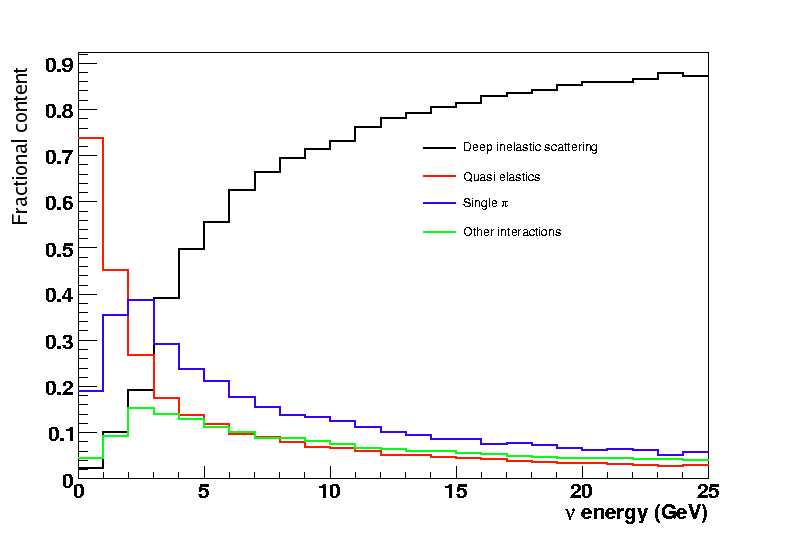}
    \end{array}$
  \end{center}
  \caption{Proportion of total number of interactions of different $\nu$ interaction processes for events generated using NUANCE and passed to the G4MIND simulation. (top) $\overline{\nu}_\mu$ (left) and $\nu_e$ CC (right), (middle) $\nu_\mu$ (left) and $\overline{\nu}_e$ CC (right) and (bottom) NC from anti-neutrinos (left) and neutrinos (right).}
\label{fig:intprops}
\end{figure}

Generation of all types of interaction was performed using the NUANCE
framework~\cite{Casper:2002sd}. 
The relative proportions of the interaction types generated by NUANCE
are shown in figure \ref{fig:intprops} where `other' interactions
include the resonant, coherent and diffractive processes.
NUANCE also includes a treatment to simulate the effect of
re-interaction within the participant nucleon, which is particularly
important for low energy interactions in high-$Z$ targets such as
iron. 

From the flux spectrum used to generate the events and the resultant interaction spectrum, the cross section as a function of energy can be estimated by normalising the highest energy bin to the high-energy limit: $0.67\times 10^{-38}$~cm$^2$~GeV$^{-1}$ for $\nu_\mu$ and $0.34\times 10^{-38}$~cm$^2$~GeV$^{-1}$ for $\overline{\nu}_\mu$~\cite{Amsler20081}. 

A new simulation of MIND using the GEANT4 toolkit G4MIND was developed to give as much flexibility to the geometry as possible so that an optimisation of all aspects of the detector could be carried out. The dimensions and spacing of all scintillator and iron pieces, as well as all external dimensions of the detector, can be controlled. This will allow for easy comparison of the simulation itself and the subsequent analyses to other potential Neutrino Factory detectors.

The detector transverse dimensions ($x$ and $y$ axes) and length in the beam direction ($z$ axis), transverse to the detector face, are controlled from a parameter file. A fiducial cross section of 14~m$\times$14~m, including 3~cm of iron for every 2~cm of polystyrene extruded plastic scintillator (1~cm of scintillator per view), was assumed. A constant magnetic field of 1~T is oriented in the positive $y$ direction throughout the detector volume. Events generated using NUANCE in iron nuclei and scintillator (polystyrene) are selected in the relative proportion of the two materials. An event vertex is then generated within one of the slabs of the appropriate material randomly positioned in three dimensions. Physics processes are modelled using the QGSP\_BERT physics lists provided by GEANT4 \cite{geant4phys}. 

Secondary particles are required to travel at least 30~mm from their production point or to cross a material boundary between the detector sub-volumes to have their trajectory fully tracked. Generally, particles are only tracked down to a kinetic energy of 100~MeV. However, gammas and muons are excluded from this cut. The end-point of a muon track is important for muon pattern recognition. 

A simplified digitisation model was considered for this simulation. Two-dimensional boxes -- termed voxels -- represent view-matched $x$ and $y$ readout positions. Any deposit which falls within a voxel has its energy deposit added to the voxel total raw energy deposit. The thickness of two centimetres of scintillator per plane assumes 1~cm per view. 

\begin{figure}
  \begin{center}$
    \begin{array}{cc}
      \includegraphics[width=0.45\textwidth]{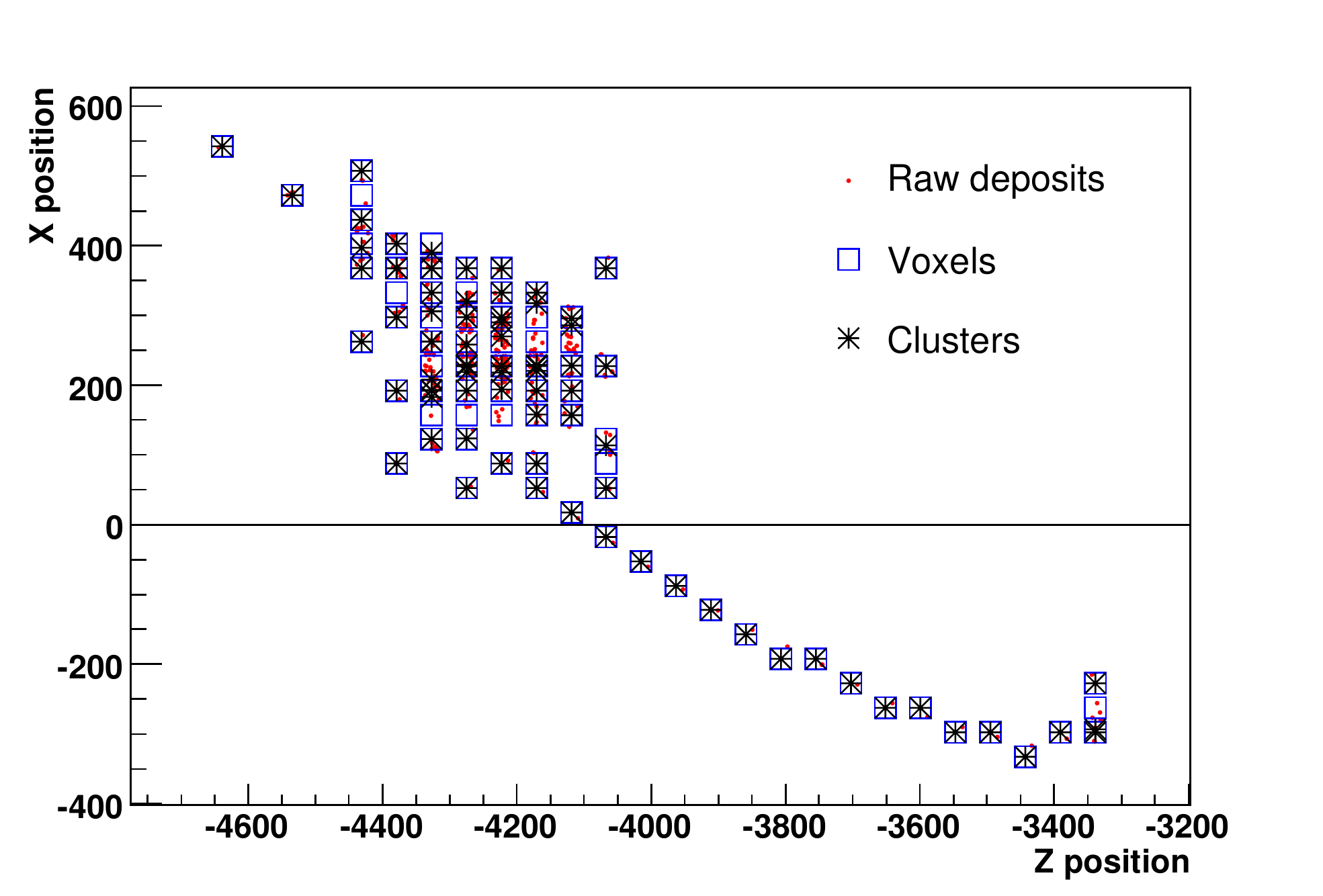} &
      \includegraphics[width=0.45\textwidth]{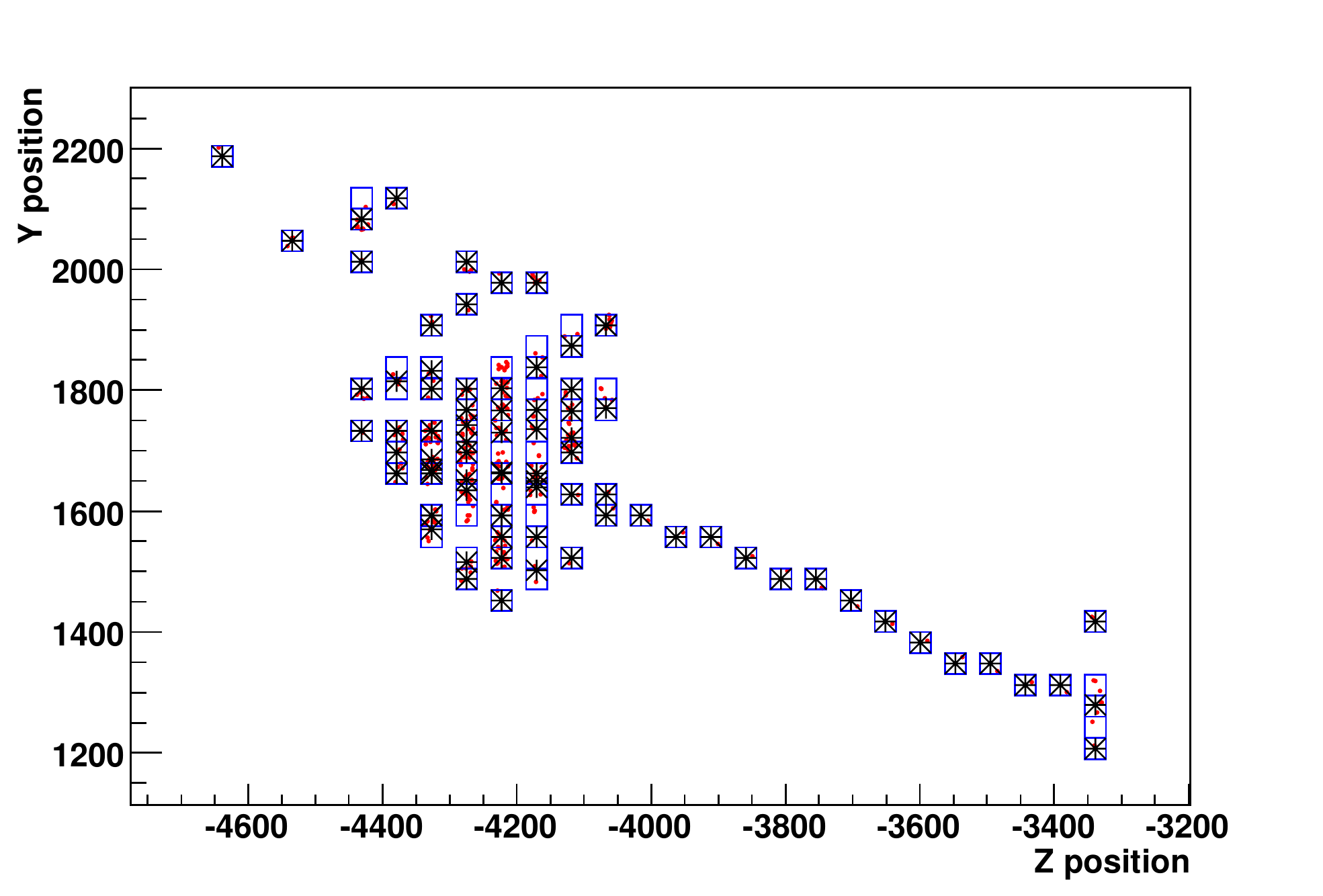} \\
    \end{array}$
    \includegraphics[width=0.45\textwidth]{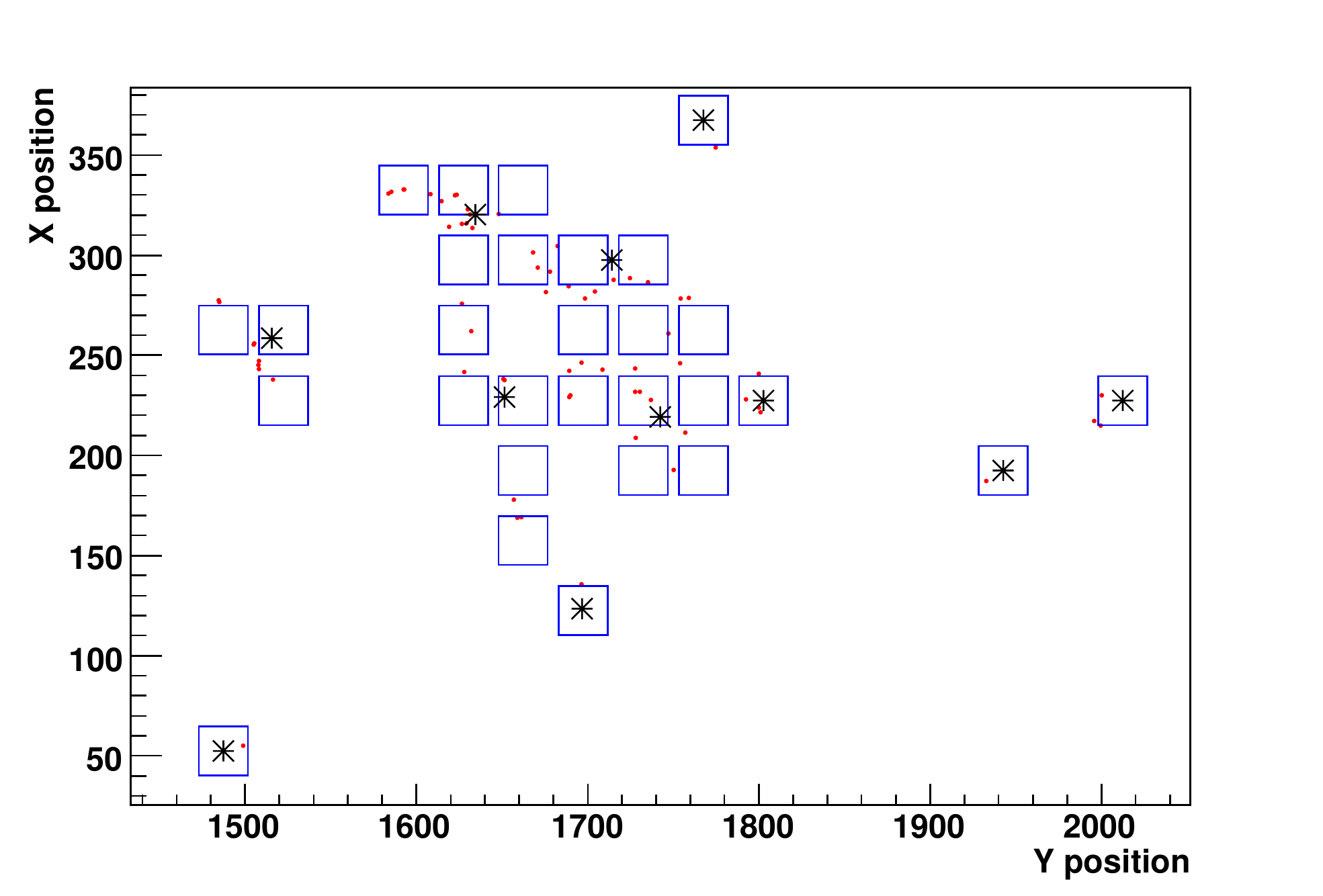}
  \end{center}
  \caption{The digitisation and voxel clustering of an example event: (top left) bending plane view, (top right) non-bending plane, (bottom) an individual scintillator plane. The individual hits are small dots (in red), the blue squares are the voxels and the black asterisks represent the centroid positions of the clusters.}
  \label{fig:voxclust}
\end{figure}

Voxels with edge lengths of 3.5~cm were chosen, which results in a position resolution of $3.5/\sqrt{12} \cong 1$~cm. The response of the scintillator bars is derived from the raw energy deposit in each voxel, read out using wavelength shifting (WLS) fibres with attenuation length $\lambda = 5$~m, as reported by the MINERvA collaboration~\cite{PlaDalmau:2005dp}. Assuming that approximately half of the energy will come from each view, the deposit is halved and the remaining energy at each edge in $x$ and $y$ is calculated. This energy is then smeared according to a Gaussian with $\sigma/E = 6\%$ to represent the response of the electronics and then recombined into $x$, $y$ and total = $x+y$ energy deposit per voxel. Assuming an output wavelength of 525~nm, a photo-detector quantum efficiency of  $\sim$30\% can be achieved (see figure \ref{fig:pde_mppc} in section \ref{sec:MIND_conceptual_design}). A threshold of 4.7 photo electrons (pe) per view, as in MINOS~\cite{Michael:2008bc}, was assumed. Any voxel in which the two views do not make this threshold is cut. If only one view is above threshold, then only the view below the cut is excluded (see section~\ref{Sec:RecG4}). The digitisation of an example event is shown in figure \ref{fig:voxclust}.

\paragraph{Reconstruction of events}
\label{Sec:RecG4}

The reconstruction and analysis packages are described in detail in a previous publication \cite{Cervera:2010rz}. We present here an update of the analysis based on the MIND simulation generated using NUANCE and GEANT4. The optimisation of cuts and the extraction of efficiencies were performed using a flux profile designed to maximise statistics in all energy bins of interest. 

Many traversing particles, particularly hadrons, deposit energy in more than one voxel. Forming clusters of adjacent voxels reduces event complexity and can improve pattern recognition in the region of the hadron shower. The clustering algorithm is invoked at the start of each event. The voxels of every plane in which energy has been deposited are considered in sequence. Where an active voxel is in contact with no other active voxel, this voxel becomes a measurement point. If there are adjacent voxels, the voxel with the largest total deposit (at scintillator edge) is sought and all active voxels in the surrounding 3$\times$3 area are considered to be part of the cluster. Adjacent deposits which do not fall into this area are considered separate.
The cluster position is calculated independently in the $x$ and $y$ views as the energy-weighted sum of the individual voxels.

The clusters formed from the hit voxels of an event are then passed to the reconstruction algorithm. 
The separation of candidate muons from hadronic activity is achieved using two methods: a Kalman filter algorithm provided by RecPack~\cite{CerveraVillanueva:2004kt} and a cellular automaton method 
(based on \cite{Emeliyanov_otr/itr-cats:tracking}).

The Kalman filter propagates the track parameters back through the planes using a helix model which takes into account multiple scattering and energy loss.  Since, in general, a muon is a minimum ionising particle (MIP) and will travel further in the detector than hadronic particles, those hits furthest downstream are assumed to be muon hits and used as a seed for the Kalman filter. The seed ``state vector'' is then propagated back to each plane with multiple hits and the matching $\chi^2$ to each of the hits is computed. 
Hits with matching $\chi^2$ below 20 are considered and in each plane the one with the best matching among these is added to the trajectory and filtered (i.e. the track parameters are updated with the new information). All accepted hits constitute the candidate muon and are presented for fitting 
with the remaining hits being considered as hadronic activity. 

Events with high $Q^2$ or low neutrino energy can be rejected by the first method, since in general the muon will not escape the region of hadronic activity.   
The cellular automaton method uses a neighbourhood function first to rank all the hits and then to form all viable combinations into possible trajectories. A ``neighbour'' is defined as a hit in an adjacent plane within a pre-defined transverse distance of the projection into that plane of the straight line connecting hits in the previous two planes. Starting from the plane with lowest $z$ position,
 hits are given incremental ranks higher than neighbours in the previous plane. Trajectories are then formed 
from every possible combination of one hit per plane starting with those of highest rank using the neighbourhood function. 

Those trajectories formed using this method are then categorised according to length and $\chi^2$. The candidate muon is then selected as the longest remaining trajectory with the lowest $\chi^2$. All other hits in the event are considered to be from hadronic activity. Even if three planes in a row have no associated candidate hits, the event can still be accepted if 70\% of the planes are considered to belong to the candidate-muon trajectory. This algorithm improves the efficiency and purity of muon candidates close the hadronic jet, since it ranks the likelihood that a given hit belongs to the muon candidate or the hadronic jet.

An additional step was added to take into account that fully-contained muons (particularly $\mu^-$) can have additional deposits at their endpoint due to captures on nuclei or due to decays. Long, well defined tracks can be rejected if there is additional energy deposited at the muon end point, since the cellular automaton interprets this as hadronic activity or a decay.  Once the cellular automaton has completed the pattern recognition, a Kalman fit is used to estimate the track parameters.  Additional energy deposits at the end of the muon track may cause confusion in determining the seed of the Kalman filter. Therefore, after sorting clusters into increasing \emph{z} position, a pattern-recognition algorithm is used to identify the section of the track associated with the long muon section, used for momentum measurement via the Kalman filter, and the endpoint.

\begin{figure}
  \begin{center}$
    \begin{array}{cc}
      \includegraphics[width=8cm, height=6cm]{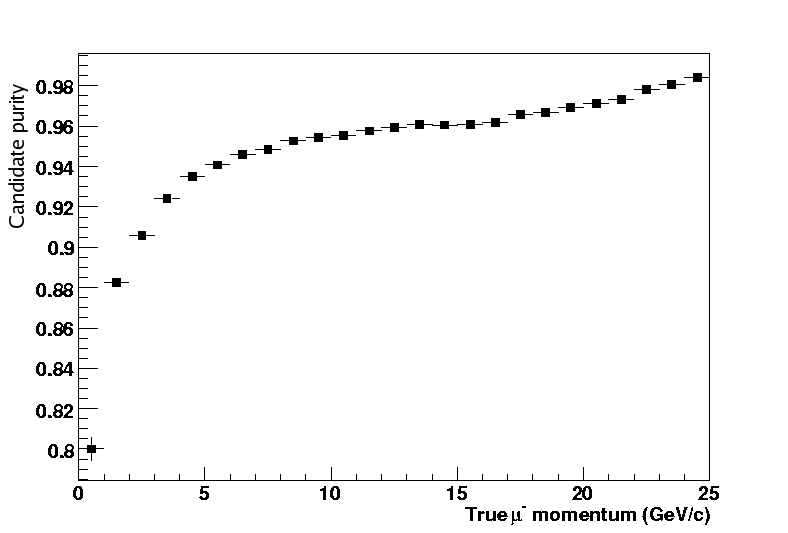} &
      \includegraphics[width=8cm, height=6cm]{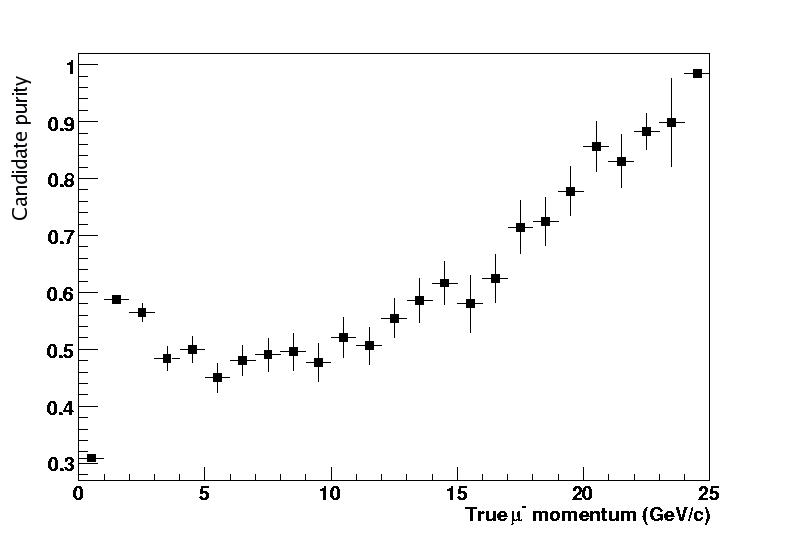}\\
      \includegraphics[width=8cm, height=6cm]{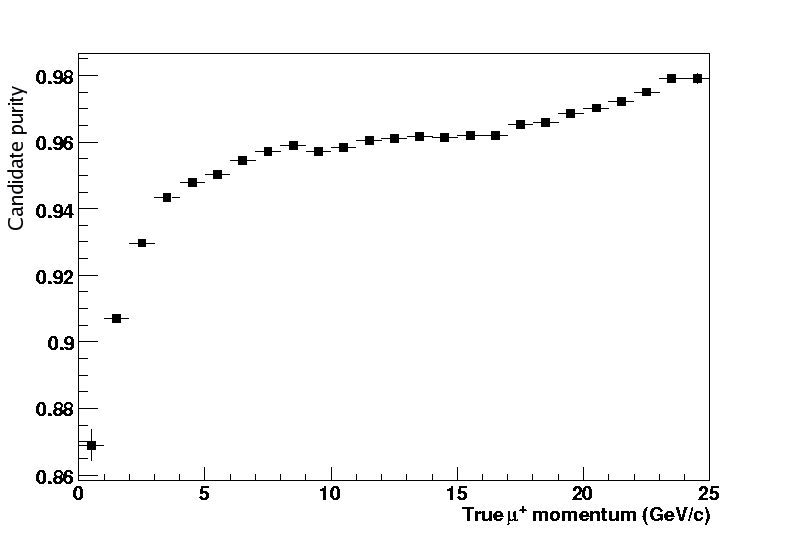} &
      \includegraphics[width=8cm, height=6cm]{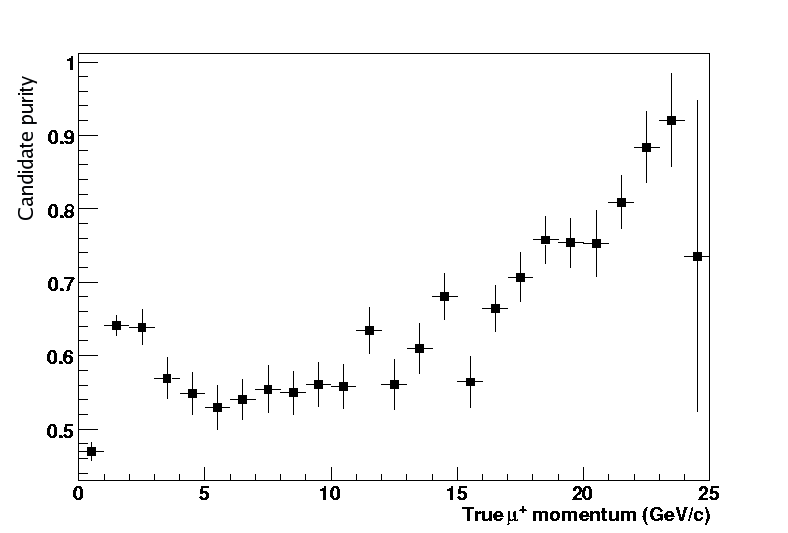}
    \end{array}$
  \end{center}
  \caption{Muon candidate hit purity for $\nu_\mu$ CC (top) and $\overline{\nu}_\mu$ CC (bottom) interactions extracted using (left) Kalman filter method and (right) cellular automaton method.}
  \label{fig:G4purity}
\end{figure}

The complete pattern-recognition chain using these algorithms leads to candidate purity (fraction of candidate hits of muon origin) for $\nu_\mu~(\overline{\nu}_\mu)$ CC events as shown in figure \ref{fig:G4purity}. A cluster is considered to be of muon origin if greater than 80\% of the raw deposits contained within the cluster were recorded as muon deposits.  Overall, the candidate purity is in excess of 90\% for true muon energies larger than $\sim 2$\,GeV.

\begin{figure}
  \begin{center}$
    \begin{array}{cc}
      \includegraphics[width=8cm, height=6cm]{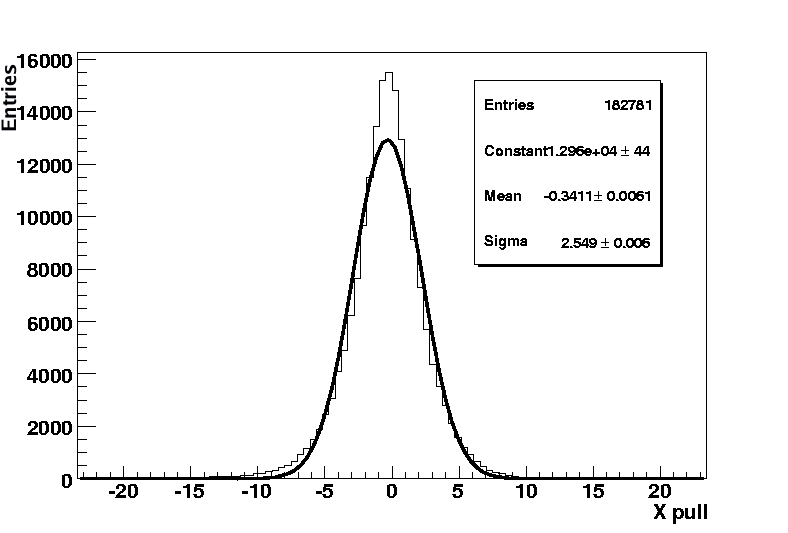} &
      \includegraphics[width=8cm, height=6cm]{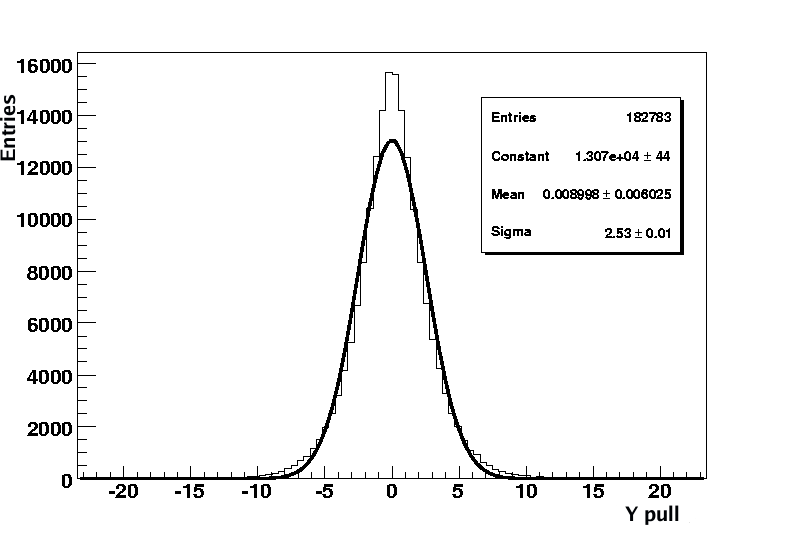} \\
      \includegraphics[width=8cm, height=6cm]{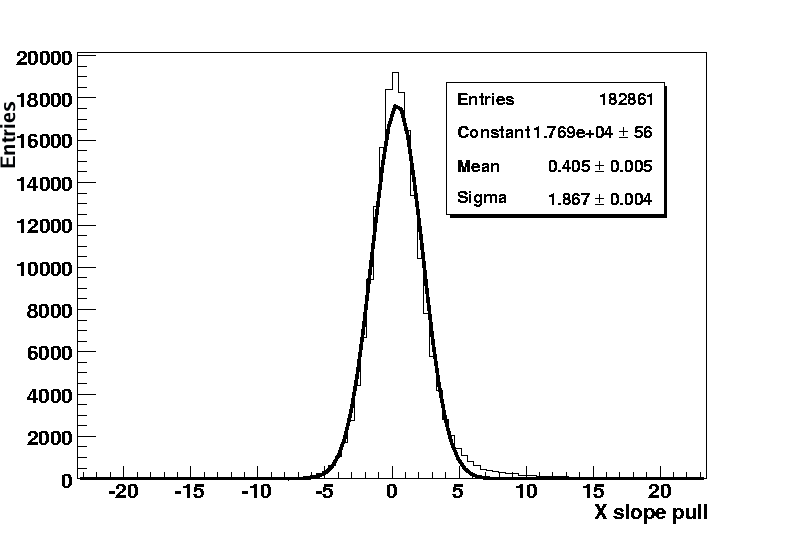} &
      \includegraphics[width=8cm, height=6cm]{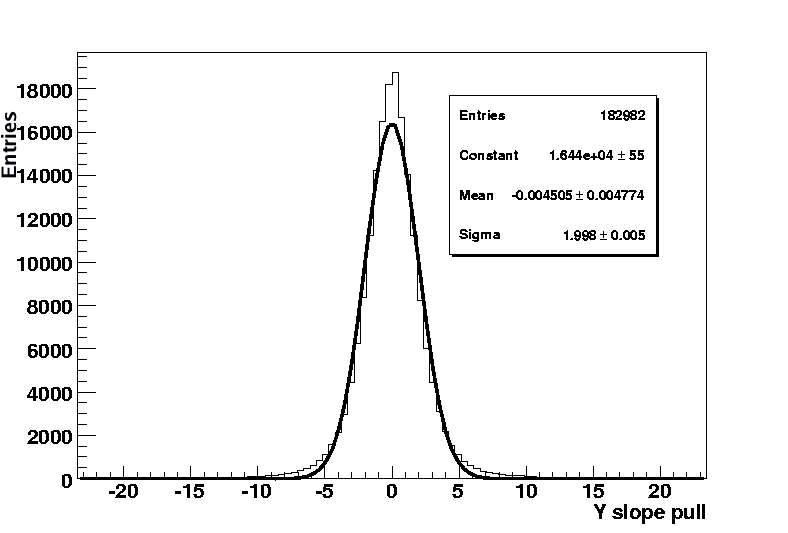} \\
      \includegraphics[width=8cm, height=6cm]{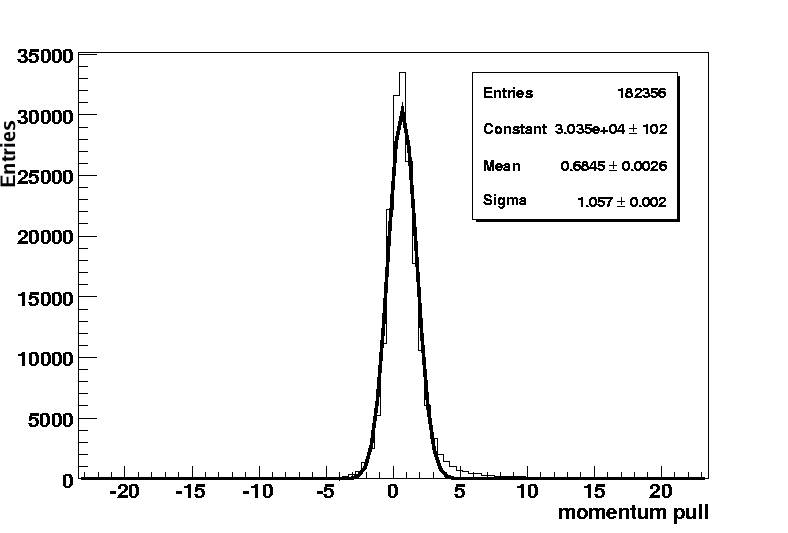} &
      \includegraphics[width=8cm, height=6cm]{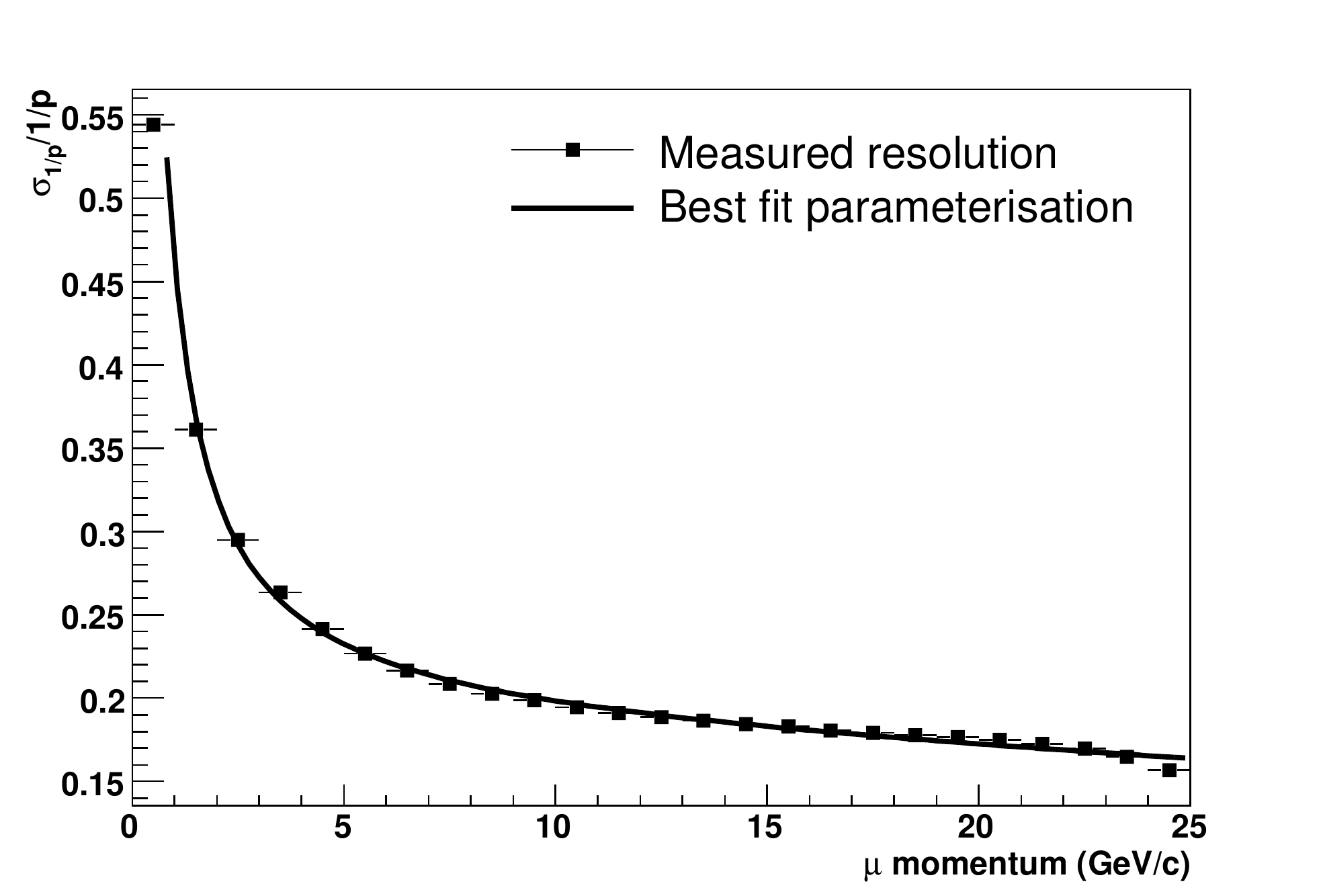}
    \end{array}$
  \end{center}
  \caption{Pulls on the five Kalman filter vector parameters and the reconstructed momentum resolution: (top) X (left) and Y (right) position, (middle) X (left) and Y (right) slopes and (bottom) momentum pull (left) and momentum resolution (right).}
  \label{fig:G4pulls}
\end{figure}

Fitting of the candidates proceeds using a Kalman filter to fit a helix to the candidate, using an initial seed estimated by a quartic fit, and then refitting any successes. Projecting successful trajectories back to the true vertex \emph{z} position, the quality of the fitter can be estimated using the pulls of the different parameters in the fit vector (see figure \ref{fig:G4pulls}).

The pulls on the position and direction exhibit a larger sigma than expected indicating that the position resolution is underestimated (i.e. that the resolution assumed in the fit is a little too small). This is attributed to clusters being made up of many particles in the high-occupancy vertex region. This effect will be better understood as the digitisation algorithms are improved. The momentum pull, however, behaves as expected but for a slight bias due to the fact that the energy loss is not updated along the track in the version of RecPack used for the study. In this version the energy loss is not changed as the filter progresses along the track and can result in a bias. The muon-momentum resolution is slightly improved from previous studies, particularly at low energies, and its resolution is parametrised as follows:
\begin{equation}
  \label{eq:G4resol}
  \displaystyle\frac{\sigma_{1/p}}{1/p} = 0.18 + \frac{0.28}{p} - 1.17\times 10^{-3}p.
\end{equation}

Hadronic reconstruction is predominantly performed using a smear on
the true quantities as described in \cite{Cervera:2010rz}. 
However, the presence of QE interactions in the sample allows for the
reconstruction of certain events using the formula:
\begin{equation}
  \label{eq:quasiEng}
  E_\nu = \displaystyle\frac{m_NE_\mu + \frac{m_{N_X}^2 - m_\mu^2 - m_N^2}{2}}{m_N - E_\mu + |p_\mu|\cos\vartheta} \, ;
\end{equation}
where $\vartheta$ is the angle between the muon momentum vector and
the beam direction, $m_N$ is the mass of the initial state nucleon,
and $m_{N_X}$ is the mass of the outgoing nucleon for the interactions
$\nu_\mu + n \rightarrow \mu^- + p$ and $\overline{\nu}_\mu + p
\rightarrow \mu^+ + n$~\cite{Blondel:2004cx}. 
QE interactions can be
selected using their distribution in $\vartheta$ and their event-plane
occupancy among other parameters. 
Should the use of
equation~\ref{eq:quasiEng} result in a negative value for the energy,
it is recalculated as the total energy of a muon using the
reconstructed momentum. 

\paragraph{Analysis of potential signal and background}
\label{par:G4ana}
There are four principal sources of background to the wrong sign muon search: incorrect charge assignment (charge misidentification), wrong sign muons from hadron decay in $\overline{\nu}_\mu$ charged current (CC), neutral current (NC) and $\nu_e$ CC events wrongly identified as $\nu_\mu$ CC. In order to reduce these backgrounds while maintaining good efficiency a number of 
offline cuts were employed. They can be organised in four categories: 1) $\nu_\mu$ CC selection cuts; 2) NC rejection cuts; 3) kinematic cuts; and 4) muon quality cuts.
Cut levels and likelihood distributions have been defined using a test statistic of $\nu_\mu$ and $\overline{\nu}_\mu$ CC and NC events.

\begin{figure}
  \begin{center}$
    \begin{array}{cc}
      \includegraphics[width=7.5cm, height=5.5cm]{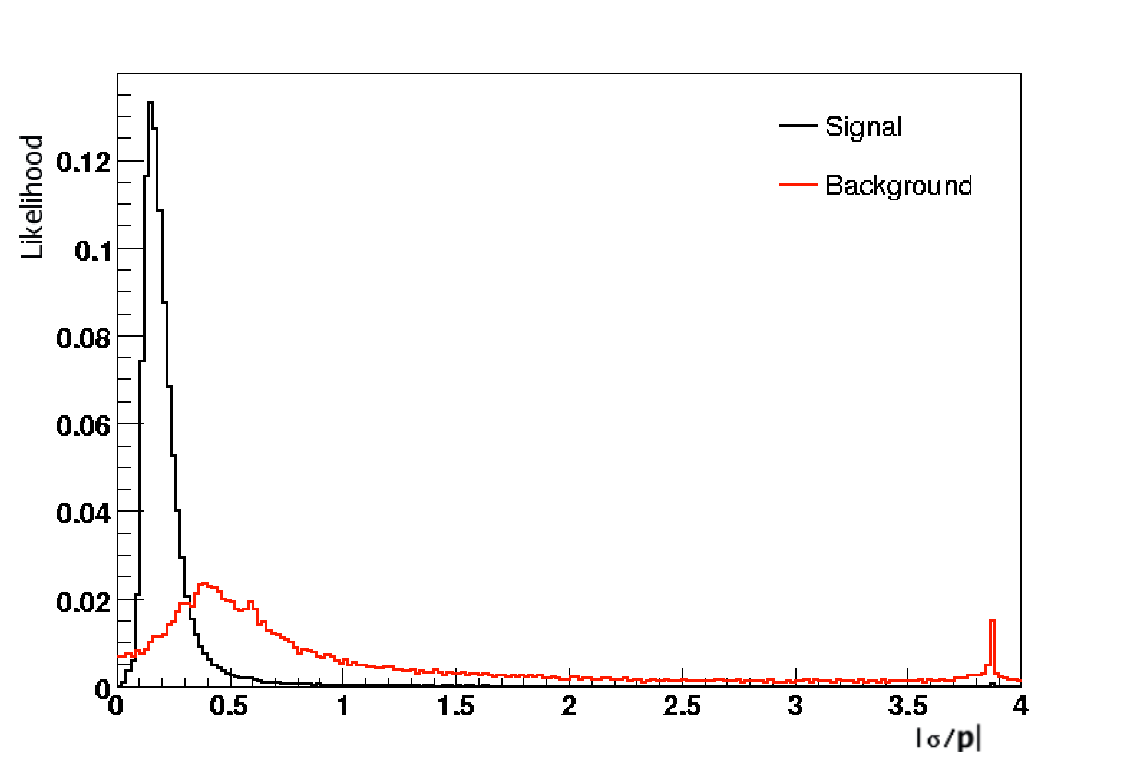} &
      \includegraphics[width=7.5cm, height=5.5cm]{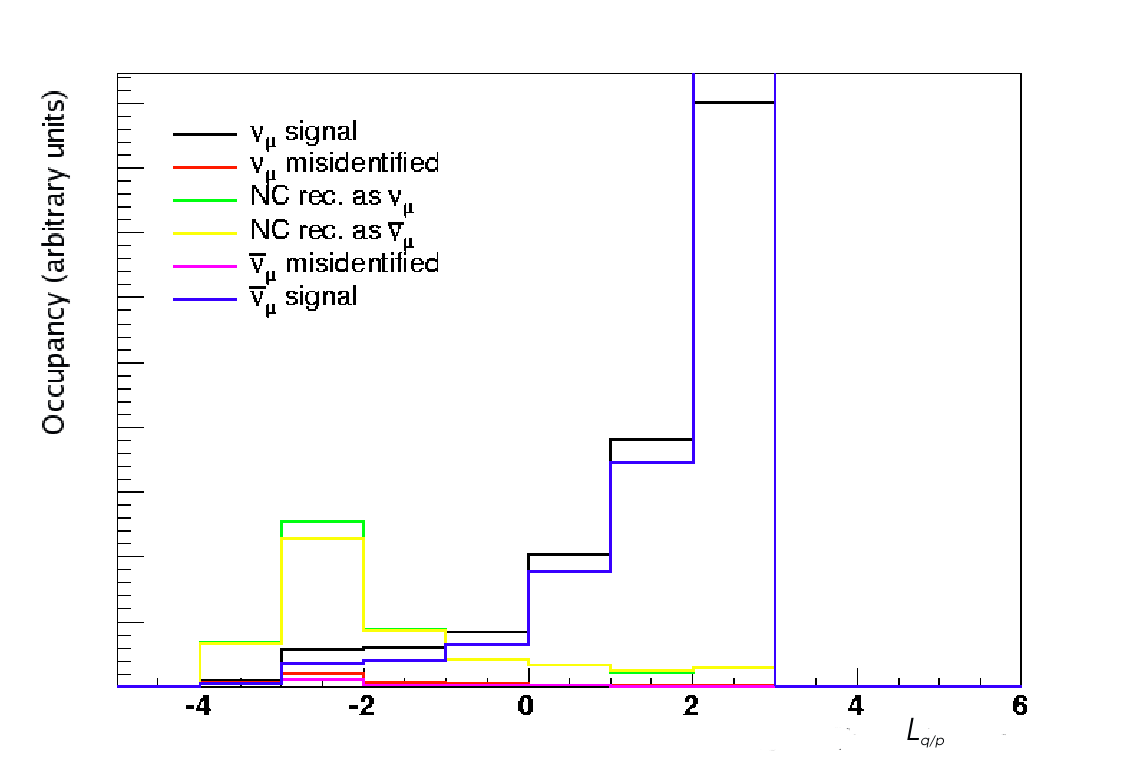}
    \end{array}$
  \end{center}
  \caption{PDF of $\displaystyle\frac{\sigma_{q/p}}{q/p}$ (left) and the resulting log likelihood distribution ($\mathcal{L}_{q/p}$) for a test statistic (right).}
  \label{fig:qPlike}
\end{figure}

\subparagraph*{q/p relative error \\}
\label{subpar:qP}
The error on the momentum parameter (charge divided by momentum, q/p) of the Kalman filter is a powerful handle in the rejection of background. The likelihood distributions for the combination of the two signals and of the neutral current and charged-current backgrounds (re-normalising in the latter case to take account of the different data-set sizes) are used to assess events as signal or background (shown in figure \ref{fig:qPlike}-left).
Signal events are selected as those with a log-likelihood parameter $\mathcal{L}_{q/p} > -0.5$. As can be seen from figure \ref{fig:qPlike}-right this cut-level effectively rejects much of the background.

\begin{figure}
  \begin{center}$
    \begin{array}{cc}
      \includegraphics[width=8cm, height=6cm]{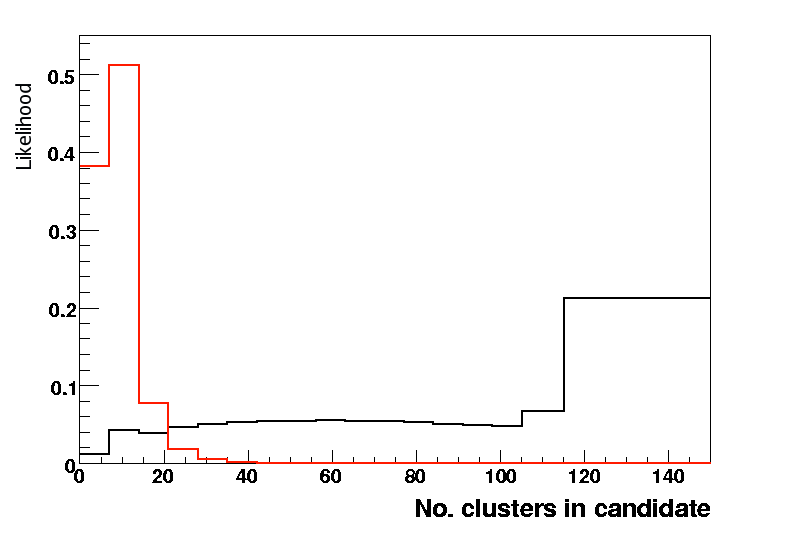} &
      \includegraphics[width=8cm, height=6cm]{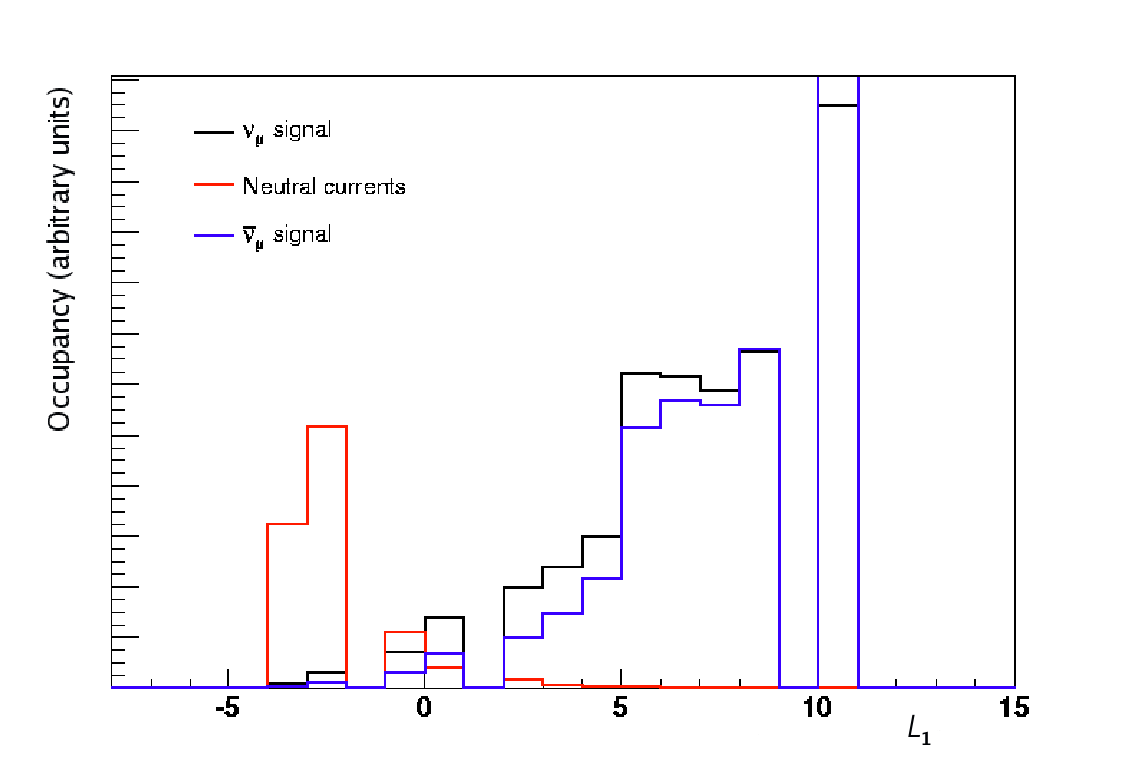}
    \end{array}$
  \end{center} 
  \caption{Number of clusters likelihood distribution function (left) and log-likelihood ratio (right) used to reject NC background. CC in black and NC in red.}
  \label{fig:NCpdfs}
\end{figure}

\subparagraph*{Neutral current rejection \\}
\label{subpar:NCrej}
Rejection of neutral current events is most efficiently performed using a likelihood analysis of trajectory quantities, as was carried out in MINOS. The three parameters used by MINOS were: number of hits in the candidate, fraction of the total visible energy in the candidate and the mean deposit per plane of the candidate. No improved background rejection was observed by using the latter two parameters, probably due to correlations between them. Only the number of hits, $l_{hit}$, was used to generate particle distribution functions (PDFs) for charged and neutral current events (see figure \ref{fig:NCpdfs}). Candidates with greater than 150 clusters are considered signal and for those with less than or equal 150 clusters, a log likelihood rejection parameter:
\begin{equation}
  \mathcal{L}_1 = \log \left( \frac{l_{hit}^{CC}}{l_{hit}^{NC}} \right) \, ;
\end{equation}
is used, which is also shown in figure \ref{fig:NCpdfs}. Allowing only those candidates where the log parameter is $\mathcal{L}_1 > 1.0$ to remain in the sample ensures that the sample is pure.

\begin{figure}
  \begin{center}$
    \begin{array}{cc}
      \includegraphics[width=7.5cm, height=4.75cm]{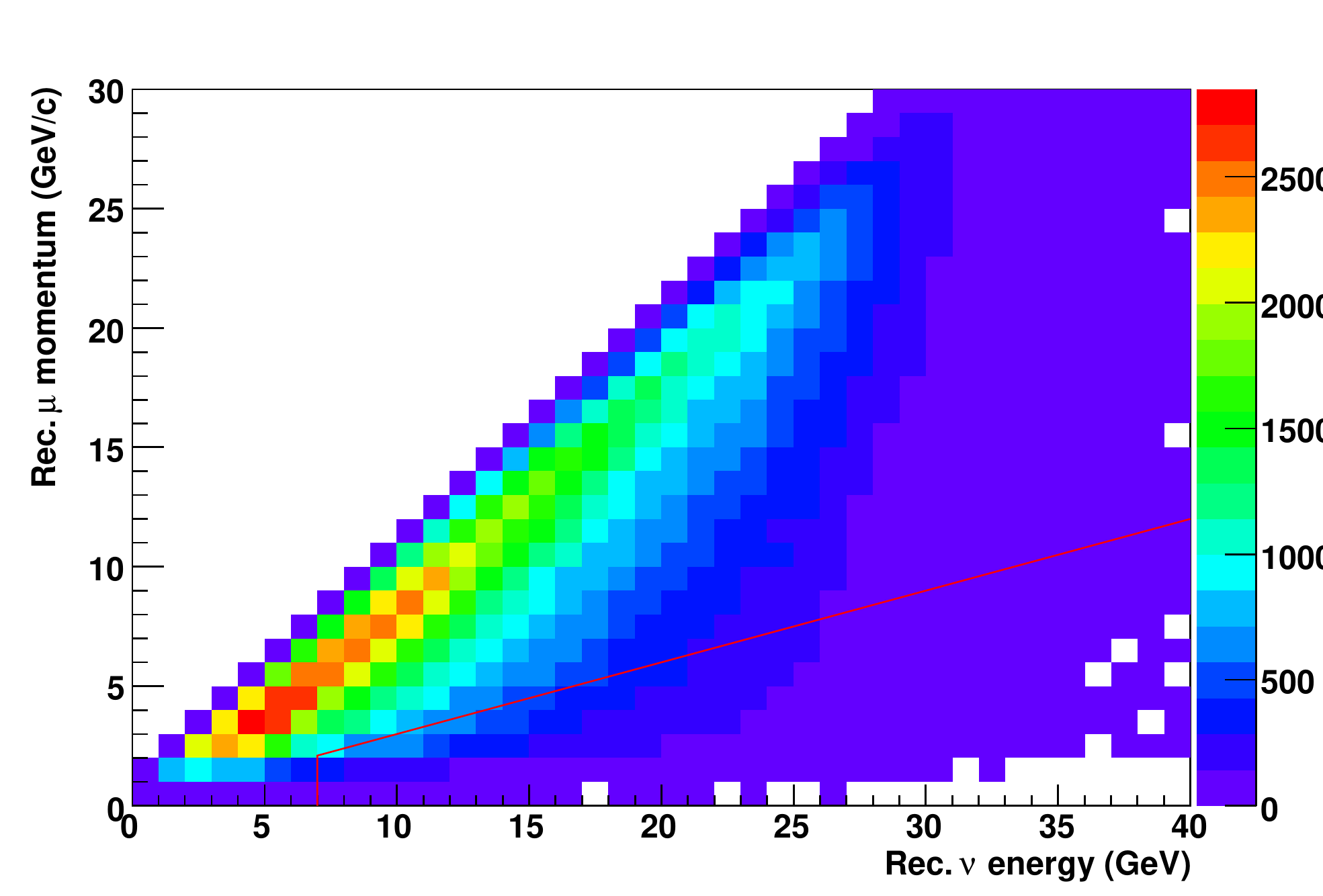} &
      \includegraphics[width=7.5cm, height=4.75cm]{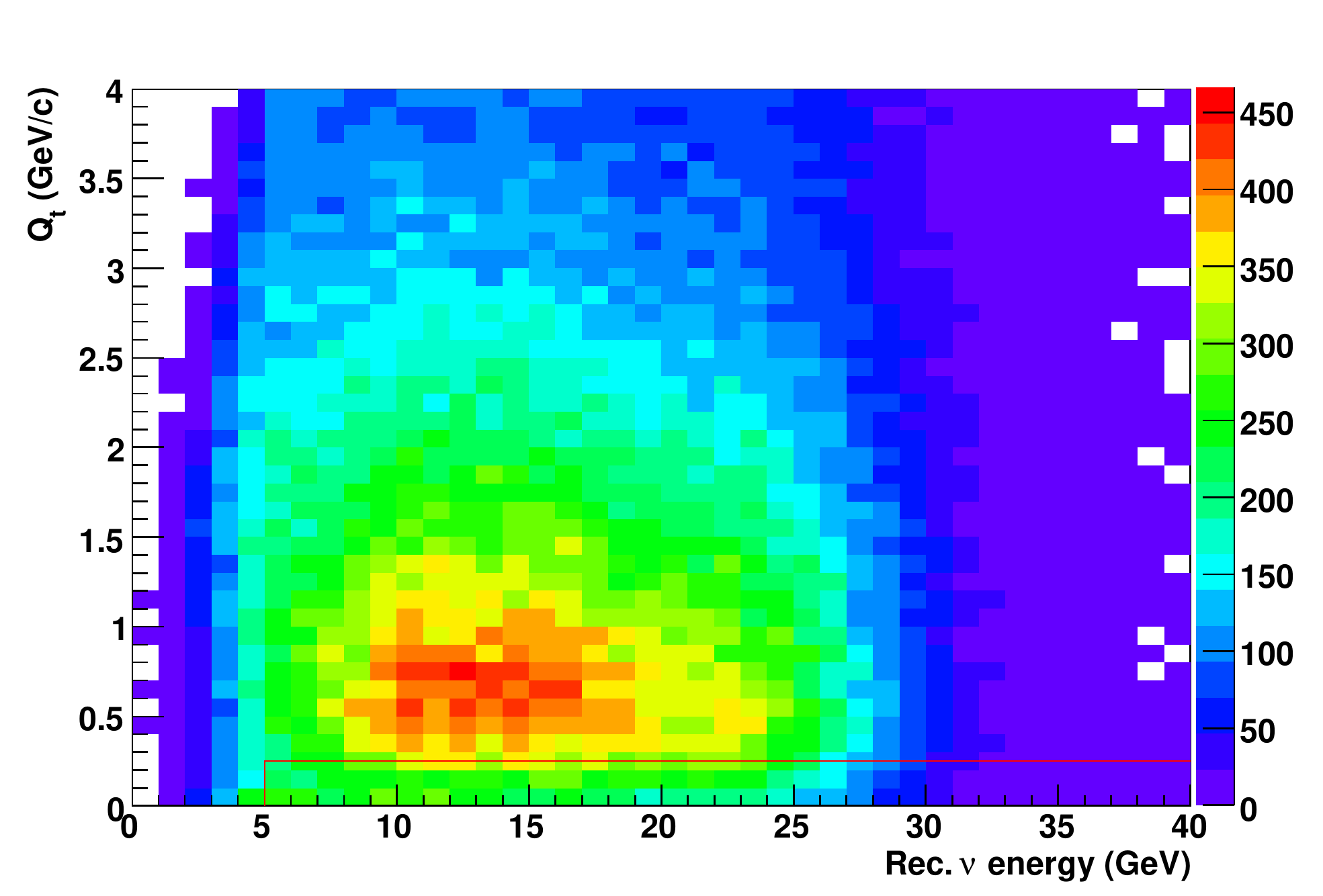} \\
      \includegraphics[width=7.5cm, height=4.75cm]{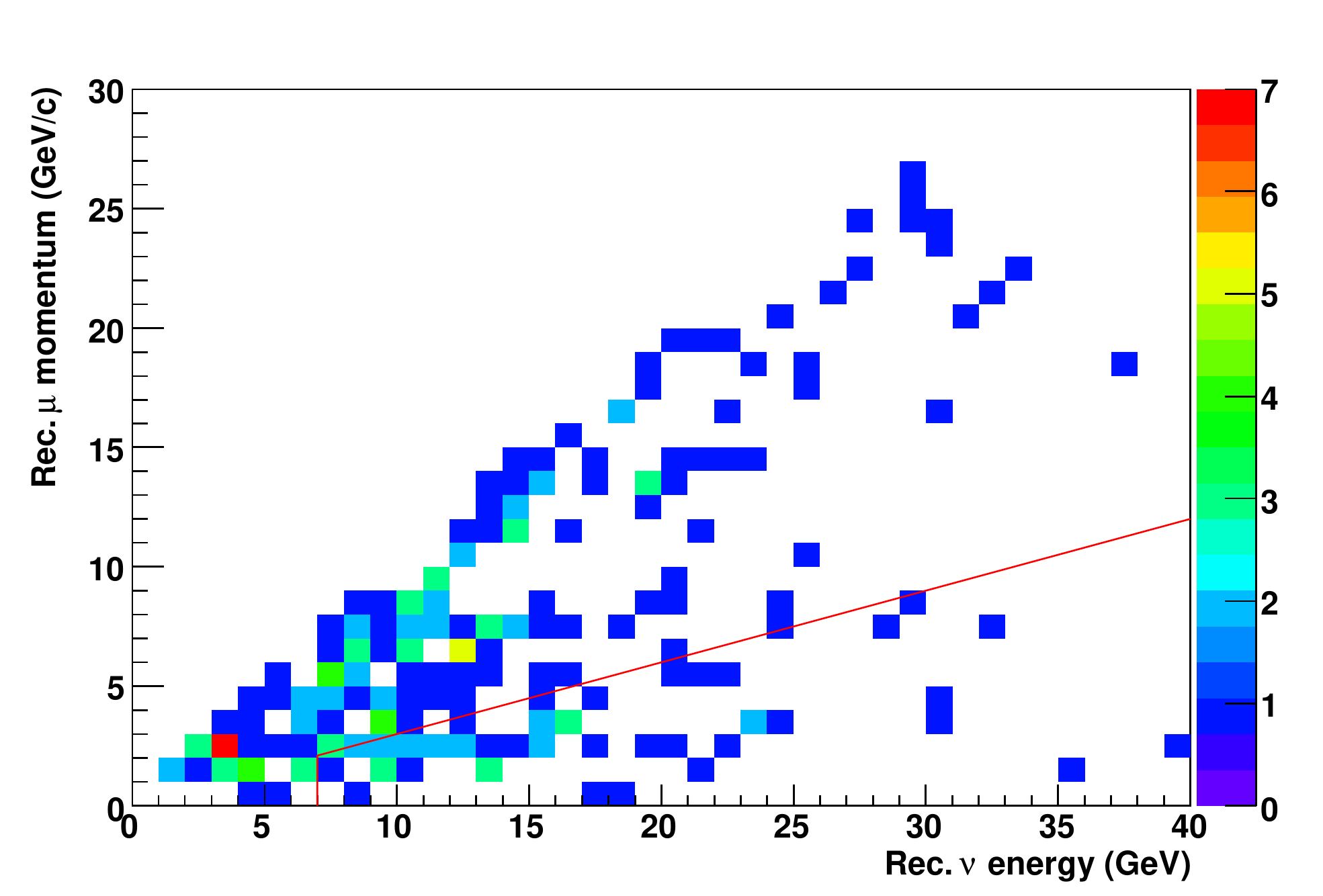} &
      \includegraphics[width=7.5cm, height=4.75cm]{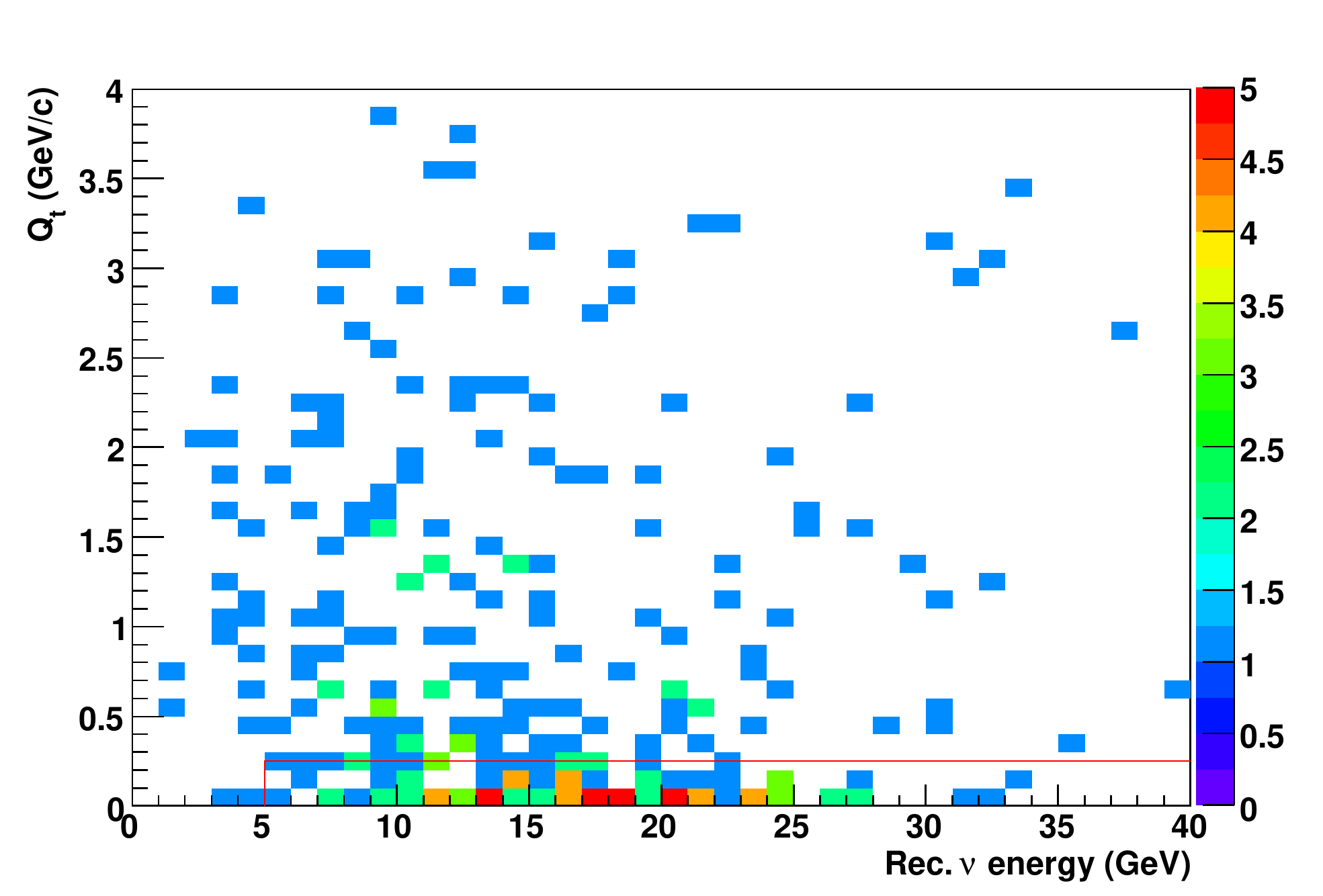} \\
      \includegraphics[width=7.5cm, height=4.75cm]{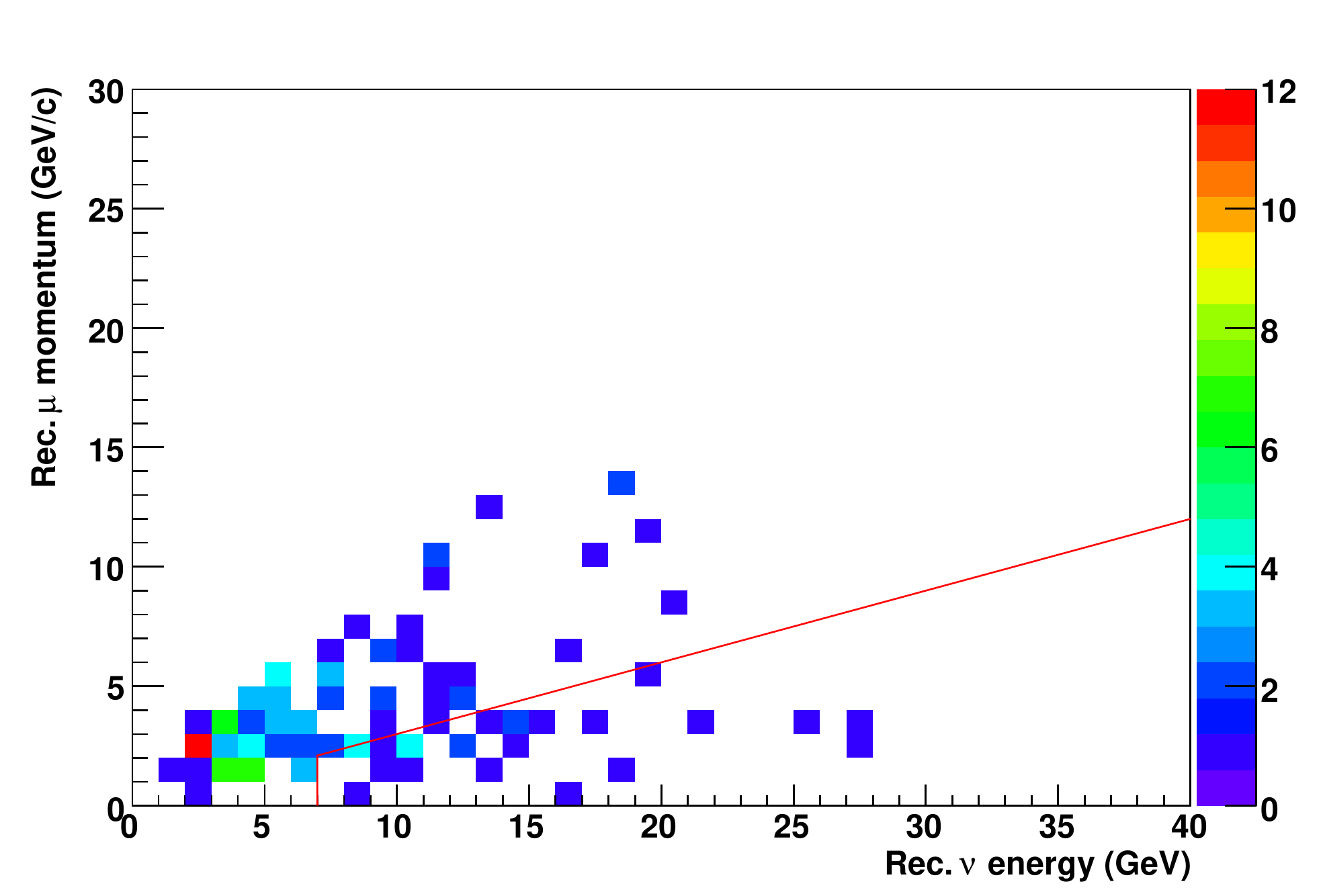} &
      \includegraphics[width=7.5cm, height=4.75cm]{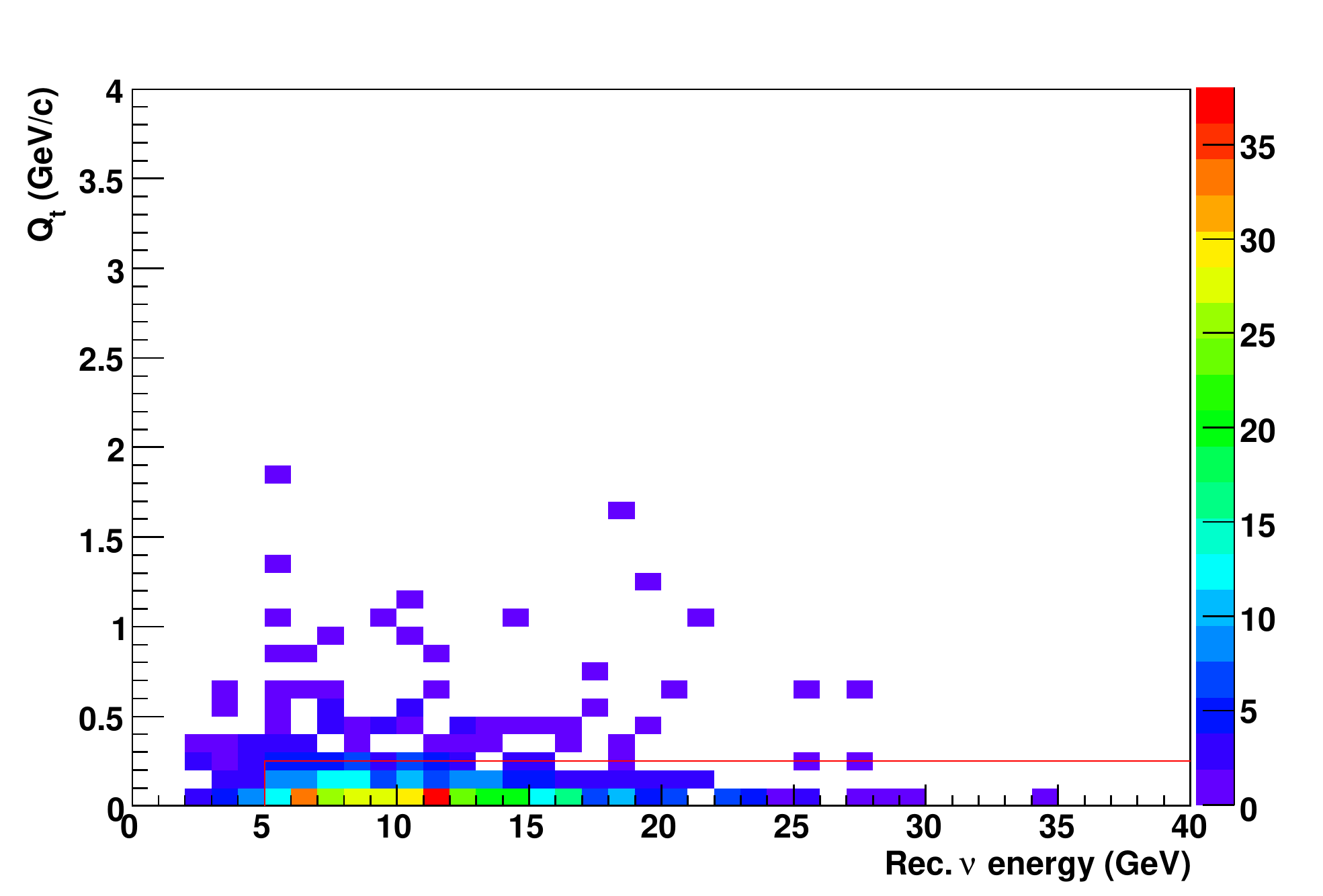} \\
      \includegraphics[width=7.5cm, height=4.75cm]{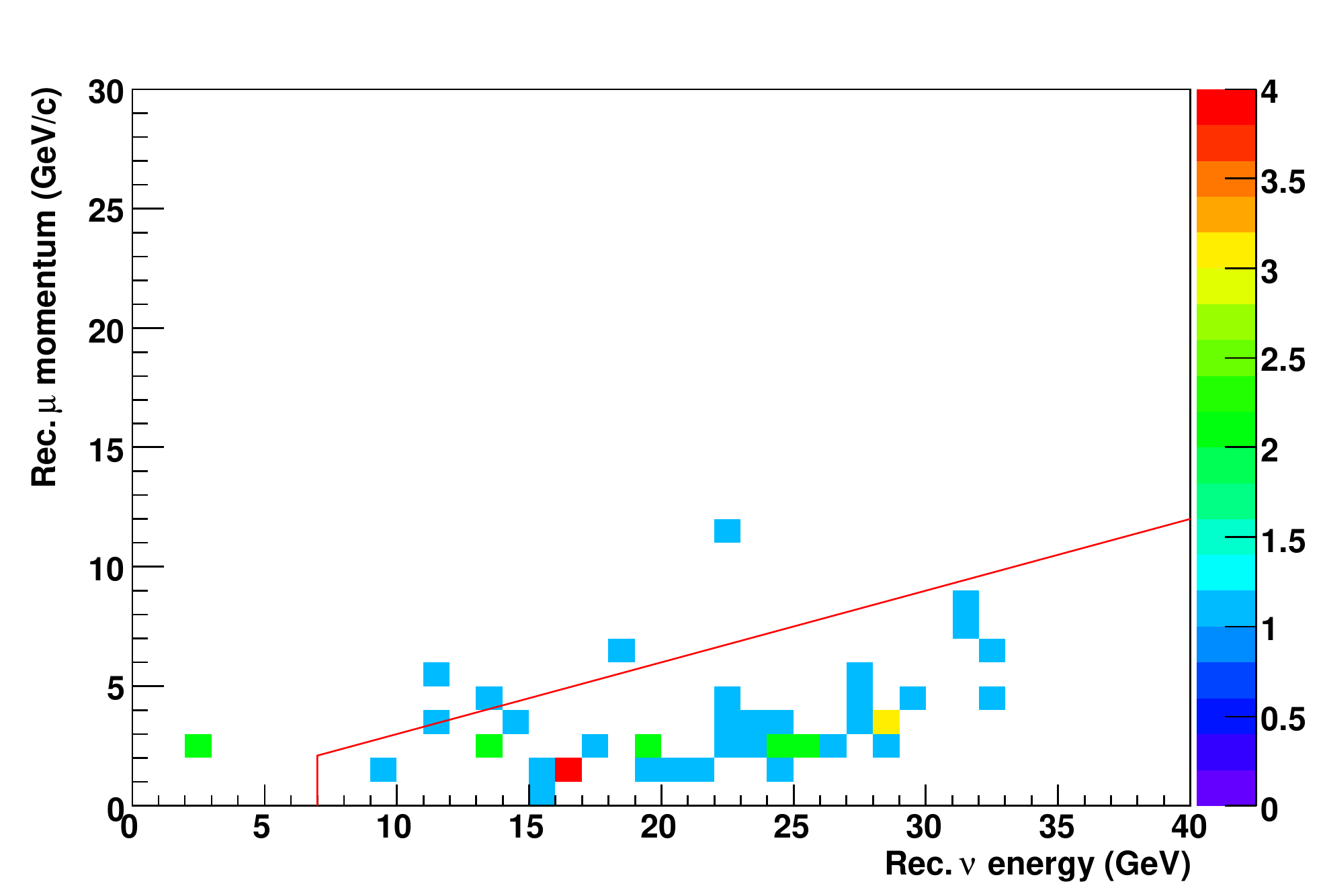} &
      \includegraphics[width=7.5cm, height=4.75cm]{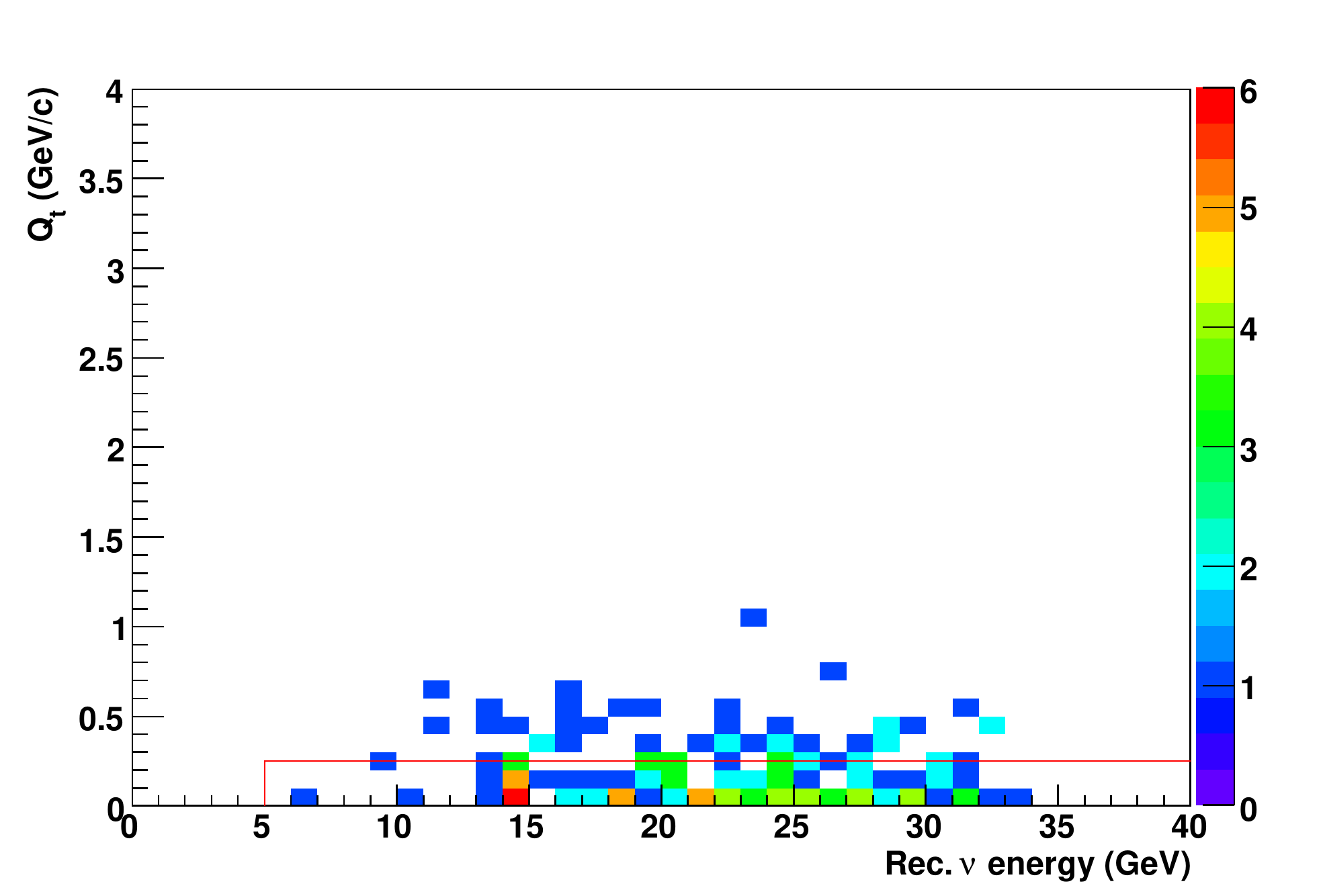} \\
    \end{array}$
  \end{center}
  \caption{Distributions of kinematic variables: (left) Reconstructed muon momentum with reconstructed neutrino energy for (top$\rightarrow$bottom) $\nu_\mu~(\overline{\nu}_\mu)$ signal, $\nu_\mu~(\overline{\nu}_\mu)$ CC background, NC background, $\nu_e~(\overline{\nu}_e)$ CC background and (right) $Q_t$ variable (in the same order).}
  \label{fig:G4kin}
\end{figure}

\subparagraph*{Kinematic cuts \\}
\label{subpar:G4kinCut}
Kinematic cuts based on the momentum and isolation of the candidate, as related to the reconstructed energy of the event $E_{rec}$, can be used to reduce backgrounds from decays. The main isolation variable is defined as $Q_t=p_\mu \sin^2\theta$, where $\theta$ is the angle between the muon candidate and the hadronic-jet vector. Cuts based on these variables are an effective way to reduce all of the relevant beam related backgrounds:
\begin{eqnarray}
  \label{eq:G4kin1}
  E_{rec}~\leq~5\mbox{~GeV or } &Q_t~>~0.25\mbox{~GeV/c} \, {\rm ; and}\\
  \label{eq:G4kin2}
  E_{rec}~\leq~7\mbox{~GeV or } &p_\mu~\geq~0.3\cdot E_{rec} \, .
\end{eqnarray}
The distributions after the application of the preceding cuts are those shown in figure \ref{fig:G4kin}. 

The acceptance level is described in equations~\ref{eq:G4kin1} and~\ref{eq:G4kin2}. With the exception of those events for which the energy is reconstructed using the quasi-elastic formula and hence do not have a hadronic shower the separation of which can be calculated, events are subject to the $Q_t$ cut if their reconstructed energy is greater than 5~GeV. Those events fulfilling the condition of equation~\ref{eq:G4kin1} and those reconstructed using the quasi-elastic formula are then subject to the second cut if they have $E_{rec} > 7$~GeV. All remaining events which fulfil the conditions of equation~\ref{eq:G4kin2} (those above and to the left of the red lines in figure \ref{fig:G4kin}) are kept in the data set for the next series of cuts.

\subparagraph*{Additional quality cuts \\}
\label{subpar:addition}
A number of additional cuts have been developed to reduce the background from opposite sign $\nu_\mu~(\overline{\nu}_\mu)$ CC events at high neutrino energy. A set of new cuts in two categories are used: 1) Those related to the range in \emph{z}, displacement in the bending plane, and reconstructed momentum of the candidate; and 2) a cut using the results of a re-fit of the remaining events to a quadratic.

\begin{figure}
  \begin{center}
    \includegraphics[width=8cm, height=6cm]{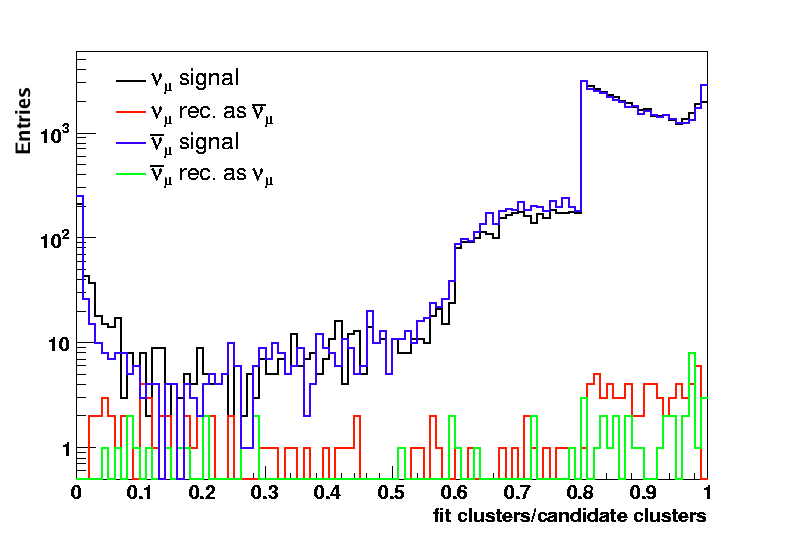}
  \end{center}
  \caption{Distribution of the proportion of clusters fitted in the trajectory.}
  \label{fig:fitnode}
\end{figure}

Additionally, a fiducial cut requiring that the first cluster in a
candidate be at least 2~m from the end of the detector is employed to
reduce the misidentification of candidates originating at high
\emph{z}. 
Moreover, a high proportion of the misidentified candidates are not
fully fitted. 
The distribution of the ratio of the candidate clusters which are
fitted with respect to the total number of candidate clusters for
signal and background is shown in figure \ref{fig:fitnode}. 
Accepting only those events with a candidate with 60\% of its clusters
fitted further reduces the background levels.

\begin{figure}
  \begin{center}$
    \begin{array}{cc}
      \includegraphics[width=8cm, height=6cm]{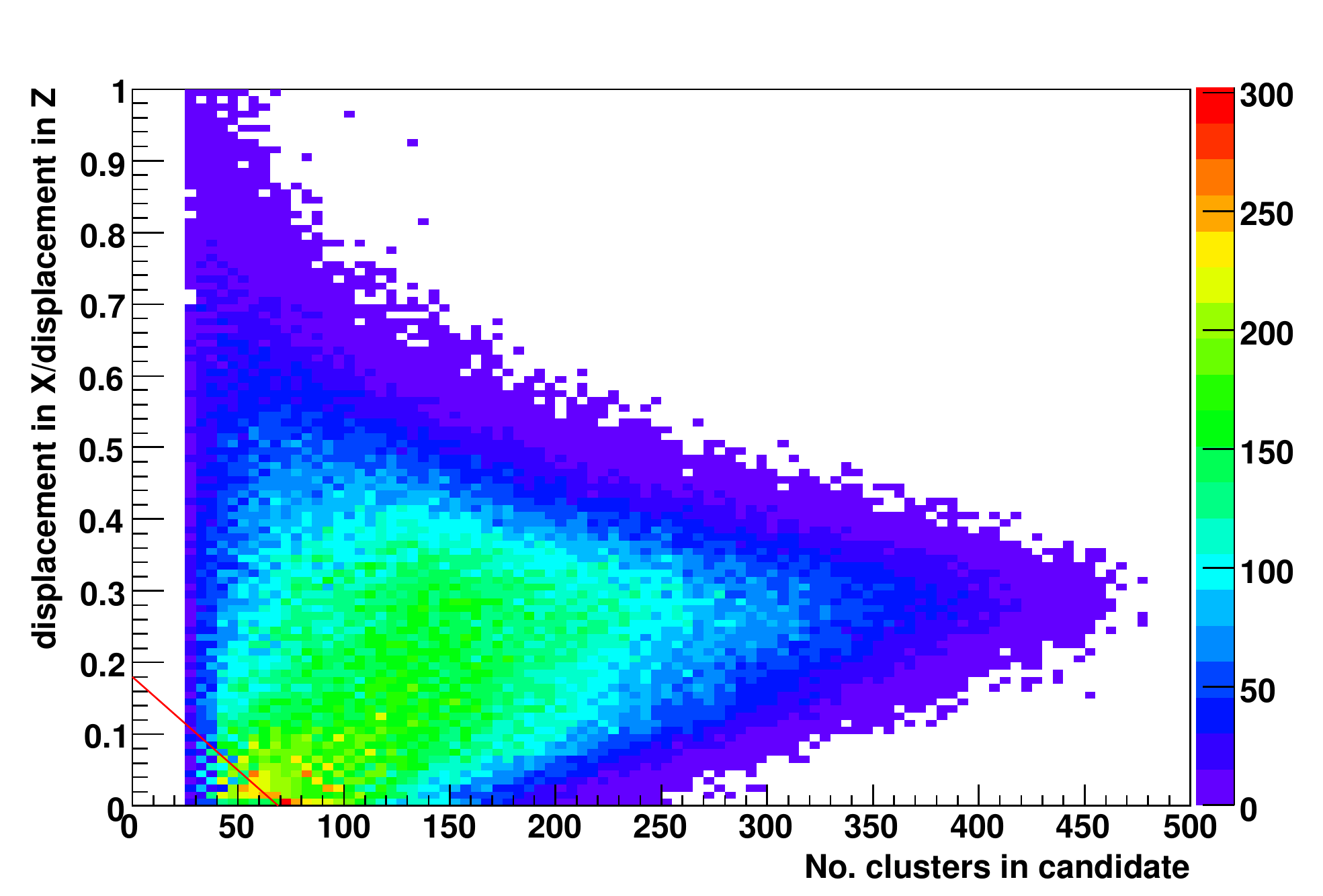} &
      \includegraphics[width=8cm, height=6cm]{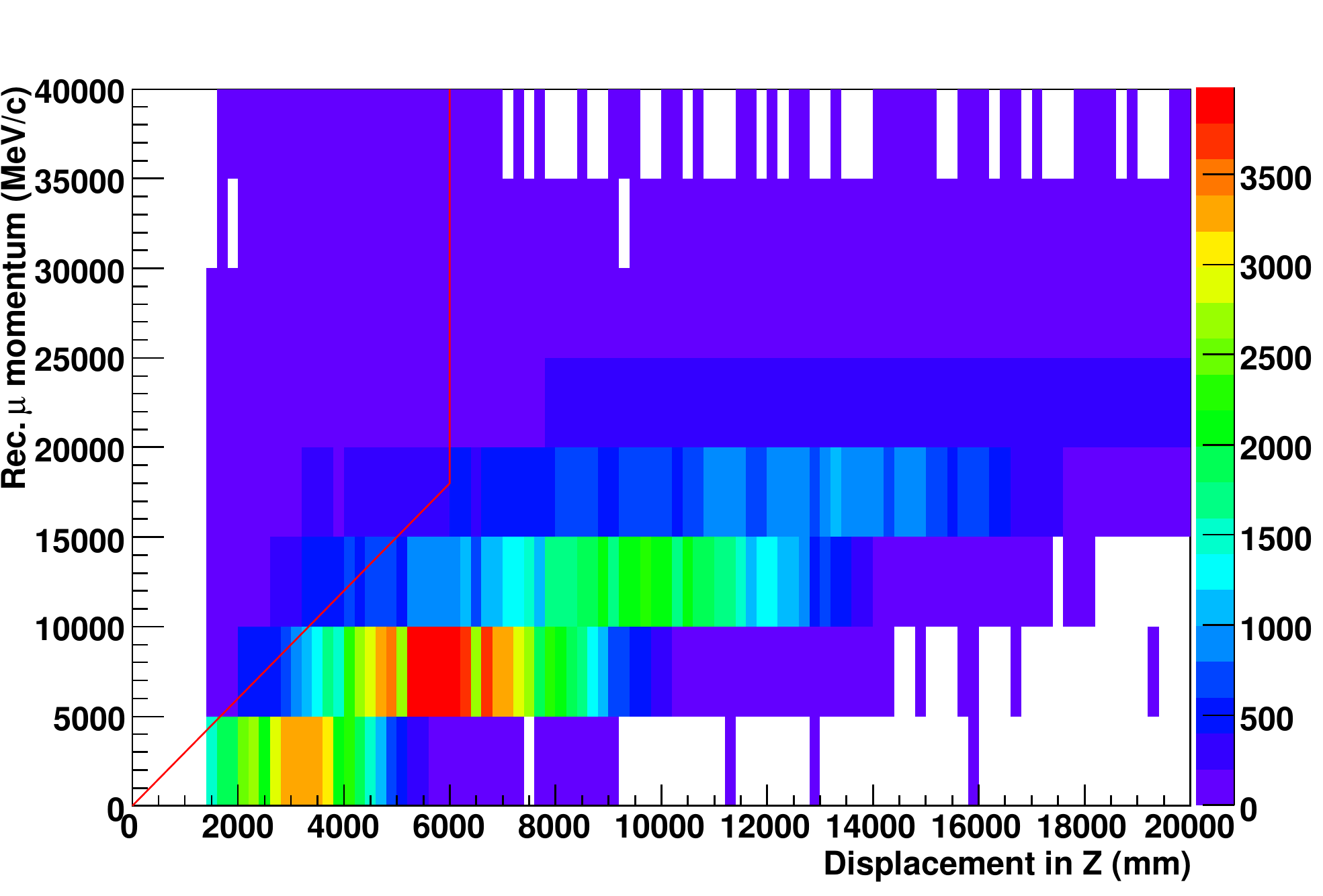} \\
      \includegraphics[width=8cm, height=6cm]{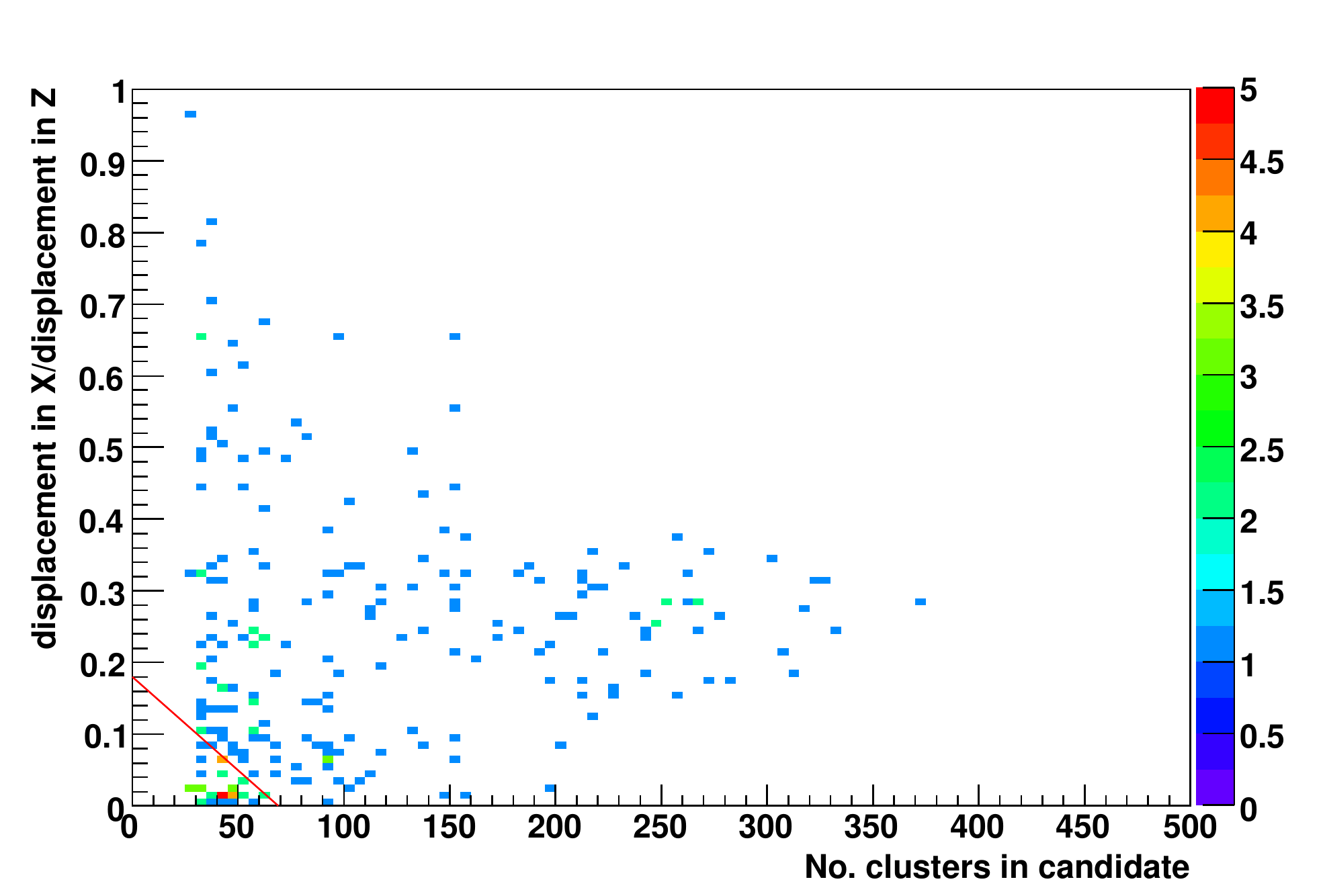} &
      \includegraphics[width=8cm, height=6cm]{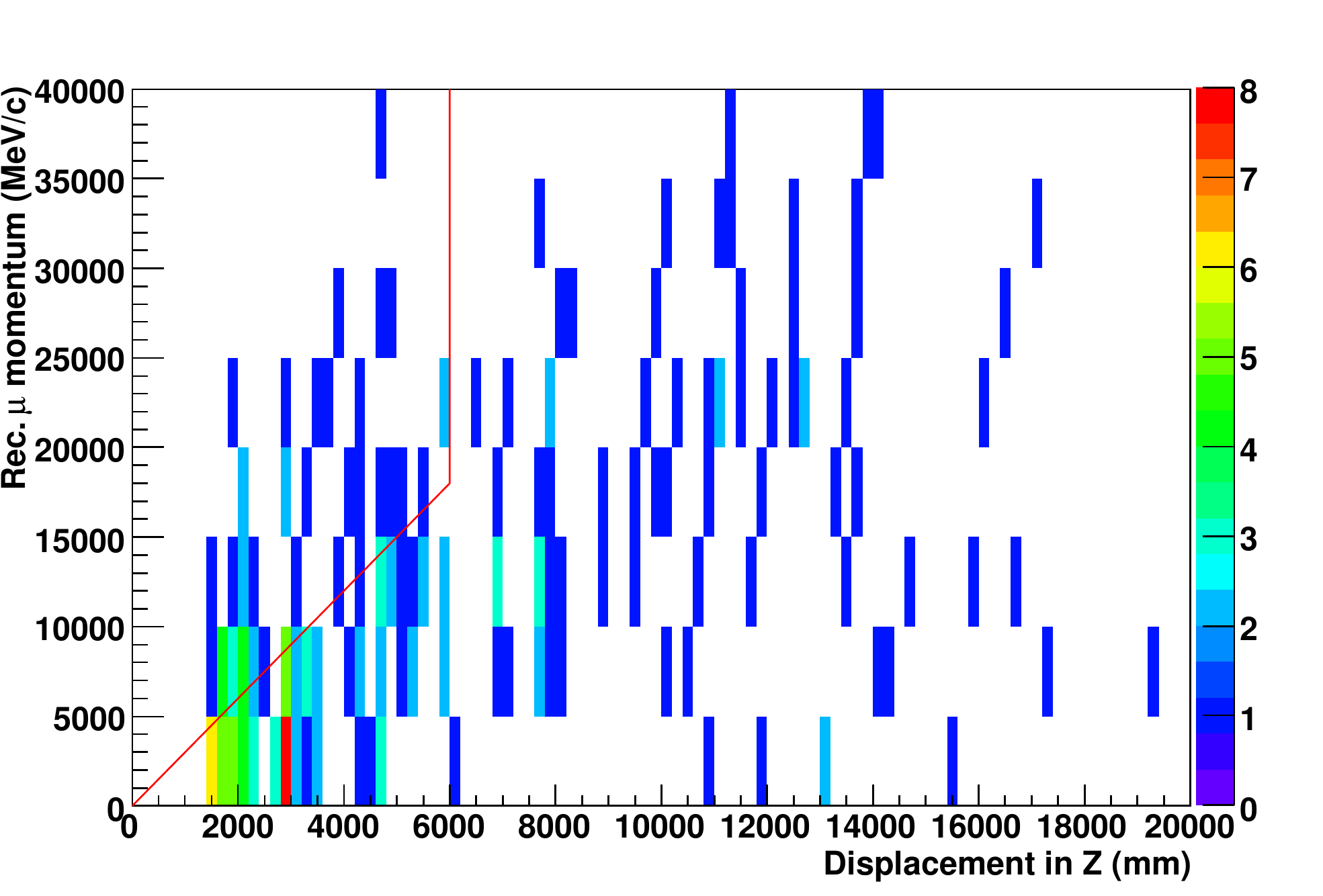} \\
    \end{array}$
  \end{center}
  \caption{Distributions of momentum and displacement with cut levels: (top left) relative displacement in the bending plane to the $z$ direction against candidate hits for signal events, (top right) reconstructed momentum against displacement in $z$ for signal events and (bottom) as top for $\nu_\mu~(\overline{\nu}_\mu)$ CC backgrounds.}
  \label{fig:dispMom}
\end{figure}

The momentum of the muon candidate can be badly reconstructed due to multiple scattering, badly assigned hits, and other effects. These badly reconstructed events contribute significantly to the overall background. In order to reduce this effect, three separate, but related cuts are employed. Firstly, a maximum reconstructed momentum, due to the finite resolution of the detector, is enforced at 40~GeV. Plotting the reconstructed momentum against the displacement in the longitudinal direction $dispZ$ of the candidate allows the candidate to be separated from the background (see figure \ref{fig:dispMom}). Additionally, the ratio between the lateral displacement in the bending plane $dispX$ and $dispZ$ against the number of hits in the candidate (see figure \ref{fig:dispMom}) provides further separation. The background events tend to be concentrated at low relative displacement and low number of hits. Events are accepted if they meet the conditions described in equation~\ref{eq:dispMom} and illustrated by the red lines in figure \ref{fig:dispMom}:
\begin{eqnarray}
  \label{eq:dispMom}
  \displaystyle \frac{dispX}{dispZ} &>& 0.18 - 0.0026\cdot N_h \, ; {\rm and} \\
  \label{eq:dispMom2}
  dispZ~&>&~6000\mbox{ mm} \mbox{~~~or~~~} p_\mu~\leq~3\cdot~dispZ \, ;
\end{eqnarray}
where $N_h$ is the number of clusters in the candidate, $dispZ$ is in units of mm, and $p_\mu$ in units of MeV/c.

\begin{figure}
  \begin{center}
    \includegraphics[scale=0.6]{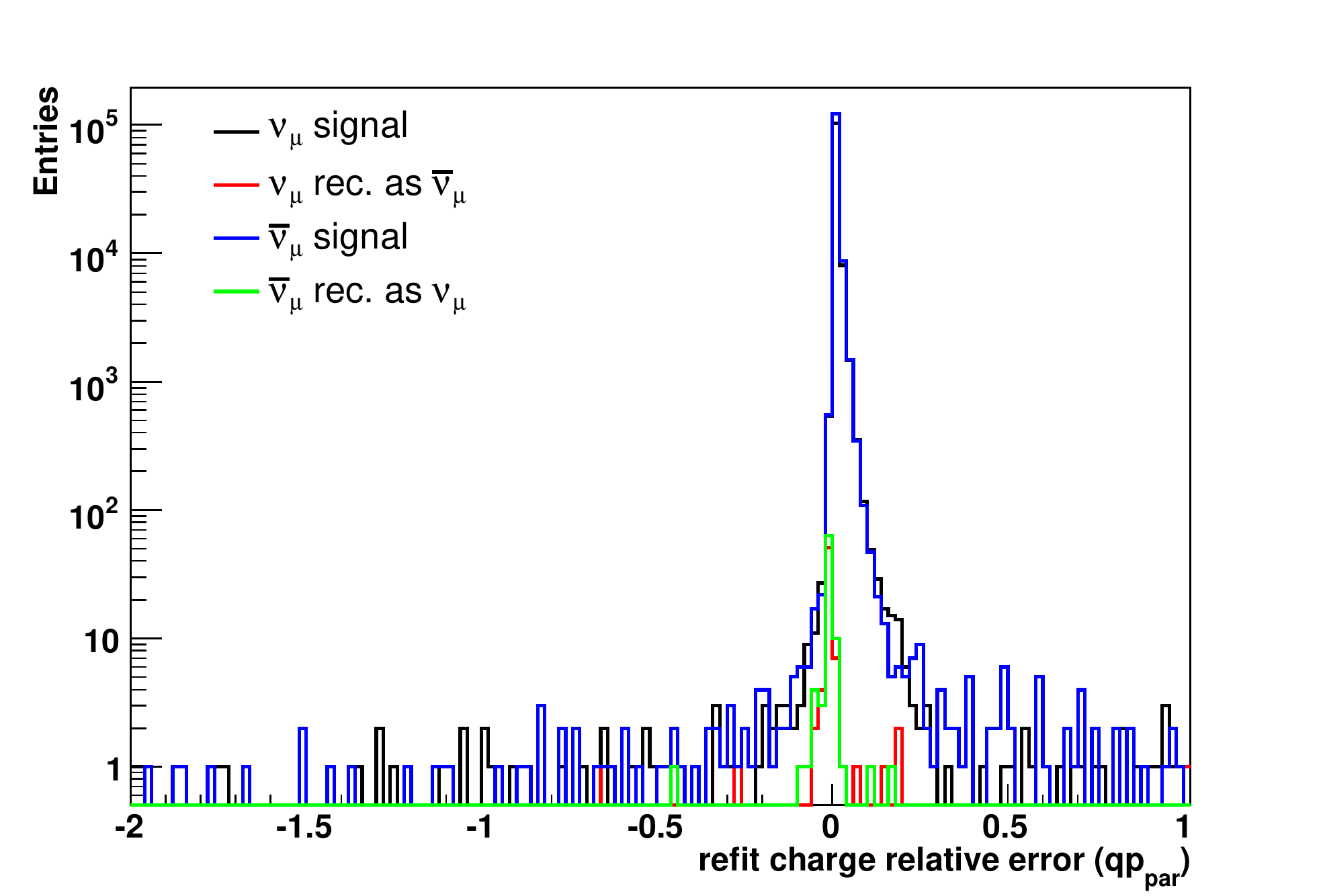}
  \end{center}
  \caption{Distribution of the $qp_{par}$ variable with the region where the parameter is $<$ 0 representing those candidates fitted with charge opposite to the initial Kalman filter (region of highest occupancy shown).}
  \label{fig:qcharge}
\end{figure}

The final cut involves fitting the candidate's projection onto the bending plane to a parabola. If the charge fitted is opposite to that found by the Kalman filter (in the current simulation a negatively-charged muon bends upwards so that for a parabola defined as $a + bz + cz^2$ the parameter $c$ would be positive and the charge of the muon is $Q_{par} = -sign(c)$) the quality of the fit is assessed using the variable:
\begin{equation}
  \label{eq:qCharge}
  \begin{array}{ll}
  qp_{par} &= \left\{ \begin{array}{c} \left|\displaystyle\frac{\sigma_c}{c}\right|\mbox{, if } Q_{par} = Q_{kal} \, ;\\ \\
      -\left|\displaystyle\frac{\sigma_c}{c}\right|\mbox{, if } Q_{par} = -Q_{kal} \, ;\end{array}\right.
  \end{array}
\end{equation}
where $Q_{kal}$ is the charge fitted by the Kalman filter fit. Defining the parameter in this way ensures that the cut is independent of the initial fitted charge. There are two types of events that are accepted: those with $qp_{par} > 0.0$, in which there is no change in charge, and those candidates with $qp_{par} < -1.0$, in which there is a change in charge but the quality of the fit is poor, so the fit cannot be trusted. This $qp_{par}$ cut effectively reduces the background level from CC mis-identification, as can be seen in figure \ref{fig:qcharge}.

\subparagraph*{Cut summary \\}
\label{subpar:Csumm}
In  summary, the test statistic leads to the chain of cuts described above and in table~\ref{tab:G4cutSum}. Application of this analysis and the resulting efficiencies and background suppression are described in the following section.
\begin{table}
\caption{Summary of cuts applied to select the golden channel appearance signals. The level of absolute efficiency and the main background level (with the dominant neutrino species) after each cut are also shown.}
  \label{tab:G4cutSum}
  \begin{center}
    \begin{tabular}{|c|c|c|c|c|c|}
      \hline
      {\bf Cut} & {\bf Acceptance level} & \multicolumn{2}{c|}{{\bf Eff. after cut}}& \multicolumn{2}{c|}{{\bf Main back ($\times 10^{-3}$)}} \\
                &                        & {\bf $\nu_\mu$}                           & {\bf $\overline{\nu}_\mu$} & {\bf $\nu_\mu$} & {\bf $\overline{\nu}_\mu$} \\
      \hline
      \hline
      Fiducial & \emph{z1} $\leq$ 18000~mm & 0.85 & 0.91 & 120($\nu_e$) & 100($\overline{\nu}_e$)\\
      & {\tiny where \emph{z1} is the lowest \emph{z} cluster in the candidate} & & & &  \\
      \hline
      Track quality & $\mathcal{L}_{q/p} > -0.5$ & 0.76 & 0.85 & 20($\nu_e$) & 20($\overline{\nu}_e$)\\
      \hline
      Max. momentum & $P_\mu \leq 40$~GeV & 0.76 & 0.84 & 20($\nu_e$) & 20($\overline{\nu}_e$)\\
      \hline
      CC selection & $\mathcal{L}_1 > 1.0$ & 0.74 & 0.83 & 0.49($\nu_e$) & 1.6($\nu_\mu$)\\
      \hline
      Fitted proportion & $N_{fit}/N_h \geq 0.6$& 0.73 & 0.83 & 0.46($\nu_e$) & 1.2($\nu_\mu$)\\
      \hline
      Kinematic & $E_{rec} \leq 5~GeV \mbox{ or } Q_t > 0.25$ & 0.63 & 0.77 & 0.65($\overline{\nu}_\mu$) & 0.59($\nu_\mu$)\\
      & $E_{rec} \leq 7~GeV \mbox{ or } P_\mu \geq 0.3E_{rec}$&  &  &  &  \\
      \hline
      Displacement & $dispX/dispZ > 0.18 - 0.0026N_h$ & 0.59 & 0.72 & 0.38($\overline{\nu}_\mu$) & 0.38($\nu_\mu$) \\
      & $dispZ > 6000~mm \mbox{ or } P_\mu \leq 3dispZ$ &  &  &  & \\
      \hline
      Quadratic fit & $qp_{par} < -1.0$ or $qp_{par} > 0.0$ & 0.58 & 0.71 & 0.07($\overline{\nu}_\mu$) & 0.07($\nu_\mu$) \\
      \hline
    \end{tabular}
  \end{center}
\end{table}

\paragraph{MIND response to the golden channel}
\label{par:responseMat}
Using a large data set of $3\times 10^6$ events each of $\nu_\mu$ CC, $\overline{\nu}_\mu$ CC, $\nu_e$ CC, $\overline{\nu}_e$ CC and $7\times 10^6$ NC interactions from neutrinos and anti-neutrinos generated using NUANCE and tracked through the GEANT4 representation of MIND, the expected efficiency and background suppression for the reconstruction and analysis of the golden-channel appearance for both polarities of stored muons has been carried out. In addition, an estimate of the likely systematic error of the analysis has been made.

\subparagraph*{Analysis efficiency \\}
\label{subpar:baseline}
The resultant efficiencies for both polarities and the corresponding background levels expected for the appearance channels are summarised in Figs.~\ref{fig:CCback}~--~\ref{fig:G4Eff}. Numeric response matrices for each of the channels may be found in Appendix~\ref{app:response}.
As can be seen in figure \ref{fig:CCback}, the expected level of background from CC misidentification is significantly below $10^{-3}$ at all energies for the new simulation and re-optimised analysis.
\begin{figure}
  \begin{center}$
    \begin{array}{cc}
      \includegraphics[width=8cm, height=6cm]{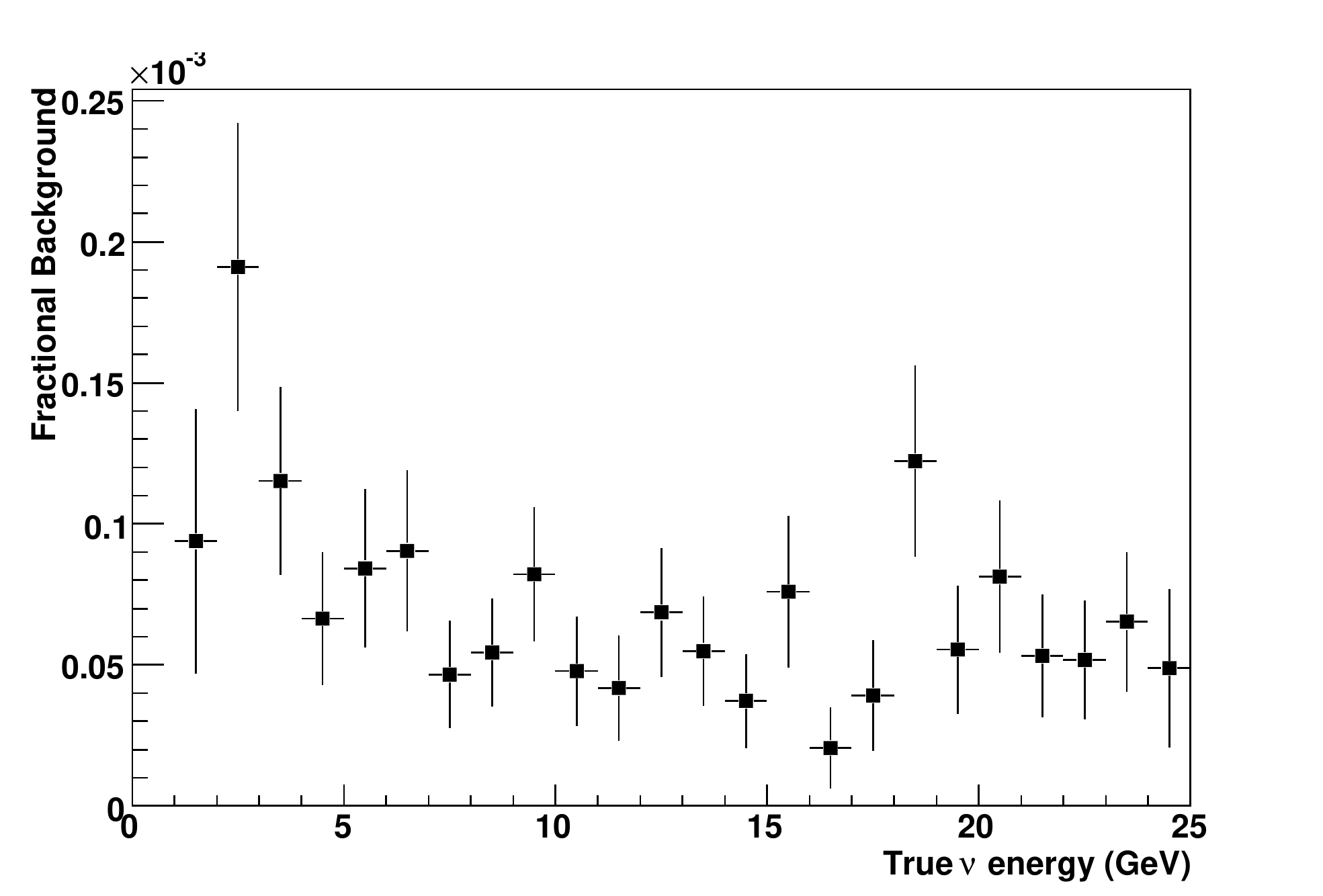} &
      \includegraphics[width=8cm, height=6cm]{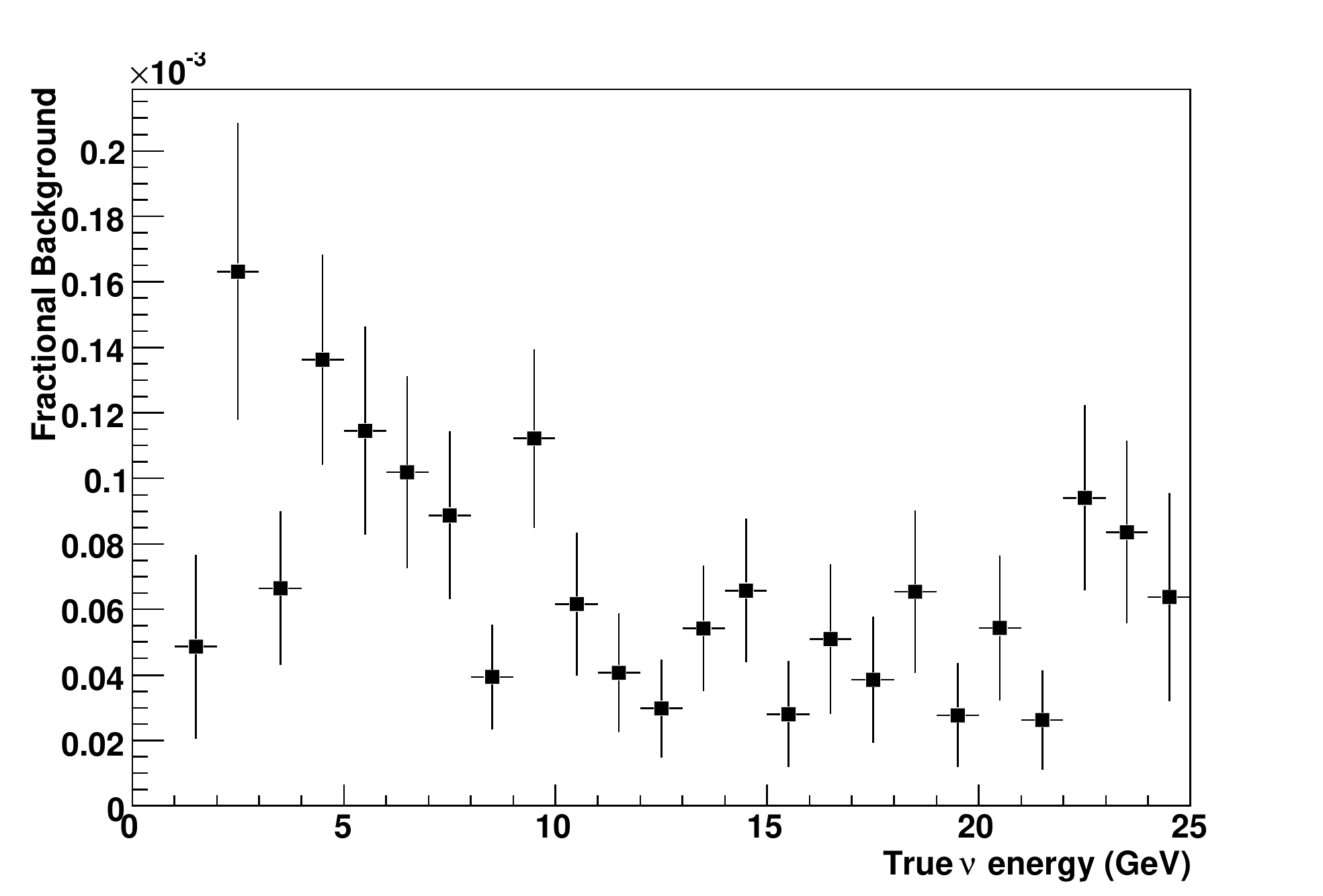} \\
    \end{array}$
  \end{center}
  \caption{Background from mis-identification of $\nu_\mu~(\overline{\nu}_\mu)$ CC interactions as the opposite polarity. (left) $\overline{\nu}_\mu$ CC reconstructed as $\nu_\mu$ CC, (right) $\nu_\mu$ CC reconstructed as $\overline{\nu}_\mu$ CC as a function of true energy.}
  \label{fig:CCback}
\end{figure}

The background from neutral current interactions also lies at or below the $10^{-4}$ level, with the high-energy region exhibiting a higher level than the low-energy region due to the dominance of DIS interactions because of increased visible energy and particle multiplicity. As expected, the NC background tends to be reconstructed at low energy due to the missing energy.
\begin{figure}
  \begin{center}$
    \begin{array}{cc}
      \includegraphics[width=7.5cm, height=5.5cm]{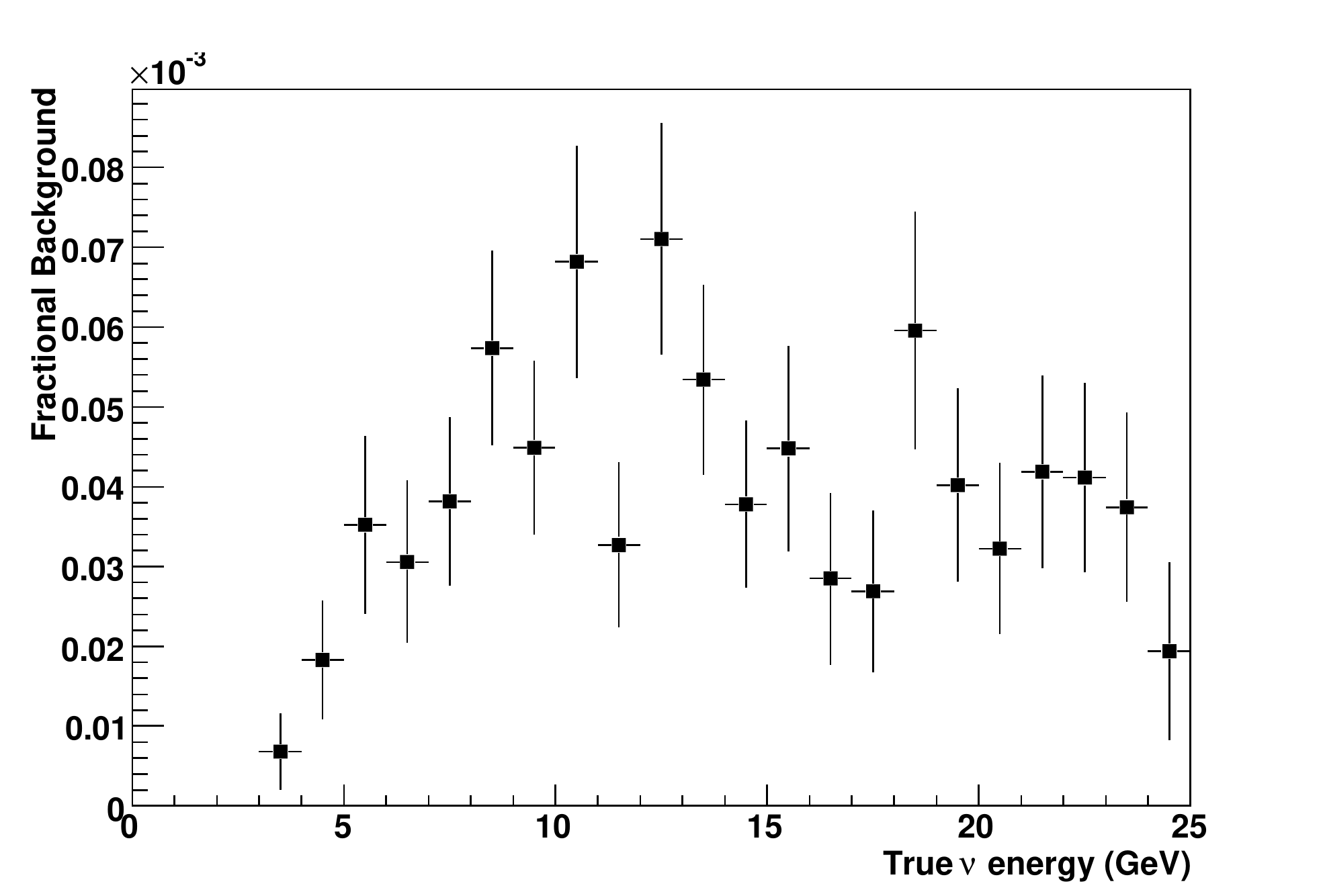} &
      \includegraphics[width=7.5cm, height=5.5cm]{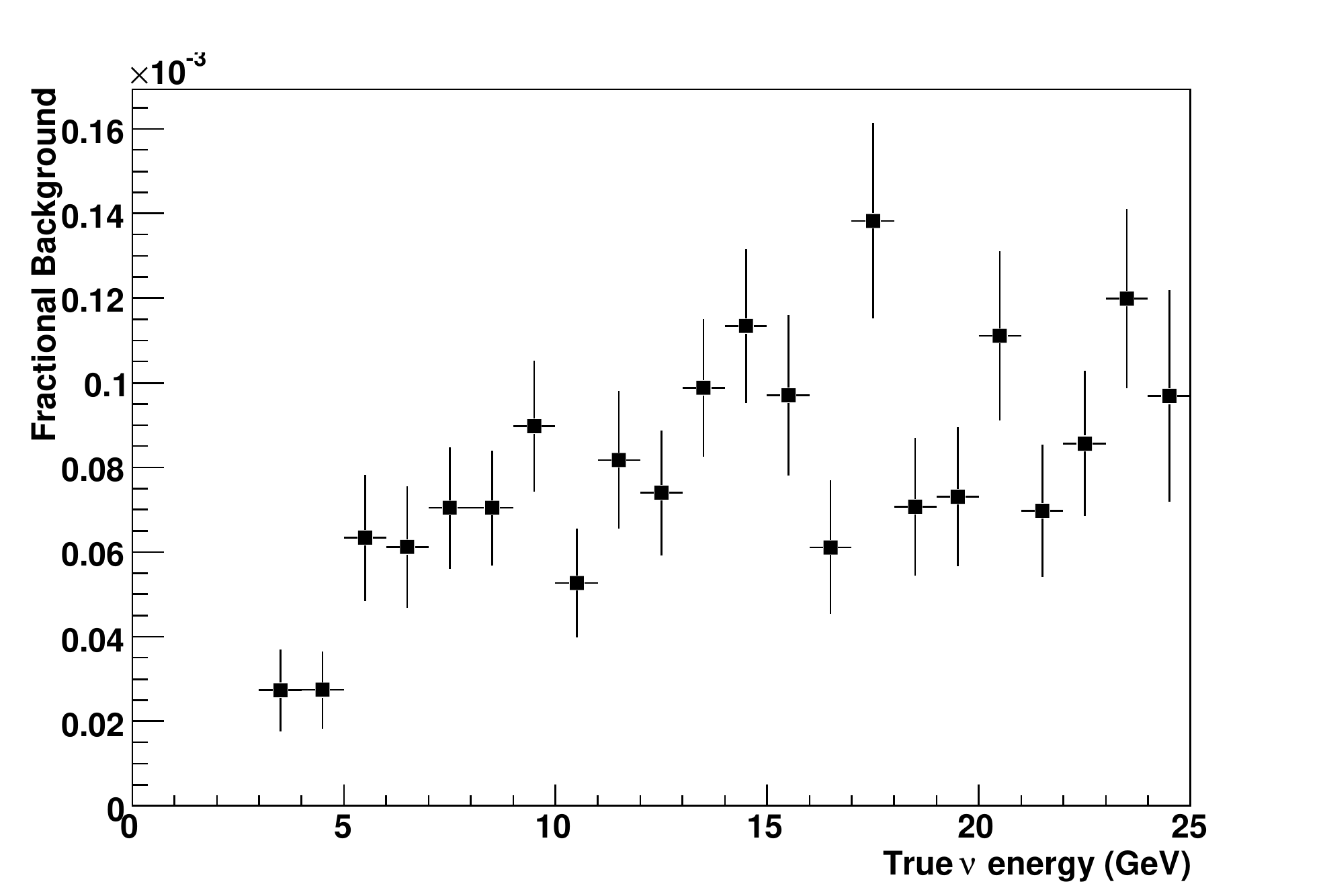} \\
    \end{array}$
  \end{center}
  \caption{Background from mis-identification of NC interactions as $\nu_\mu~(\overline{\nu}_\mu)$ CC interactions. (left) NC reconstructed as $\nu_\mu$ CC, (right) NC reconstructed as $\overline{\nu}_\mu$ CC. (top) as a function of true energy.}
  \label{fig:NCback}
\end{figure}

The background from $\nu_e~(\overline{\nu}_e)$ CC interactions is once again expected to constitute a very low level addition to the observed signal. This background is particularly well suppressed due to the thickness of the iron plates and the tendency for the electron shower to overlap with any hadronic activity.
\begin{figure}
  \begin{center}$
    \begin{array}{cc}
      \includegraphics[width=7.5cm, height=5.5cm]{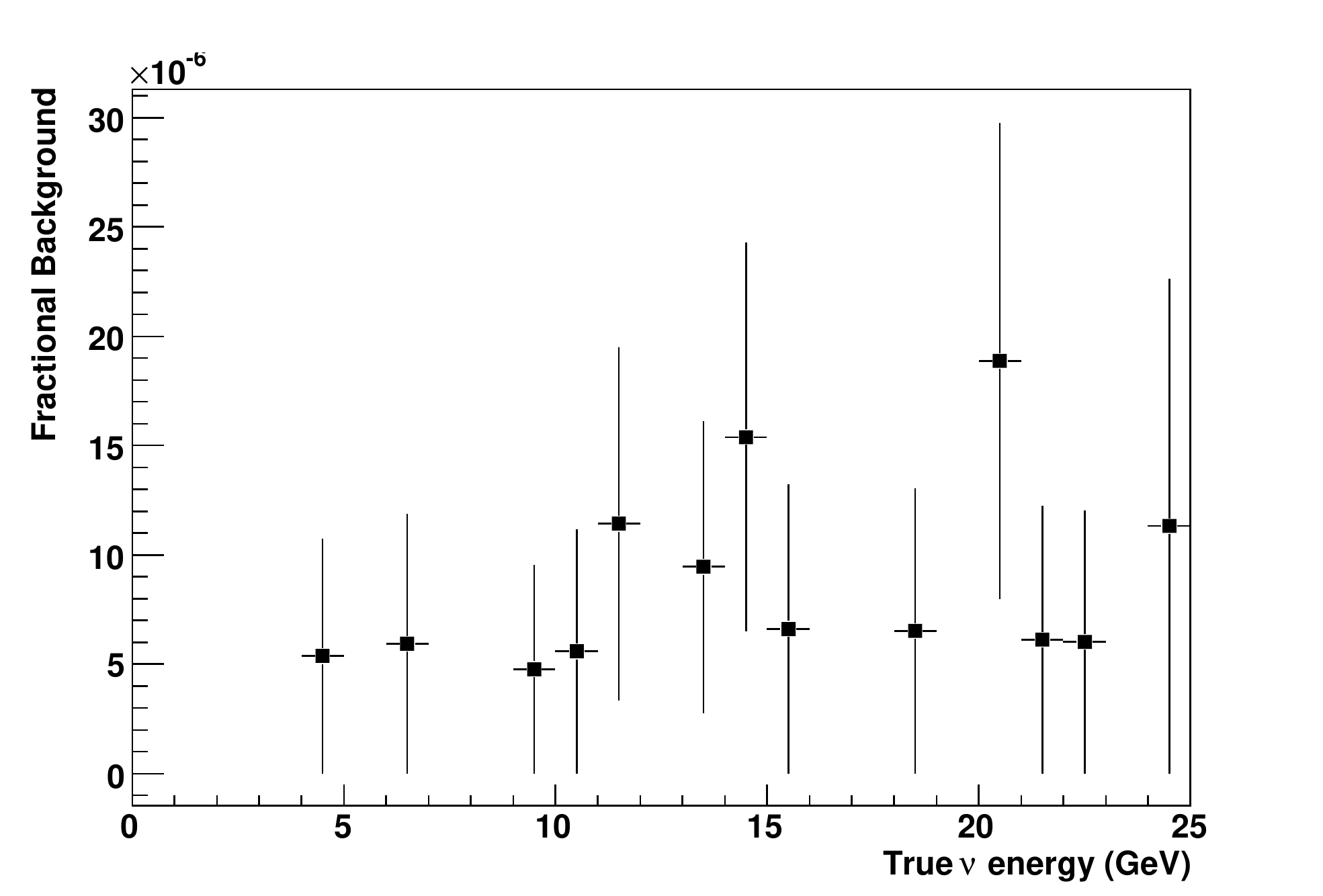} &
      \includegraphics[width=7.5cm, height=5.5cm]{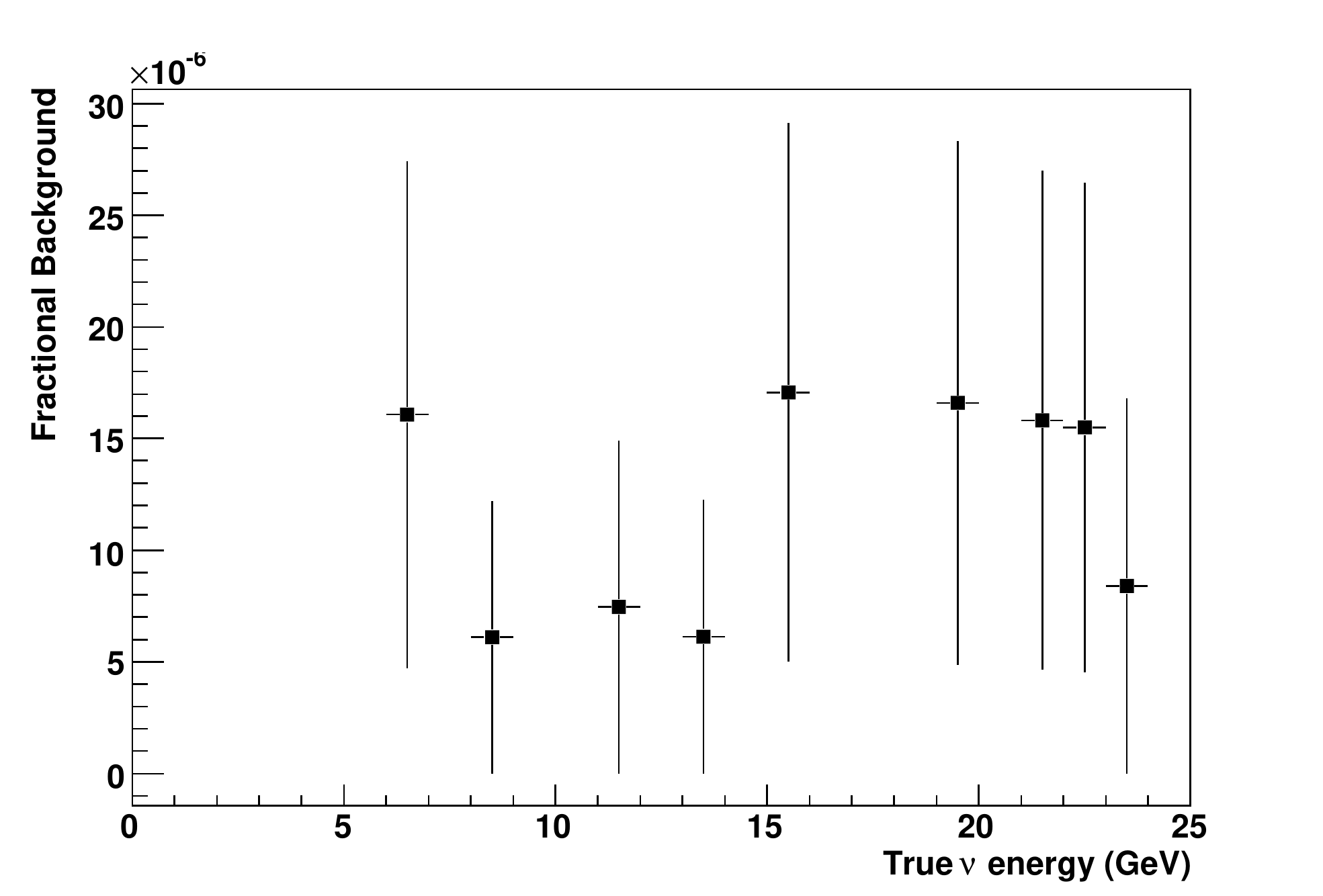} \\
    \end{array}$
  \end{center}
  \caption{Background from mis-identification of $\nu_e~(\overline{\nu}_e)$ CC interactions as $\nu_\mu~(\overline{\nu}_\mu)$ CC interactions. (left) $\nu_e$ CC reconstructed as $\nu_\mu$ CC, (right) $\overline{\nu}_e$ CC reconstructed as $\overline{\nu}_\mu$ CC as a function of true energy.}
  \label{fig:eCback}
\end{figure}

The efficiency of detection of the two $\nu_\mu$ polarities was expected to have a threshold lower than that seen in previous studies due to the presence of non-DIS interactions in the data sample. The efficiencies expected for the current analysis are shown in figure \ref{fig:G4Eff}.
\begin{figure}
  \begin{center}$
    \begin{array}{cc}
      \includegraphics[width=7.5cm, height=5.5cm]{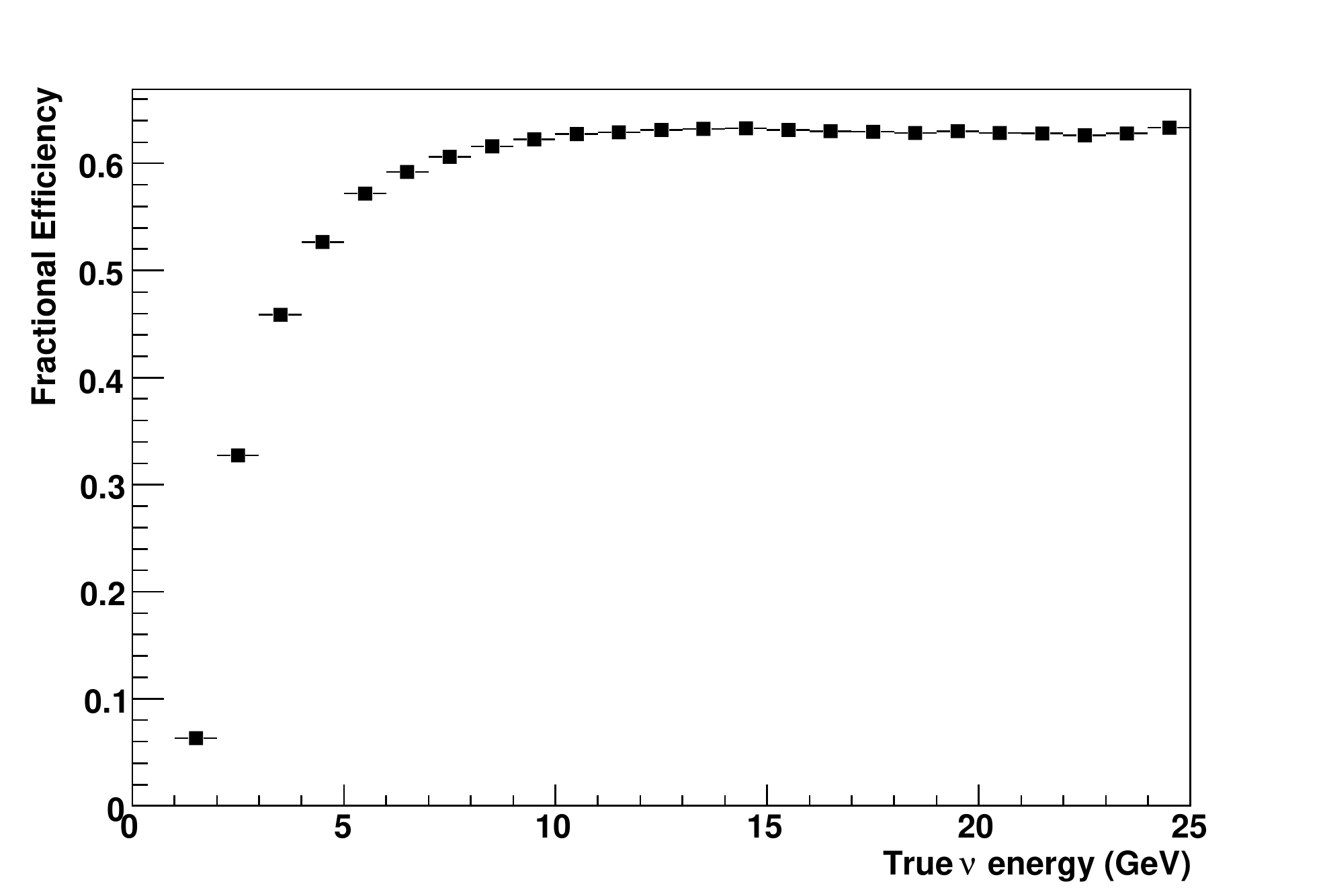} &
      \includegraphics[width=7.5cm, height=5.5cm]{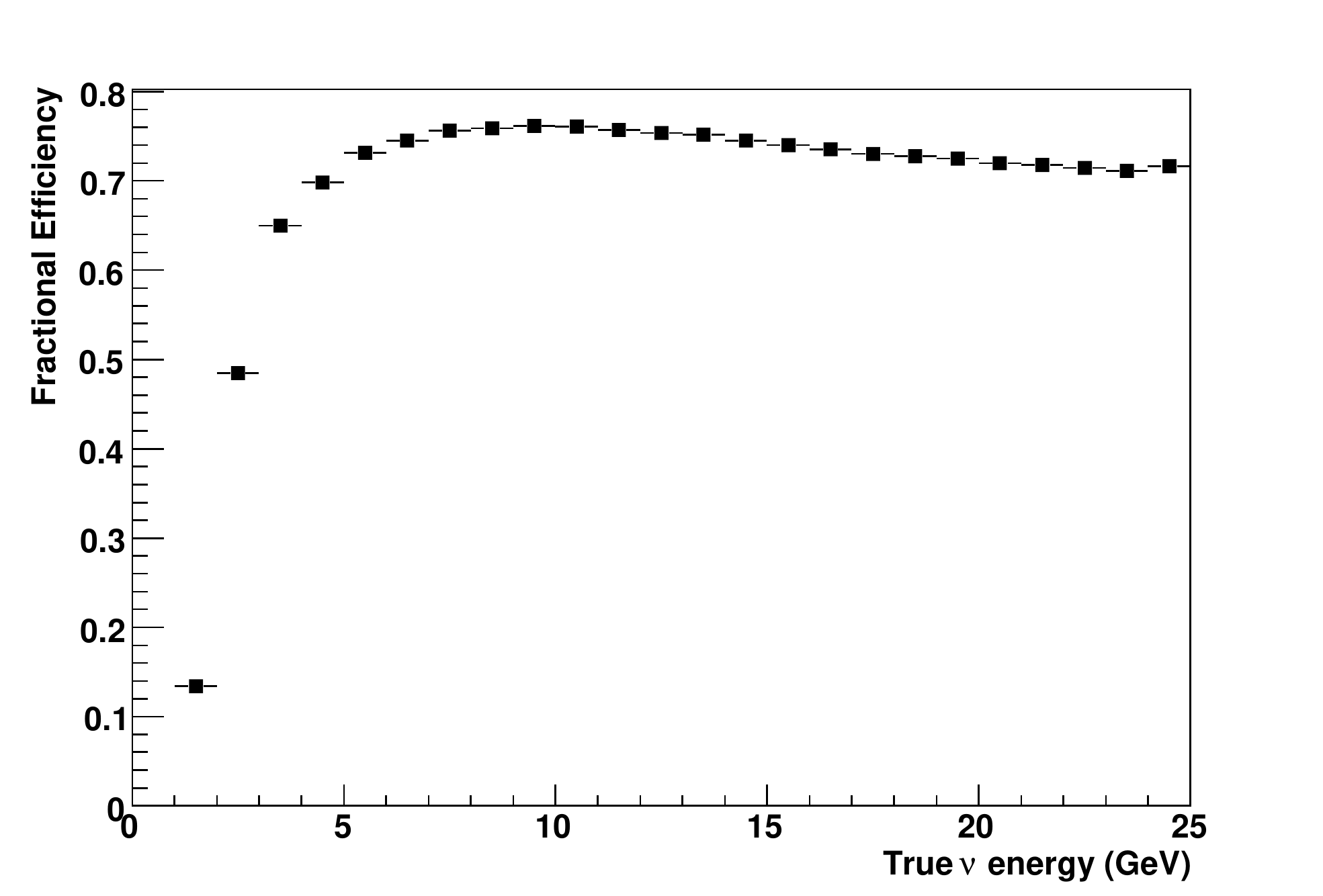} \\
    \end{array}$
  \end{center}
  \caption{Efficiency of reconstruction of $\nu_\mu~(\overline{\nu}_\mu)$ CC interactions. (left) $\nu_\mu$ CC efficiency, (right) $\overline{\nu}_\mu$ CC efficiency as a function of true energy.}
  \label{fig:G4Eff}
\end{figure}

\begin{figure}
  \begin{center}
    \includegraphics[width=9.5cm, height=7.5cm]{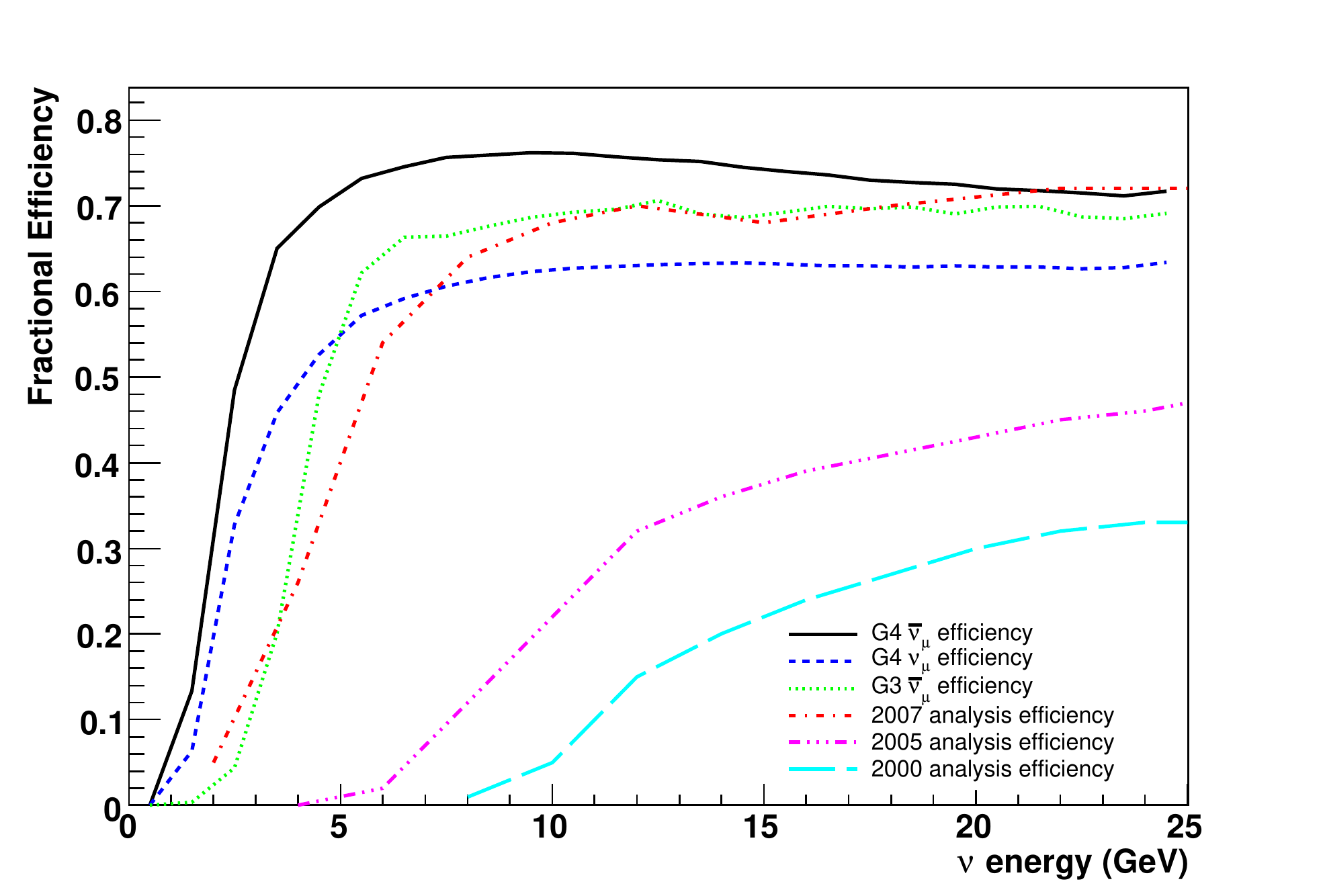}
  \end{center}
  \caption{Evolution of the MIND $\overline{\nu}_\mu$ CC detection efficiency. For the 2000 analysis see~\cite{Cervera:2000kp}, for 2005 see~\cite{CerveraVillanueva:2005ym} and for 2007 see~\cite{CerveraVillanueva:2008zz}.}
  \label{fig:EffEvolv}
\end{figure}

Figure \ref{fig:EffEvolv} shows a comparison of the resultant
$\overline{\nu}_\mu$ efficiency with that extracted in previous
studies.  
In the studies described in
\cite{Cervera:2000kp,CerveraVillanueva:2008zz,CerveraVillanueva:2005ym},
the improvement in threshold with the development of the analysis and
the introduction of the full spectrum of interactions is clear. 
The thresholds of the newest results, between 2\,GeV and 3\,GeV,
correspond well with the expected saturation point of sensitivity to
the extraction of the oscillation parameters. 
The inclusion of non-DIS events is responsible for reducing the
threshold, manifested by comparing the black and blue curves (G4
efficiency) with the green curve (G3 efficiency) in figure
\ref{fig:EffEvolv} that only included DIS events. 

\begin{figure}
  \begin{center}$
    \begin{array}{cc}
      \includegraphics[width=7.5cm, height=5.5cm]{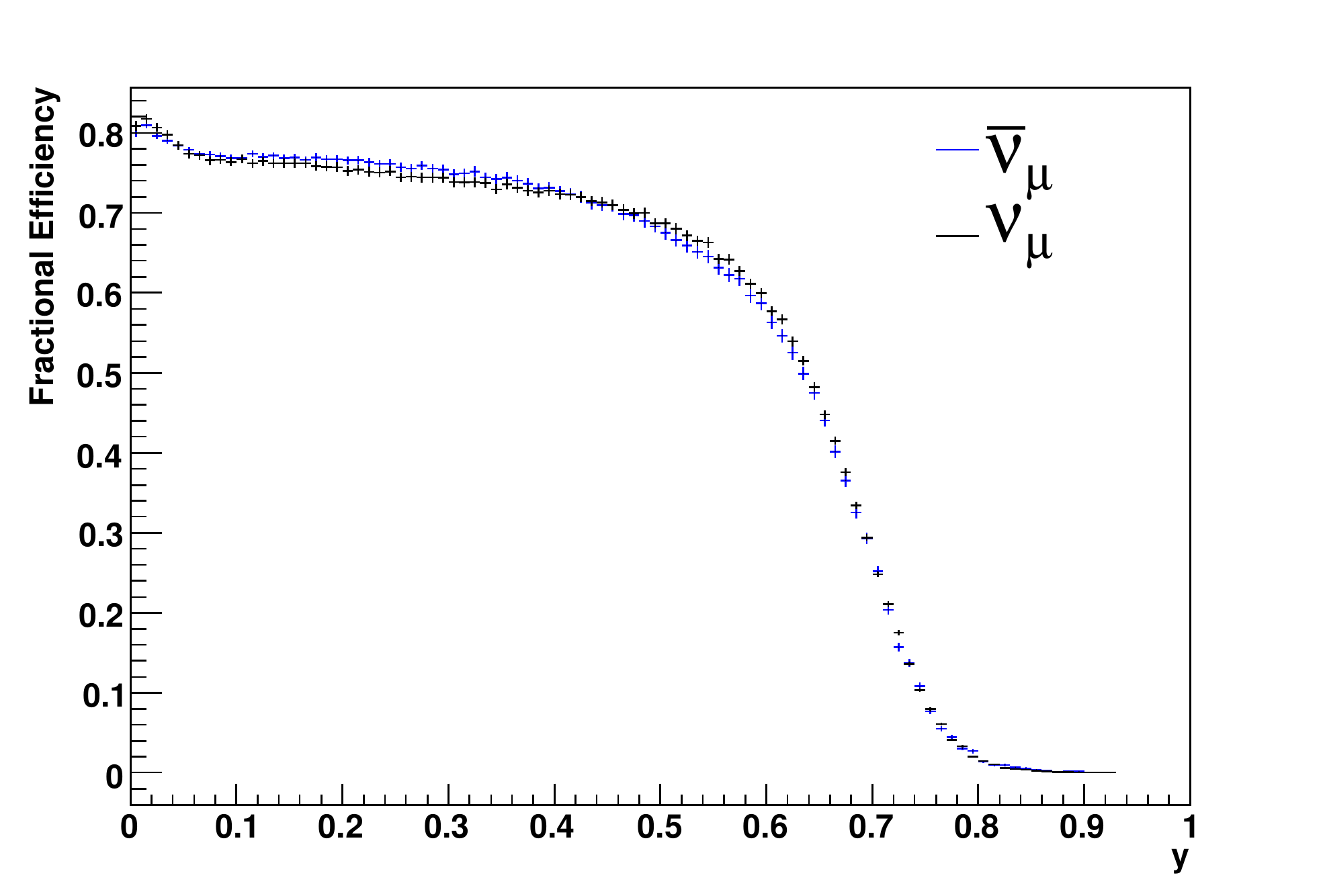} &
      \includegraphics[width=7.5cm, height=5.5cm]{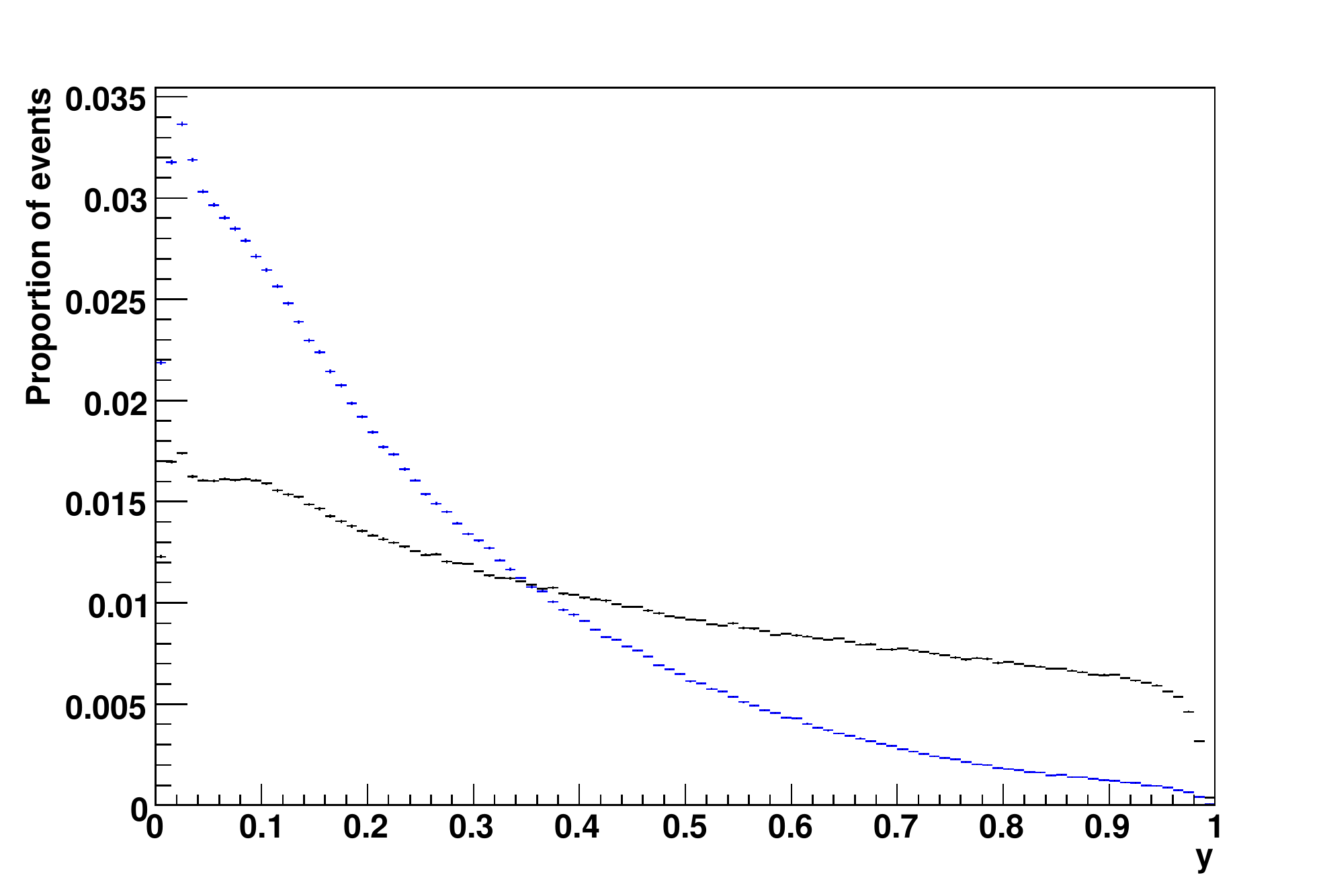}\\
    \end{array}$
  \end{center}
  \caption{$\nu_\mu$ CC and $\overline{\nu}_\mu$ CC signal detection efficiency as a function of $y$ (left) and the normalised distribution of all events considered in each polarity as a function of $y$ (right).}
  \label{fig:Effbjorken}
\end{figure}

The difference in efficiency between the two appearance channels is effectively described by the difference in the inelasticity of neutrino and anti-neutrino CC interactions. Neutrino DIS interactions with quarks have a  flat distribution in the Bjorken variable:
\begin{equation}
  y = \displaystyle\frac{E_\nu - E_l}{E_\nu} \, ;
\end{equation}
with $E_l$ being the scattered-lepton energy. However, anti-neutrinos interacting with quarks follow a  distribution $\propto~(1-y)^2$ ~\cite{Zuber:2004nz}. For this reason, neutrino interactions generally involve a greater energy transfer to the target. As can be seen from figure \ref{fig:Effbjorken}-(top left), the efficiencies for the two species, as a function of $y$ after all cuts, are the same to within the uncertainties over the full $y$ range. Hence, the difference in neutrino and anti-neutrino efficiencies, when translated into true neutrino-energy, can be explained by the greater abundance of neutrino events at high $y$. However, since the cross section for the interaction of neutrinos is approximately twice that for anti-neutrinos it is not expected that this reduced efficiency will affect the fit to the observed spectrum significantly.

\subparagraph*{Study of systematic uncertainties \\}
\label{subSec:syst}
The efficiencies described above will be affected by several systematic effects. There will be many contributing factors including uncertainty in the determination of the parameters used to form the cuts in the analysis, uncertainty in the determination of the hadronic shower energy and direction resolution, uncertainty in the relative proportions of the different interaction types and any assumptions in the representation of the detector and electronics. While exact determination of the overall systematic error in the efficiencies is complicated, an estimate of the contribution of different factors can be obtained by setting certain variables to the extremes of their errors.

\begin{figure}
  \begin{center}$
    \begin{array}{cc}
      \includegraphics[width=8cm, height=6cm]{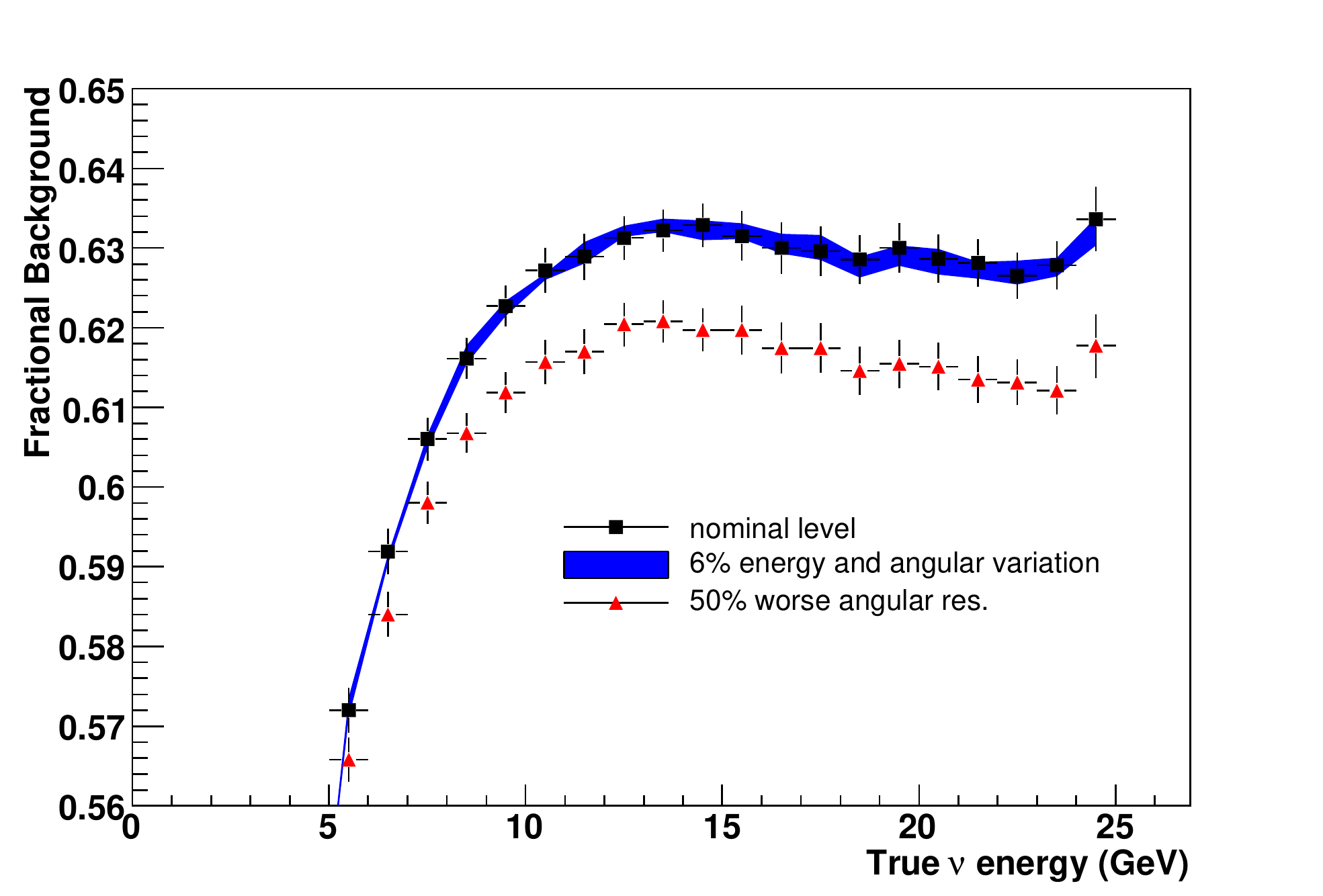} &
      \includegraphics[width=8cm, height=6cm]{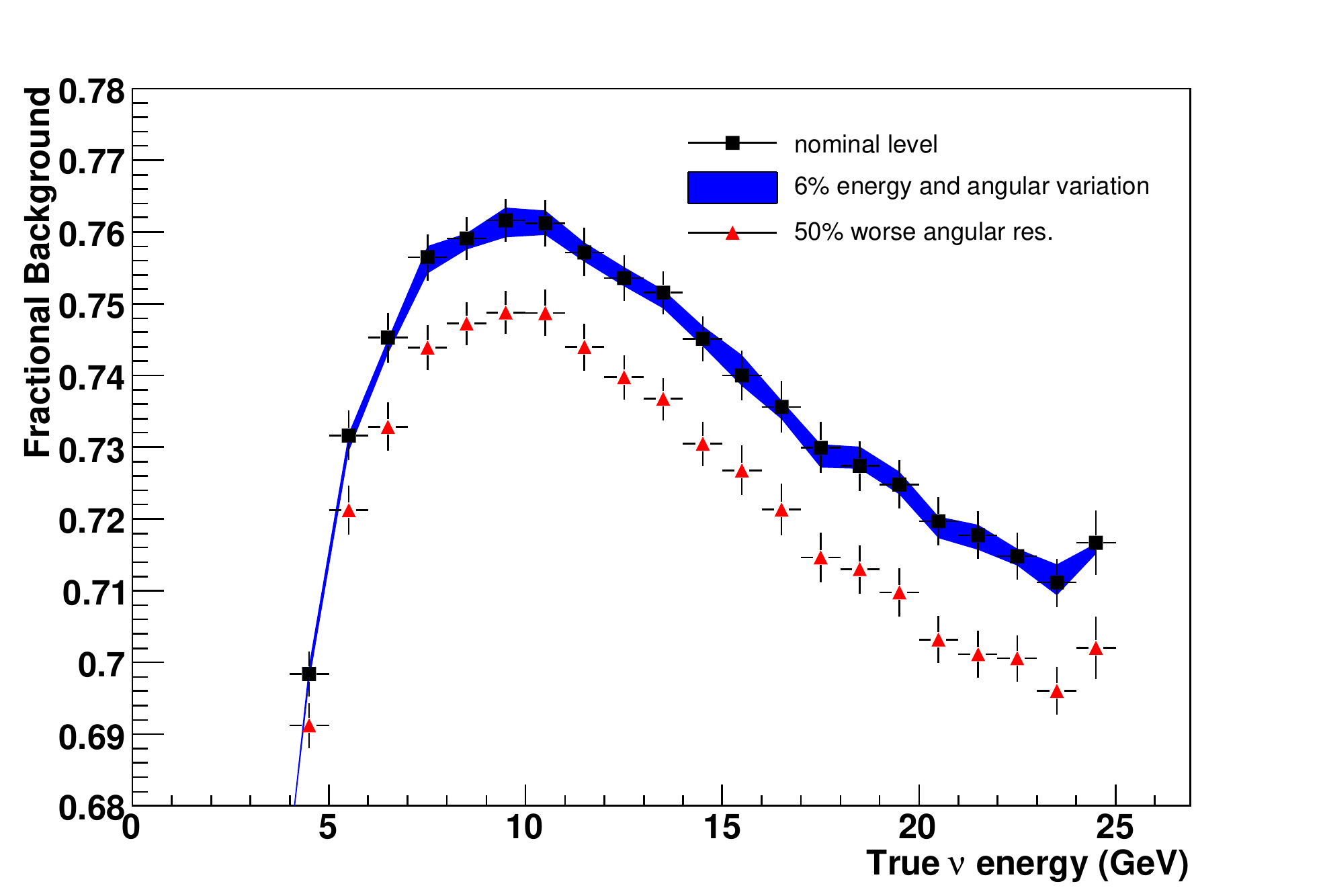}\\
      \includegraphics[width=8cm, height=6cm]{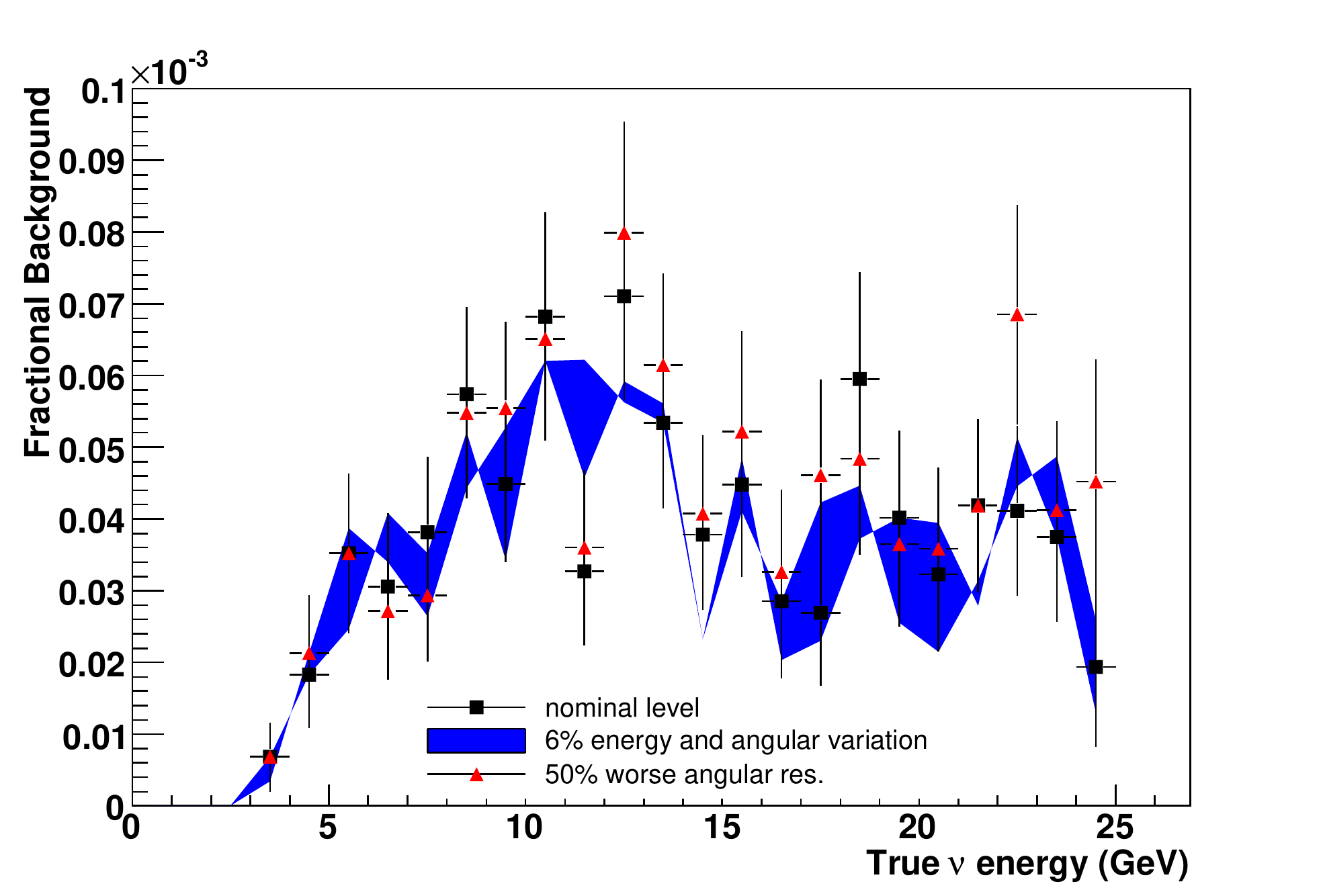} &
      \includegraphics[width=8cm, height=6cm]{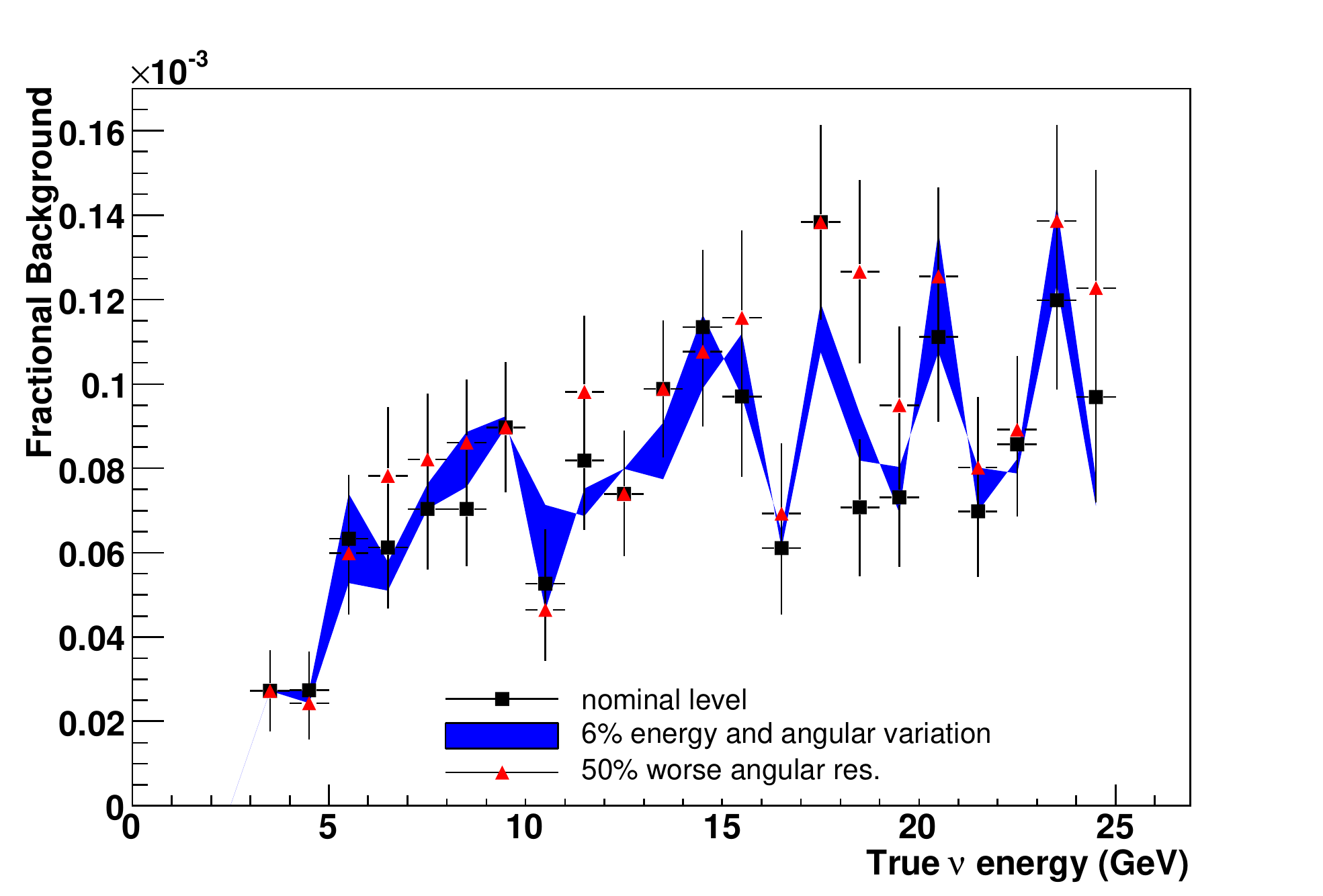}
    \end{array}$
  \end{center}
  \caption{Variation of signal efficiency (top) and NC backgrounds (bottom) due to a 6\% variation in the hadron shower energy and direction resolution and a more pessimistic 50\% reduction in angular resolution (focused on region of greatest variation).}
  \label{fig:hadresSyst}
\end{figure}

Since the neutrino energy is reconstructed generally from the sum of the hadron shower and primary muon energies, and this reconstructed quantity is used in the kinematic cuts, uncertainty in the energy and direction resolution of the detector could contribute significantly to the systematic error in the efficiencies. Taking a 6\% error as quoted for the energy scale uncertainty assumed by the MINOS collaboration~\cite{Michael:2008bc} and varying the constants of the energy and direction smears by this amount, it can be seen (figure \ref{fig:hadresSyst}) that, to this level, the hadronic resolutions have little effect on the true neutrino-energy efficiencies. However, the hadronic direction resolution is likely to have far greater uncertainty and would be very sensitive to noise in the electronics. Also shown in figure \ref{fig:hadresSyst} are the efficiencies when the hadronic energy resolution parameters are 6\% larger but with a 50\% increase in the angular resolution parameters. Even at this level the observed difference in efficiency is only at the level of 1\%.

\begin{figure}
  \begin{center}$
    \begin{array}{cc}
      \includegraphics[width=7.5cm, height=4.5cm]{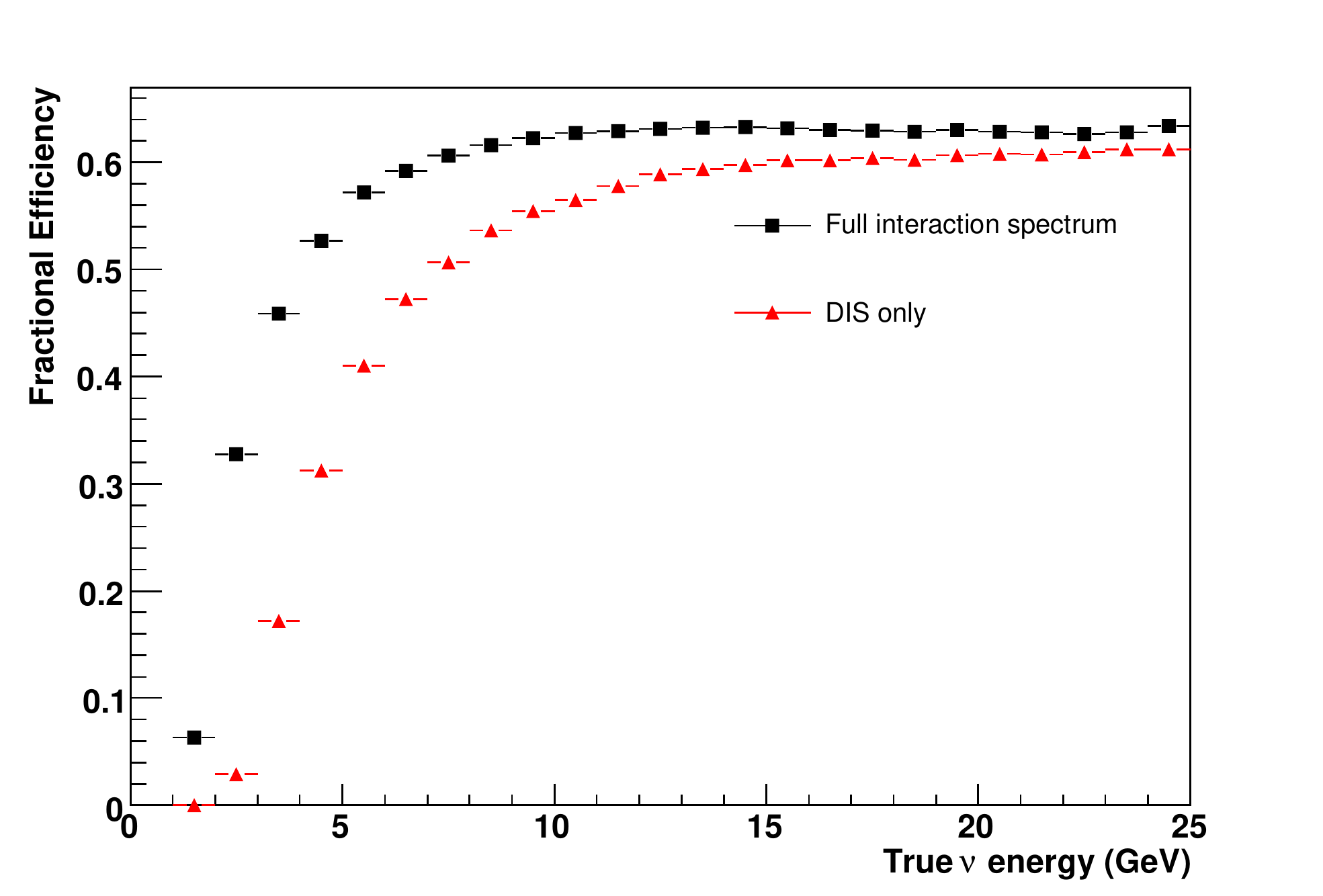} &
      \includegraphics[width=7.5cm, height=4.5cm]{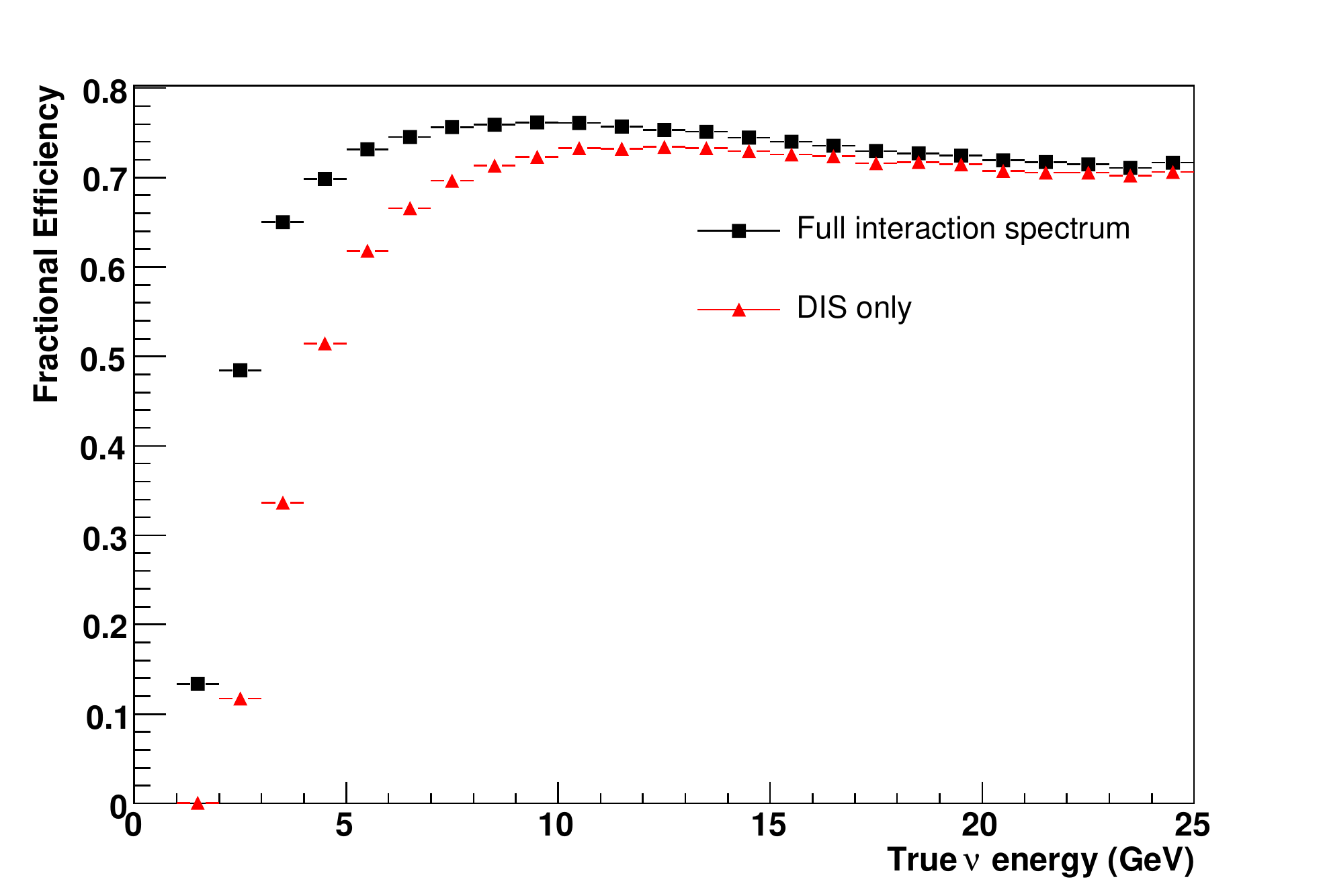}\\
      \includegraphics[width=7.5cm, height=4.5cm]{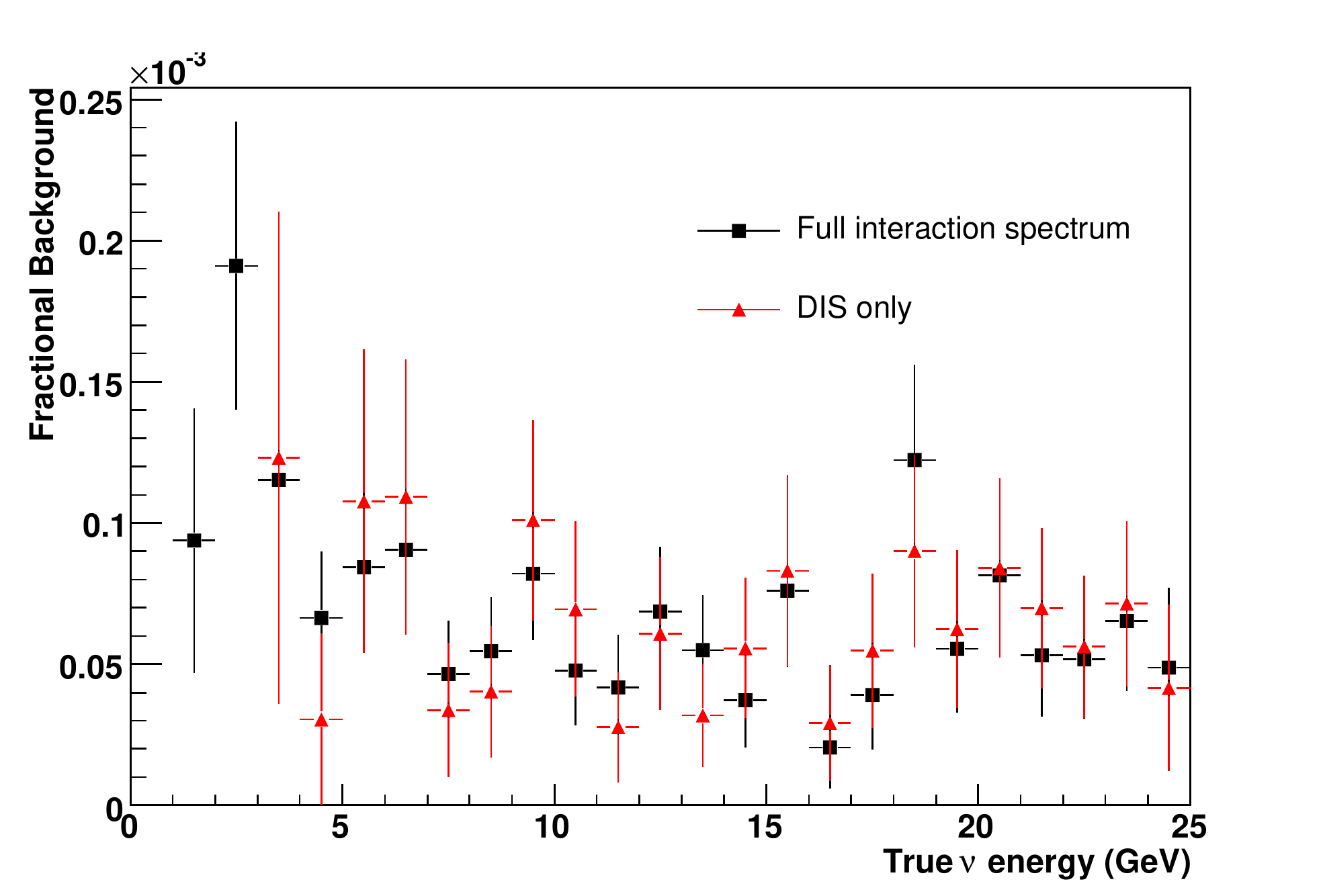} &
      \includegraphics[width=7.5cm, height=4.5cm]{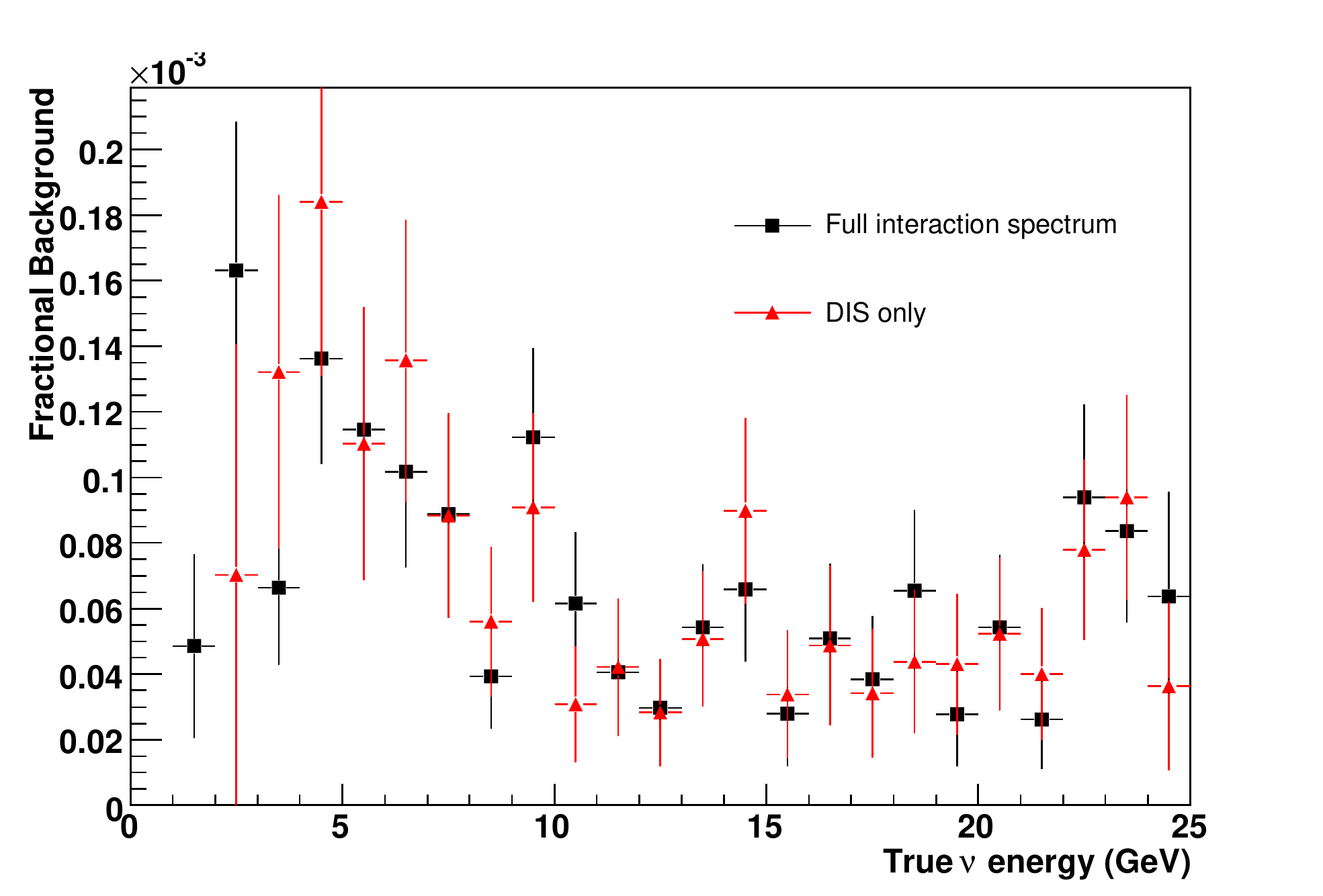}\\
      \includegraphics[width=7.5cm, height=4.5cm]{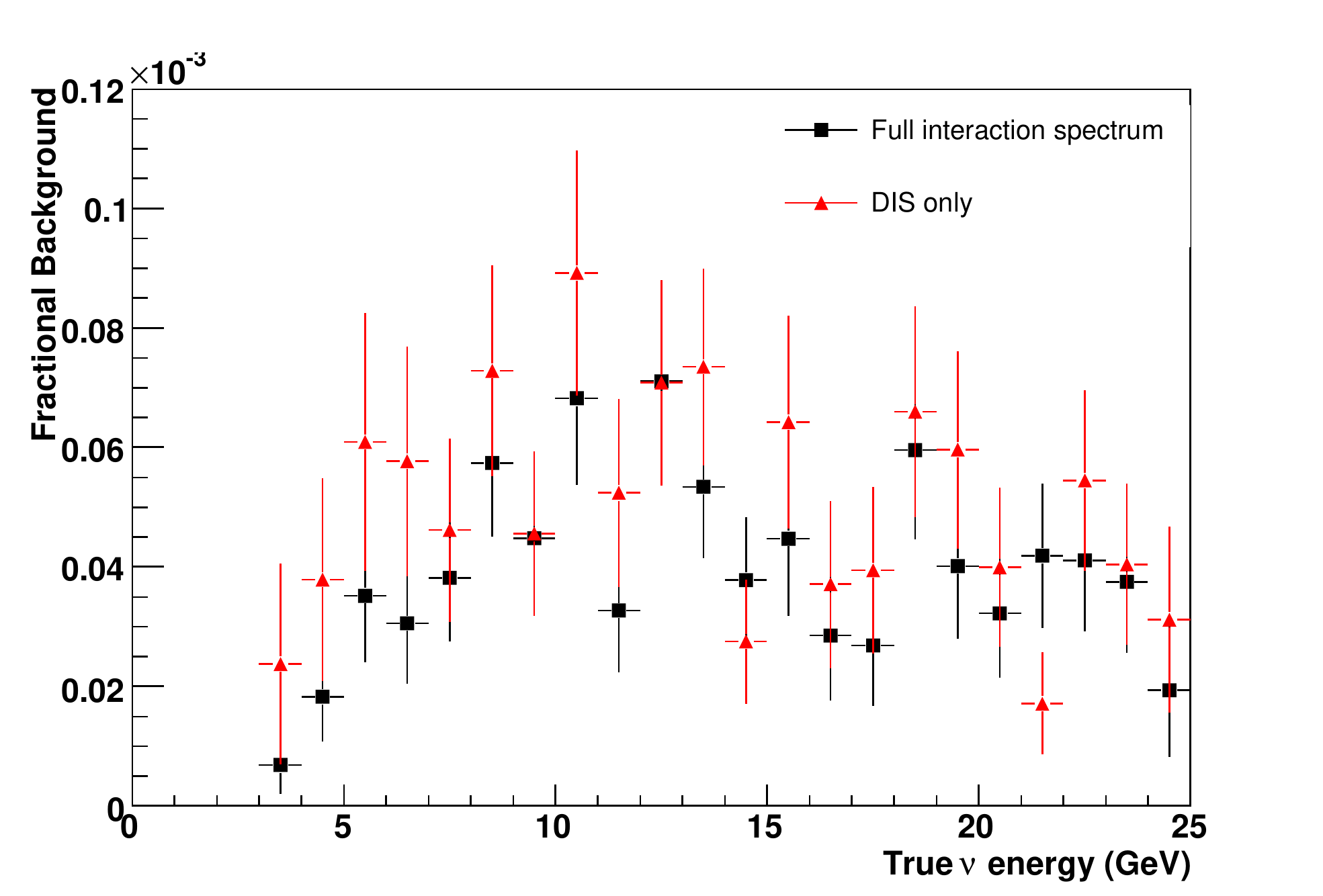} &
      \includegraphics[width=7.5cm, height=4.5cm]{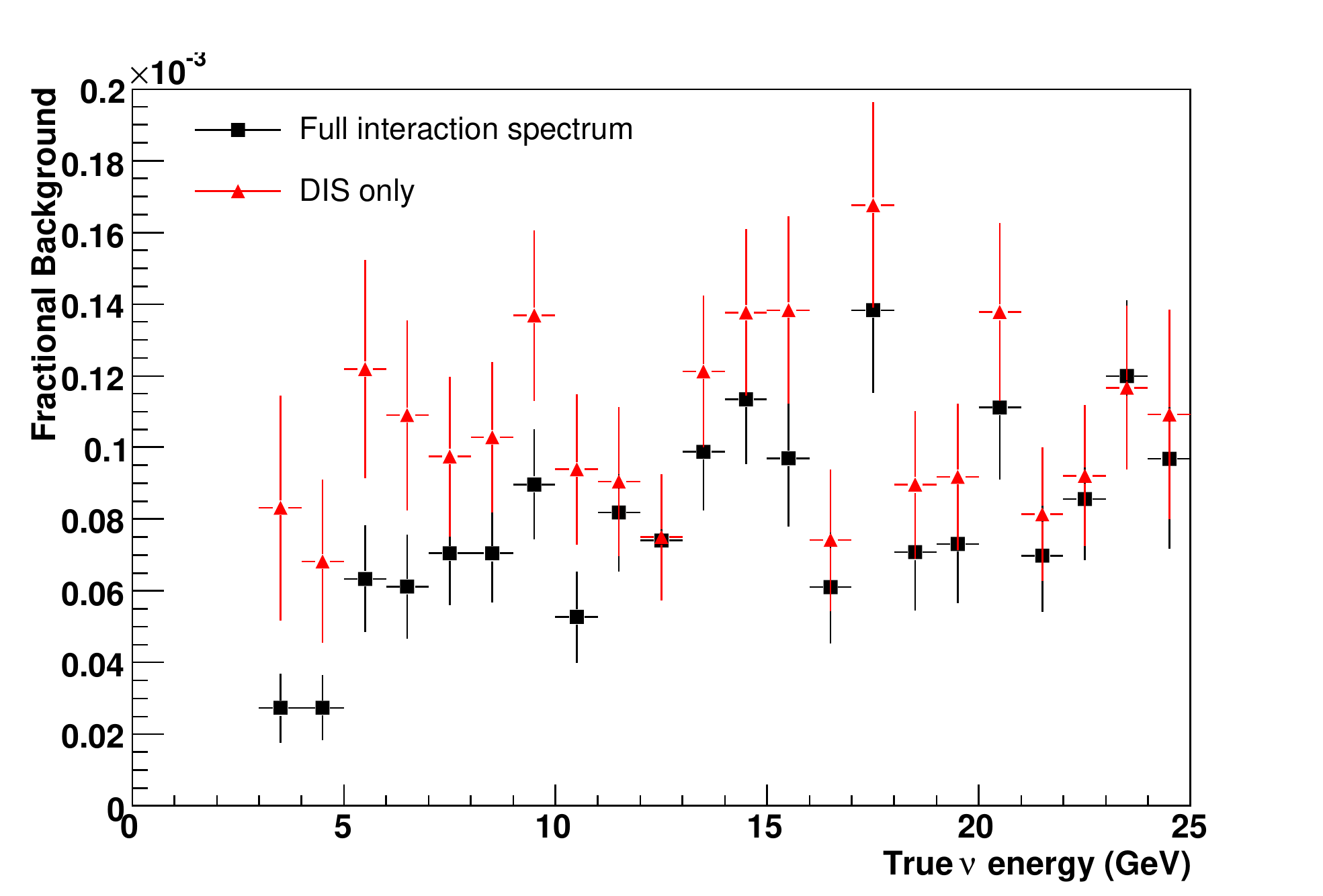}\\
      \includegraphics[width=7.5cm, height=4.5cm]{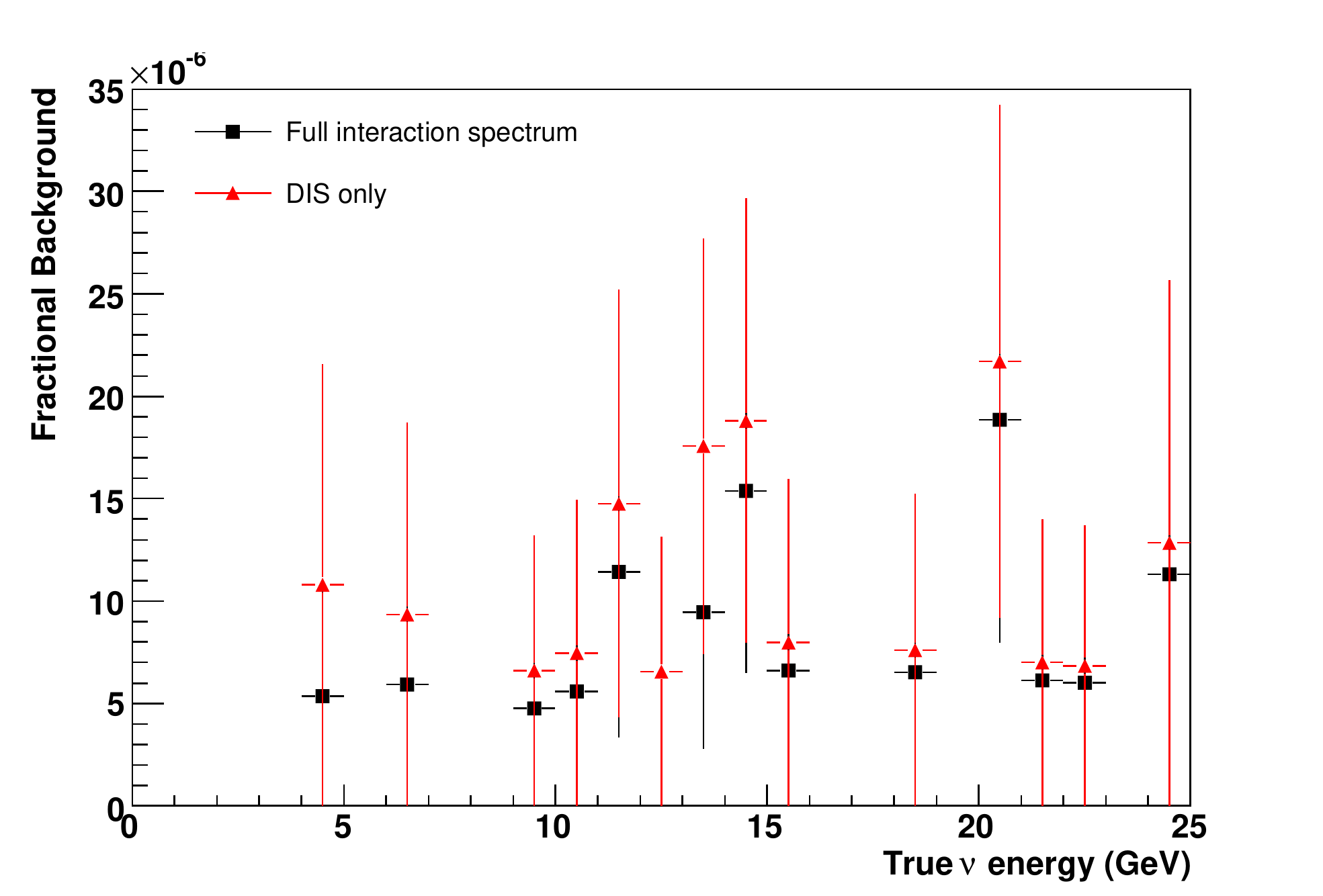} &
      \includegraphics[width=7.5cm, height=4.5cm]{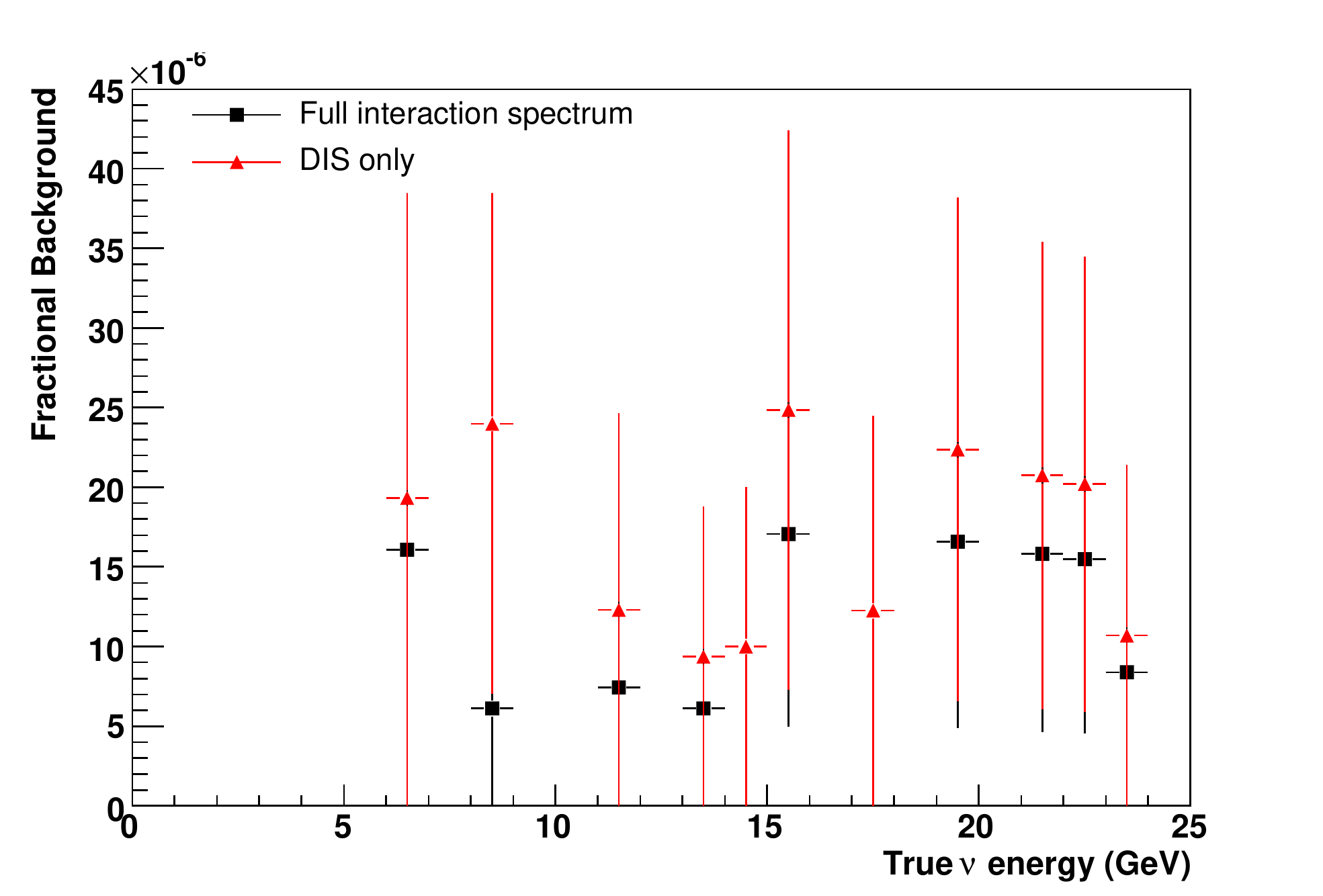}
    \end{array}$
  \end{center}
  \caption{Efficiencies for a pure DIS sample compared to the nominal case. (top) Signal efficiency, (second line) $\nu_\mu~(\overline{\nu}_\mu)$ CC background, (third line) NC background and (bottom) $\nu_e~(\overline{\nu}_e)$ CC background. $\nu_\mu$ appearance on the left and $\overline{\nu}_\mu$ appearance on the right.}
  \label{fig:DISonly}
\end{figure}

\begin{figure}
  \begin{center}$
    \begin{array}{cc}
      \includegraphics[width=7.5cm, height=5.5cm]{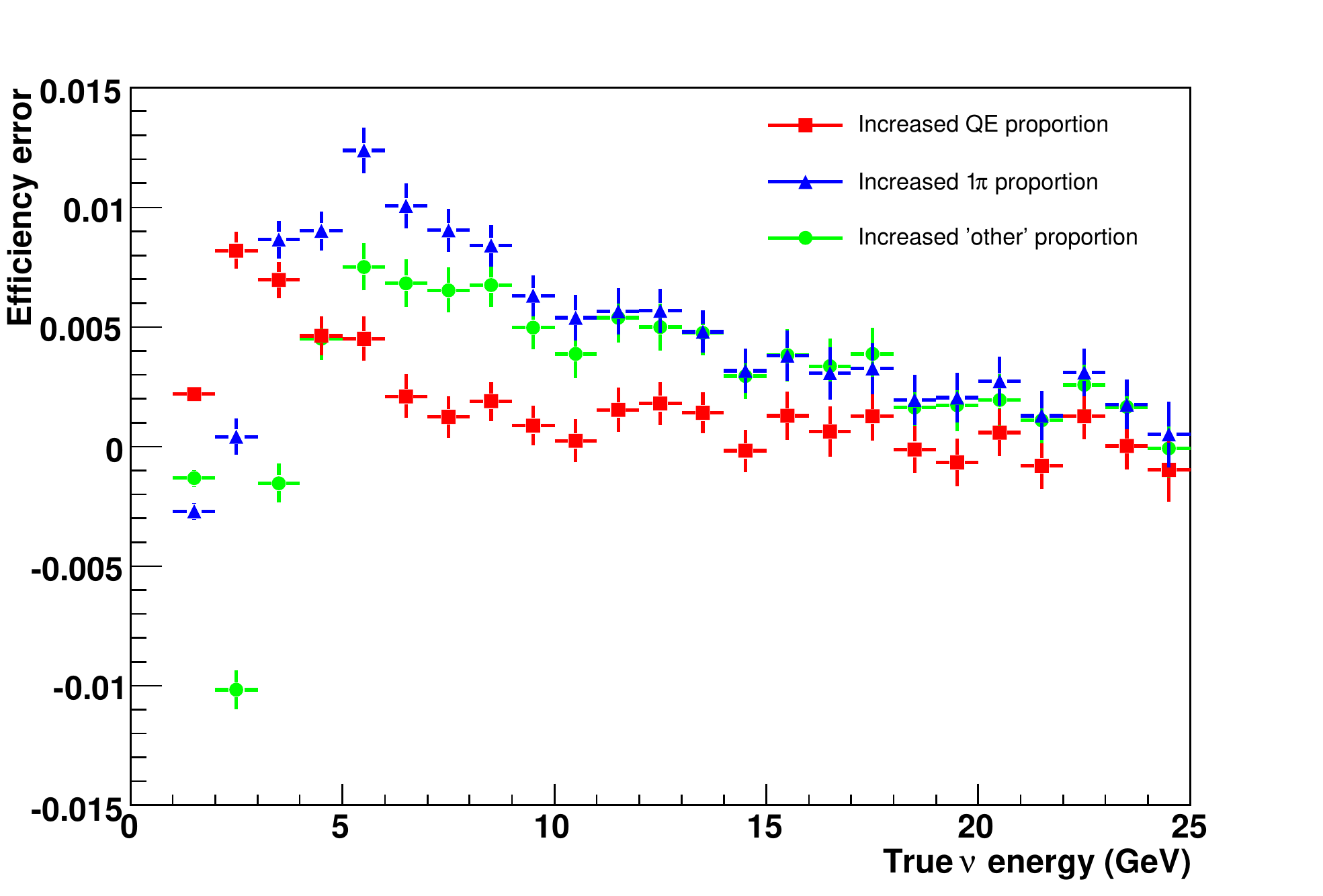} &
      \includegraphics[width=7.5cm, height=5.5cm]{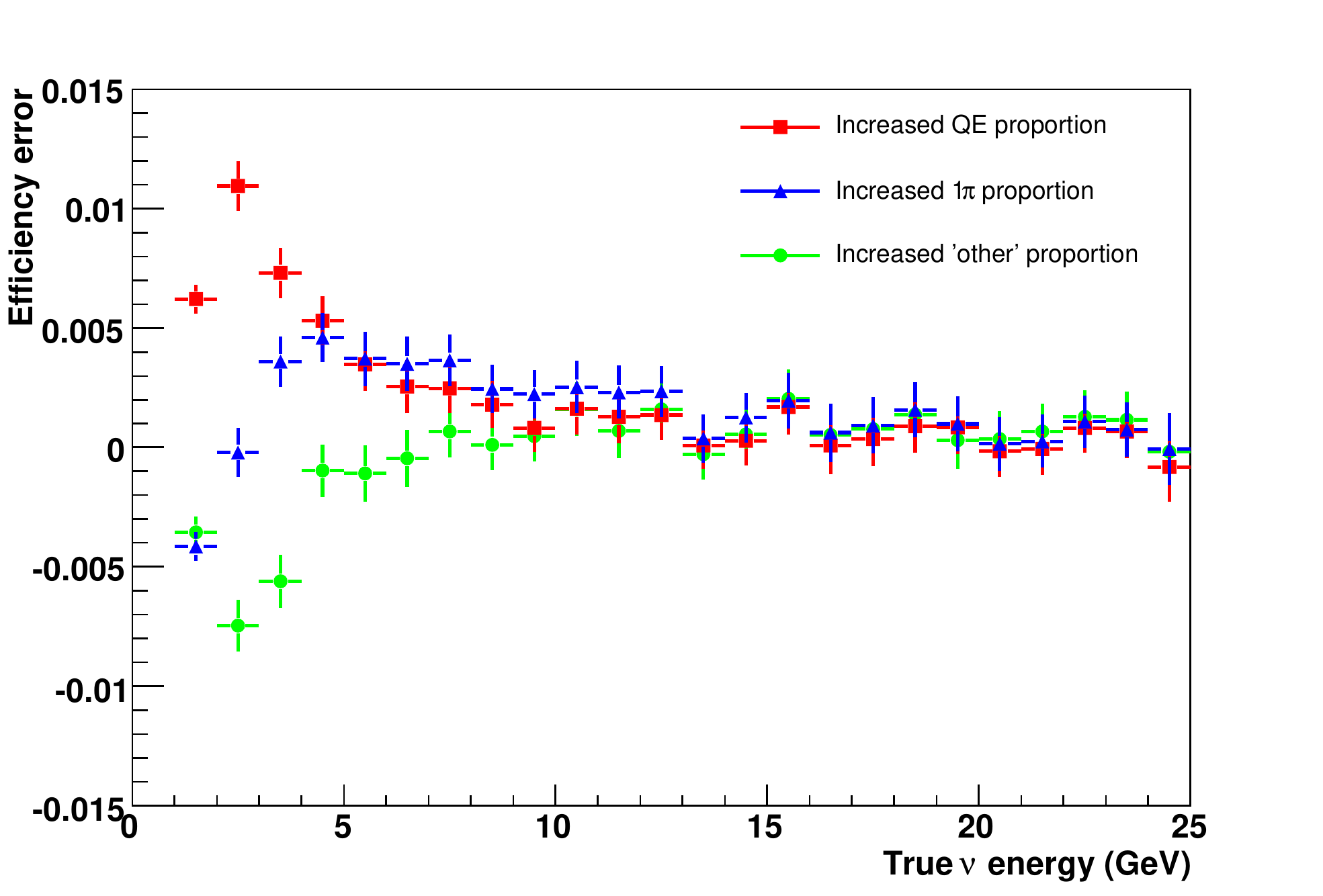}\\
      \includegraphics[width=7.5cm, height=5.5cm]{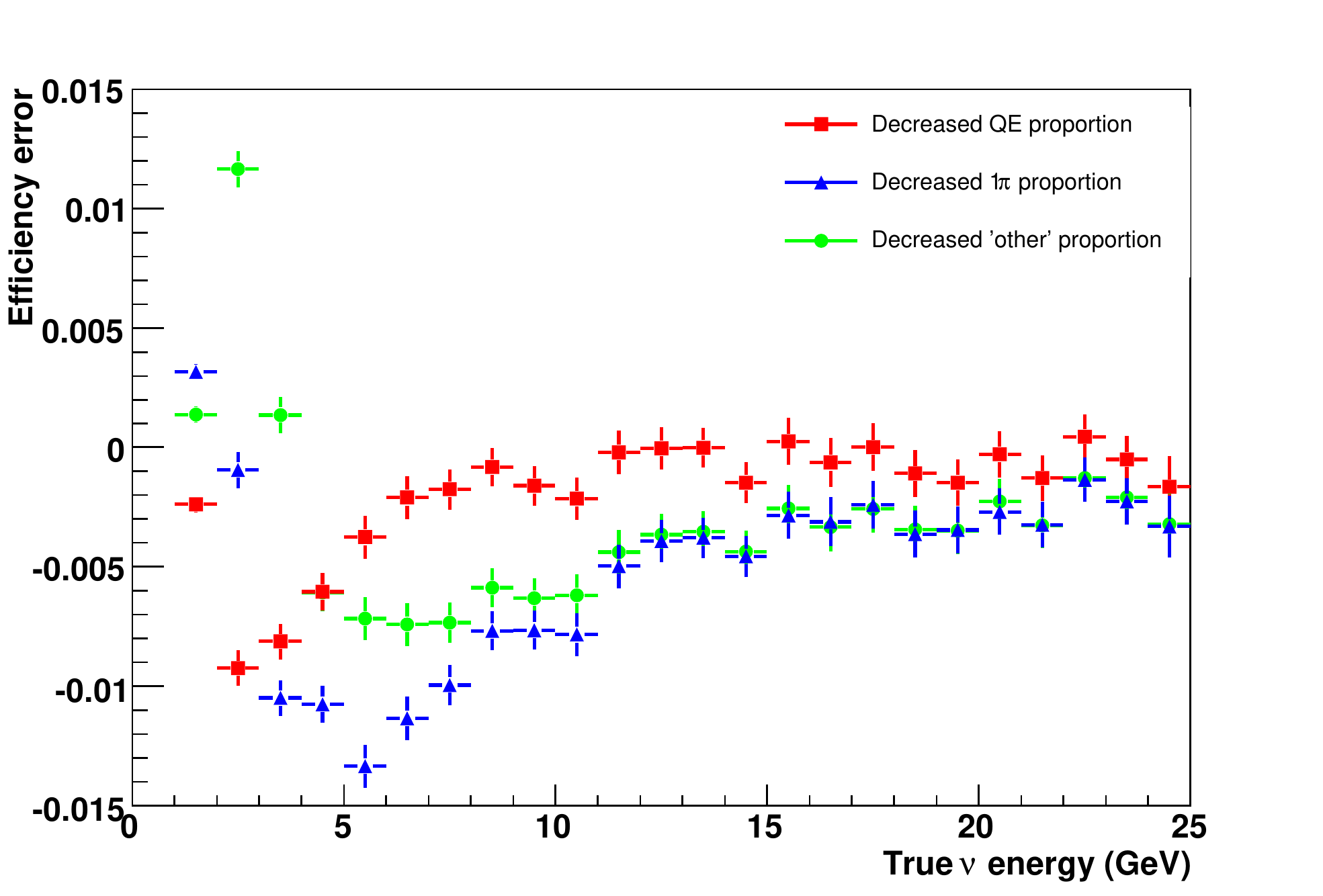} &
      \includegraphics[width=7.5cm, height=5.5cm]{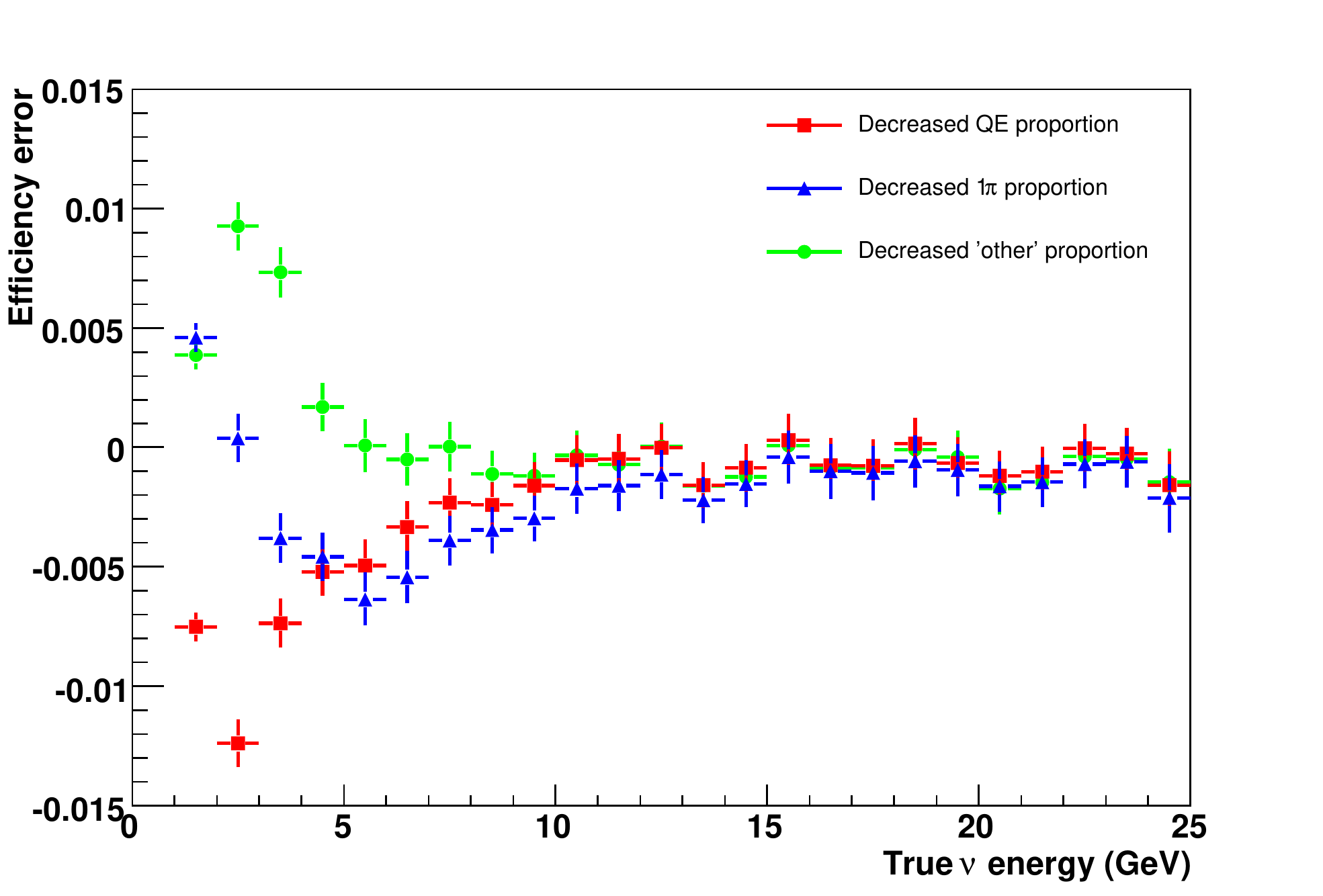}
    \end{array}$
  \end{center}
  \caption{Calculated error on signal efficiencies on increasing (top) and decreasing (bottom) the proportion of non-DIS interactions in the data-set. (left) Errors on true energy $\nu_\mu$ CC efficiency and (right) errors on true energy $\overline{\nu}_\mu$ CC efficiency}
  \label{fig:intvar}
\end{figure}

The relative proportions of QE, DIS and other types of interaction in the data sample could have a significant effect on the signal efficiencies and backgrounds. The change in efficiency if the data sample were purely DIS is shown in figure \ref{fig:DISonly}.
Although experimental data are available confirming the presence of non-DIS interactions in the energy region of interest, there are significant errors in the transition regions (see for example~\cite{Lyubushkin:2008pe,AguilarArevalo:2010zc}). These errors lead to an uncertainty in the proportion of the different types of interaction that can affect the efficiencies. In order to study the systematic error associated with this effect, many runs over the data-set were performed using different random seeds to exclude events of a particular type from the data-set or alternatively to exclude an equivalent proportion of the ``rest''. As an illustration of the method, consider the contribution from QE interactions. Taking the binned errors on the cross-section measurements from~\cite{Lyubushkin:2008pe,AguilarArevalo:2010zc}, a run to reduce the QE content would exclude a proportion of events in a bin so that instead of contributing a proportion:
\begin{equation}
  \label{eq:nomQE}
  \displaystyle\frac{N_{QE}}{N_{tot}} \, ;
\end{equation}
where $N_{QE}$ is the total number of QE interactions in the bin of interest and $N_{tot}$ is the total number of interactions in the bin, it would instead contribute:
\begin{equation}
  \label{eq:redQE}
  \displaystyle\frac{N_{QE} - \sigma_{QE}N_{QE}}{N_{tot} - \sigma_{QE}N_{QE}} \, ;
\end{equation}
where $\sigma_{QE}$ is the proportional error on the QE cross section for the bin. The equivalent run to increase the QE contribution reduces the contribution of the ``rest'' by an amount calculated to give the corresponding proportional increase in QE interactions:
\begin{equation}
  \label{eq:incQE}
  \displaystyle\frac{N_{QE} + \sigma_{QE}N_{QE}}{N_{tot} + \sigma_{QE}N_{QE}} = \displaystyle\frac{N_{QE}}{N_{tot} - \epsilon N_{rest}} \, ;
\end{equation}
where $N_{rest}$ is the total number of non-QE interactions in the bin and $\epsilon$ is the required proportional reduction in the `rest'. Solving for $\epsilon$ yields the required reduction:
\begin{equation}
  \label{eq:redRest}
  \epsilon = \displaystyle\frac{\sigma_{QE}}{1 + \sigma_{QE}} \, .
\end{equation}
Sampling randomly and repeating runs ensures that any observed change in efficiency is not solely due to the particular events excluded. The 1$\sigma$ systematic error can be estimated as the mean difference between the nominal efficiency and the increase due to a higher QE proportion or decrease due to exclusion. The errors in the true $\nu_\mu$ and $\overline{\nu}_\mu$ efficiencies extracted using this method to vary the contribution of QE, single pion (1$\pi$) and other non-DIS interactions are shown in figure \ref{fig:intvar}. Errors for 1$\pi$ resonant reactions are estimated to be $\sim$20\% below 5~GeV (as measured by the K2K near detector~\cite{:2008eaa}) and at 30\% above. Due to the large uncertainty, both theoretically and experimentally, on the models describing other resonances, coherent, diffractive and elastic processes, a very conservative error of 50\% is taken when varying the contribution of the ``others''. As can be seen, the systematic effect is at the level of 1\% in the efficiency threshold region with increased QE and 1$\pi$ interactions generally increasing the efficiency and increased contribution of the ``other'' interactions having the possibility to decrease efficiency. This last effect is likely to be predominantly due to resonances producing multiple tracks. The effect on backgrounds, while expected to be minimal, has not yet been studied due to the statistical limitations in the data sample, making any calculation using this method unreliable.

\subsubsection{MIND conceptual design}
\label{sec:MIND_conceptual_design}

The Magnetised Iron Neutrino Detector (MIND) is an iron and scintillator sampling calorimeter which is essentially a larger version of the MINOS detector \cite{Lang:2001rw}.  We have chosen a cross section of 14~m~$\times$~14~m in order to maximise the ratio of the fiducial mass to the total mass.  The magnetic field will be toroidal as in MINOS and MIND will also use extruded scintillator for the readout material.  Details on the iron plates, magnetisation, scintillator, photo-detector and electronics are given below.  

\paragraph{Iron plates}
\label{par:iron_plates}

For the iron plates in MIND, we are following the design strategy that was used for MINOS.  The plates are octagons with overall dimension of 14 m $\times$ 14~m and 3.0 cm thick.  They are fabricated from strips that are 1.5 cm thick, 2~m (or 3~m)  wide and up to 14 m long.  Two layers of crossed strips  are plug welded together to form the full plate.  MINOS used 2~m wide strips and we know that fabrication of iron components of this width and with lengths of up to 14 m is possible.  Depending on the final plate fabricator, strip widths greater than 2~m are likely to be possible, so both 2~m and 3~m strip-width models were investigated.  Initially it was thought that the much larger weight of the MIND plates (40~T versus 10~T for MINOS) would preclude the concept of hanging the plates on a rail system due to excessive stress in the ears.  However, our analysis of the expected stress (see section \ref{par:MIND_FEA} below) has shown that this is not the case.  Essentially, no R\&D on the MIND iron plates is needed.  Final specification of the plate mosaic structure will be determined once a plate fabricator is chosen.

\begin{figure*}
  \centering{
    \includegraphics[width=0.6\textwidth]%
      {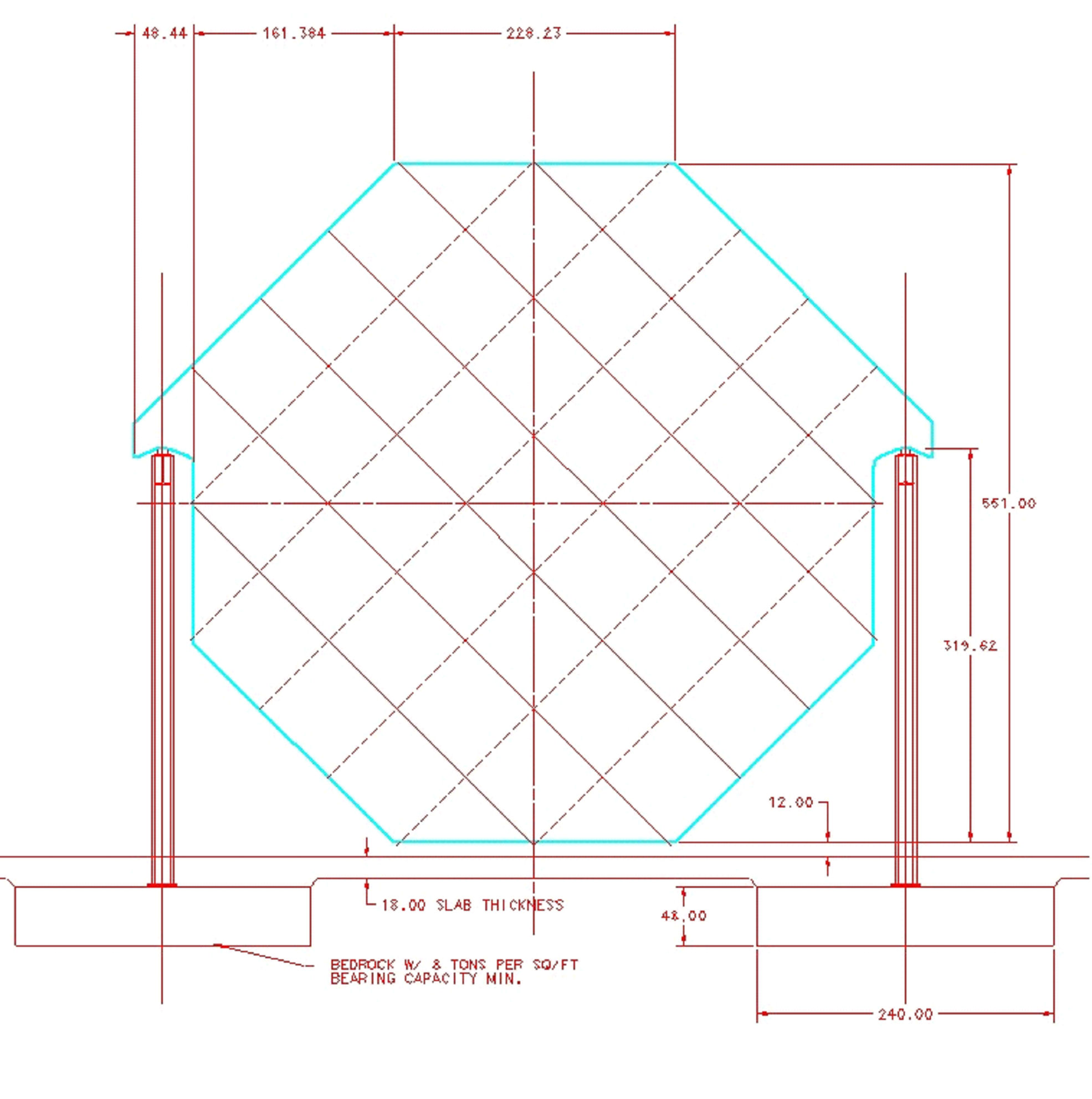}
  }
  \caption{2D diagram of MIND plate.}
  \label{Fig:MIND-2D}
\end{figure*}

\begin{figure*}
  \centering{
    \includegraphics[width=0.6\textwidth]%
      {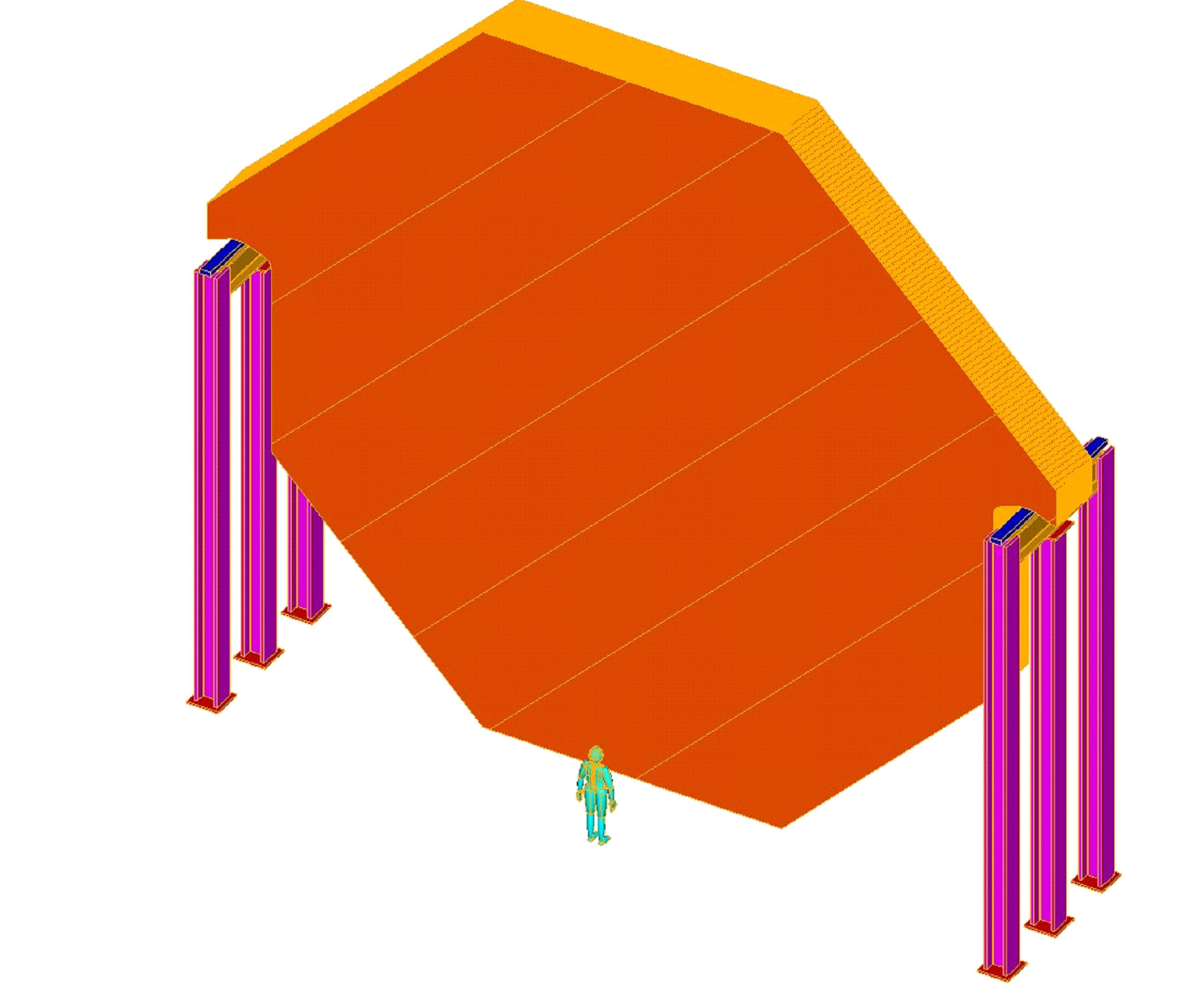}
  }
  \caption{3D diagram of MIND plate.}
  \label{Fig:MIND-3D}
\end{figure*}

For the 2~m strip model, seven 2~m strips will be required to make up
a whole layer. 
The layout of the top layer will be perpendicular to the bottom layer
in each plane.  
For the 3~m strip model, the 14~m long strips will be both 3~m wide
and 2~m wide. 
Four 3~m strips and one 2~m strip will be required to make up a whole
layer. 
The layout of the top layer and the bottom layer remains perpendicular
to one-another in each plane. 
The individual plates will be held together by plug welding the top
layer to the bottom layer.  
Each ear of the plane is supported by a structural-steel rail which is
in turn supported by structural-steel columns. 
The section of the MIND detector plane and the support structures are
shown in figure \ref{Fig:MIND-2D} and figure \ref{Fig:MIND-3D}.

A `book-end' will be used to provide lateral support for the planes as
the MIND detector is constructed. 
The plane will be attached to the book-end at the minimum of each
vertex point in the octagon and each midpoint between the vertexes. 
This book-end will consist of a framework of structural-steel members
and will be vertically supported by the floor slab and horizontally
supported by the wall of the enclosure.

\paragraph{The finite-element model}
\label{par:MIND_FEA}

\begin{figure*}
  \centering{
    \includegraphics[width=0.9\textwidth]%
      {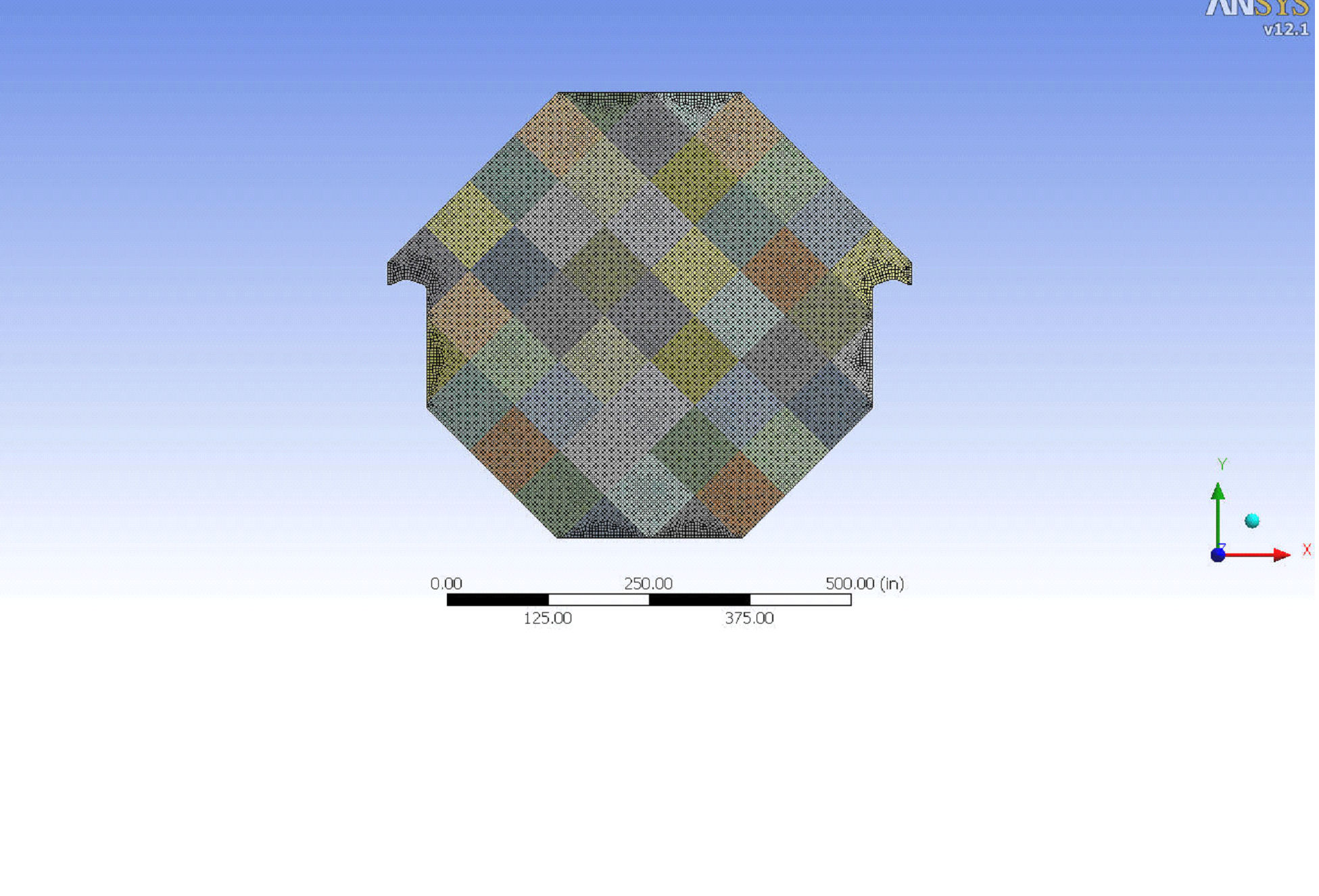}
  }
  \caption{Finite-element model of plate with 2~m strips.}
  \label{Fig:2M-model}
\end{figure*}

\begin{figure*}
  \centering{
    \includegraphics[width=0.9\textwidth]%
      {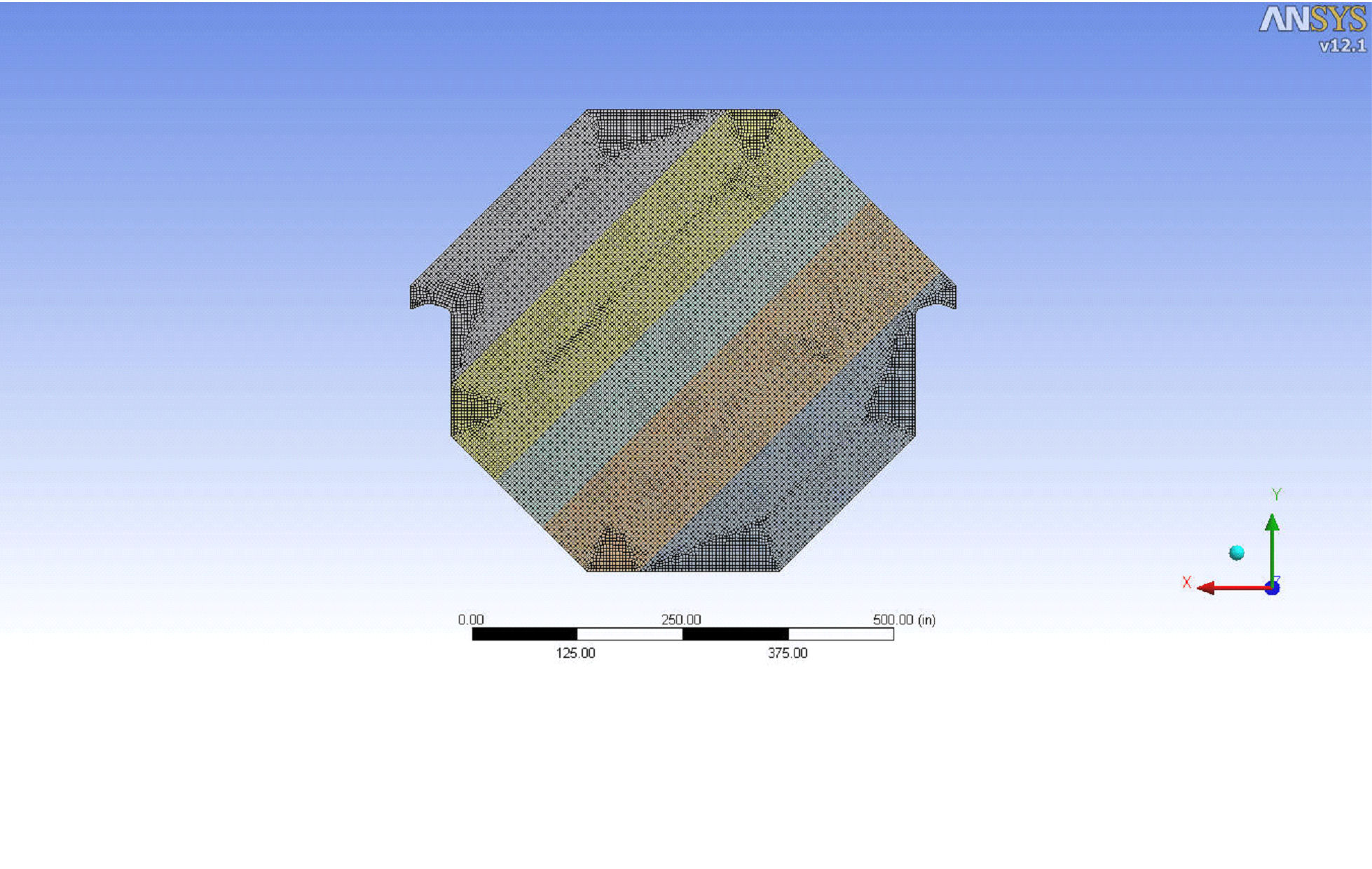}
  }
  \caption{Finite-element model of plate that utilises both 3~m and 2~m strips.}
  \label{Fig:3M-model}
\end{figure*}

\begin{figure*}
  \centering{
    \includegraphics[width=0.9\textwidth]%
      {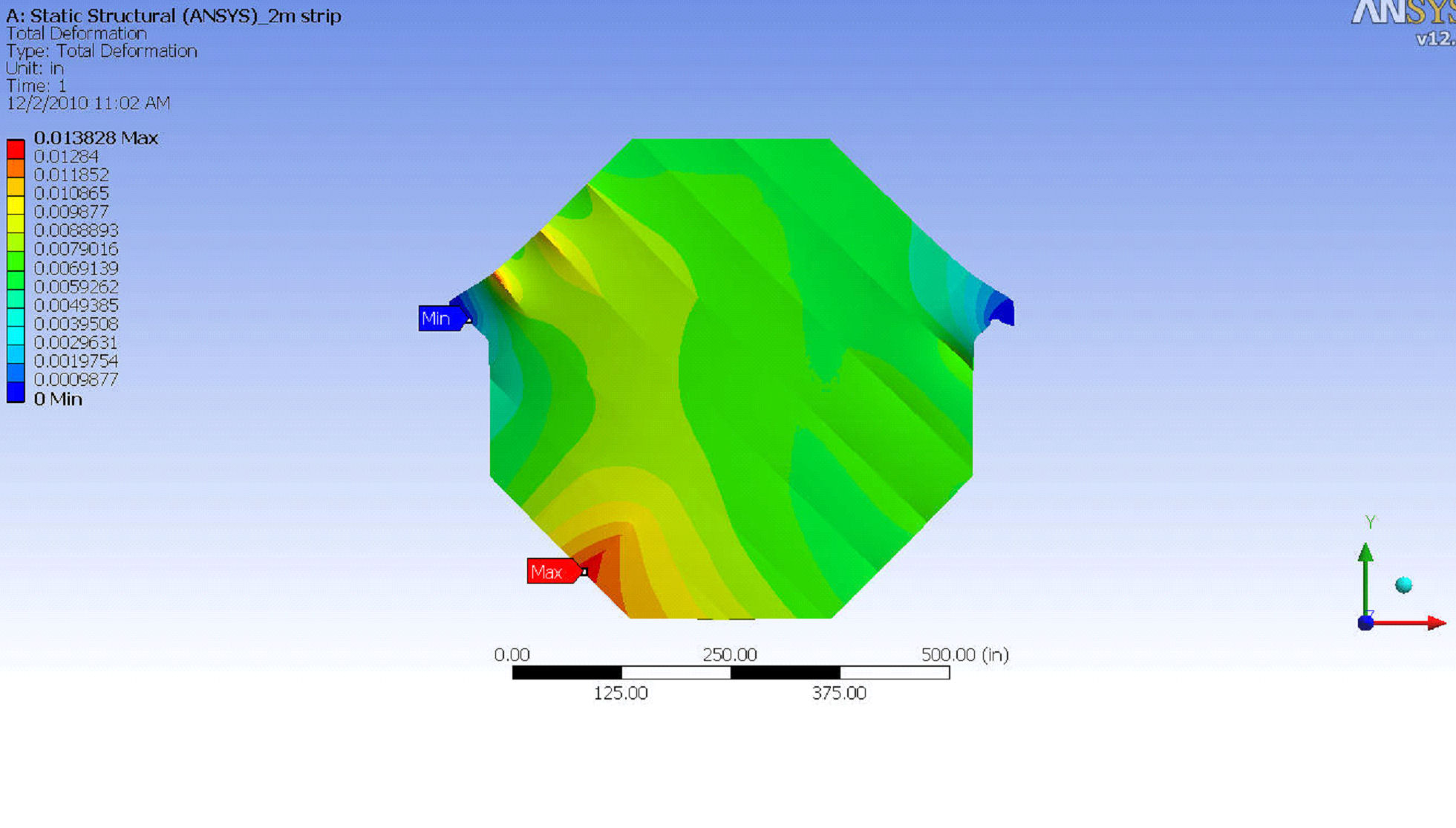}
  }
  \caption{Total deformed shape of plate fabricated with 2~m strips}
  \label{Fig:2M-deform}
\end{figure*}

\begin{figure*}
  \centering{
    \includegraphics[width=0.9\textwidth]%
      {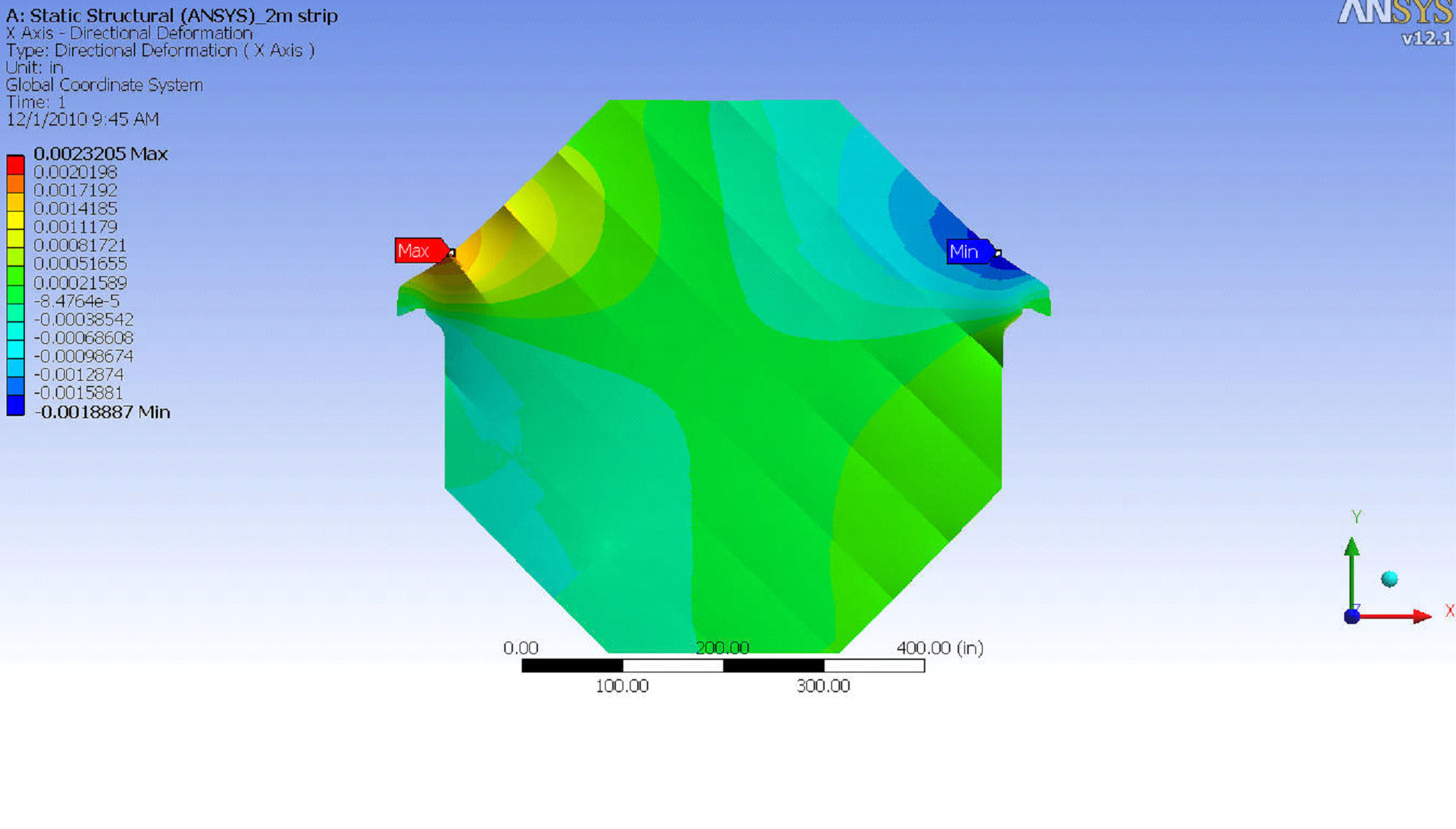}
  }
  \caption{Deformation in $x$ for plate using 2~m strips}
  \label{Fig:2M-X-deform}
\end{figure*}

\begin{figure*}
  \centering{
    \includegraphics[width=0.9\textwidth]%
      {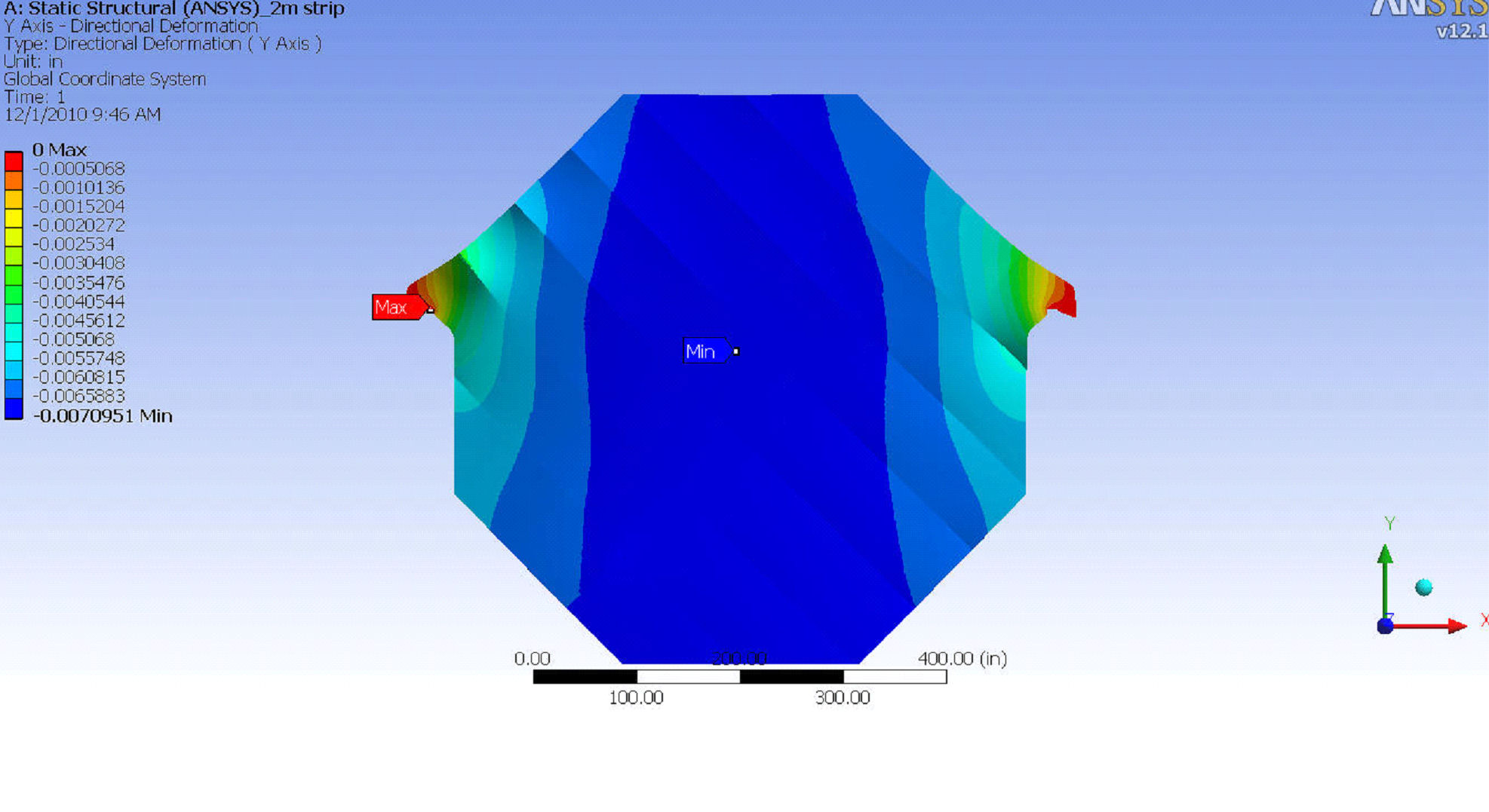}
  }
  \caption{Deformation in $y$ for plate using 2~m strips}
  \label{Fig:2M-Y-deform}
\end{figure*}

\begin{figure*}
  \centering{
    \includegraphics[width=0.9\textwidth]%
      {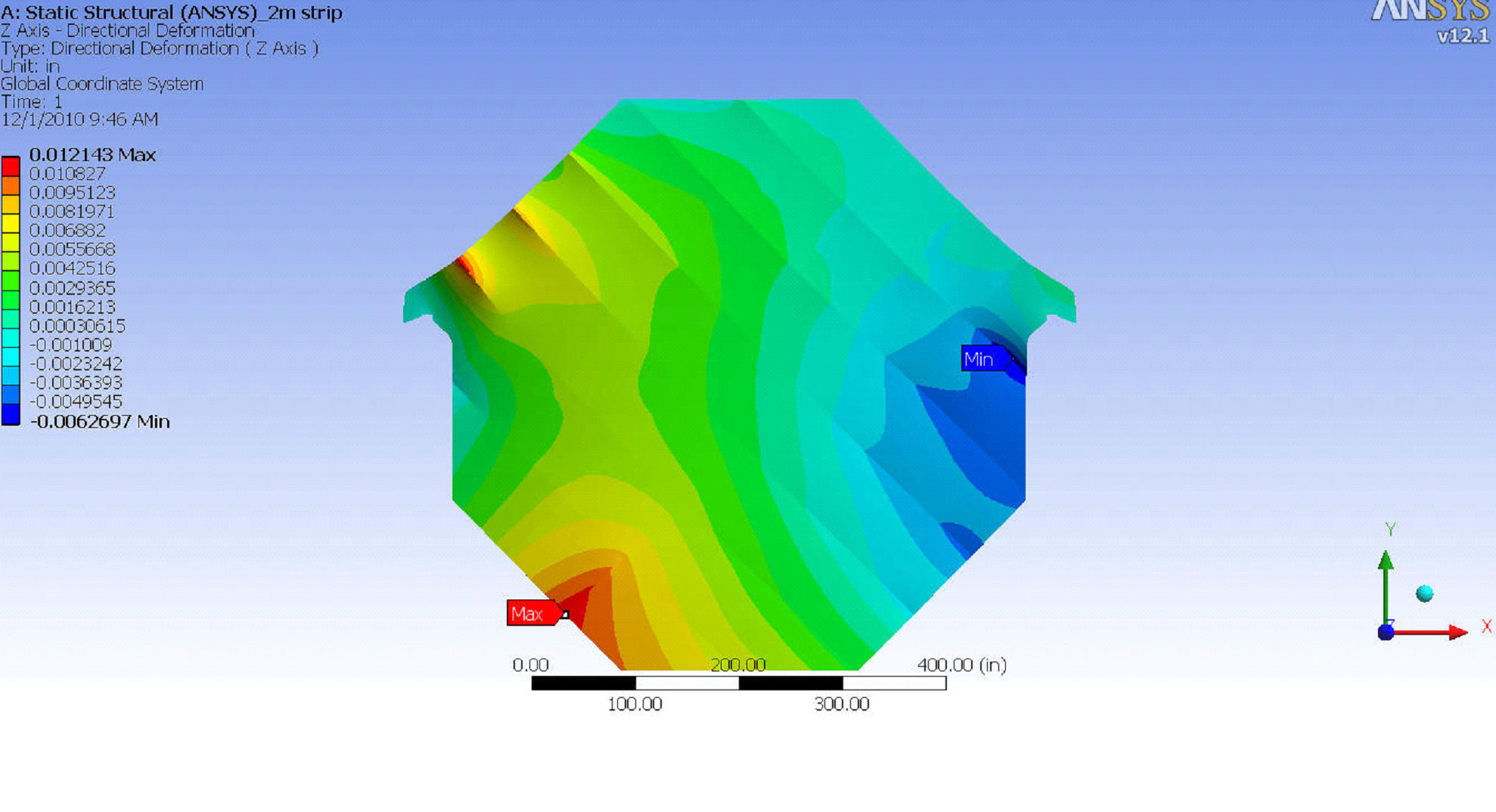}
  }
  \caption{Deformation in $z$ for plate using 2~m strips}
  \label{Fig:2M-Z-deform}
\end{figure*}

Two finite-element models of the detector plane were created using higher order solid elements. The 2~m strip model is shown in figure \ref{Fig:2M-model} and the 3~m strip model is shown in figure \ref{Fig:3M-model}.  Loading was simulated by the gravity load of the plane. The plane was fixed at the bottom of the ears and the two top vertexes are fixed in the $z$ direction to resist plane buckling. The linear buckling of the plane was also investigated. The total deformed shape of the 2~m strip plane is shown in figure \ref{Fig:2M-deform}. The directional deformations are shown in figures \ref{Fig:2M-X-deform}, \ref{Fig:2M-Y-deform} and \ref{Fig:2M-Z-deform} respectively. The maximum deflections occur at the ear and the bottom of the plane.

\begin{figure*}
  \centering{
    \includegraphics[width=0.9\textwidth]%
      {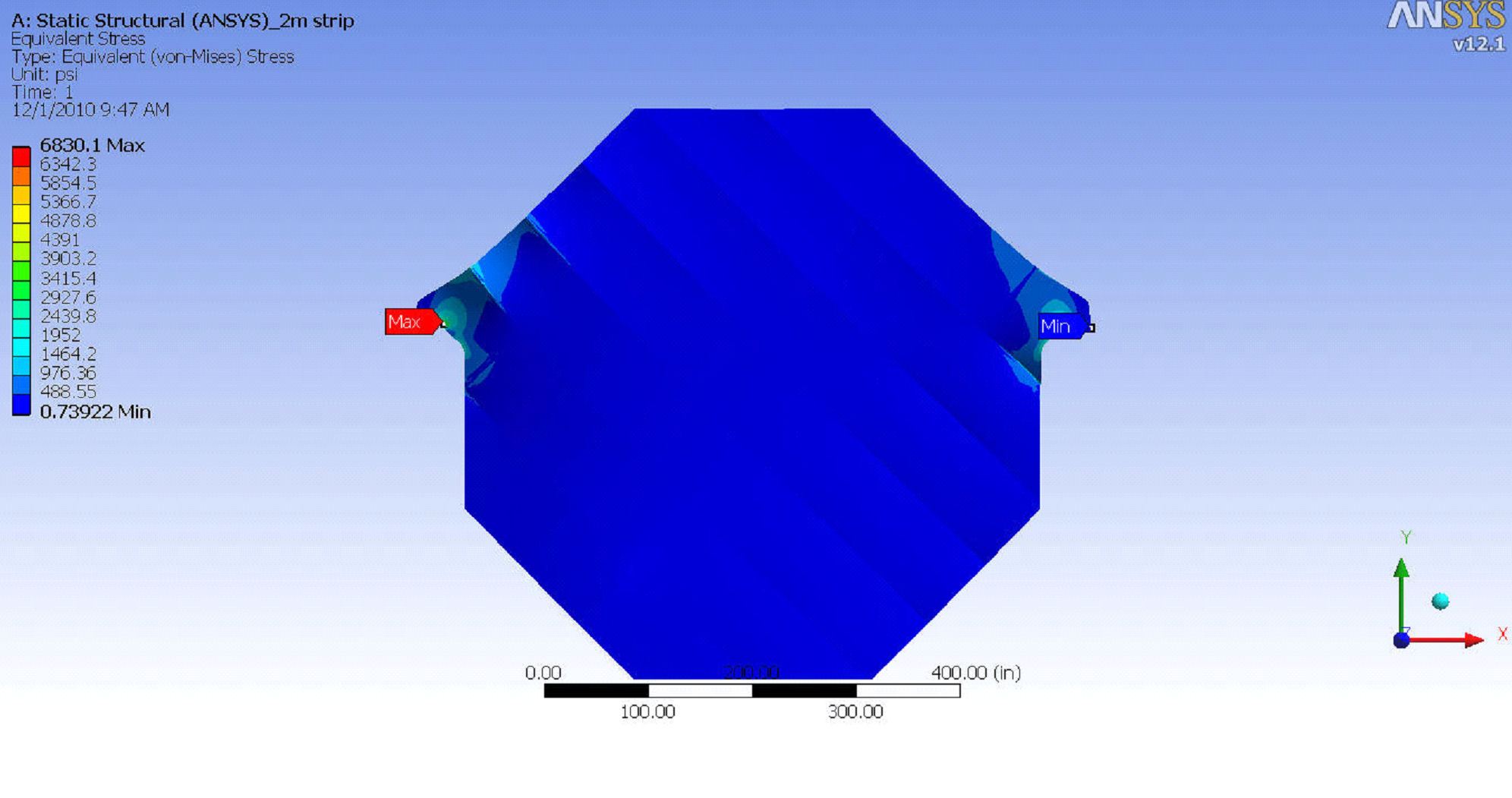}
  }
  \caption{von Mises stress in plate using 2~m strips}
  \label{Fig:2M-stress}
\end{figure*}

The stresses in the 2~m strip plane are shown in
figure \ref{Fig:2M-stress}. 
The maximum von Mises stress is 6.8\,ksi at the ear. 
In the regions away from that, all stresses are below the 12\,ksi
limit for AISI 1006 low carbon steel.  
The welded connections were examined by extracting nodal forces and
moments from the 2\,m strip plane model at 45 locations. 
The maximum load in the plane is in the ear area and is approximately
15\,000 pounds per inch.  
In the linear buckling analysis of the 2\,m strip plane, the results
show the first buckling mode has a load safety factor of 4.7. 

\begin{figure*}
  \centering{
    \includegraphics[width=0.9\textwidth]%
      {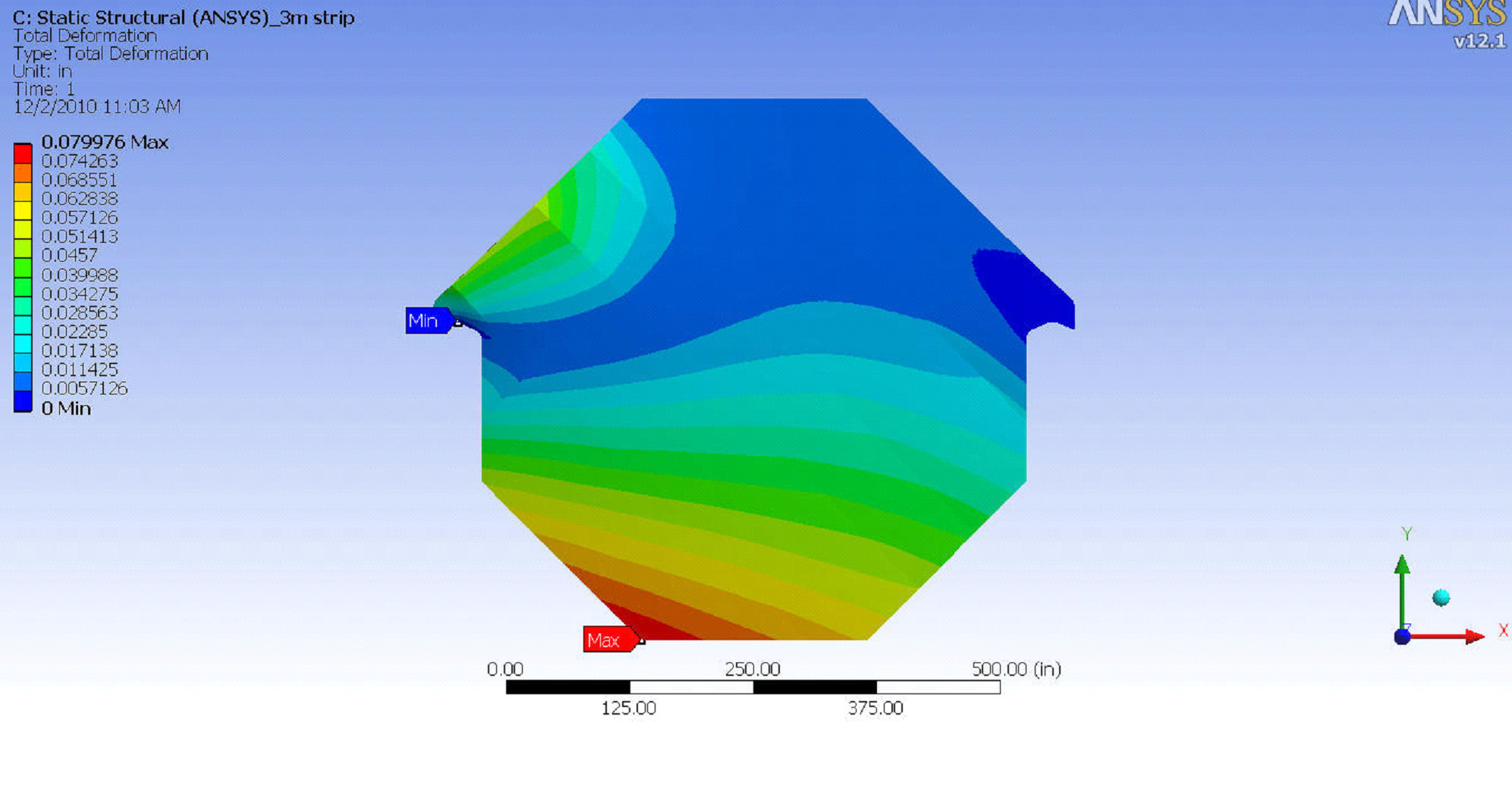}
  }
  \caption{Total deformation for plate using 3~m and 2~m strips}
  \label{Fig:3M-deform}
\end{figure*}

\begin{figure*}
  \centering{
    \includegraphics[width=0.9\textwidth]%
      {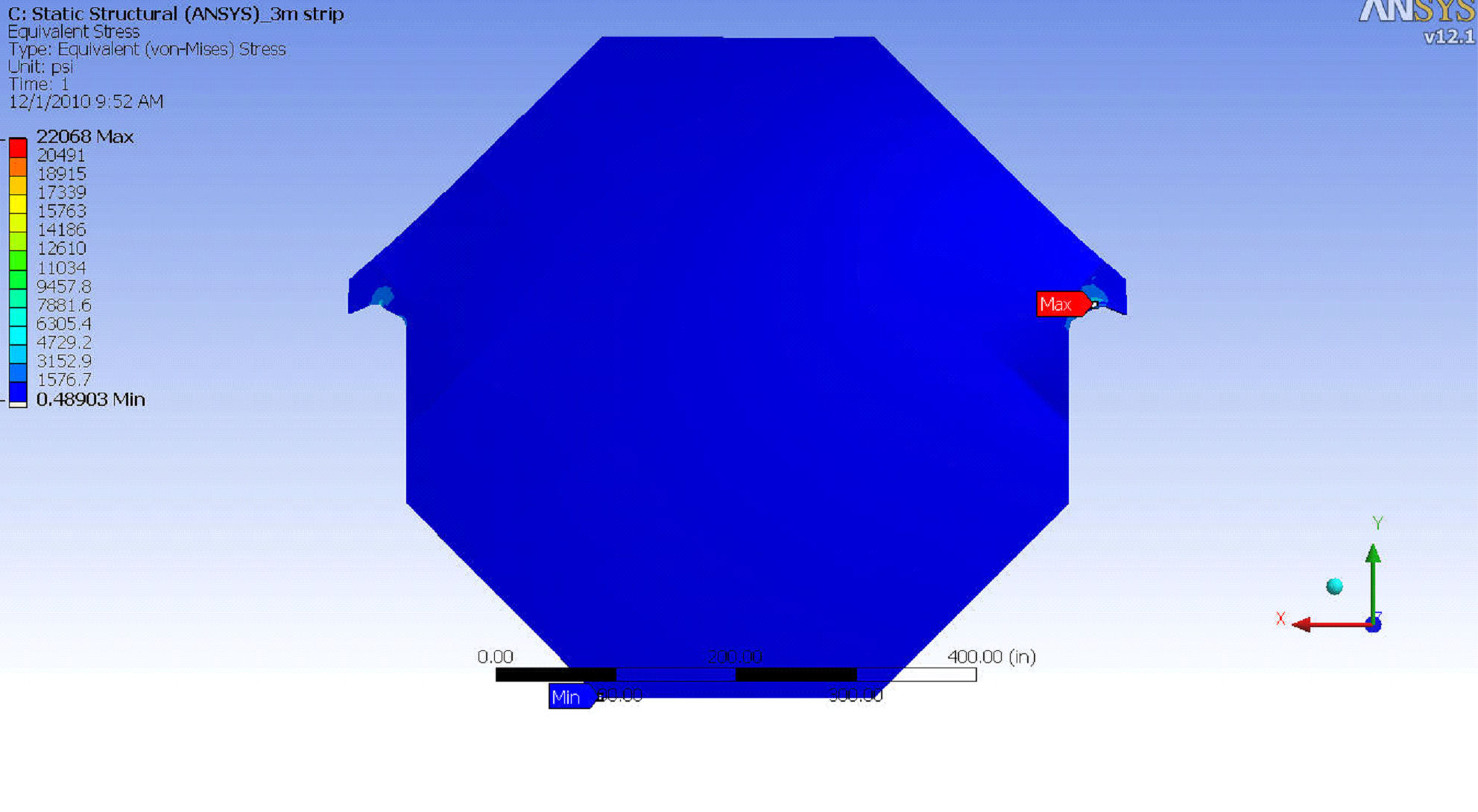}
  }
  \caption{von Mises stress in plate using 3~m and 2~m strips}
  \label{Fig:3M-stress}
\end{figure*}

For comparison, the total deformed shape of the 3~m strip plane is
shown in figure \ref{Fig:3M-deform}. 
Again, the maximum deflections occur at the ear and the bottom of the
plane. 
The stresses in the 3~m strip plane are shown in
figure \ref{Fig:3M-stress}. 
The maximum von Mises stress is 22 ksi at the ear. 
In the regions away from the concentration, all stresses are below the
12 ksi limit for AISI 1006 low carbon steel. 
In the linear buckling analysis of the 3~m strip plane, the results
show the first buckling mode has a load safety factor of 4.5, somewhat
worse than in the case of the 2~m wide strip case.

\paragraph{Magnetisation}

\begin{figure*}
  \centering{
    \includegraphics[width=0.6\textwidth]%
      {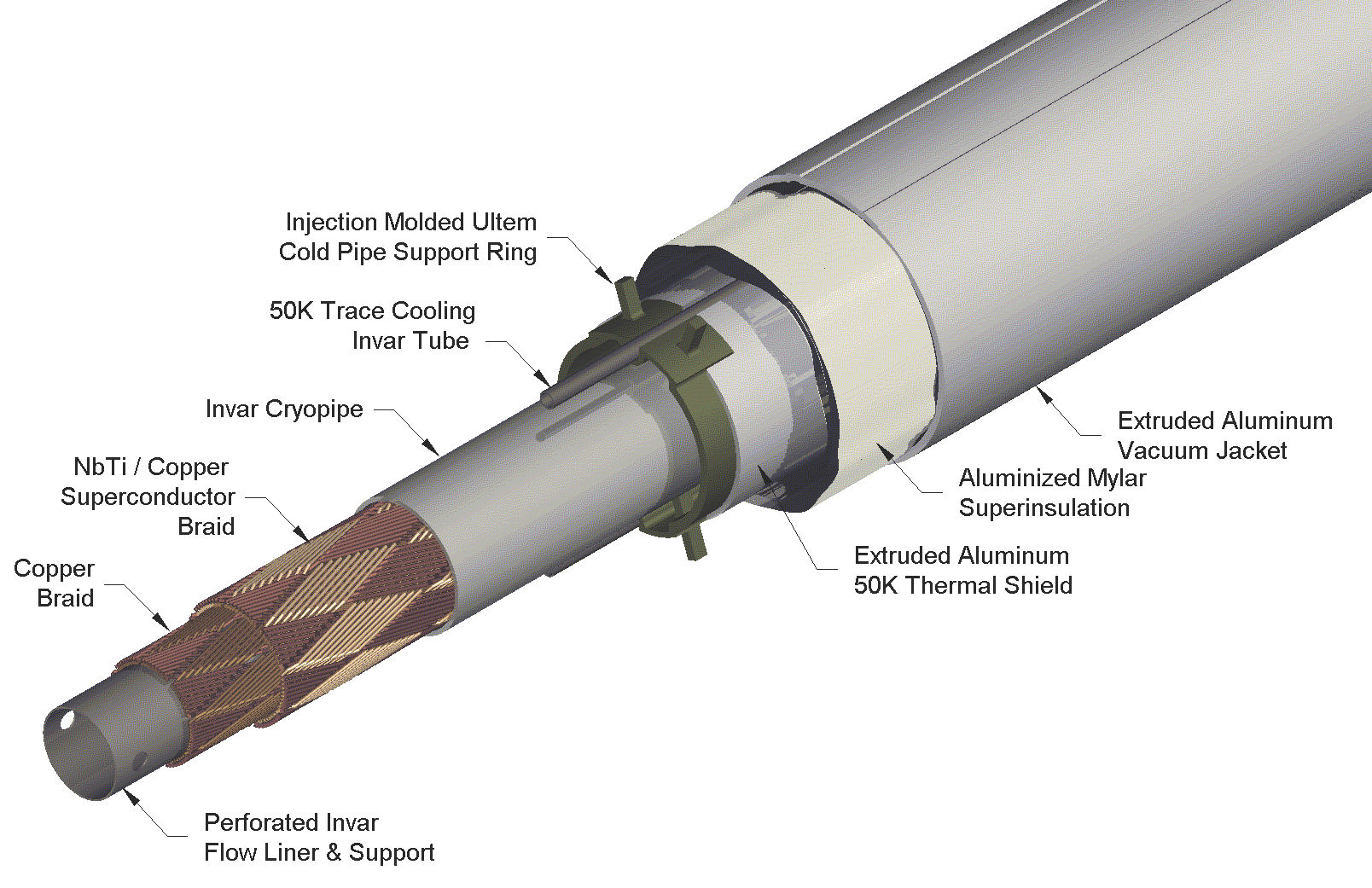}
  }
  \caption{
    Schematic of superconducting transmission line showing
    construction details.
  } 
  \label{Fig:STL}
\end{figure*}

As mentioned above, MIND will have a toroidal magnetic field like that
of MINOS.  
For excitation, however, we plan to use the concept of the
superconducting transmission line (STL) developed for the Design Study
for a staged Very Large Hadron Collider (VLHC) \cite{Ambrosio:2001ej}.  
In order to obtain good field uniformity in a 14\,m\,$\times$\,14\,m
plate, MIND requires a much larger excitation current-turn than the
15\,kA-turn that is used in the MINOS room temperature Cu coils.  
100\,kA-turn is possible using the STL.  
The STL consists of a cylindrical superconducting braid inside a pipe
cooled by super-critical helium.  
The superconductor and cryo-pipe are co-axial to a cylindrical
cryostat/vacuum vessel, figure \ref{Fig:STL}.  
Figure \ref{Fig:STL} shows the constructions details for the STL that
was prototyped and tested for the VLHC study and consisted of: 1) a
perforated Invar flow liner and support; 2) a copper stabiliser braid;
3) superconductor cable braid; 4) an Invar pipe that contains the
helium; 5) the cold-pipe support; 6) cryo-shield; 7) super-insulation;
and 8) the vacuum jacket/pipe.

\subparagraph*{Field Map}

\begin{figure*}
  \centering{
    \includegraphics[width=0.4\textwidth]%
      {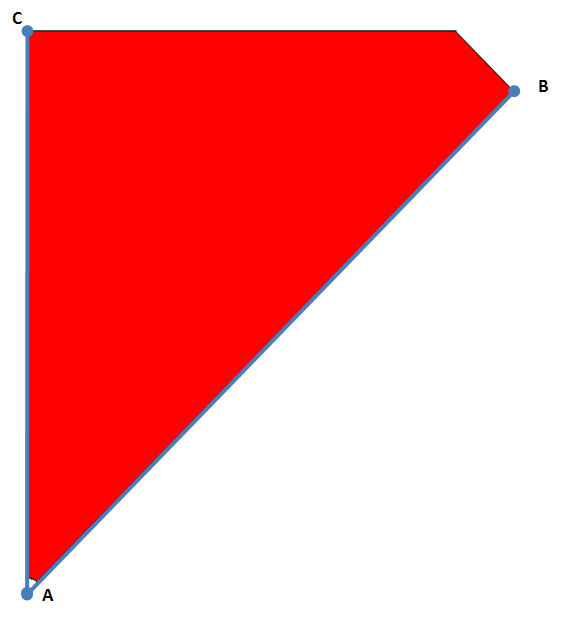}
  }
  \caption{1/8th model used for magnetic analysis.}
  \label{Fig:one-eight-plate}
\end{figure*}

\begin{figure*}
  \centering{
    \includegraphics[width=0.76\textwidth]%
      {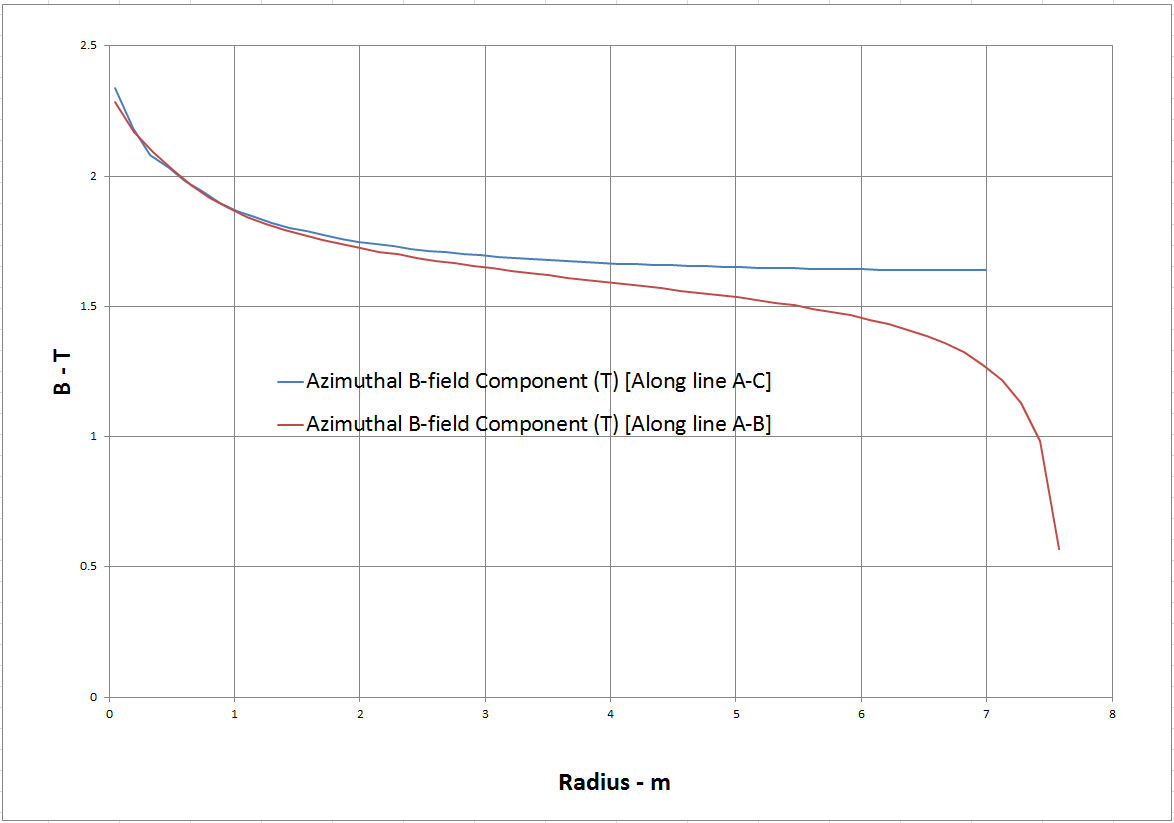}
  }
  \caption{
    Magnetic field along the lines A-B and A-C in figure
    \ref{Fig:one-eight-plate}
  }
  \label{Fig:AB-field}
\end{figure*}

\begin{figure*}
  \centering{
    \includegraphics[width=0.8\textwidth]%
      {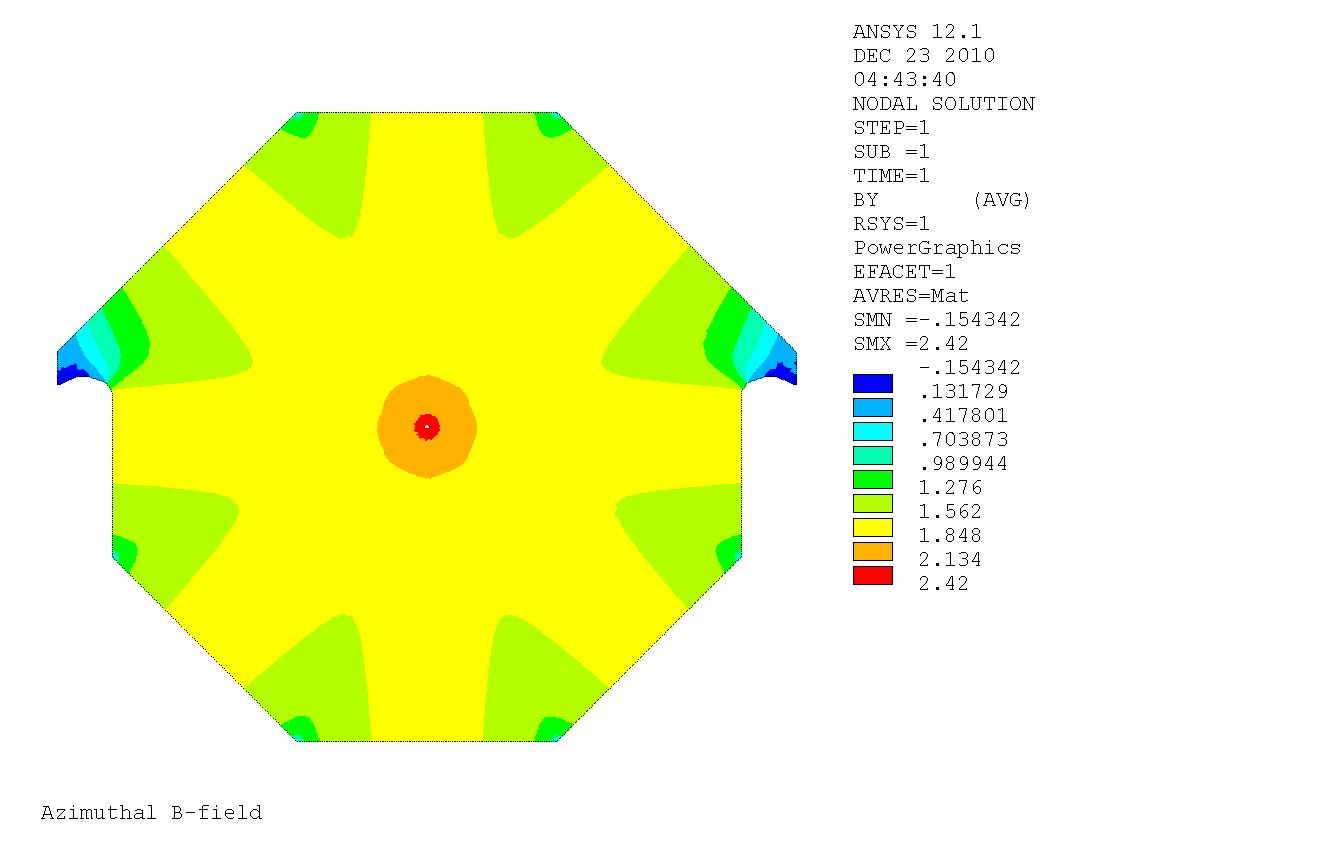}
  }
  \caption{Contours of constant magnetic B-field.}
  \label{Fig:Field-Map-2D}
\end{figure*}

Using the MIND plate geometry shown in figure \ref{Fig:MIND-2D}, a 2D
magnetic analysis of the plate was performed.  
Figure \ref{Fig:one-eight-plate} shows the model (1/8th) that was used
in the analysis.  
A 100\,cm diameter hole for the STL was assumed and the MINOS
steel \cite{Michael:2008bc} BH curve was assumed.  
For this analysis, an excitation current of 100 kA was used.  
This was the critical current achieved at 6.5\,K in the STL test stand
assembled for the VLHC proof-of-principle.  
In figure \ref{Fig:AB-field} we give the azimuthal B field along the
two lines (A-B and A-C) shown in figure \ref{Fig:one-eight-plate}.  
Figure \ref{Fig:Field-Map-2D} gives the 2D contour lines of constant
B.


\paragraph{R\&D Requirements}

The STL described above was optimised for the 37 km radius of the
VLHC.  
The loop length needed for MIND is quite a bit shorter than the 233 km
needed for the VLHC.  
The optimised design of an STL for MIND is therefore likely to be
different.  
The 2.5\,cm He-flow region can most likely be reduced, for example.  
In addition, we would like to consider an STL with multiple internal
current loops so that a 100 kA room temperature driving source is not
needed. 
The total amount of superconductor would be roughly the same;  5-10
turns with a current of 20-10 kA each would be more manageable from
the viewpoint of the power source. 

\paragraph{Detector planes}

\subparagraph*{Scintillator \\}

Particle detection using extruded scintillator and optical fibres is a
mature technology.  
MINOS has shown that co-extruded solid scintillator with embedded
wavelength shifting (WLS) fibres and PMT readout produces adequate
light for MIP tracking and that it can be manufactured with excellent
quality control and uniformity in an industrial setting.  
Many experiments use this same technology for the active elements of
their detectors, such as the K2K
Scibar \cite{RodriguezMarrero:2007zz}, the T2K INGRID, the T2K P0D,
the T2K ECAL \cite{Kudenko:2008ia} and the Double-Chooz
detectors \cite{Greiner:2007zzd}.

Our initial concept for the readout planes for MIND is to have both an
$x$ and a $y$ view between each plate. 
The simulations  done to date have assumed a scintillator extrusion
profile that is 3.5\,$\times$\,1.0\,cm$^2$.  
This gives both the required point resolution and light yield.  
We are also considering an option where we use triangular extrusions
similar to those used in Minerva \cite{McFarland:2006pz}. 

\subparagraph*{Rectangular extrusions}

\begin{figure*}[htb]
  \centering{
    \includegraphics[width=0.4\textwidth]%
      {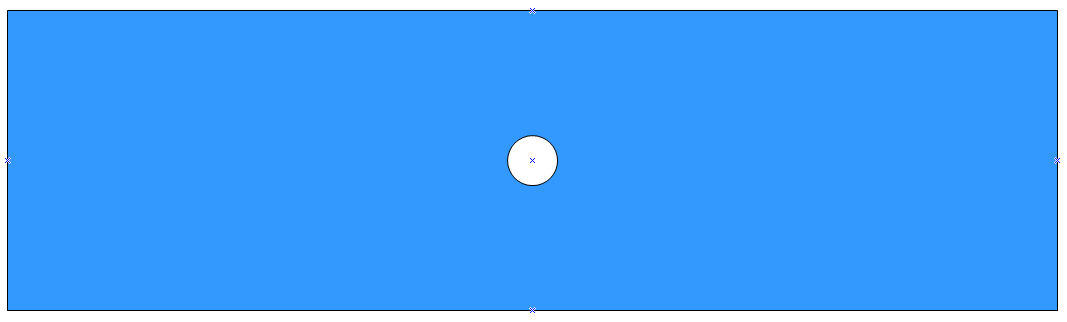}
  }
  \caption{Schematic of rectangular scintillator extrusion.}
  \label{Fig:rectanex}
\end{figure*}

The existing MIND simulations have assumed that the readout planes
will use a rectangular extrusion that is 3.5\,$\times$\,1.0\,cm$^2$,
see figure \ref{Fig:rectanex}.  
A 1 mm hole down the centre of the extrusion is provided for insertion
of the wavelength shifting fibre.  
This is a relatively simple part to manufacture and has already been
fabricated in a similar form for a number of small-scale applications.  
The scintillator strips will consist of an extruded polystyrene core
doped with blue-emitting fluorescent compounds, a co-extruded TiO$_2$
outer layer for reflectivity, and a hole in the middle for a WLS
fibre.  
Dow Styron 665\,W polystyrene pellets are doped with PPO (1\% by
weight) and POPOP (0.03\% by weight). 
The strips have a white, co-extruded, 0.25 mm thick TiO$_2$ reflective
coating.  
This layer is introduced in a single step as part of a co-extrusion
process.  
The composition of this capstocking is 15\% TiO$_2$ (rutile) in
polystyrene.  
In addition to its reflectivity properties, the layer facilitates the
assembly of the scintillator strips into modules. 
The ruggedness of this coating enables the direct gluing of the strips
to each other and to the module skins which results in labour and time
savings for the experiment.  
This process has now been used in a number of experiments.

\subparagraph*{Minerva extrusions \\}

\begin{figure*}
  \centering{
    \includegraphics[width=0.5\textwidth]%
      {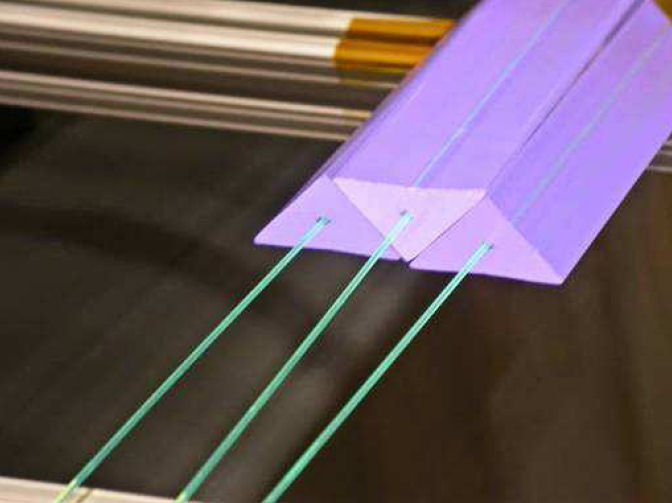}
  }
  \caption{Minerva extrusions showing partial readout plane and wavelength shifting fibres.}
  \label{Fig:Minerva-triangle}
\end{figure*}

\begin{figure*}
  \centering{
    \includegraphics[width=0.8\textwidth]%
      {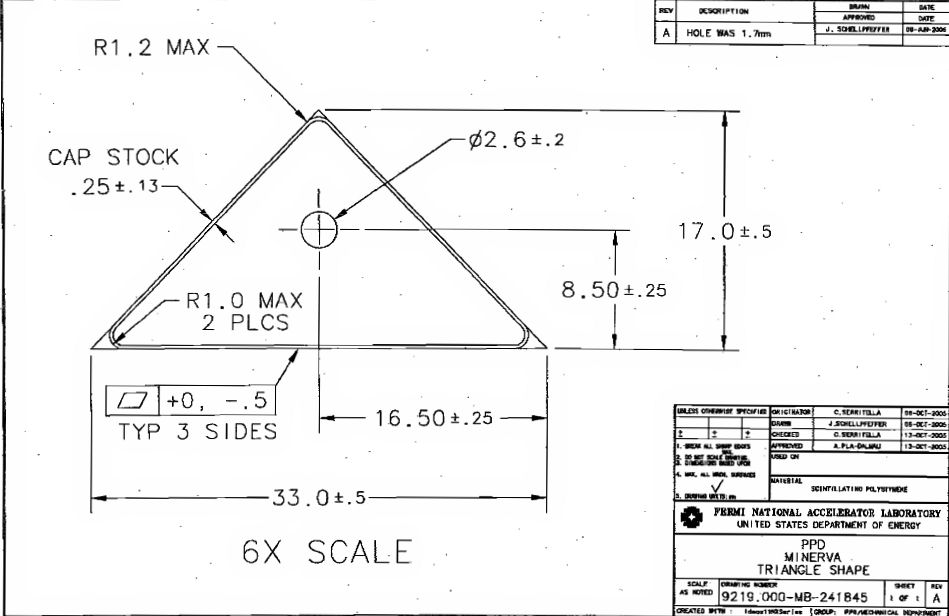}
  }
  \caption{Minerva extrusion engineering drawing.}
  \label{Fig:Minerva-triangle-line}
\end{figure*}

We are also considering using the Minerva extrusion (see
figure \ref{Fig:Minerva-triangle}) for MIND.  The triangle has a 3.3-cm
base and a 1.7-cm height, and a 2.6 mm hole for a WLS fibre (figure \ref{Fig:Minerva-triangle-line}). 


\subparagraph*{Photo-detector \\}

Given the rapid development in recent years of solid-state photo-detectors based on Geiger mode operation of silicon avalanche photodiodes, we have chosen this technology for MIND.  Although various names are used for this technology, we will use silicon photomultiplier or SiPM.

\subparagraph*{SiPM Overview \\}

\begin{figure*}
  \centering{
    \includegraphics[width=0.5\textwidth]%
      {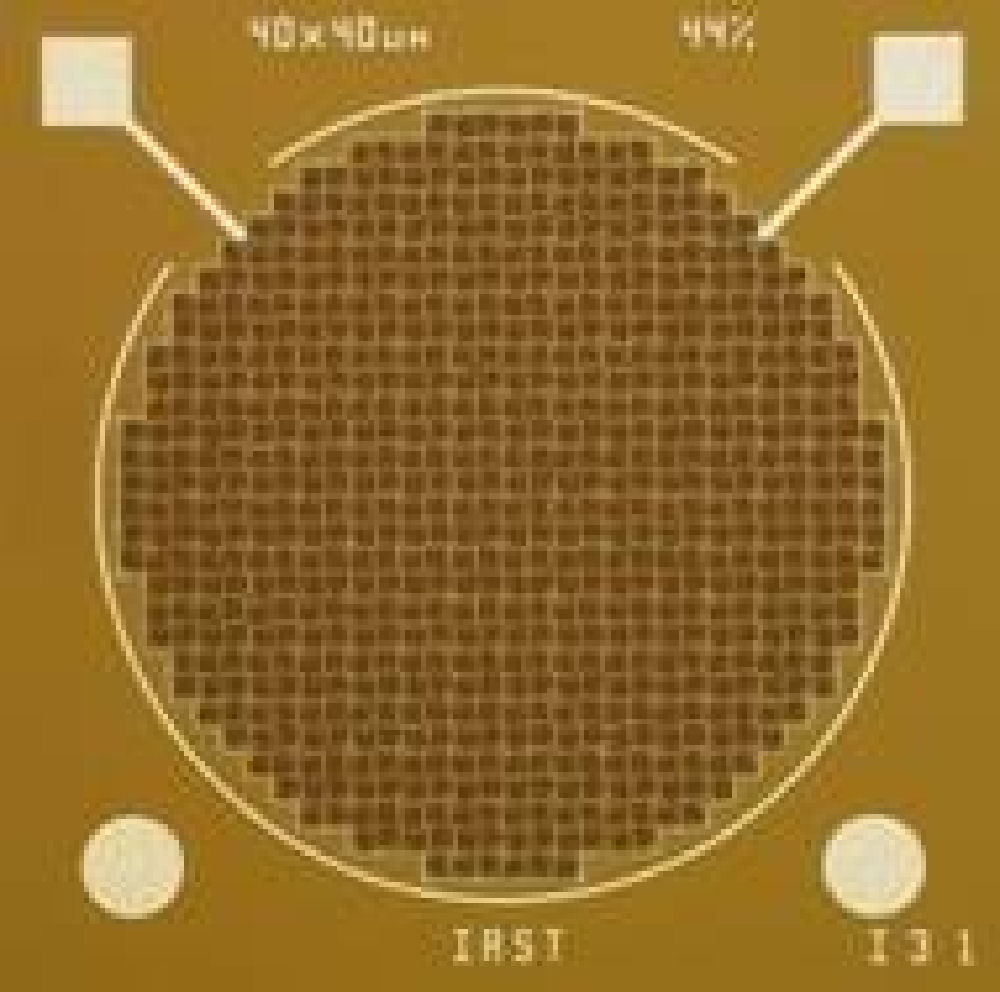}
  }
  \caption{Photograph of SiPM}
  \label{Fig:SiPM}
\end{figure*}

SiPM is the often-used name for a type of photo detector formed by
combining many small avalanche photodiodes operated in the Geiger mode
to form a single detector \cite{Sadygov:1996,Bacchetta:1996dc}.
Detailed information and basic principles of operation of these
``multi-pixel"  photodiodes can be found  in a recent review paper and
the references therein \cite{Renker:2009zz}. The first generation of
these detectors use a polysilicon resistor connected to each avalanche
photodiode forming  a pixel. Pixels usually vary in size from 10
$\times 10 \, \mu$m$^2$ to 100 $\times 100 \, \mu$m$^2$ (see
figure \ref{Fig:SiPM}). 
All the diodes are connected to a common electrical point on one side,
typically through the substrate, and all the resistors are connected
to a common grid with metal traces on the other side to form a
two-node device. 
A typical SiPM will have from 100 to 10\,000 of these pixels in a
single device, with the total area from 1 to 10 mm$^2$.  
Because all the diode and the individual quenching resistors are
connected in parallel, the SiPM device as a whole appears as a single
diode. 
In operation, the device appears to act somewhat like a conventional
APD, but in detail it is radically different. 
Because the diodes are operated in the Geiger mode, and because every
pixel of the SiPM device is nearly identical, the sum of the fired
pixels gives the illusion of an analog signal that is proportional to
the incident light, but it is, essentially, a digital device.  

SiPMs have a number of advantages over conventional photomultiplier
tubes, including very high photon-detection efficiency, complete
immunity to magnetic fields, excellent timing characteristics, compact
size, and physical robustness. 
They are particularly well suited to applications with fibres, as the
natural size of the SiPM is comparable to that of
fibres \cite{Balagura:2005gh,Yokoyama:2010qa}. 
But the most important single feature of the SiPM is that it can be
manufactured in standard microelectronics facilities using nearly
standard CMOS processing. 
This means that huge numbers of devices can be produced without any
manual labour, making the SiPMs very economical. 
Furthermore, it is possible to integrate the electronics into the SiPM
itself, which reduces cost and improves performance. 
Initial steps have been taken in this direction, though most current
SiPMs are not manufactured in this way. 
The advantages of the SiPM are that it improves performance, reduces
cost and can be tailored to a specific application.
As the use of SiPMs spreads, so will the use of custom SiPMs with
integrated electronics, just as ASICs have superseded standard logic
in electronics.  

\begin{figure}[htb]
\centering\includegraphics[width=10cm,angle=0]{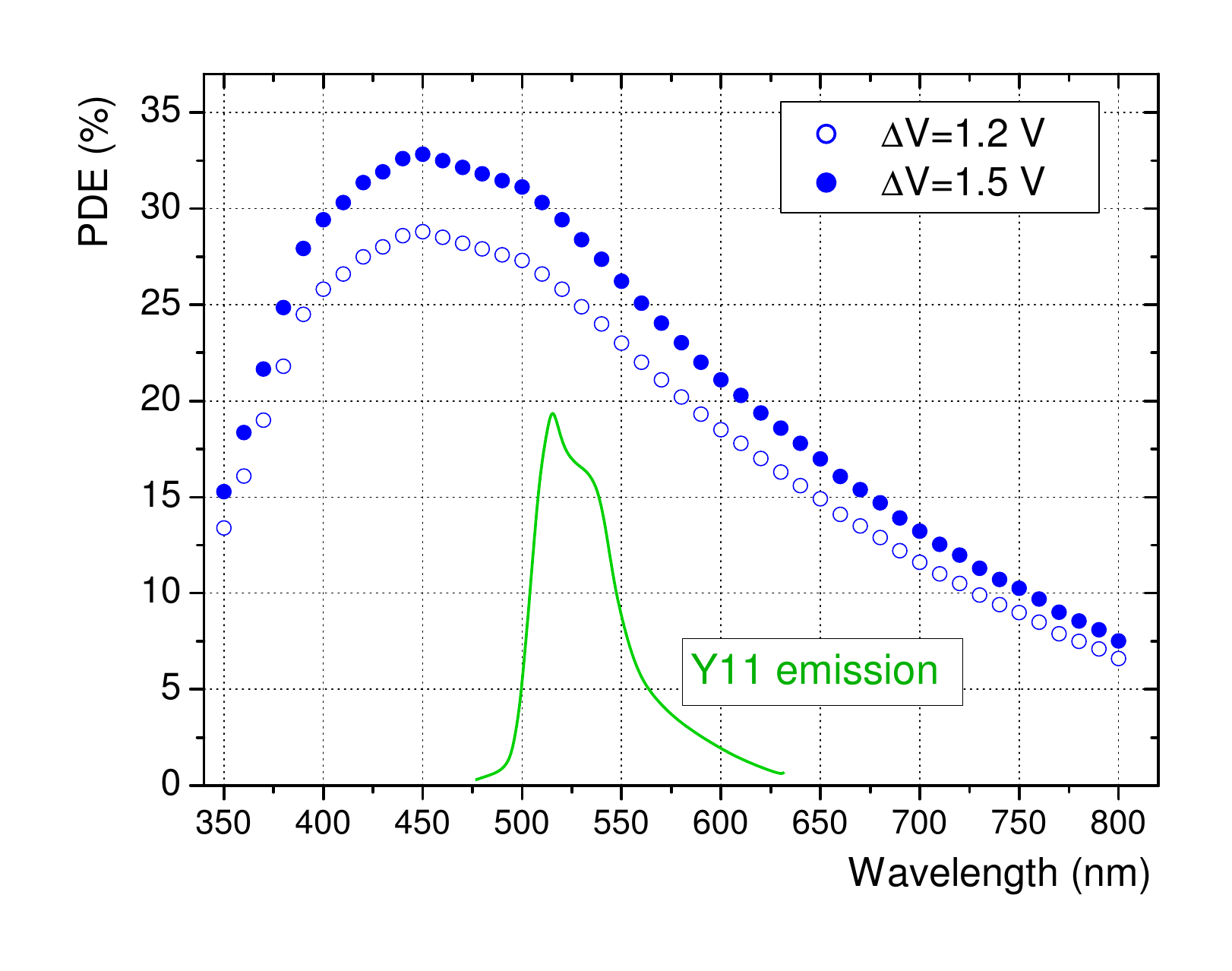}
\caption{Photon detection efficiency  of a Hamamatsu MPPC as a function of wavelength of the detected light at  $\Delta V$ of 1.2 and 1.5~V at 25$^{\circ}$C.  The Y11(150 ppm) Kuraray fibre emission spectrum (in a. u.) for fibre length of 150 cm (from Kuraray specification) is also shown.}
\label{fig:pde_mppc}
\end{figure}

The photon-detection efficiency (PDE) of a SiPM is the product of 3
factors:
\begin{equation}
{\rm PDE} = QE\cdot\varepsilon_{Geiger}\cdot\varepsilon_{pixel} \, ;
\label{eq:pde}
\end{equation}
where $QE$ is the wavelength-dependent quantum efficiency,
$\varepsilon_{Geiger}$, is the probability  to initiate the Geiger
discharge by a photo-electron, and $\varepsilon_{pixel}$ is the
fraction of the total photodiode area occupied by sensitive pixels.
The bias voltage affects one parameter in the
expression \ref{eq:pde}, $\varepsilon_{Geiger}$.   
The geometrical factor $\varepsilon_{pixel}$  is  completely 
determined by the photodiode topology, and is in the range 50-70\%.    
The PDE of a device manufactured by Hamamatsu \cite{HAMAMATSU}
(Hamamatsu uses the name multi-pixel photon counter, MPPC) as function
of the wavelength of the detected light is shown in
figure~\ref{fig:pde_mppc}.

\subparagraph*{Implementation for a very large system \\}

Although SiPMs with integrated electronics are in their infancy, we
can say quite a bit about what such a device might look like in the
case of instrumenting a very large system of extruded scintillator
with wavelength shifting fibre readout, with a channel count of many
million channels. 
In a system of that size, there is no question that an application
specific approach, which can reduce the unit costs by factors of five
to ten, will easily justify the additional non-refundable expense
inherent in a custom development. 
In this application, the SiPM would have an area of about  one square
mm, with about 100 pixels and electronics around the periphery of the
device. 
Each cell would have a few transistors along one of the sides to
provide active quenching---this differs from first-generation SiPMs,
which use a passive resistor as the quenching element. 
In the case of active quenching, each pixel has essentially a digital
latch associated with it. 
When a pixel fires, the latch is set, until an external reset signal
arms the latch once again. 
This mechanism should not be confused with devices such as vertex
pixel detectors. 
In this case, because the pixel capacitance is very small, the signal
voltage is typically between 1~V and 2~V. 
Unlike vertex detectors, there is no amplification needed, and
therefore, there is almost no standing current required in the
transistors and the power dissipation is very small.  
It is only slightly larger than in the case of passive SiPMs and, in
any case, much smaller than in vertex detectors. 
Active quenching has many advantages over passive quenching. 
For the application we are considering here, one of the main
advantages is being able to control precisely when the pixels of the
SiPM are re-armed. 
This greatly improves the dynamic range available for a given number
of pixels because each pixel can fire only once during the
signal-collection period and therefore it is simple to correct for the
probability that some pixels were hit more than once. 
In the passive quenching case, the pixels will start recharging while
there are still signal photons arriving, allowing some pixels to fire
when the pixel is not fully charged, violating the rule that every
pixel gives the same signal. 
This makes saturation correction complicated and unreliable, as it
depends on many details of the signal and the SiPM. 
For the application we are considering here, an active quenching SiPM
with 100 pixels gives about the same resolution as a passively
quenched SiPM with 250 pixels. 
The smaller number of pixels reduces the ratio of the active area to
the area lost to routing, resistors, and optical isolation trenches
between the pixels and makes up for the added dead area associated
with the active-quench circuitry. 
 
\subparagraph*{Readout Electronics \\}

On the periphery of the chip there is circuitry that latches the
number of fired pixels in a FIFO, adds a time stamp and issues
periodic latch resets to the quenching circuits. 
The communication with the chip is serial, with an input, an output
and a clock, all of which are differential signals. 
SiPM bias and ground complete the connections for a total of eight.  
The power for the digital circuitry on the chip is extracted from the
clock lines. 
A number of chips would be connected with flex cables, in a ring
topology, to a data-concentrator module which would service a large
number of SiPMs.  
A reasonable bandwidth available on differential lines over a flex
cable for distances up to a few meters is between 10~Mbps and
100~Mbps. 
Depending on the rate of signals in the detector, and including things
such as protocol overhead and data redundancy, a reasonable estimate
of the number of SiPM chips that can be serviced by a single
data-concentrator is a few thousand.
We expect that in this application, the number of SiPMs in a module
associated with a single data-concentrator would be limited by
mechanical and operational considerations to something like 250 to
500. 
From the data concentrator, the data travel over optical links to
higher-level data collectors.

\subsubsection{Options for Far Detectors}

The Magnetised Iron Neutrino Detector described above is the baseline
for the distant neutrino detectors.
Alternatives to the MIND which, though currently less mature, may
offer advantages, are under consideration and will be discussed briefly
below.

\paragraph{Totally Active Scintillating Detector (TASD)}

One alternative for the MIND for the Neutrino Factory detector is to
build a totally active scintillator detector (TASD). 
Although one could consider using TASD for the baseline Neutrino
Factory with 25\,GeV muon storage rings, TASD has been primarily
studied as an option for a machine of lower energy (4\,GeV to
5\,GeV).
For this scenario, the detector must have excellent neutrino
event-detection efficiency down to event energies of
roughly 500\,MeV.  
Recently, new ideas have made plausible large, cost-effective
detectors \cite{lenf1,lenf2} for lower muon energy, and initial
studies of the corresponding physics reach suggest that an experiment
with a 4\,GeV to 5\,GeV Neutrino Factory would provide an impressive
sensitivity to the oscillation parameters. 
Furthermore, compared with high-energy facilities, a low-energy
Neutrino Factory would require a less expensive acceleration scheme
and a cheaper storage ring that presents fewer underground-engineering
issues.

\subparagraph{Overview}

The Neutrino Factory TASD being considered consists of plastic
scintillator bars with a triangular cross-section arranged in planes
which make x and y measurements.  
Optimisation of the cell cross section still needs further study, but
for now we have chosen the MINER$\nu$A extrusion
design \cite{McFarland:2006pz}.  
The scintillator bars have a length of $15$\,m and the triangular
cross-section has a base of $3$\,cm and a height of $1.5$\,cm.  

\begin{figure}[ht]
\begin{center}
\includegraphics[width=0.79\textwidth]{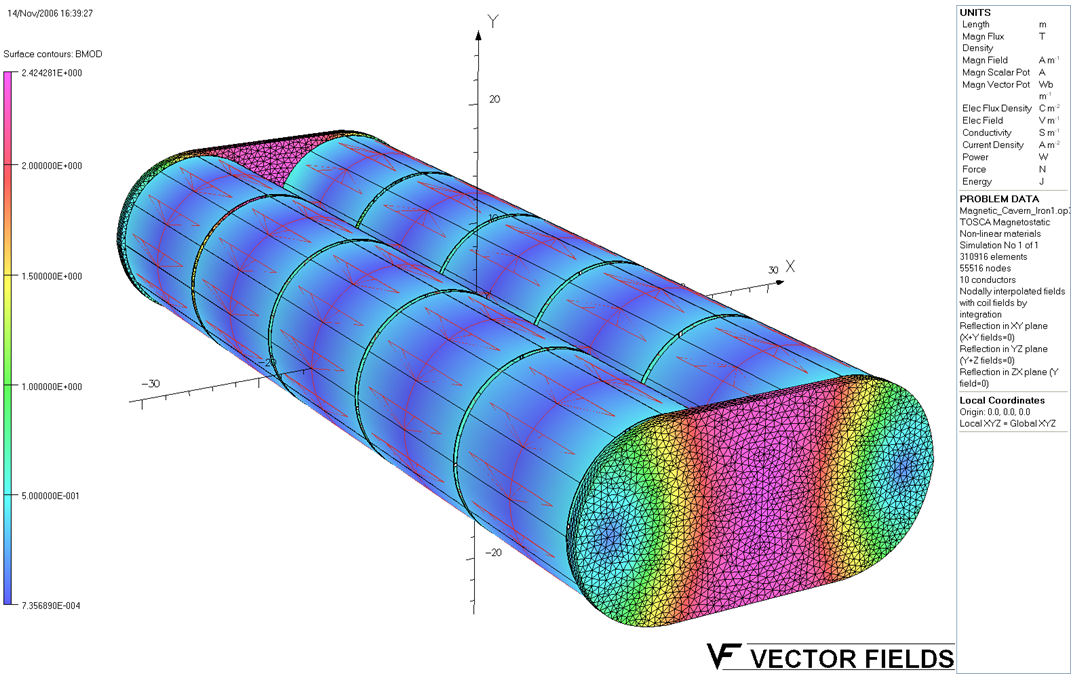}
\end{center}
\caption{Schematic of the Magnetic Cavern concept.}
\label{fig:MagCav}
\end{figure}

Magnetising the large detector volume presents a significant technical
challenge for a Neutrino Factory TASD.  
Conventional room temperature magnets are ruled out due to their
prohibitive power consumption, and conventional superconducting
magnets are too expensive, due to the cost of the enormous cryostats
needed in a conventional superconducting magnet design. 
In order to eliminate 
the cryostat, we have investigated a concept based on the
superconducting transmission line (STL), the technology chosen for
field excitation in MIND. 
The solenoid windings now consist of this superconducting cable which
is confined in its own cryostat (figure \ref{Fig:STL}). 
Each solenoid ($10$ are required for the full detector) consists of
$150$ turns and requires $7\,500$ m of STL. 
There is no large vacuum vessel and thus none of the large vacuum
loads which make the cryostats for large conventional superconducting
magnets so expensive. 
The 10 solenoids forming what we have called a ``Magnetic Cavern" is
shown in figure \ref{fig:MagCav}.

\begin{figure}[ht]
\begin{center}
\includegraphics[width=0.7\textwidth]{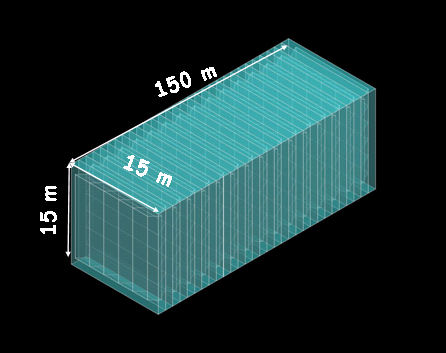}
\end{center}
\caption{Schematic of the Totally Active Scintillator Detector.}
\label{fig:tasd}
\end{figure}

The Neutrino Factory TASD basic response has been simulated with
GEANT4 version 8.1.  
The GEANT4 model (figure~\ref{fig:tasd}) of the detector included each
of the individual scintillator bars, but did not include edge effects
on light collection or the effects of a central wavelength-shifting
fibre. 
A uniform 0.5 Tesla magnetic field was simulated.

Samples of isolated muons in the  momentum range between $100$~MeV$/c$
and $15$~GeV$/c$ were simulated to allow for the determination of the
detector's momentum resolution and charge identification
capabilities. 
The NUANCE \cite{Casper:2002sd} event generator was used to simulate 1
million $\nu_e$ and 1 million $\nu_\mu$ interactions. 
Events were generated in $50$ mono-energetic neutrino-energy bins
between $100$~MeV and $5$~GeV. 
The results that follow only have one thousand events processed
through the full GEANT4 simulation and reconstruction. 
 
The detector response was simulated assuming a light yield consistent
with MINER$\nu$A measurements and current photo-detector
performance \cite{Paley:2008zza}. 
In addition, a 2 photo-electron energy resolution was added through
Gaussian smearing. 
Since  a complete pattern recognition algorithm was beyond the scope
of this initial study, the Monte Carlo information was used to aid in
pattern recognition. 
All digitised hits from a given simulated particle, where the
reconstructed signal was above 0.5 photo electrons, were collected. 
When using the isolated particles, hits in neighbouring $x$ and $y$
planes were used to determine the 3 dimensional position of the
particle.
The position resolution was found to be approximately $4.5$~mm RMS
with a central Gaussian of width $2.5$~mm. 
These space points were then passed to the RecPack Kalman
track-fitting package \cite{CerveraVillanueva:2004kt}.  
 
\begin{figure}[ht]
\begin{center}
\includegraphics[width=0.5\textwidth]{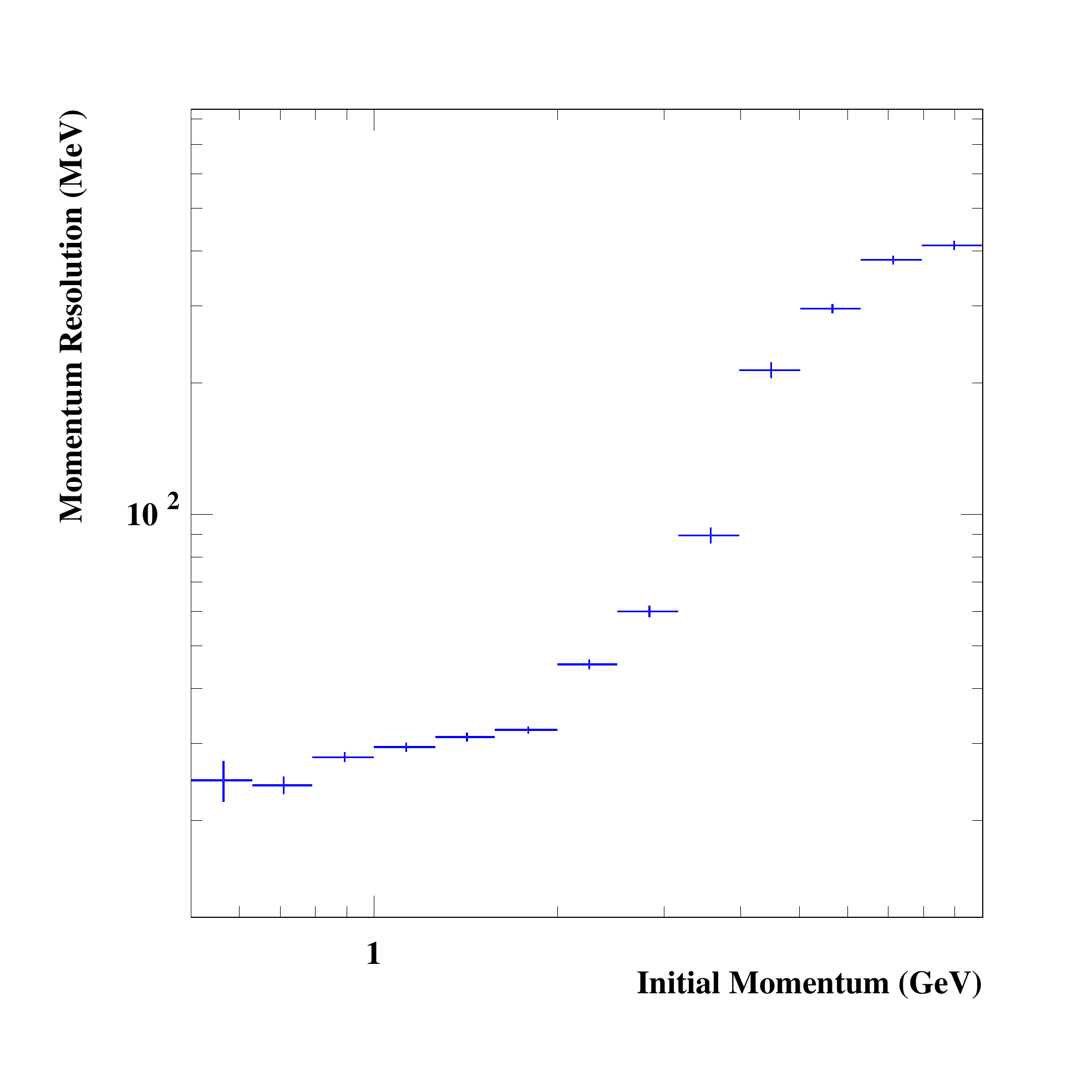}
\end{center}
\caption{TASD momentum resolution as a function of the muon momentum.}
\label{fig:tasd-mom}
\end{figure}

For each collection of points, the track fit was performed with an assumed positive and negative charge. The momentum resolution and charge misidentification rates were determined by studying the fitted track in each case which had the better $\chi^2$ per degree of freedom. Figure~\ref{fig:tasd-mom} shows the momentum resolution as a function of muon momentum. 

\begin{figure}[ht]
  \begin{center}
    \includegraphics[width=0.49\textwidth]{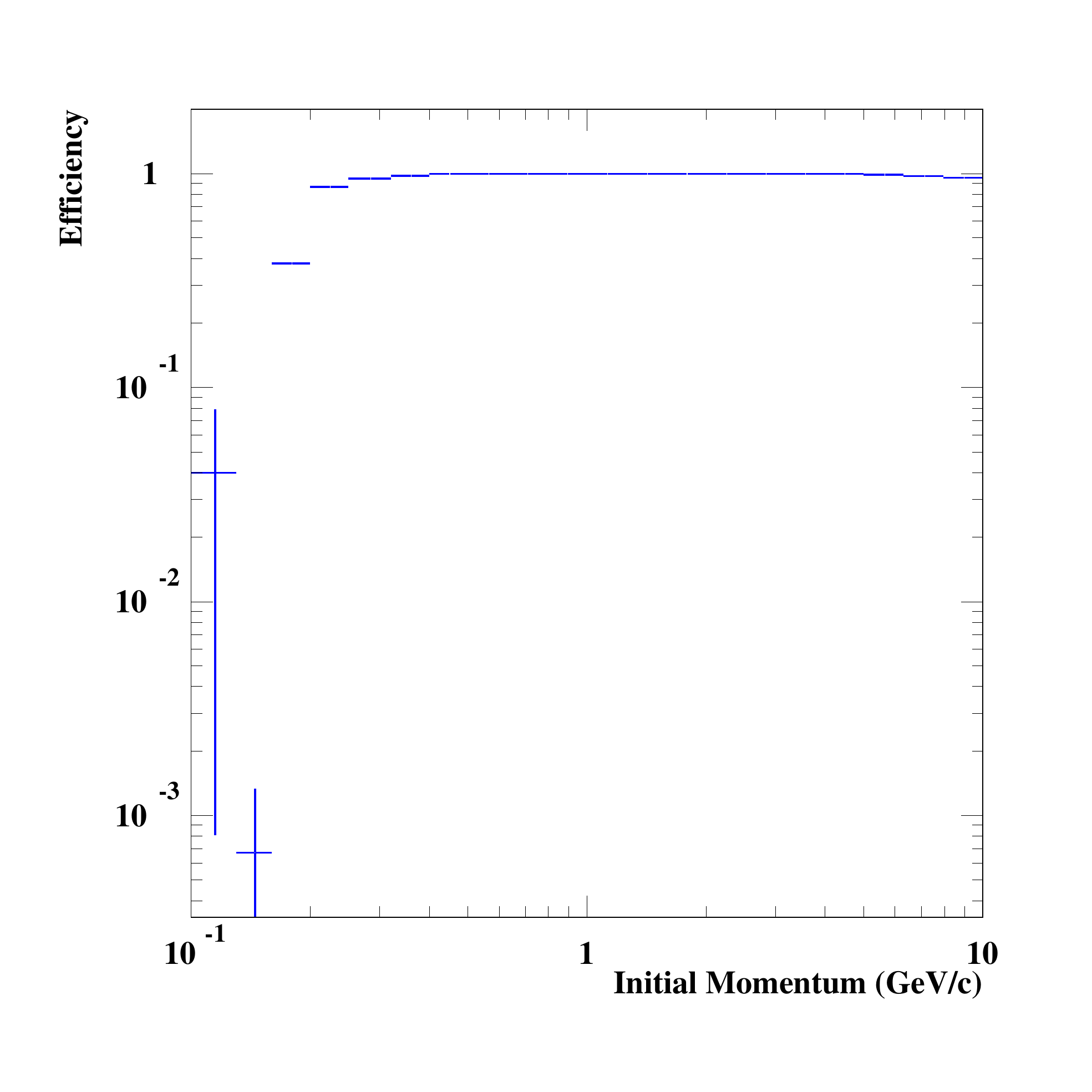}
    \includegraphics[width=0.49\textwidth]{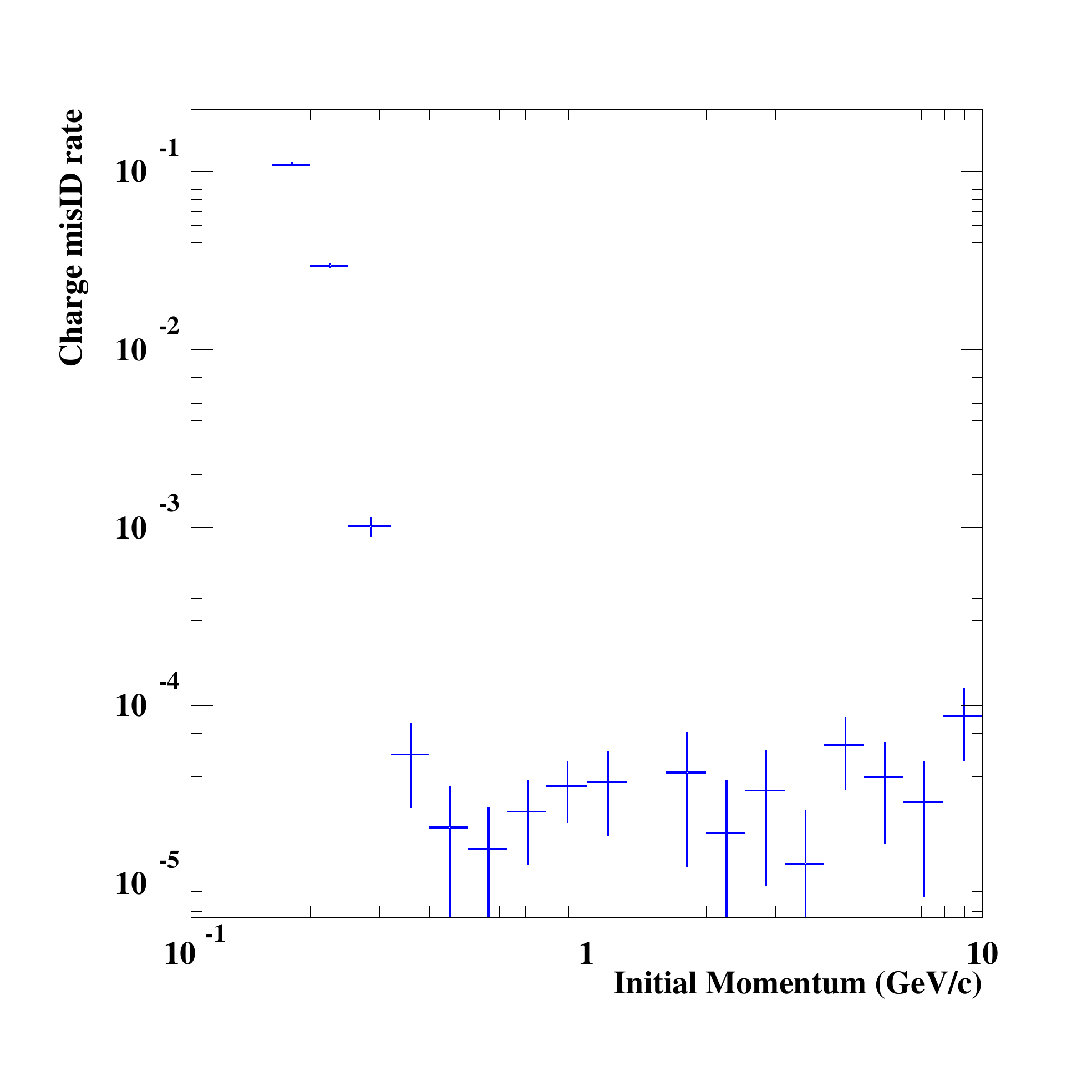}
  \end{center}
  \caption{
    Left: Efficiency for reconstructing positive muons. 
    Right: Muon charge mis-identification rate as a function of the
    initial muon momentum.
  } 
  \label{fig:Track}
\end{figure}

As a tracker, TASD achieves a resolution of better than $10\%$ over
the momentum range studied. Figure \ref{fig:Track} (left) shows the
efficiency for reconstructing positive muons as a function of the
initial muon momentum. The detector becomes fully efficient above
$400$~MeV.  The charge mis-identification rate was determined by
counting the rate at which the track fit with the incorrect charge had
a better $\chi^2$ per degree of freedom than that with the correct
charge. Figure \ref{fig:Track} (right) shows the charge mis-identification rate as a function of the initial muon momentum.

Based on these initial studies, TASD meets the required muon charge
mis-identification rate criterion needed for physics at a Neutrino
Factory and exhibits exceptional tracking and momentum resolution.  
In order to understand fully the potential performance of TASD at a
Neutrino Factory, a Monte Carlo study with full pattern recognition
reconstruction and full neutrino event reconstruction must be
performed, as has been done for MIND.  
This will allow for a detailed and quantitative understanding of
backgrounds and thus present a realistic estimate of the detector's
full potential at a Neutrino Factory. 

\subparagraph{Performance}

In this section we will show the sensitivity to the neutrino
oscillation parameters which can be obtained by a 20 kton Totally
Active Scintillating Detector (TASD) used in conjunction with a
low-energy Neutrino Factory. 
This setup was originally considered in references~\cite{lenf1,lenf2}
and later in references \cite{LENF,mythesis,mynsi}, in which full
details can be found. 
As explained in the main text, we considered the possibility of the
addition of the platinum channels ($\nu_{\mu}\rightarrow\nu_{e}$ and
$\bar{\nu}_{\mu}\rightarrow\bar{\nu}_{e}$) in addition to the usual
golden channels ($\nu_{e}\rightarrow\nu_{\mu}$ and
$\bar{\nu}_{e}\rightarrow\bar{\nu}_{\mu}$). 
The main assumptions used to simulate the TASD are as follows: we
assume an energy threshold of 0.5 GeV with a detection efficiency of
$94\%$ above 1 GeV and $74\%$ below 1 GeV for $\nu_{\mu}$ and
$\bar{\nu}_{\mu}$ with a background of $10^{-3}$. 
For $\nu_{e}$ and $\bar{\nu}_{e}$ we assume an efficiency of $47\%$
above 1 GeV and $37\%$ below 1 GeV with a background of $10^{-2}$. 
The main sources of background are assumed to arise from charge
mis-identification and neutral-current events. 
The energy resolution is assumed to be $10\%$ for all channels. 
A 100\,kton liquid argon (LAr) detector was also simulated, with the
assumptions used to simulate the optimistic performance being the same
as for the TASD, so that the optimistic performance of a 100\,kton LAr
detector coincides roughly with that of a 100\,kton TASD.  

\begin{figure}[htp]
     \centering
     \subfigure[~$\theta_{13}$ discovery potential.]{
          \includegraphics[scale=0.5]{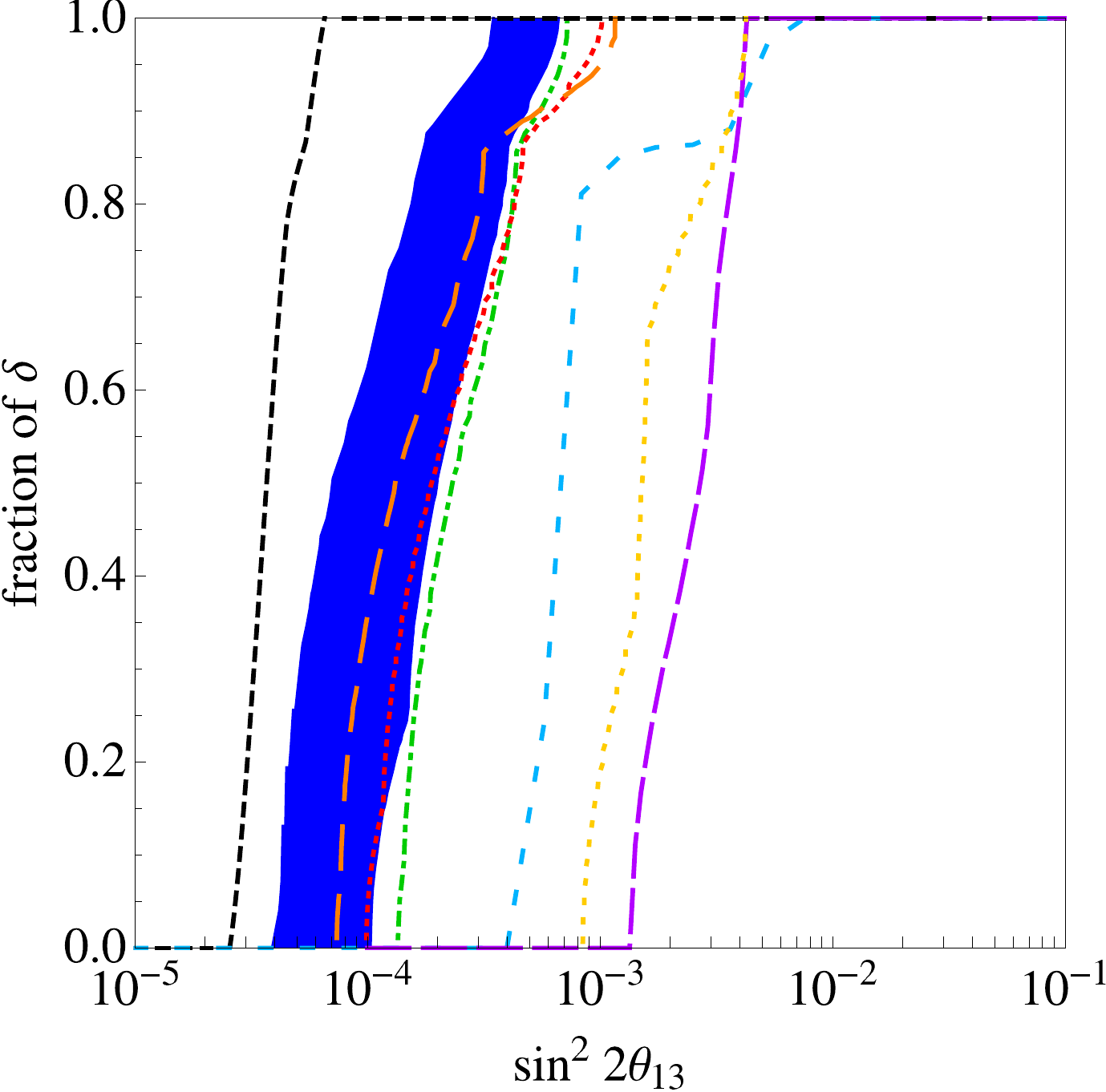}}
     \subfigure[~CP discovery potential.]{
          \includegraphics[scale=0.5]{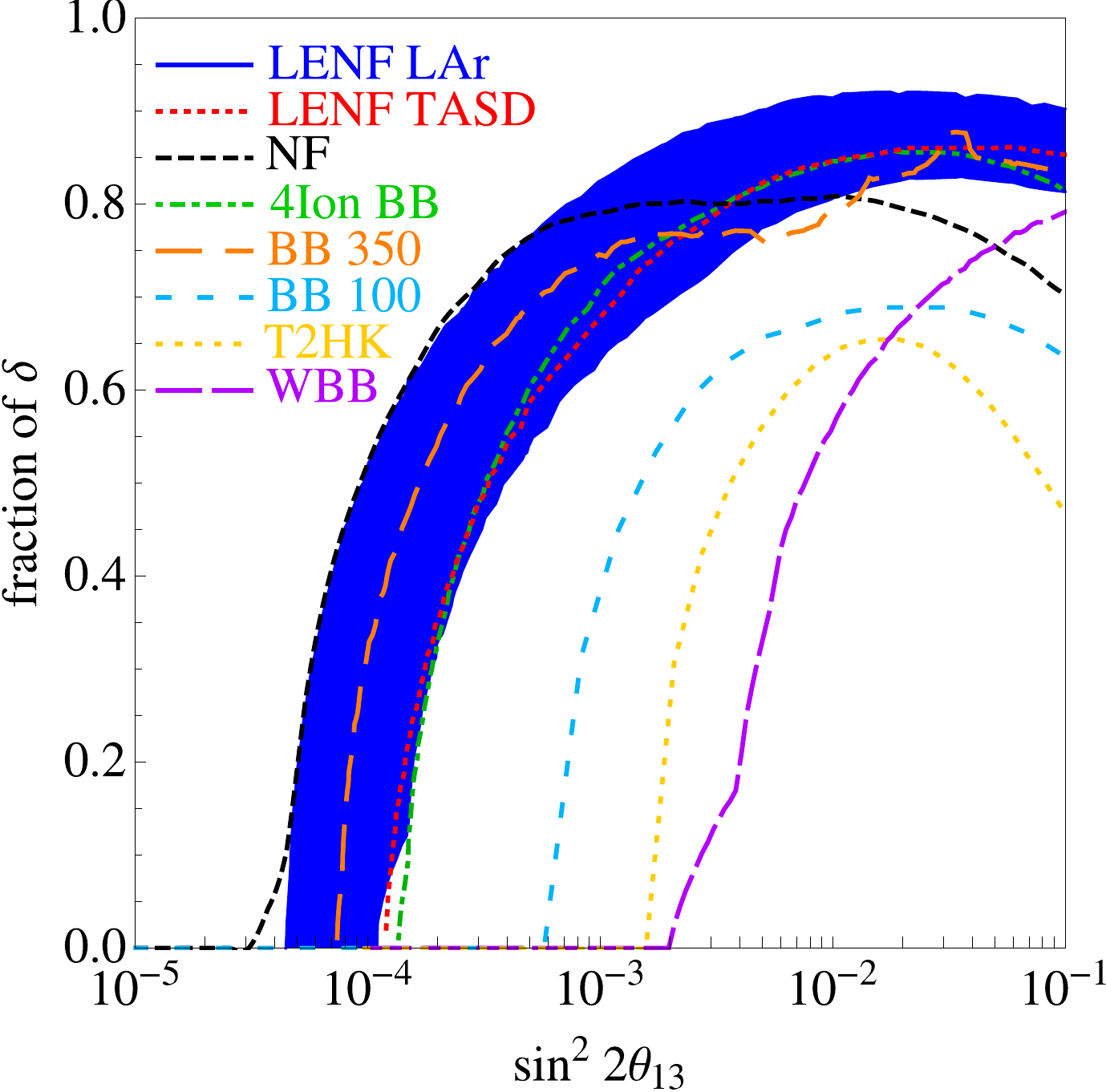}}\\
     \subfigure[~Hierarchy sensitivity.]{
          \includegraphics[scale=0.5]{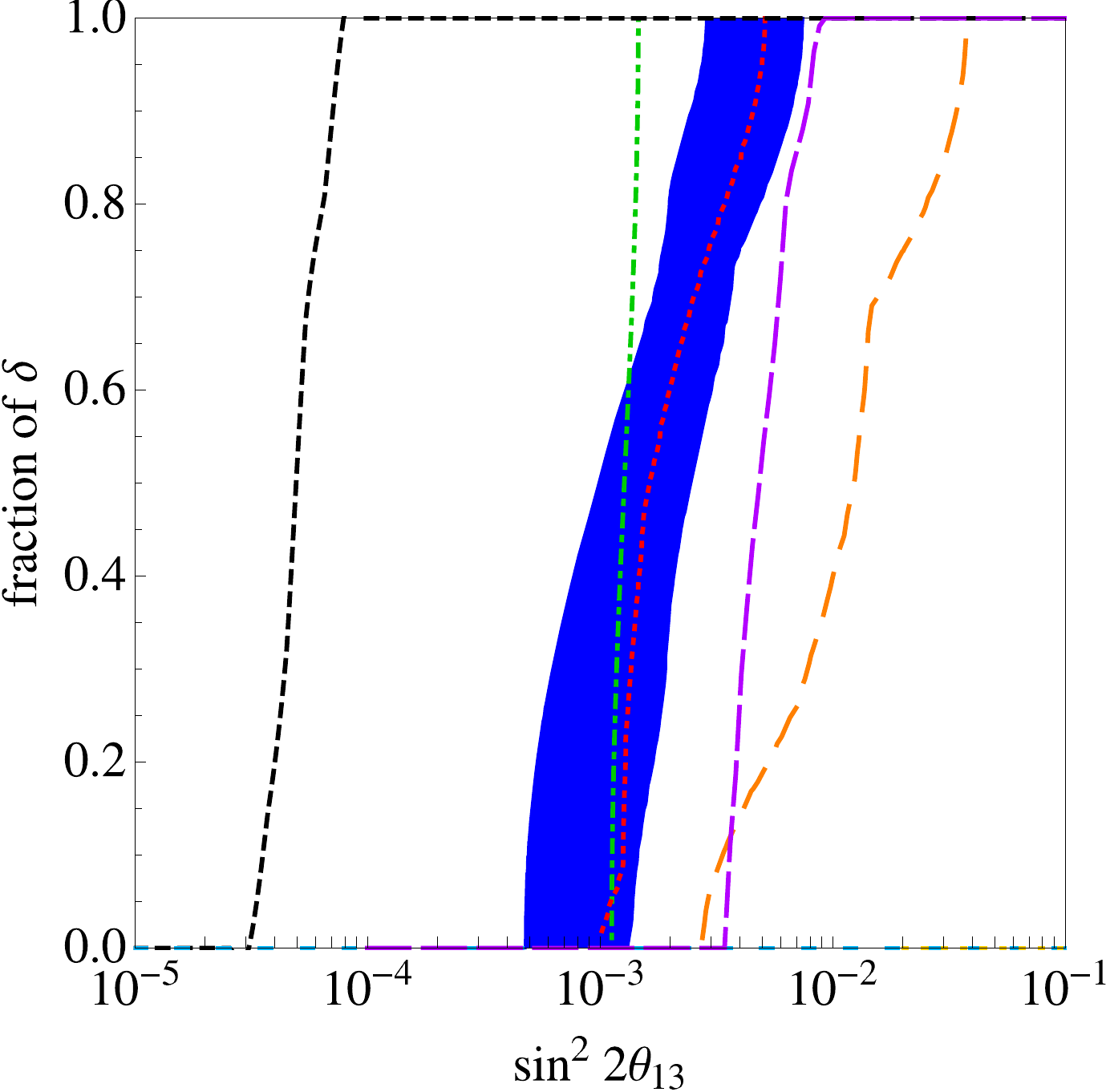}}
     \caption{$3\sigma$ allowed contours for the LENF with 20 kton
     TASD (red line) and 100 kton LAr detector (blue band), the HENF
     (black line), T2HK (yellow line), the wide-band beam (purple
     line) and three $\beta$-beams (green, orange, light blue lines)
     for a) $\theta_{13}$ discovery potential, b) CP discovery
     potential and c) hierarchy sensitivity. From reference \cite{LENF}.}
\label{fig:exps}
\end{figure}

In figure \ref{fig:exps} we show the performance of the 20\,kton TASD
to standard oscillation parameters when used with a low-energy
Neutrino Factory with a single baseline of 1\,300 km and assuming a
muon flux of $1.4\times10^{21}$ useful muon decays per year per
polarity, running for 5 years per polarity. 
The $3\sigma$ $\theta_{13}$ discovery potential, CP discovery
potential and sensitivity to the mass hierarchy as a function of
$\sin^{2}2\theta_{13}$ in terms of the CP fraction are shown, for the
low-energy Neutrino Factory with a 20\,kton TASD (red line), the
low-energy Neutrino Factory with a 100\,kton LAr detector (blue band -
the left-hand edge corresponds to the optimistic assumptions and the
right-hand edge to the conservative assumptions), and other
experiments - the ISS baseline Neutrino Factory \cite{ISS} (black
line), various $\beta$-beams (green \cite{betabeam},
orange \cite{gamma350_1,gamma350_2}, blue \cite{4ion} lines),
T2HK \cite{ISS} (yellow line) and the wide-band beam \cite{WBB}
(purple line).
In these simulations we used the same oscillation parameters as in
reference \cite{baseline}, and have assumed a normal hierarchy.

It can be seen that a low-energy Neutrino Factory with TASD has
sensitivity to CP violation which is roughly equal to that of the
pessimistic performance of the 100 kton LAr detector. 
The sensitivity to $\theta_{13}$ is also very good and again is
roughly equal to that of the pessimistic LAr detector, whilst the
sensitivity to the mass hierarchy is better than that of the
pessimistic LAr detector.

\begin{figure}[htb]
  \includegraphics[width=0.49\textwidth]{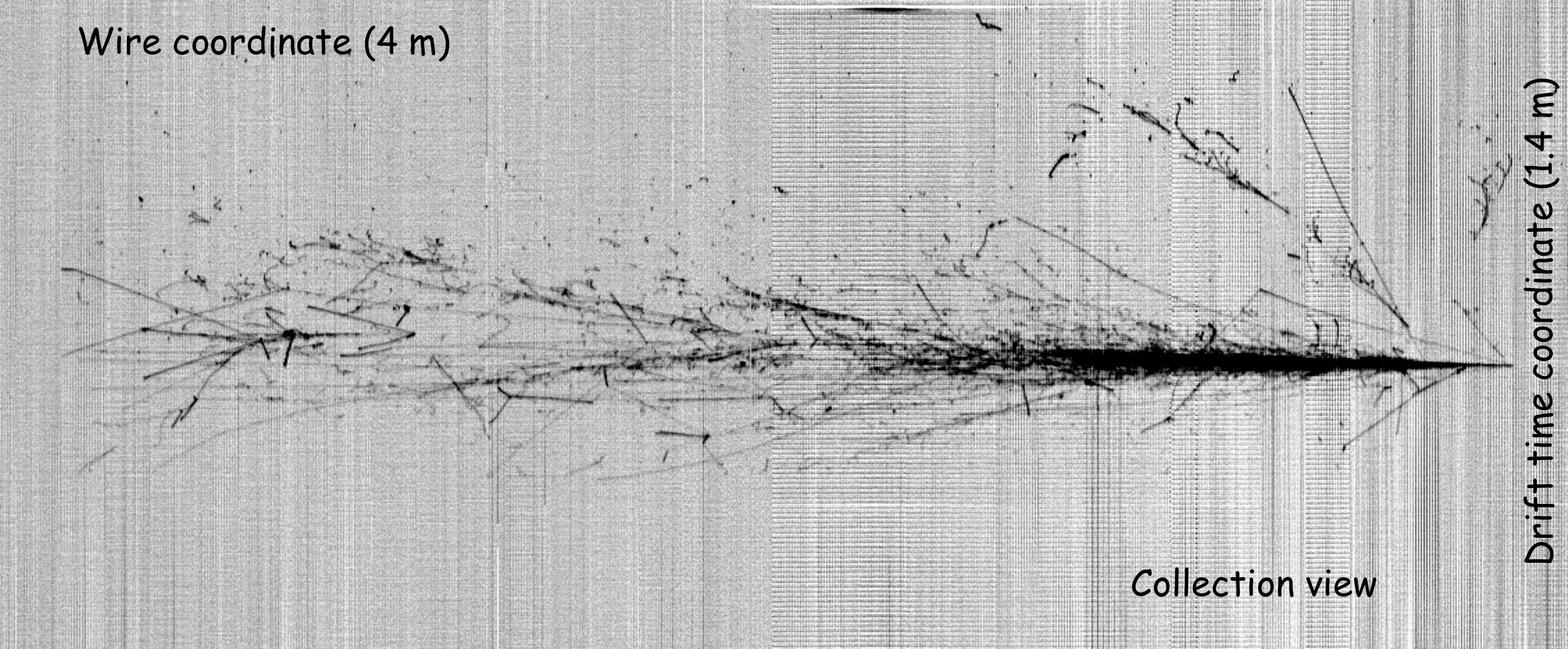}
  \includegraphics[width=0.49\textwidth]{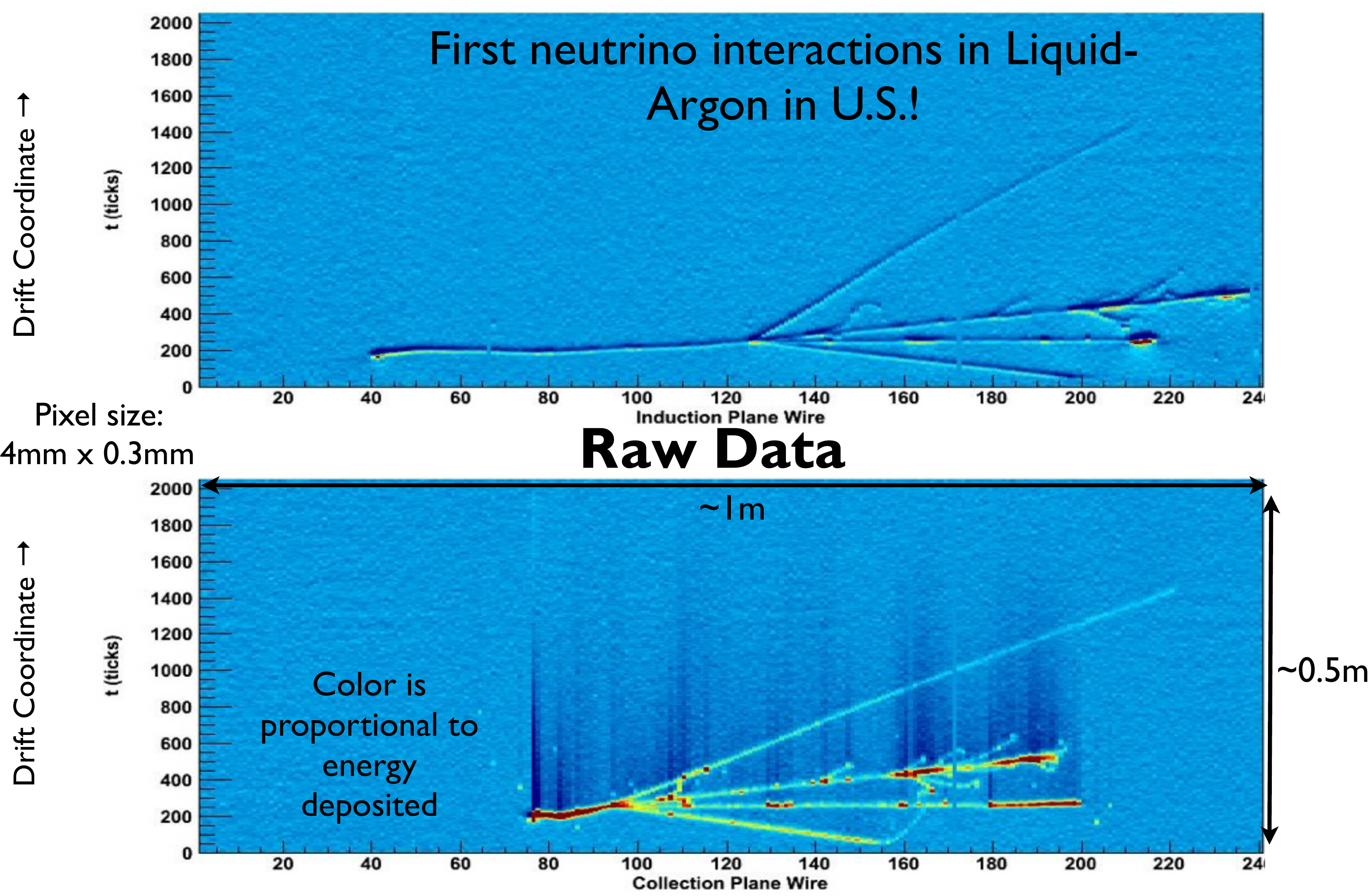} 
  \caption{
    Neutrino interactions as seen in the (left) ICARUS (A. Guglielmi,
    Neutrino'10) and (right) ArgoNeut detectors (M. Soderberg,
    Neutrino'10) 
  }  
\label{Events} 
\end{figure}

\paragraph{Liquid Argon Detectors}
\subparagraph*{Overview\\} 
The Liquid Argon Time Projection Chamber (LArTPC) offers the possibility of
truly isotropic tracking and calorimetry with fine sampling at the level of a
few-percent of a radiation length. 
Pioneered by the ICARUS project~\cite{Amerio:2004ze}, extensive
studies and R\&D projects have proven the potential of the  LArTPC to
provide a clean identification of $\nu_\mu,\nu_e$ charged-current
events  with the efficient rejection of neutral-current backgrounds,
excellent energy resolution and sensitivity over a wide energy band
including down to very low momentum thresholds. 
These properties make the LArTPC an interesting candidate for a
large-scale far detector for a Neutrino Factory and a number of groups
worldwide are actively pursuing LArTPC concept development. Fully
contained neutrino interactions have recently been reported by ICARUS
illuminated by the CNGS beam (see figure \ref{Events}(left)) and from the
ARGONEUT collaboration situated in the NUMI beam at FNAL (see
figure \ref{Events}(right)). 

The largest LArTPC constructed to date is the $2 \times 300$ ton
ICARUS T600 detector which has a maximum drift length of $1.5$ m. 
R\&D programmes underway  in Europe, USA and Japan are focused on
the issues surrounding scaling up the 
LArTPC technology to the multi-kiloton scales demanded by a Neutrino
Factory in addition to  next generation projects for particle
astrophysics and proton decay.

\begin{figure}[htb] 
  \includegraphics[height=3.cm]{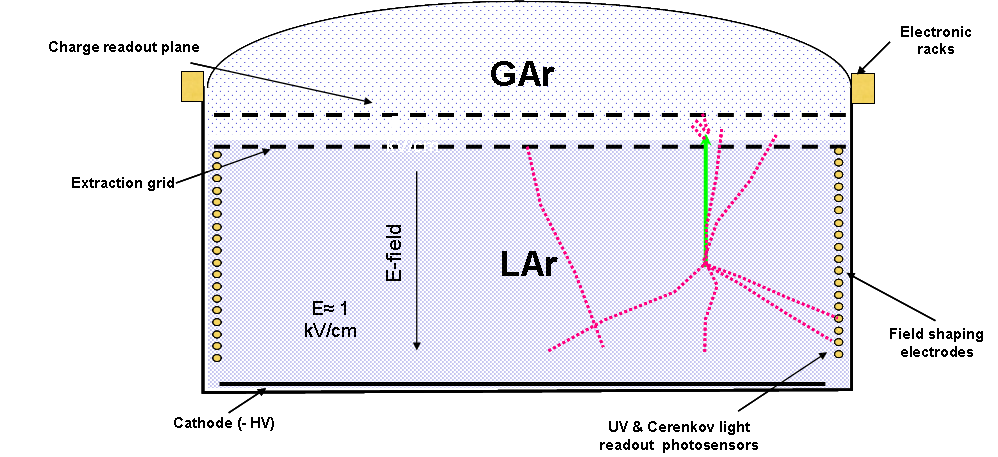} 
  \includegraphics[height=3.cm]{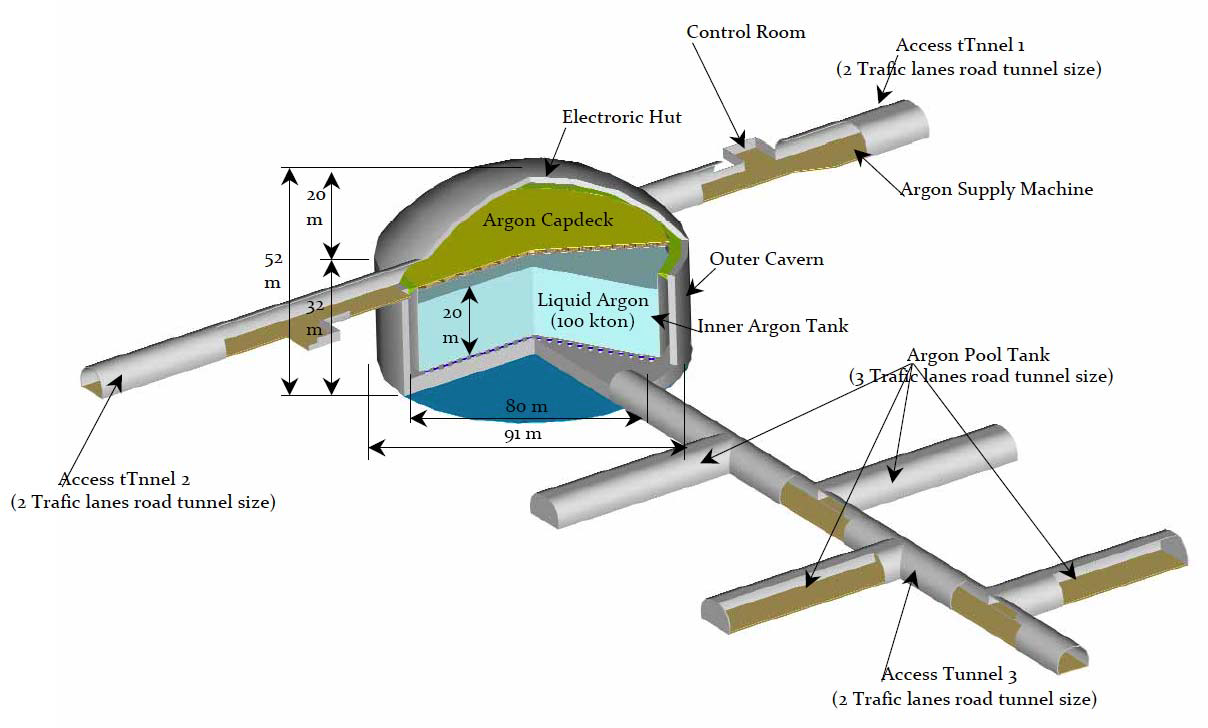} 
  \includegraphics[height=4.cm]{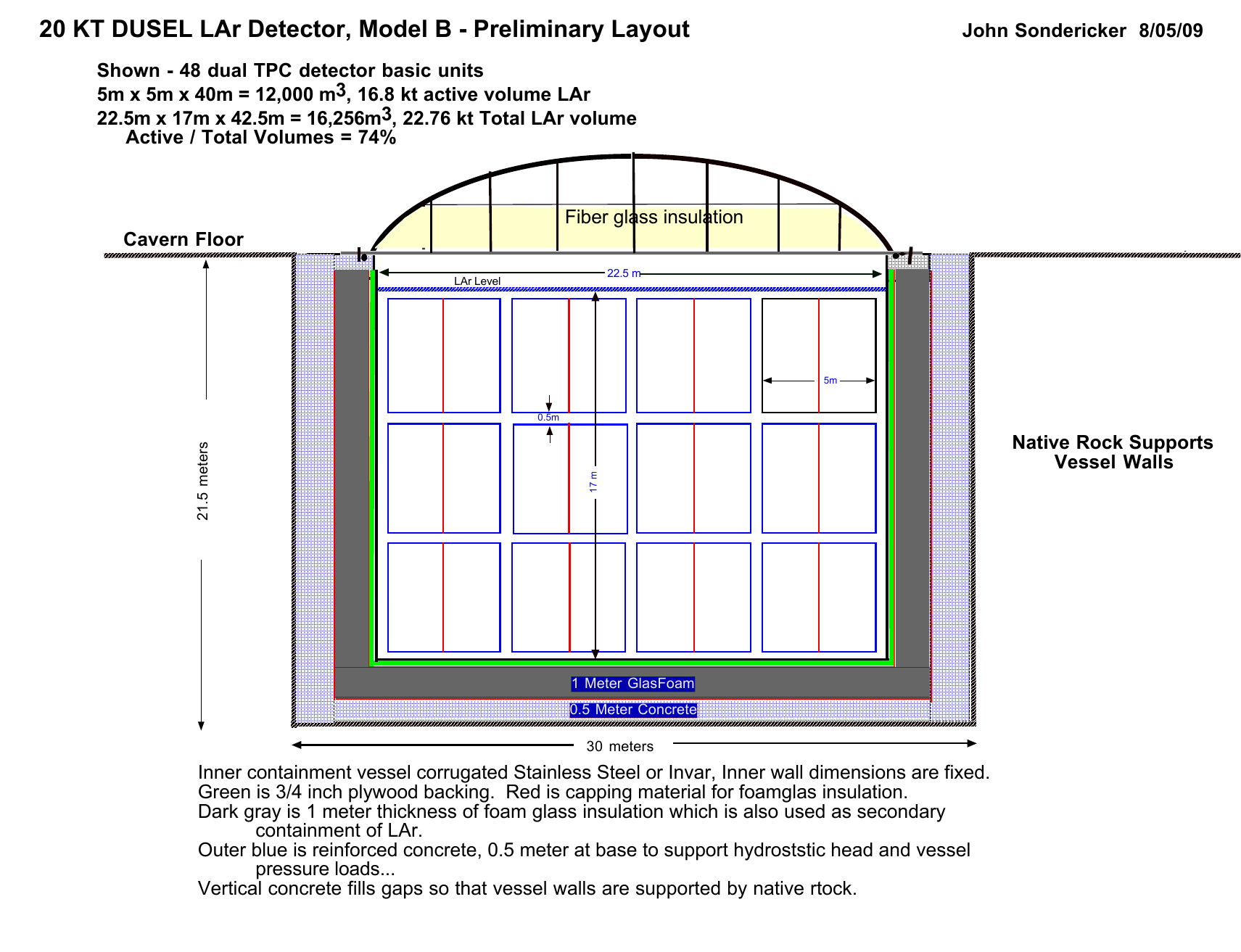} 
  \caption{
    (left) GLACIER, (middle) JPARC-to-Okinoshima 100 kt concept and
    (right) 20 kt detector for DUSEL.
  } 
  \label{Concepts} 
\end{figure}

Several design concepts have emerged for large scale LArTPC detectors
up to $\sim 100$ kt scale 
\cite{Rubbia:2009md,Bartoszek:2004si,Cline:2006st,Angeli:2009zza,Hasegawa-talkNeutrino2010,Fleming-talkNNN09}.
These designs achieve the large mass scale by being made up either of
a number of smaller identical modules or by using a single volume
based on Liquid Natural Gas (LNG) cryogenic tank technology. 
Examples of projects from Europe, Japan and the US are shown in
figure \ref{Concepts}. 
GLACIER and the JPARC-to-Okinoshima project are both based on a single $100$
kton LNG tank with an electron drift distance of up to $20$m whereas the DUSEL
design achieves large mass scales by stacking 
$20$ kton modules thereby keeping the drift distance down to $2.5$ m. 

With all of these large scale proposals, there are three main technical areas
where R\&D solutions need to be found:
\begin{itemize}
\item Engineering issues related to constructing large LAr volumes with very
  low thermal losses, preferably in an underground location and operation over
  many years in a safe and stable environment;
\item Drift lengths of  more than $10$ times the $1.5$\,m ICARUS T600
  drift length will only be achievable if drift fields of $0.5-1$\,kV/cm can be
  applied and if the level of electro-negative impurities, which degrade the
  released ionisation charge, can be controlled at the level of a few tens of
  parts per trillion (ppt); and
\item Reconstructing the $\nu_e \rightarrow \nu_\mu$ and $\bar{\nu}_e
  \rightarrow \bar{\nu}_\mu$~`golden' oscillation channel of the
  Neutrino Factory requires that the entire LAr volume be magnetised.
\end{itemize}

\subparagraph*{R\&D Activity\\}
In this section we summarise the status of R\&D activity in liquid
argon technology. 
\begin{itemize}
  \item {\bf Readout and Electronics:}
  The 2D reconstruction of charge in ICARUS modules is read out by
  planes that cannot easily scale in size, due to their large
  capacitance.  
  An alternative proposal for single volume designs with
  drift distances $>10$ m (e.g. GLACIER) is to operate in double-phase
  liquid and gaseous Ar; ionisation charge amplification occurring in
  the gas phase. 
  A $3$ litre double-phase TPC with thick GEM readout (macroscopic hole
  multipliers fabricated in standard
  PCBs) \cite{Badertscher-IEEE,Badertscher-Pisa} has been successfully
  operated at CERN.  
  More recent developments involve the production of an $X-Y$~2D
  projective anode for a Japanese $250$ litre prototype, studies with
  bulk-micromegas \cite{Rubbia-NNN10} and imaging the secondary
  scintillation light in a single phase LAr
  volume \cite{Lightfoot:2009zz}.
  ASIC developments using cold electronics (preamplifiers, digitisers
  and multiplexers) that work inside the cryostat are under development
  both in Europe \cite{Centro-CERN} and the
  USA \cite{Fleming-talkNNN09};

\item {\bf Liquid-argon vessels:}
LAr storage vessels are based on a stainless steel vacuum-insulated
Dewar. One of the 
large-scale designs \cite{Cline:2006st} scales this technology by
installing support structures within the LAr volume. Studies of the
suitability of industrial Liquid Natural Gas (LNG) technology for
large-scale LAr storage (up to $2\times 10^5$\,m$^3$) \cite{Technodyne}
have concluded that, with only passive insulation, boil-off losses of
only $0.04 \%$/day for a $100$\,kton vessel are expected. 
A stainless
steel/Invar membrane is also being considered for the DUSEL $20$\,kton design
\cite{Fleming-talkNNN09}.  Issues surrounding the feasibility of constructing and running
safely such detectors, particularly in underground sites has been studied as
part of the LAGUNA design study \cite{Rubbia:2010zz};

\item {\bf High Voltage Systems:}
The drift time of ionisation electrons over a distance of $\sim 10$\,m is around
$10$ ms for a drift velocity of $1\, {\rm mm/\mu s}$. It is desirable for the drift velocity to be as high as possible, so drift fields in the region of a few kV/cm are needed (in the MV-range for 10\,m). The ArDM experiment \cite{Kaufmann:2007zz} are generating the high voltages required inside the LAr volume via a submerged 
Cockcroft-Walton voltage multiplier. Drift fields of $\sim 1$ kV/cm have
already been demonstrated  over a  $1.2$ m drift distance \cite{Rubbia-Strasb}.

\item {\bf Liquid-argon purity:}
Attaining and maintaining sufficient purity of LAr is a 
challenge given that electronegative impurities (mainly $O_2$) must be kept at a level of a
few tens of ppt in order to ensure electron lifetimes of the order of $10$ ms.
Proof-of-principle tests for purging with argon gas before
filling with LAr have been performed at
FNAL \cite{Jaskierny:2006sr}, KEK \cite{Maruyama-talkNNN09}, and
CERN \cite{Rubbia-Strasb} (a 6\,m$^3$ volume 
achieved $3$ ppm $O_2$ contamination after purging). 
A closed gaseous Ar purification stage (via cartridges) in addition to a LAr 
purification stage has been used in ICARUS. Options for purification cartridges  are being studied
as part of the ArDM project at FNAL \cite{Andrews:2009zza}. Direct measurements of long electron drifts are underway at CERN ($5$ m horizontal drift \cite{Cline-talkNNN09}) and at the  University of Bern
(ARGONTUBE: $5$m vertical drift);

\item {\bf Magnetisation:} 
The magnetisation of a multi-kiloton LAr volume is one of the key issues 
surrounding the evaluation of a LAr TPC as a potential Neutrino Factory
detector. Small-scale prototype  LAr TPC's have been operated in
magnetic fields; e.g. a $10$~lt, $150$~mm maximum drift length volume in a
field of $0.55$~T \cite{Badertscher:2005te}, with no significant degradation of the imaging
properties. Conceptual studies of magnetising a  multi-kiloton LAr volume
have concluded that conventional warm coils would dissipate MW's of heat
~\cite{Cline:2001pt}, so more recent ideas include immersing a
super-conducting solenoid directly into the LAr volume~\cite{Ereditato:2005yx}, or
using superconducting transmission lines ~\cite{Bross-talkNeutrino2010}. 
Studies are on-going regarding magnetisation issues or the potential of a `hybrid' detector consisting of a LAr TPC in close
proximity to a magnetised MIND-type module or measuring  $\nu/\bar{\nu}$~QE CC reactions with nuclei on a statistical basis 
\cite{Huber:2008yx} in the context of a low energy Neutrino Factory;
    
\item {\bf Algorithmic reconstruction:}
With the increased interaction rates associated with neutrino beam
intensity upgrades, it is essential that automatic reconstruction algorithms for neutrino
interaction events in LAr are developed. Studies in Europe \cite{Barker} and via the LArSoft collaboration
in the USA \cite{Spitz-talkNPC} are actively developing algorithms to reconstruct
hit clusters, tracks, event vertex points and shower objects. The resulting
`toolbox' of algorithms will be used to validate the performance assumptions for
LAr detectors in common use for neutrino oscillation physics studies (see e.g. \cite{Barger:2007jq,FernandezMartinez:2010zza}).     
These tools will also be useful in estimating reliably the feasibility of non
magnetic field solutions to $\nu-\bar{\nu}$ separation in the golden channel
discussed above; and

\item {\bf Roadmaps:}
The  R\&D programmes in Europe and Japan for future large-scale LAr detectors are closely coordinated. Work is ongoing in delivering the large electron multiplier (LEM) readout technologies, demonstration of long drift lengths (5 m), culminating with the construction of a $1$ kton device, which is widely accepted to be the largest possible 
affordable detector that will allow reliable extrapolations to $100$ kton. 
The roadmap in the USA includes the exploitation of the ArgoNeuT in a
neutrino beam, the construction of the 0.1 kton MicroBooNE detector
and delivering a 20 kton detector for DUSEL.  
\end{itemize}

\paragraph{Large-volume liquid scintillator detectors}

Large-volume liquid scintillator detectors (Borexino\cite{Alimonti:2008gc}, KamLAND \cite{Suekane:2004ny}) are presently used for low-energy neutrino physics. Even larger scintillator detectors have been proposed, such as the 50 kton LENA \nocite{Wurm:2007cy,Oberauer:2006cd,Hochmuth:2006gz,MarrodanUndagoitia:2006qs,MarrodanUndagoitia:2006qn,MarrodanUndagoitia:2006rf,MarrodanUndagoitia:2006re,Undagoitia:2005uu,Autiero:2007zj,Wurm:2010ny,Lachenmaier:2010zz} \cite{Wurm:2007cy}--\cite{Lachenmaier:2010zz} detector or the 10 kton HanoHano\cite{Batygov:2008mr,Learned:2007zz} detector. Although these are totally-active scintillator detectors (TASD), these single-volume detectors are conceptually different from segmented scintillator detectors.

The performance of a conventional large volume liquid scintillator detector for high-energy neutrinos (1--5 GeV) has been studied recently \cite{Learned:2009rv,Peltoniemi:2009xx,Smith:2010zzc,Moellenberg2010} and a more detailed study is in preparation \cite{LENA-GeV}. It was found that, if the detector, particularly the read-out electronics, is well designed, one can reconstruct the simplest events using photon-arrival times in phototubes. The lepton-flavour identification was determined to be almost perfect and the energy resolution was found to be better than 5\%. Hence a large-volume liquid-scintillator detector can be used in conventional neutrino beam experiments \cite{Peltoniemi:2009hv} and high-energy beta beam \cite{Peltoniemi:2009zk} experiments, as well as in atmospheric neutrino experiments. If it can be magnetised, it could be also be considered for the detector at a Neutrino Factory \cite{Peltoniemi:2009zb}. 

\begin{figure}[tbhp]
\begin{center}
\includegraphics[width=0.6\textwidth]{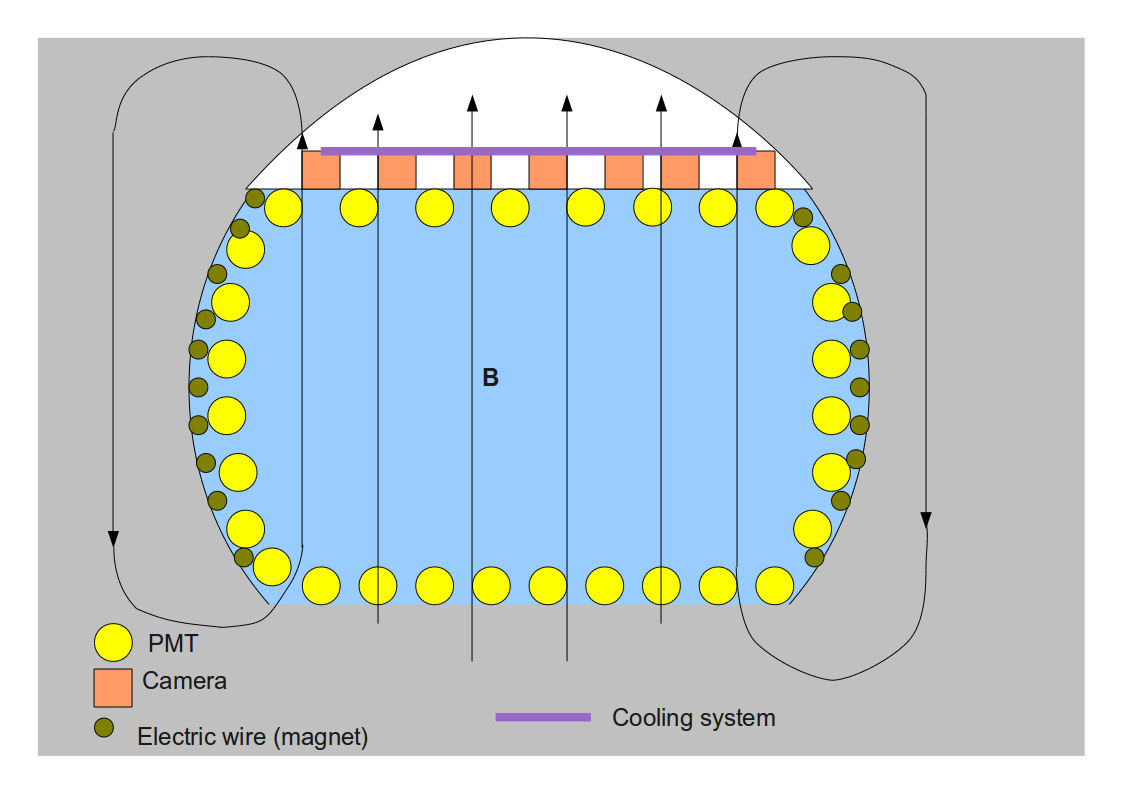}
\caption{A possible layout of the detector. This shows the transverse cross section. Photomultipliers (yellow circles) are located at all sides of the detector and cameras at the upper surface. In this case the vertical magnetic field is generated by longitudinal wires.}
\label{cross}
\end{center}
\end{figure}

Here we consider a magnetised single-volume liquid scintillator detector of approximately 100 kton. A phenylxylylethane (PXE) scintillator \cite{Back:2004zn, Quirk:2008zz, MarrodanUndagoitia:2009kq, Wurm:2010ad} doped with PPO (up to 10 g PPO/1 kg PXE) is able to produce fast scintillation decay times, essential for high energy applications (such as the Neutrino Factory). Additionally, we may not need a separate tank and water shielding or buffer as in LENA, but can use rock walls with a suitable liner, thus reducing the cost and optimising the detector volume.

The required shape and volume could be achieved with a horizontal cavern of 30 m diameter and a length of 150 m. The diameter is mostly limited by the transparency of the liquid, typical attenuation lengths being of the order of 15 m \cite{MarrodanUndagoitia:2009kq, Wurm:2010ad}. Previous studies performed for LENA hint that at least 30 m width is accessible without excessive cost \cite{Nuijten2010}. 

It is assumed that photomultiplier tubes are installed in all sides of the detector volume, including the bottom, the surface and both ends. Photomultiplier tubes (PMT) designed to work in magnetic fields, possibly embedded in $\mu$-metal shields and aligned relative to the field with mirrors for light collection, will be required. As an alternative, semiconductor-based (or hybrid) photo-sensors are also under study. They may be faster and immune to magnetic fields, but may suffer from noise and are currently expensive.

Previous studies \cite{LENA-GeV} suggest that a 3\% photo-cathode coverage is sufficient for high-energy events. The devices should be able to record multiple photons reliably, with a time resolution in the ns scale, to be achieved by 16 bit Flash ADCs. We may optionally install an array of one-photon capable cameras on the upper surface, when such technology is available and affordable (see e.g. \cite{Learned:2009rv}). The liquid-air interface can be used as a lens (objective), because of the high refractive index (1.6 - 1.8) of the oil used as the scintillator. To measure the very faint scintillation light we need single-photon capable pixelized photo-sensors. Potential devices exist based on a number of different technologies such as: ICCD (CCD with image intensifier), EMCCD (Electron multiplier CCD), SPADA (Single photon avalanche diode array) or MCP (micro-channel plate) varieties with binary pixel readout. 

A severe problem with all semiconductor photodetectors is their very high noise rate. A typical noise rate is 100~kHz per pixel at room temperature. With cooling, we may expect 10 kHz per pixel at $0^\circ$ C, and no less than 1~kHz at the freezing point of the liquid, but somewhat lower at liquid nitrogen temperatures. Because of the noise, the camera readout must be externally triggered with the PMT coincidence. A local rolling memory for at least 100 ns, or the full readout time window, is required. The analysis of the PMT data determines the vertex and the end of the muon track with a precision of 20~cm. 

It is reasonable to assume 50\% quantum efficiency and 70\% geometric efficiency. If we also assume 90\% transmission through the lens, this then gives a total photo-efficiency of 30\%. Assuming a photocell composed of $64\times64$ pixels (4096 channels), a lens with a 15 cm diameter and an aperture of 0.30, we obtain a mean resolution of 10 cm at 15 m distance. With a lens diameter of 15 cm and 10\% coverage of the upper surface area, we will measure about 10 photons for a 10 cm path-length along the trajectory of a muon (assuming a typical photon yield of $10^4$ photons/MeV and 20\% transparency at a distance of 15~m). 

The detector performance with PMT time signals was simulated with the prototype code ``Scinderella". The code simulates the passage of muons and scintillation light emission, individual hadrons and electromagnetic showers and tries to make a fit to a test event using the simulated recorded data of the photo-sensors. For a more quantitative picture, a more detailed simulation applying GEANT4 and physical event generators should be used. Here we considered only the bending of muon tracks. Much more detail about the tracking capability is given elsewhere \cite{LENA-GeV}. 

It was found that the muon charge was reliably measured using the time profile information of the photo-sensors. The simulations indicate that a magnetic field strength between 0.02 T for ideal electronics and 0.05 T for current standard technology might be sufficient. Older technology designed only for low-energy events, like that used in Borexino \cite{Alimonti:2008gc}, is not sufficient. The decay time of the scintillator affected the performance. For high-energy events, the lower transparency of higher-doped scintillators is adequate, if we are not aiming for performance at low energies.

Using the additional camera setup reduces the 20 cm resolution
achieved by the PMTs down to a statistical accuracy for an individual
track to better than 5 cm. With this resolution, the charge of the
muon can be identified for $B > 0.02$ T  and a 10 MeV energy
resolution can be achieved. The camera setup, in particular, improves
the capability to distinguish multiple tracks, such as charged pions
and recoil protons. The identification of 10--80 cm long tracks is
enhanced with the addition of the cameras when compared to a PMT only
setup. The camera setup also makes recoil neutron tracks visible. The
neutrons scatter from free protons, giving the proton recoil energies
up to half of the neutron energy, within a radius of $\sim
1$~m. Moreover, the neutron may cause the spallation of additional
neutrons that are absorbed giving gammas of a few MeV. The combination
of the camera and the PMT measurement may help to identify the
absorption signals.  
 
The camera readout will improve the measurement of the energy of the hadron showers produced in multi-GeV neutrino interaction. Electrons and photons causing electromagnetic showers can be identified by observing the gap between the vertex and the shower, which is 42~cm on average. This is important for rejection of neutral-current background events (and also for proton decay $p \to e^ + \pi^ 0$). 
 
Identifying a pion (from a muon) is important for neutral current
background rejection. The pion decays with a mean life-time at rest of 26\,ns. The resulting muon (4 MeV for a decay at rest) has a path-length of about 2 cm. In the case of decay in flight, the decay is recognised by an energy bump in the time profile and a sharp change in the direction of the path. To see the angle, we need a timing resolution better than 1 ns. Negative pions may be absorbed before decaying (giving photons).
	
We can summarise that a liquid scintillator detector setup with a magnetic field $B>0.02$~T, good PMTs and adequate electronics is sufficient to determine the muon charge. Combining PMT readout with a camera system gives additional tracking capability, with very good event identification and background rejection. Further studies to quantify these effects at a neutrino beam will be carried out in the future.

\subsection{Near detectors}

\subsubsection{Introduction}

As mentioned in section~\ref{sec:Baseline_near}, the baseline is to
have four near detectors, one at the end of each of the four
straights of the two storage rings. 
This allows the measurement of the neutrino flux of each neutrino
beam and reduces the systematic error in the oscillation parameters
\cite{Tang:2009na}. 
Apart from the flux measurement, the near detector will also be able
to perform precision measurements of neutrino-nucleon,
neutrino-electron, and charm-production cross-sections.
Each of these measurements is necessary to reduce the systematic errors
of the neutrino oscillation analysis. 
Additionally, the near detector will embark on an extended programme
of precision neutrino physics and searches for Non Standard
Interactions (NSI). 

Section~\ref{sec:ND_performance} describes the performance required of
a near detector at a Neutrino Factory. 
It includes an analysis using neutrino-electron scattering to extract
the neutrino flux with a near detector, a description of how the near
detector flux measurement can be used to extrapolate to the far
detector to constrain the parameters of the neutrino oscillation
signal, a description of the neutrino scattering physics that can be
achieved with a near detector (including cross-sections, QCD and other
electroweak physics topics) and finally a section on the measurement
of charm production from neutrino interactions (one of the dominant
backgrounds in the oscillation signal in the far detector) and the tau
search, which can be used to constrain NSI.  

Section~\ref{sec:ND_design} describes the requirements of the near
detector to achieve the above physics goals, including two possible
options that are being considered, one with a scintillating fibre
tracker and the other using a high resolution straw-tube tracker.
In addition, two other options will be considered to perform 
charm and tau measurements; one of which includes an emulsion
detector, the other a silicon vertex detector.

\subsubsection{Performance requirements}
\label{sec:ND_performance}

\paragraph{Neutrino flux measurement; inverse muon decay}

\subparagraph*{Introduction \\}

In order to perform measurements of neutrino oscillations at a
neutrino facility, it is necessary to establish the rate of neutrino
interactions. 
The aim of the near detector of the Neutrino Factory is to
measure precisely the absolute neutrino flux, the neutrino cross
sections, and to estimate the backgrounds in the far detector. 
Hence, careful design of a near detector is crucial for the reduction
of systematic uncertainties in the long-baseline neutrino-oscillation
experiment. 
The paragraphs which follow demonstrate that quasi-elastic
neutrino-electron scattering can be used to measure the neutrino flux
coming from the Neutrino Factory storage ring with a systematic
uncertainty of $ \sim 1 \%$. 

\subparagraph*{Quasi-Elastic scattering off electrons in the near detector \\}

Quasi-elastic neutrino-electron scattering is suitable for the measurement of the neutrino flux because its absolute cross-section can be calculated theoretically with confidence. The two process of interest for neutrinos from $\mu^-$ decays are:
\begin{eqnarray}
 \nu_{\mu} + e^{-} & \rightarrow & \nu_{e} + \mu^{-} \, ; 
   {\rm and}  \label{p1} \\
 \bar{ \nu_{e} } + e^{-} & \rightarrow & \bar{ \nu_{\mu} } + \mu^{-} \, .
   \label{p2}
\end{eqnarray}
For the processes in equation \ref{p1}, also known as inverse muon
decay, the differential cross section is isotropic in the
centre-of-mass system. The total cross section is given by:  
\begin{equation}
\sigma=\frac{G^2_F}{\pi}\frac{ (s - m^2_{\mu})^{2} }{s} \, .
\end{equation}
For the process in equation \ref{p2}, also known as muon production
through annihilation, the differential cross section in the
centre-of-mass system is given by:
\begin{equation}
 \frac{d\sigma}{dcos\theta}=\frac{G^2_F}{\pi}\frac{(s-m^2_{\mu})^{2}}{s} \times 
\left(1+\frac{s-m^2_{\mu}}{s+m^2_{\mu}}cos\theta\right)\left(1+\frac{s-m^2_{e}}{s+m^2_{e}}cos\theta\right) \, ;
\end{equation}
and the total cross section is:
\begin{equation}
 \sigma=\frac{G^2_F}{\pi}\frac{ (s - m^2_{\mu})^{2} }{s^2}(E_{e}E_{\mu}+\frac{1}{3}E_{\nu 1}E_{\nu 2}) \, ;
\end{equation}
where $E_{\nu 1}$ and $E_{\nu 2}$ are the energies of the neutrinos. $E_{\nu 1}$ and $E_{\nu 2}$ depend, in turn, only on  $s$. Both processes have a threshold at $\sim 11$~GeV.

\subparagraph*{Simulation of the near detector \\}

A Monte Carlo simulation of muon decay in flight along the length of the 600\,m straight section of the decay ring has been developed. The two leptonic processes in equations \ref{p1} and \ref{p2} have been simulated in order to determine the detector requirements and to select the best criteria for the suppression of the background. After this, the GENIE \cite{Andreopoulos:2009zz} event generator and the GEANT4 simulation tool were used to simulate the entire spectrum of neutrino interactions and the response of the detector.

\subparagraph*{Near detector requirements \\}

If we want to measure the neutrino flux by using the quasi-elastic neutrino-electron scattering for earlier measurements of these processes see \cite{Vilain:1996yf,Mishra:1990yf}), the detector has to be able to distinguish between the leptonic events and inclusive charged current (CC) neutrino-nucleon scattering:
\begin{equation}
 \nu_{\mu} + N \rightarrow \mu^{-} + X \, ;
 \label{p3}
\end{equation}
which has a cross section a few orders of magnitude larger. Some of the events from the charged-current processes (equation \ref{p3}) can mimic leptonic events from quasi-elastic neutrino-electron scattering, but instead of a single muon in the final state there will be also a hadronic system $X$. The measured recoil energy of this hadronic system can be used as a good criterion for the suppression of the background.

\begin{figure}
\begin{center}
  \includegraphics[width=0.5\textwidth]{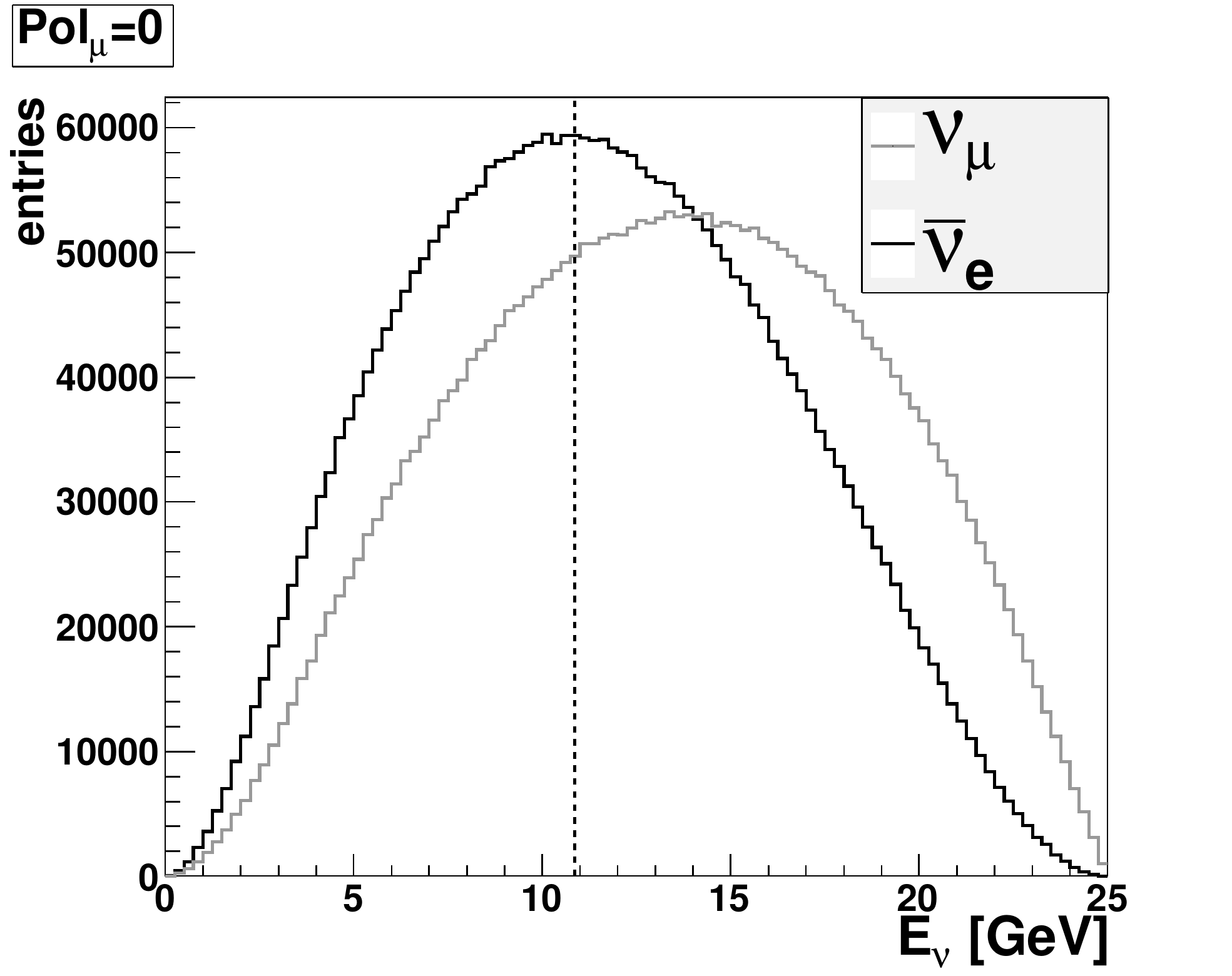}
\end{center}
\caption{Distributions over the neutrino energy on the plane perpendicular to the straight section and 
100~m away from its end. Dotted lines indicate the threshold for the process of quasi-elastic scattering off electrons.}
\label{NuE}
\end{figure}

The energy spectrum of the beam of neutrinos 100\,m from the end of
the straight section are shown in figure \ref{NuE}; the
thresholds for the two processes of interest are also shown.

\begin{figure}
\begin{center}
  \includegraphics[width=.49\textwidth]{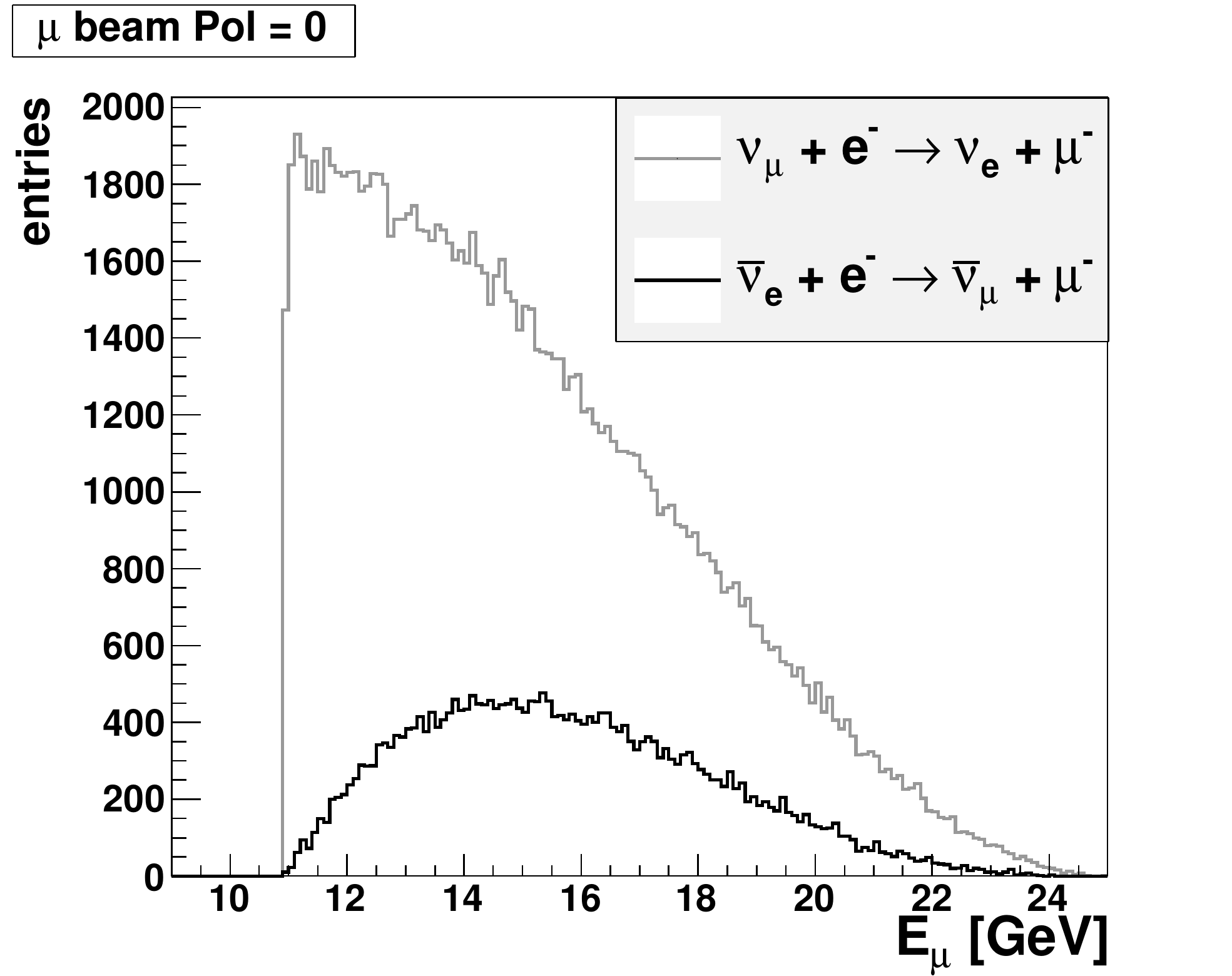} 
  \includegraphics[width=.49\textwidth]{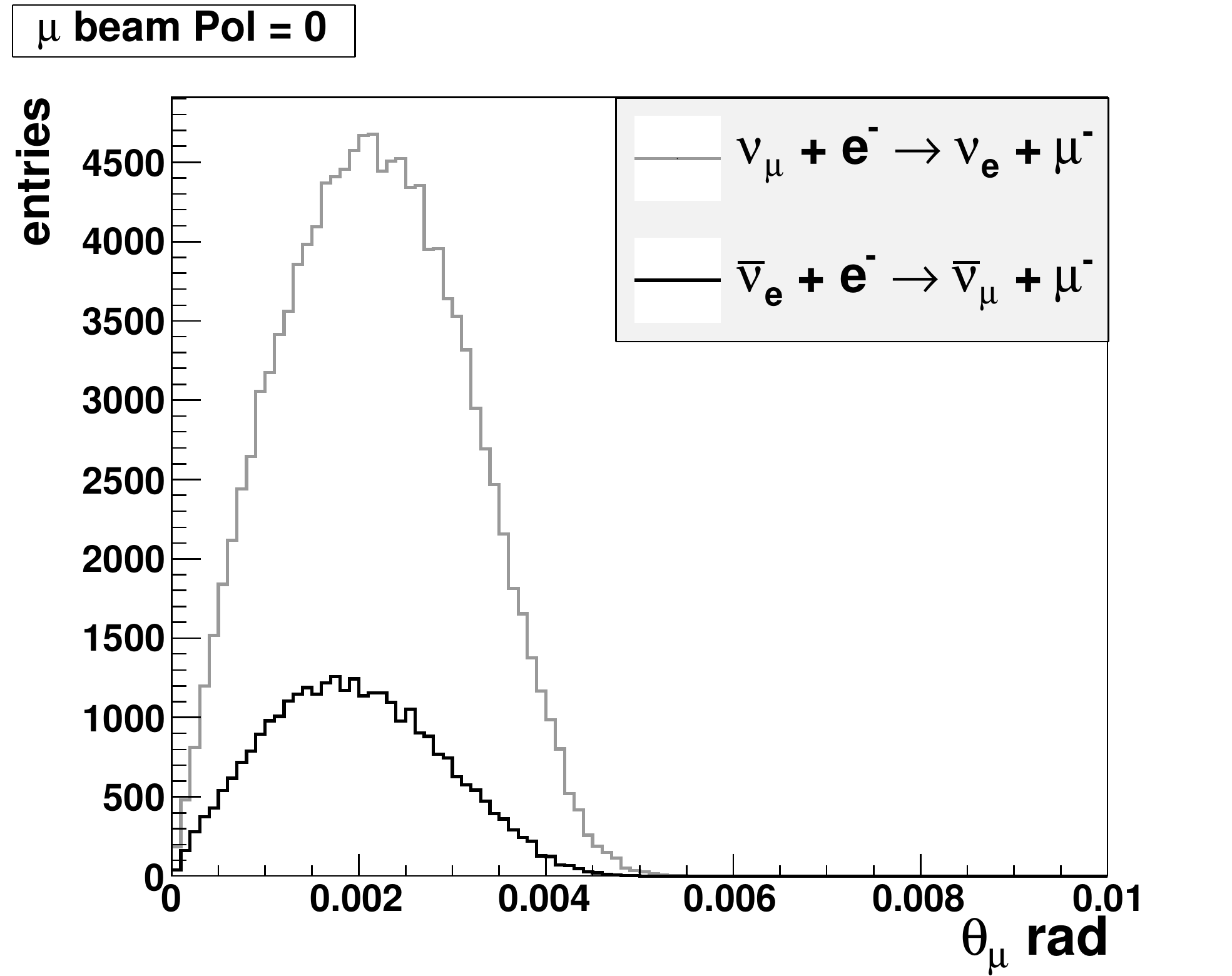}
\end{center}
  \caption{
    The energy distributions (left) and the polar angle distributions
    (right) of the muons generated in the detector by the reactions
    $\nu_{\mu} + e^{-} \rightarrow \nu_{e} + \mu^{-}$ and
    $\bar{\nu}_{e}  + e^{-} \rightarrow \bar{\nu}_{\mu} + \mu^{-}$.
  }
  \label{Mu_Th_E}
\end{figure}

Figure \ref{Mu_Th_E} shows the distributions over the energy and the
polar angle of the muons from quasi-elastic neutrino-electron
scattering at that plane. One can see that all these muons have very
small polar angles $\theta_{\mu} < 5$~mrad. This angular spread comes
mainly from the muon beam divergence as the intrinsic scattering angle
in processes \ref{p1} and \ref{p2} is much smaller.  
We use this as another criterion for suppression of the background.

The properties of the quasi-elastic scattering off electrons described
above impose specific requirements on the near detector. 
The detector has to provide an interaction rate sufficient for the
measurement despite the very small cross sections of the leptonic
processes, $\sim  4\times 10^{-41}$~cm$^2$ at 15~GeV (for comparison,
the total $\nu_{\mu}N$ CC cross section is  $\sim 10^{-37}$~cm$^2$ at
this energy). 
This requires a solid detector. 
Also, the detector must be able to reconstruct the polar angle of the
scattered muon with maximum precision. 
This requires a low-$Z$ tracker. 
At the same time, the near detector has to be able to measure the
hadronic-recoil energy in the background events down to values of
several MeV. 
This requires a precise calorimeter.

\subparagraph*{Neutrino event generation \\}

In order to test different criteria for the suppression of the background from CC reactions and to determine the strict requirements for the near detector we make use of the neutrino event generator GENIE \cite{RefGenie1,Andreopoulos:2009rq} to simulate the interactions of the neutrinos with a detector made of polystyrene ($\rho = 1.032$~g/cm$^3$). The GENIE simulation uses as an input the neutrino flux created by the simulation of muon decay in flight.

The following neutrino interaction processes are included in the GENIE event generator:
\begin{itemize}
 \item Quasi-elastic scattering;
 \item Elastic NC scattering;
 \item Baryon resonance production in charged and neutral current interactions;
 \item Coherent neutrino-nucleus scattering;
 \item Non-resonant deep inelastic scattering (DIS);
 \item Quasi-elastic charm production;
 \item Deep-inelastic charm production;
 \item Neutrino-electron elastic scattering; and
 \item Inverse muon decay.
\end{itemize}
The process  $\bar{\nu}_{e}  + e^{-} \rightarrow \bar{\nu}_{\mu} +
\mu^{-}$ is not included in GENIE. This is not crucial for our
simulation since, for unpolarised muons, it has a rate that is $\sim
10$ times less than that of inverse muon decay.

\subparagraph*{GEANT4 simulation of the near detector \\}

In the GENIE data files all final state particles are recorded using the GHEP library. These particles are then read through an interface by the GEANT4 simulation and treated as primary particles for tracking through the volume of the detector.

The detector that has been simulated is a low-$Z$, high-resolution scintillating-fibre tracker. It consists of consecutive modules placed perpendicular to the beam axis. Each module consists of five planes made of plastic-scintillator slabs 1~cm thick and a fibre station. The scintillator slabs serve to absorb and measure the energy of recoil particles in the interaction. The fibre station consists of four layers of fibres with horizontal orientation and four layers with vertical orientation. Signals from individual fibres are used to construct space points (hits) from charged particles crossing the station.

\begin{figure}
\begin{center}
 \includegraphics[width=.45\textwidth]{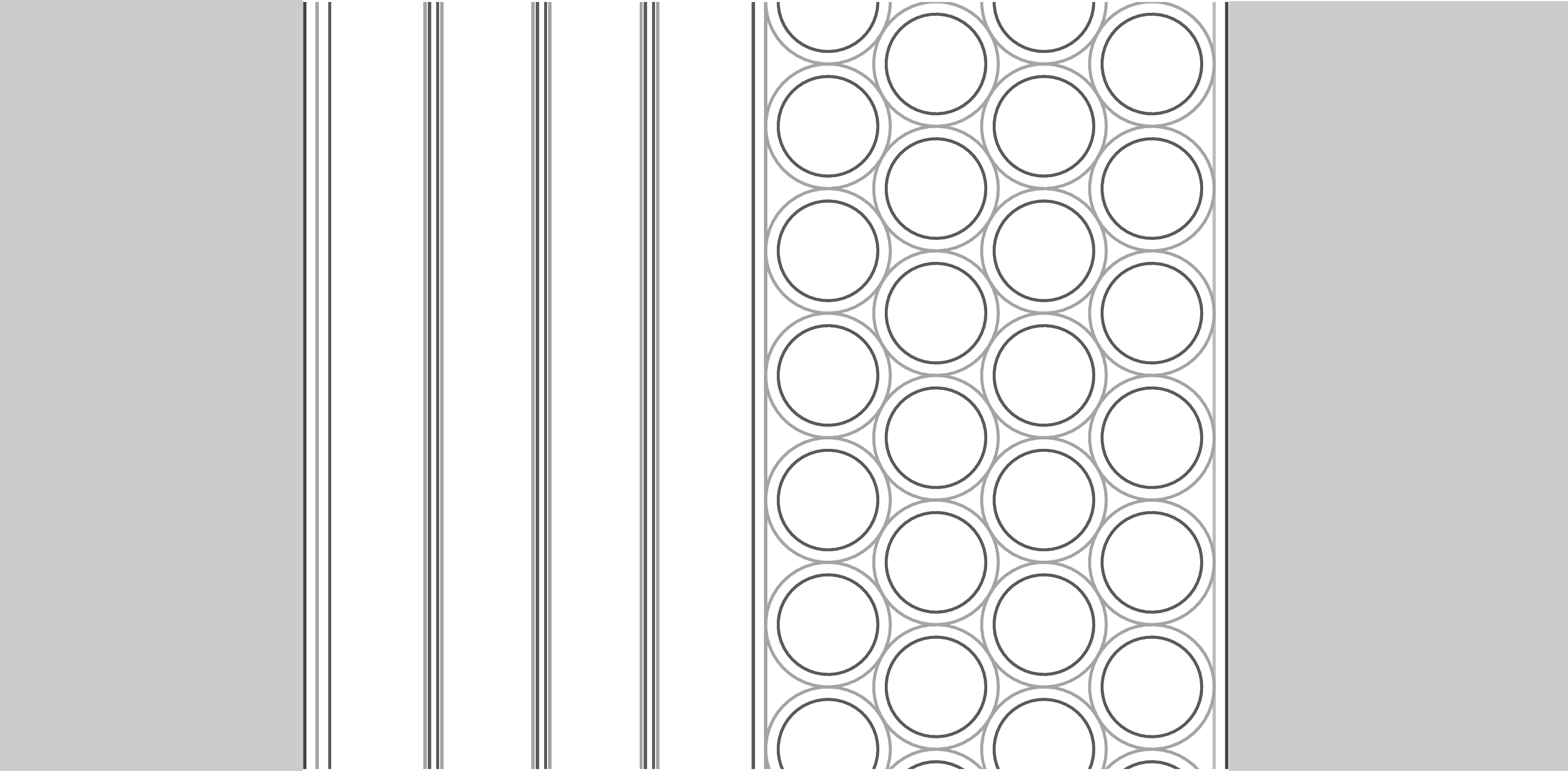}
 \includegraphics[width=.45\textwidth]{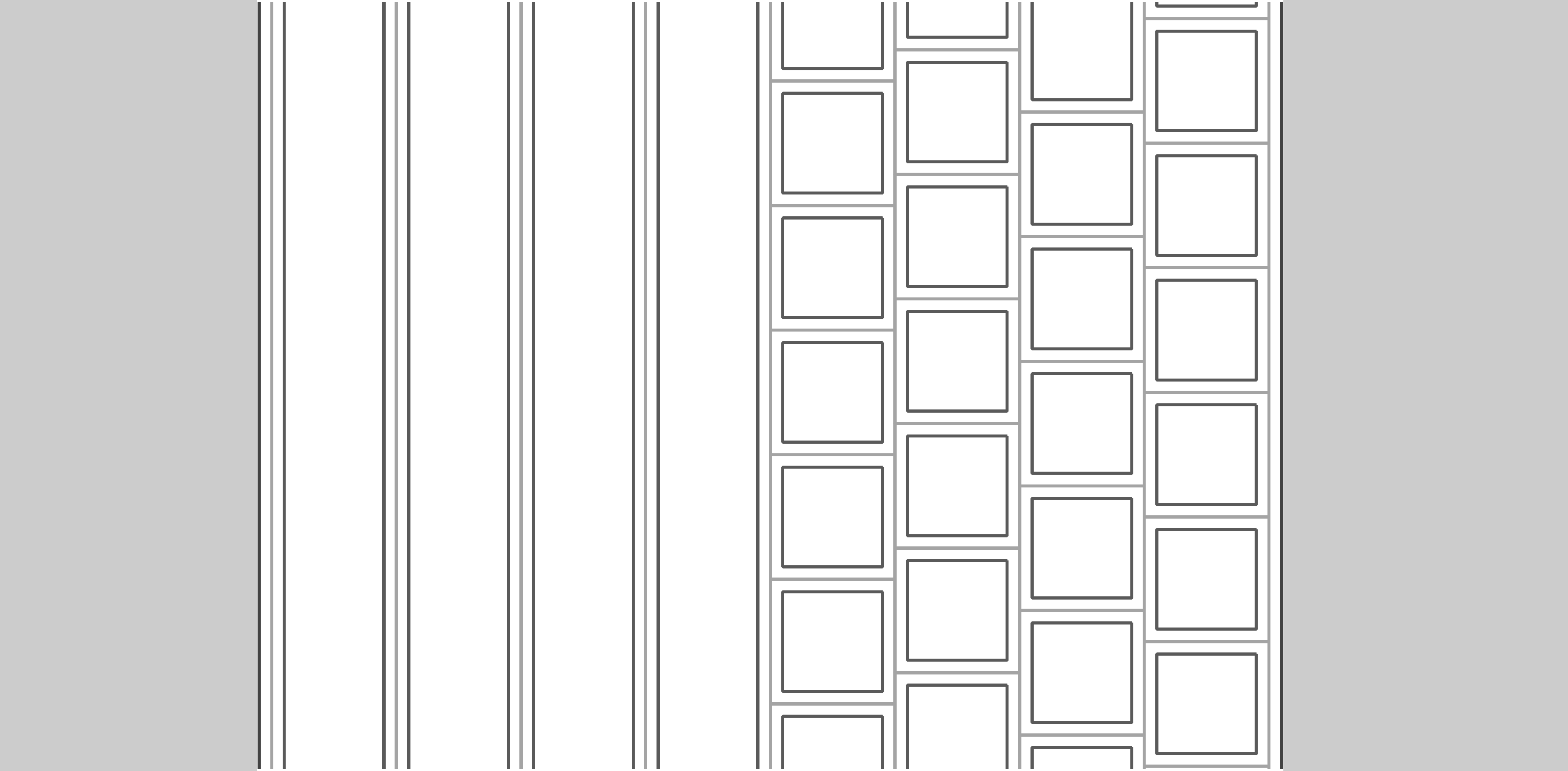}
\end{center}
\caption{Schematic drawing of a fibre station made of cylindrical (left) and squared (right) fibres.}
\label{fibres1}
\end{figure}

Three different fibre-station conceptual designs have been simulated. 
The first option is a station made of cylindrical fibres with radius
of 0.5\,mm.
For this design, the position of the fibre centres in a given layer is
shifted by 0.25\,mm with respect to centres of the fibres in the
neighbouring layer (figure \ref{fibres1} - left). 
The other two options consider a station made of square fibres with
0.5\,mm side  and displacement between the neighbouring layers of
0.25\,mm and 0.125~mm, respectively (figure \ref{fibres1}, right panel). 
In each design the station thickness is 4~mm. 

\begin{figure}
\begin{center}
  \includegraphics[width=.6\textwidth]{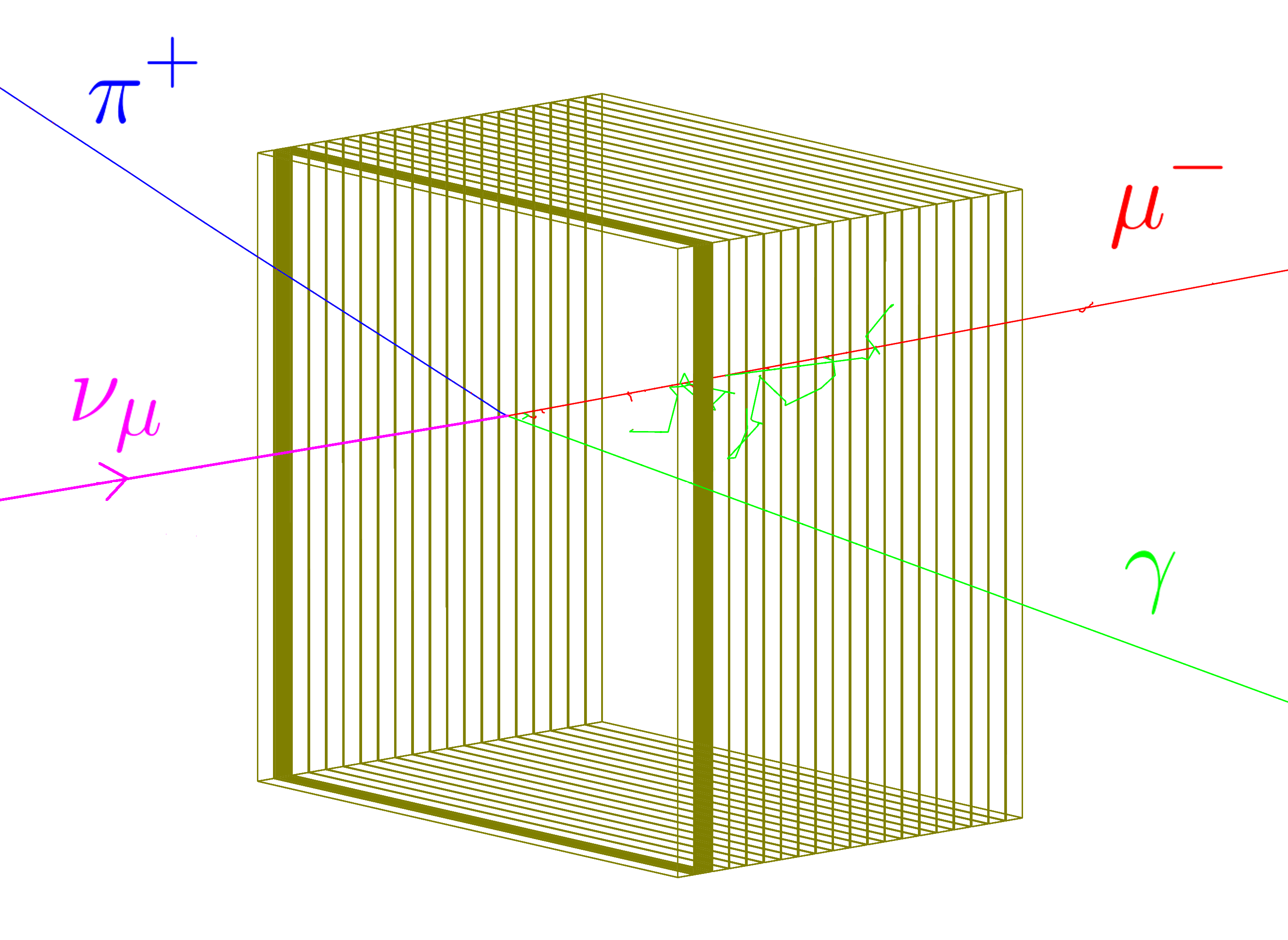}
\end{center}
\caption{Schematic drawing of the detector with a $\nu_{\mu}N$ interaction in its second module.}
\label{detector}
\end{figure}

The overall dimensions of the detector are $1.5 \times 1.5 \times
1.08$~m$^3$ which corresponds to a mass of $\sim 2.5$~Ton. 
A sketch of the detector with a $\nu_{\mu}N$ interaction in its second
module is shown in figure \ref{detector}.

\subparagraph*{Digitisation \\}  

The fibre signal is taken to be proportional to the energy deposition in the corresponding fibre. It is then corrected for the attenuation of the light in its path between the hit and the fibre end. The signal is then smeared with a Gaussian with $\sigma /E = 25\% $. 
A simplified parametrisation is used for the absorber blocks. The signal in a given slab is taken to be proportional to the total energy deposition and is smeared with a Gaussian with $\sigma /E = 5\%$.

\subparagraph*{Reconstruction and analysis \\}

\emph{\noindent Reconstruction of the muon track:}

{\noindent 
For the purpose of these studies a simplified reconstruction of the events has been used.  No pattern recognition is implemented and we simply use hits in the fibres belonging to the muon track.
A space point measurement is then created for each fibre station. Its $x$ and $y$ coordinates are calculated independently as the weighted average of the coordinates of the centres of the hit fibres. The weights are proportional to the digitised signals. The space points obtained this way are used to reconstruct the muon track via the Kalman filter (e.g. taking multiple scattering into account).
}

\begin{figure}
\begin{center}
 \includegraphics[width=.49\textwidth]{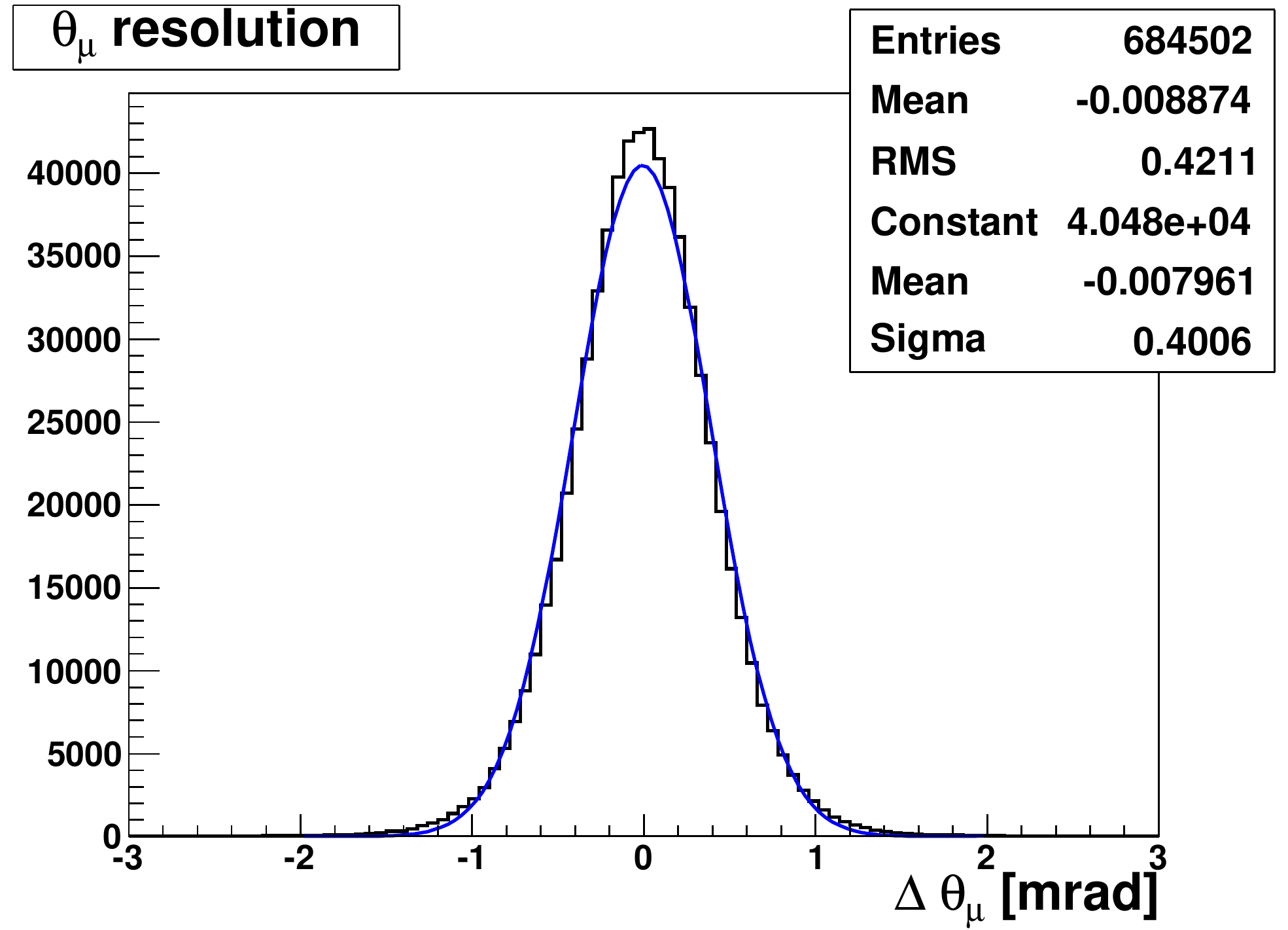}
 \includegraphics[width=.49\textwidth]{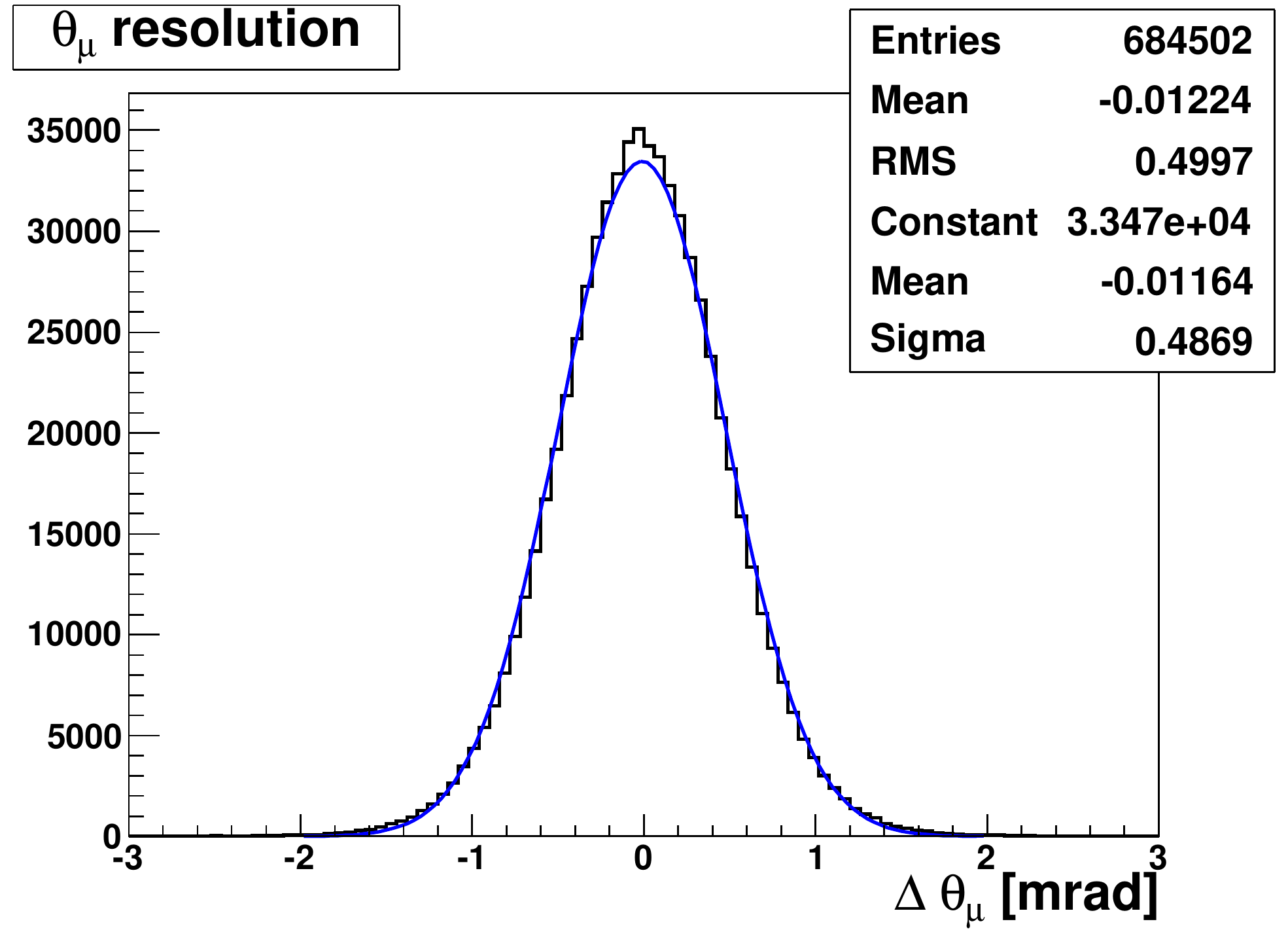}
 \includegraphics[width=.49\textwidth]{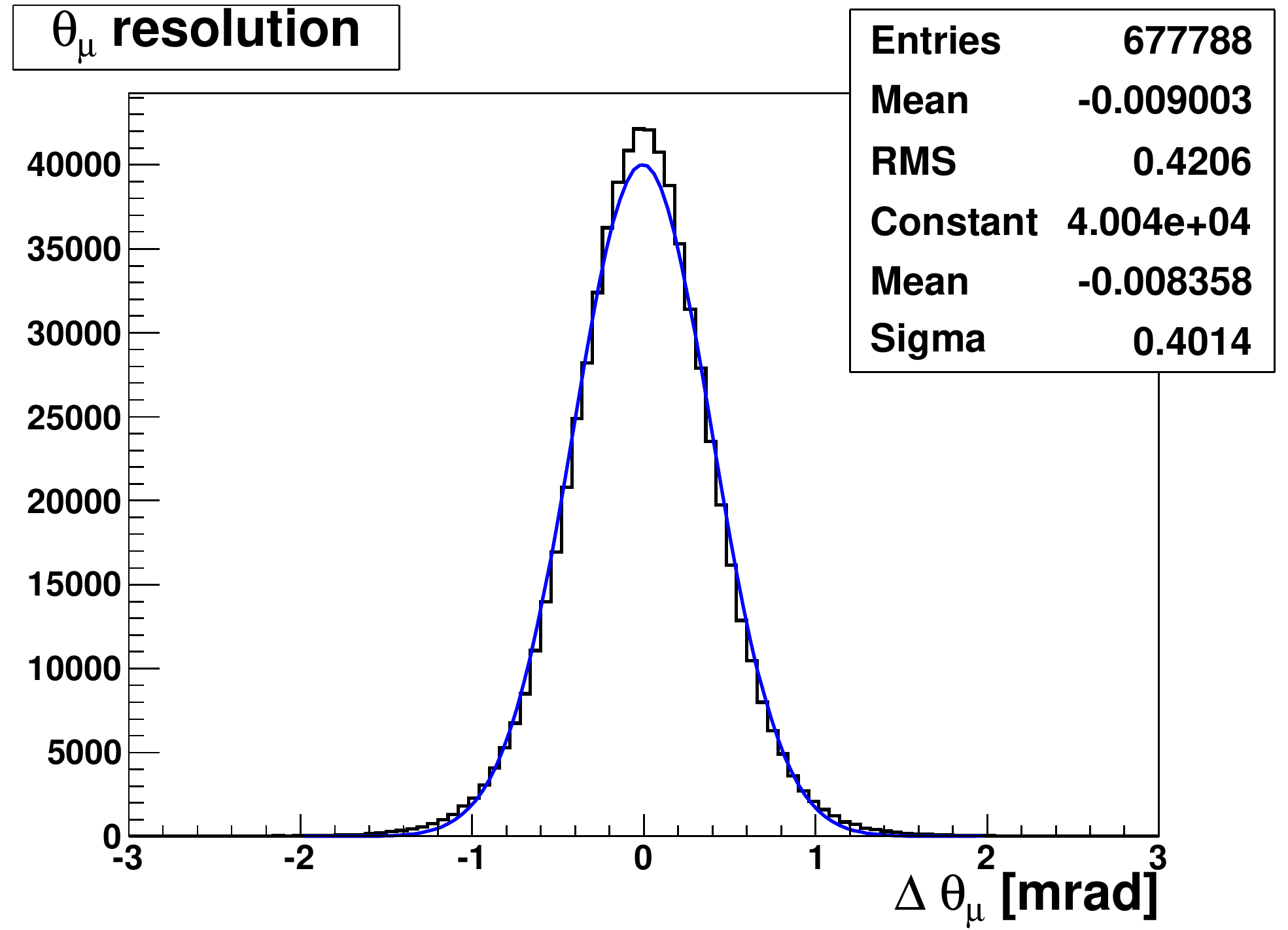}
\end{center}
\caption{The difference between the reconstructed polar angle of the muons and its true value  from the simulation for a fibre station made of cylindrical staggered fibres (top left), squared fibres staggered by 0.25~mm (top right) and squared  fibres staggered by 0.125~mm  (bottom). Gaussians are fitted to all distributions.}
\label{resol}
\end{figure}

The difference between the reconstructed polar angle of the muon and its true value from the simulation is used to measure the resolution. Figure \ref{resol} shows the distributions over this difference for the three different conceptual designs of the fibre stations. It is seen that the resolution is $\sim0.5$~mrad in all three cases.

{\noindent 
\emph{Background rejection exploiting energy deposition in the
  absorber blocks}
}

\begin{figure}
\begin{center}
 \includegraphics[width=.6\textwidth]{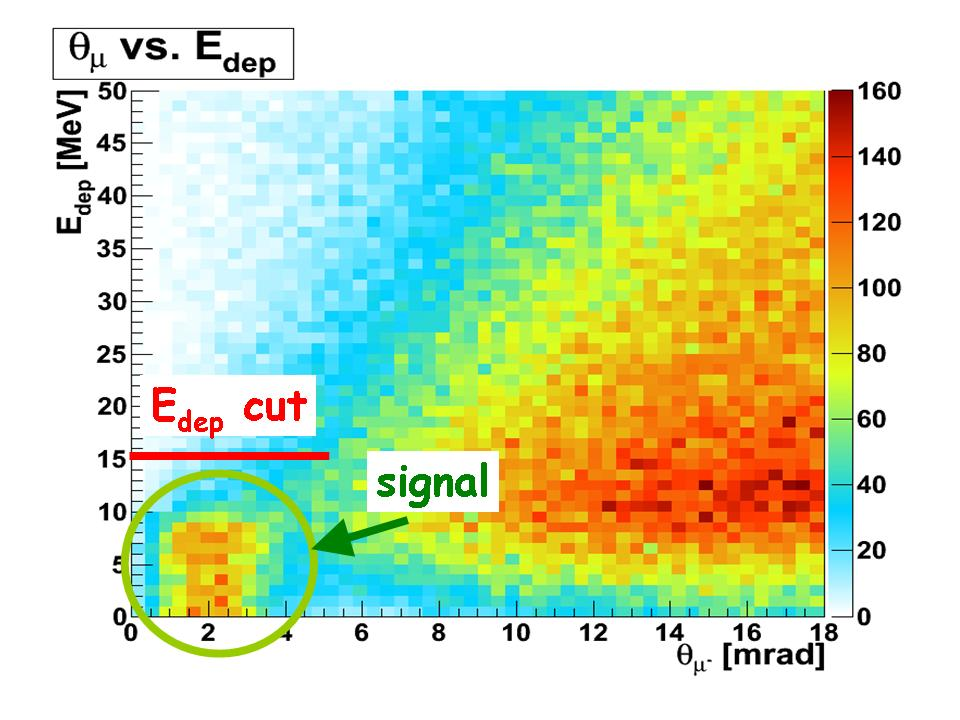} 
\end{center}
\caption{Energy deposition in the first illuminated scintillating slab compared to the reconstructed muon scattering angle.}
\label{fit1}
\end{figure}

\begin{figure}
\begin{center}
 \includegraphics[width=.49\textwidth]{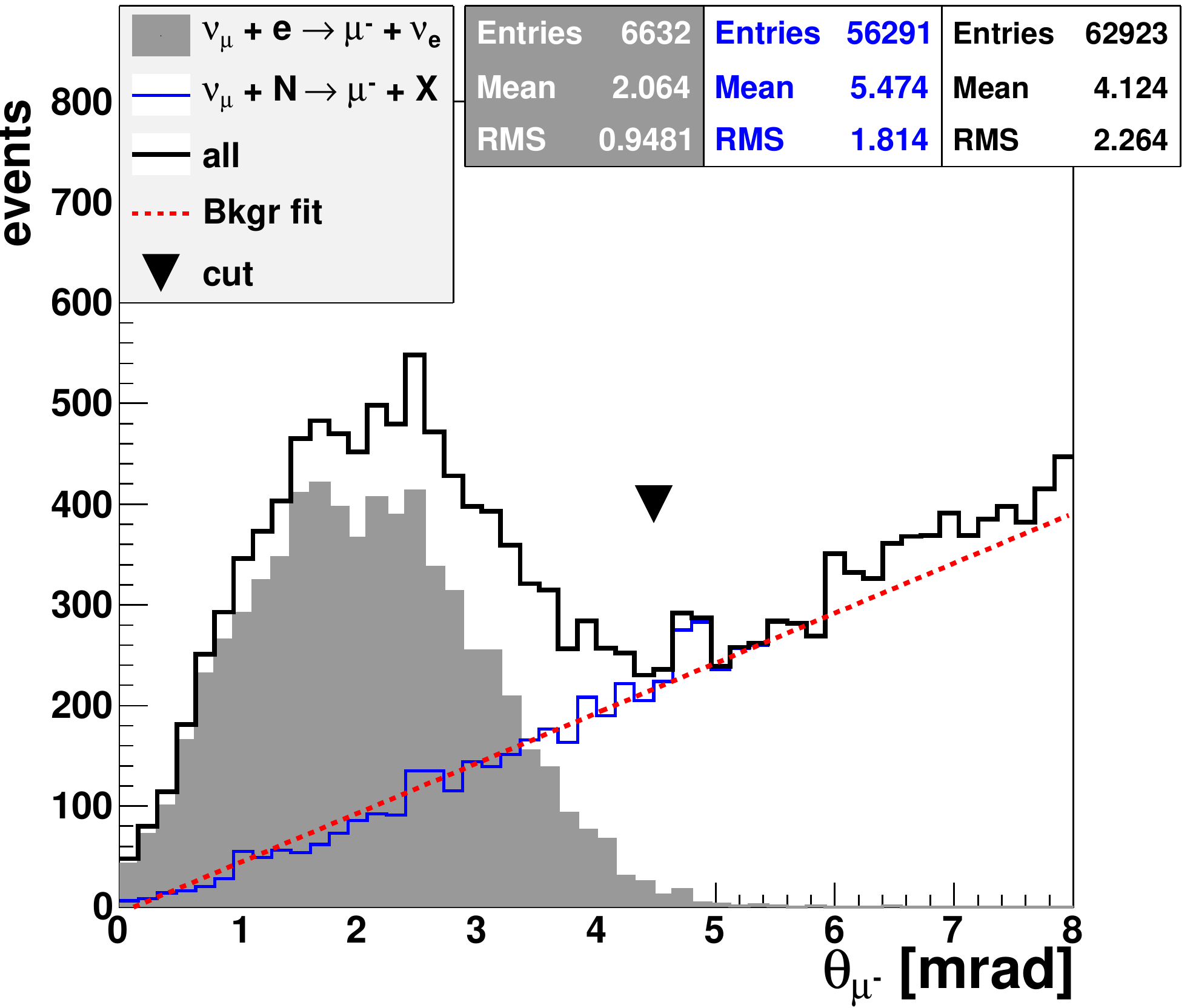}
 \includegraphics[width=.49\textwidth]{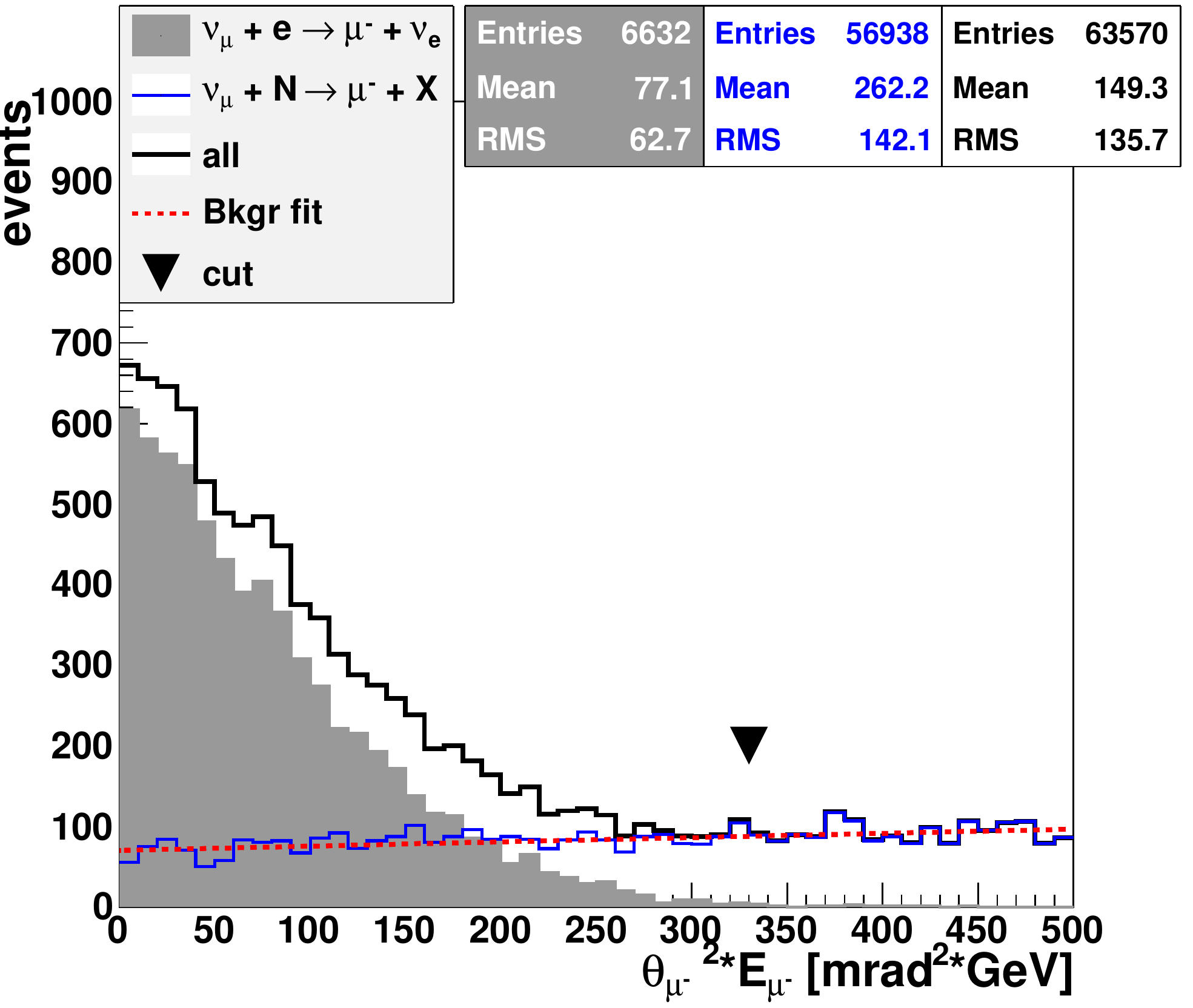}
\end{center}
\caption{Distributions over the reconstructed polar angle $\theta_{\mu}$ (left) and the \emph{inelasticity} $\theta^2_{\mu}E$ (right) of the outgoing muons. The leptonic events are filled with grey, the hadronic events are plotted in blue and the total spectrum is in black. The cut value is denoted by a black inverted triangle. The red lines indicate the background extrapolation.}
\label{fit2}
\end{figure}

{\noindent 
We propose to use the total energy deposition in the first illuminated detector module as a first tool to reject background events.
Figure \ref{fit1} shows the energy deposition in the first illuminated detector module plotted as a function of the reconstructed-muon scattering angle. It is seen that the leptonic events and the CC background are well
separated. Figure \ref{fit2} shows the distributions of the outgoing muons over the reconstructed polar angle $\theta_{\mu}$
and the variable $\theta^2_{\mu} \times E$. The variable $\theta^2_{\mu} \times E$ is proportional to the event inelasticity $1-y = E_{had}/E_{\nu}$. Only events with a total energy
deposition in the first illuminated absorber block less than 15~MeV are selected. A linear
extrapolation of the CC background toward $\theta_{\mu} = 0$ is used to evaluate the number of
background events under the leptonic peak.
}

The results obtained exploiting both suppression variables are summarised in Table \ref{ND_IMB_results}.
It is seen that by imposing a suitable cut on the recoil energy and subtracting the fitted inclusive background with a muon in the final state under the leptonic peak, it is possible to evaluate statistically the number of events due to pure leptonic scattering with a precision of $\sim 1\%$. Both variables ${\theta}_{\mu}$ and $\theta^2_{\mu} \times E$ may be used for this task.
\begin{table}
\caption{Results for background extrapolation exploiting two different suppression variables. The true number of leptonic events is 6632.}
\label{ND_IMB_results}
\begin{center}
\begin{tabular}{|c|c|c|c|c|c|}
\hline
{\bf Cut on} & {\bf All events} & {\bf Bkgr events} & {\bf Bkgr events}  & {\bf Simulated} & {\bf Leptonic}    \\
             & {\bf below the}  & {\bf below the}   & {\bf from the fit} & {\bf leptonic}  & {\bf events}        \\
             & {\bf cut}        & {\bf cut}         &                    & {\bf events}    & {\bf from the fit}  \\
\hline
${\theta}_{\mu}$                            & 9450        & 2860           & $ 2865 \pm 57 $  & 6632     & $6585 \pm 57$ \\
$\theta^2_{\mu} \times E$ & 9284       & 2666           & $ 2596 \pm 74 $  & 6632     & $6688 \pm 74$ \\
\hline
\end{tabular}
\end{center}
\end{table}

\subparagraph*{Conclusions \\}

The quasi-elastic neutrino-electron scattering can be used to measure
the neutrino flux coming from the Neutrino Factory storage ring. 
The angle ${\theta}_{\mu}$ and the \emph{inelasticity} 
$1-y \sim \theta^2_{\mu} \times E$ have similar discriminating power. 
The latter variable seems to have a flatter distribution when
${\theta}_{\mu}\rightarrow 0$. 

\paragraph{Neutrino flux measurement; neutral current elastic
  neutrino-electron scattering}

The experimental determination of the absolute neutrino flux below
11~GeV will rely upon the measurement of neutral current elastic
scattering off electrons: $\nu e^- \rightarrow \nu e^-$. 
The total cross section for NC elastic scattering off electrons is
given by~\cite{Marciano:2003eq}: 
\begin{eqnarray}
\sigma (\nu_l e \to \nu_l e) & = & \frac{G_\mu^2 m_e E_\nu}{2\pi} \left[ 1 -4 \sin^2 \theta_W + \frac{16}{3} \sin^4 \theta_W \right] \\
\sigma (\bar{\nu}_l e \to \bar{\nu}_l e) & = & \frac{G_\mu^2 m_e E_\nu}{2\pi} \left[ \frac{1}{3} -\frac{4}{3} \sin^2 \theta_W + \frac{16}{3} \sin^4 \theta_W \right]
\end{eqnarray}
where $\theta_W$ is the weak mixing angle. 
For $\sin^2 \theta_W \simeq~0.23$ the cross sections are very
small $\sim 10^{-42} (E_\nu/\gev)$\,cm$^2$. 
Neutral current elastic scattering off electrons can be used to determine the
absolute flux normalisation since the cross sections only depend upon
the the knowledge of $\sin^2 \theta_W$. 
The value of $\sin^2 \theta_W$ at the average momentum transfer
expected at Neutrino Factory near detector $Q\sim 0.07 \gev$ can be extrapolated down from the
LEP/SLC measurements with a precision of $\sim 0.2\%$ within the
Standard Model (SM). 
However, in order to take into account potential deviations from the
SM predictions, in the flux extraction we must initially consider a
theoretical uncertainty $\leq 1\%$, obtained from direct measurements
of $\sin^2 \theta_W$ at momentum scales comparable those that will
pertain at the Neutrino Factory near detector.
As discussed in Section \ref{SBL:sec:sin2thetaW}, precision
electroweak measurements with the near-detector data at a Neutrino
Factory can determine the value of $\sin^2 \theta_W$ to
better than 0.3\%.  
The theoretical uncertainty on the absolute flux normalisation can
therefore be improved substantially by a combined analysis with the
electroweak measurements. 

\begin{figure}[ht]
\centering\includegraphics[width=.85\textwidth]{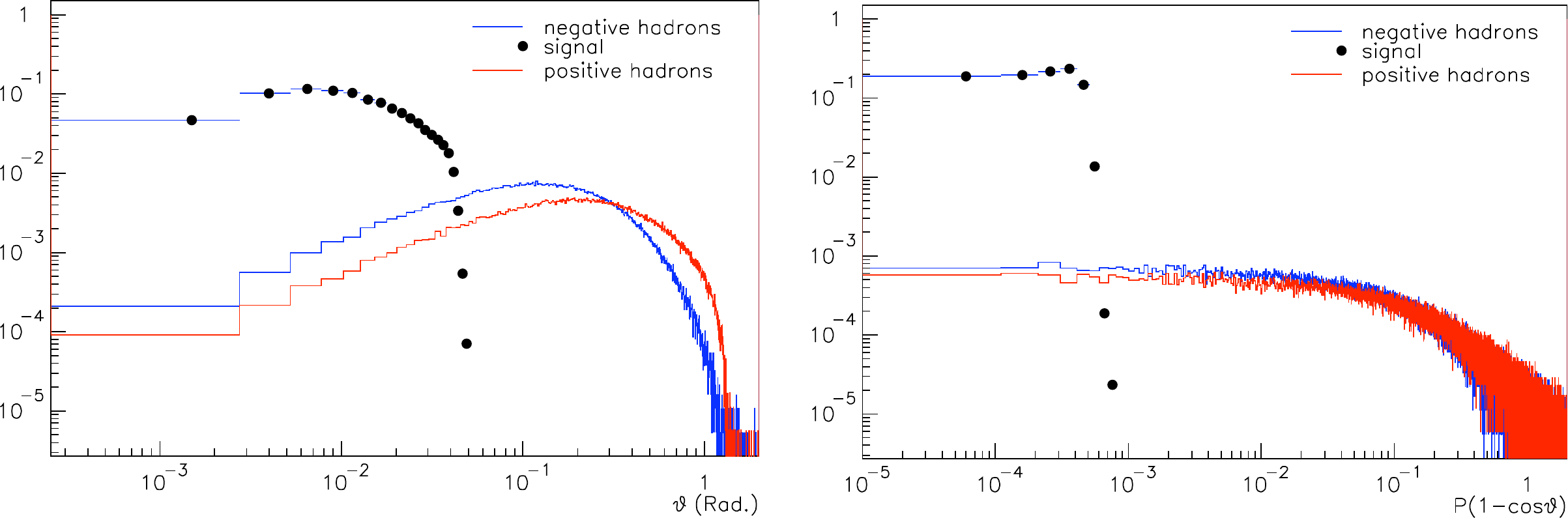}
\caption{Distributions of the angle of the electron with respect to the beam direction
(left) and of the discriminating variable $P(1 -\cos \theta)$ (right) for NC elastic
scattering off electrons and for the corresponding backgrounds in the
near detector.
}
\label{fig:NCnu-e}
\end{figure}

The signature of the process $\nu_l (\bar{\nu}_l) e \rightarrow \nu_l (\bar{\nu}_l) e$ is a single
electron in the final state, emitted almost collinearly with the beam
direction ($\theta \sim $ mrad). The dominant backgrounds are given by
NC $\pi^0$ production and single photon production in which one
photon fakes a single electron. A smaller background contribution is
given by $\nu_e$ quasi-elastic scattering events in which the
proton is not visible. This measurement
requires a detector which can distinguish between photons and electrons
efficiently. 
The low density magnetised tracker proposed for the Neutrino Factory near detector can identify electrons and positrons
and reconstruct the corresponding track parameters, allowing a background rejection
$\leq 10^{-6}$ .
Figure~\ref{fig:NCnu-e} shows the distributions of kinematic variables for signal
and background.
It is thus possible to select a sample of NC elastic scattering events off electrons
with small background in the Neutrino Factory near detector. The main limitation of such a measurement
is the statistics of the selected sample, which, for the Neutrino Factory  near detector will not be a problem.
It must be noted that in a low density magnetised design the
background originates 
from asymmetric $\gamma$ conversions in which the positron is not reconstructed.
This type of background is expected to be charge-symmetric and this fact gives a
powerful tool to calibrate the $\pi^0/\gamma$ background in-situ.

\paragraph{Effects of High $\Delta m^2$ Oscillations on the Flux Extraction}
\label{SBL:sec:fluxosc}

The results described in the previous sections were obtained under the
assumption that the events observed in the near detector 
originate from the same 
(anti)-neutrino flux produced by the decay of the parent muons.  
The recent results from the MiniBooNE experiment might suggest the
possibility of relatively high $\Delta m^2$ anti-neutrino oscillations
consistent with the LSND signal. 
This effect, if confirmed, seems to indicate a different behaviour
between neutrinos and anti-neutrinos, which would imply CP or CPT
violation. 
The MINOS experiment also reported different oscillation parameters
between $\nu_\mu$ and $\bar{\nu}_\mu$ from the disappearance analysis,
although this result is as yet not statistically compelling.

The presence of high $\Delta m^2$ oscillations with characteristic
oscillation length comparable to the near detector baseline at the
Neutrino Factory energies, would imply that the spectra observed in
the near detector could be already distorted by neutrino oscillations. 
The main effect expected on the flux extraction from a MiniBooNE/LSND
oscillation is that a deficit is induced in the $\bar{\nu}_\mu$ CC
spectrum from a significant disappearance rate.
Any in situ determination of the fluxes would then require the
unfolding of the oscillation effect from the measured spectra. The
measurement strategy in the near detector should necessarily include a
combined oscillation and flux analysis. 
Since the near detector cannot be easily moved, different
complementary measurements are needed. 

Several follow-up experiments have been proposed to investigate the
MiniBooNE/LSND effects: move MiniBooNE to a near detector location,
OscSNS at the ORNL neutron spallation source or a two-detector LAr
experiment at the CERN PS. 
Each experiment is expected to cover the region in the oscillation
parameter space consistent with MiniBooNE/LSND data, so that, by the
time the Neutrino Factory experiments will take data, the high 
$\Delta m^2$ oscillation hypothesis may be confirmed or disproved.
However, the precision which will ultimately be achieved
in the determination of the fluxes at the Neutrino Factory near
detector is directly connected to the high $\Delta m^2$ oscillation
parameters.
If the oscillation is confirmed, we will then need dedicated precision
measurements in the near detector at the Neutrino Factory.

\paragraph{Influence of near detector flux data on far detector sensitivities}

Any neutrino near-detector facility will have many possible
functions. Among these will be the measurement of the absolute flux in
the direction of the far detectors. There is the possibility that this
measurement could be used, in addition to the determination of the
absolute normalisation, to project a non-oscillation flux prediction
to the far detector site to be used in the determination of the
oscillation parameters. This would require the reconstruction of the
whole flux spectrum at the near detector so that the projection could
be carried out reliably. The initial studies of the power of a
technique for the projection of the near detector flux spectrum as a
means of extracting oscillation parameters are described in the
paragraphs which follow.

\subparagraph*{Flux measurement \\}
\label{subSec:Flux}

\begin{figure}
  \begin{center}$
    \begin{array}{cc}
      \includegraphics[height=5.5cm,width=7.5cm]{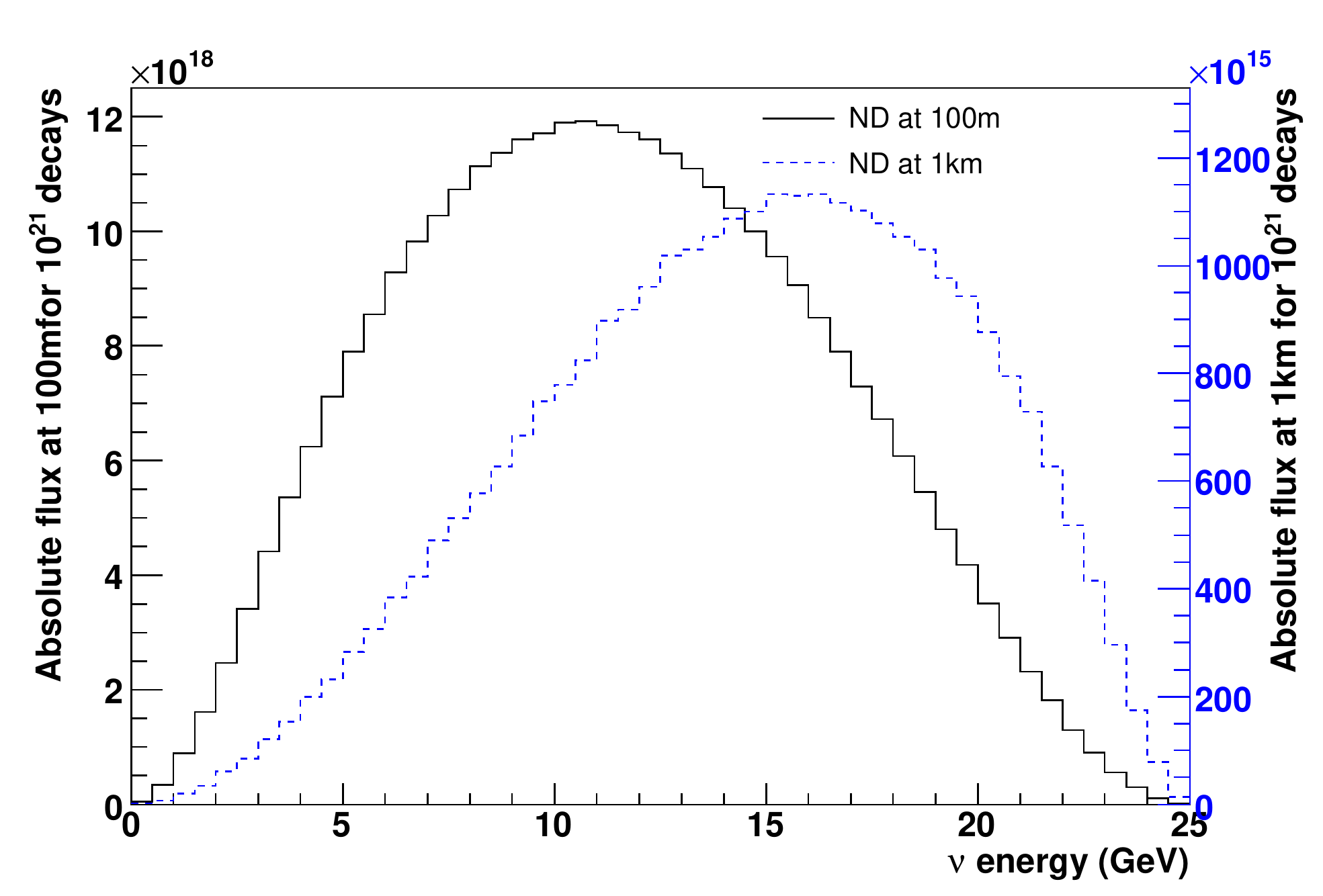} &
      \includegraphics[height=5.5cm,width=7.5cm]{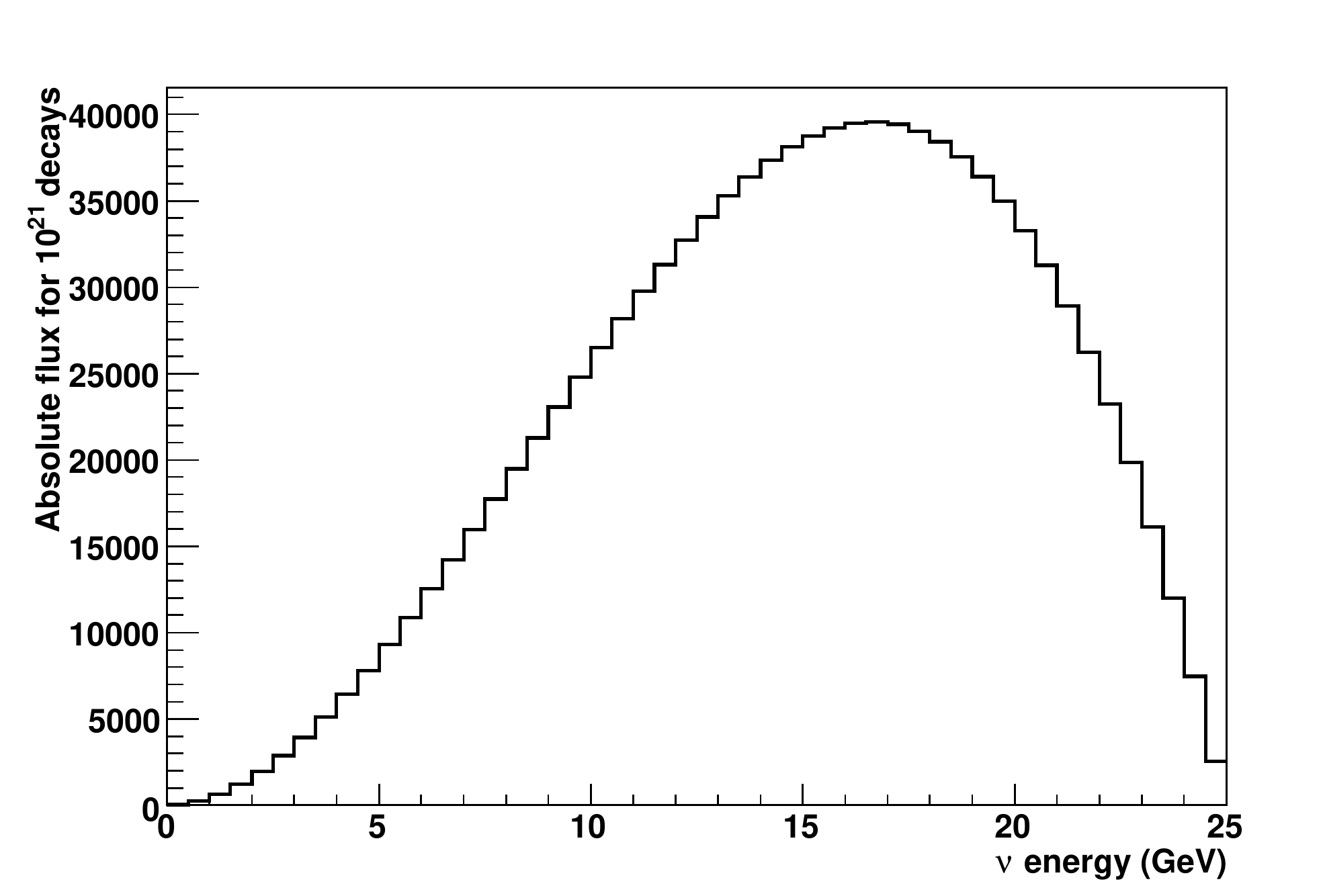}\\
      \includegraphics[height=5.5cm,width=7.5cm]{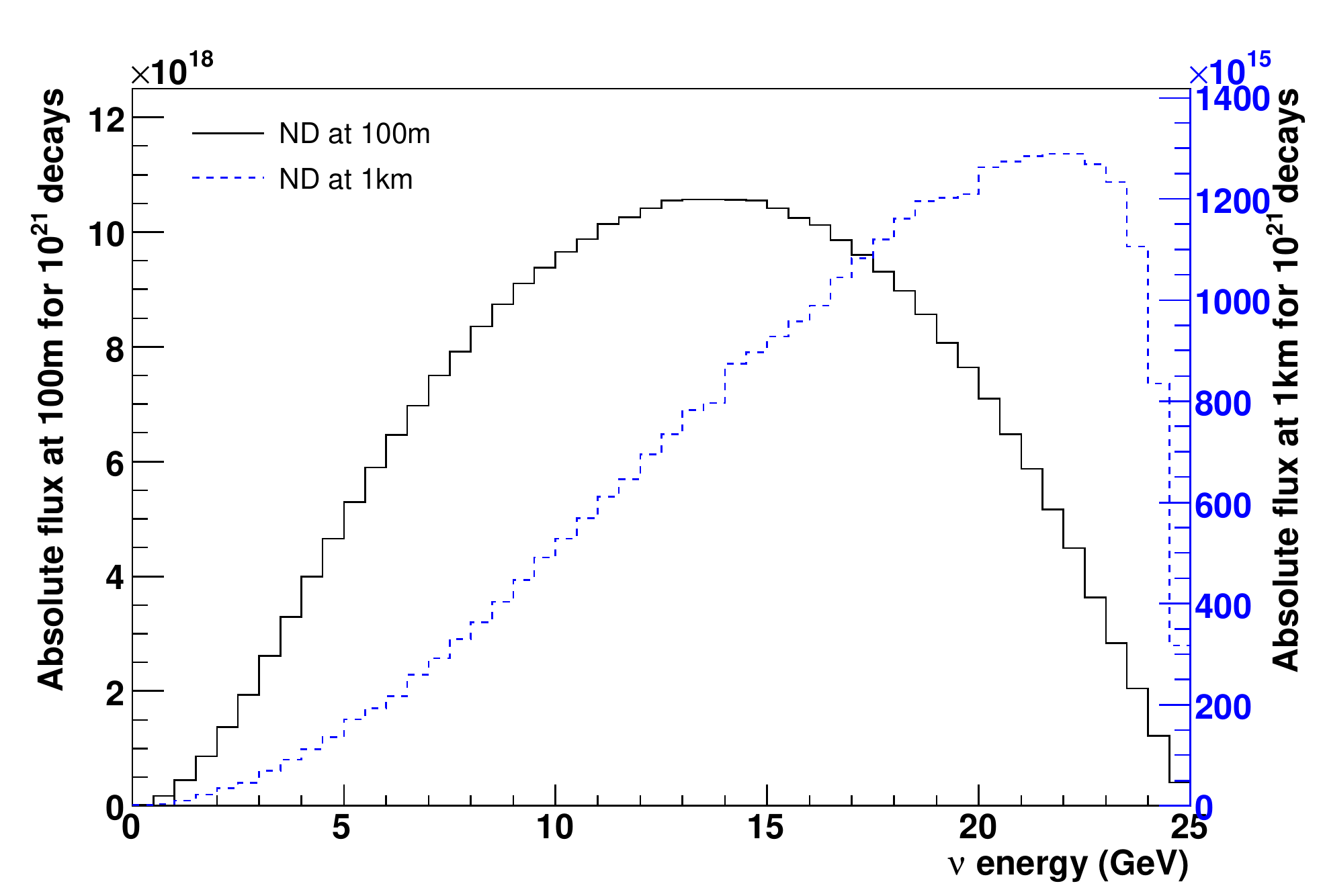} &
      \includegraphics[height=5.5cm,width=7.5cm]{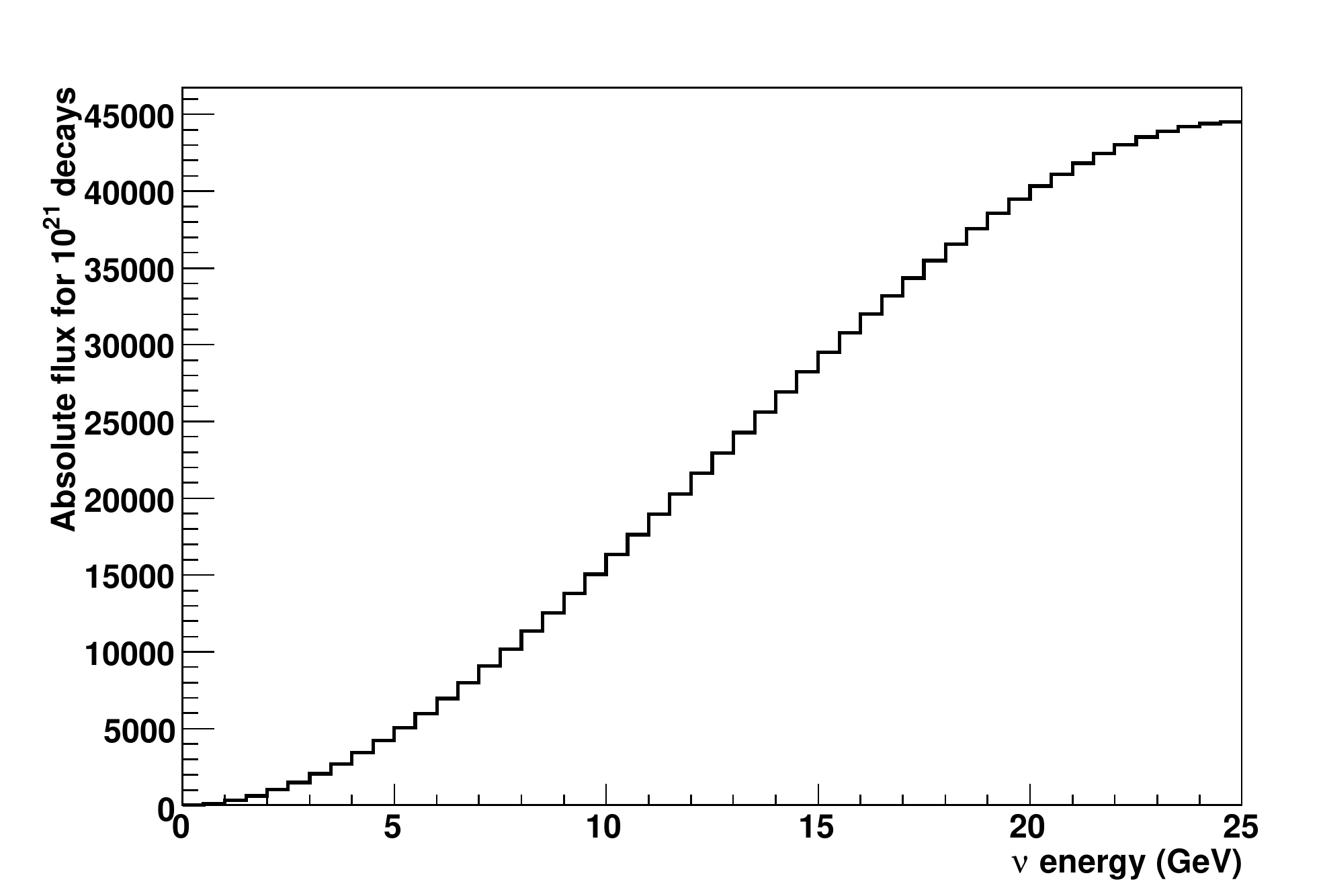}\\
    \end{array}$
  \end{center}
  \caption{Difference in expected fluxes at near and far sites: (top) $\nu_e$ flux through a 1~m radius detector 100~m and 1~km from a 600~m decay straight (left) and at a 7~m radius detector 4000~km from the Neutrino Factory; (bottom) $\nu_\mu$ flux for the same detectors (using unpolarised muon expectation).}
  \label{fig:NDNFflux}
\end{figure}

Comparison of the neutrino flux at the near and far sites is
problematic for various reasons. 
A detector within $\sim$1~km of the beam pipe subtends
a far greater solid angle than a large scale detector positioned
thousands of kilometres from the source. 
In addition, the solid angle subtended by the near detector as seen
from different parts of the decay pipe varies considerably. 
Considering a 1~m radius near detector placed 100~m from the end of a
600~m long straight decay length and the $14\times 14$~m$^2$ cross
section MIND at 4\,000~km from the same decay pipe it can be seen that
the solid angle of the near detector as seen from the two ends of the
decay pipe ranges from $6.4\times 10^{-6}$\,sr to 
$3.1\times 10^{-4}$~sr, a range of two orders of magnitude, whereas
for the far detector the solid angle ranges between 
$1.22\times 10^{-11}$\,sr to $1.23\times 10^{-11}$\,sr, up to 7 orders
of magnitude smaller and varying by only 1\% of the larger value.
As can be seen in figure \ref{fig:NDNFflux}, this results in a
different energy spectrum for near detector distances up to a distance
of $\sim$1~km. 
Due to the steep angle of the beam direction, positioning the near
detector facility at 1~km would be restrictively expensive, as well as
problematic from an engineering standpoint. 
Therefore, one of the main focuses of this study must be to determine
whether it is still possible to extract a prediction for the far
detector flux, even if the near detector flux differs considerably
from the far detector spectrum, when the near detector is positioned
close to the end of the decay straight. 

Direct comparison of the expected fluxes at the two sites allows for a
clear determination of the change in the energy spectrum. 
This technique would require any near detector to have excellent
energy resolution to reduce the errors in the determination of the
near detector flux before projection. 
However, the proximity of the detector to the beam source means that a
large mass is not a requirement from a statistical point of view and
hence a high resolution detector constructed of multiple sub-detectors
could be built with designs based on upgrades to the
NOMAD~\cite{Barichello:2003gu} or Minerva~\cite{Kafka:2010zza}
detectors as possible candidates. 
The studies described below focus on the determination of oscillation
parameters via the Neutrino Factory golden channel using a matrix
representation of all aspects of the set-up and is inspired by the
technique used by the MINOS collaboration~\cite{Michael:2006rx}.

\subparagraph*{Flux projection for non-oscillation prediction \\}
\label{sec:matMeth}

The technique essentially involves three matrices describing the set-up: near detector response, flux projection and far detector response; in addition to cross-section matrices for the relevant processes and a parametrisation of the oscillation probability which is ultimately used in the determination of the sensitivity to oscillation parameters. Through purely mathematical arguments one can prove that the oscillation probability is related to the two observed signals via the relationship:
\begin{equation}
  \label{eq:OscProbNeFa}
  P_{osc} = M_{FD}^{-1}M_{dat}M_{ND}M_{nOsc}^{-1} \, ;
\end{equation}
where $M_{FD}$ and $M_{ND}$ are matrices representing the combination
of cross-section and response for $\nu_\mu~(\overline{\nu}_\mu)$ at
the far detector and $\nu_e~(\overline{\nu}_e)$ at the near detector
respectively, $M_{nOsc}$ relates the expected far detector
$\nu_e~(\overline{\nu}_e)$ flux without oscillations to the expected
$\nu_e~(\overline{\nu}_e)$ flux at the near detector and $M_{dat}$ is
the ratio of the observed $\nu_\mu~(\overline{\nu}_\mu)$ interaction
spectrum at the far detector to the observed
$\nu_e~(\overline{\nu}_e)$ interaction spectrum at the near detector. 
The extracted function would then be fit to the oscillation
probability formul\ae to find the best fit values of the $\theta_{13}$
and $\delta$. 
This technique formed the basis of the first study of the matrix
method which was presented in~\cite{Laing:2008zzb}.

There are, however, a number of problems with this direct method. The finite resolution of both detectors would mean the direct ratio of observed signals could not be used without some correction but, more importantly, fitting such a complex function as the oscillation formula, particularly in the low energy region, can be problematic. Inversion of the response of a detector, particularly a coarse grained calorimeter like the far detector can lead to statistical instabilities or bias in the prediction of the true energy interactions (see~\cite{Cowan:1998ji} for a more detailed discussion of this inverse problem). For these reasons the next step in the study involved a re-definition of the fitted quantity.

This updated method attempts to fit the observed far detector spectrum directly by using the projection of the observed near detector spectrum. That is, the predicted spectrum for a given grid point on the ($\theta_{13}$,$\delta$) plane is calculated as:
\begin{equation}
  \label{eq:nf2008}
  N_{FD} = M_{FD}P_{osc}(\theta_{13},\delta)M_{nOsc}M_{ND}^{-1}N_{ND} \, ;
\end{equation}
where $N_{FD}$ and $N_{ND}$ are the observed far and near detector
spectra respectively, with other symbols defined as in
equation~\ref{eq:OscProbNeFa}. 
In this way only the higher resolution near-detector response needs to
be inverted. 
The prediction here would, however, use binned data as opposed to
using the integration of the flux calculations, so the binning used at
the near detector would have to be optimised for the projection. 
Calculation of the expected appearance spectrum from the
no-oscillation flux spectrum is aided by fine binning, however,
statistical significance and detector resolution limit how fine the
binning can be.
An initial study using the golden-channel oscillation without
backgrounds was performed and presented in~\cite{Laing:2008zz} using a
near detector with $\nu_e~(\overline{\nu}_e)$ detection threshold of
5~GeV and energy resolution of 35\%/$\sqrt{E}$. 

In order to predict the interaction spectra at the far detector, the near detector would be required to measure both the $\nu_\mu~(\overline{\nu}_\mu)$ and $\overline{\nu}_e~(\nu_e)$ interaction spectra. The only background to the $\nu_\mu~(\overline{\nu}_\mu)$ measurement is likely to be from neutral current interactions. Using a combination of missing $p_T$ and vertex reconstruction, both of which should be measured with high resolution at a near detector, this could be suppressed to at least the level in the far detector. The $\overline{\nu}_e~(\nu_e)$ measurement would require a more sophisticated analysis to achieve a pure sample. Both channels could benefit from the study of low-rate background processes such as electron scattering but these processes are limited by statistics and take place often in restricted energy ranges and as such could not be used alone to perform this analysis. Any final analysis is likely to use a combination of channels and analyses to maximise statistics and purity.

Projection of the predicted flux is carried out using scaling matrices calculated using a comparison of the true fluxes at the near and far sites. The ultimate analysis would benefit from the use of a Monte Carlo study either to quantify the additional neutrinos expected in the beam from neutral current interactions and the corrections required to take into account the near-detector resolution or to construct a probability matrix relating directly near-detector interactions to unoscillated far-detector interactions, which is the method favoured in the MINOS analysis~\cite{Michael:2006rx}. However, a simple scaling argument allows for the understanding of the near detector effects and can be directly compared to existing studies which do not yet take into account the effect of additional neutrinos scattered into the beam by interactions.

\subparagraph*{Comparison of near detector projection to standard fit method \\}

An initial study of the power of this technique has been carried out
assuming a 100~kg cylindrical detector of 1~m radius placed 100~m from
the end of a 600~m straight decay section. The flux expected at the
near detector site is predicted by generating muon decays randomly 
along a straight line with an appropriate beam divergence and
calculating the expected spectrum from the detector acceptance as
calculated for each decay position. The detector is modelled using a
conservative estimate of the $\overline{\nu}_e~(\nu_e)$ energy
resolution of 35\%/$\sqrt{E}$(GeV) with efficiency rising linearly
from 0\% at 0~GeV up to 70\% above 4~GeV. The
$\nu_\mu~(\overline{\nu}_\mu)$ resolution is set at
20\%/$\sqrt{E}$(GeV), with efficiency of 80\% for $\overline{\nu}_\mu$
and 60\% for $\nu_\mu$ above 4~GeV (similar to the far detector). A
full detector simulation is not yet available to test the validity of
these assumptions or  to estimate the background levels, as such the
current study assumes the two signals are separated without
background. 

\begin{figure}
  \begin{center}$
    \begin{array}{cc}
    \includegraphics[height=5.5cm,width=7.5cm]{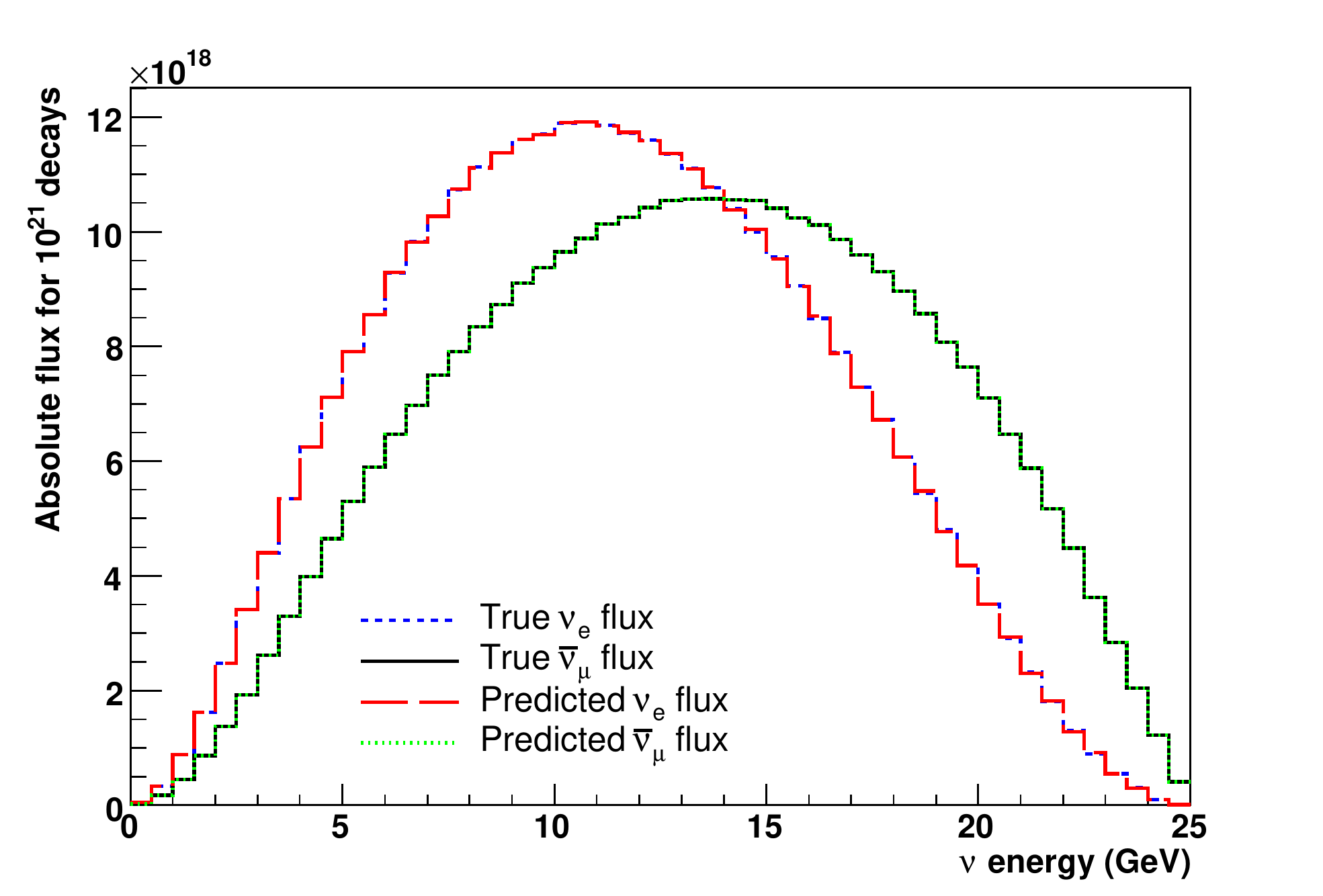} &
    \includegraphics[height=5.5cm,width=7.5cm]{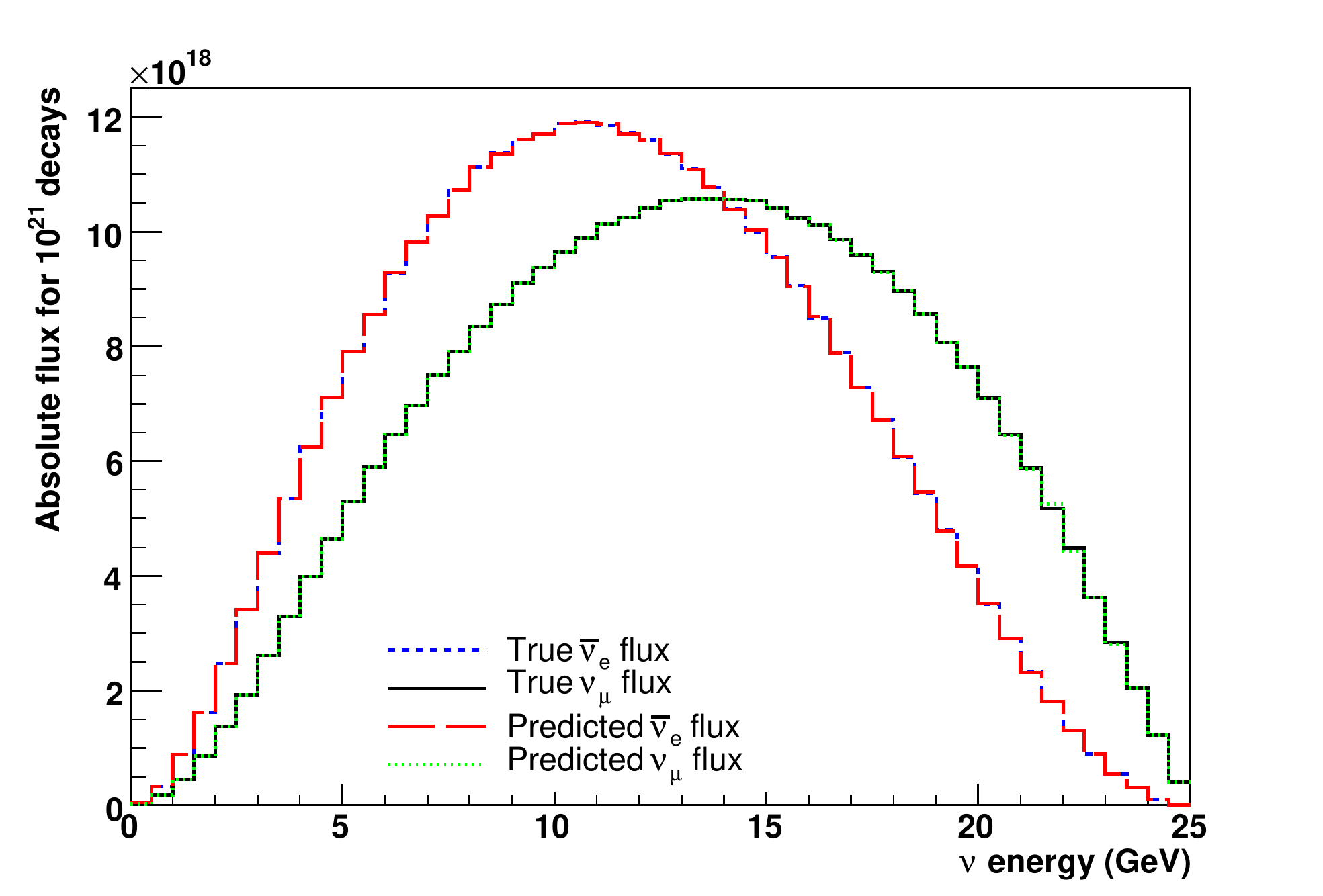}\\
    \includegraphics[height=5.5cm,width=7.5cm]{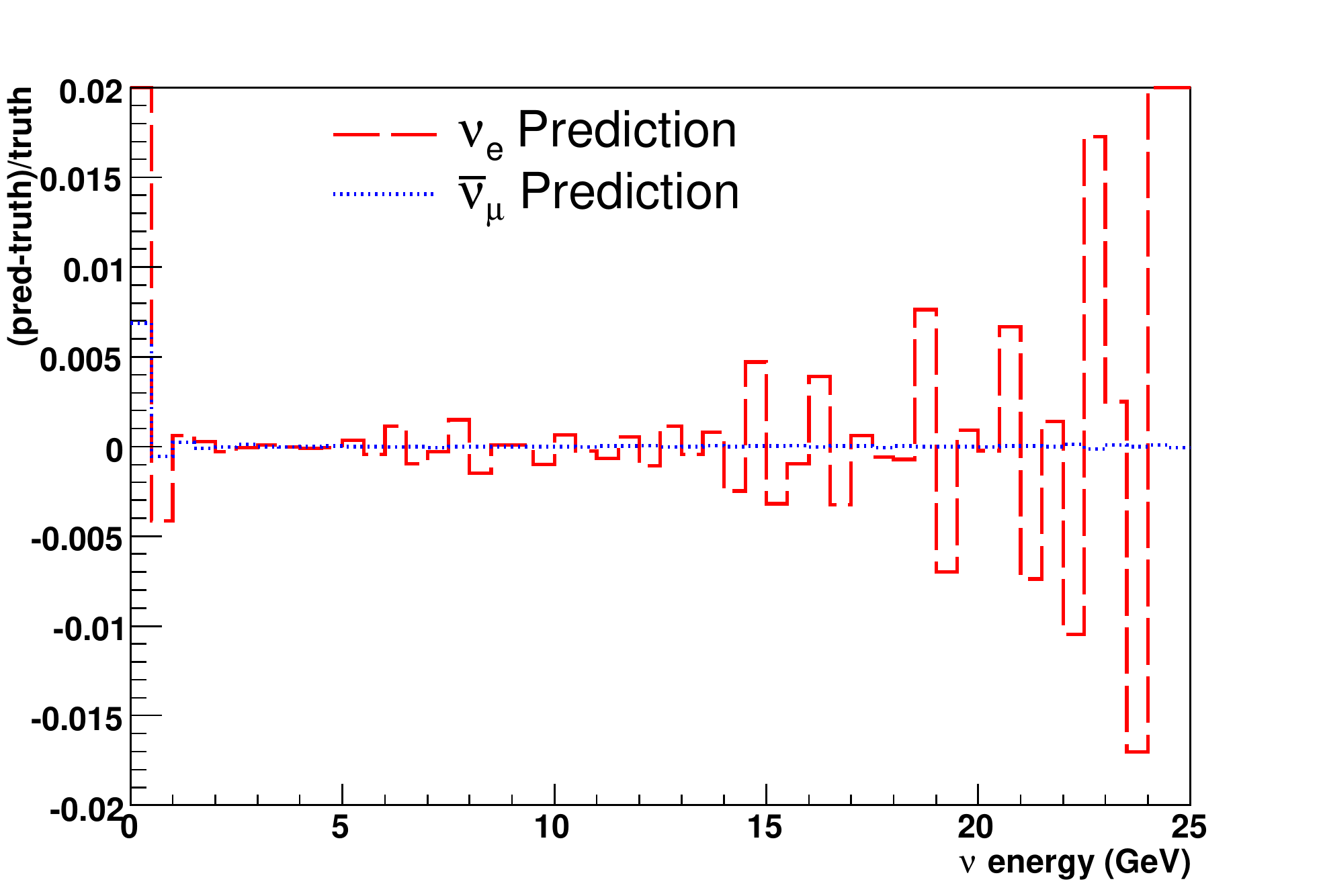} &
    \includegraphics[height=5.5cm,width=7.5cm]{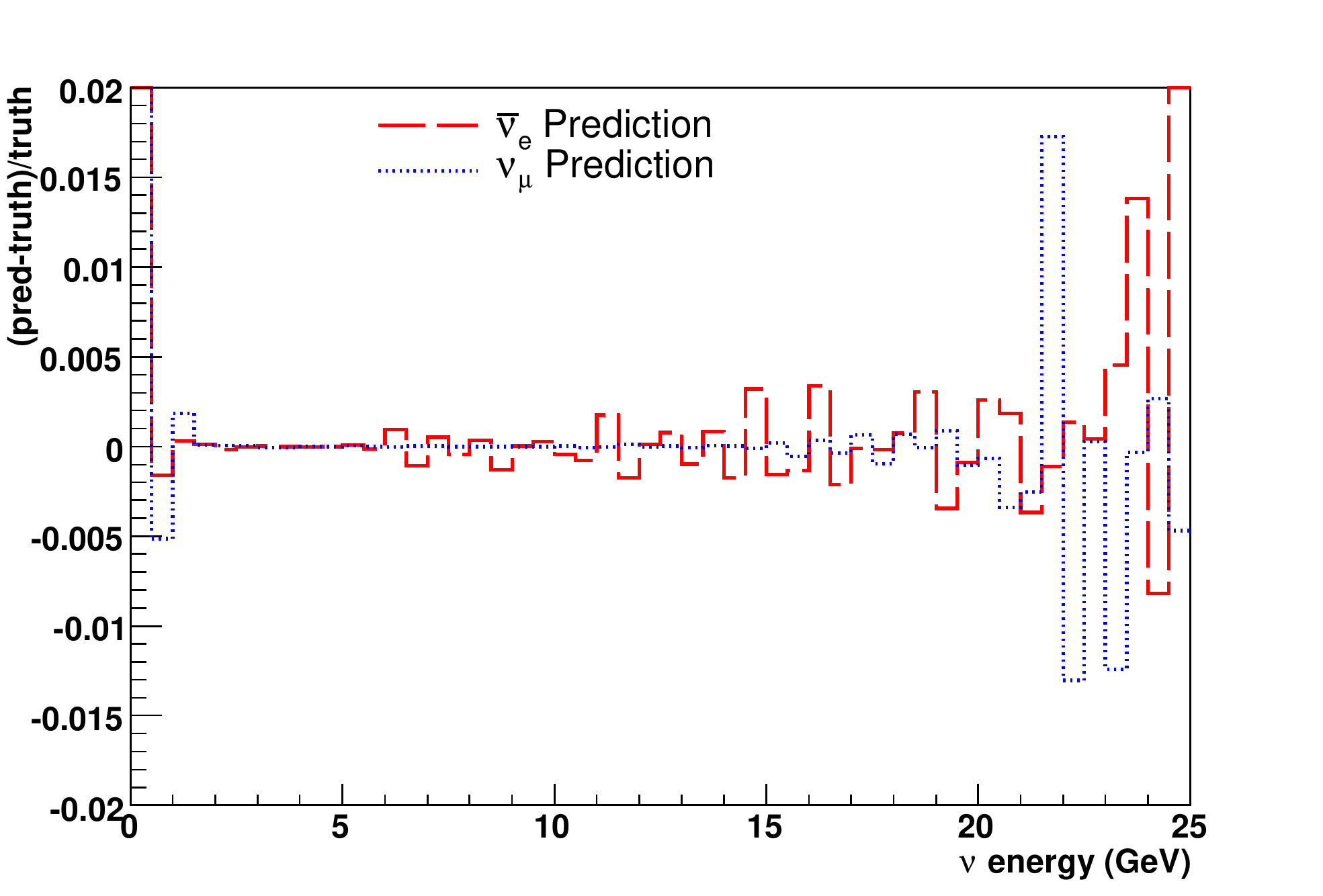}
    \end{array}$
  \end{center}
  \caption{Comparison of predicted to true flux through the near detector for stored $\mu^+$ (left) and stored $\mu^-$(right). (top) Actual spectra and (bottom) fractional residuals.}
  \label{fig:predfluxND}
\end{figure}

A smear is performed on the calculated interactions at the near
detector and the flux and correlation matrices for each channel and
the far detector are then projected to the appropriate far detector in
bins of width 0.5\,GeV. 
The total integrated interaction spectra expected at a near detector
over a data taking run of 5 years allows the flux through the detector
to be predicted to a high level of accuracy using the inversion of the
response matrices (see figure \ref{fig:predfluxND}). 
Only slight oscillation of the predicted values is visible at higher
energies. 
However, the quality of the prediction is likely to be worse if the
technique were to be used for a limited data-set, or if the
efficiencies or energy resolutions were reduced. 
In the future a more sophisticated unfolding will be developed.

\begin{figure}
  \begin{center}$
    \begin{array}{cc}
    \includegraphics[width=7cm,height=6cm]{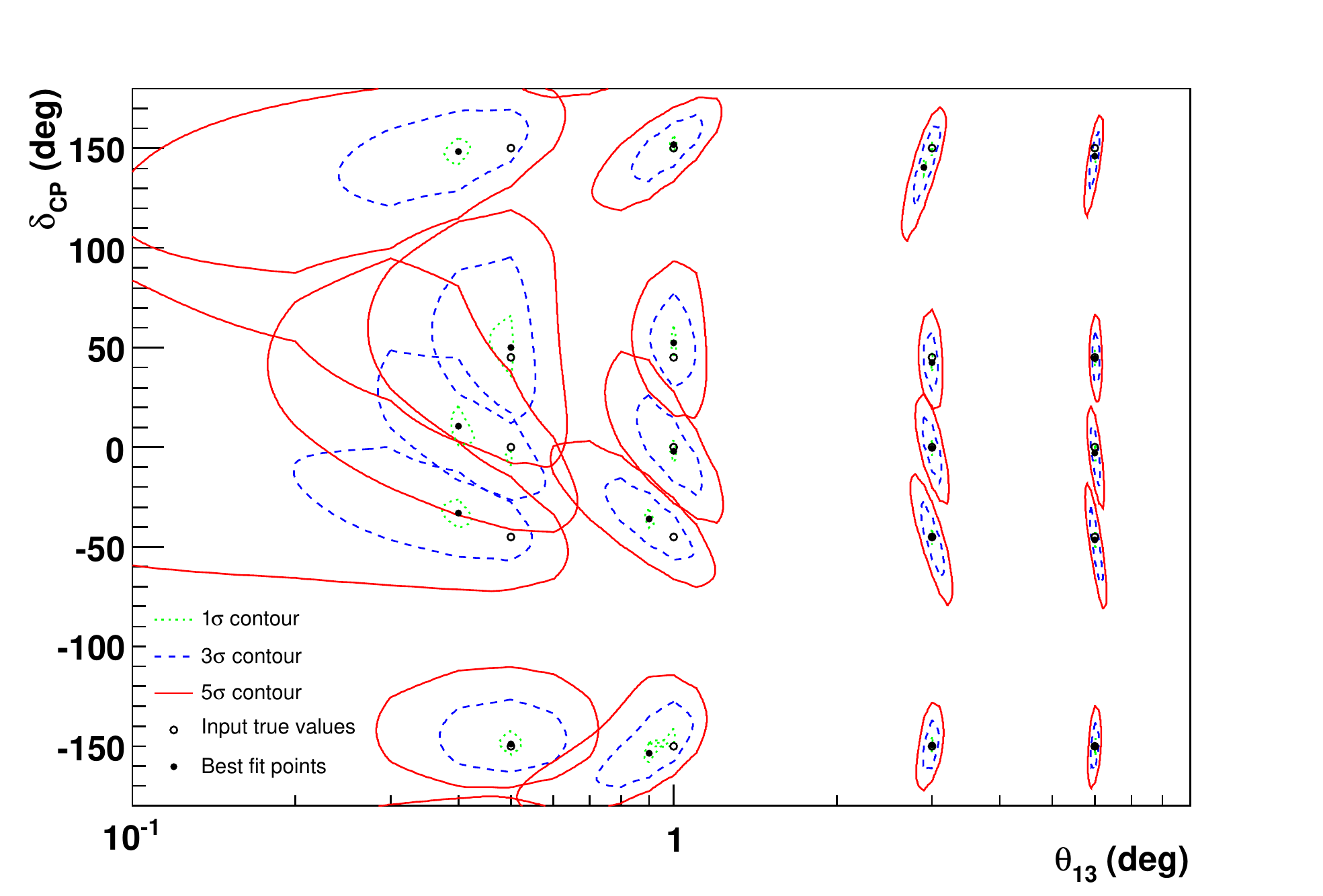} &
    \includegraphics[width=7cm,height=6cm]{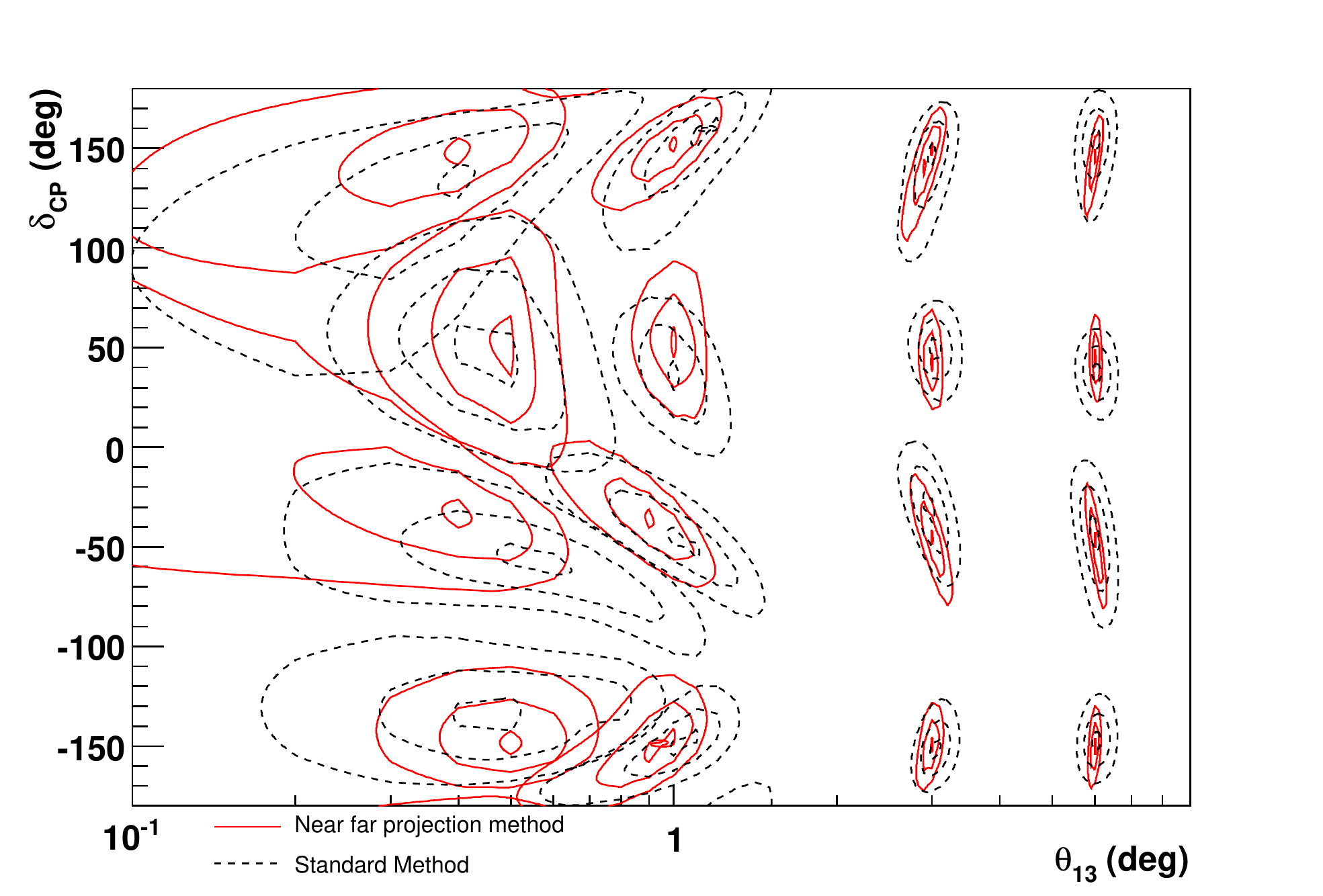}
    \end{array}$
  \end{center}
  \caption{Fits to simulated data using the near-far prediction (left) and compared to equivalent fits performed allowing the flux to be determined as part of the fit (right). All fits assume normal hierarchy.}
  \label{fig:NeFfits}
\end{figure}

The far detector spectra are calculated with the non-oscillation predictions from the near detector. The far-detector spectra obtained are then used to perform a fit using the function:
\begin{equation}
  \label{eq:chicorr}
  \chi^2 = \displaystyle\sum_i\sum_j(N_{i,j}-n_{i,j})V^{-1}_{i,j}(N_{i,j}-n_{i,j})^T \, ;
\end{equation}
where $i$ is the detector baseline, $j$ the polarity, $N$ the predicted spectrum, $n$ the data spectrum and $V$ the correlation matrix, composed of the projected matrix of the prediction and the expected errors on the far detector measurement. Figure \ref{fig:NeFfits} shows the results of fits to a range of $\theta_{13}$ and $\delta$ values using this technique (left) compared to similar fits performed in which the neutrino flux is allowed to be part of the fit (right), rather than be constrained by the near-detector data. As can be seen, the resolution of $\theta_{13}$ is as good or better using the projection technique. Resolution of $\delta$ is generally better for large $\theta_{13}$ but reduces to a similar level for $\theta_{13} \leq 1^{\circ}$.

\begin{figure}
  \begin{center}$
    \begin{array}{cc}
    \includegraphics[width=7.5cm,height=6cm]{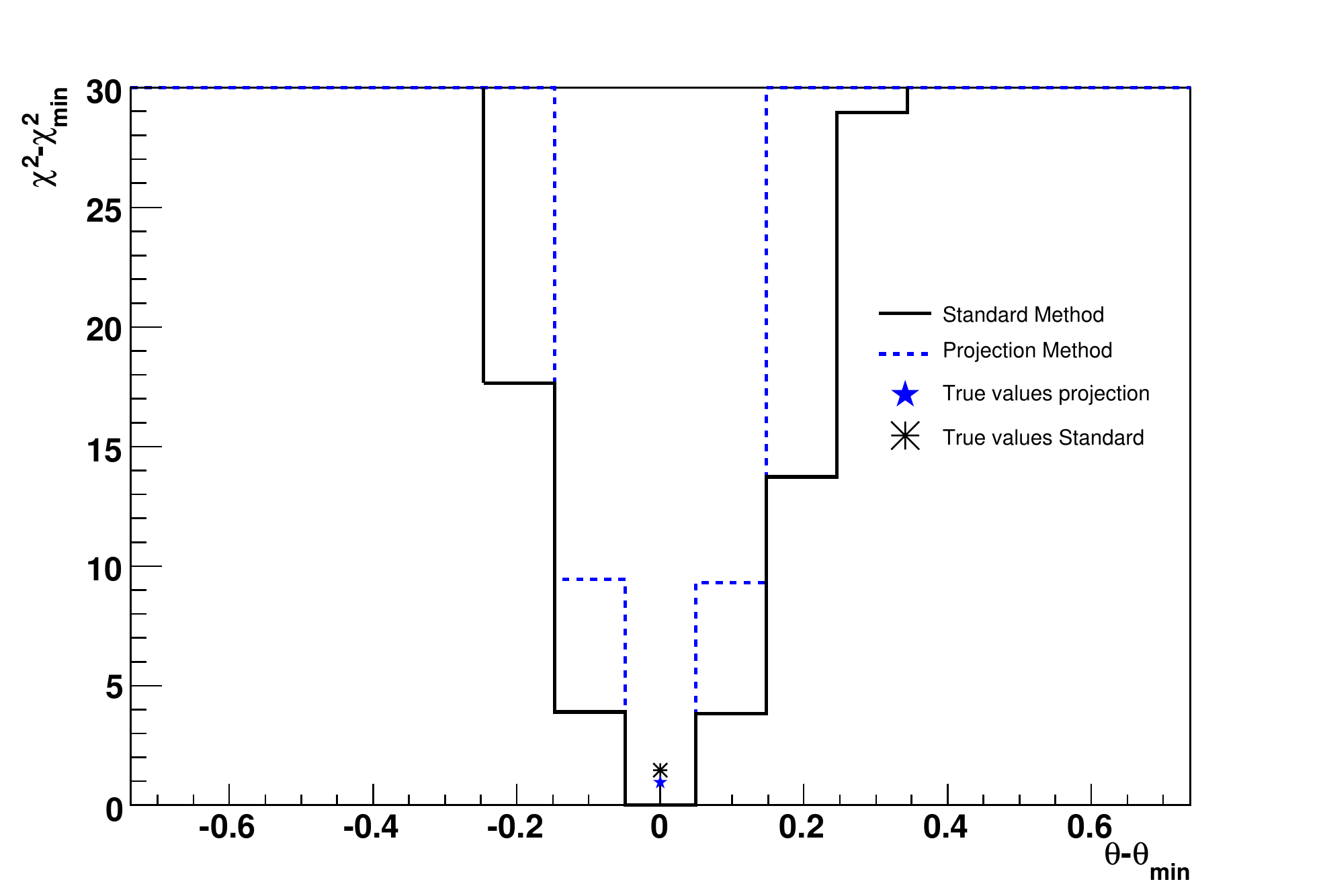} &
    \includegraphics[width=7.5cm,height=6cm]{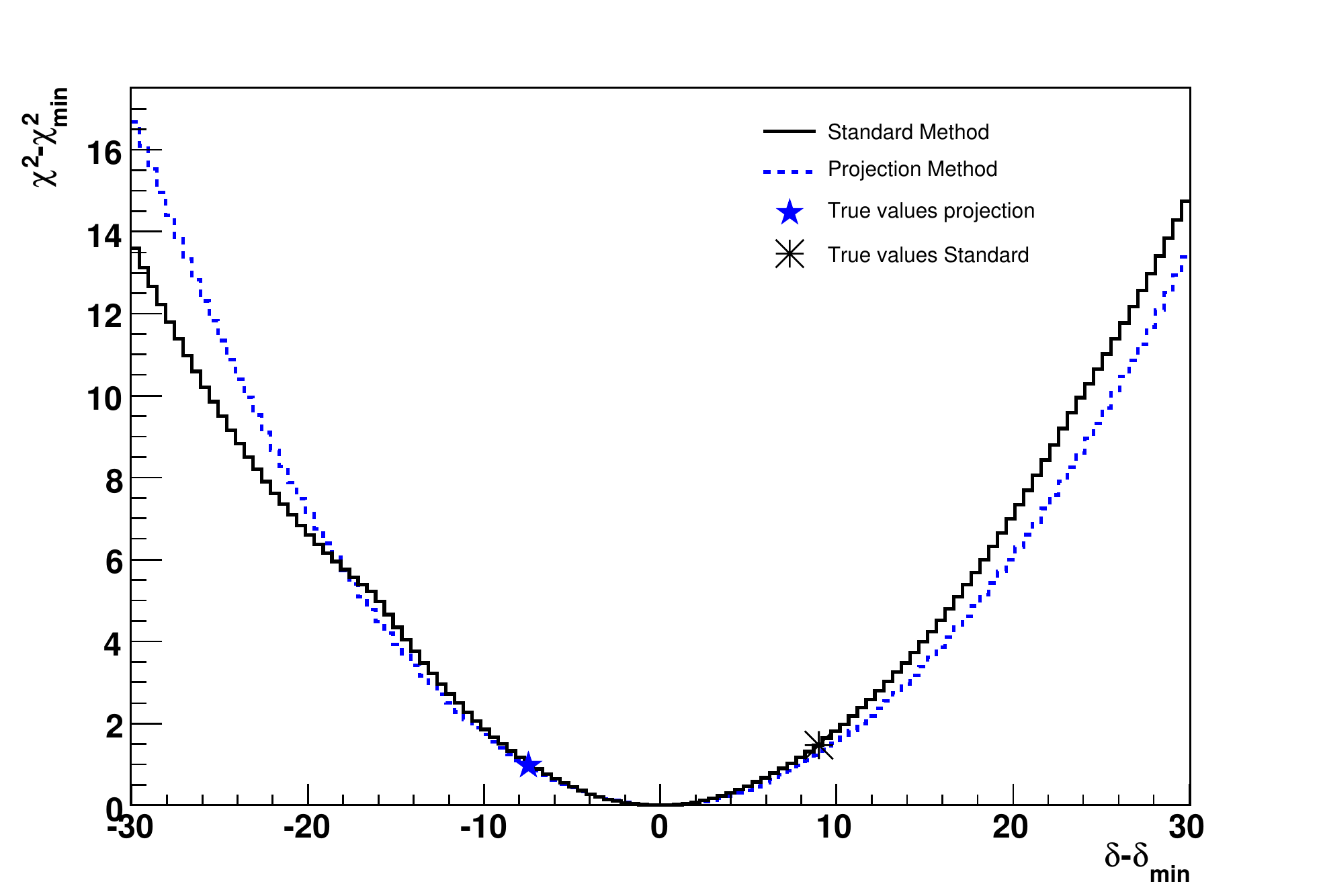}
    \end{array}$
  \end{center}
  \caption{Projection of the minimum $\chi^2$ onto the $\theta_{13}$ axis (left) and the $\delta$ axis (right) for $\theta_{13}=1^\circ$ and $\delta=45^\circ$.}
  \label{fig:projNF}
\end{figure}

This tendency is illustrated in figure \ref{fig:projNF} for a set of true values of $\theta_{13} = 1^{\circ}$ and $\delta = 45^{\circ}$ where the minimum $\chi^2$ is projected onto each axis in turn.
At the 1$\sigma$ level ($\chi^2-\chi^2_{min}=1$) the fit to $\theta_{13}$ is very similar for both methods, however, the near-far projection is already better at the 3$\sigma$ level ($\chi^2-\chi^2_{min}=9$). The projection onto the $\delta$ axis for both methods is very similar, although for this example case the true value is a better fit in the case of the near-far-projection method.
\begin{figure}
  \begin{center}$
    \begin{array}{cc}
    \includegraphics[width=7.5cm,height=6cm]{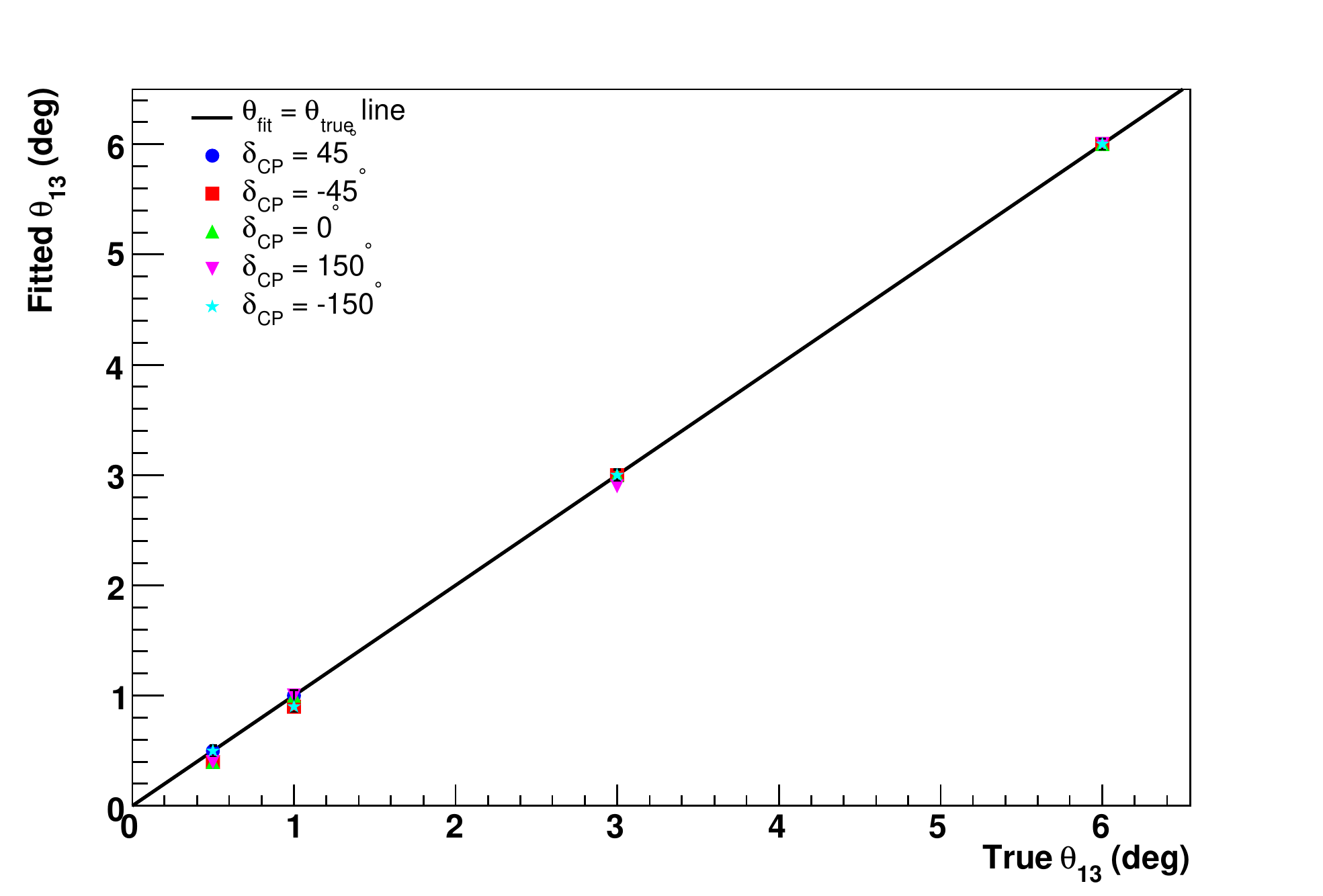} &
    \includegraphics[width=7.5cm,height=6cm]{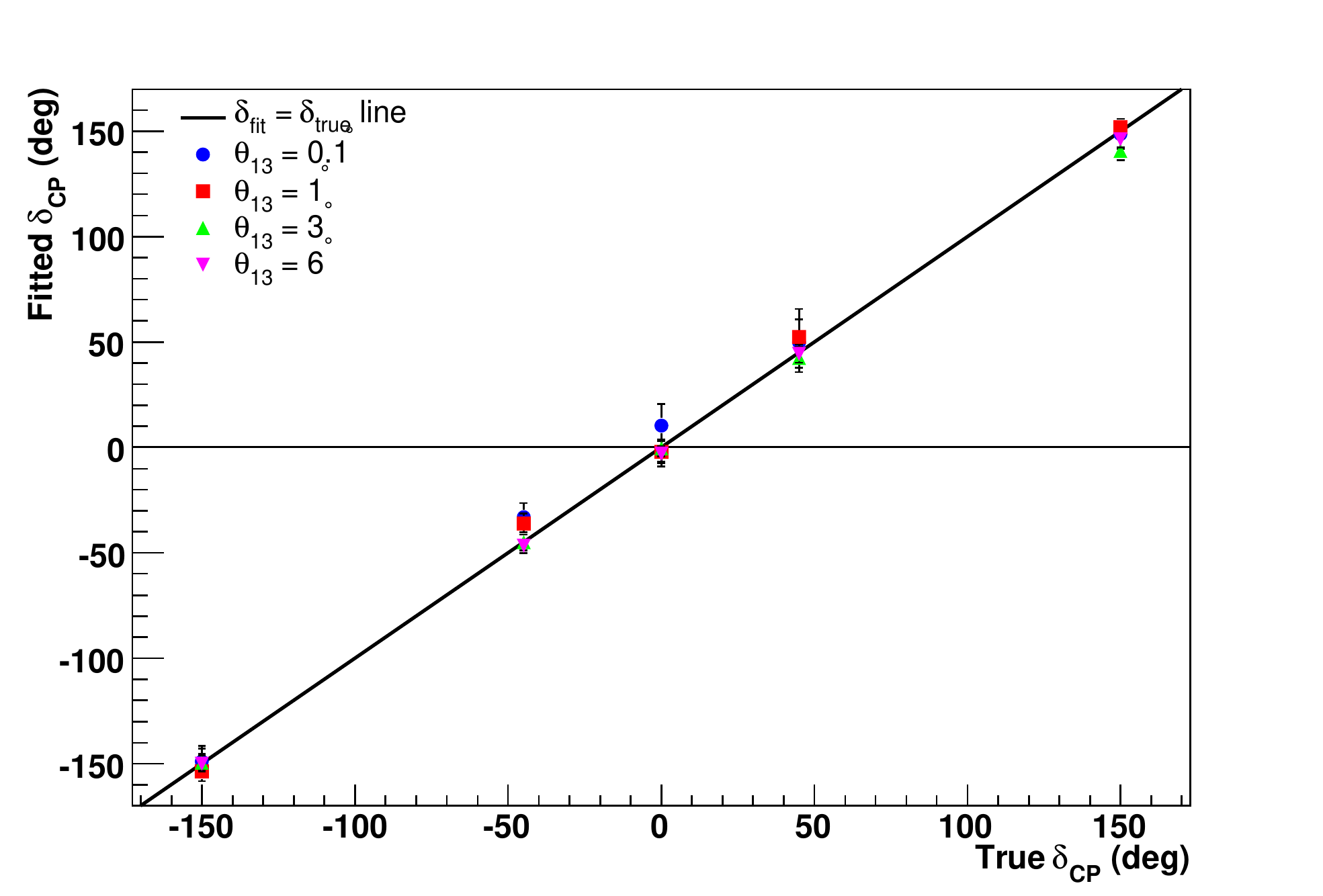}\\
    \includegraphics[width=7.5cm,height=6cm]{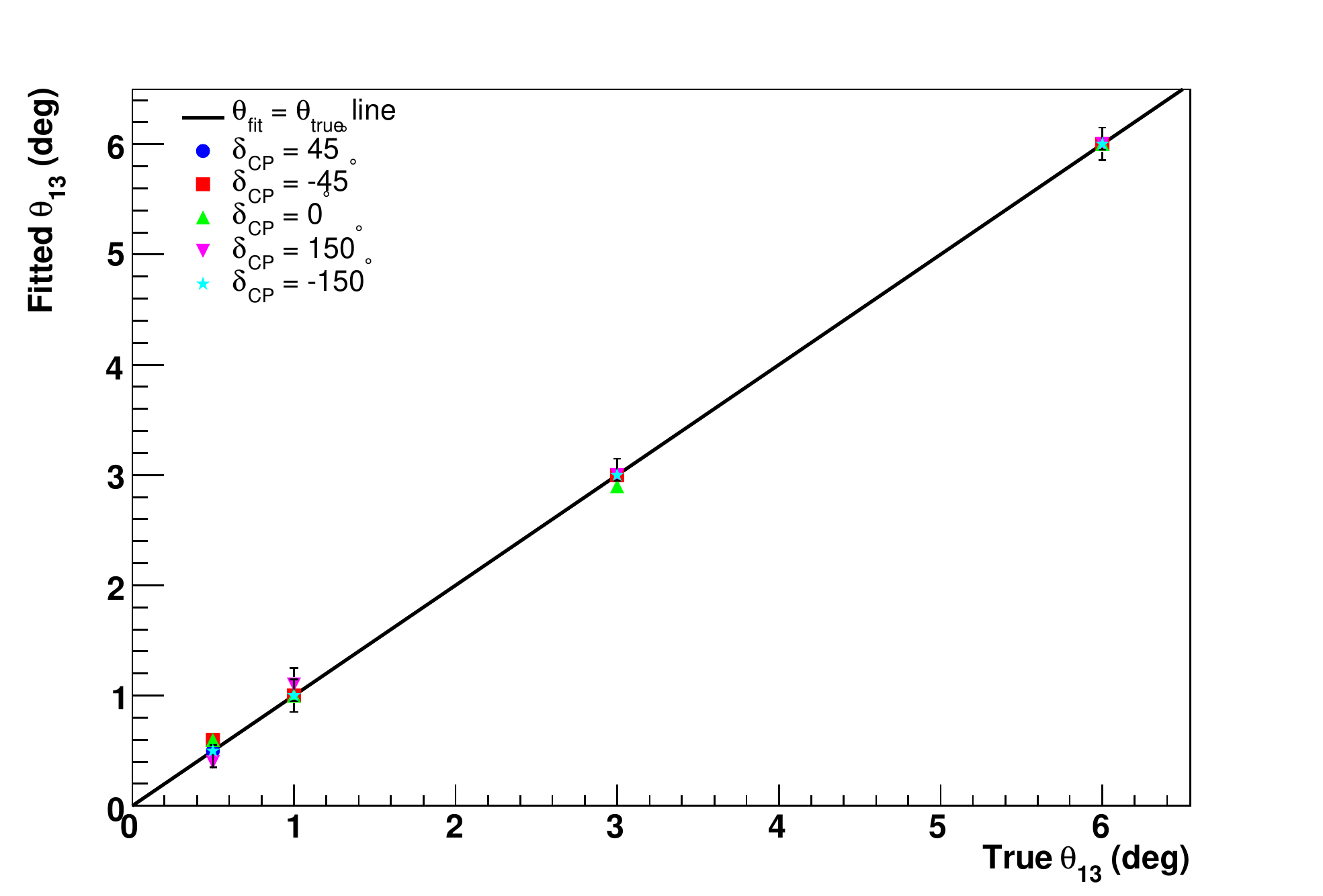} &
    \includegraphics[width=7.5cm,height=6cm]{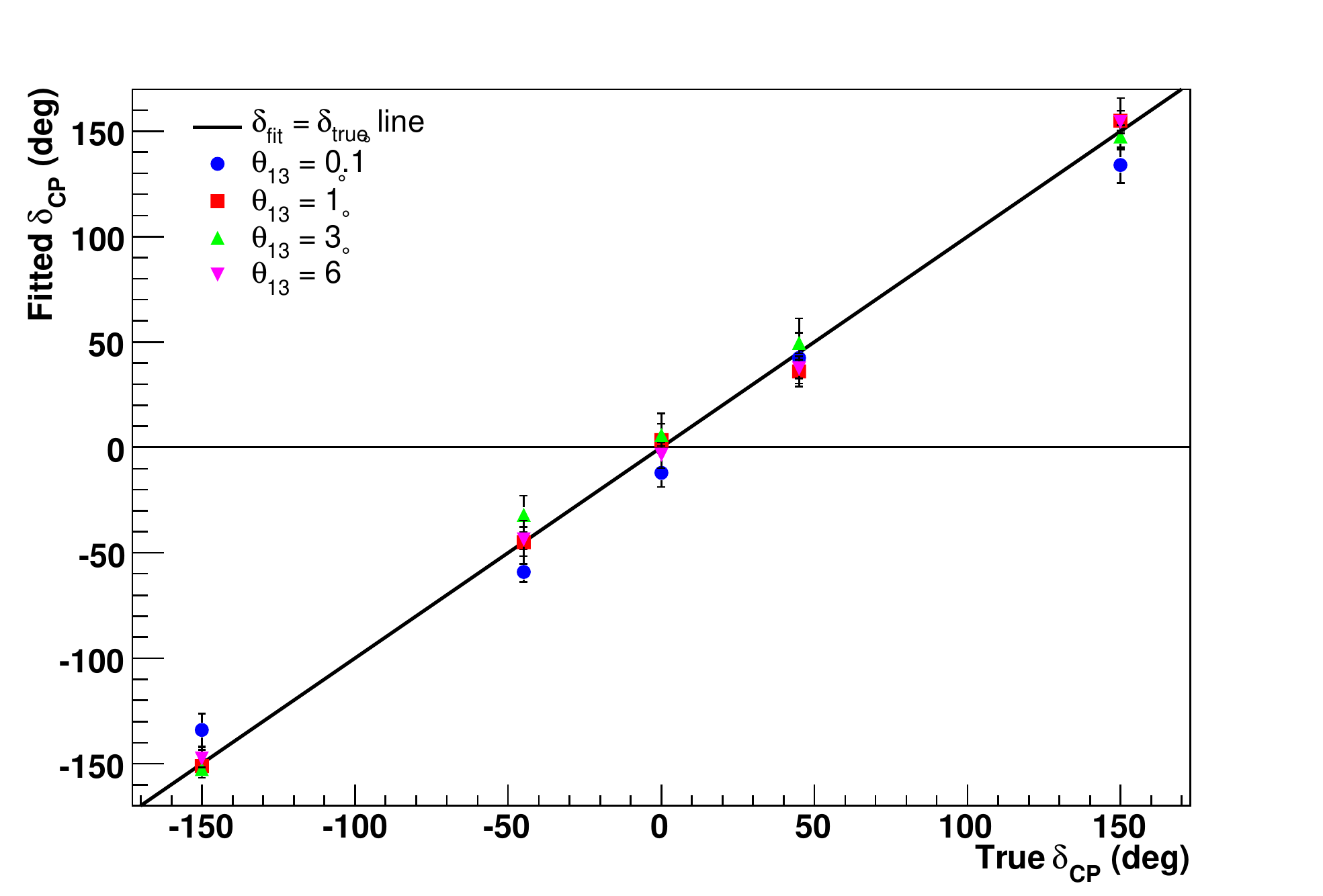}
    \end{array}$
  \end{center}
  \caption{Quality of fit to $\theta_{13}$ (left) and $\delta$ (right) for a range of values. Using the near-far projection method (top) and a method in which the neutrino flux is part of the fit (bottom).}
  \label{fig:trendNF}
\end{figure}

Figure \ref{fig:trendNF} shows the trend for the measurement of the oscillation parameters for both the near-far projection method and the fitted-flux method. In both cases, 1$\sigma$ errors are garnered from the projection of the minimum $\chi^2$ onto the appropriate axis and $\theta_{13}$ is determined very precisely, with the true value being the best fit value in most cases. Determination of $\delta$ is not as precise, with the precision of the near-far projection method being somewhat better. Considering the deviation of the best fit delta value from the true value, it can be seen that the mean difference for the standard method is $\sim0.9\sigma$ whereas for the projection method it is $\sim0.6\sigma$.

\subparagraph*{Conclusions \\}

A basic method for the projection of the observed near detector
spectrum at the Neutrino Factory to the far-detector sites has been
developed. 
Under the assumption of pure near detector signals it has been shown
that this projection can be used to predict accurately the far
detector spectrum in the absence of oscillations. 
While in its current form this method has no clear advantage at the
$1\sigma$ level over a method in which the flux is extracted from the
fit directly, there seems to be evidence of improved sensitivity at
$3\sigma$. 
These preliminary results show the potential for a high-resolution
near detector to improve the precision with which the oscillation
parameters can be measured.

\paragraph{Study of neutrino interactions}
\label{SBL:sec:nuint}

The unprecedented neutrino fluxes available for the Neutrino
Factory program will allow the collection of a large number of
inclusive neutrino charged current (CC) interactions. 
As discussed above, the reduction of systematic uncertainties for the
neutrino oscillation program requires a highly segmented near
detector, thus providing excellent resolution in the reconstruction of
neutrino events. 
The combination of this substantial flux with a finely segmented near
detector offers a unique opportunity to produce a range of neutrino
scattering physics measurements, in addition to those needed by the
long base line oscillation program. 
The combined statistics and precision expected in the near detector
will allow precise tests of fundamental interactions and better
understanding of the structure of matter. 
Given the broad energy range of the beam, a diverse range of physics
measurements is possible in the Neutrino Factory near detector.
To provide a flavour for the outstanding physics potential, we give a
short description of the studies that can be performed at a Neutrino
Factory near detector for a few selected topics. 
A more detailed and complete discussion of the short baseline physics
potential will appear in a separate physics working group paper and in
subsequent Neutrino Factory reports. 

\subparagraph*{Electroweak Physics}
\label{SBL:sec:sin2thetaW}

Neutrinos are a natural probe for the investigation of
electroweak physics.  Interest in a precise determination of the weak mixing angle ($\sin^2 \theta_W$)
at Neutrino Factory energies via neutrino scattering is twofold: a) it provides a direct measurement of
neutrino couplings to the $Z$ boson and b) it probes a different scale of momentum transfer
than LEP by virtue of not being on the $Z$ pole. The weak mixing angle can be extracted
experimentally from three main NC physics processes:
\begin{enumerate}
  \item Deep inelastic scattering off quarks inside nucleons: 
        $\nu N \to \nu X$; 
  \item Elastic scattering off electrons: $\nu e^- \to \nu e^-$; and
  \item Elastic scattering off protons: $\nu p \to \nu p$.
\end{enumerate}

The most precise measurement of $\sin^2 \theta_W$ in neutrino deep
inelastic scattering (DIS) comes from the NuTeV experiment, which
reported a value that is $3\sigma$ from the Standard
Model~\cite{Zeller:2001hh}.
The Neutrino Factory near detector can perform a similar analysis in
the DIS channel by measuring the ratio of NC and CC interactions
induced by neutrinos: 
\begin{equation}
{\cal R}^\nu \equiv \frac{\sigma^\nu_{\rm NC}}{\sigma^\nu_{\rm CC}}
 \simeq \rho^2 \left( \frac{1}{2} - \sin^2 \theta_W +\frac{5}{9} \left(1 + r \right) \sin^4 \theta_W  \right)
\end{equation}
where $\rho$ is the relative coupling strength of the neutral to charged current interactions
($\rho =1$ at tree level in the Standard Model) and $r$ is the ratio of anti-neutrino to
neutrino cross section ($r \sim 0.5$).

The measurement of $\sin^2 \theta_W$ from DIS interactions can be
performed with a low density magnetised tracker since an accurate
reconstruction of the NC event kinematics and of the $\nu_e$ CC
interactions are crucial to keep the systematic uncertainties on the
event selection under control. 
The analysis selects events in the near detector after imposing a cut on the
visible hadronic energy of $E_{\rm had} > 3$ GeV, as in the NOMAD
$\sin^2 \theta_W$ analysis (the CHARM analysis had $E_{\rm had} > 4$
GeV). 

The use of a low density magnetised tracker can substantially reduce systematic uncertainties
with respect to a massive calorimeter. The largest experimental systematic uncertainty in NuTeV is related to the subtraction
of the $\nu_e$ CC contamination from the NC sample. Since the low density tracker at the Neutrino Factory
can efficiently reconstruct the electron tracks, the $\nu_e$ CC interactions can be
identified on an event-by-event basis, reducing the corresponding uncertainty to a
negligible level. Similarly, uncertainties related to the location of the interaction vertex,
noise, counter efficiency etc. are removed by the higher resolution and by the different
analysis selection. 

A second independent measurement of $\sin^2 \theta_W$ can be obtained from NC
$\nu_\mu e$ elastic scattering. This channel has lower systematic uncertainties since it does
not depend upon the knowledge of the structure of nuclei, but has limited statistics
due to its very low cross section. The value of $\sin^2 \theta_W$ can be extracted from
the ratio of neutrino to anti-neutrino interactions~\cite{Marciano:2003eq}:
\begin{equation} \label{eqn:NCel}
{\cal R}_{\nu e} (Q^2) \equiv \frac{\sigma(\bar{\nu}_\mu e \to \bar{\nu}_\mu e)}{\sigma(\nu_\mu e \to \nu_\mu e)} (Q^2)
\simeq \frac{1 - 4 \sin^2 \theta_W + 16 \sin^4 \theta_W}{3 -12 \sin^2 \theta_W + 16 \sin^4 \theta_W}
\end{equation}
in which systematic uncertainties related to the selection and
electron identification cancel out. 

The extraction of the weak mixing angle is dominated by the systematic
uncertainty on the $\bar{\nu}_\mu / \nu_\mu$ flux ratio, which enters
equation \ref{eqn:NCel}.
At a Neutrino Factory this systematic uncertainty will be considerably
reduced over conventional beam facilities used for previous studies.
Together, the DIS and the NC elastic scattering channels
involve substantially different scales of momentum transfer, providing a tool to test the
running of $\sin^2 \theta_W$ in a single experiment. To this end, the study of NC elastic scattering
off protons can provide additional information since it occurs at a momentum scale which is
intermediate between the two other processes. Furthermore, in the two NC elastic processes
off electrons and protons it is possible to reconstruct the $Q^2$ on an event-by-event
basis, providing additional sensitivity. 

\subparagraph*{Strange Content of the Nucleon}
\label{sec:deltas}

The role of the strange quark in the
proton remains a central investigation in hadronic physics. The interesting question is
to what extent the strange quarks contribute substantially to the nucleon vector and axial-vector
currents. A large observed value of the strange quark contribution to the nucleon spin
(axial current), $\Delta s$, would require further theoretical speculations with respect to present
assumptions. The nucleon spin structure also affects the couplings of axions and
supersymmetric particles to dark matter. To better understand this, experiments at
MIT/Bates, Mainz, and Jefferson Lab probed and continue to probe the contribution of
strange quarks to the electromagnetic (vector) current. However, the only reliable
measurement of $\Delta s$ can be obtained from the detection of neutrino-proton NC elastic
scattering, $\nu p \to \nu p$.
In the limit $Q^2 \to 0$, the differential cross section is
proportional to the square of the iso-vector axial-vector form factor plus/minus the strange
axial form factor, $(G_A \pm G_s)^2$, where $G_s^2(Q^2=0) = \Delta s$.
Unfortunately, previous neutrino
scattering experiments have not been precise enough to provide a definitive statement on
the contribution of the strange sea to either the axial or vector form factor.
The Neutrino Factory near detector neutrino beam will be sufficiently intense that
a measurement of NC elastic scattering 
can provide a definitive statement on
the contribution of the strange sea to either the axial or vector form factor.
Systematic uncertainties can be reduced by measuring the NC/CC ratios for both neutrinos and
anti-neutrinos as a function of $Q^2$:
\begin{equation}
{\cal R}_{\nu p} (Q^2) \equiv \frac{\sigma(\nu_\mu p \to \nu_\mu p)}{\sigma(\nu_\mu n \to \mu^- p)}(Q^2); \;\;\;\;\;
{\cal R}_{\bar{\nu} p} (Q^2) \equiv \frac{\sigma(\bar{\nu}_\mu p \to \bar{\nu}_\mu p)}{\sigma(\bar{\nu}_\mu p \to \mu^+ n)}(Q^2)
\end{equation}

\subparagraph*{Structure of the Nucleon}
The following have been identified as important physics topics to address the structure of the nucleon at a near detector of a Neutrino Factory:
\begin{itemize}
\item Measurement of form factors and structure functions;
\item QCD analysis of parton distribution functions;
\item $d/u$ parton distribution functions at large Bjorken-$x$;
\item GLS sum rule and $\alpha_s$;
\item Non-perturbative contributions and higher twists;
\item Quark-hadron duality; and
\item Generalised parton distributions.
\end{itemize}

\subparagraph*{Parton Distribution Functions}
\label{SBL:sec:PDFs}

A QCD analysis of the near detector data in the framework of global fits to extract
parton distribution functions is a crucial step to constrain systematic uncertainties
on the electroweak measurements.  In addition,
precision measurements of (anti)-neutrino structure functions and differential cross sections
would directly affect the long-baseline oscillation searches,
providing an estimate of all background processes which are dependent
upon the angular distribution of the outgoing particles in the far
detector.

For quantitative studies of inclusive deep-inelastic lepton-nucleon
scattering, it is vital to have precise $F_3$ structure functions,
which can only be measured with neutrino and antineutrino beams, as
input into global PDF fits.  Because it depends on weak axial quark
charges, the $F_3$ structure function is unique in its ability to
differentiate between the quark and antiquark content of the nucleon.
On a proton target, for instance, the neutrino and antineutrino $F_3$
structure functions (at leading order in $\alpha_s$) are given by:
\begin{eqnarray}
xF_3^{\nu p}(x)
&=& 2 x \left( d(x) - \bar u(x) + \bar s(x) + \cdots \right)\, {\rm ; and} \\
xF_3^{\bar\nu p}(x)
&=& 2 x \left( u(x) - \bar d(x) - \bar s(x) + \cdots \right)\, .
\end{eqnarray}
In contrast, electromagnetic probes are sensitive only to a sum of
quark and antiquark PDFs.  Unfortunately, the neutrino scattering
cross sections have considerably larger uncertainties than the
electromagnetic inclusive cross sections at present.
The high intensity Neutrino Factory facility offers
the promise to reduce the gap between the uncertainties on the weak and electromagnetic
structure functions, and would have a major impact on global PDF
analyses.
In addition to data in the DIS region, there is considerable interest
in obtaining data at low $Q^2$ (down to $Q^2 \sim 1$~GeV$^2$) and low
$W$ ($W < 2$~GeV).

Global PDF fits show that at large values of (Bjorken) $x$ ($x >
0.5-0.6$) the $d$ quark distribution (or the $d/u$ ratio) is very
poorly determined. 
The main reason for this is the
absence of free neutron targets.  
Because the electric charge on the $u$ quark is larger than that on
the $d$, the electromagnetic proton $F_2$ structure function data
provide strong constraints on the $u$ quark distribution, but are
relatively insensitive to the $d$ quark distribution. 
To constrain the $d$ quark distribution as a function of the hadronic four-momentum
transfer squared, $t$, a precise knowledge
of the corresponding neutron $F_2^n$ structure functions is required, which in
practice is extracted from inclusive deuterium $F_2$ data.
At large values of $x$ the nuclear corrections in deuterium become
large and, more importantly, strongly model dependent, leading to
large uncertainties on the resulting $d$ quark distribution.

Perhaps the cleanest and most direct method to determine the $d/u$
ratio at large $x$ is from neutrino and antineutrino DIS on hydrogen.
Should a hydrogen target at the Neutrino Factory near detector become
available, a new measurement of 
neutrino and antineutrino DIS from {\it hydrogen} would offer significantly
improved uncertainties and could therefore make an important discovery
about the $d/u$ behaviour as $x \to 1$. To be competitive with the
proposed JLab 12~GeV experiments, the kinematic reach would need
to be up to $x \sim 0.85$ and with as large a $Q^2$ range as possible
to control the higher twist and other sub-leading effects in $1/Q^2$.

\subsubsection{Near detector baseline design}
\label{sec:ND_design}

\paragraph{Option A - scintillating fibre tracker}

\begin{figure}
\begin{center}
 \includegraphics[width=.95\textwidth]{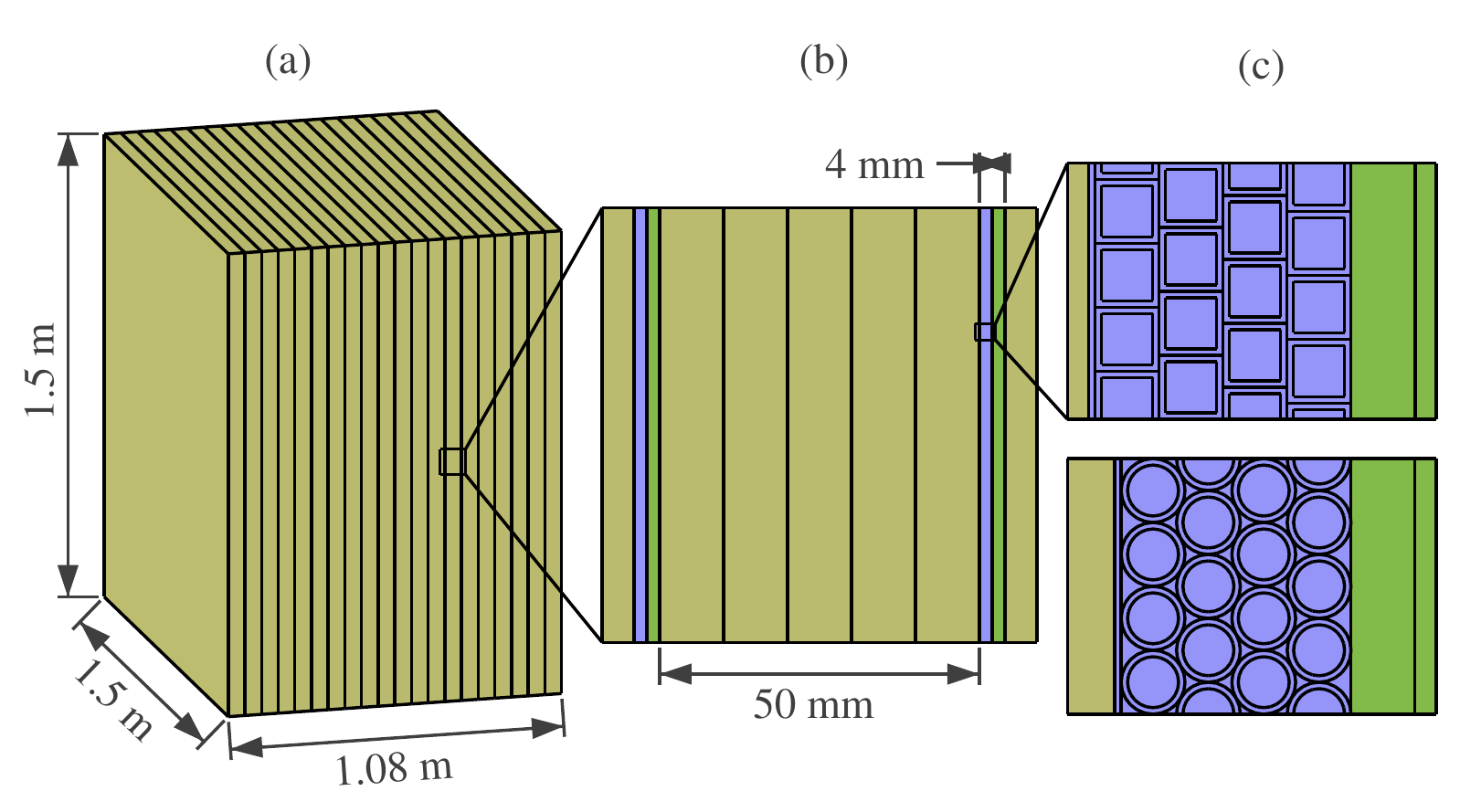}
\end{center}
\caption{Schematic drawing of the scintillating fibre tracker (a). In each fibre station (b,c) the layers of fibres with horizontal orientation are in blue and the layers with vertical orientation are in green. The absorbers are shown in brown.}
\label{fibres2}
\end{figure}

This option considers a low $Z$, high resolution scintillating fibre
tracker. 
It is composed of consecutive modules placed perpendicular to the beam
axis (figure \ref{fibres2}a). Each module (figure \ref{fibres2}b) has
two volumes: an absorber made of plastic scintillator slabs and a fibre
station made of thin scintillating fibres. The recoil
hadronic energy is measured in the absorber, while the fibre station reconstructs
particle tracks with high angular resolution.  

The overall thickness of the absorber is 5\,cm. 
It is subdivided into five layers perpendicular to the beam axis. 
The light attenuation in the material of the slabs introduces a
dependence of the measured signal on the position of the particle
hit, which can be eliminated by reading the signal from the absorber scintillator
slabs at both edges of a slab. 

The fibre station (figure \ref{fibres2}c) consists of 4 layers of
fibres with horizontal orientation and 4 layers of fibres with
vertical orientation. 
Three different conceptual designs for the fibre station are under
consideration. 
The first one is a station made of cylindrical 0.5\,mm thick fibres
(figure \ref{fibres2}c-bottom). 
For this design, the position of the fibres in a given layer is
shifted by 0.25\,mm relative to the fibres in the neighbouring layer. 
The other two options consider a station made of squared 0.5\,mm thick
fibres with a displacement between neighbouring layers of 0.25\,mm and
0.175\,mm respectively (figure \ref{fibres2}c-top). 
In all three options the fibre-station thickness is 4\,mm and the
station consists of  $\sim 12\,000 $ fibres. 
Our baseline option is a detector made of 20 consecutive
modules. Thus, the overall detector dimensions are 
$1.5 \times 1.5 \times 1.08$\,m$^3$ that corresponds to a total mass
of $\sim 2.5$\,Ton.

In order to achieve maximal resolution on the muon scattering angle
the detector volume should be free of magnetic field. 
Such a choice implies an additional magnetised detector, placed
downstream of the scintillating fibre tracker to measure the sign,
momentum and trajectory of the outgoing muons.

\paragraph{Option B - High resolution straw tube tracker}
We propose here a possible high resolution straw tube tracker inspired
by the HiResM$\nu$ near detector \cite{Mishra:2008nx} being considered
for the LBNE project at Fermilab \cite{Fermilab_P5,Fermilab_SG}. 
In this section we describe the detector being considered for LBNE,
which would have identical features to the one at a Neutrino Factory. 

Building upon the NOMAD-experience \cite{Altegoer:1997gv}, we propose
a low-density tracking detector with a fiducial mass of 7.4\,Ton as a
neutrino target. 
The active-target tracker will have a factor of two more sampling
points along the $z$-axis ($\nu$ -direction) and a factor of six more
sampling points in the plane transverse to the neutrino compared to
the NOMAD experiment. 
The proposed detector will further enhance the resolving power by
having an order of magnitude more tracking points and coverage for
side-exiting neutrals and muons. 

We are proposing straw-tube trackers (STT) for the active neutrino
target, similar to the ATLAS Transition Radiation Tracker
\cite{Akesson:2004nj,Akesson:2004nh,Abat:2008zza} and the COMPASS
detector \cite{Bychkov:2002xw}. 
The tracker will be composed of straw tubes with 1 cm diameter, in the
vertical ($y$) and horizontal ($x$) directions. 
In front of each module a plastic radiator made of many thin foils
allows the identification of electrons through their transition
radiation. The nominal fiducial volume for CC analysis is: 
$350\times 350\times 600$~cm$^3$, corresponding to 7.4 tons of mass
with an overall density $\rho < 0.1$~g/cm$^3$. 

\begin{figure}
\begin{center}
 \includegraphics[width=.85\textwidth]{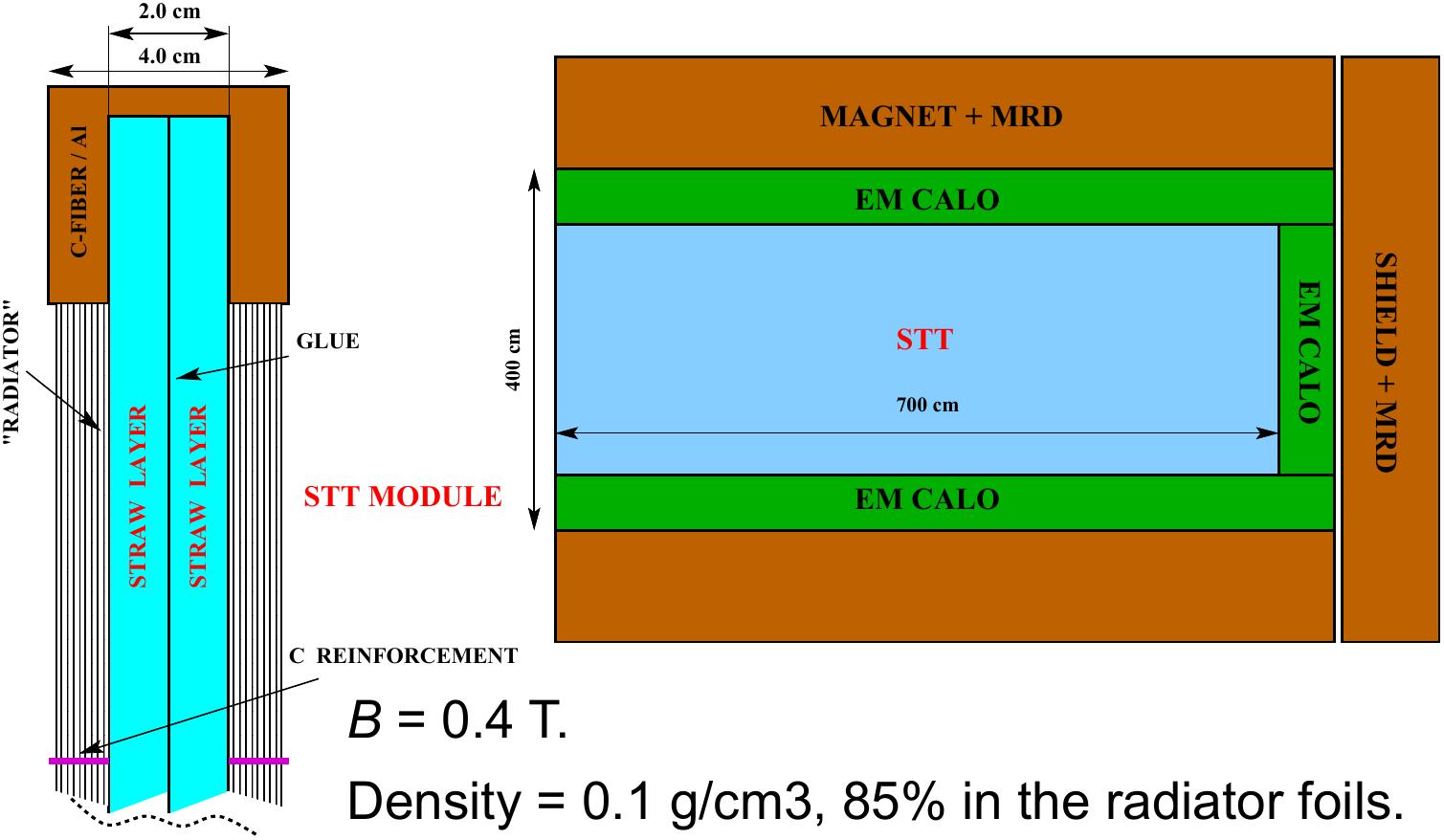}
\end{center}
\caption{Sketch of the proposed HiRes detector showing the inner straw tube tracker (STT), the electromagnetic calorimeter (EM CALO) and the magnet with the muon range detector (MRD). Also shown is one module of the proposed straw tube tracker (STT). Two planes of straw tubes are glued together and held by an aluminium frame.}
\label{fig:HiRes}
\end{figure}

The STT will be surrounded by an electromagnetic calorimeter (sampling Pb/scintillator) covering
the forward and side regions. Both sub-detectors will be installed inside a dipole magnet providing
a magnetic field of $\sim 0.4$~T. An external muon detector based upon Resistive Plate Chambers (RPC)
will be placed outside of the magnet (see figure~\ref{fig:HiRes}).

The neutrino target will be mainly composed of carbon, with a radiation length of about
5~m and space-point resolution around 200~$\mu$m. The momentum resolution is dominated by multiple scattering for tracks 1\,m long ($\Delta p/p = 0.05$), while the measurement error for $p = 1$\,GeV tracks would be $\Delta p/p = 0.006$. The proposed
detector will measure track position, $dE/dx$, and transition radiation (with Xe filling) over the
entire instrumented volume. The unconverted photon energy will be measured in the calorimeters
with a target energy resolution of $\sim 10\%/\sqrt{E}$. The detector will provide:
\begin{itemize}
\item Full reconstruction of charged particles and gammas;
\item Identification of electrons, pions, kaons, and protons from
      $dE/dx$;
\item Electron (positron) identification from transition radiation 
      ($\gamma > 1\,000$);
\item Full reconstruction and identification of protons down to
      momenta of 250\,MeV; and
\item Reconstruction of electrons down to momenta of 80\,MeV from
      curvature in the B-field. 
\end{itemize}

\begin{figure}
\begin{center}
 \includegraphics[width=.80\textwidth]{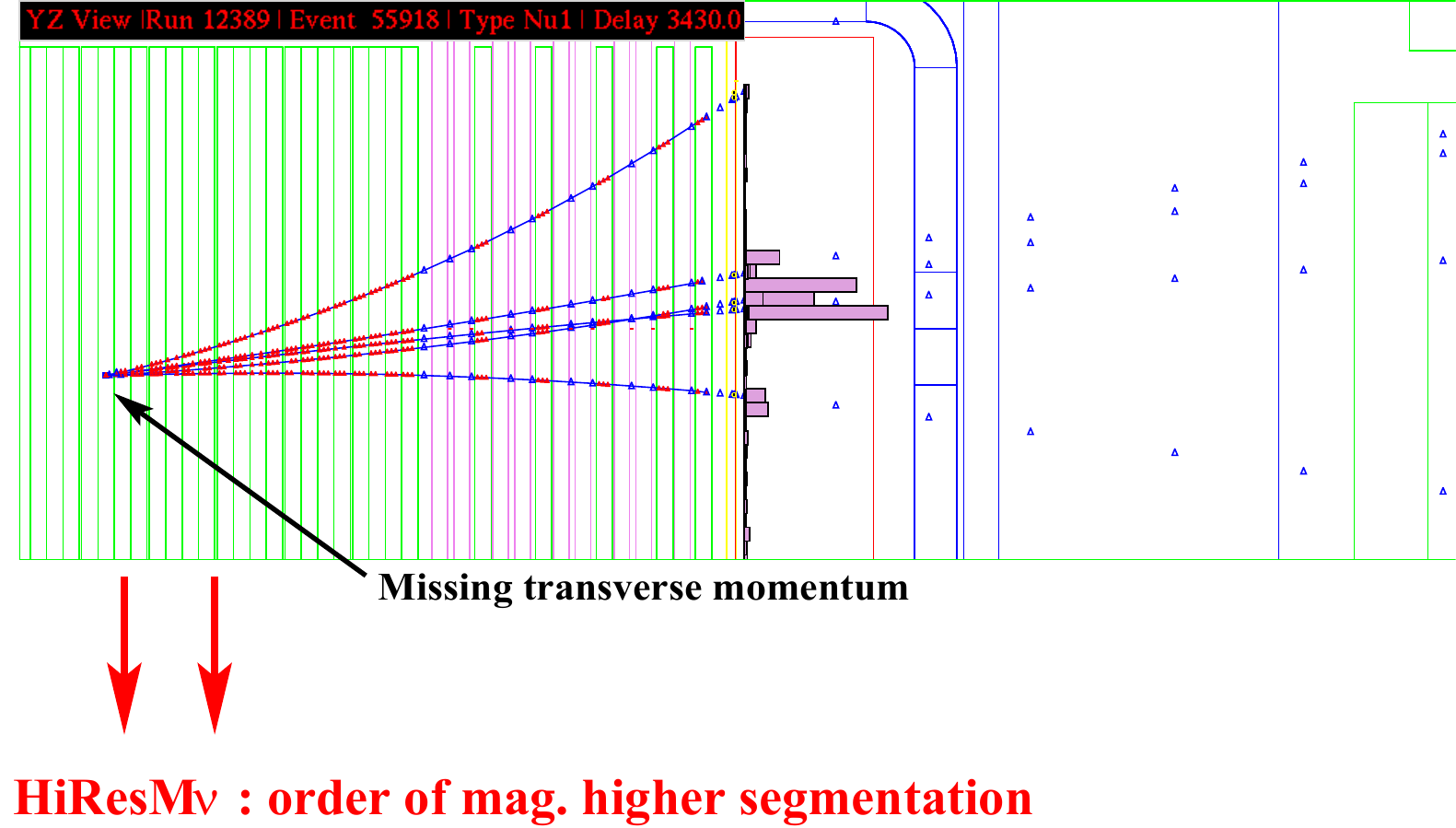}
\end{center}
\caption{Candidate NC Event in NOMAD. Tracking charged particles in the HiRes detector
will provide a factor of two higher segmentation along z-axis and a factor of six higher segmentation
in the transverse-plane compared to NOMAD.}
\label{fig:NC_NOMAD}
\end{figure}

The proposed near detector will measure the relative abundance, the
energy spectrum, and the detailed topologies for $\nu_\mu$,
$\overline{\nu}_\mu$, $\nu_e$ and $\overline{\nu}_e$ induced
interactions, including the momentum vectors of negative, positive and
neutral ($\pi^0$, $K_s^0$, $\Lambda$ and $\overline{\Lambda}$)
particles composing the hadronic system. 
A NC event candidate in NOMAD, shown in figure~\ref{fig:NC_NOMAD},
gives an idea of the precision with which the charged-particles and
the forward gammas were measured. 
Detailed simulations of this detector have been carried out in the
context of the LBNE proposals \cite{LBNE}. These simulations will be
adapted to the neutrino spectra at a Neutrino Factory to derive the
performance parameters of this detector in this context. 
We expect to present such performance parameters in the Reference
Design Report.   

\paragraph{Charm and Tau Detector} 
A near detector at a Neutrino Factory must measure the charm
cross-section to validate the size of the charm background in the far
detector, since this is the main background to the wrong-sign muon
signature (see section~\ref{sec:MIND}). The charm cross-section and
branching fractions are poorly known, especially close to threshold,
so a near detector would need to be able to detect charm
particles. Data available for charm cross-sections from neutrino
interactions result from dimuon events in large mass neutrino
experiments or from high resolution emulsion detectors operating at
neutrino beams (for a comprehensive review see~\cite{Lellis:2004yn}
and references therein). Recent results come from the dimuon data
in the CHORUS \cite{:2008xp} (8\,910 neutrino induced and 430
antineutrino induced dimuon events) and NOMAD \cite{nomad:2010dimu}
experiments ($\sim 15\,000$ dimuon events). The theoretical uncertainty
in the charm cross-section arises from the errors in the strange sea
content of the nucleon, the semi-leptonic charm to muon branching
fraction (with a 10\% relative error) and the longitudinal structure
function ($F_L$) and higher-twist effects. For this reason, it is
paramount to make an independent near detector measurement of the
charm cross-section with of order $10^6$ charm events and make the
error in the charm cross-section negligible in the estimation of the
neutrino oscillation background.

Since tau events have a similar signature to charm events, any
detector that can measure charm should be able to measure taus as
well. This is important to explore couplings of Non Standard
Interactions (NSI) at source $\epsilon_{\tau\mu}^s$, $\epsilon_{\tau
  e}^s$ or detection $\epsilon_{\tau\mu}^d$, $\epsilon_{\tau e}^d$(see
section~\ref{sec:nsi} for a comprehensive treatment). Either an
emulsion based detector or a semiconductor vertex detector for charm
and tau detection could be used for this purpose.  

A near detector at about 100\,m from the muon decay ring needs to
operate in a high intensity environment ($\sim 10^9~\nu_\mu$ CC events
per year in a detector of mass 1 ton). The ideal detector to identify
the decay topologies of tau leptons and charmed hadrons should be able
to cope with the neutrino event rate and any background from the
facility. While in principle, the muons are bound inside the storage
ring, a calculation of the expected muon escape probability has not
been performed yet and would likely be due to rare processes (for
example, stray muons either escaping early in one of the arcs or the
early part of the straight section and end up scattered towards the
near detector). 
Should this be a problem, active shielding with
magnetised iron toroids to sweep away stray muons could be
used. Another background problem will be due to photons. 
Passive shielding, for example, for example using 30--50\,$X_0$ of
high-$Z$ material, will be required to remove electromagetic radiation
arising from radiative muon decay (1.4\%) and radiation from the decay
electrons.  
However, this shielding will
in turn cause ``neutrino radiation'' from neutrino interactions in the
shielding itself to create a large muon flux upstream of the near
detector. The shielding could be made active and could be used as a
beam profile monitor and, if sufficiently segmented, could be used to
measure the beam divergence. A solution to shielding in the near
detector has not been found yet and will be studied in time for the
Reference Design Report.

Notwithstanding the above mentioned considerations for background
shielding, the near detector is required to: 
\begin{itemize}
\item Have high resolution to identify the short-lived charm hadrons
  and tau  leptons; 
\item Measure the momentum and the charge of decay particles; and
\item Perform a complete and accurate kinematic reconstruction of
  neutrino events. 
\end{itemize}
In the next two sections we identify two different ways of achieving
these goals, one with an emulsion-based detector and the other with a
silicon vertex detector. 

\subparagraph*{Emulsion Detector}

Nuclear emulsions are a very well proven technology operating on
neutrino and charged-particle beams since the 1950s. 
Emulsion technology, pure nuclear emulsion and Emulsion Cloud Chamber
(ECC) targets, has already demonstrated that it is a superb technique
for the study of decay topologies
\cite{Niu:1971xu,Ushida:1980rr,Ushida:1980rs,Albanese:1985wk,:2007ny,Kodama:2000mp,Agafonova:2010dc}. 
A discussion of the performance that has been achieved in previous
emulsion-based experiments is beyond the scope of this section. 
We recall only the outstanding accuracy of this technology in detecting
short-lived particles with an excellent signal to background ratio. 
For details we refer to \cite{9783642036057,Arrabito:2007td} and
references therein.

So far the largest emulsion film production for a high-energy physics
experiment is the one for the OPERA detector \cite{Nakamura:2006xs}. 
Therefore, in the following we consider the emulsion films used for
the OPERA target \cite{Acquafredda:2009zz}. 
Each film has transverse dimensions $10 \times 13$\,cm$^2$ and
consists of a 44\,$\mu$m thick emulsion layer on both sides of a
205\,$\mu$m thick triacetyl cellulose (TAC) base. 
The radiation length of the nuclear emulsions is 5.5\,cm and of the
TAC base 31~cm, while the density is 2.84\,g/cm$^3$ and 1.35\,g/cm$^3$
respectively.  

The possibility of exploiting the emulsion technology for a near
detector has been already discussed in \cite{Abe:2007bi}. 
It was stressed that the main issue is whether it can cope with the
high rate that will be observed at a Neutrino Factory. 
Here we estimate the practical limits and the performance that could
be envisaged for a detector based on the emulsion technology in 
a very intense neutrino beam. 
We consider a pure emulsion target, followed by a magnetic
spectrometer, exploiting nuclear emulsions as a high-precision
(sub-micron) tracker.

The target proposed consists of a sequence of 150 films for a total
length of about 4.6 cm. 
Therefore, the proposed target weight is 1\,kg and has a thickness
of about $0.2\,X_0$. 
An important issue is the number of interactions
that can be stored in the target whilst preserving the capability of
connecting them unambiguously with the hits recorded by the
electronic detectors. 
Experience with OPERA bricks exposed to a neutrino beam, and
integrating over thousands of interactions, shows that up to 10
neutrino interactions per cm$^3$ can be stored \cite{Aoki:2010zz}.
Therefore, in a target (about 500 cm$^3$ of
nuclear emulsions) we can collect up to 5\,000 neutrino interactions.

Downstream of the target, we consider a spectrometer: consisting of a
sandwich of nuclear emulsions and very light material that we call a
``spacer''. 
This material provides a long lever arm between two consecutive
emulsions (tracking devices) with a stable mechanical structure. A
Rohacell plate, a few centimetres thick, fulfils this
requirement. The trajectory measured with the emulsions which precede
and follow the spacer provides the measurement of the charge and
momentum of the particle. The total length of the target plus
spectrometer is of the order of 10 cm. 

Emulsion spectrometers have also been exposed to charged particle
beams of 0.5, 1.0 and 2.0\,GeV \cite{Fukushima:2008zzb}. 
The spectrometer was composed of three emulsion sheets interleaved
with two spacers, each of which produces an air gap of 15\,mm, and 
immersed in a dipole magnetic field of 1.06\,T.
The measured momentum resolution in the range 0.5 to 2.0\,GeV is
$13.3\pm 0.3$\%.  

Monte Carlo simulations have been performed in order to compute the
momentum resolution and the charge identification efficiency of the
spectrometer. Depending on the magnetic field, on the relative
alignment of the emulsion plates in the spectrometer and on the
spectrometer geometry, the momentum resolution for a 10 GeV muon is
better than 25\%, with a charge misidentification better than
0.2\%. 
As far as the electrons are concerned, the momentum resolution
is as good as in the muon case, while the charge misidentification is
much worse due to showering. Very preliminary results show that the
electron charge misidentification is of the order of 40\%. However,
further studies are needed before we can draw firm quantitative
conclusions on the electron charge misidentification.

Although it is not the goal of this study, it is worth noting that
downstream of the spectrometer we have to place an electronic detector
with the aim of providing the time stamp for the events. We plan to
perform the scanning of the events without any electronic detector
prediction. Therefore, time information is essential in order to
match the emulsion information to that of the electronic
detector to allow separation of the charged-current and
neutral-current events.
Furthermore, this electronic detector is also needed to 
identify primary electrons. Given the expected event density in the
target and its dimension, the electronic detector will be able to
provide the time stamp only if it has a position accuracy of the order
of 50~$\mu$m. Therefore, one could imagine synergy between the
emulsion detector and an electronic detector (either a silicon or
scintillating fibre detector) operating on the same beam-line. 

The final question that needs to be addressed is whether an emulsion
based detector can operate close to the storage ring of a Neutrino
Factory. A dedicated study to answer this question will be carried out
in time for the Reference Design Report. 

\subparagraph*{Silicon vertex detector\\}

The second possible detector capable of identifying short lived
particles at a near detector of a Neutrino Factory is a silicon vertex
tracker. The advantage of this type of detector is that it is able to
operate at a high event rate and still have very good spatial
resolution. This is necessary to distinguish the primary neutrino
interaction vertex from the secondary vertex due to the short lived
charm hadron or the tau lepton. Downstream of the vertex detector, we
need a tracking detector capable of distinguishing electrons from
muons in a magnetic field. So, a possible configuration could consist
of a silicon strip or pixel detector, followed by either the
scintillating fibre tracker or the straw tube tracker mentioned in the
previous section. 

\begin{figure}[htbp]
\begin{center}
  \raisebox{-0.5\height}{\includegraphics[width=.65\linewidth]{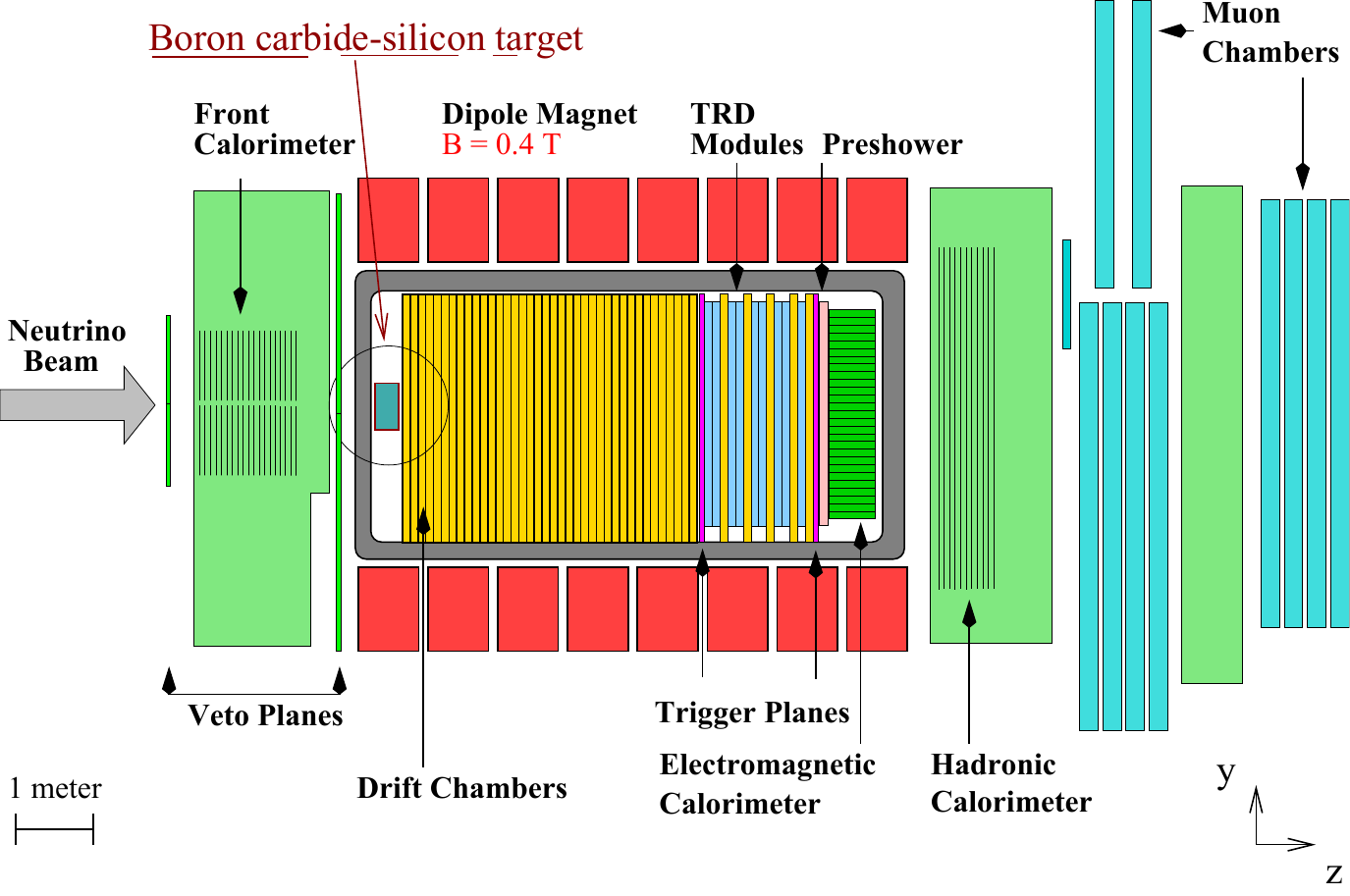}}%
  \hspace{0.05\linewidth}%
  \raisebox{-0.5\height}{\includegraphics[width=.30\linewidth]{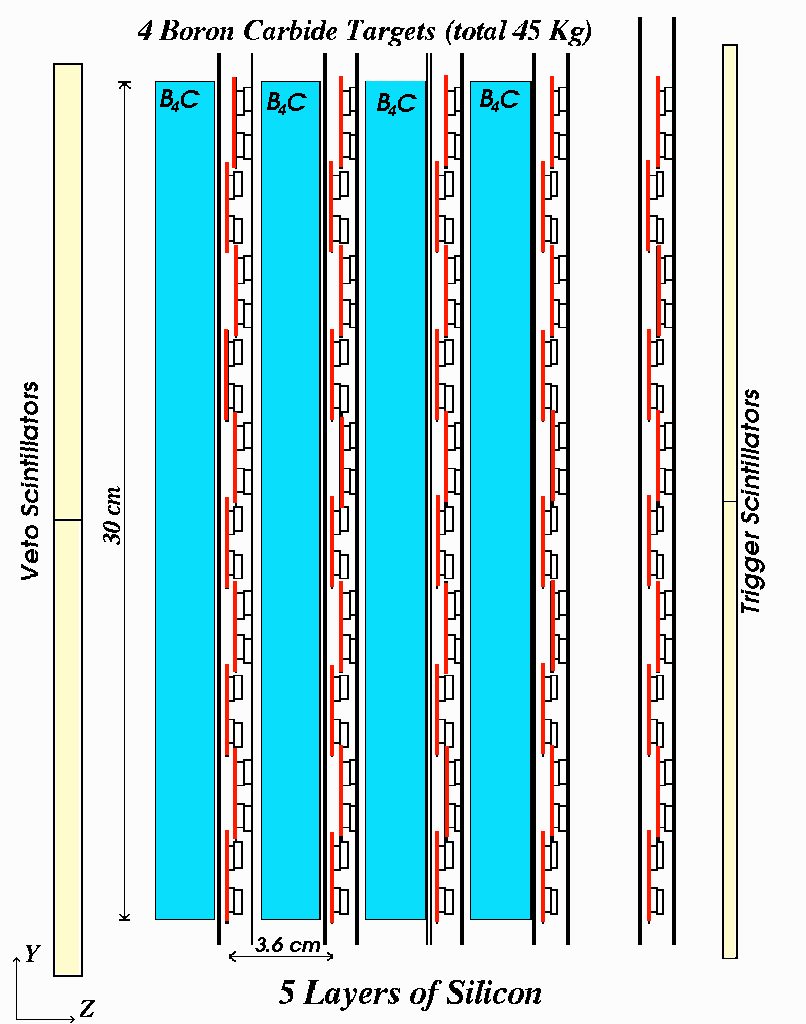}}
\end{center}
\caption{Left: The NOMAD detector with the Silicon TARget (NOMAD--STAR).Right: Side view of NOMAD-STAR.}
\label{fig:nomad_star}
\end{figure}

Such a vertex detector could be similar to the NOMAD--STAR detector
that was installed upstream of the first drift chamber of the NOMAD
neutrino oscillation experiment \cite{Altegoer:1997gv} (see figure
\ref{fig:nomad_star}).
The main aim
of this detector was to test the capabilities of silicon detectors
for short-baseline neutrino oscillation searches
\cite{GomezCadenas:1995ij,GomezCadenas:1996jx}. However, this set-up
can mimic a possible design  for a near detector at a Neutrino Factory
\cite{Ellis:2006wj}. It was used to measure the impact parameter and
double-vertex resolution to determine the charm-detection efficiency. 


The flux of neutrinos at a near detector 80\,m downstream of a
muon storage ring \cite{Tang:2009na} will produce  
$2.3\times 10^8 \nu_\mu$ CC events per year in a detector of similar
mass as NOMAD--STAR (50\,kg). With the measured efficiencies from the
NOMAD-STAR analysis, this would correspond to about $7\times 10^5$
charm events reconstructed per year. 
This would allow a sensitive measurement of the charm cross-section
close to the charm threshold.


The reconstruction of taus from an impact parameter signature with a
dedicated silicon vertex detector was studied in the NAUSICAA proposal
\cite{GomezCadenas:1995ij}. A  silicon vertex detector with a $B_4 C$
target was proposed as an ideal medium to identify taus. Standard
$\nu_\mu$ CC interactions have an impact parameter resolution of
28~$\mu$m, while tau decays have an impact parameter resolution of
62~$\mu$m. By performing a cut on the impact parameter significance
($\sigma_{IP}/IP$) one can separate  one prong decays of the tau from
the background. For three prong decays of the tau, a double-vertex
signature is used to separate signal from background. The total net
efficiency of the tau signal in NAUSICAA was found to be 12\%. With
this efficiency, one could have a sensitivity of  $P_{\mu\tau}<
3\times 10^{-6}$ at 90\% C.L. on the $\mu-\tau$ conversion
probability.  

Another idea proposed in 1996 was to use a hybrid detector
emulsion-silicon tracking to improve the tau-detection efficiency
\cite{GomezCadenas:1996jx}. A Letter of Intent (called TOSCA) was
submitted to the CERN SPSC in 1997 \cite{TOSCA} with a detector based
around this idea.  Tau detection efficiencies of 42\%, 10.6\% and 27\%
were determined for the muon, electron and one charged hadron decays
of the tau, yielding a net probability of $P_{\mu\tau}< 0.75\times
10^{-5}$ for the CERN SPS beam at 350 GeV.  

Assuming 12\% efficiency from the NAUSICAA proposal, and assuming that
charm production is about 4\% of the $\nu_\mu$ CC rate between 10 and
30 GeV (CHORUS measured $6.4\pm 1.0\%$ at 27 GeV)
\cite{Lellis:2004yn}, would imply a signal of $1.2\times 10^8$ tau
events and $4\times 10^7$ charm events. 
Charm events from anti-neutrinos (for example 
$\overline{\nu}_e$) mimic the potential signal. The identification of
the positron can reduce the background, but electron and positron
identification normally has a lower efficiency than muon
identification. It is very important to have a light detector (i.e., a
scintillating fibre tracker) behind the vertex detector inside a
magnetic field to identify the positron with high efficiency (in the
best scenarios $\sim 80\%$ would be the maximum achievable). A further
way to separate the charm background from signal is to use the
kinematic techniques of NOMAD. Assuming the NOMAD net efficiency
yields a $P_{\mu\tau}< 2\times 10^{-6}$. These considerations are
preliminary and a full study needs to be carried out to validate these
assumptions.

\clearpage
\section{Towards the Reference Design Report and R\&D requirements}
\label{Sect:FuturePlans}

The IDS-NF baseline Neutrino Factory (2010/2.0) has been presented in
this Interim Design Report.
An internationally coordinated R\&D programme designed to address the
key technological issues that underpin the Neutrino Factory has been
underway for a number of years.
The details of this programme will not be repeated here.
However, the studies that have been carried out to prepare the
material presented in this report have led to the identification of a
number of issues that must be addressed through an R\&D programme.
In addition, the conceptual-design and engineering tasks that must be
carried out before the Reference Design Report (RDR) can be prepared
have also been identified.
The paragraphs which follow summarise the R\&D programme that is
required to reduce the technical risks that the Neutrino Factory
project presents and the steps that the IDS-NF collaboration plans to
take to deliver the RDR.

In parallel to the programme outlined below, work will continue on the
evaluation of alternatives to some of the systems that are presently
included in the baseline specification.
Examples of such considerations include the possibility that an FFAG
may be a cost-effective replacement for one or more of the RLAs.
The IDS-NF has established a ``change control'' process
\cite{IDS-NF-002} by which a baseline change will be considered if the
proposed alternative can be demonstrated to have substantially reduced
technical risk or give a performance, or a cost, advantage.
The work required to develop such alternatives to systems that are
presently included in the baseline is not presented in the paragraphs
that follow.

\subsection{Accelerator systems}
\subsubsection{R \& D tasks}

In this section, an outline of the the R\&D programme required to
reduce the technical risks that the Neutrino Factory project presents
is given.
Since detailed descriptions of the R\&D programme are available
elsewhere, the R\&D programme will be presented as a concise list of
the tasks that must be accomplished.
In the paragraphs which follow, important design and R\&D topics are
identified, a number of which are printed in boldfaced type to
indicate that their completion is essential to the production of the
RDR. 

\paragraph{Proton driver}

The IDS-NF will not have a specific proton driver in its baseline
design.  
Instead, multiple laboratories will describe how they could
construct a proton driver that meets our specifications. 
R\&D that will be important to the Neutrino Factory project will have
two goals: to demonstrate that the facility in question can meet our
requirements, and to provide an estimate of the cost, over and above
existing or planned facilities, of constructing an proton driver for a
Neutrino Factory.   
Particular challenges of a Neutrino Factory proton driver are the high
power and consequently high currents required, the very short bunches
that must be delivered, and injecting such an intense beam into the
various rings that are needed.

\paragraph{Target}

The principal R\&D tasks that must be carried out to complete the
specification of the Neutrino Factory target are:
\begin{itemize}
  \item \textbf{Re-design of the solenoid capture system and its 
                shielding:}\\
    Two areas of concern necessitate this:
    \begin{itemize}
    \item {\it Cryogenic thermal loads:}\\
      The heat load on the superconducting
      magnets is too high in the current design.  Specifications
      must be defined for the heat load per proton-driver pulse
      (temperature rise on each pulse), the local maximum,
      time-averaged heat load (required local cryogen flow rate),
      and the total time-averaged heat load (thermal capacity
      of cryogenic system).  The shielding system must be re-designed to
      meet these requirements;
    \item {\it Mechanical forces:}\\  
      The forces on the solenoids are extremely
      high.  Specifications for acceptable forces need to be defined,
      and the shielding re-design must be constrained to meet these
      specifications;
    \end{itemize}
  \item Coolant flow in the internal shield:\\  
    A system needs to be designed and simulated to ensure adequate
    coolant flow everywhere. 
    It would be prudent to test the coolant flow patterns in a full
    size mock-up, for which inexpensive low-\textit{Z} beads will
    suffice;
  \item \textbf{Definition of the full infrastructure for a target
    station}. \\
    This must include the outer shielding and containment, the remote
    handling systems, and the mercury loop.  These need to be defined
    to sufficient detail that a cost estimate can be made and any
    significant technical issues identified;
  \item Nozzle design and tests: \\
    The performance of the 1~cm diameter nozzle for the mercury jet in
    the MERIT experiment was poorer than desired at jet velocities of
    15--20\,m/s.  
    A program of simulation and design is under way with the goal of
    developing a better nozzle.  
    This issue should not, however, be left only to design, but should
    be addressed in laboratory tests once a revised design is
    developed, on the time scale of 2~years;
  \item \textbf{The beam dump:}  \\
    A complete design of the beam dump must be
    produced.  This must allow for the possibility of a failure
    of the mercury flow, in which case the entire proton beam would
    be incident on the dump.  Since the beam dump would contain mercury,
    the hydrodynamics of the jet and the beam hitting this pool
    must be understood, and systems to mitigate the effects of the
    splash must be designed and simulated.  It would also be useful
    to conduct a laboratory test of the system in the presence of
    the mercury jet;
  \item Pion-yield calculations and measurements: \\
    Different particle production simulation codes have produced
    results that, while in broad agreement, sometimes disagree in
    detail for predicting quantities that are important to the
    design of a Neutrino Factory~\cite{Strait:2009}.
    Studies must continue to compare
    the results from different codes, compare their results with
    experimental data, and assess the uncertainty in our predictions.
    Collecting additional experimental data to
    improve the models in the codes, as proposed in \cite{Sadler:2010}
    would also be desirable; and
  \item Beam window designs:\\
    Complete designs of the windows must be performed.
\end{itemize}

\paragraph{Muon front-end}

The R\&D tasks required to complete the specification of the muon
front-end are:
\begin{itemize}
  \item RF cavities in magnetic fields: \\
    Our understanding of the limitations on RF gradients caused by
    magnet fields must be improved through experiments such as
    MuCool.  
    As the limitations are better understood, lattice designs will
    need to be revised accordingly, either by modifying the baseline
    design or switching to one of the alternative designs;
  \item \textbf{Particle losses:}  \\
    A system for managing particle losses needs to be defined.
    Three items have been identified for investigation: a proton
    absorber to remove low momentum protons; a chicane to remove
    high-momentum protons; and transverse collimators to remove
    high-amplitude particles. 
    A Wien filter has also been suggested as a method to filter
    off-momentum particles.  
    Energy deposition in various systems, in particular
    superconducting magnets, must be computed. 
    Appropriate shielding will need to be designed;
  \item \textbf{A full engineering design:} \\
    Designs for superconducting magnets,
    RF cavities, and their associated services are required.
    Heat deposition in lithium hydride absorbers need to be studied,
    and an active cooling system needs to be designed if it is
    needed;
  \item Further optimisation of lattice optics:\\  
    In particular, between
    the phase rotation and ionisation cooling sections, there is a
    longitudinal mismatch, and the transverse matching can also
    be improved; and
  \item Ionisation cooling experiments:\\  
    In particular, the results from MICE are critical.
\end{itemize}

\paragraph{Linac and RLAs}

To complete the specification of the linac and RLA systems, the
following tasks must be completed:
\begin{itemize}
\item \textbf{Full lattice design:}\\
  While the lattice design is nearly complete, there are a few sections
  (such as matching sections in the arcs and the second injection
  chicane) which still need to be designed;
\item \textbf{Magnet designs:}\\
  First-pass designs of all magnets need to be made;
  \item \textbf{Physical layout:}\\
    In particular this is necessary
    around injection chicanes, separators, and arc crossings, to
    ensure that the theoretical lattice design can be constructed; and
  \item \textbf{Full particle tracking:}\\  
    Particles need to be tracked through the entire system.  
    This needs to include realistic magnet designs. 
    Matching sections designs need to be adjusted to reduce losses.
\end{itemize}

\paragraph{FFAG}

The principal R\&D tasks required to complete the specification of the
muon FFAG are:
\begin{itemize}
  \item \textbf{Determine the optimal amount of, and method for, the
    chromaticity correction;}
  \item \textbf{Compute the matched beam in longitudinal phase space:}\\
    Computing the optimal phase-space shape for the longitudinal
    distribution is complex~\cite{Berg:2006ih}.  This needs to be
    done for this design, taking into account the asymmetric
    dependence of time-of-flight on energy and the dependence of
    time-of-flight on transverse amplitude;
  \item \textbf{Design matching to other systems:}\\
    In particular for longitudinal phase space.
  \item \textbf{6-D tracking through the system;}
  \item \textbf{Determine error tolerances;}
  \item \textbf{Perform approximate cost comparison of FFAG design for an RLA
      replacement;}
  \item \textbf{Produce single-layer combined-function designs for main ring
      magnets;} and
  \item \textbf{Hardware studies of kicker magnets and their power
      supplies.} 
\end{itemize}

\paragraph{Decay ring}

To complete the specification of the decay ring, the following issues
must be addressed:
\begin{itemize}
  \item \textbf{Design injection system;}
  \item \textbf{Decide on inclusion of sextupoles;}
  \item \textbf{Polarimeter design;}
  \item Design of OTR system for angular divergence measurement;
  \item \textbf{Decide on optimal system for measuring or computing
    neutrino flux spectrum going to far the detectors;} and
  \item Compute particle spectrum based on storage ring tracking: \\
    An accurate neutrino spectrum is needed to simulate the far
    detectors' performance.
    The spectrum and trajectories of lost muons, electrons, and
    synchrotron radiation photons are needed
    to compute heat loads in the storage
    ring and undesired particles in the near detector.
\end{itemize}

\subsubsection{Plan to produce the RDR}

\begin{figure}
  \includegraphics[width=\linewidth]%
    {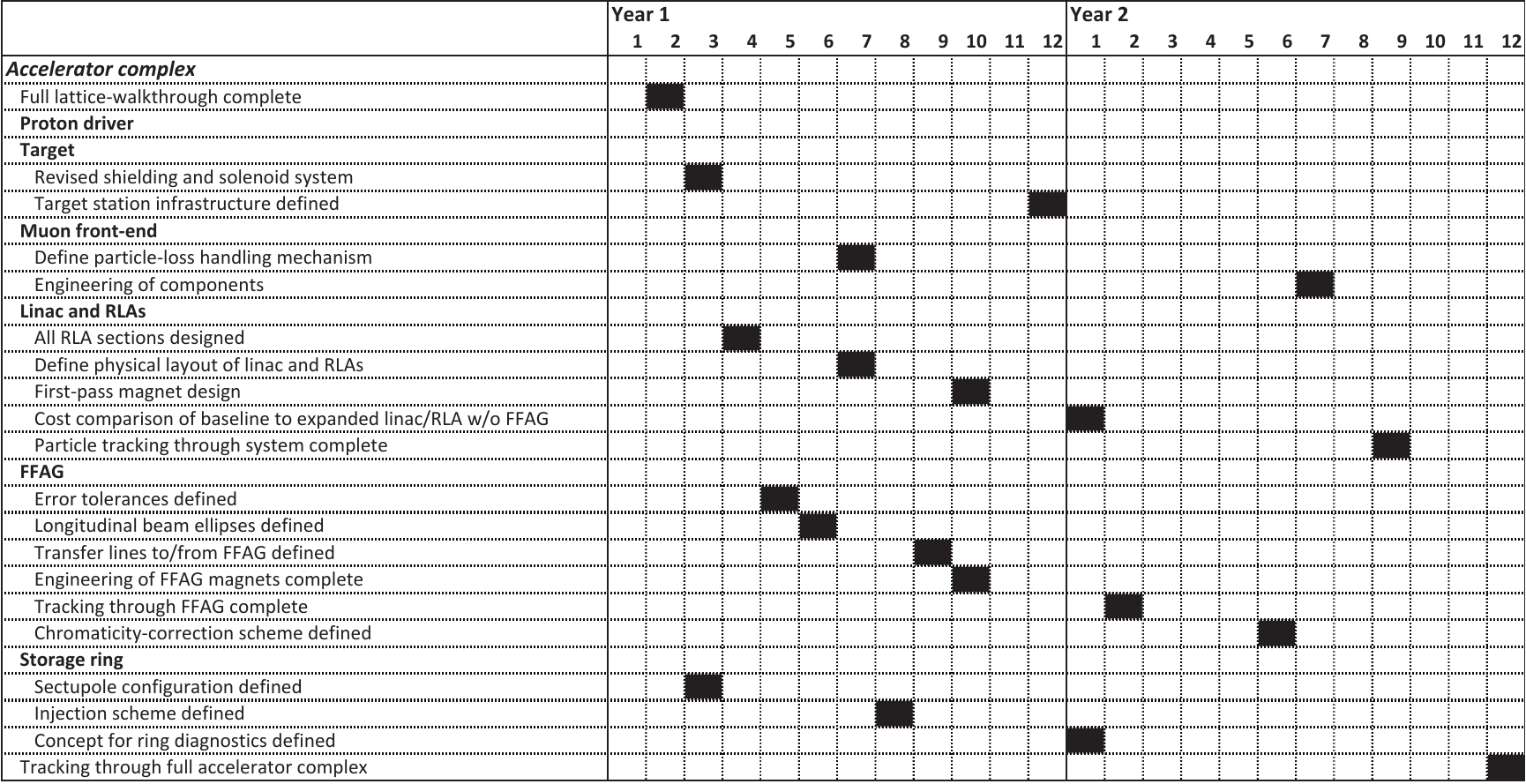}
  \caption{
    Approximate time-line for completing important tasks needed
    for the RDR.
  }
  \label{fig:acc:rdr}
\end{figure}

Our primary goals for the accelerator portion of the RDR are to have:
\begin{itemize}
\item A complete design for a Neutrino Factory accelerator facility that
  we believe is technologically feasible and would have the required
  performance;
\item An estimate of the cost of such a facility.  This will require
  designs of the system components at a sufficiently detailed level to
  produced such a cost estimate; and
\item A complete end-to-end tracking result for a realistic particle
  distribution through the system to verify that the system works as
  expected. 
\end{itemize}
An approximate time-line for completing these tasks is shown in figure
\ref{fig:acc:rdr}.   
These tasks give the required input for attaining the primary goals
for the RDR.  
The time-line does not include common tasks such as supplying
component designs and making cost estimates, except when they are
emphasised because, for instance,
component designs are more ambiguous than usual.

\subsection{Detector Systems}

\subsubsection{Simulation and Analysis Work towards RDR}
\label{sec:detector_RDR}
While there has been considerable progress in the definition of a
baseline detector configuration, including an improved analysis of
MIND and defining the specifications of the near detector, there is
still a significant amount of work that needs to be performed before
the performance of the near and far detectors at the Neutrino Factory
can be finalised for the Reference Design Report (RDR).
 
The efficiency of identification of $\nu_\mu$ and $\overline{\nu}_\mu$
CC interactions, along with the associated backgrounds, has been
studied for MIND. 
All efficiencies and backgrounds have been calculated using full
pattern recognition and reconstruction of neutrino events, that now
include QEL and RES events, in addition to DIS events. 
As a consequence, the efficiencies obtained demonstrate an extension
of the performance of MIND at low energies that allows the oscillation
maximum to be within the efficiency plateau, while maintaining backgrounds
at the $10^{-4}$ level. 
The systematic error on the signal efficiencies obtained is about
$\sim 1\%$ over the whole neutrino-energy range. 
However, there are still a number of simplifications that have been
carried out in the MIND simulations that need to be addressed before
we can quote a final sensitivity for the RDR. 

Similarly, the requirements of the near detector have been identified,
but we have yet to perform a full near detector simulation and
analysis to determine the realistic prospects to measure the flux over
the full energy range, to measure cross-sections, to measure the
divergence of the neutrino beam, to measure charm production and to
search for NSI from tau events.  
We will itemise the next steps needed to complete these and other
studies in time for the RDR in the following sections.

\paragraph{Steps towards RDR for MIND analysis}  
\subparagraph*{Multi-variate likelihood analysis}

\begin{figure}
  \begin{center}$
    \begin{array}{cc}
      \includegraphics[width=8.5cm, height=6.5cm]{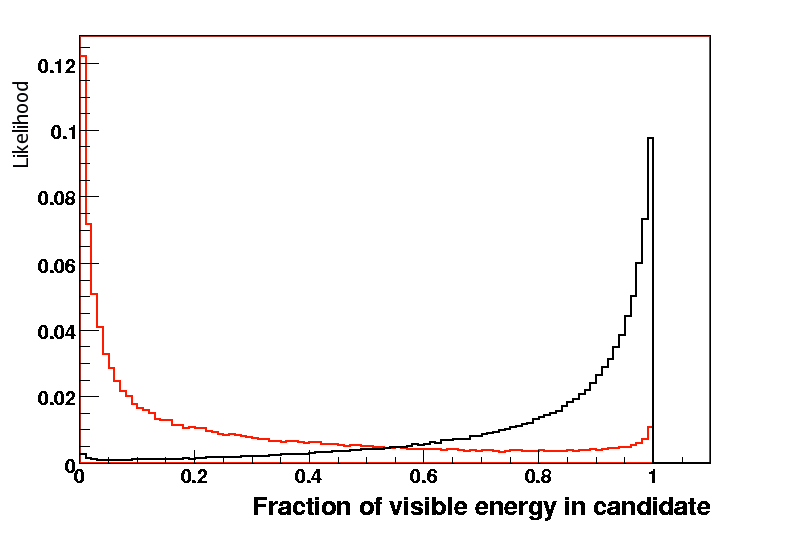} &
      \includegraphics[width=8.5cm, height=6.5cm]{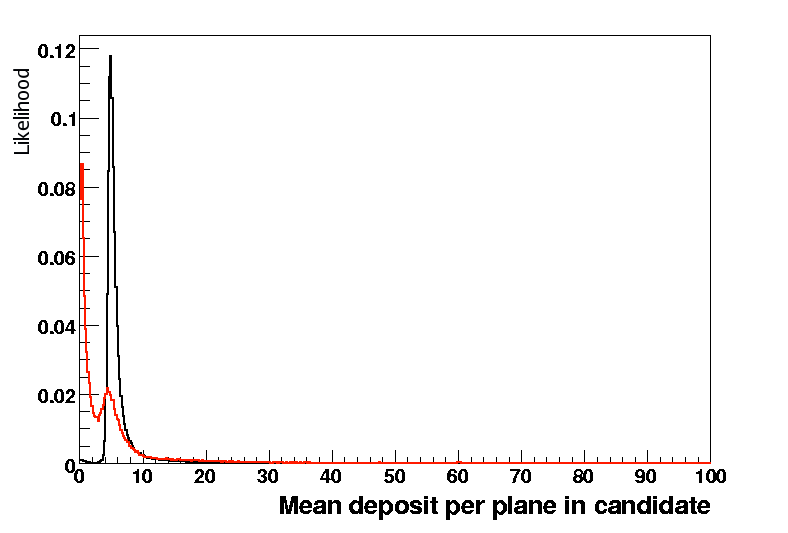}
    \end{array}$
  \end{center} 
  \caption{Additional PDFs for the NC rejection parameters. (left) fraction of visible energy in candidate and (right) candidate mean deposit per plane. CC in black and NC in red.}
  \label{fig:NCpdfs_2}
\end{figure}

A number of additional steps will be required to fully benchmark the
performance of MIND. 
This section presents some possible improvements to the analysis which
have not yet been fully exploited. 
As mentioned in section~\ref{subpar:NCrej}, two energy-deposit based
parameters used by MINOS were studied but not included in the analysis
presented here. 
The extra parameters include the total energy in the candidate and the
mean deposit per plane of the candidate. 
The distributions of these variables for a test set of neutral-current
and charged-current events were used to form PDFs, named $l_{frac}$
and $l_{mean}$ respectively, shown in figure \ref{fig:NCpdfs_2}. 
Samples are taken from the NC and CC PDFs to form the four
log-likelihood rejection parameters:
\begin{eqnarray}
  \label{eq:lik21}
  \mathcal{L}_1 &= &\log \left( \frac{l_{hit}^{CC}}{l_{hit}^{NC}} \right)\\
  \label{eq:lik22}
  \mathcal{L}_2 &= &\log \left( \frac{l_{hit}^{CC}\times l_{frac}^{CC}}{l_{hit}^{NC}\times l_{frac}^{NC}} \right)\\
  \label{eq:lik23}
  \mathcal{L}_3 &= &\log \left( \frac{l_{hit}^{CC}\times l_{mean}^{CC}}{l_{hit}^{NC}\times l_{mean}^{NC}} \right)\\
    \label{eq:lik24}
    \mathcal{L}_4 &= &\log \left( \frac{l_{hit}^{CC}\times l_{frac}^{CC}\times l_{mean}^{CC}}{l_{hit}^{NC}\times l_{frac}^{NC}\times l_{mean}^{NC}} \right)
\end{eqnarray}

\begin{figure}
  \begin{center}$
    \begin{array}{cc}
      \includegraphics[width=7.5cm, height=5.5cm]{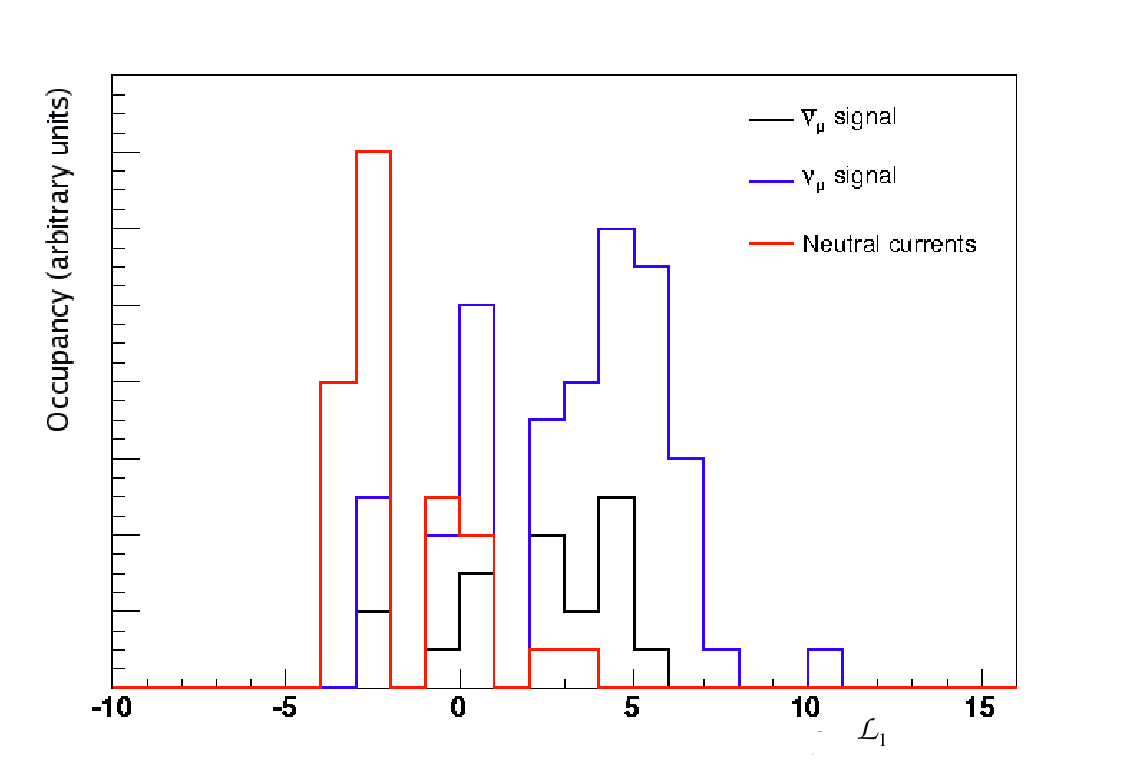} &
      \includegraphics[width=7.5cm, height=5.5cm]{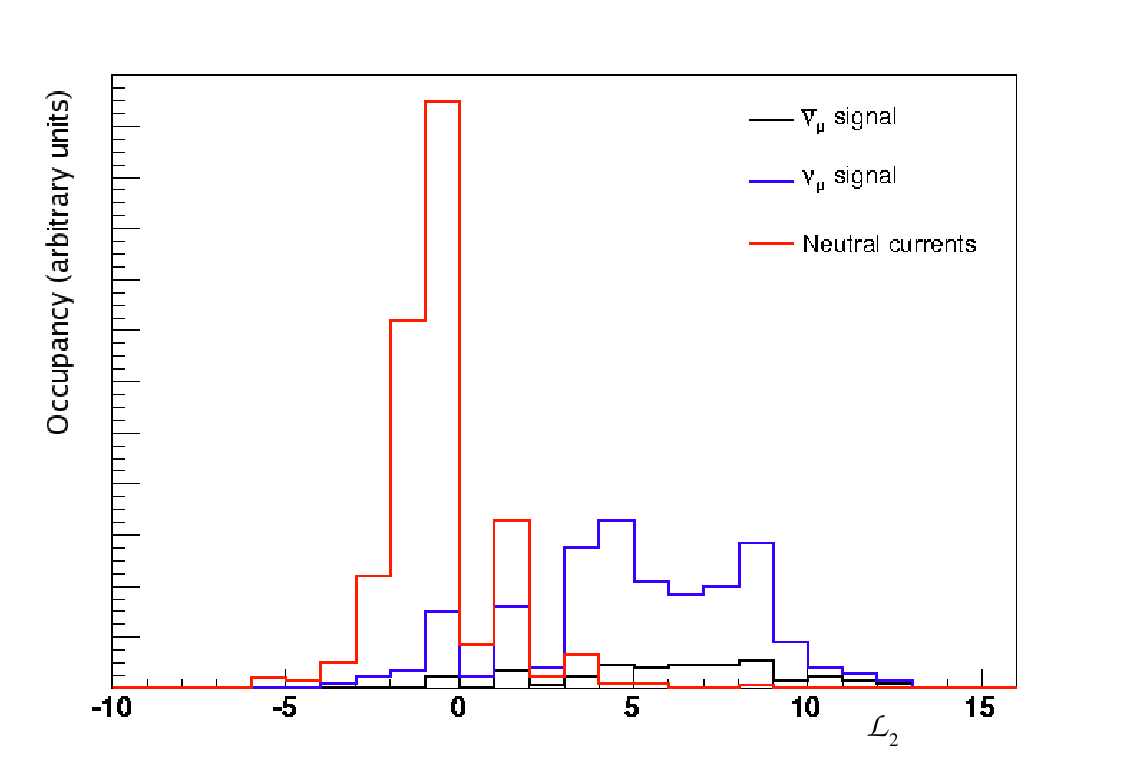} \\
      \includegraphics[width=7.5cm, height=5.5cm]{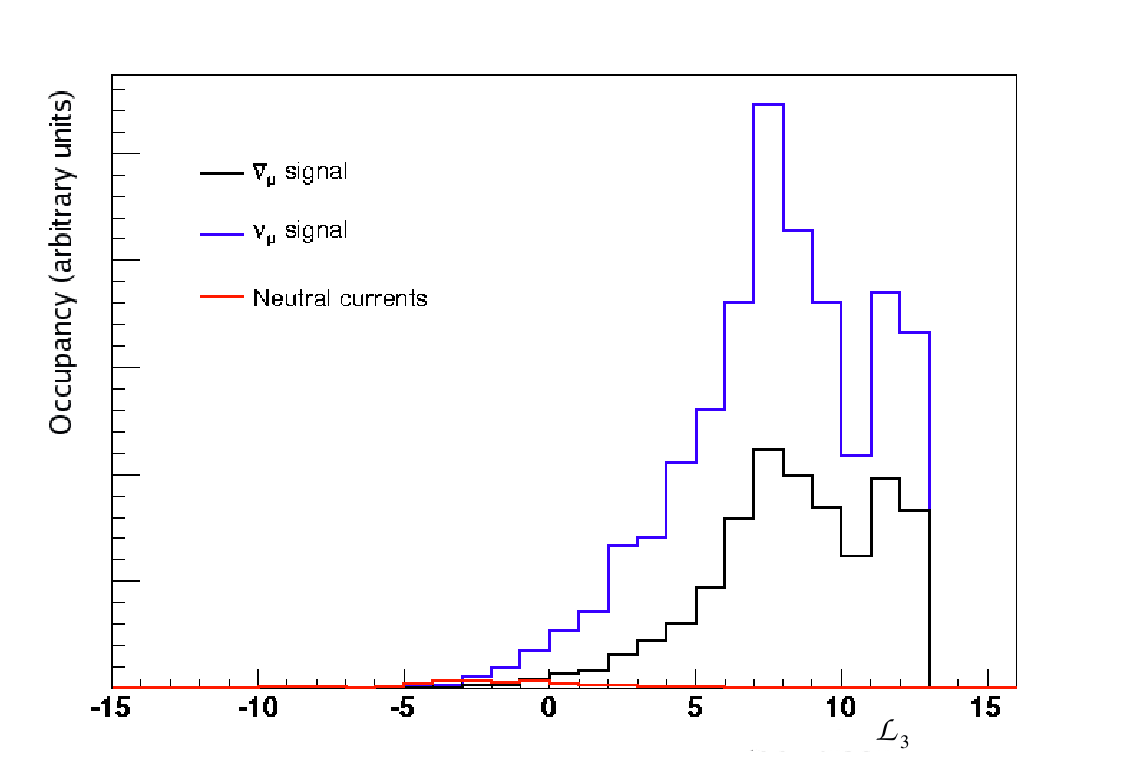} &
      \includegraphics[width=7.5cm, height=5.5cm]{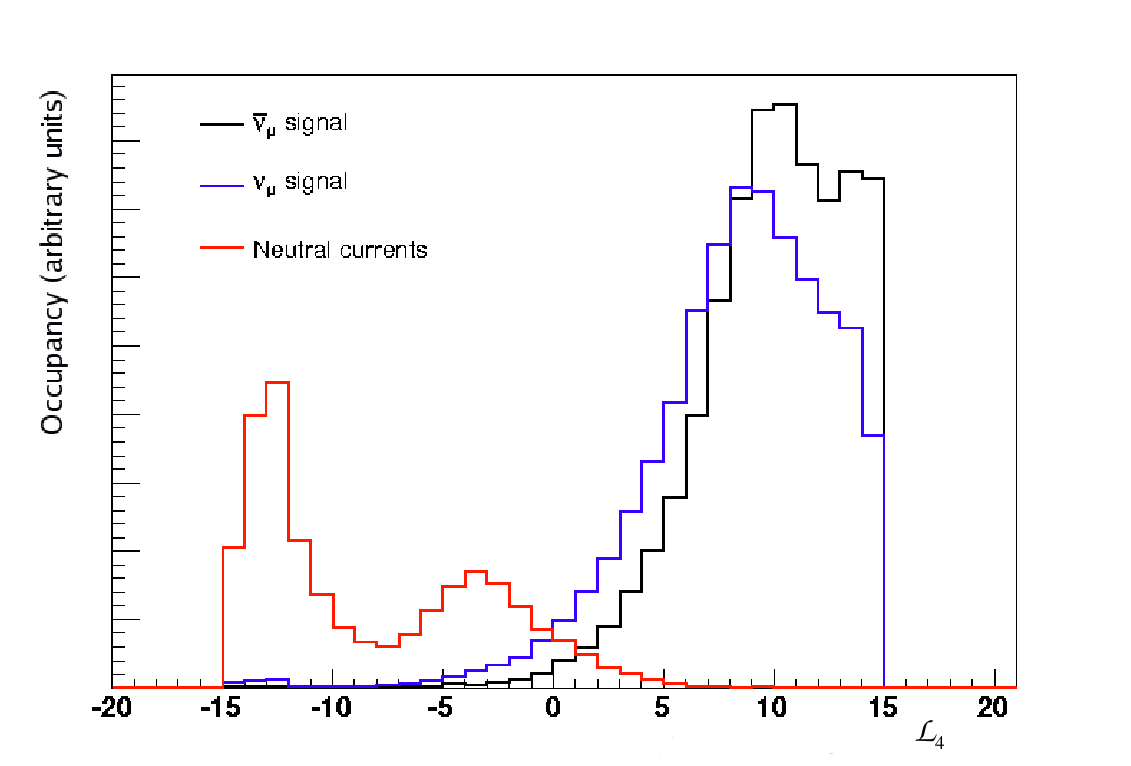}
    \end{array}$
  \end{center} 
  \caption{Log likelihood distributions for the four possible scenarios. (top left) $\mathcal{L}_1$, (top right) $\mathcal{L}_2$, (bottom left) $\mathcal{L}_3$ and (bottom right) $\mathcal{L}_4$}
  \label{fig:an2logs}
\end{figure}

An analysis using the likelihood functions described in equations~\ref{eq:lik21}~--~\ref{eq:lik24} was implemented but no improvement in performance was obtained with respect to using $\mathcal{L}_1$. Distributions of these functions for a test statistic are shown in figure \ref{fig:an2logs}. There is significant correlation between these parameters. Development of a multivariate analysis or neural net based on these variables, which take their correlations into account, will be studied for future analyses.

\subparagraph*{Muon momentum measurement by range}

Measurement of the muon momentum using the range of the candidate in
the detector was shown to improve resolution, particularly at low
momenta, by MINOS~\cite{Adamson:2007gu}. 
An independent measure of the momentum could aid in the MIND
analysis. 
The cuts currently implemented via equations~\ref{eq:dispMom}
and~\ref{eq:dispMom2} are highly correlated with the range of the muon
and their power to remove background is likely to be improved with the
inclusion of the muon range measurement. 

\subparagraph*{Hadronic reconstruction}

The studies presented use parameterisations based on results presented
by Monolith and MINOS to estimate the quality of reconstruction of the
hadronic energy and direction.  
These parameters are important both for the reconstruction of neutrino
energy and for the formation of the kinematic cuts. 
In the ultimate analysis, reconstruction will be performed on an
event-by-event basis using deposited charge and distribution
information from that part of the event not associated to the
candidate muon. 

A dedicated study will be required to understand fully the best method
to perform the hadronic reconstruction. 
Using the observable quantities from the event, we could form a seed
for a jet fitting algorithm which could reconstruct both the energy
and momentum of the hadronic shower. 
Using the known structure of neutrino interactions, we can also use
the combination of the muon momentum and hadronic energy to calibrate
the total energy of the neutrino interaction.  

\subparagraph*{Cosmic backgrounds}

The background from cosmic rays and cosmic neutrinos has not been
studied yet. 
This will be important to determine the final detector design and its
location, either in an underground laboratory or in a green field site
on the surface with minimal overburden. 
While understanding the timing of the accelerator system and
directional arguments will enable the rejection of many of these
events it is clear that some level of rock overburden or active
vetoing will be required, especially when considering the surface area
of a 50--100\,kTon MIND. 
The expected Neutrino Factory duty cycle---the proportional time
window when beam related interactions are possible---of $\sim$10\%
will need to be taken into account to determine the level of
overburden that will be required.  

\subparagraph*{Tau background}

Another important study that is required is the determination of the
contribution to the reconstructed spectrum from $\nu_e \rightarrow
\nu_\tau~(\overline{\nu}_e \rightarrow \overline{\nu}_\tau)$
interactions. 
This oscillation channel would be expected to produce a similar
absolute flux of $\nu_\tau~(\overline{\nu}_\tau)$ to that produced by
the golden channel $\nu_\mu~(\overline{\nu}_\mu)$ oscillation
\cite{Indumathi:2009hg}. 
While the high interaction threshold for these species means that
fewer interactions will take place in the detector, tau interactions
become significant above $\sim$8\,GeV. 
Since the $\tau$ produced in CC interactions will decay promptly and
has a branching fraction for $\mu$ containing channels of (17.36 $\pm$
0.05)\%~\cite{Amsler20081} it is possible that a significant fraction
of these events may survive the analysis. 
However, understanding this background, if treated with care, can
contribute information to the fit instead of detracting from
it~\cite{Donini:2010xk}. 
The distortion caused is highly sensitive to the oscillation
parameters so fitting for this distortion could help in the removal of
ambiguities in the determination of $\theta_{13}$ and $\delta_{CP}$. 
To include this study in the full simulation of MIND, we will need to
migrate from NUANCE to GENIE, as NUANCE does not contain tau
production and decays. 
Migration of the full simulation to GENIE has already commenced.

\subparagraph*{Realistic MIND Geometry with Toroidal Magnetic Field Map}

The MIND simulation described in section~\ref{sec:MIND} includes an
idealised geometry with square iron plates 14\,m$\times$14\,m and
3\,cm thick, with two layers of scintillator, each of 1~cm thickness. 
The magnetic field assumed was a simple uniform dipole field of 1~T. 
However, a design that can be realised in practice, with achievable
engineering constraints consisting of octagonal plates of
14\,m$\times$14\,m and with a toroidal field map between 1\,T and
2.2\,T, was presented in \ref{sec:MIND_conceptual_design}. 
While we do not expect that the change in geometry will affect the
MIND performance significantly, it is necessary to perform a new
simulation with these new parameters. 
These changes will be carried out for the RDR.

\paragraph{Steps towards RDR for Near Detector}

We have two options for the near detector baseline: one with a
scintillating-fibre tracker and the other with a straw-tube tracker. 
These detectors need to be immersed in a magnetic field to be able to
take full advantage of the detector resolution to measure the momentum
of the particles produced in the neutrino interaction. 
Either option would have to be followed by a muon spectrometer, which
could be a smaller version of MIND. 
We need to benchmark both of these options and determine which of the
detectors provides  better performance. 
Additionally, upstream of the near detector, we would place a high
resolution charm and tau detector. 
We need to establish whether an emulsion detector is feasible at a
Neutrino Factory or whether the only way we can measure charm and tau
events is using a silicon vertex detector.  

We also need to determine for the RDR whether we need four near
detectors or whether two detectors will suffice (for example, this
could be achieved if the near detector were on rails, and it could be
moved to either of the straight sections of the storage ring,
depending on the charge of the muon). 
Moreover, we need to determine the optimum distance from the near
detector to the end of the decay straight and what shielding is
required. 
All of these considerations need to be resolved through detailed
simulations before the publication of the RDR. 
Full response matrices, similar to those that have been prepared for
the MIND (see Appendix \ref{app:response}), for signal and background
at the final near detector configuration, will also need to be
determined so that the physics performance may be quantified. 

To achieve these goals, the following simulations still need to be carried out:
\begin{itemize}
\item Neutrino electron scattering: Simulations for Inverse Muon Decay
  (IMD) have already been produced as a way of calibrating the
  neutrino flux (section~\ref{sec:ND_performance}). 
  However, IMD has a threshold of 11~GeV, and we need a method of
  determining the flux at lower energies as well. 
  Neutrino-electron $\nu_e e^- \rightarrow \nu_e e^-$ and
  $\overline{\nu}_e e^- \rightarrow \overline{\nu}_e e^-$ elastic
  scattering can provide such a flux measurement over the whole
  neutrino energy range and can be used to cross-check the IMD
  results;
\item Flux extrapolation from near to far detector: Preliminary
  results on the flux extrapolation method have been presented in
  section~\ref{sec:ND_performance}. 
  A method to deduce and take into account the systematic errors from
  the near detector data storage must be developed;
\item Tau and charm analysis: A full
  simulation with a silicon vertex detector and an emulsion detector
  needs to be carried out to determine the tau and charm detection
  efficiency and relevant backgrounds of the tau/charm detector at a
  neutrino factory;
\item Near detector shielding: This is an important topic that needs
  to be addressed at the interface between the storage ring and the
  near detector. 
  From the accelerator end, we need to understand potential muon
  losses in the storage ring and photon contamination from radiative
  decays. 
  From the near detector point of view, we need to ascertain the
  impact of these backgrounds and what type of shielding can be
  tolerated. 
  The flux of neutrino induced muons from the shielding itself will
  also need to be calculated; and
\item Performance of near detector for neutrino scattering physics
  topics: Once the near detector facility has been defined, the
  performance of a near detector to determine neutrino cross-sections,
  parton distribution functions and other topics in neutrino
  scattering physics will be carried out.    
\end{itemize}

\subsubsection{Detector R\&D Plan}

\paragraph{Magnetisation}

\subparagraph{STL optimisation for MIND}

Since the STL implementation we envision for MIND is essentially
identical to that planned for the VLHC and what was actually
prototyped during the R\&D program for the VLHC, there is a relatively
small amount of work that needs to be done prior to developing the
engineering design document for MIND.  
The one area of investigation would be to use multiple superconductor
loops within the STL cryostat.  
The overall size of the STL would remain the same, the total amount of
the superconductor would remain the same, the forces would be
essentially the same, but now the external excitation current would be
reduced from 90-100kA to that number divided by the number of separate
circuits in the STL.  
Configurations with between 5 and 10 circuits will be investigated.  
This makes the external power supply and the room temperature current
leads much more straightforward.

\paragraph{Photo-detectors}

Given the enormous amount of R\&D currently underway in Europe, the
United States, Russia and  Japan on various Geiger-mode, multi-pixel
avalanche photo-diodes (and the rapid progress on performance and
cost), we do not plan to initiate any separate R\&D in this area.  
We will monitor the progress on these types of devices world wide and
stay abreast of advances by communicating with our
colleagues actively working in the field and by staying current with
the literature on the devices. 

\paragraph{Scintillator}

Extruded scintillator is an advanced and mature technology. 
There are a number of areas that we will devote R\&D to, however.
These are summarised below.

\subparagraph*{Specify final extrusion profile through simulation (square or triangle)}

Simulation studies will advise us on the relative merits of square
versus triangle cross-section extrusions.  
These studies will lead to a final choice regarding the scintillator
cross section and dimensions.  
Once the extrusion profile design is finalised, we will initiate discussions with extrusion die manufacturers to optimise the die design for:
\begin{itemize}
  \item Part uniformity;
  \item Production speed; and
  \item Tooling lifetime.
\end{itemize}

\subparagraph*{Engineer detector plane mechanics}

The iron plate design currently is quite mature.  
R\&D will be needed to integrate the detector planes with the steel.  
We will start with the concepts that were used by MINOS and then
extend them to address the larger cross section of the MIND and the
fact that the photo-detectors will be mounted directly on the
scintillator extrusions, thus eliminating the need for fibre
manifolds.  
This simplifies the detector plane tremendously and also allows for
the additional bolts needed for the iron plate support.  
The work will include 2D and 3D modelling with associated ANSYS
analyses. 

 \subparagraph*{Investigate the possibility of co-extruding the fibre with the scintillator.}

Possibly the most manpower-intensive step in the detector plane
fabrication is the insertion (and gluing) of the WLS fibre into the
scintillator extrusion.  
Preliminary investigations have been done at Fermilab to develop
extrusion die tooling that would allow commercial WLS fibre (from
Kuraray most likely) to be co-extruded with the scintillator (inserted
into the hot melt zone and pulled along with the polymer flow).  
Although this is a very tricky process, initial studies were
successful with post-cladding Kuraray WLS fibre with various thin (few
hundred micron) layers of polyethylene, Kapton and Kynar.  
Industrial experts in this field have been contacted and believe it is
possible to develop the process tooling to accomplish this.  
We will work with the Fermilab Scintillator Detector group and outside
vendors to develop process die tooling for this application. 

\subparagraph*{Alternate detector plane possibility}

Although we believe that solid scintillator is the optimum choice for
the detection planes, we will monitor the progress the INO
collaboration is making with RPC R\&D and production for their
detector.  
INO is a candidate far detector for a Neutrino Factory located in
Europe or Japan.

\paragraph{Fibre}

Although we do not feel that R\&D is needed to improve the performance
of the best available wavelength shifting (WLS) fibre that is 
currently available, this product only comes from a single-source
supplier, Kuraray.  
We will investigate with university groups versed in polymer science
and optical-fibre fabrication to see if an alternate technology base
can be developed that could be transferred to industry in order to
present experimenters with a choice of vendors.  
This could also have a positive impact on cost.

\paragraph{Prototyping}

Prototyping a scale version of MIND will allow the technical
feasibility of the MIND design, including the use of extruded
scintillator and SiPM readout options. 
There will be an added benefit that this prototype can be exposed to a
test beam to validate the performance of the detector under realistic
conditions. 

AIDA (Advanced European Infrastructures for Detectors at Accelerators)
\cite{AIDA} is a European project to develop infrastructures for
particle physics detector R\&D. 
As part of this project, a MIND prototype will be assembled and placed
at the end of the H8 beam-line by a consortium of European
institutions. 
The main purpose will be to benchmark and validate simulations to
determine the muon charge identification performance, but also to
measure hadron energy reconstruction and test different reconstruction
algorithms. 
After studying its performance the module will be available as a muon
spectrometer for future users of the beam. 
Also, as part of AIDA, a small Totally Active Scintillating Detector
(TASD) will be constructed and placed inside the Morpurgo magnet of
the H8 beam line. 
The device will be used to measure the electron charge identification
efficiency in a test beam with a realistic detector.  

\paragraph{Detailed Costing and time-line for the RDR}

\begin{figure}
  \includegraphics[width=\linewidth]{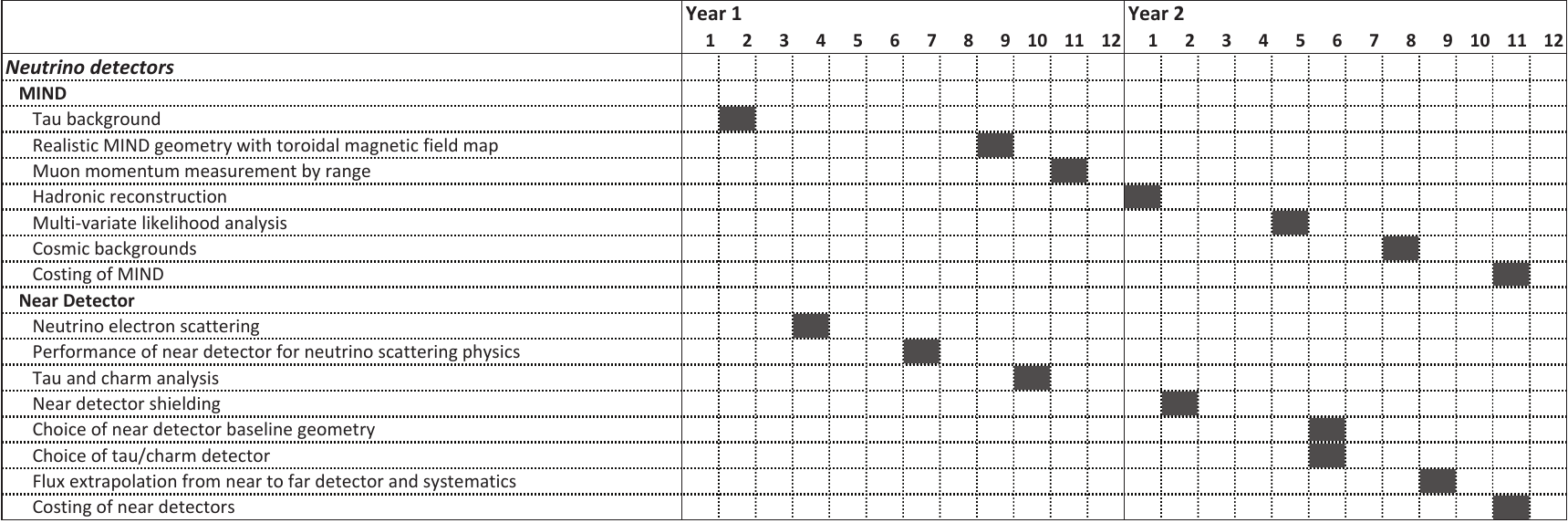}
  \caption{
    Approximate time-line for completing important near and far 
    detector tasks needed for the RDR.
  }
  \label{Fig:DetRDRPlan}
\end{figure}

For the RDR, we will provide a detailed costing for MIND.  
We will start with the WBS structure for MINOS and then develop our
own costing model based on the MIND components, fabrication techniques
and siting issues.  
At present, we do not see any big unknowns that would
result in large cost uncertainties.
An approximate time-line for the completion of the near and far detector
tasks that are required to deliver the RDR is shown in figure
\ref{Fig:DetRDRPlan}. 

\subsection{Estimation of the cost of the facility}
\label{Sect:Costing}

The cost of providing the accelerator complex and the neutrino
detectors must be evaluated and presented in the Reference Design
Report (RDR) to allow the value of the outstanding physics reach to be
evaluated.
The RDR will contain an estimate of the capital cost of the proposed
accelerator complex and neutrino detectors together with an evaluation
of the uncertainty on the costing.
The estimation will need to identify items such as expenses related
to the final design, installation and commissioning.  
Consideration needs to be given to items such as: 
\begin{itemize}
  \item The cost of a programme of R\&D that will address the
        remaining technical issues and mitigate the various risks.
        This programme will include the construction and test of
        prototypes; 
  \item The cost that will be incurred in the design and manufacture
        of specialised tooling;
  \item The construction of the facility, including materials,
        equipment, and labour.  In addition, the cost of installation
        and the field-supervision of the construction will be estimated;
  \item The cost of inspection, commissioning, and testing of
        components and systems; and
  \item The cost implications that arise in the development of designs
        that meet safety and radiation requirements.  
\end{itemize}
The estimation of the cost of the facility as complex of the Neutrino
Factory is challenging, requires a significant engineering effort,
and must be carefully organised to ensure consistency of approach
across the systems that make up the facility.
The costing methodology that has been adopted by the IDS-NF
collaboration is outlined in the paragraphs which follow.

The cost estimate to be presented in the RDR will be based on a
Project Breakdown Structure (PBS) for the accelerator complex and the
neutrino detectors.
The PBS is a hierarchical breakdown of the elements of the accelerator
complex and detector systems.
The top level of the PBS is shown in table \ref{Tab:PBS}.
The Neutrino Factory facility is broken down into the accelerator
complex and the neutrino detectors at level 2.
At level 3, the principal systems are identified.
The breakdown then continues until a level of detail is reached at
which a cost can be determined for the component or sub-system.
The costing tool developed at CERN \cite{Ref:CERNcostingTOOL} will be
used for the management of the PBS and the costing data.
The tool allows the costs relating to each component to be entered and
provides full functionality for indexation and reporting at various
levels and in various formats and currencies.  
\begin{table}
  \caption{
    Top-level Project Breakdown Structure for the Neutrino Factory
    that will be elaborated to prepare the cost estimate to be
    presented in the RDR.
  }
  \label{Tab:PBS}
  \begin{center}
    \begin{tabular}{|l|l|l|}
      \hline
      {\bf Level 1}    & {\bf Level 2}       & {\bf Level 3}                  \\
      \hline
      Neutrino Factory & Accelerator complex & Proton driver                  \\
                       &                     & Target                         \\
                       &                     & Muon front-end                 \\
                       &                     & Linac and RLAs                 \\
                       &                     & FFAG                           \\
                       &                     & Storage ring                   \\
                       & Neutrino Detectors  & Near detector                  \\
                       &                     & Intermediate baseline detector \\
                       &                     & Magic baseline detector        \\
      \hline
    \end{tabular}
  \end{center}
\end{table}

The tool allows data related to different options to be entered as
parallel sub-PBS structures, and provides an easy way to combine such
options in a project report. 
This functionality is particularly interesting as for some
sub-systems, for example the proton driver, a number of options will
be carried to the RDR.
The tool also has facilities to take various financial factors, for
example indexation, price escalation, and costs that relate to the
specific example site under consideration, to be taken into account.
Finally the tool provides ``version management'', archive
functionality, cost monitoring for each element or sub-system so that
the evolution of the cost can be followed over the lifetime of the
project.
Example sites (CERN, FNAL, and RAL) have been chosen to allow
site-specific aspects of various choices to be evaluated.

Preparing an estimate of the cost for a large project such as the
Neutrino Factory is a collective effort.
Each sub-system (PBS level 3) will have an individual responsible for
the collection of the cost estimates for all sub-nodes in the PBS and
for entering and maintaining the information in the costing tool.
A costing panel composed of the sub-system conveners and the
individuals responsible for coordinating the cost data, will be
convened to ensure that a coherent and uniform approach is maintained
across all the subsystems.
At the component or sub-system level a pragmatic approach will be
taken to determine the contribution to the overall cost. 
In some cases it may be possible to provide an analytical approach to
determine the cost of the element from manufacturers, while in other
cases costing formul\ae or scaling from previous experience will be
applied. 
In each case an ``assumption data sheet'' will be prepared in which
the list of components will be specified together with the technical
and economic assumptions that have been made to derive the cost.
The assumption data sheets will also contain a change record and be
stored in the costing tool.

The complexity of the Neutrino Factory facility is such that the
implementation of a number of the sub-systems carry known technical
risks. 
Examples of such sub-systems include the muon front-end, where the
reduction in the break-down potential of cavities in the presence of
magnetic field  may lead to the need to revise the design as discussed
in Appendix \ref{sec:acc:fe:FrontEnd_Alternatives}.
Such cases will be dealt with by developing a ``risk register'' in
which such risks are identified and the cost of mitigation is
presented.
A ``risk score'' will be assigned to each element of the risk table. 
The risk score is defined as the product of the probability that the
risk will occur and the impact on the Neutrino Factory project should
the un-mitigated risk occur.
The risk score will be used to inform the definition of the R\&D
programme that must be carried out as the first phase of the Neutrino
Factory project.  
The cost of the R\&D programme will also be evaluated and presented as
part of the total cost of the Neutrino Factory project.

The Neutrino Factory project will be carried out by a large,
international collaboration funded by a variety of stakeholders.
The cost of the facility that will be presented in the RDR must
therefore be accessible to all the different stakeholders.
While it is recognised that the costing presented in the RDR must
address the issues of the different costing models that have been
adopted by the various stakeholders in a clear and concise manner, the
details of the paradigm that will be adopted have yet to be defined.

\clearpage
%
\section*{Acknowledgements}

During the course of the IDS-NF to date, we have been welcomed at a
number of laboratories across the world and therefore thank the
CERN, FNAL, and RAL laboratories and the Tata Institute
of Fundamental Research for hosting the IDS-NF plenary meetings.
The authors acknowledge the support of the European Community under
the European Commission Framework Programme 7 Design Study: EUROnu,
Project Number 212372. 
The authors acknowledge the support of Grants-in-Aid for Scientific
Research, Japan Society for the Promotion of Science and the World
Premier International Research Center Initiative (WPI Initiative),
MEXT, Japan.
The work was supported by the Science and Technology Facilities
Council under grant numbers PP/E003192/1, ST/H001735/1, ST/H003142/1
and through SLAs with STFC supported laboratories.
This research was partially supported by the Director, Office of
Science, Office of High Energy Physics, of the U.S. Department of
Energy, under contract numbers DE-AC02-07CH11359, DE-AC02-05CH11231, 
DE-AC02-98CH10886, DE-AC05-06OR23177, and DE-AC05-00OR22725.

%
\cleardoublepage
\section*{Appendices}
\appendix
\section{FNAL Proton Driver}
\label{sec:acc:fnal}

\begin{figure}
  \includegraphics[angle=0,width=0.95\textwidth]%
    {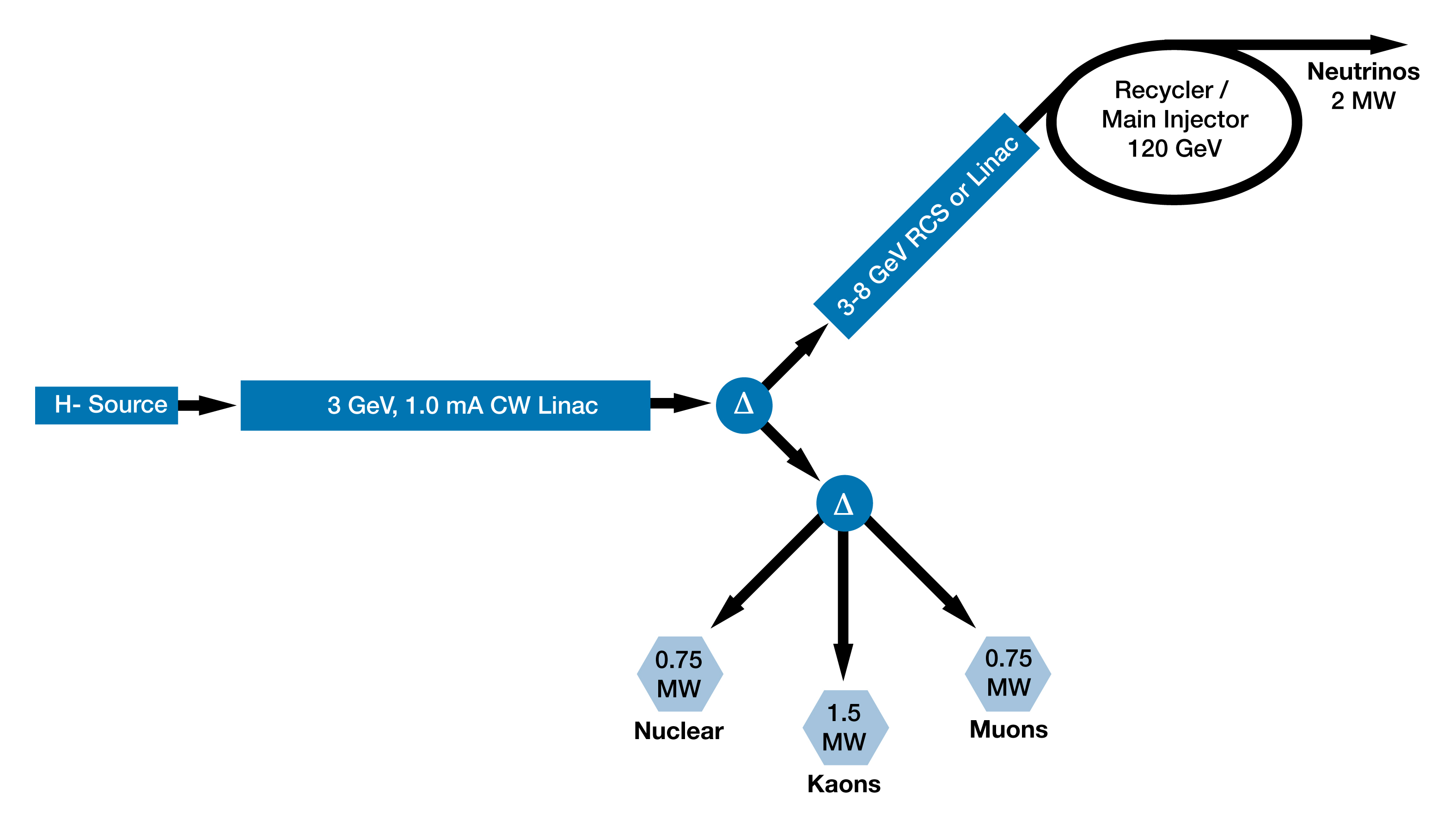}
  \caption{Schematic of FNAL Project~X. 3~to 8~GeV acceleration
    will be done using a pulsed linac.
  }
  \label{fig:acc:fnal:schem}
\end{figure}

FNAL proposes to build Project~X~\cite{Holmes:2010} as a new high
intensity proton source.
Project~X has three physics goals: 
\begin{enumerate}
\item To provide neutrino beams for long baseline neutrino oscillation
  experiments;
\item To provide intense kaon and muon beams for precision
  experiments; and 
\item To develop a path toward a muon source for a possible
  Neutrino Factory 
\end{enumerate}
The third goal provides the beam needed for a Neutrino Factory but additional accelerator rings 
will be needed to provide the correct time structure for the beam. 
A schematic layout of the Project~X reference design is shown in figure \ref{fig:acc:fnal:schem}.
 
Project~X will accelerate H$^{-}$ ions in two linacs. 
The first linac will be CW and accelerate beam to $\approx$3~GeV. 
The beam will then be directed either to a switchyard area to perform
precision experiments or to the second linac. 
A pulsed linac will accelerate the beam from the CW linac to $\approx$8~GeV, the injection energy of the Recycler Ring. 
After converting the H$^{-}$ beam, protons are accumulated in the Recycler Ring before transfer to 
the Main Injector where the beam is accelerated for the long baseline neutrino program. 

The beam originates from a 1---10~mA DC H$^{-}$ source.  
The beam is bunched and accelerated by a CW normal-conducting RFQ to 2.5~MeV. 
The RFQ is followed by a Medium Energy Beam Transport (MEBT) section, which includes a chopper
following a pre-programmed time-line formatting the bunch pattern. 
Since the linac average beam current is 1~mA and the beam current at
the ion source can be as high as 10~mA, up to 90\% of the beam has to
be removed by a chopper in the MEBT section.
The bunch spacing will be 3.1\,ns with a maximum intensity of
$1.9\times10^{8}$ protons per bunch exiting the front end. 
However, the CW linac will have an average beam current of 1\,mA
(averaged over 1\,$\mu$s). 

\begin{figure}
  \includegraphics[width=\textwidth]{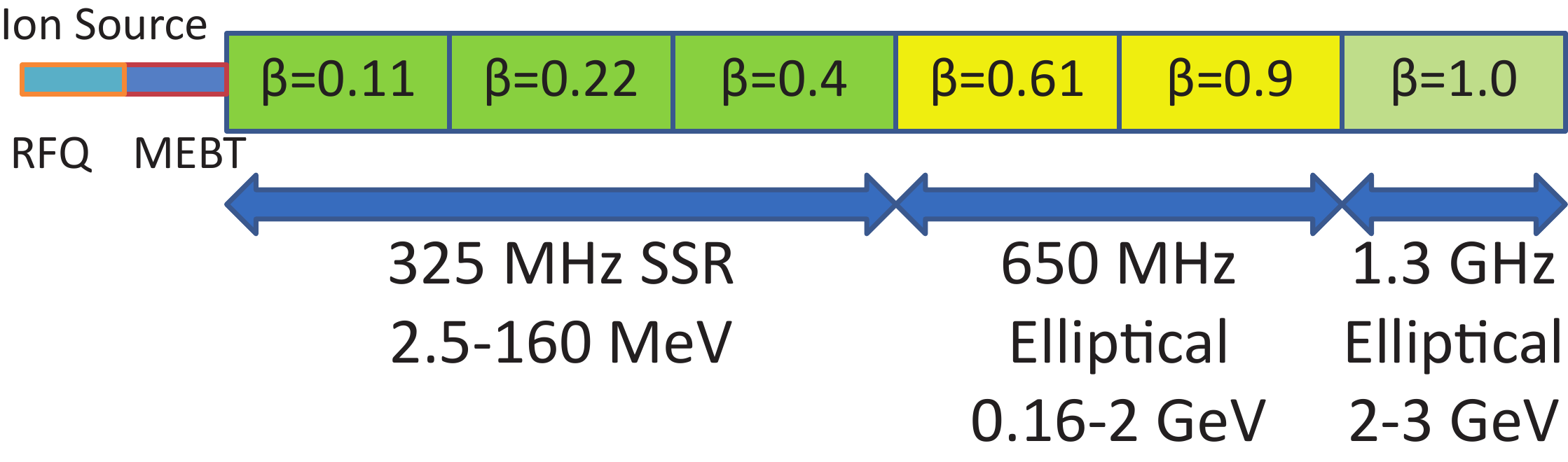}
  \caption{Front-end and CW linac where the energy range and
    geometrical phase velocity are shown for each section.}
  \label{fig:acc:fnal:linac}
\end{figure}

The CW linac will use several types of superconducting RF cavities. 
Three types of single-spoke cavities operating at 325~MHz will be used to 
accelerate beam from 2.5~MeV to 160~MeV. 
To increase the energy to 2~GeV, two types of 650~MHz elliptical
cavities will be used. 
1300~MHz elliptical cavities complete the CW linac with a final energy
of 3~GeV.  
Figure \ref{fig:acc:fnal:linac} shows a schematic of the front end and
CW linac.
Table \ref{tab:acc:fnal:linac} contains the CW linac cavity parameters. 
Cavities and focusing elements are grouped within cryomodules. 
In the 325~MHz section of the linac focusing is provided by solenoids.  
In the 650~MHz section a standard focusing/defocusing (FD)
quadrupole doublet lattice is used, followed by a FODO lattice in the
1300~MHz section.  
All magnets are superconducting with built-in dipole correctors for
beam steering.  
\begin{table}
  \caption{
    Accelerating cavity specifications for the CW linac. $\beta_{G}$ is
    cavity geometrical phase velocity.
    The type of accelerating structure used in the various sections of
    the linac are noted on the figure: SSR refers to single spoke
    resonators; the 650\,MHz elliptical cavities are optimised for two
    geometrical phase velocities in the low energy (LE) and high
    energy (HE) sections of the linac; and the 1.3\,GHz elliptical
    cavities are referred to as ``ILC'' cavities.
  }
  \centering
  \begin{tabular}{|c|c|c|c|c|c|c|c|} \hline
    {\bf Section}&{\bf $\beta_{G}$}&{\bf Freq}&{\bf Cavity}&{\bf Number of}&{\bf Gradient}&{\bf Q$_{0}$}&{\bf Energy}\\
    ~&~&(MHz)&{\bf type}&{\bf cavities}&(MV/m)&($10^{10}$)&(MeV)\\ \hline
    SSR0&0.114&325&Single&26&6&0.6&2.5---10\\ 
    ~&~&~&Spoke&~&~&~&~\\ \hline
    SSR1&0.215&325&Single&18&7&1.1&10---32\\
    ~&~&~&Spoke&~&~&~&~\\ \hline
    SSR2&0.42&325&Single&44&9&1.3&32---160\\ 
     ~&~&~&Spoke&~&~&~&~\\ \hline
   LE650&0.61&650&Elliptic&42&16&1.7&160---500\\ \hline
    HE650&0.9&650&Elliptic&96&19&1.7&500---2000\\ \hline
    ILC&1&1300&Elliptic&72&17&1.5&2000---3000 \\ \hline
  \end{tabular}
  \label{tab:acc:fnal:linac}
\end{table}

The CW linac accelerates H$^{-}$ ions having the base bunch frequency
of 325~MHz set by the RFQ.
The beam may be steered toward the high-energy linac (pulsed linac), 
to the experimental area, or to the linac dump. 
The injection to the pulsed linac is controlled by a pulsed
switching-magnet.
If the switch magnet is off, then the beam encounters another
selection dipole magnet to steer it to the 3~GeV experimental
area or the dump.  
In the experimental area, an RF beam separator is used to split the
beam.

The 3--8~GeV pulsed linac is also a superconducting linac. 
Six 2.2~ms pulses of beam are provided at a rate of 10~Hz.
These pulses of beam are accelerated in the pulsed linac and
transferred to the Recycler/Main Injector in support of the
long-baseline neutrino program.  
The pulsed linac will deliver 26~mA-ms of charge in less than
0.75~s to the Recycler Ring;  
Project~X will deliver 345~kW at 8~GeV. 
The pulsed linac is based on 1.3~GHz, 9-cell cavities optimised for
$\beta = 1$ and International Linear Collider (ILC) type cryomodules,
providing a FODO focusing structure.
A cavity gradient of 25~MV/m, which is readily achieved with current
superconducting RF technology, means that 224 cavities are needed to
accelerate the H$^{-}$ beam from 3~to 8~GeV.

\begin{figure}
  \includegraphics[width=\textwidth,height=98.4mm]{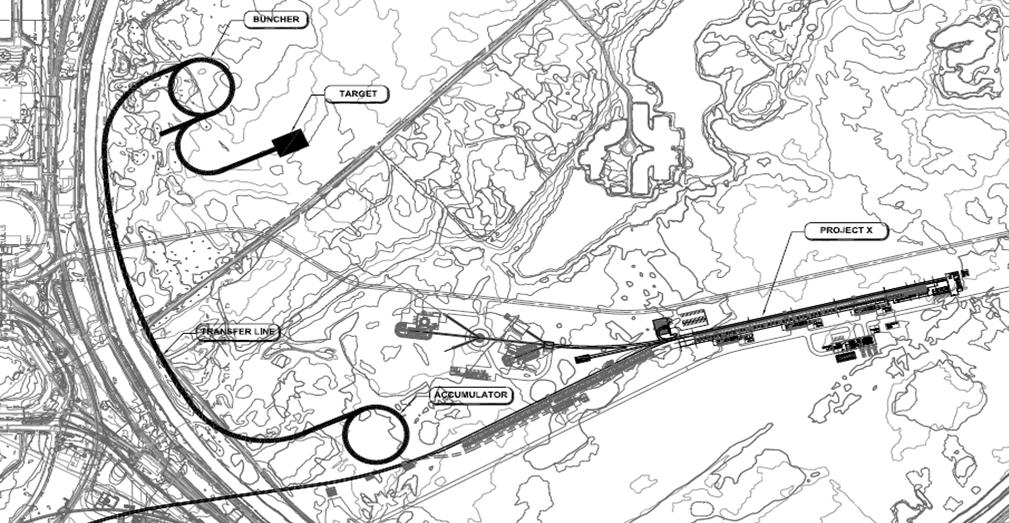}
  \caption{Project~X siting within the Tevatron ring and possible
    locations of accumulator and proton ring to form a proton driver
    for a Neutrino Factory}
  \label{fig:acc:fnal:pd}
\end{figure}

The two linacs fit comfortably within FNAL's Tevatron Ring. 
The basic idea is that, after upgrading Project~X, an accumulation
ring will be put at the end of the second linac.  
The accumulated protons would then be transferred to a separate bunch 
compression ring. 
From that ring, beam will be sent to the Neutrino Factory target
station.  
The target station and ensuing muon collection, acceleration and decay
ring will fit within the Tevatron Ring as well.  
Figure \ref{fig:acc:fnal:pd} shows a possible layout of the additional
rings and target station. 

To achieve 4~MW at 8~GeV, the pulsed linac will need to deliver
10~mA-ms of charge in less than 20~ms (50~Hz).
With the initial 1~mA average CW linac current, half of the CW linac
beam would have to be accelerated by the pulse linac to achieve 4~MW.
This would mean that the high energy linac would have to pulse 50\% of
the time.  
Instead of converting the pulsed linac to CW, the average injection
current can be raised.
Provisions are being designed into Project~X to support an 
upgrade of the CW linac to deliver a current of 4--5~mA. 

The transfer line from the pulsed linac to the accumulator ring will
have the same characteristics as the Project~X transfer line to the
Recycler that is being designed to avoid loss of the H$^{-}$ ions due
to Lorentz stripping, black-body radiation stripping, and stripping
through collisions with residual gas in the beam pipe.
Dipole fields will be limited to 0.05\,T to prevent stripping of the
weakly bound second electron. 
The vacuum of the transfer line will be required to be 
$\sim 5\times10^{-9}$\,Torr.
To mitigate black-body-radiation stripping, a liquid-nitrogen cryogenic
shield will surround the beam pipe. 
The transport line will include a transverse collimation scheme for
capturing large amplitude particles, a momentum-collimation system for
the protection of off-energy particles, and a passive phase-rotator
cavity to compensate for energy jitter.

The preliminary design for the accumulator ring has a circumference of
$\approx 250$\,m.  
Injection will incorporate a stripping system to convert H$^{-}$-ions
to protons.
Foil or laser stripping of electrons from a beam of such power will
need to be developed. 
The stripping processes are quantum mechanical which could leave 
$\sim 1$\% of the H$^{-}$ ions unaffected. 
The non-stripped beam results in a need for a $\approx 50$\,kW beam dump
integrated into the proton accumulation ring.

The major concern with a foil stripping system is the survival of the
foil.
Current systems rely upon short H$^{-}$ beam pulses.  
Pulses much longer than 1~ms will deposit enough energy to melt/damage
the foil.
The foil stripping-injection process will have to employ
transverse/longitudinal phase-space painting to spread the energy
deposition over the foil.
There must be enough time between pulses ($\approx 10$~ms) such that
the foil can radiate.
The bulk of the energy deposited in the foil comes from multiple
passes of protons.  
To reduce the number of proton hits, the circulating beam is moved
away from the foil when there is no beam being injected into the
linac. 
Development of cooled foils or a rotating foil may also increase the
stripping foil survivability.

A demonstration of high efficiency laser stripping system is needed.
A laser based stripping system depends upon two abrupt magnetic fields
and a bright-broad laser intersecting the beam. 
The H$^{-}$ beam is subjected to an abrupt exposure to a high-field dipole magnet 
where an electron will immediately Lorentz strip.
After the dipole, the hydrogen beam is exposed to a laser that excites
the remaining electron.  
A second high-field dipole magnet Lorentz strips the excited electron. 
The laser beam is divergent and at an angle with respect to the hydrogen beam 
so that the spread of Lorentz transformed photon energies matches the spread in hydrogen-beam energies. 
The laser must be bright enough to ensure a high efficiency for
exciting the hydrogen. 
The laser could be high-powered, pulsing synchronously with the beam.  
A moderate-power laser with a resonant cavity across the beam could also be used. 
Either arrangement needs development to survive in the radiation area of an accelerator enclosure. 

The CW linac front-end will be programmed to give short bursts,
injecting beam repeatedly into the same RF buckets of the proton
accumulator ring.
When $2\times10^{13}$ protons have been loaded into each of three RF buckets, 
the RF voltage is increased to shorten the bunch lengths. 
The proton beam is then transferred to the bunching ring. 
A preliminary design for both rings has been developed; beam
instabilities remain to be studied.
Space charge, as well as other conditions, will influence the high
intensity bunches. 
The bunching ring will accept the proton bunches and immediately perform 
bunch rotation of the beam to the final bunch length.  
The beam is then extracted within 240\,$\mu$s; extractions are
separated by 120~$\mu$s.  
The extracted beam is then directed to the target. 
The last transfer line magnet element will be outside the target
station shielding and equipment; the last focusing and steering
elements will be at least 3~m from the target.

\section{RAL proton driver}
\label{sec:acc:ral}

\subsection{Introduction}
\label{sec:acc:ral:Introduction}

\begin{figure}
  \includegraphics[width=0.95\textwidth]{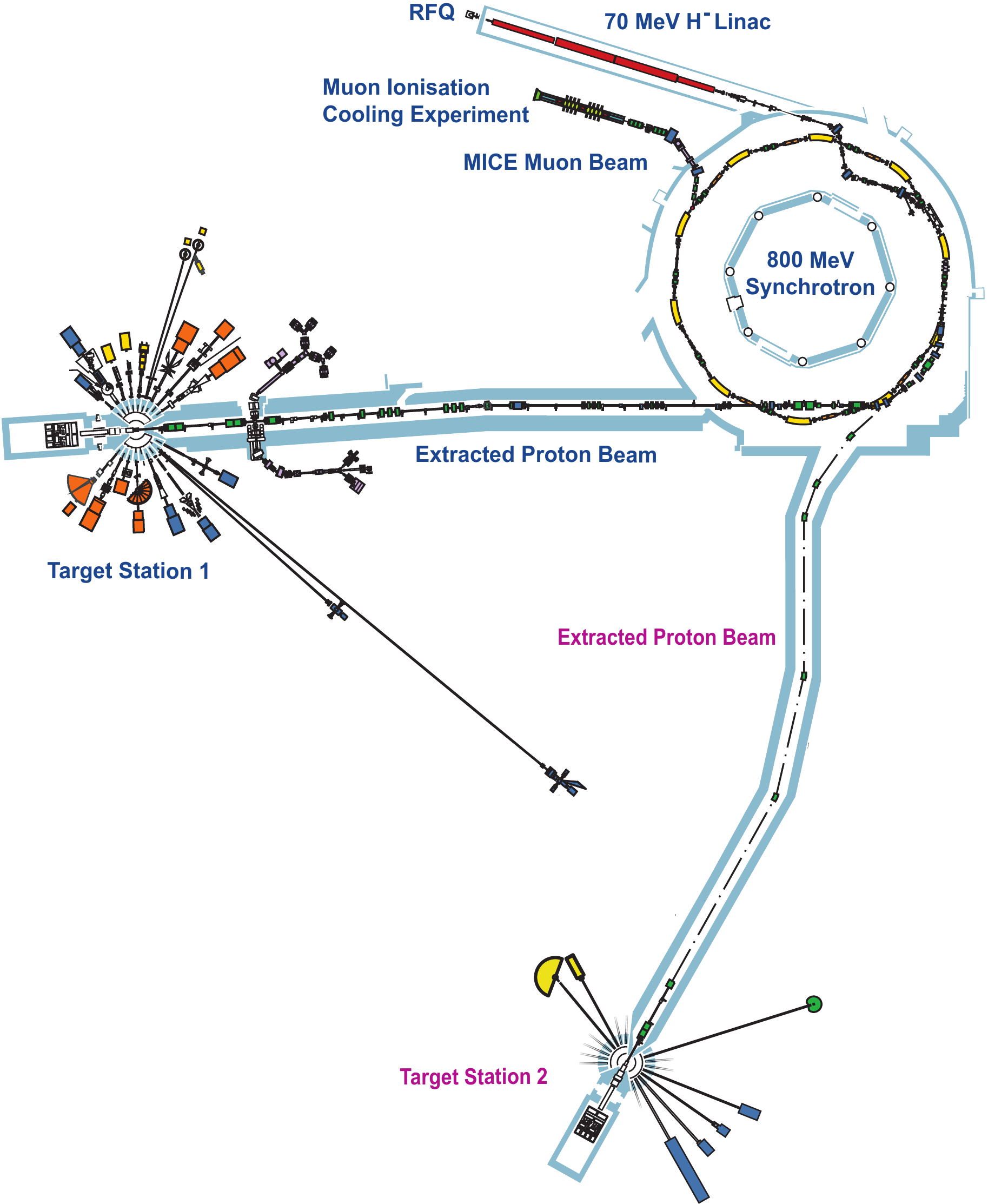}
  \caption{ISIS schematic layout.}
  \label{fig:acc:ral:isis} 
\end{figure}

The Rutherford Appleton Laboratory (RAL) is home to ISIS, the world's
most productive spallation neutron source. ISIS has two neutron
producing target stations (TS-1 and TS-2), driven at 40~Hz and 10~Hz
respectively by a 50~Hz, 800~MeV proton beam from a rapid cycling
synchrotron (RCS), which is fed by a 70~MeV H$^-$ drift tube linac
(DTL)~\cite{Findlay:2007}. The schematic layout of the ISIS facility
is shown in figure~\ref{fig:acc:ral:isis}.

Potential upgrades of the ISIS accelerators to provide beam powers of
2--5~MW in the few GeV energy range could be envisaged as the starting
point for a proton driver shared between a short-pulse spallation
neutron source and the Neutrino Factory \cite{Pasternak:2009}. The
concept of sharing a proton driver between other facilities and the
Neutrino Factory is an attractive, cost-effective solution which is
already being studied in site-specific cases at
CERN~\cite{Garoby:2009a} and FNAL~\cite{Holmes:2010zza}. Although
in the RAL case the requirements for the Neutrino Factory baseline
proton energy and time structure are different from those for a
spallation neutron source, an additional RCS or FFAG booster bridging
the gap in proton energy and performing appropriate bunch compression
seems feasible. 

\subsection{ISIS megawatt upgrades}
\label{sec:acc:ral:ISIS MW}

A detailed comparison of reasonable upgrade routes for ISIS that will
provide a major boost in beam power has been carried out in order to
identify optimal upgrades. Designs are to be developed primarily for
an optimised neutron facility, and will include the provision of an
appropriate proton beam to the existing TS-2 target station. This
forms part of the ongoing research programme into high intensity
proton beams at ISIS~\cite{Warsop:2008, Warsop:2010}, based on
understanding, optimising and upgrading both the existing ISIS RCS and
putative new upgrade synchrotrons at ISIS.  Development and
experimental testing of simulation codes is under way using the SNS
code ORBIT~\cite{orbit} and also with the in-house code
SET~\cite{Pine:2008}.  The latter is presently being expanded to cover
3-D particle motion, exploiting the parallel computing facilities
available at RAL.  The aim is to adapt models being verified on the
present ISIS synchrotron to proposed new running regimes.

The recommended first stage of the upgrade path is to replace parts or
all of the 70~MeV H$^-$ injector.  
Replacement with a new or partly new linac of the same energy could
address obsolescence issues with the present linac, and ensure
reliable operation for the foreseeable future.  The more exciting, but
more challenging, option is to install a higher energy linac (up to
$\approx$\,180~MeV), with a new optimised injection system into the
present ring.  This could give a substantial increase in beam power
(up to a factor of 2), but there are numerous issues to be considered,
and these are currently being studied~\cite{Thomason:2010}. Until the
study is complete, it will not be possible to confirm the viability of
such an upgrade.  However, the calculations, simulation models and
experimental comparisons with the existing machine required in the
course of the work will form an essential baseline for any further
ISIS upgrades. 

The next stage is a new $\approx$\,3.2~GeV RCS that can be employed to increase the energy of the existing ISIS beam to provide powers of $\approx$\,1~MW. This new RCS would require a new building, along with a new $\approx$\,1~MW target station. The new RCS could be built with minimal interruptions to ISIS operations, would give predictable increases in power at reasonable estimated costs, and would have well-defined upgrade routes. RCS designs will include the features required for fast injection directly from the existing ISIS RCS, together with the option for optimised multi-turn injection from a new 800~MeV H$^-$ linac.

\begin{figure}
  \includegraphics[width=0.95\textwidth]{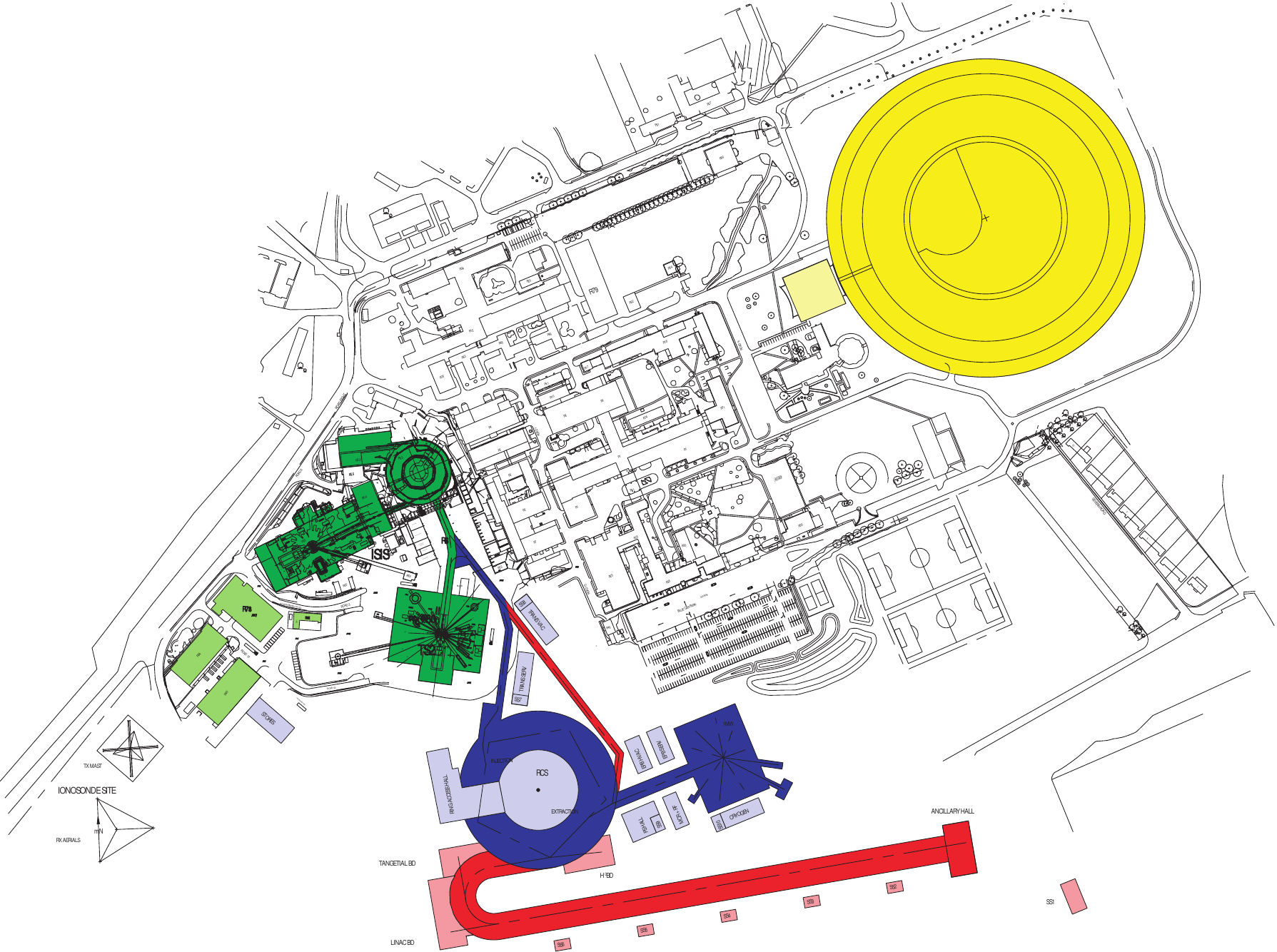}
  \caption{
    Schematic showing the RAL site with ISIS (dark green), the Diamond 
    light source (yellow), $\approx$\,3.2~GeV RCS (blue) and 800~MeV 
    linac (red).
    The light green shading indicates the workshop and other ancillary
    buildings.
  }
  \label{fig:acc:ral:mwisis} 
\end{figure}

 The final upgrade stage is to accumulate and accelerate beam in the $\approx$\,3.2~GeV RCS from a new 800~MeV linac for 2--5~MW beams. It should be noted that a significant collimation section or achromat would be required after the linac to provide a suitably stable beam for injection into the RCS. The new RCS and 800~MeV linac would need to be located some distance from the present accelerators. A schematic layout of these upgrades to the ISIS facility is shown in figure~\ref{fig:acc:ral:mwisis}.
 
Studies and simulations will assess the key loss mechanisms that will impose intensity limitations. Important factors include injection, RF systems, instabilities, loss control and longitudinal and transverse space charge.

\subsubsection{$\approx$\,3.2~GeV RCS studies}

\begin{figure}
  \includegraphics[width=0.7\textwidth]{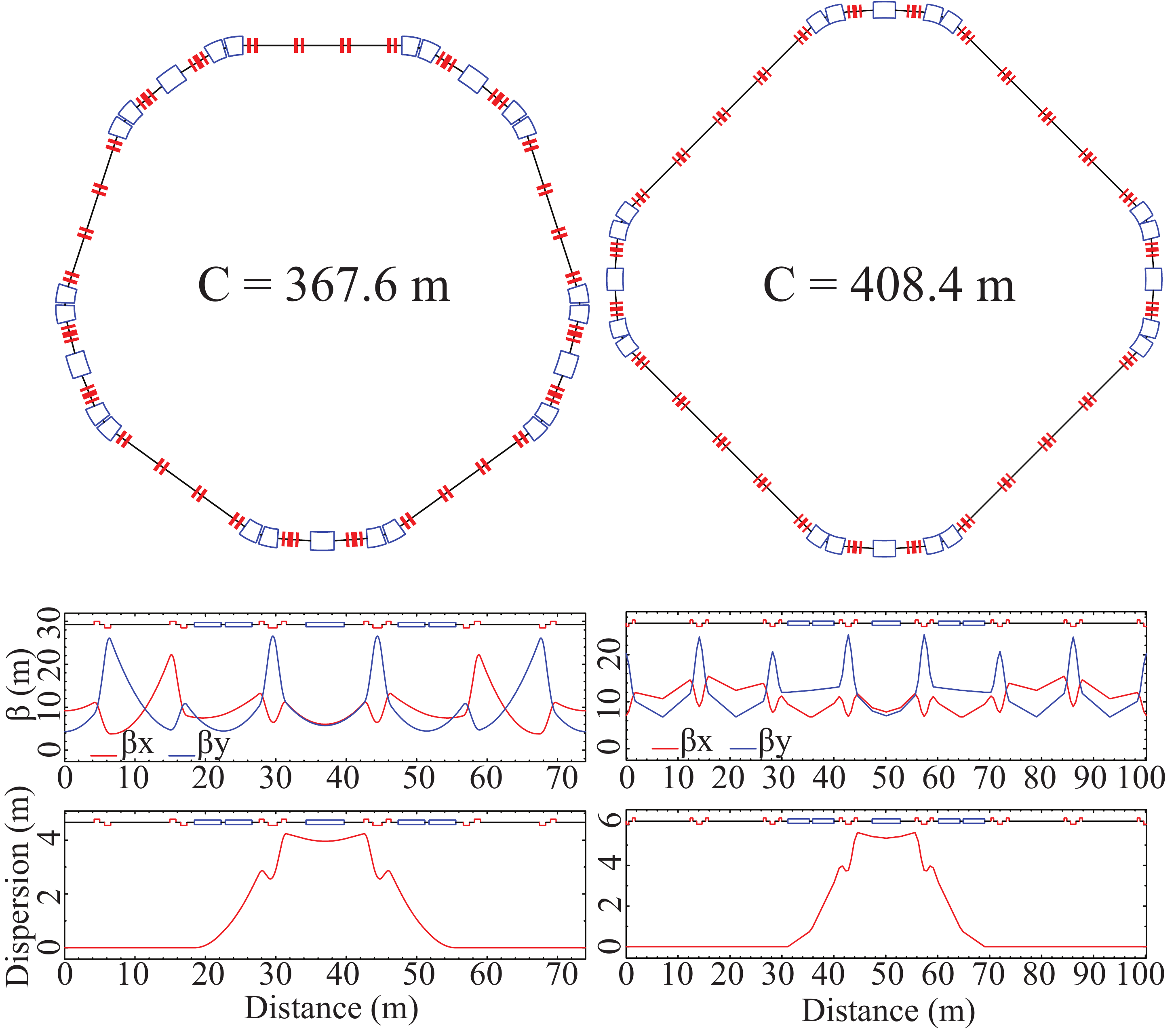}
  \caption{
    Schematic layouts for 5SP (left) and 4SP (right) RCS designs, with
    their respective beta and dispersion functions (below).
  } 
  \label{fig:acc:ral:mwlattices} 
\end{figure}

There are a number of possible candidates for the $\approx$\,3.2~GeV, 50~Hz RCS, but studies are presently focused on a 3.2~GeV doublet-triplet design with five super-periods (5SP) and a 3.2~GeV triplet design with four super-periods (4SP)~\cite{Thomason:2008}, as shown in figure~\ref{fig:acc:ral:mwlattices}.

The 5SP ring has a mean radius ($R$) of 58.5\,m ($R/R_0 = 9/4$, where 
$R_0 = 26.0$\,m is the mean radius of the ISIS 800\,MeV synchrotron)
and RF cavities running at harmonic number $h = 9$, i.e. at nine times
the ring revolution frequency (6.18--7.15\,MHz). 
This ring is optimised to give small dipole apertures and therefore to
minimise the magnet power supply requirements. 
Meanwhile, the 4SP ring has a mean radius of 65.0\,m ($R/R_0 = 5/2$)
and RF cavities running at harmonic number $h = 5$, i.e. at five times
the ring revolution frequency (3.09---3.57\,MHz). 
This ring is optimised to make fast injection from ISIS easier, but
has larger apertures. 
Both of these ring designs (and appropriate variations) will be
studied in detail in order to assess their suitability for the
recommended upgrades. 
Initial work, however, has concentrated mostly on the 5SP design. 

Work is now under way to study the key issues for the $\approx$\,3.2~GeV ring designs, underpinned by extensive development of the relevant codes and benchmarking during machine physics studies on ISIS. The main topics include space charge, injection, provision for RF, beam stability and the requirement to keep beam losses below about 0.01\%.

\subsubsection{800~MeV linac studies}

An 800~MeV H$^-$ linac design has been produced~\cite{Rees:2009},
following the designs of the SNS~\cite{Campisi:2007} and European
Spallation Source (ESS) \cite{ESS:2002} linacs. The initial 74.8~MeV stage is based around the 324~MHz frequency of the 2.5~MW peak power Toshiba klystron used in the J-PARC linac~\cite{JPARC:2003}. The design includes an ion source, low energy beam transport (LEBT), 3~MeV radio frequency quadrupole (RFQ) and medium energy beam transport (MEBT), all based on the Front End Test Stand (FETS) at RAL~\cite{Letchford:2010}, followed by a 74.8~MeV DTL. An intermediate energy beam transport (IEBT) collimation section follows the DTL.

Three options have been considered for acceleration from 74.8 to $\approx$\,200~MeV: a room temperature coupled cavity linac (CCL) at 648~MHz, and superconducting cavity linacs at 648~MHz (ScL1) or 324~MHz (ScLa), both with geometric $\beta_G$ values of 0.45. The first two options require a high power klystron development at 648~MHz, but are preferred to the 324~MHz ScLa option for reasons of practicality and beam dynamics. The CCL design option could also be adopted for the $\approx$\,180~MeV linac to replace the ISIS 70~MeV H$^-$ injector, although in this instance the IEBT would not be required.

\begin{figure}
  \includegraphics[width=0.7\textwidth]{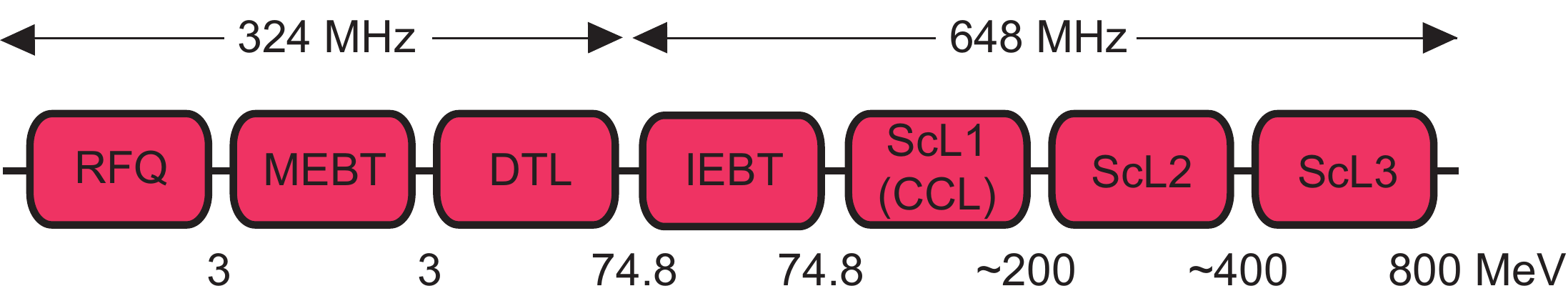}
  \caption{Schematic arrangement for the 800~MeV linac, showing the option of ScL1 or CCL.}
  \label{fig:acc:ral:linac} 
\end{figure}

In the 800~MeV linac after $\approx$\,200~MeV, new superconducting structures are used with a $\beta_G$ value of 0.62 at energies up to $\approx$\,400~MeV and a $\beta_G$ value of 0.76 for $\approx$\,400 to 800~MeV.  The preferred options continue to use 648~MHz cavities, first in ScL2 stage to $\approx$\,400~MeV and then in ScL3 to 800~MeV. The schematic arrangement for the 800~MeV linac is shown in figure~\ref{fig:acc:ral:linac}.

Beam transport lines between the linacs and rings need achromatic bending sections and also added cavities to ramp the output energy of the beam and to control the beam momentum spread for H$^-$ charge-exchange injection into the new $\approx$\,3.2~GeV RCS.

\subsection{Common proton driver}
\label{sec:acc:ral:Common PD}

In a common proton driver for a neutron source and the Neutrino Factory, based on a
2--5~MW ISIS upgrade with an 800~MeV linac and a $\approx$\,3.2~GeV
RCS, both facilities have the same ion source, RFQ, MEBT, linac, H$^-$
injection, accumulation and acceleration to
$\approx$\,3.2~GeV. Bunches of protons are shared between the two
facilities at $\approx$\,3.2~GeV, and a dedicated RCS or FFAG
booster must then accelerate the Neutrino Factory bunches to meet the requirements
for the Neutrino Factory baseline (4~MW and 5--15~GeV). Taking the optimistic case
of a total power of 4--5~MW at $\approx$\,3.2~GeV, some possible bunch
sharing scenarios are outlined in table~\ref{tab:acc:ral:PulseSharing}.

\begin{table*}
  \caption{
    Scenarios for bunch sharing between an upgraded ISIS neutron
    source and a Neutrino Factory, assuming bunches will be
    transferred from the $\approx$\,3.2~GeV RCS at 50~Hz with a total
    power of 4--5~MW and that 4~MW is required for the Neutrino
    Factory target.} 
  \begin{center}
    \begin{tabular}{|c|c|c|c|c|c|c|c|c|}
      \hline
      {\bf $\approx$\,3.2~GeV} & {\bf Power}      & {\bf Total}      & {\bf Bunch}   & {\bf Protons}               & {\bf Number of} & {\bf Power}   & {\bf Number of} & {\bf NF booster} \\
      {\bf RCS}                & {\bf at 3.2~GeV} & {\bf number}     & {\bf spacing} & {\bf per bunch}             & {\bf bunches}   & {\bf to ISIS} & {\bf bunches}   & {\bf energy} \\
      {\bf design}             & (MW)             & {\bf of bunches} & (ns)          & ($\times$10$^{13}$)          & {\bf to ISIS}   & (MW)          & {\bf to NF}     & (GeV)  \\
      \hline
      4SP                & 5          & 5          & 280     & 3.9                   & 2         & 2       & 3         & 4.3    \\
      4SP                & 5          & 5          & 280     & 3.9                   & 3         & 3       & 2         & 6.4    \\
      5SP                & 5          & 9          & 140     & 2.2                   & 6         & 3.33    & 3         & 7.7    \\
      5SP                & 4          & 9          & 140     & 1.76                  & 6         & 2.66    & 3         & 9.6    \\
      \hline
    \end{tabular}
  \end{center}
  \label{tab:acc:ral:PulseSharing}
\end{table*}

\begin{figure}
  \includegraphics[width=0.7\textwidth]{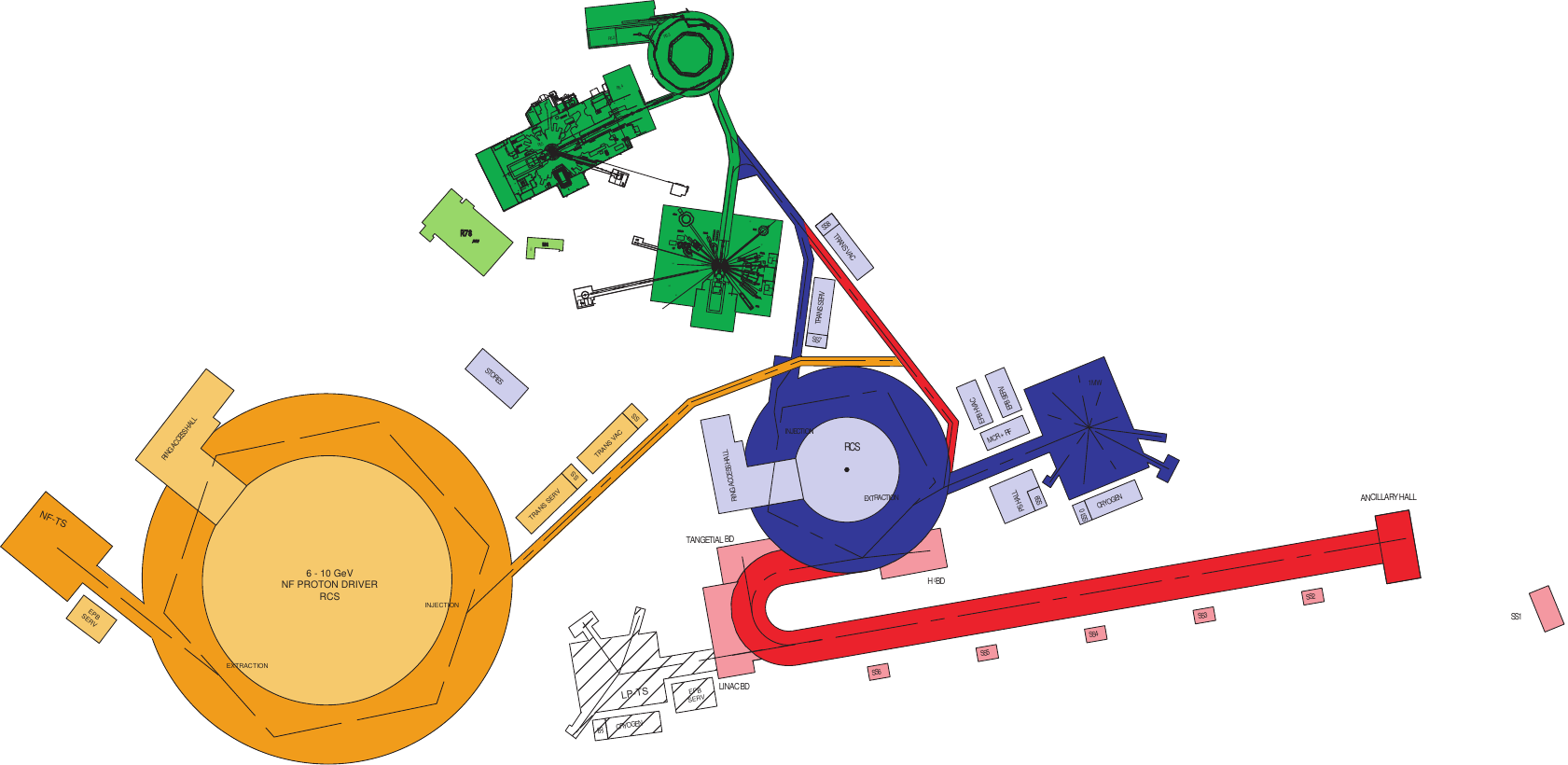}
  \caption{
    Conceptual layout of the Neutrino Factory proton driver, with the dedicated Neutrino Factory
    booster shown in orange. 
  }
  \label{fig:acc:ral:mwisis+nf} 
\end{figure}

Assuming that at least half of the power at $\approx$\,3.2~GeV should
be delivered to the neutron source, both the 4SP and 5SP
$\approx$\,3.2~GeV RCS designs could meet the power and energy needs
of the Neutrino Factory (although for the 4SP design only two bunches are delivered
rather than the Neutrino Factory baseline of three). It would appear that the 5SP
design is most suitable, as it meets all the requirements of the Neutrino Factory
baseline and provides more beam power to the neutron source, but its
merits need to be established by thorough beam dynamics studies.
In order to give some flexibility in
case the total power at $\approx$\,3.2~GeV is somewhat less than 5~MW,
6.4--10.3~GeV RCS and FFAG booster designs will be considered. 
Figure \ref{fig:acc:ral:mwisis+nf} shows the conceptual 
layout of the common proton driver. 

\subsubsection{Dedicated Neutrino Factory booster}

\begin{figure}
  \includegraphics[width=0.7\textwidth, angle=0]{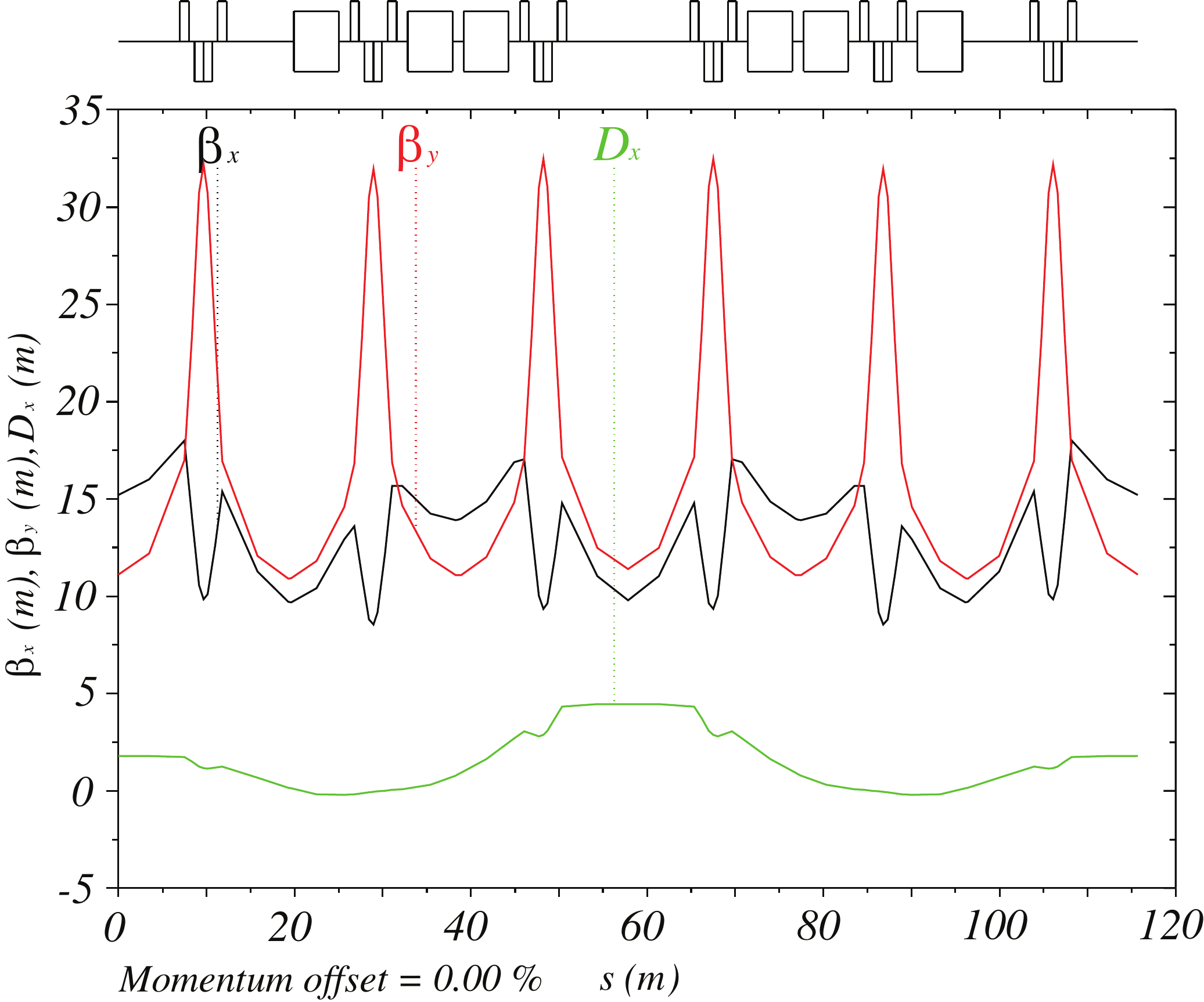}
  \caption{Optical functions in the dedicated Neutrino Factory booster RCS ring for 3.2--9.6~GeV.}
  \label{fig:MainRCStwiss} 
\end{figure}

Based on the time structure and longitudinal dynamics of the ISIS upgrade $\approx$\,3.2~GeV RCS, a further RCS or an FFAG can be considered as a booster ring to reach the required Neutrino Factory baseline. Preliminary RCS designs~\cite{Pasternak:2009} have concentrated on achieving the necessary acceleration and bunch compression with present-day, cost-effective RCS technology, e.g., dipole magnets with a maximum field of 1.2~T, an RF system similar to that used at ISIS~\cite{Seville:2008} and long straight sections for injection, extraction, RF and collimation.  
An RCS design based on injection from the 5SP 3.2~GeV booster with
harmonic number 9 and 4~MW total beam power (the last entry in
Table~\ref{tab:acc:ral:PulseSharing}) has been investigated. This case
dictates a rather large final proton energy of 9.6~GeV, but allows
delivery of the required beam parameters to both facilities with
minimal impact on the neutron source performance.  The RCS has six
super-periods with six FDF triplet cells each, uses only three
quadrupole families and allows for a flexible choice of gamma
transition. The main RCS parameters for this design are summarised in
Table~\ref{Tab:MainRCS} and the optical functions are shown in
figure~\ref{fig:MainRCStwiss}. Although the preliminary lattice design
has been produced, a great deal of work remains to be done to produce
a full conceptual scenario.  

\begin{table*}
  \caption{Parameters of the dedicated Neutrino Factory booster RCS ring for 3.2--9.6~GeV.}
  \begin{center}
    \begin{tabular}{|l|r|}
      \hline
      {\bf Parameter}                  & {\bf Value}        \\
      \hline
      Number of super-periods           & 6                  \\
      Circumference (m)                & 694.352            \\
      Harmonic number                  & 17                 \\
      RF frequency (MHz)               & 7.149--7.311       \\
      Injection energy (GeV)           & 3.2                \\
      Extraction energy (GeV)          & 9.6                \\
      Maximum dipole field (T)         & 1.2                \\
      Tune                             & 8.72 (h), 7.82 (v) \\
      Long straight section length (m) & 14~m               \\
      Gamma transition                 & 13.37 (flexible)   \\
      RF voltage per turn (MV)         & $\approx$\,3.7     \\
      \hline
     \end{tabular}
  \end{center}
  \label{Tab:MainRCS}
\end{table*}

FFAG options are yet to be explored, and would be based on technology which remains to be fully tested, but in principle would offer the advantage of allowing all the bunches to be extracted to the Neutrino Factory target with the same energy (unlike the RCS where the 120~$\mu$s sequential extraction delay required by the Neutrino Factory baseline would give time for the main magnet field to vary between bunches).

\subsubsection{Neutrino Factory bunch compression}

Optimised longitudinal muon capture in the muon front end of the Neutrino Factory requires compression of the proton bunch length from the $\approx$\,100~ns for the neutron source to 1--3~ns at the Neutrino Factory target. Several methods have been proposed in order to reach this goal~\cite{Prior:2008zzc}, based on either adiabatic compression during acceleration or fast phase rotation at the end of acceleration (or in an additional dedicated compressor ring).

Adiabatic compression during acceleration requires relatively high RF voltage (V) because the bunch length scales as V$^{-1/4}$. Variations of this method apply higher harmonic RF systems or lattices just below transition at the end of compression. Compression by fast phase rotation allows a lower RF gradient, but requires earlier bunch stretching to reduce the momentum spread just before the rotation and
does not allow the compressed bunches to be held for many turns. Manipulations close to transition may also be applied in this scheme. Fast phase rotation in an additional dedicated compressor ring, possibly based on the CERN design~\cite{Garoby:2009a}, could provide an alternative solution if RF manipulation in the booster itself proves impractical.  

Bunch compression is clearly of vital importance to the success of a
common proton driver and future studies must address the longitudinal
dynamics and the space charge forces in detail. 

\subsection{Summary}
\label{sec:acc:ral:Summary}

A common proton driver for neutrons and neutrinos compatible with an
ISIS upgrade is an attractive solution to create a cost-effective,
multi-user facility, but careful attention must be given to potential
sharing conflicts between the neutron and neutrino communities. A
conceptual design has been produced, in which it appears to be
feasible that the Neutrino Factory baseline can be met, as shown in
Table~\ref{tab:acc:ral:PDParams}, although a lot of the detailed beam
dynamics remains to be done and no consideration has yet been given to
beam transport to the pion-production target. 
\begin{table*}
  \caption{
    Baseline proton beam parameters at the Neutrino Factory pion-production
    target compared with expected parameters from a proton driver based on an ISIS MW upgrade at RAL.}
  \begin{center}
    \begin{tabular}{|l|r|r|}
      \hline
      {\bf Parameter}                                              & {\bf Baseline} & {\bf RAL}         \\
      \hline
      Beam power (MW)                                              & 4         & 4          \\
      Pulse repetition frequency (Hz)                              & 50        & 50         \\
      Proton kinetic energy (GeV)                                  & 5--15     & 6.4--10.3  \\
      Proton rms bunch length (ns)                                 & 1--3      & 1--3       \\
      Number of proton bunches per pulse                           & 3         & 2 or 3     \\
      Sequential extraction delay ($\mu$s)                         & 120       & 120        \\
      RMS transverse bunch size (rms bunch radius) at target (mm)  & 1.2       &            \\
      Geometric transverse emittance at target ($\mu$m)            & $< 5$     &            \\
      $\beta^*$ of proton beam at target (cm)                      & $\geq 30$ &            \\
      \hline
    \end{tabular}
  \end{center}
  \label{tab:acc:ral:PDParams}
\end{table*}

\begin{figure}
  \includegraphics[width=0.7\textwidth]{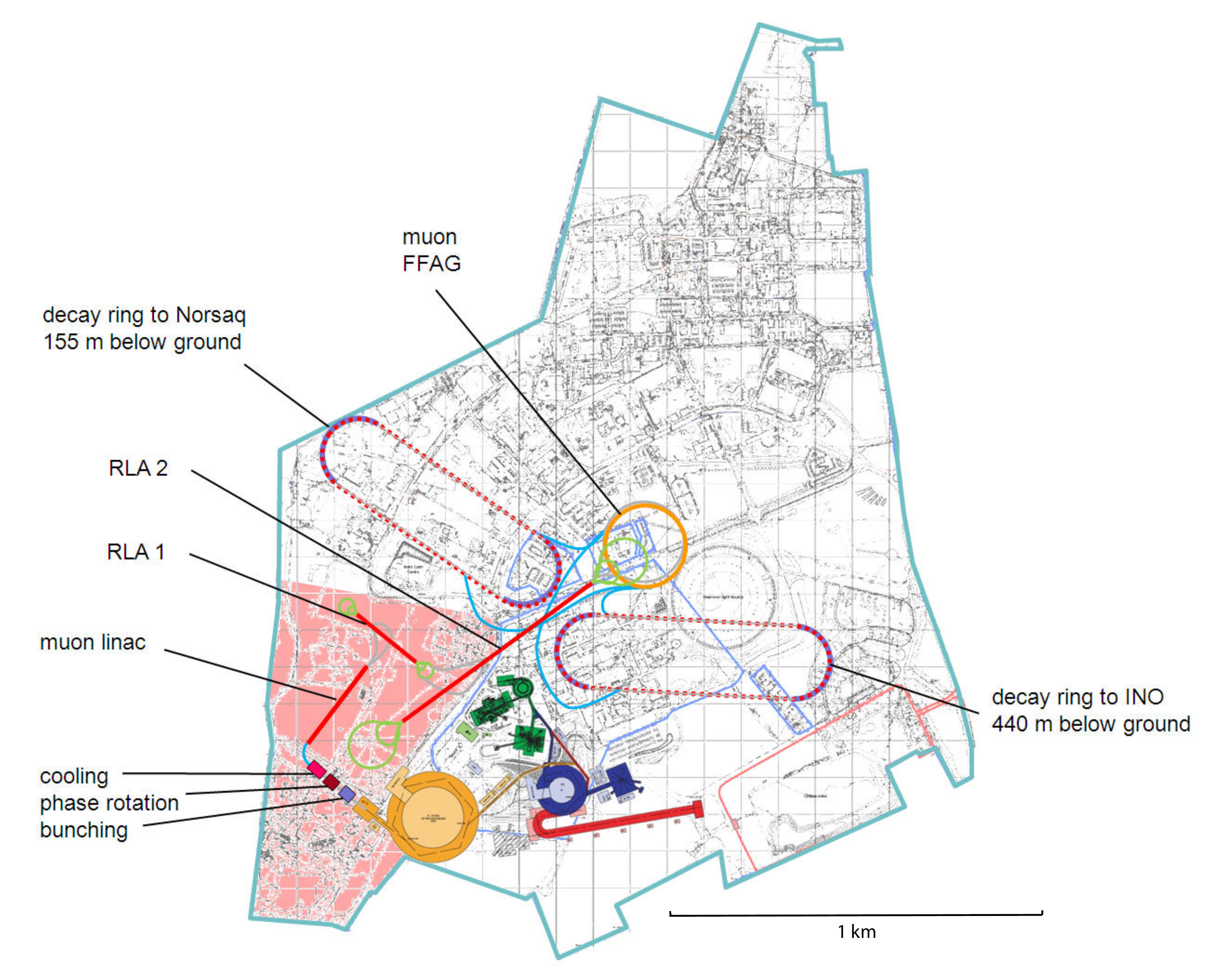}
  \caption{
    Schematic layout of the Neutrino Factory on the HSIC site. The components of the
    proton driver are shown in blue, red and orange as in
    figures~\ref{fig:acc:ral:mwisis} and~\ref{fig:acc:ral:mwisis+nf},
    and the HSIC site boundary is shown in light blue. The area shown
    in pink is former UKAEA land which would need to be
    decommissioned.
    Norsaq (in Greenland) and the Indian Neutrino Observatory (INO)
    are possible far detector sites.
  } 
  \label{fig:acc:ral:hsic} 
\end{figure}

The site-specific design at RAL is clearly at a preliminary stage, and will require extensive effort on beam dynamics and accelerator engineering (and strategic research and development in a number of key areas) before it can be regarded as viable. The common proton driver could fit onto the RAL site, on land already set aside for large facilities and research expansion, but the complete Neutrino Factory would require the use of part of the Harwell Science and Innovation Campus (HSIC), where some former UK Atomic Energy Authority (UKAEA) land would need to be decommissioned before any building or tunnelling work could begin. A possible schematic layout of the Neutrino Factory on the HSIC site is shown in figure~\ref{fig:acc:ral:hsic}.

\section{Alternative Target Materials}
\subsection{Powder Jet}
\label{App:PowderJet}

A fluidised tungsten-powder target is being developed at Rutherford
Appleton Laboratory \cite{Densham:2010} as an alternative to the
baseline mercury jet. 
The motivation for studying a flowing-powder target is to investigate
whether such a technology can combine some of the advantages of a
solid target with those of a liquid while avoiding some of the
disadvantages of either. 
The configuration of such a target in a neutrino factory would be
similar to that for the mercury jet, i.e., it is broadly compatible
with the baseline solenoid arrangement.  
As for the mercury jet, a return path for the tungsten powder would
need to be devised, possibly by engineering a longitudinal gap between
the coils, although other configurations may be possible. 
A tungsten powder target of 50\% material fraction has been computed
to deliver the same meson production performance as the mercury jet,
by increasing the length to compensate for the reduced material fraction.

The potential attractions of such a scheme include: intrinsic
resilience of grains to beam induced stress waves and radiation
damage; no propagation of stress waves to pipe walls; possibility to
recirculate material in an external cooling loop; continual reforming
of target-material geometry; short conduction path and consequently
good heat transfer characteristics; no cavitation or associated
phenomena; and negligible interaction with magnetic fields.
Some potential issues that would need to be addressed for a fluidised powder target system are expected to include:
erosion of pipe and container walls, particularly for recirculation at high velocity in a lean phase;
secondary heating and radiation damage of a pipe wall for the
preferred geometry of a contained flowing powder rather than a free
jet; and
integration of a powder-recirculation system within the confines of the target station and solenoid system.

\begin{figure}
  \includegraphics[width=0.5\linewidth]{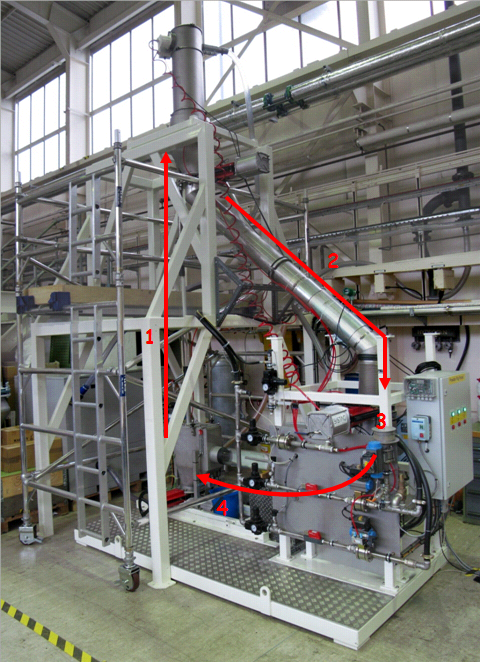}
  \caption{
    Fluidised powder test rig showing (1) Suction/lift, (2)
    Chute to hopper, (3) Pressurised hopper and (4) Powder ejection to
    receiver hopper.
  } 
  \label{fig:acc:tgt:powrig}
\end{figure}

\begin{figure}
  \includegraphics[width=\linewidth]{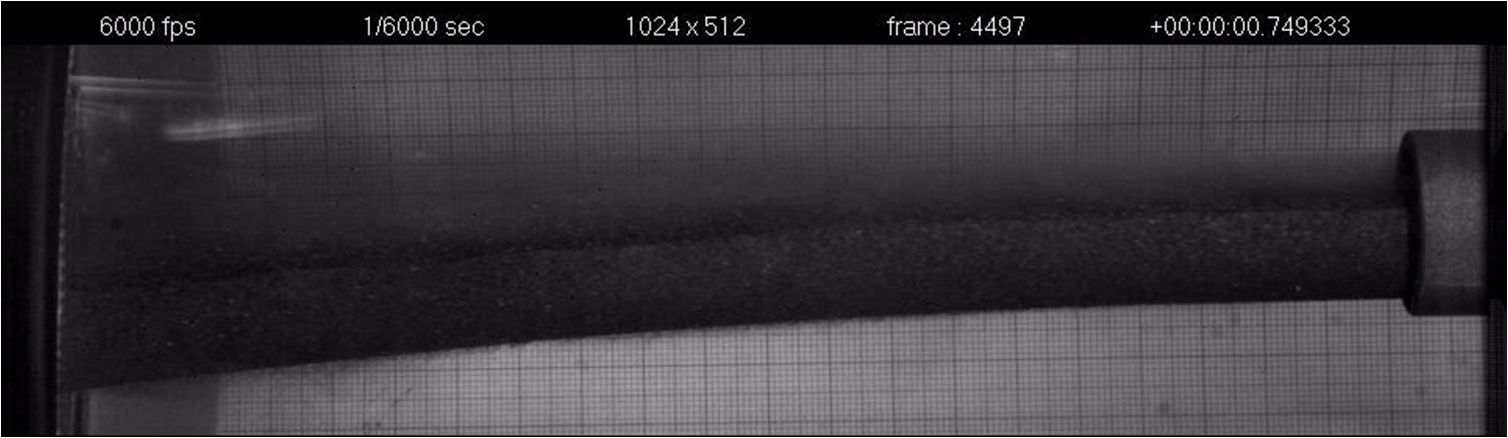}
  \caption{
    Free tungsten powder jet demonstrating stable dense flow at
    3.7 m/s and 40-45\% material fraction.
  } 
  \label{fig:acc:tgt:powjet}
\end{figure}

Most of these issues can be studied in an off-line experimental programme,
which has already begun~\cite{Densham:2010}.
The RAL test rig is shown in figure~\ref{fig:acc:tgt:powrig}.
The test programme has demonstrated the feasibility of generating a stable flow regime of tungsten powder at around 40--45\% material fraction 
in both contained and open jet configurations. 
The long dimension of the target is horizontal, suitable for a neutrino factory.
An example of an open, 2~cm diameter, tungsten powder jet with a
velocity of 3.7 m/s generated by the rig operating at a 2 bar driving
pressure is shown in figure~\ref{fig:acc:tgt:powjet}.
A
vertical lift vacuum recirculation system has lifted the powder at the rate
required for the rig to be able to operate in CW mode for the optimised
target radius of 0.47 cm. In a neutrino factory,
this arrangement would require a longitudinal gap
between the solenoid elements to accommodate the vertical lift pipe. It may
be possible to devise a recirculation path that is closer to that proposed
for the mercury loop, at the cost of increased complexity.

A free jet requires an even higher velocity flow than the mercury jet to
generate the necessary target interaction length. This high velocity
generates concerns over erosion of the downstream components.
Consequently, the preferred configuration is for low velocity (2--3~m/s)
flow within a contained pipe, which should be tolerant to multiple beam
pulses interacting with the same material as it traverses the discharge
pipe. This configuration has received the focus of the test effort to date.
No erosion has been observed in the pipes used for experiments so far,
although this has only been for about 30 minutes of integrated operation. The
main concern with this configuration is with secondary heating and radiation
damage of the pipe walls; external cooling of the pipe wall, possibly by
high pressure helium, is expected to be necessary. The design of such a
circuit should permit rapid replacement of such pipework.

\subsection{Solid Target}
\label{App:SolidTrgt}

\begin{figure}
  \includegraphics[width=0.5\linewidth]{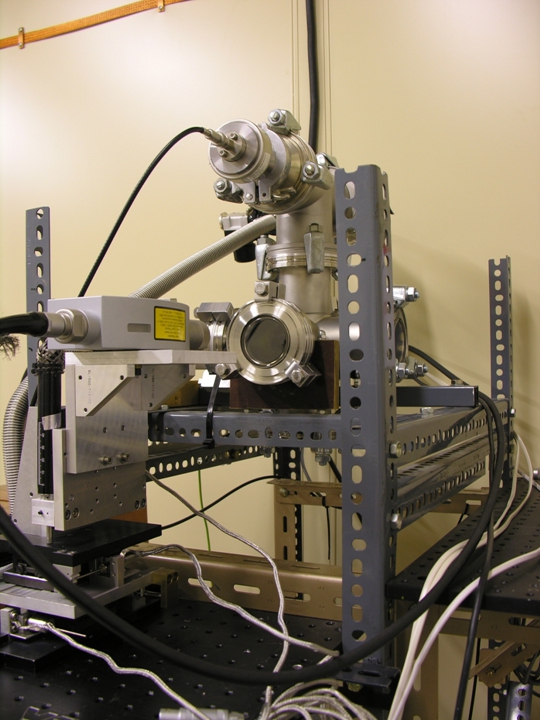}
  \caption{
    The solid target test rig. 
    The cable from the pulsed power supply enters from the left. 
    The test wire can be viewed through a number of windows. The head
    of the Doppler laser vibrometer used to measure the shock waves
    for comparison with modelling can be seen in the foreground.
  }
  \label{fig:acc:tgt:solid-test}
\end{figure}

\begin{figure}
  \begin{center}
    \includegraphics[width=\linewidth]{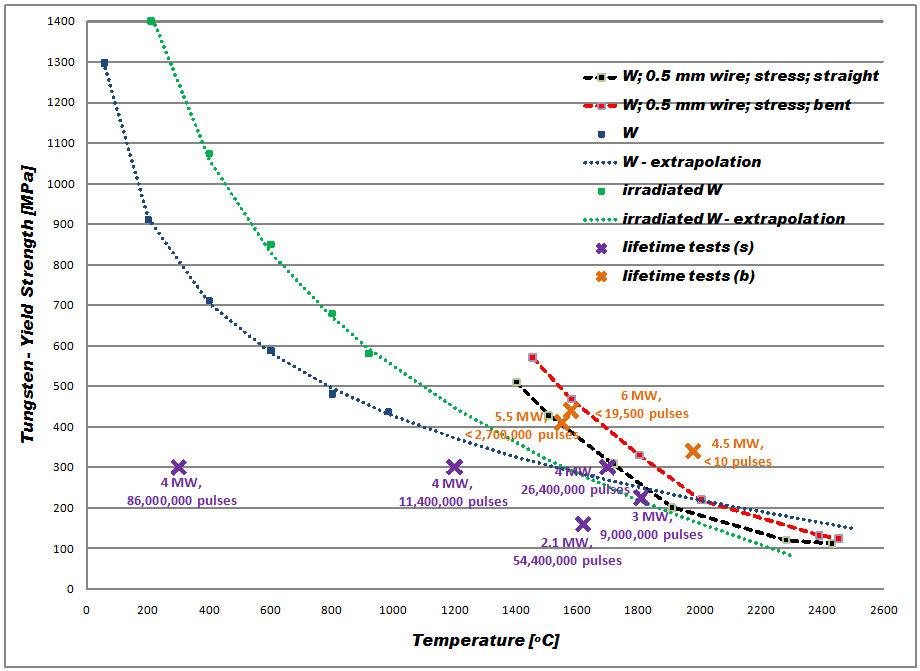}\\[-1mm]
  \end{center}
  \caption{
    The measured yield strength of tungsten. 
    The square points show earlier measurements of the yield strength
    of tungsten as a function of temperature both before and after
    irradiation. 
    The dotted lines are a fit to these points, used to extrapolate
    them. 
    The dashed lines are measurements made by us, which constrain the
    yield strength to lie between them. 
    Note that there is a difference at lower temperature because the
    strain rate we use is much larger than in the existing
    measurements. 
    The crosses show measured lifetimes for the targets. 
    For yellow crosses, the target broke, for purple it did not. 
    The working temperature for a Neutrino Factory target will be less
    than $1\,500\,^\circ$C and one year's operation corresponds to 1.25M
    pulses with 400 bars.
  }
  \label{fig:acc:tgt:solid-lifetime}
\end{figure}

A solid-target option is also being studied for a Neutrino Factory. 
This is motivated by the many years of experience in the use of high
atomic-number materials as targets, for example at ISIS, and the clear
benefits for radio-protection. 
Two materials have been investigated, tantalum and tungsten, and
initial studies have focused on their lifetime under the thermal
shock created by the high-energy-density proton beam. 
A test facility (see figure \ref{fig:acc:tgt:solid-test}) employing a
current pulse of up to 8\,mA and with a rise time of 100\,ns has been
used to emulate in wires, ranging from 0.35\,mm to 1\,mm in diameter,
the thermal shock in the target. 
Some results of this work are shown in figure
\ref{fig:acc:tgt:solid-lifetime}. 
At the working temperature of the target ($\sim 1\,200$~$^\circ$C),
tantalum is not strong enough.
However, it has been demonstrated that the lifetime of tungsten is
more than sufficient \cite{Bennett:2008}. 
To verify this outcome, it is planned to test real tungsten targets in
a proton beam of high energy density at CERN in 2011. 
During this work, the Young's modulus of tantalum and tungsten at
temperatures and strain-rates much larger than measurements presented
in the existing literature have been made.
This work on tantalum and tungsten outlined above is being published in 
appropriate journals. 

After shock, the second most important issue for solid targets is the
heat generated by the beam. 
The only practical solution to this is to change the target between
beam pulses. 
This must be done in a way that minimises the impact on useful pion
production and the heat load to the superconducting capture coils. 
Various schemes, employing 200---500 targets, are being considered and 
detailed engineering studies of this are now starting. 
In addition, we must also worry about the effect of radiation on the
targets. 
With 200 target bars, the activation rate is already slower than a single
ISIS target.  
Several ISIS targets have been used and, while they have not been
examined in detail, there are no visible signs of the expected effects
of radiation damage. 
Further, earlier studies of irradiated tungsten indicate that it
remains strong enough for use after significant neutron irradiation. 
Nevertheless, we are investigating making detailed measurements of an
ISIS target that has been irradiated to verify these observations.  
Finally, MARS simulations show that the pion yield is comparable to
the yields from the mercury-jet system \cite{Back:2008sr}. 

\section{Alternative lattices for the muon front-end}
\label{sec:acc:fe:FrontEnd_Alternatives}
\begin{figure}
  \begin{center}
    \includegraphics[width=49mm,height=44mm]{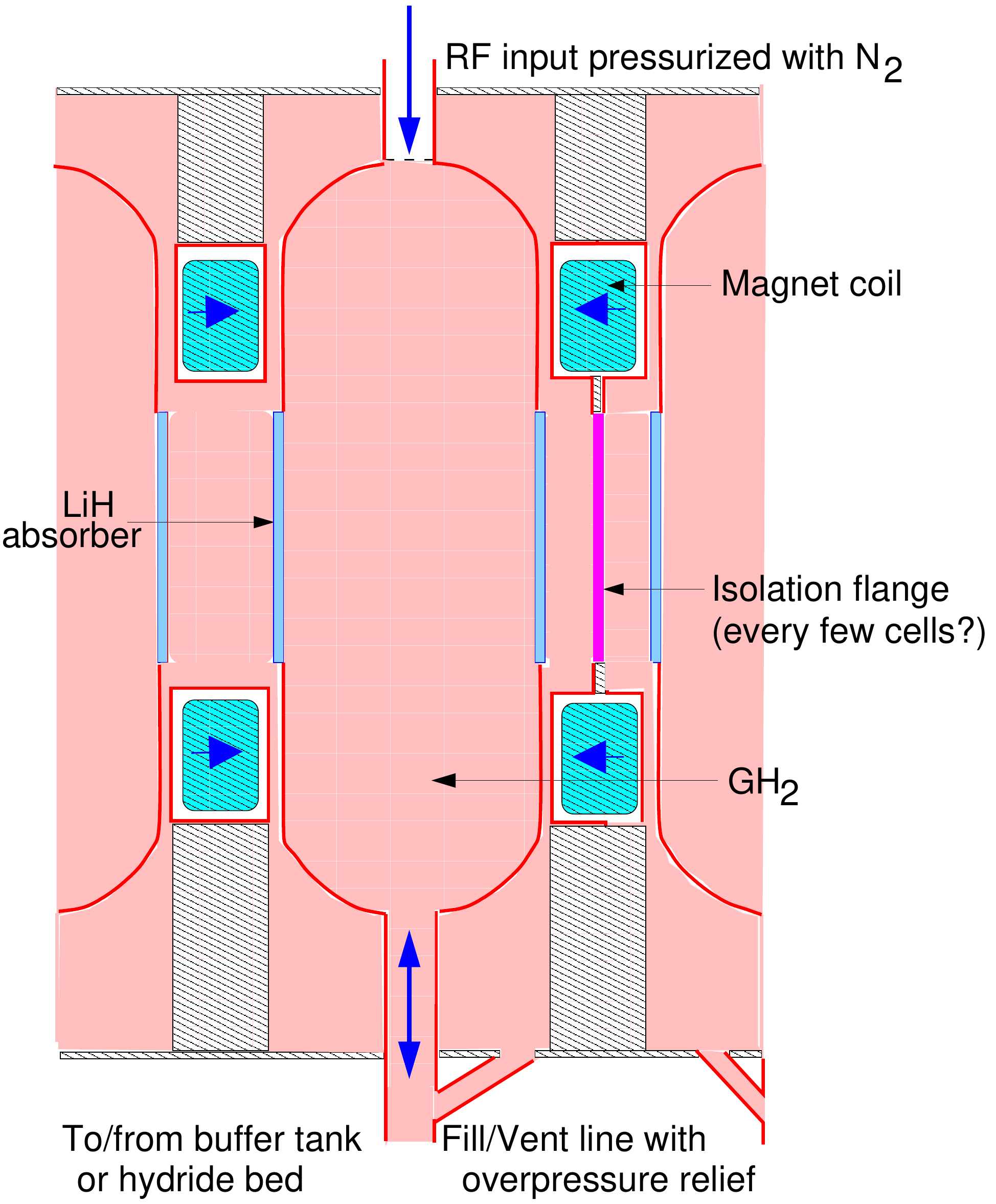}\\[12pt]
    \includegraphics[width=0.44\linewidth]{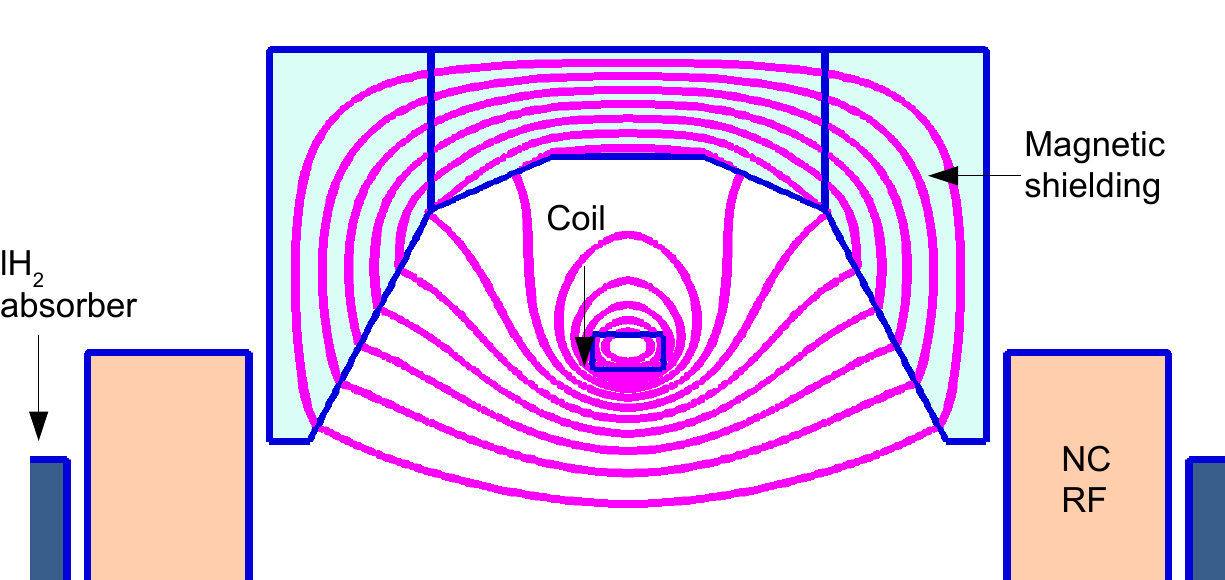}\\[12pt]
    \includegraphics[width=0.33\linewidth]{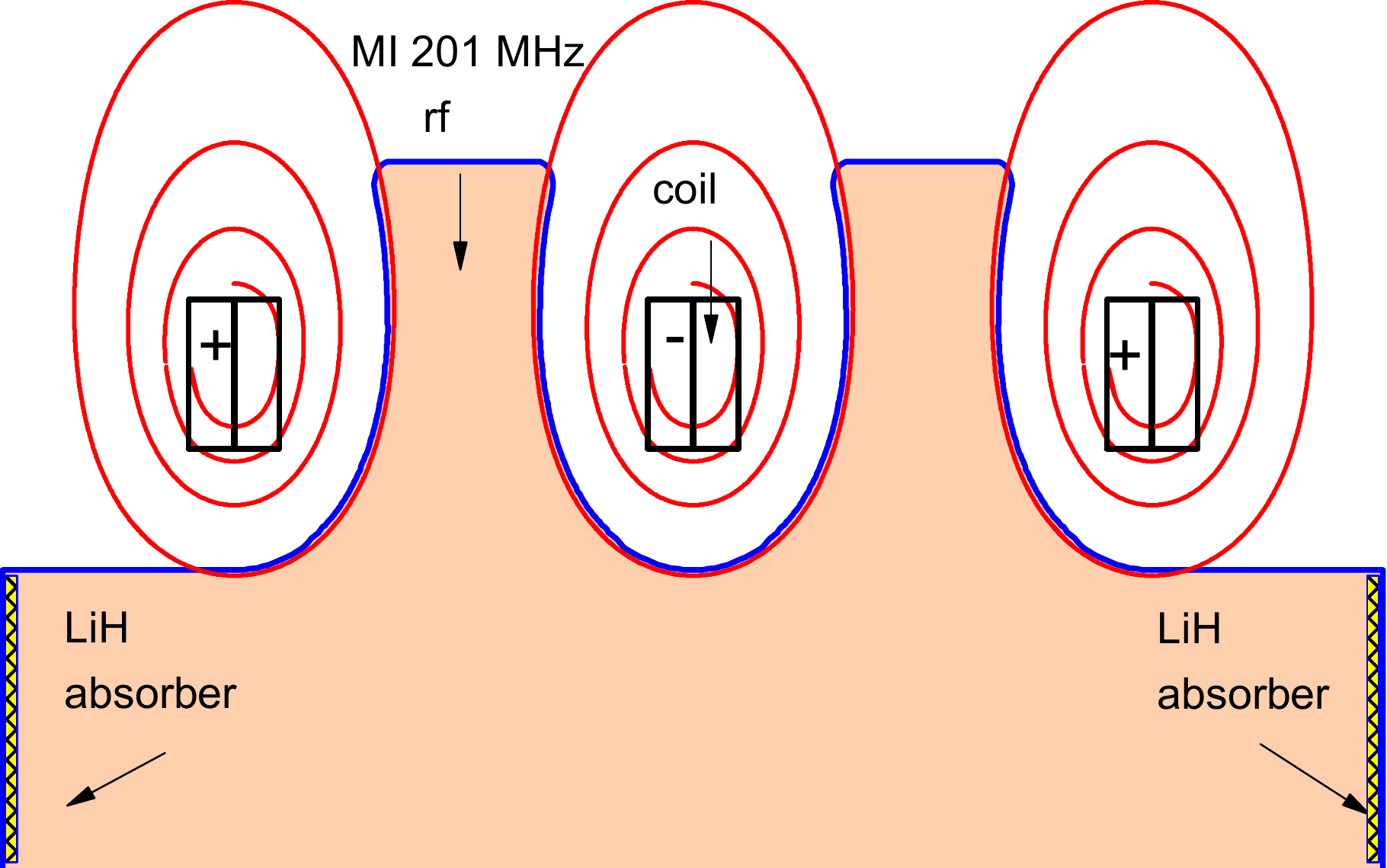}
  \end{center}
\caption{Schematic drawings of the proposed alternative lattices: (top)
  high pressure gas-filled hybrid lattice; (middle) shielded RF cavities;
  (bottom) magnetically insulated lattice.}
\label{fig:acc:fe:alternate-lattice-schematics}
\end{figure}

\begin{figure}
\includegraphics[width=0.7\textwidth]{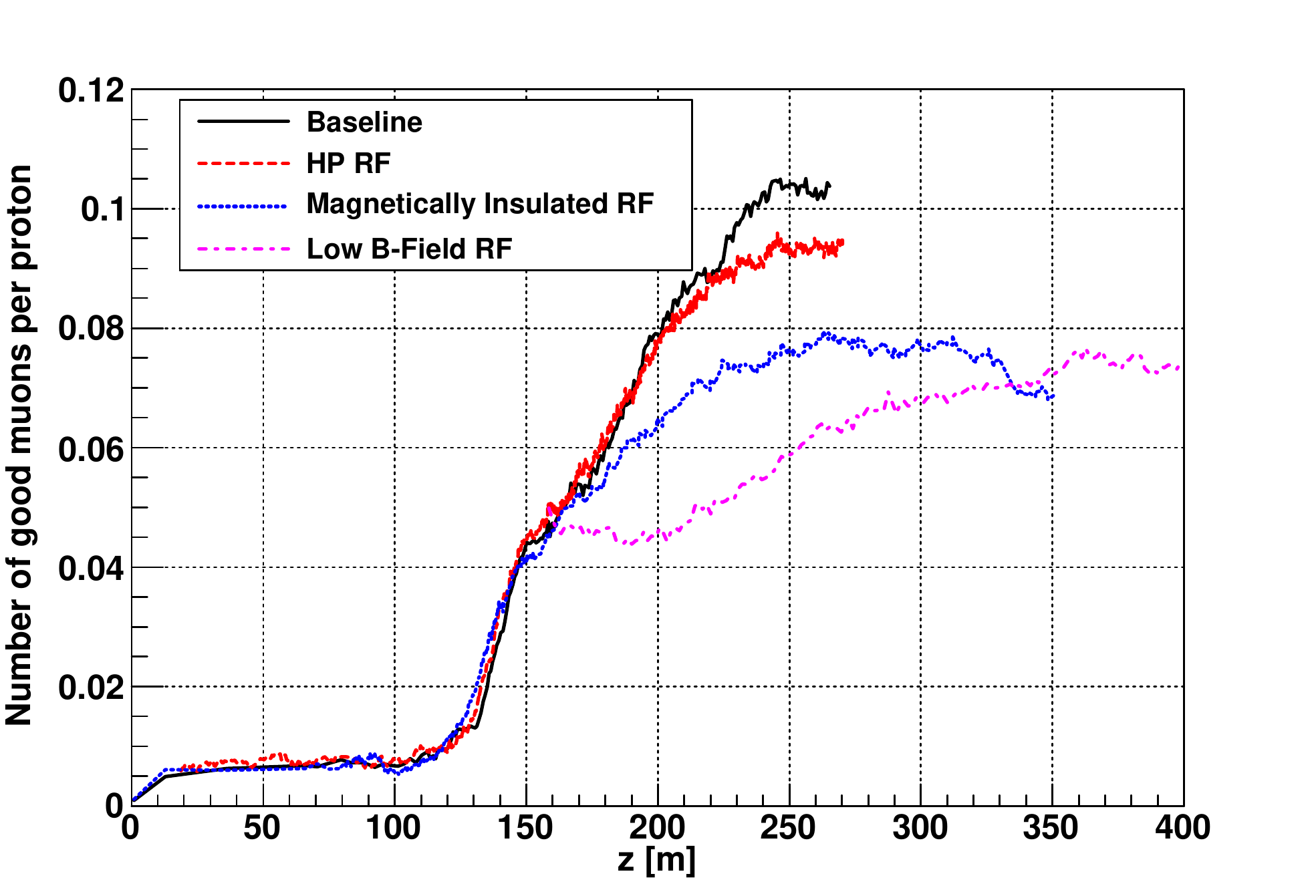}
\caption{Comparison of muon capture performance of the proposed
  alternative lattices and the baseline outlined above. A `good' muon is
  one that falls within a momentum range of $\pm$ 100 MeV/c of the
  nominal momentum, that has longitudinal amplitude $A_L^2 <$ 0.15 m
  and has transverse amplitude squared $A_\perp^2 < $ 0.030 m.}
\label{fig:acc:fe:alternate-lattice-performance}
\end{figure}

The presence of magnetic fields overlapping RF cavities has been
identified as a technical risk that may reduce the capture efficiency
of the muon front-end due to a reduction in the peak gradient that can
be achieved in the RF cavities~\cite{Moretti:2005zz,Palmer:2009zza}.
Three alternative lattices have been proposed
for the cooling section to mitigate this risk. A schematic of the
three lattices and the number of muons accepted as a function of
distance along the channel is shown in
figure \ref{fig:acc:fe:alternate-lattice-schematics} and
\ref{fig:acc:fe:alternate-lattice-performance}~\cite{Alekou:2010zza}.
The principal characteristics of the three alternative lattices are:
\begin{enumerate}
\item RF cavities filled with high pressure (HP) gas have been
  demonstrated to operate without degradation even in the presence of
  strong magnetic fields. A study of a hybrid system using high
  pressure gas has shown that the baseline performance can be achieved
  even in the presence of strong magnetic fields
  \cite{Gallardo:2010td}.
  The cavities will be tested in the presence of beam and the
  gas-ionisation effect on RF breakdown will be studied in the near
  future;
\item Magnetically insulated cavities are designed such that magnetic
  field lines are parallel to the RF cavity surface, so that field
  emission electrons are redirected onto the RF cavity surface before
  they are accelerated by the RF field. It is hoped that this may
  prevent breakdown~\cite{Stratakis:2010zz}; and
\item In the shielded lattice, the magnetic field on the RF cavities
  is reduced by stretching the lattice and adding shielding around the
  coils. This reduces the acceptance and performance of the system but
  much of the performance can be recovered by reducing the geometric
  emittance of the beam by means of an initial acceleration after the
  phase rotation system~\cite{Rogers:2009zzc}.
\end{enumerate}

\section{Replacement of the RLA arcs by a compact FFAG lattice}

\begin{figure}
  \begin{center}
    \includegraphics[width=0.85\textwidth]{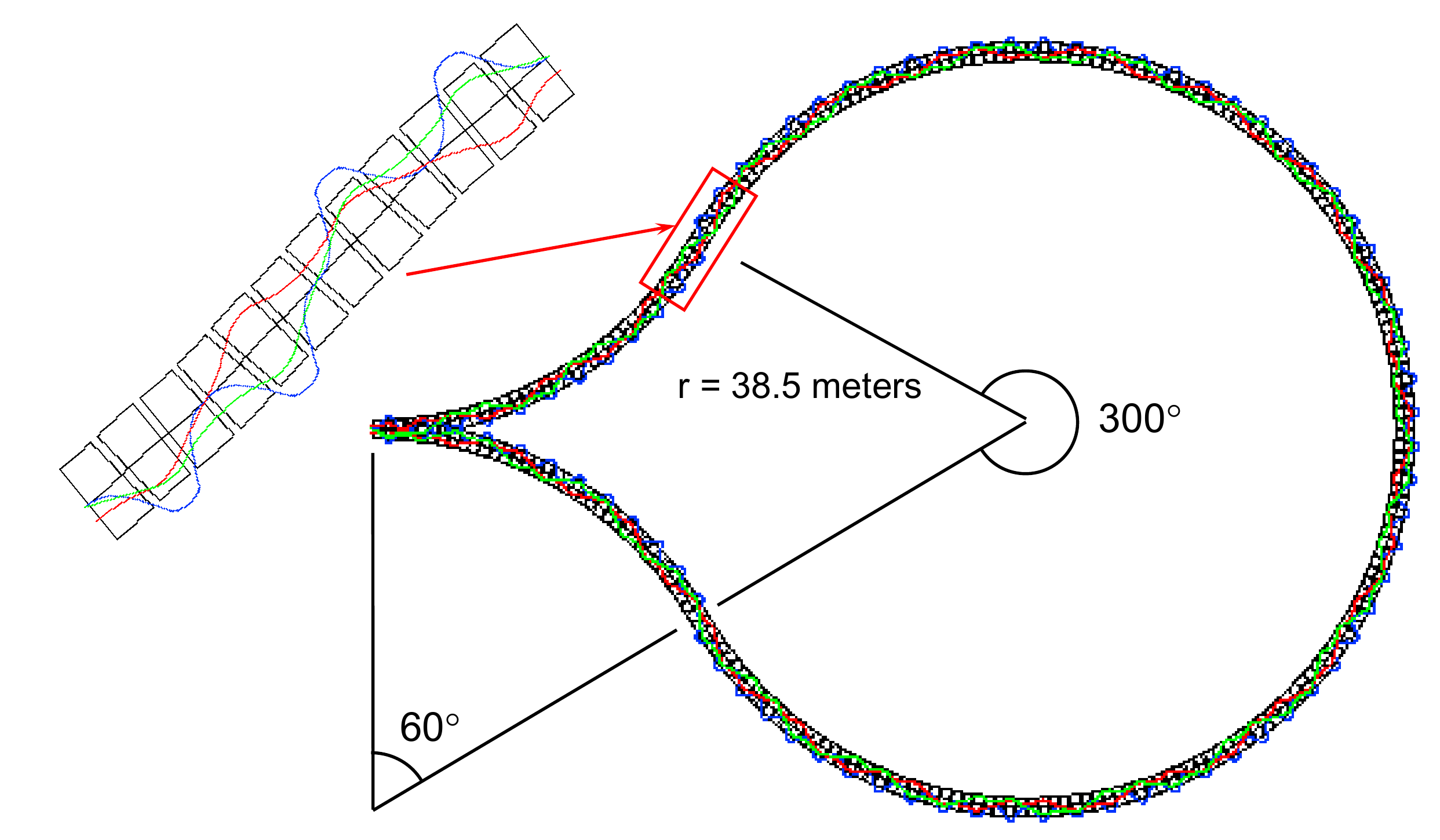}
  \end{center}
\caption{Layout and building-block-cell of a droplet-shaped return arc
  based on ns-FFAG optics.}
\label{fig:acc:rla:ffag_ab1}
\end{figure}

We developed a complete linear optics for a 4-pass muon dog-bone RLA
consisting of a single linac and two droplet-shaped return arcs at
each end of the linac.  
In order to reduce the 
number of required return arcs, we employ the  
non-scaling fixed-field alternating-gradient (NS-FFAG) arc lattice 
design, which allows transport of two 
consecutive passes with very different energies through the same string 
of arc magnets. 
As illustrated in figure \ref{fig:acc:rla:ffag_ab1}, each droplet arc
consists of a 60$^\circ$ outward bend, a 300$^\circ$ inward bend and
another 60$^\circ$ outward bend so that the net bend is 180$^\circ$. 
Such an arc geometry has the advantage that the outward and inward
bends are made up of similar cells and the geometry automatically
closes without the need for any additional straight sections, thus
making it simpler and more compact.
In the scheme presently under consideration, a 0.9~GeV/c muon beam is
injected at the middle of a linac with 0.6 GeV energy gain per
pass. 
Therefore, with a four-pass scheme, there are 1.2 and 2.4 
GeV/c passes through one arc and 1.8 and 3.0 GeV/c passes through the 
other.  In addition to accommodating 
the appropriate momenta, each arc must satisfy the following
requirements:
(a) for each momentum transported, the offset of the periodic orbit
must be zero at the entrance and the exit to the arc to ensure that
the beam goes through the centre of the linac; (b) the arc must be
achromatic for each  momentum to guarantee matching to the linac; (c)
the arc must be mirror symmetric, so that $\mu^+$ and $\mu^-$ can pass
through the same lattice in opposite directions; (d) the arc must be
nearly isochronous for both energies to ensure the proper phasing with
the linac RF; and (e) the size of the orbit offsets, beta functions,
and dispersion should be small enough to keep the aperture size
reasonable.

We used a ns-FFAG triplet magnet arrangement as the basic cell of the
arc design.  
The inward-bending triplet 
cell has an outward-bending combined-function magnet with positive gradient 
(horizontally focusing) at the 
centre and two inward-bending magnets located on either side with equal 
negative gradients. The outward-bending triplet-cell has the same
structure but reversed dipole fields.
The cell symmetry ensures that their 
periodic solutions have $\alpha_x$ = $\alpha_y$ = 0 and $D^\prime_x$ = 0 
at the beginning and the 
end. Since perturbative-method codes 
are not suitable for large-momentum-offset cases, we studied the NS-FFAG 
optics using the Polymorphic 
Tracking Code (PTC) module of the MAD-X program \cite{Schmidt:2006sa}, which
allows symplectic integration through all elements with user control
over the precision (with the full or expanded Hamiltonian).

Since the two arc designs are very similar, we focus on the details 
of the optics of the 1.2\,GeV/c and the 2.4\,GeV/c arcs. 
We choose the lower 1.2 GeV/c momentum as the nominal momentum going
through the magnet centres. 
The constraint that the lower-energy periodic orbit must have zero
offset coming in and out of the cell is then automatically
satisfied. Besides, once the 1.2 GeV/c linear optics is adjusted using
quadrupole gradients, introduction of the sextupole and octupole
magnetic field components required for accommodating the 2.4~GeV/c
momentum does not change it. This decouples the 1.2 GeV/c linear
optics from the 2.4~GeV/c optics 
and ensures that, once the 
$D_x$ = 0 and $D^\prime_x$ = 0 
conditions are met for 1.2~GeV/c, they 
are not affected by the tuning 
of the 2.4~GeV/c optics.

\begin{figure}
  \begin{center}
    \includegraphics[width=0.85\textwidth]{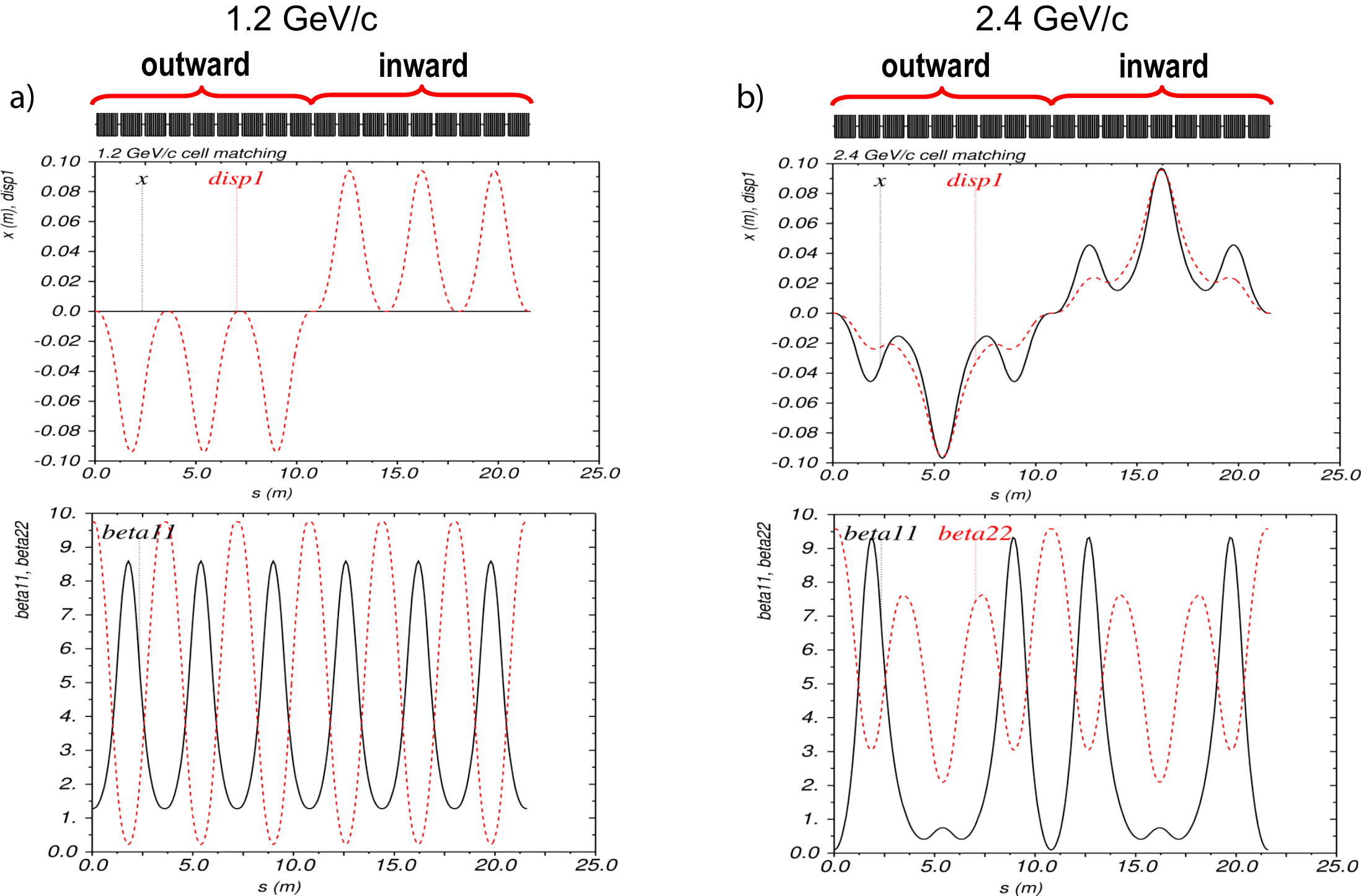}
  \end{center}
  \caption{
    Periodic orbit, dispersion, and beta functions for the
    inward-outward bending super-cell transition at 1.2\,GeV/c and
    2.4\,GeV/c, the two momenta of interest.
  }
  \label{fig:acc:rla:ffag_ab2}
\end{figure}

The triplet is composed of 1 m long magnets separated by 20~cm 
gaps with each magnet's bending angle being 5$^\circ$. 
The quadrupole gradients are adjusted to make the cell 
achromatic. 
Figure \ref{fig:acc:rla:ffag_ab2}a shows the 1.2~GeV/c 
periodic orbit and dispersion, as well as the beta functions for 
the inward- outward-bending cells. In this case, 
the Bogacz-Lebedev 
$\beta_{11}$ and $\beta_{22}$
are equal to the usual horizontal 
and vertical beta-functions, respectively. 
The outward-bending cell has the same beta functions but its
dispersion has opposite sign. 
With a single triplet, one is not able to satisfy, at the same time,
the zero orbit-offset and achromatic conditions for 2.4~GeV/c optics. 
Therefore, three triplet cells are combined into a ``super-cell''. 
Sextupole and octupole field components are 
introduced in the middle magnets of the triplets while maintaining 
the super-cell's overall mirror symmetry. 
By tuning the sextupole components, we simultaneously satisfy the
conditions of the super-cell being achromatic and having zero incoming
and outgoing periodic-orbit offsets. 
The octupole components are adjusted, keeping their strengths the same
until transverse stability is reached in both dimensions. 
Figure \ref{fig:acc:rla:ffag_ab2}b shows the 2.4~GeV/c periodic orbit,
dispersion, and the beta functions for the inward- and outward-bending
super-cells. 
In the outward-bending super-cell, the dipole and sextupole fields are 
reversed. 
The beta functions in the outward-bending super-cell remain the same
while the periodic orbit and dispersion change sign. Since both the
inward and outward-bending super-cells are achromatic and have zero 
incoming and outgoing periodic orbit offsets, it is 
clear that the super-cells are automatically matched at both 
energies. 
Since the net bend of each super-cell is 
15$^\circ$, they can be combined easily to form the 60$^\circ$ outward
and 300$^\circ$ inward bends of the droplet arc.

Finally, the linac optics needs to be matched to both arcs for 
all passes simultaneously by adjusting the strengths of the linac
fixed-field quadrupoles. 
The new solution was obtained by modifying the bi-sected linac profile
where the quadrupole strengths increase linearly from the linac's
centre toward the ends. 
Investigation of the complete non-linear dynamics 
aiming to optimise the RLA's dynamic aperture is 
under study, as well as phase control to maintain RF synchronisation.

\section{Scaling FFAG replacing the final RLA stage}
\label{App:FFAG}

The possibility of using the scaling type of FFAG ring for muon
acceleration has already been proposed \cite{Kuno:2001tb}, but it was
done assuming very low RF frequencies, incompatible with a frequency
of the order of 200~MHz or higher. 
Here we present a scheme based on stationary-bucket acceleration
\cite{Mori:2006zn} to accelerate muon beams in a scaling FFAG ring
with an RF frequency of 200~MHz. 

\subsection{Muon Beam Acceleration}

The example of a 3.6~GeV to 12.6~GeV ring with parameters given in
table \ref{tab:acc:sffag:muonrpara} is considered. 
It is assumed that scaling FFAG magnets with a maximum field of
about 4~T are used. 
This is a reasonable assumption once superconducting magnets with a
left-right asymmetric coil distribution \cite{Nakamoto:2004nj} are
employed to realise the scaling-field law.  
In order to allow the simultaneous acceleration of $\mu^+$ and $\mu^-$
beams, the path length per cell of the synchronous particle is
adjusted to be a multiple of $\frac{1}{2}\beta_s \lambda_{RF}$, with
$\beta_s$ the ratio of the synchronous particle velocity to the speed
of light, and $\lambda_{RF}$ the RF wavelength. 
\begin{table}
    \caption{Ring parameters.}
    \label{tab:acc:sffag:muonrpara}
    \begin{center}
    \begin{tabular}{|l|r|}
\hline
	Lattice type		& scaling FFAG FDF triplet\\
	Injection energy & 3.6 GeV\\
	Extraction energy & 12.6 GeV\\
	RF frequency & 200 MHz\\
	Mean radius		& $\sim$ 160.9 m \\
	Synchronous energy (kinetic) & 8.04 GeV\\
	Harmonic number $h$ & 675\\
	Number of cells 	& 225\\
	Field index $k$		& 1\,390\\
	RF peak voltage (per turn)& 1.8 GV\\
	Number of turns & 6\\
	$B_{max}$ (at 12.6 GeV)	& 3.9~T\\
	Drift length & $\sim$ 1.5~m\\
        	Horizontal phase advance per cell & 85.86 deg.\\
	Vertical phase advance per cell &  33.81 deg.\\
	Excursion & 14.3 cm\\ 
\hline 
    \end{tabular}
    \end{center}
\end{table}

We use stepwise particle-tracking in a geometrical field model with
Enge-type fringe field \cite{Enge:1967} to study the beam dynamics. 
Results of single-particle tracking at fixed energy show a normalised
transverse acceptance larger than 30~$\pi$.mm.rad for both horizontal
and vertical planes.

\subsubsection{6D Simulation of a Whole Acceleration Cycle}
\label{sec:acc:sffag:6Dmu}

At the beginning of 6D tracking, the bunch of particles is prepared as
follows: 1\,000 particles are uniformly distributed inside a
transverse, 4D ellipsoid (Water-bag distribution) and uniformly inside
an ellipse in the longitudinal plane. 
The initial normalised bunch emittances are 30~$\pi$.mm.rad in both
horizontal and vertical planes and 150~mm in the longitudinal plane.

\begin{figure}
  \begin{center}
    \includegraphics[width=8cm]{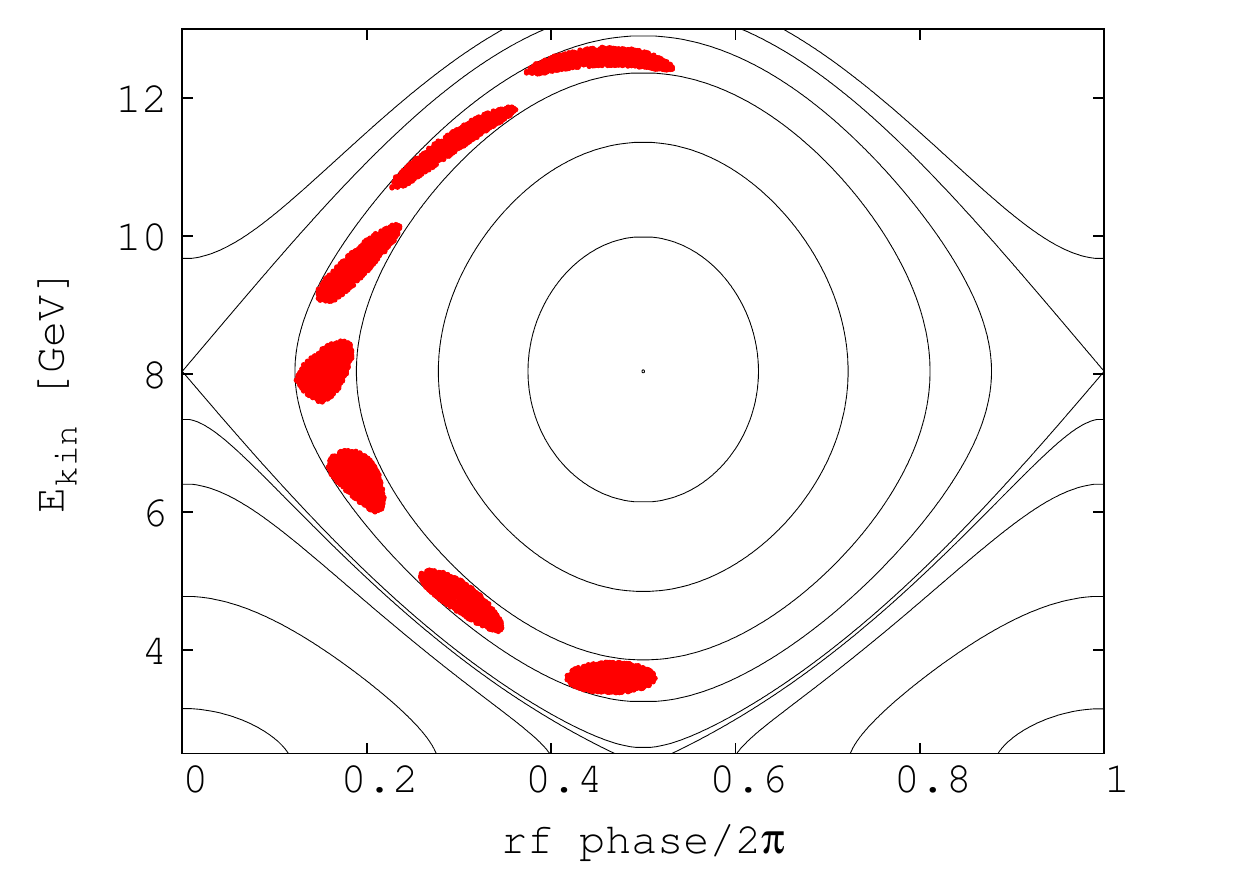}
  \end{center}
  \caption{
    6-turn acceleration cycle plotted in the longitudinal phase
    space. Hamiltonian contours are superimposed.
  } 
  \label{fig:acc:sffag:longit_mu_6d}
\end{figure}

\begin{figure}
  \begin{minipage}[b]{.5\linewidth}
    \includegraphics[width=8cm]{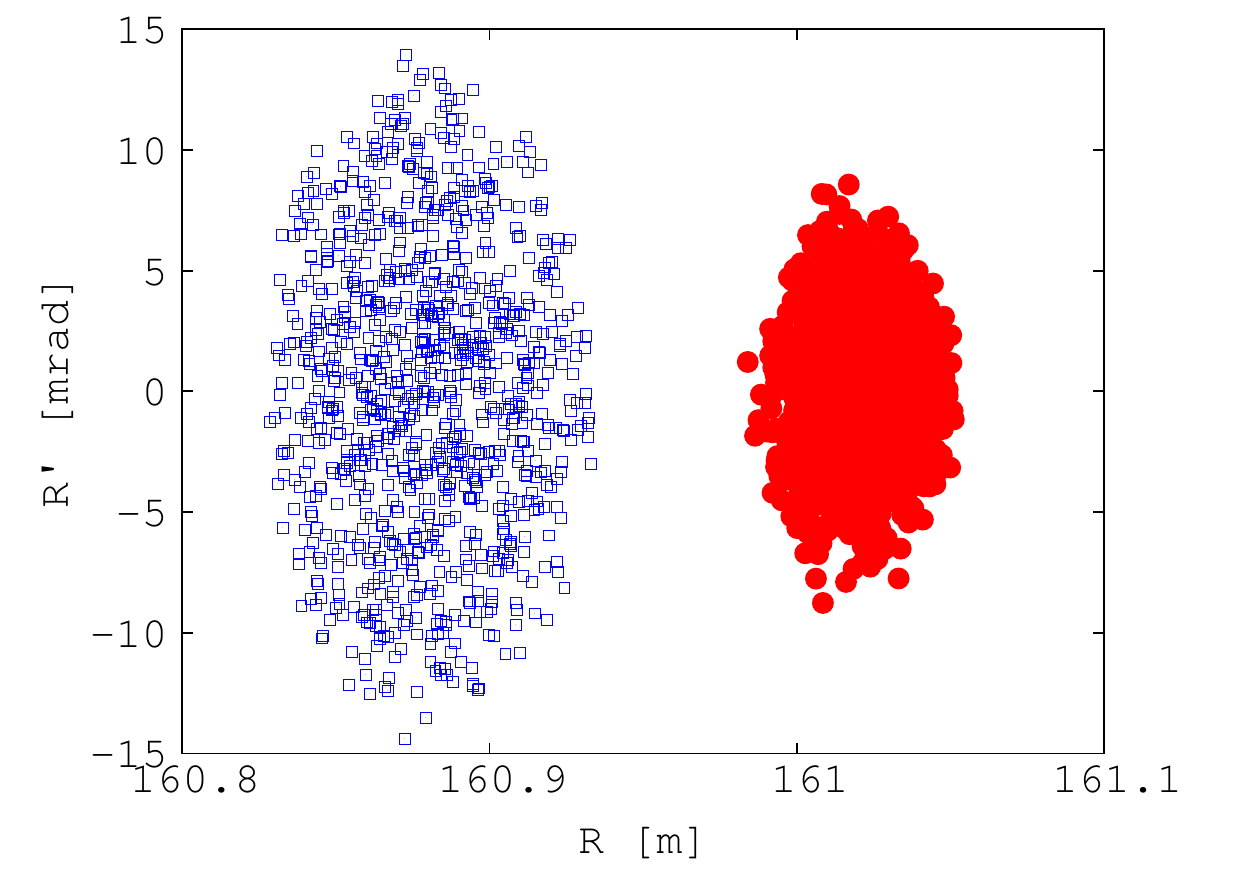}
  \end{minipage} \hfill
  \begin{minipage}[b]{.5\linewidth}
    \includegraphics[width=8cm]{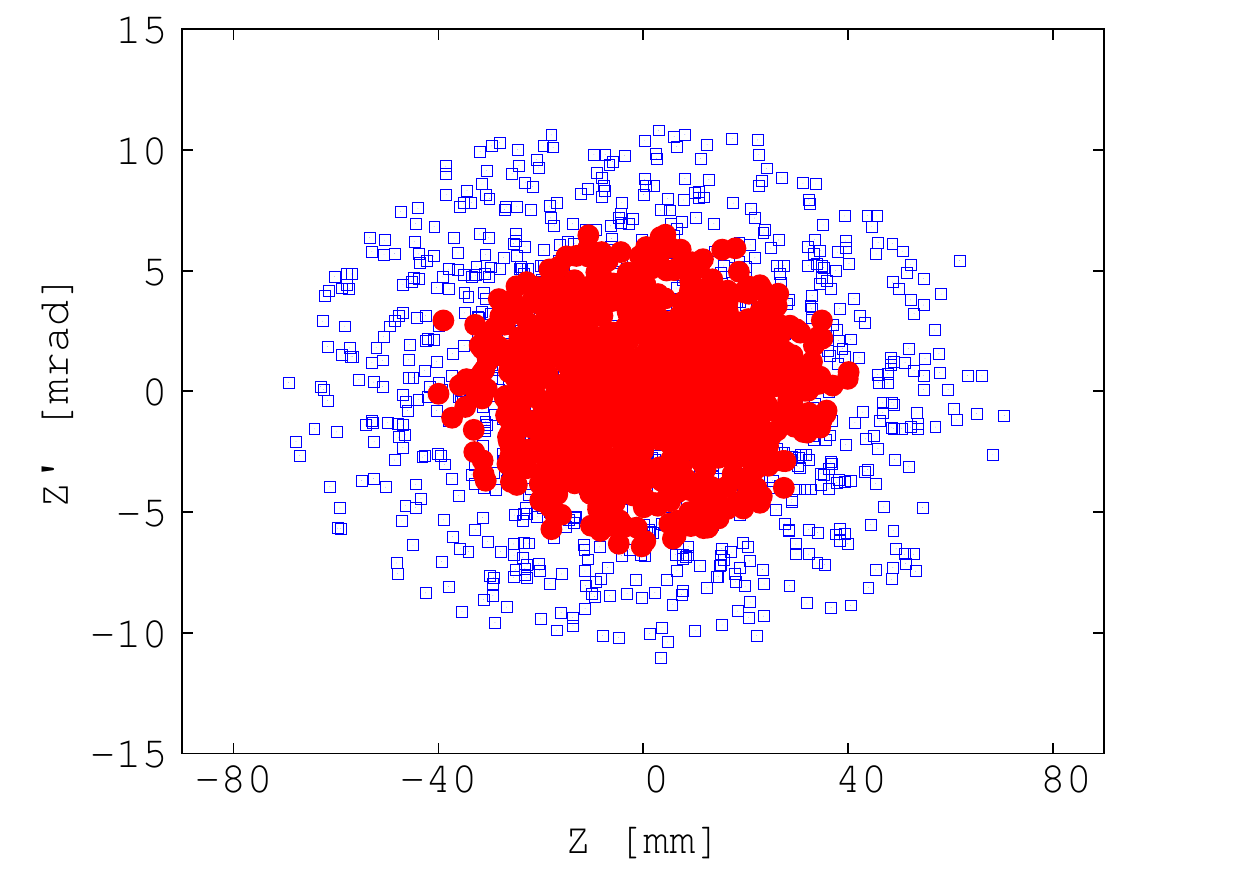}
  \end{minipage}
  \caption{
    Initial (blue squares) and final (red dots) particle positions
    plotted in the horizontal (left figure) and vertical (right
    figure) phase spaces.
  } 
  \label{fig:acc:sffag:trans_mu_6d}
\end{figure}

Tracking results show no beam lost during the acceleration cycle.
No significant emittance blow-up is observed in either the
longitudinal (see figure \ref{fig:acc:sffag:longit_mu_6d}) or the
transverse planes (see figure \ref{fig:acc:sffag:trans_mu_6d}).  

Tracking introducing random alignment errors in the lattice has been
performed. 
No beam loss has been observed with an rms alignment error smaller
than 1~mm. 

\subsection{Summary}

A scheme to accelerate muon beams inside the stationary RF bucket of
scaling FFAG rings using 200~MHz RF cavities has been proposed. 
The ring can accelerate both $\mu^+$ and $\mu^-$ beams simultaneously. 
Acceleration is performed within 6 turns, the RF is used in an efficient
manner. 
A detailed tracking-study of the example of a 3.6~GeV to 12.6~GeV muon
ring has been presented.  
6D particle tracking in a soft-edge-field model shows that the
acceptance of this scheme is larger than 30~$\pi$.mm.rad in both
horizontal and vertical planes, and larger than 150~mm in the
longitudinal plane.  
No beam loss and no significant emittance blow-up was observed in
neither the transverse nor the longitudinal planes. 
This scheme also shows a good tolerance to errors.

\section{Chromaticity correction for the linear non-scaling FFAG}

A particle with large betatron amplitude has a different
time-of-flight because the path length is longer. 
The effect exists in any ring accelerator, but turns out to be the
crucial issue in a muon FFAG for the following reasons.
Firstly, the transverse emittance of a muon beam is huge, even after
ionisation cooling. 
Secondly, a muon FFAG accelerates the beam outside the RF bucket
with an almost isochronous lattice.  
Energy gain is sensitive to the RF phase at each turn since there are
no synchrotron oscillations which average the different energy
gains at each turn.  
As a result, the longitudinal emittance blows up significantly after
acceleration in a linear, ns-FFAG \cite{Machida:2006sj}.  
Furthermore, a scheme to use an FFAG cascade to boost the muon energy
is not possible. 

\begin{figure}
  \centering
  \includegraphics[width=0.5\linewidth]{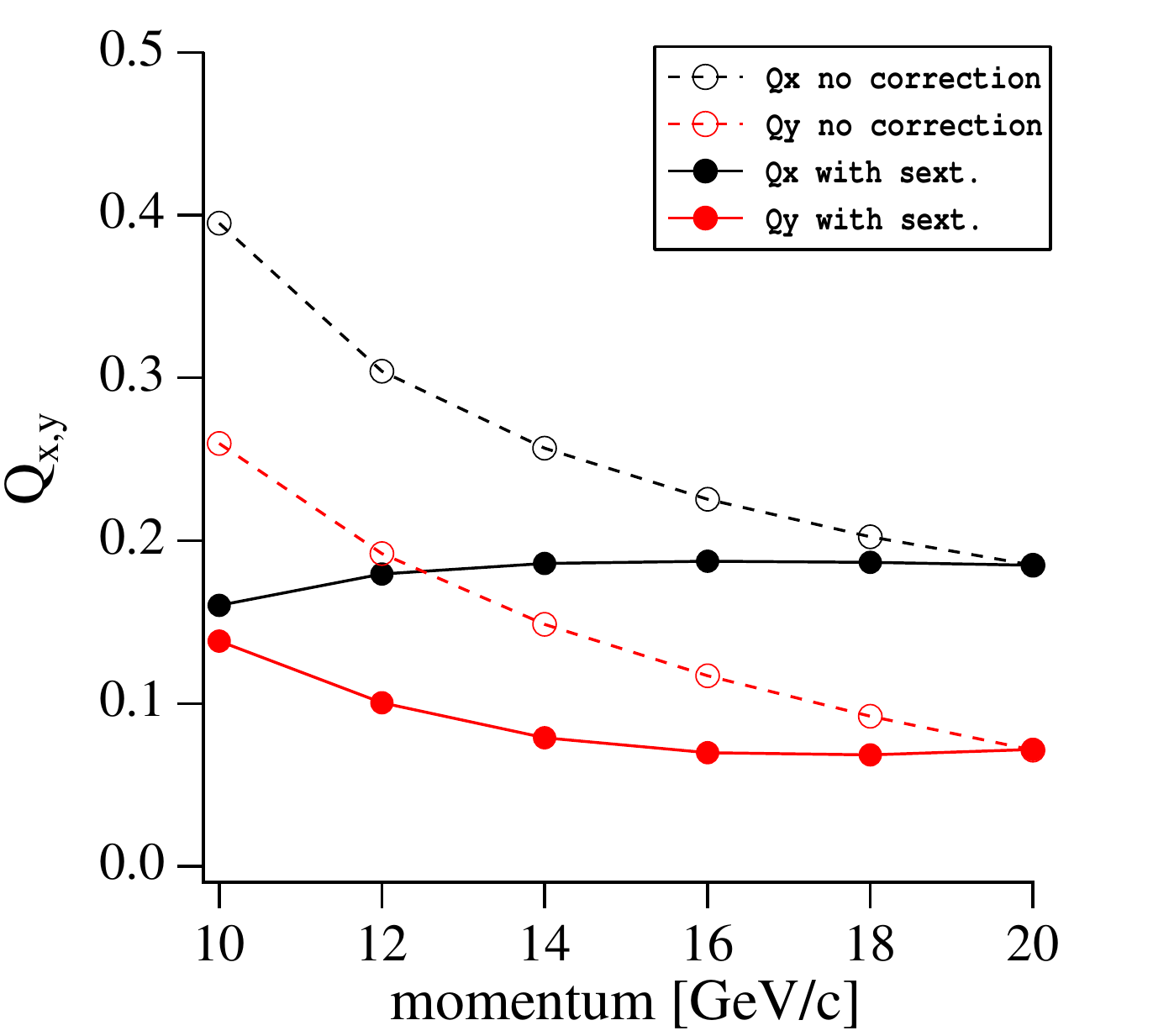}
  \caption{Tune as a function of momentum for a ns-FFAG
    without (dashed lines/open circles) and with (solid lines/filled circles)
    sextupole components to correct chromaticity.}
  \label{fig:app:acc:ffag:1}
\end{figure}

\begin{figure}
  \centering
  \includegraphics[width=0.5\linewidth]{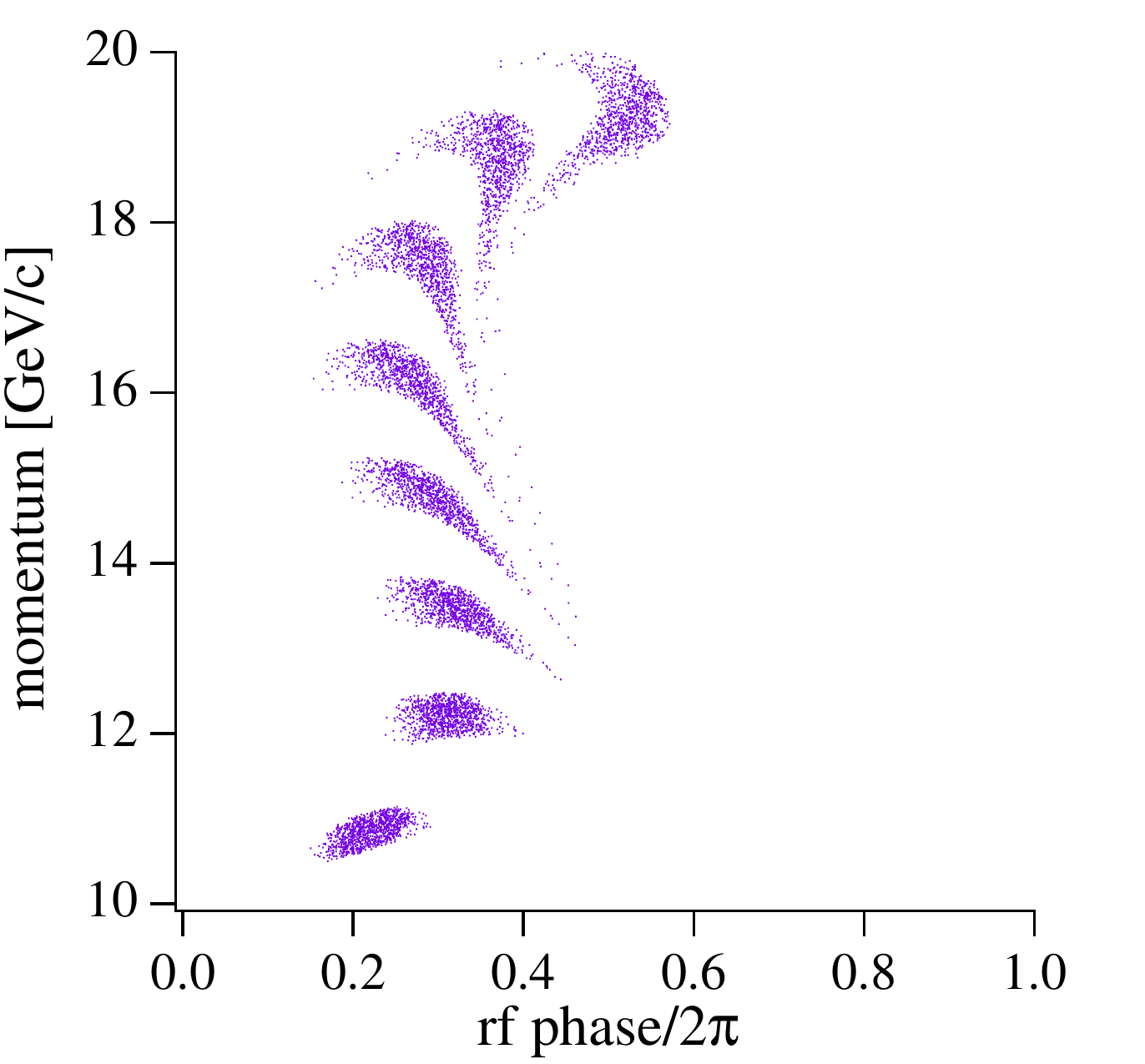}%
  \includegraphics[width=0.5\linewidth]{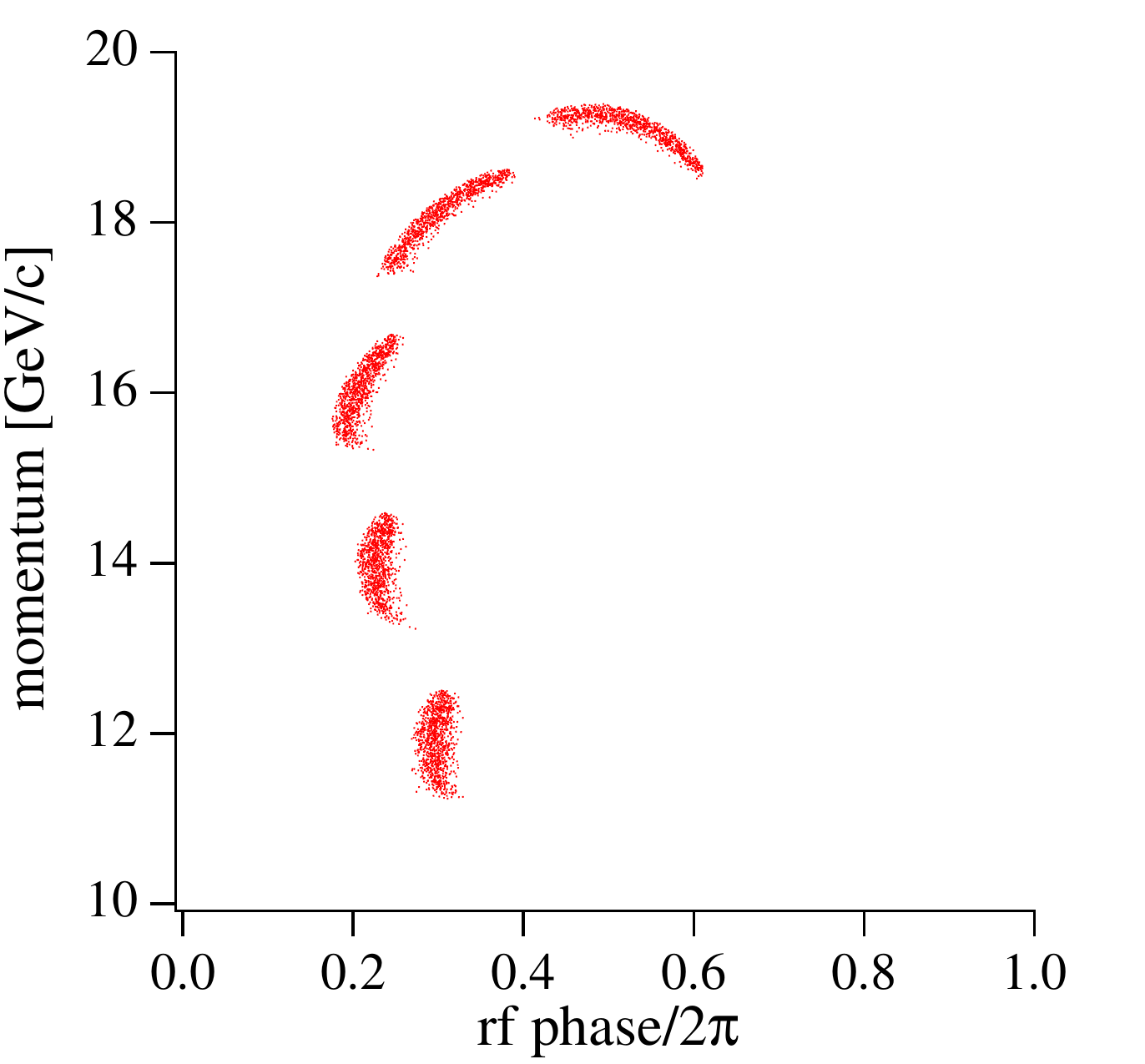}
  \caption{Longitudinal phase space on successive turns with (left) and
    without (right) chromaticity correction.}
  \label{fig:app:acc:ffag:2}
\end{figure}

It was found that chromaticity correction mitigates the
problem \cite{Berg:2007}. 
Sextupole fields introduce amplitude-dependent revolution orbits, which
cancel the increase of time-of-flight experienced by large amplitude
oscillations. 
In a chromaticity-corrected lattice as shown in figure
\ref{fig:app:acc:ffag:1}, longitudinal-emittance blowup can be
eliminated as shown in  figure \ref{fig:app:acc:ffag:2}.

\begin{figure}
  \centering
  \includegraphics[width=0.5\linewidth]{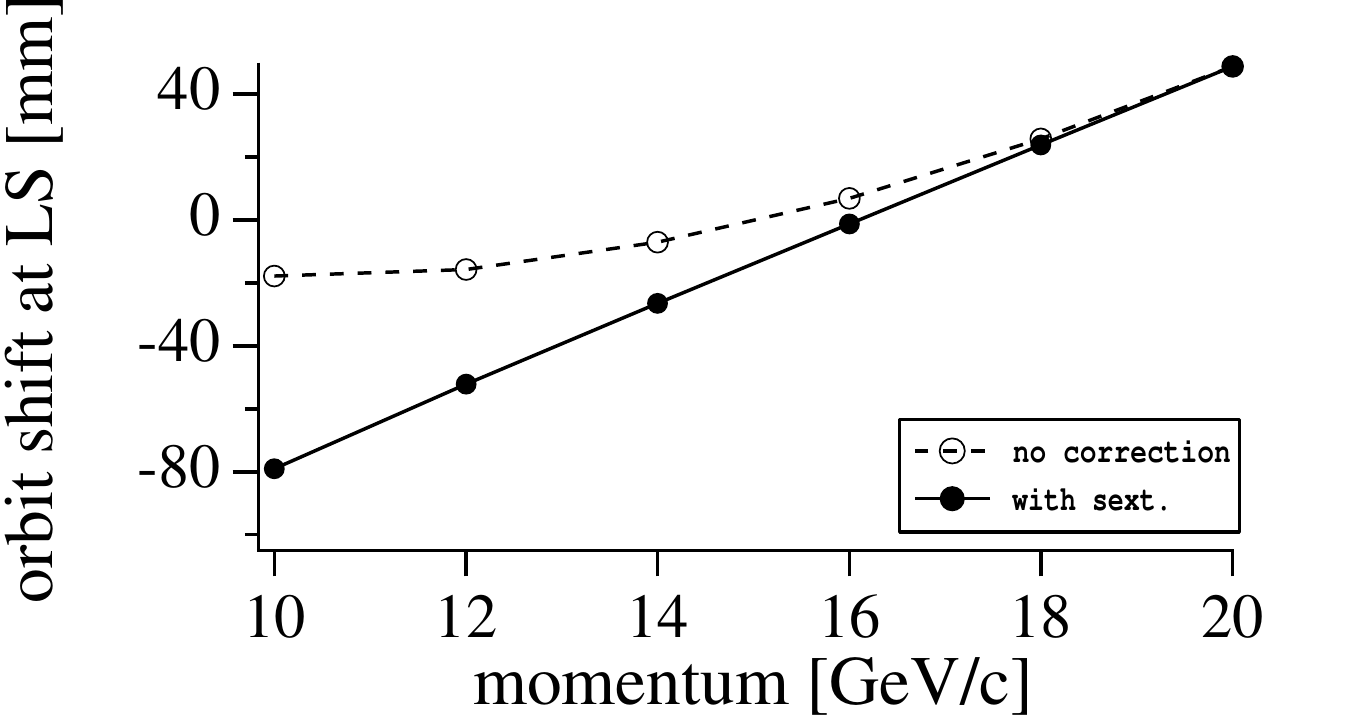}%
  \includegraphics[width=0.5\linewidth]{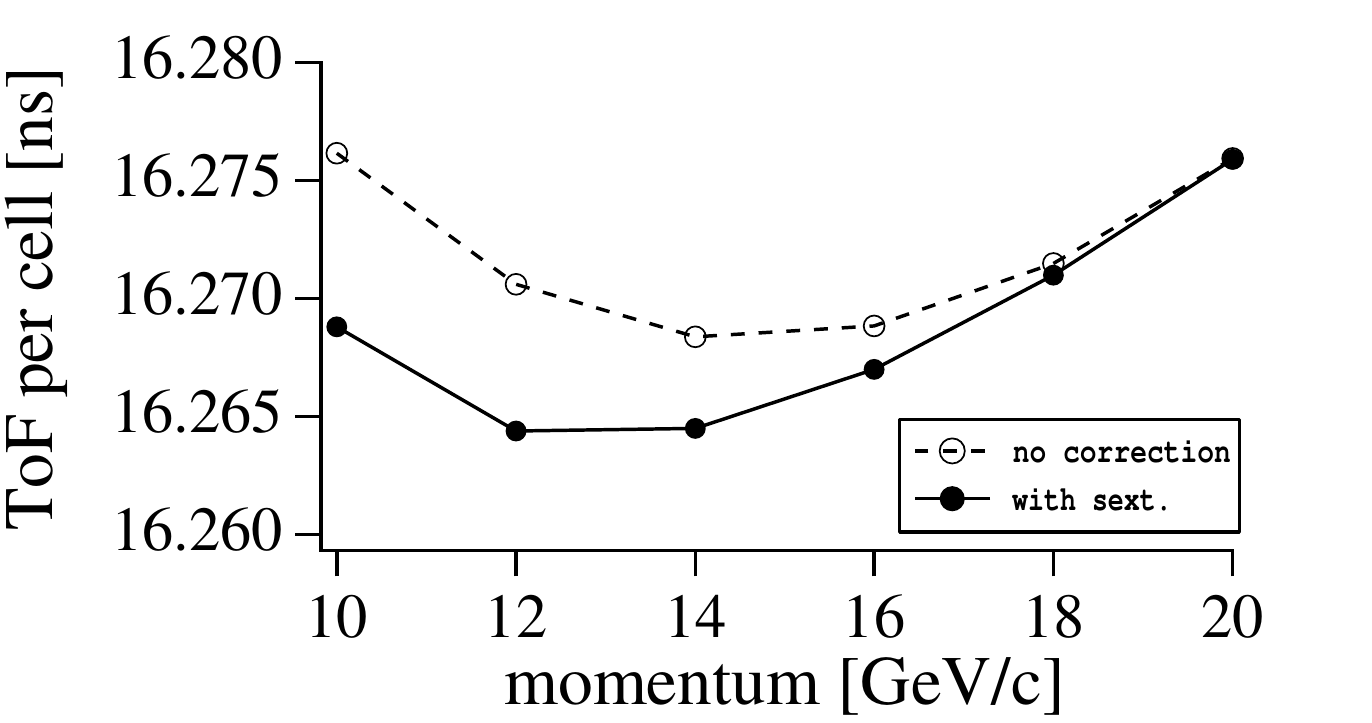}
  \caption{
    Left: closed position in the long straight (LS) versus momentum
    without (dashed lines, open circles) and with (solid lines, filled
    circles) chromaticity correction.  
    Right: time of flight as a function of momentum without and with
    chromaticity correction.
  } 
  \label{fig:app:acc:ffag:4}
\end{figure}

On the other hand, there is a price to be paid.
In the chromaticity corrected lattice, the orbit shift due to
acceleration is large, as shown in figure \ref{fig:app:acc:ffag:4}.
This directly increases the magnet aperture and the cost. 
Another side-effect is the asymmetry of the time-of-flight curve shown
in figure \ref{fig:app:acc:ffag:4}. 
This implies an increase of total RF voltage.  
Finally, the most significant impact of the chromaticity correction is
the reduction of the dynamic aperture. 
Non-linearity introduced to correct chromaticity makes the dynamic
aperture smaller than 30~mm (dynamic aperture is defined the maximum
transverse amplitude which can survive during acceleration).
As a result of these adverse effects, chromaticity corrections are
not included in the present muon FFAG lattice.

\section{Response Matrices for MIND analysis}
\label{app:response}
This appendix summarises the response matrices of signal (wrong sign $\nu_\mu$ and $\overline{\nu}_\mu$ appearance) and all backgrounds ($\overline{\nu}_\mu$, $\nu_\mu$ CC, $\overline{\nu}_e$, $\nu_e$ CC and NC) in bins of true and reconstructed neutrino energy, relevant to an oscillation analysis. Each entry in the table is the survival probability for each species. In all tables, columns represent the true neutrino energy in GeV and rows the reconstructed energy, also in GeV. The overflow bin in reconstructed energy represents all events with a reconstructed energy greater than the known maximum.
\vspace{0.25cm}

{\large {\bf $\nu_\mu$ appearance} }
\vspace{0.25cm}

\hspace{1.0cm}    {\large Signal Efficiency }

\begin{table}[!ht]
  \caption{
    $\nu_\mu$ appearance detection efficiency. 
    All values $\times 10^{-4}$.
  }
  \begin{center}
  \begin{tabular}{|l||c|c|c|c|c|c|c|c|c|c|c|}
    \hline
    & {\tiny 0-2.0} & {\tiny 2.0-3.0} & {\tiny 3.0-4.0} & {\tiny 4.0-5.0} & {\tiny 5.0-6.0} & {\tiny 6.0-8.0} & {\tiny 8.0-10.0} & {\tiny 10.0-12.5} & {\tiny 12.5-15.0} & {\tiny 15.0-20.0} & {\tiny 20.0-25.0}\\
    \hline
    \hline
    {\tiny 0-2.0} & {\scriptsize211.8} & {\scriptsize331.0} & {\scriptsize48.55} & {\scriptsize13.92} & {\scriptsize5.73} & {\scriptsize2.371} & {\scriptsize0.790} & {\scriptsize0.377} & {\scriptsize0.142} & {\scriptsize0.076} & {\scriptsize0.039}\\
\hline
{\tiny 2.0-3.0} & {\scriptsize154.4} & {\scriptsize1379} & {\scriptsize522.0} & {\scriptsize118.0} & {\scriptsize42.03} & {\scriptsize15.18} & {\scriptsize4.903}
& {\scriptsize1.916} & {\scriptsize1.049} & {\scriptsize0.572} & {\scriptsize0.156}\\
\hline
{\tiny 3.0-4.0} & {\scriptsize45.17} & {\scriptsize906.6} & {\scriptsize1408} & {\scriptsize554.5} & {\scriptsize145.6} & {\scriptsize41.82} & {\scriptsize13.36}
& {\scriptsize4.334} & {\scriptsize2.211} & {\scriptsize0.934} & {\scriptsize0.429}\\
\hline
{\tiny 4.0-5.0} & {\scriptsize17.93} & {\scriptsize422.1} & {\scriptsize1377} & {\scriptsize1352} & {\scriptsize544.5} & {\scriptsize105.4} & {\scriptsize26.26}
& {\scriptsize8.951} & {\scriptsize3.288} & {\scriptsize1.449} & {\scriptsize0.722}\\
\hline
{\tiny 5.0-6.0} & {\scriptsize7.509} & {\scriptsize155.6} & {\scriptsize744.7} & {\scriptsize1476} & {\scriptsize1233} & {\scriptsize318.5} & {\scriptsize49.26}
& {\scriptsize14.82} & {\scriptsize4.535} & {\scriptsize2.307} & {\scriptsize0.761}\\
\hline
{\tiny 6.0-8.0} & {\scriptsize2.017} & {\scriptsize70.17} & {\scriptsize421.2} & {\scriptsize1469} & {\scriptsize2729} & {\scriptsize2010} & {\scriptsize447.4} & {\scriptsize71.04} & {\scriptsize18.31} & {\scriptsize7.302} & {\scriptsize3.025}\\
\hline
{\tiny 8.0-10.0} & {\scriptsize0.112} & {\scriptsize7.406} & {\scriptsize46.89} & {\scriptsize230.1} & {\scriptsize850.6} & {\scriptsize2371} & {\scriptsize1823} & {\scriptsize380.8} & {\scriptsize56.63} & {\scriptsize13.65} & {\scriptsize5.212}\\
\hline
{\tiny 10.05-12.5} & {\scriptsize0} & {\scriptsize2.134} & {\scriptsize12.61} & {\scriptsize37.15} & {\scriptsize134.9} & {\scriptsize963.9} & {\scriptsize2756} & {\scriptsize2011} & {\scriptsize416.5} & {\scriptsize48.96} & {\scriptsize11.50}\\
\hline
{\tiny 12.5-15.0} & {\scriptsize0} & {\scriptsize0.126} & {\scriptsize2.739} & {\scriptsize11.20} & {\scriptsize27.05} & {\scriptsize127.4} & {\scriptsize886.8} & {\scriptsize2519} & {\scriptsize1788} & {\scriptsize242.5} & {\scriptsize22.84}\\
\hline
{\tiny 15.0-20.0} & {\scriptsize0} & {\scriptsize0.126} & {\scriptsize0.498} & {\scriptsize2.346} & {\scriptsize7.930} & {\scriptsize36.72} & {\scriptsize173.2} & {\scriptsize1207} & {\scriptsize3621} & {\scriptsize2840} & {\scriptsize340.5}\\
\hline
{\tiny 20.0-25.0} & {\scriptsize0} & {\scriptsize0} & {\scriptsize0.083} & {\scriptsize0.076} & {\scriptsize0.088} & {\scriptsize1.186} & {\scriptsize11.52} & {\scriptsize61.27} & {\scriptsize380.2} & {\scriptsize2665} & {\scriptsize2562}\\
\hline
{\tiny overflow} & {\scriptsize0} & {\scriptsize0} & {\scriptsize0} & {\scriptsize0} & {\scriptsize0} & {\scriptsize0.514} & {\scriptsize0.757} & {\scriptsize5.685} & {\scriptsize31.86} & {\scriptsize477.2} & {\scriptsize3338}\\
\hline
  \end{tabular}
  \end{center}
\end{table}
\newpage

\hspace{1.0cm}       {\large $\overline{\nu}_\mu$ CC background }

\begin{table}[!ht]
  \begin{center}
  \begin{tabular}{|l||c|c|c|c|c|c|c|c|c|c|c|}
    \hline
    & {\tiny 0-2.0} & {\tiny 2.0-3.0} & {\tiny 3.0-4.0} & {\tiny 4.0-5.0} & {\tiny 5.0-6.0} & {\tiny 6.0-8.0} & {\tiny 8.0-10.0} & {\tiny 10.0-12.5} & {\tiny 12.5-15.0} & {\tiny 15.0-20.0} & {\tiny 20.0-25.0}\\
    \hline
    \hline
    {\tiny 0-2.0} & {\scriptsize0} & {\scriptsize0} & {\scriptsize0.096} & {\scriptsize0} & {\scriptsize0} & {\scriptsize0.041} & {\scriptsize0} & {\scriptsize0} & {\scriptsize0} & {\scriptsize0} & {\scriptsize0}\\
\hline
{\tiny 2.0-3.0} & {\scriptsize0.346} & {\scriptsize0.546} & {\scriptsize0} & {\scriptsize0.083} & {\scriptsize0} & {\scriptsize0} & {\scriptsize0} & {\scriptsize0} & {\scriptsize0} & {\scriptsize0.039} & {\scriptsize0.020}\\
\hline
{\tiny 3.0-4.0} & {\scriptsize0.173} & {\scriptsize0.273} & {\scriptsize0.096} & {\scriptsize0} & {\scriptsize0} & {\scriptsize0.084} & {\scriptsize0} & {\scriptsize0.065} & {\scriptsize0.029} & {\scriptsize0.077} & {\scriptsize0.020}\\
\hline
{\tiny 4.0-5.0} & {\scriptsize0.173} & {\scriptsize0.546} & {\scriptsize0.192} & {\scriptsize0.166} & {\scriptsize0} & {\scriptsize0.042} & {\scriptsize0.068} & {\scriptsize0.065} & {\scriptsize0} & {\scriptsize0} & {\scriptsize0.020}\\
\hline
{\tiny 5.0-6.0} & {\scriptsize0} & {\scriptsize0.273} & {\scriptsize0.384} & {\scriptsize0.083} & {\scriptsize0.187} & {\scriptsize0.042} & {\scriptsize0.034} & {\scriptsize0} & {\scriptsize0.029} & {\scriptsize0.039} & {\scriptsize0.039}\\
\hline
{\tiny 6.0-8.0} & {\scriptsize0} & {\scriptsize0.273} & {\scriptsize0.288} & {\scriptsize0.249} & {\scriptsize0.468} & {\scriptsize0.209} & {\scriptsize0.068}
& {\scriptsize0.097} & {\scriptsize0.115} & {\scriptsize0.096} & {\scriptsize0.039}\\
\hline
{\tiny 8.0-10.0} & {\scriptsize0} & {\scriptsize0} & {\scriptsize0.096} & {\scriptsize0} & {\scriptsize0} & {\scriptsize0.084} & {\scriptsize0.137} & {\scriptsize0.129} & {\scriptsize0.086} & {\scriptsize0.116} & {\scriptsize0.020}\\
\hline
{\tiny 10.0-12.5} & {\scriptsize0} & {\scriptsize0} & {\scriptsize0} & {\scriptsize0} & {\scriptsize0.094} & {\scriptsize0.042} & {\scriptsize0.205} & {\scriptsize0.032} & {\scriptsize0.086} & {\scriptsize0.019} & {\scriptsize0.138}\\
\hline
{\tiny 12.5-15.0} & {\scriptsize0} & {\scriptsize0} & {\scriptsize0} & {\scriptsize0.083} & {\scriptsize0.094} & {\scriptsize0} & {\scriptsize0.102} & {\scriptsize0}
& {\scriptsize0.058} & {\scriptsize0.058} & {\scriptsize0.079}\\
\hline
{\tiny 15.0-20.0} & {\scriptsize0} & {\scriptsize0} & {\scriptsize0} & {\scriptsize0} & {\scriptsize0} & {\scriptsize0.084} & {\scriptsize0.068} & {\scriptsize0.065}
& {\scriptsize0.029} & {\scriptsize0.116} & {\scriptsize0.079}\\
\hline
{\tiny 20.0-25.0} & {\scriptsize0} & {\scriptsize0} & {\scriptsize0} & {\scriptsize0} & {\scriptsize0} & {\scriptsize0.042} & {\scriptsize0} & {\scriptsize0.032} & {\scriptsize0.086} & {\scriptsize0.058} & {\scriptsize0.059}\\
\hline
{\tiny overflow} & {\scriptsize0} & {\scriptsize0} & {\scriptsize0} & {\scriptsize0} & {\scriptsize0} & {\scriptsize0} & {\scriptsize0} & {\scriptsize0} & {\scriptsize0} & {\scriptsize0.019} & {\scriptsize0.098}\\
\hline
  \end{tabular}
  \end{center}
  \caption{\emph{$\overline{\nu}_\mu$ CC background. All values $\times 10^{-4}$.}}
\end{table}
\vspace{0.25cm}

\hspace{1.0cm}       {\large $\nu_e$ CC background }

\begin{table}[!ht]
  \begin{center}
  \begin{tabular}{|l||c|c|c|c|c|c|c|c|c|c|c|}
    \hline
    & {\tiny 0-2.0} & {\tiny 2.0-3.0} & {\tiny 3.0-4.0} & {\tiny 4.0-5.0} & {\tiny 5.0-6.0} & {\tiny 6.0-8.0} & {\tiny 8.0-10.0} & {\tiny 10.0-12.5} & {\tiny 12.5-15.0} & {\tiny 15.0-20.0} & {\tiny 20.0-25.0}\\
    \hline
    \hline
    {\tiny 0-2.0} & {\scriptsize0} & {\scriptsize0} & {\scriptsize0} & {\scriptsize0} & {\scriptsize0} & {\scriptsize0} & {\scriptsize0} & {\scriptsize0} & {\scriptsize0.020} &
{\scriptsize0.013} & {\scriptsize0}\\
\hline
{\tiny 2.0-3.0} & {\scriptsize0} & {\scriptsize0} & {\scriptsize0} & {\scriptsize0} & {\scriptsize0} & {\scriptsize0} & {\scriptsize0.023} & {\scriptsize0.045} & {\scriptsize0.040} & {\scriptsize0} & {\scriptsize0}\\
\hline
{\tiny 3.0-4.0} & {\scriptsize0} & {\scriptsize0} & {\scriptsize0} & {\scriptsize0} & {\scriptsize0} & {\scriptsize0} & {\scriptsize0} & {\scriptsize0} & {\scriptsize0} & {\scriptsize0.013} & {\scriptsize0.055}\\
\hline
{\tiny 4.0-5.0} & {\scriptsize0} & {\scriptsize0} & {\scriptsize0} & {\scriptsize0} & {\scriptsize0} & {\scriptsize0.028} & {\scriptsize0} & {\scriptsize0} & {\scriptsize0}
& {\scriptsize0} & {\scriptsize0.014}\\
\hline
{\tiny 5.0-6.0} & {\scriptsize0} & {\scriptsize0} & {\scriptsize0} & {\scriptsize0} & {\scriptsize0} & {\scriptsize0} & {\scriptsize0} & {\scriptsize0} & {\scriptsize0} & {\scriptsize0} & {\scriptsize0.014}\\
\hline
{\tiny 6.0-8.0} & {\scriptsize0} & {\scriptsize0} & {\scriptsize0} & {\scriptsize0} & {\scriptsize0} & {\scriptsize0} & {\scriptsize0} & {\scriptsize0} & {\scriptsize0.020}
& {\scriptsize0} & {\scriptsize0}\\
\hline
{\tiny 8.0-10.0} & {\scriptsize0} & {\scriptsize0} & {\scriptsize0} & {\scriptsize0} & {\scriptsize0} & {\scriptsize0} & {\scriptsize0} & {\scriptsize0} & {\scriptsize0} & {\scriptsize0} & {\scriptsize0}\\
\hline
{\tiny 10.0-12.5} & {\scriptsize0} & {\scriptsize0} & {\scriptsize0} & {\scriptsize0.054} & {\scriptsize0} & {\scriptsize0} & {\scriptsize0} & {\scriptsize0} & {\scriptsize0} & {\scriptsize0} & {\scriptsize0}\\
\hline
{\tiny 12.5-15.0} & {\scriptsize0} & {\scriptsize0} & {\scriptsize0} & {\scriptsize0} & {\scriptsize0} & {\scriptsize0} & {\scriptsize0} & {\scriptsize0} & {\scriptsize0} & {\scriptsize0} & {\scriptsize0}\\
\hline
{\tiny 15.0-20.0} & {\scriptsize0} & {\scriptsize0} & {\scriptsize0} & {\scriptsize0} & {\scriptsize0} & {\scriptsize0} & {\scriptsize0} & {\scriptsize0.022} & {\scriptsize0.020} & {\scriptsize0} & {\scriptsize0}\\
\hline
{\tiny 20.0-25.0} & {\scriptsize0} & {\scriptsize0} & {\scriptsize0} & {\scriptsize0} & {\scriptsize0} & {\scriptsize0} & {\scriptsize0} & {\scriptsize0} & {\scriptsize0} & {\scriptsize0} & {\scriptsize0}\\
\hline
{\tiny overflow} & {\scriptsize0} & {\scriptsize0} & {\scriptsize0} & {\scriptsize0} & {\scriptsize0} & {\scriptsize0} & {\scriptsize0} & {\scriptsize0} & {\scriptsize0} & {\scriptsize0} & {\scriptsize0}\\
\hline
  \end{tabular}
  \end{center}
  \caption{\emph{$\nu_e$ CC background. All values $\times 10^{-4}$.}}
\end{table}
\newpage

\hspace{1.0cm}       {\large NC background }

\begin{table}[!ht]
  \begin{center}
  \begin{tabular}{|l||c|c|c|c|c|c|c|c|c|c|c|}
    \hline
    & {\tiny 0-2.0} & {\tiny 2.0-3.0} & {\tiny 3.0-4.0} & {\tiny 4.0-5.0} & {\tiny 5.0-6.0} & {\tiny 6.0-8.0} & {\tiny 8.0-10.0} & {\tiny 10.0-12.5} & {\tiny 12.5-15.0} & {\tiny 15.0-20.0} & {\tiny 20.0-25.0}\\
    \hline
    \hline
    {\tiny 0-2.0} & {\scriptsize0} & {\scriptsize0} & {\scriptsize0} & {\scriptsize0} & {\scriptsize0.035} & {\scriptsize0} & {\scriptsize0.026} & {\scriptsize0.050} & {\scriptsize0.022} & {\scriptsize0.015} & {\scriptsize0}\\
\hline
{\tiny 2.0-3.0} & {\scriptsize0} & {\scriptsize0} & {\scriptsize0.034} & {\scriptsize0.061} & {\scriptsize0.070} & {\scriptsize0.016} & {\scriptsize0.052} &
{\scriptsize0.050} & {\scriptsize0.034} & {\scriptsize0.030} & {\scriptsize0.039}\\
\hline
{\tiny 3.0-4.0} & {\scriptsize0} & {\scriptsize0} & {\scriptsize0} & {\scriptsize0.061} & {\scriptsize0.070} & {\scriptsize0} & {\scriptsize0.052} & {\scriptsize0.076} & {\scriptsize0.056} & {\scriptsize0.068} & {\scriptsize0.055}\\
\hline
{\tiny 4.0-5.0} & {\scriptsize0} & {\scriptsize0} & {\scriptsize0.034} & {\scriptsize0.030} & {\scriptsize0.106} & {\scriptsize0.157} & {\scriptsize0.210} & {\scriptsize0.088} & {\scriptsize0.146} & {\scriptsize0.106} & {\scriptsize0.086}\\
\hline
{\tiny 5.0-6.0} & {\scriptsize0} & {\scriptsize0} & {\scriptsize0} & {\scriptsize0.030} & {\scriptsize0.035} & {\scriptsize0.031} & {\scriptsize0.026} & {\scriptsize0.076} & {\scriptsize0.067} & {\scriptsize0.030} & {\scriptsize0.023}\\
\hline
{\tiny 6.0-8.0} & {\scriptsize0} & {\scriptsize0} & {\scriptsize0} & {\scriptsize0} & {\scriptsize0} & {\scriptsize0.094} & {\scriptsize0.105} & {\scriptsize0.113} & {\scriptsize0.067} & {\scriptsize0.046} & {\scriptsize0.047}\\
\hline
{\tiny 8.0-10.0} & {\scriptsize0} & {\scriptsize0} & {\scriptsize0} & {\scriptsize0} & {\scriptsize0.035} & {\scriptsize0.047} & {\scriptsize0.026} & {\scriptsize0.088} & {\scriptsize0.011} & {\scriptsize0.015} & {\scriptsize0.023}\\
\hline
{\tiny 10.0-12.5} & {\scriptsize0} & {\scriptsize0} & {\scriptsize0} & {\scriptsize0} & {\scriptsize0} & {\scriptsize0} & {\scriptsize0.013} & {\scriptsize0.025} & {\scriptsize0.034} & {\scriptsize0.046} & {\scriptsize0.055}\\
\hline
{\tiny 12.5-15.0} & {\scriptsize0} & {\scriptsize0} & {\scriptsize0} & {\scriptsize0} & {\scriptsize0} & {\scriptsize0} & {\scriptsize0} & {\scriptsize0} & {\scriptsize0.034} & {\scriptsize0.015} & {\scriptsize0.008}\\
\hline
{\tiny 15.0-20.0} & {\scriptsize0} & {\scriptsize0} & {\scriptsize0} & {\scriptsize0} & {\scriptsize0} & {\scriptsize0} & {\scriptsize0} & {\scriptsize0.013} & {\scriptsize0.011} & {\scriptsize0.030} & {\scriptsize0.016}\\
\hline
{\tiny 20.0-25.0} & {\scriptsize0} & {\scriptsize0} & {\scriptsize0} & {\scriptsize0} & {\scriptsize0} & {\scriptsize0} & {\scriptsize0} & {\scriptsize0} & {\scriptsize0} & {\scriptsize0} & {\scriptsize0.008}\\
\hline
{\tiny overflow} & {\scriptsize0} & {\scriptsize0} & {\scriptsize0} & {\scriptsize0} & {\scriptsize0} & {\scriptsize0} & {\scriptsize0} & {\scriptsize0} & {\scriptsize0} & {\scriptsize0} & {\scriptsize0}\\
\hline
  \end{tabular}
  \end{center}
  \caption{\emph{NC background. All values $\times 10^{-4}$.}}
\end{table}
\newpage
{\large {\bf $\overline{\nu}_\mu$ appearance } }

\hspace{1.0cm}       {\large Signal Efficiency }

\begin{table}[!ht]
  \begin{center}
  \begin{tabular}{|l||c|c|c|c|c|c|c|c|c|c|c|}
    \hline
    & {\tiny 0-2.0} & {\tiny 2.0-3.0} & {\tiny 3.0-4.0} & {\tiny 4.0-5.0} & {\tiny 5.0-6.0} & {\tiny 6.0-8.0} & {\tiny 8.0-10.0} & {\tiny 10.0-12.5} & {\tiny 12.5-15.0} & {\tiny 15.0-20.0} & {\tiny 20.0-25.0}\\
    \hline
    \hline
    {\tiny 0-2.0} & {\scriptsize169.5} & {\scriptsize228.4} & {\scriptsize47.07} & {\scriptsize9.966} & {\scriptsize3.558} & {\scriptsize1.170} & {\scriptsize0.648} & {\scriptsize0.485} & {\scriptsize0.173} & {\scriptsize0.077} & {\scriptsize0}\\
\hline
{\tiny 2.0-3.0} & {\scriptsize378.0} & {\scriptsize1007} & {\scriptsize392.8} & {\scriptsize103.1} & {\scriptsize24.72} & {\scriptsize7.939} & {\scriptsize2.594} & {\scriptsize1.260} & {\scriptsize0.490} & {\scriptsize0.270} & {\scriptsize0.157}\\
\hline
{\tiny 3.0-4.0} & {\scriptsize310.3} & {\scriptsize1762} & {\scriptsize1234} & {\scriptsize435.1} & {\scriptsize115.1} & {\scriptsize22.06} & {\scriptsize4.983}
& {\scriptsize2.197} & {\scriptsize0.951} & {\scriptsize0.481} & {\scriptsize0.216}\\
\hline
{\tiny 4.0-5.0} & {\scriptsize97.66} & {\scriptsize1271} & {\scriptsize2188} & {\scriptsize1306} & {\scriptsize448.7} & {\scriptsize79.27} & {\scriptsize12.66} & {\scriptsize3.521} & {\scriptsize1.729} & {\scriptsize0.828} & {\scriptsize0.452}\\
\hline
{\tiny 5.0-6.0} & {\scriptsize24.41} & {\scriptsize419.5} & {\scriptsize1627} & {\scriptsize2009} & {\scriptsize1188} & {\scriptsize258.6} & {\scriptsize30.55}
& {\scriptsize8.561} & {\scriptsize2.825} & {\scriptsize1.367} & {\scriptsize0.748}\\
\hline
{\tiny 6.0-8.0} & {\scriptsize6.926} & {\scriptsize146.9} & {\scriptsize895.2} & {\scriptsize2620} & {\scriptsize3790} & {\scriptsize2173} & {\scriptsize386.3}
& {\scriptsize51.72} & {\scriptsize12.80} & {\scriptsize4.622} & {\scriptsize2.715}\\
\hline
{\tiny 8.0-10.0} & {\scriptsize0.346} & {\scriptsize12.01} & {\scriptsize87.04} & {\scriptsize400.3} & {\scriptsize1451} & {\scriptsize3242} & {\scriptsize2012} & {\scriptsize354.2} & {\scriptsize48.22} & {\scriptsize12.81} & {\scriptsize5.371}\\
\hline
{\tiny 10.0-12.5} & {\scriptsize0.173} & {\scriptsize3.274} & {\scriptsize25.27} & {\scriptsize70.34} & {\scriptsize224.4} & {\scriptsize1460} & {\scriptsize3617} & {\scriptsize2256} & {\scriptsize393.9} & {\scriptsize49.64} & {\scriptsize15.19}\\
\hline
{\tiny 12.5-15.0} & {\scriptsize0} & {\scriptsize0.136} & {\scriptsize5.283} & {\scriptsize26.99} & {\scriptsize55.52} & {\scriptsize199.2} & {\scriptsize1243} & {\scriptsize3171} & {\scriptsize1971} & {\scriptsize249.2} & {\scriptsize29.33}\\
\hline
{\tiny 15.0-20.0} & {\scriptsize0} & {\scriptsize0.136} & {\scriptsize0.384} & {\scriptsize2.907} & {\scriptsize14.79} & {\scriptsize68.24} & {\scriptsize273.2} &
{\scriptsize1634} & {\scriptsize4485} & {\scriptsize3151} & {\scriptsize357.7}\\
\hline
{\tiny 20.0-25.0} & {\scriptsize0} & {\scriptsize0} & {\scriptsize0} & {\scriptsize0} & {\scriptsize0.281} & {\scriptsize1.713} & {\scriptsize19.83} & {\scriptsize95.11} & {\scriptsize528.3} & {\scriptsize3212} & {\scriptsize2766}\\
\hline
{\tiny overflow} & {\scriptsize0} & {\scriptsize0} & {\scriptsize0} & {\scriptsize0} & {\scriptsize0} & {\scriptsize0.209} & {\scriptsize1.058} & {\scriptsize7.980} & {\scriptsize44.73} & {\scriptsize631.9} & {\scriptsize3982}\\
\hline
  \end{tabular}
  \end{center}
  \caption{\emph{$\overline{\nu}_\mu$ appearance detection efficiency. All values $\times 10^{-4}$}}
\end{table}

\hspace{1.0cm}       {\large $\nu_\mu$ CC background }

\begin{table}[!ht]
  \begin{center}
  \begin{tabular}{|l||c|c|c|c|c|c|c|c|c|c|c|}
    \hline
    & {\tiny 0-2.0} & {\tiny 2.0-3.0} & {\tiny 3.0-4.0} & {\tiny 4.0-5.0} & {\tiny 5.0-6.0} & {\tiny 6.0-8.0} & {\tiny 8.0-10.0} & {\tiny 10.0-12.5} & {\tiny 12.5-15.0} & {\tiny 15.0-20.0} & {\tiny 20.0-25.0}\\
    \hline
    \hline
    {\tiny 0-2.0} & {\scriptsize0} & {\scriptsize0.251} & {\scriptsize0} & {\scriptsize0} & {\scriptsize0} & {\scriptsize0} & {\scriptsize0} & {\scriptsize0} & {\scriptsize0} & {\scriptsize0} & {\scriptsize0}\\
\hline
{\tiny 2.0-3.0} & {\scriptsize0.112} & {\scriptsize0.251} & {\scriptsize0} & {\scriptsize0} & {\scriptsize0} & {\scriptsize0} & {\scriptsize0} & {\scriptsize0} & {\scriptsize0} & {\scriptsize0} & {\scriptsize0.039}\\
\hline
{\tiny 3.0-4.0} & {\scriptsize0.224} & {\scriptsize0.126} & {\scriptsize0} & {\scriptsize0.076} & {\scriptsize0} & {\scriptsize0} & {\scriptsize0} & {\scriptsize0} & {\scriptsize0} & {\scriptsize0} & {\scriptsize0.020}\\
\hline
{\tiny 4.0-5.0} & {\scriptsize0} & {\scriptsize0.628} & {\scriptsize0.249} & {\scriptsize0.076} & {\scriptsize0.176} & {\scriptsize0.079} & {\scriptsize0} & {\scriptsize0} & {\scriptsize0} & {\scriptsize0} & {\scriptsize0}\\
\hline
{\tiny 5.0-6.0} & {\scriptsize0} & {\scriptsize0} & {\scriptsize0.166} & {\scriptsize0.378} & {\scriptsize0.088} & {\scriptsize0.040} & {\scriptsize0.033} & {\scriptsize0} & {\scriptsize0.028} & {\scriptsize0} & {\scriptsize0}\\
\hline
{\tiny 6.0-8.0} & {\scriptsize0} & {\scriptsize0.251} & {\scriptsize0.249} & {\scriptsize0.605} & {\scriptsize0.529} & {\scriptsize0.119} & {\scriptsize0.329}
& {\scriptsize0.126} & {\scriptsize0.057} & {\scriptsize0.095} & {\scriptsize0.098}\\
\hline
{\tiny 8.0-10.0} & {\scriptsize0} & {\scriptsize0.126} & {\scriptsize0} & {\scriptsize0.151} & {\scriptsize0.176} & {\scriptsize0.316} & {\scriptsize0.099} & {\scriptsize0.063} & {\scriptsize0.028} & {\scriptsize0.057} & {\scriptsize0.078}\\
\hline
{\tiny 10.0-12.5} & {\scriptsize0} & {\scriptsize0} & {\scriptsize0} & {\scriptsize0.076} & {\scriptsize0.088} & {\scriptsize0.277} & {\scriptsize0.066} & {\scriptsize0.063} & {\scriptsize0.142} & {\scriptsize0.076} & {\scriptsize0.098}\\
\hline
{\tiny 12.5-15.0} & {\scriptsize0} & {\scriptsize0} & {\scriptsize0} & {\scriptsize0} & {\scriptsize0.088} & {\scriptsize0.119} & {\scriptsize0.197} & {\scriptsize0.126} & {\scriptsize0.113} & {\scriptsize0.057} & {\scriptsize0.098}\\
\hline
{\tiny 15.0-20.0} & {\scriptsize0} & {\scriptsize0} & {\scriptsize0} & {\scriptsize0} & {\scriptsize0} & {\scriptsize0} & {\scriptsize0.033} & {\scriptsize0.063} & {\scriptsize0.085} & {\scriptsize0.038} & {\scriptsize0.078}\\
\hline
{\tiny 20.0-25.0} & {\scriptsize0} & {\scriptsize0} & {\scriptsize0} & {\scriptsize0} & {\scriptsize0} & {\scriptsize0} & {\scriptsize0} & {\scriptsize0.031} & {\scriptsize0.028} & {\scriptsize0.057} & {\scriptsize0.059}\\
\hline
{\tiny overflow} & {\scriptsize0} & {\scriptsize0} & {\scriptsize0} & {\scriptsize0} & {\scriptsize0} & {\scriptsize0} & {\scriptsize0} & {\scriptsize0} & {\scriptsize0.057} & {\scriptsize0.038} & {\scriptsize0.078}\\
\hline
  \end{tabular}
  \end{center}
  \caption{\emph{$\overline{\nu}_\mu$ CC background. All values $\times 10^{-4}$.}}
\end{table}
\newpage

\hspace{1.0cm}       {\large $\overline{\nu}_e$ CC background }

\begin{table}[!ht]
  \begin{center}
  \begin{tabular}{|l||c|c|c|c|c|c|c|c|c|c|c|}
    \hline
    & {\tiny 0-2.0} & {\tiny 2.0-3.0} & {\tiny 3.0-4.0} & {\tiny 4.0-5.0} & {\tiny 5.0-6.0} & {\tiny 6.0-8.0} & {\tiny 8.0-10.0} & {\tiny 10.0-12.5} & {\tiny 12.5-15.0} & {\tiny 15.0-20.0} & {\tiny 20.0-25.0}\\
    \hline
    \hline
    {\tiny 0-2.0} & {\scriptsize0} & {\scriptsize0} & {\scriptsize0} & {\scriptsize0} & {\scriptsize0} & {\scriptsize0} & {\scriptsize0.030} & {\scriptsize0} & {\scriptsize0} &
{\scriptsize0.017} & {\scriptsize0.018}\\
\hline
{\tiny 2.0-3.0} & {\scriptsize0} & {\scriptsize0} & {\scriptsize0} & {\scriptsize0} & {\scriptsize0} & {\scriptsize0.037} & {\scriptsize0} & {\scriptsize0.029} & {\scriptsize0} & {\scriptsize0} & {\scriptsize0.053}\\
\hline
{\tiny 3.0-4.0} & {\scriptsize0} & {\scriptsize0} & {\scriptsize0} & {\scriptsize0} & {\scriptsize0} & {\scriptsize0} & {\scriptsize0} & {\scriptsize0} & {\scriptsize0} & {\scriptsize0} & {\scriptsize0}\\
\hline
{\tiny 4.0-5.0} & {\scriptsize0} & {\scriptsize0} & {\scriptsize0} & {\scriptsize0} & {\scriptsize0} & {\scriptsize0} & {\scriptsize0} & {\scriptsize0} & {\scriptsize0} & {\scriptsize0.035} & {\scriptsize0}\\
\hline
{\tiny 5.0-6.0} & {\scriptsize0} & {\scriptsize0} & {\scriptsize0} & {\scriptsize0} & {\scriptsize0} & {\scriptsize0} & {\scriptsize0} & {\scriptsize0} & {\scriptsize0.026} & {\scriptsize0} & {\scriptsize0.018}\\
\hline
{\tiny 6.0-8.0} & {\scriptsize0} & {\scriptsize0} & {\scriptsize0} & {\scriptsize0} & {\scriptsize0} & {\scriptsize0} & {\scriptsize0} & {\scriptsize0} & {\scriptsize0} & {\scriptsize0} & {\scriptsize0}\\
\hline
{\tiny 8.0-10.0} & {\scriptsize0} & {\scriptsize0} & {\scriptsize0} & {\scriptsize0} & {\scriptsize0} & {\scriptsize0} & {\scriptsize0} & {\scriptsize0} & {\scriptsize0} & {\scriptsize0.017} & {\scriptsize0}\\
\hline
{\tiny 10.0-12.5} & {\scriptsize0} & {\scriptsize0} & {\scriptsize0} & {\scriptsize0} & {\scriptsize0} & {\scriptsize0.037} & {\scriptsize0} & {\scriptsize0} & {\scriptsize0} & {\scriptsize0} & {\scriptsize0}\\
\hline
{\tiny 12.5-15.0} & {\scriptsize0} & {\scriptsize0} & {\scriptsize0} & {\scriptsize0} & {\scriptsize0} & {\scriptsize0} & {\scriptsize0} & {\scriptsize0} & {\scriptsize0} & {\scriptsize0} & {\scriptsize0}\\
\hline
{\tiny 15.0-20.0} & {\scriptsize0} & {\scriptsize0} & {\scriptsize0} & {\scriptsize0} & {\scriptsize0} & {\scriptsize0} & {\scriptsize0} & {\scriptsize0} & {\scriptsize0} & {\scriptsize0} & {\scriptsize0}\\
\hline
{\tiny 20.0-25.0} & {\scriptsize0} & {\scriptsize0} & {\scriptsize0} & {\scriptsize0} & {\scriptsize0} & {\scriptsize0} & {\scriptsize0} & {\scriptsize0} & {\scriptsize0} & {\scriptsize0} & {\scriptsize0}\\
\hline
{\tiny overflow} & {\scriptsize0} & {\scriptsize0} & {\scriptsize0} & {\scriptsize0} & {\scriptsize0} & {\scriptsize0} & {\scriptsize0} & {\scriptsize0} & {\scriptsize0} & {\scriptsize0} & {\scriptsize0}\\
\hline
  \end{tabular}
  \end{center}
  \caption{\emph{$\overline{\nu}_e$ CC background. All values $\times 10^{-4}$.}}
\end{table}
\vspace{0.25cm}

\hspace{1.0cm}       {\large NC background }

\begin{table}[!ht]
  \begin{center}
  \begin{tabular}{|l||c|c|c|c|c|c|c|c|c|c|c|}
    \hline
    & {\tiny 0-2.0} & {\tiny 2.0-3.0} & {\tiny 3.0-4.0} & {\tiny 4.0-5.0} & {\tiny 5.0-6.0} & {\tiny 6.0-8.0} & {\tiny 8.0-10.0} & {\tiny 10.0-12.5} & {\tiny 12.5-15.0} & {\tiny 15.0-20.0} & {\tiny 20.0-25.0}\\
    \hline
    \hline
    {\tiny 0-2.0} & {\scriptsize0} & {\scriptsize0} & {\scriptsize0} & {\scriptsize0} & {\scriptsize0.070} & {\scriptsize0.016} & {\scriptsize0.013} & {\scriptsize0.038} & {\scriptsize0.056} & {\scriptsize0.023} & {\scriptsize0.055}\\
\hline
{\tiny 2.0-3.0} & {\scriptsize0} & {\scriptsize0} & {\scriptsize0.068} & {\scriptsize0.030} & {\scriptsize0.070} & {\scriptsize0.094} & {\scriptsize0.026} &
{\scriptsize0.038} & {\scriptsize0.079} & {\scriptsize0.061} & {\scriptsize0.063}\\
\hline
{\tiny 3.0-4.0} & {\scriptsize0} & {\scriptsize0} & {\scriptsize0.137} & {\scriptsize0.091} & {\scriptsize0.106} & {\scriptsize0.110} & {\scriptsize0.092} & {\scriptsize0.088} & {\scriptsize0.213} & {\scriptsize0.129} & {\scriptsize0.117}\\
\hline
{\tiny 4.0-5.0} & {\scriptsize0} & {\scriptsize0} & {\scriptsize0.068} & {\scriptsize0.122} & {\scriptsize0.317} & {\scriptsize0.268} & {\scriptsize0.275} & {\scriptsize0.189} & {\scriptsize0.213} & {\scriptsize0.236} & {\scriptsize0.250}\\
\hline
{\tiny 5.0-6.0} & {\scriptsize0} & {\scriptsize0} & {\scriptsize0} & {\scriptsize0.030} & {\scriptsize0} & {\scriptsize0.063} & {\scriptsize0.079} & {\scriptsize0.101} & {\scriptsize0.067} & {\scriptsize0.084} & {\scriptsize0.055}\\
\hline
{\tiny 6.0-8.0} & {\scriptsize0} & {\scriptsize0} & {\scriptsize0} & {\scriptsize0} & {\scriptsize0.070} & {\scriptsize0.094} & {\scriptsize0.171} & {\scriptsize0.076} & {\scriptsize0.180} & {\scriptsize0.137} & {\scriptsize0.156}\\
\hline
{\tiny 8.0-10.0} & {\scriptsize0} & {\scriptsize0} & {\scriptsize0} & {\scriptsize0} & {\scriptsize0} & {\scriptsize0.016} & {\scriptsize0.105} & {\scriptsize0.076} & {\scriptsize0.101} & {\scriptsize0.084} & {\scriptsize0.117}\\
\hline
{\tiny 10.0-12.5} & {\scriptsize0} & {\scriptsize0} & {\scriptsize0} & {\scriptsize0} & {\scriptsize0} & {\scriptsize0} & {\scriptsize0.039} & {\scriptsize0.06} & {\scriptsize0.056} & {\scriptsize0.053} & {\scriptsize0.055}\\
\hline
{\tiny 12.5-15.0} & {\scriptsize0} & {\scriptsize0} & {\scriptsize0} & {\scriptsize0} & {\scriptsize0} & {\scriptsize0} & {\scriptsize0} & {\scriptsize0.025} & {\scriptsize0.011} & {\scriptsize0.053} & {\scriptsize0.031}\\
\hline
{\tiny 15.0-20.0} & {\scriptsize0} & {\scriptsize0} & {\scriptsize0} & {\scriptsize0} & {\scriptsize0} & {\scriptsize0} & {\scriptsize0} & {\scriptsize0} & {\scriptsize0.011} & {\scriptsize0.023} & {\scriptsize0.047}\\
\hline
{\tiny 20.0-25.0} & {\scriptsize0} & {\scriptsize0} & {\scriptsize0} & {\scriptsize0} & {\scriptsize0} & {\scriptsize0} & {\scriptsize0} & {\scriptsize0} & {\scriptsize0} & {\scriptsize0} & {\scriptsize0.008}\\
\hline
{\tiny overflow} & {\scriptsize0} & {\scriptsize0} & {\scriptsize0} & {\scriptsize0} & {\scriptsize0} & {\scriptsize0} & {\scriptsize0} & {\scriptsize0} & {\scriptsize0} & {\scriptsize0} & {\scriptsize0.008}\\
\hline
  \end{tabular}
  \end{center}
  \caption{\emph{NC background. All values $\times 10^{-4}$.}}
\end{table}

%
\newpage
\bibliographystyle{styles/utphys}
\bibliography{./IDS-NF_IDR}
%
\end{document}